\documentclass[twoside]{tex/tufte-book} 
\pdfoutput=1
\usepackage{natbib}

\usepackage{microtype} 

\usepackage{lipsum} 

\usepackage{booktabs} 

\usepackage{graphicx} 
\graphicspath{{graphics/}} 
\setkeys{Gin}{width=\linewidth,totalheight=\textheight,keepaspectratio} 

\usepackage{fancyvrb} 
\fvset{fontsize=\normalsize} 

\usepackage{xspace} 

\newcommand{\monthyear}{\ifcase\month\or January\or February\or March\or April\or May\or June\or July\or August\or September\or October\or November\or December\fi\space\number\year} 

\usepackage{makeidx} 
\makeindex 

\newsavebox{\coverimage}
\savebox{\coverimage}{\includegraphics[width=\linewidth]{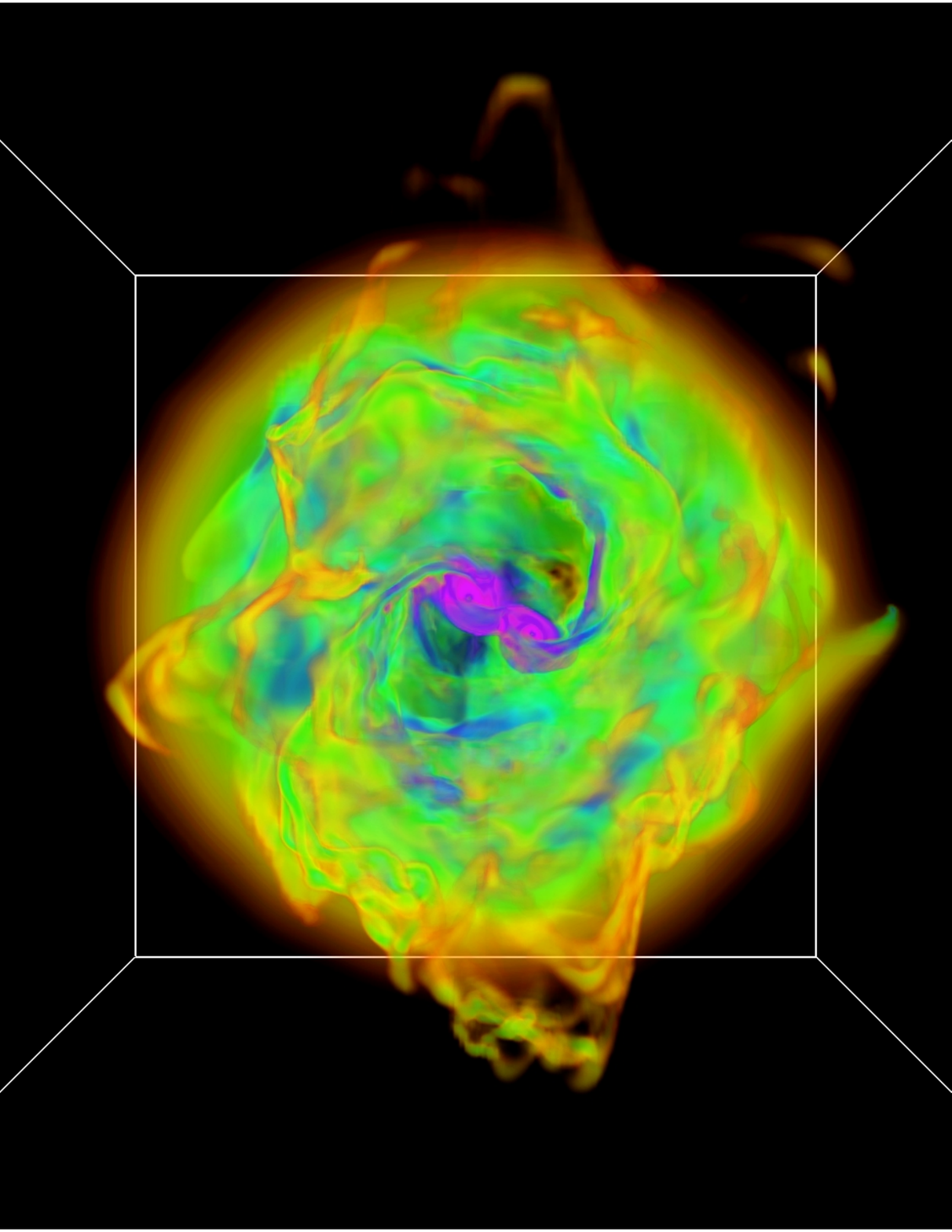}}

\usepackage{xcolor}
\definecolor{WHITE}{HTML}{FFFFFF}

\title[Notes on Star Formation]{%
\setlength{\parindent}{0pt}%
\textcolor{white}{Notes on}\par%
\textcolor{white}{Star Formation}%
\par \vspace{1cm}%
} 

\author{\textcolor{white}{Mark R.~Krumholz}} 

\publisher{\textcolor{white}{The Open Astrophysics Bookshelf}} 

\usepackage{bm}

\newcommand{\kb}{k_{\mathrm{B}}}
\newcommand{\msun}{M_{\odot}}
\newcommand{\rsun}{R_{\odot}}
\newcommand{\lsun}{L_{\odot}}
\newcommand{\avir}{\alpha_{\mathrm{vir}}}
\newcommand{\vecr}{\mathbf{r}}
\newcommand{\vecv}{\mathbf{v}}
\newcommand{\vecg}{\mathbf{g}}
\newcommand{\vecx}{\mathbf{x}}
\newcommand{\veck}{\mathbf{k}}
\newcommand{\vecB}{\mathbf{B}}
\newcommand{\vecf}{\mathbf{f}}
\newcommand{\nhat}{\hat{\mathbf{n}}}
\newcommand{\vecS}{\mathbf{S}}
\newcommand{\vecI}{\mathbf{I}}
\newcommand{\vecPi}{\boldsymbol{\Pi}}
\newcommand{\vecT}{\mathbf{T}}
\newcommand{\veco}{{\bm \Omega}}
\newcommand{\ehat}{\hat{\mathbf{e}}}
\newcommand{\alphab}{\alpha_{\mathrm{B}}}

\usepackage{amssymb}

\titleformat{\chapter}%
  [display]
  {\relax\ifthenelse{\NOT\boolean{@tufte@symmetric}}{\begin{fullwidth}}{}}
  {\itshape\huge\thechapter}
  {0pt}
  {\huge\rmfamily\itshape}
  [\ifthenelse{\NOT\boolean{@tufte@symmetric}}{\end{fullwidth}}{}]

\newcounter{problemset}
\newcommand{\problemset}{%
   \refstepcounter{problemset}
   \chapter*{Problem Set~\theproblemset}
   \addcontentsline{toc}{chapter}{Problem Set~\theproblemset}
}

\newcounter{solutionset}
\newcommand{\solutionset}{%
   \refstepcounter{solutionset}
   \chapter*{Solutions to Problem Set~\thesolutionset}
   \addcontentsline{toc}{chapter}{Solutions to Problem Set~\thesolutionset}
}

\usepackage{appendix}

\usepackage{transparent}
\usepackage{eso-pic}
\newcommand\BackgroundPic{%
\put(0,0){%
\parbox[b][\paperheight]{\paperwidth}{%
\vfill
\centering
\includegraphics[width=\paperwidth,height=\paperheight,%
keepaspectratio]{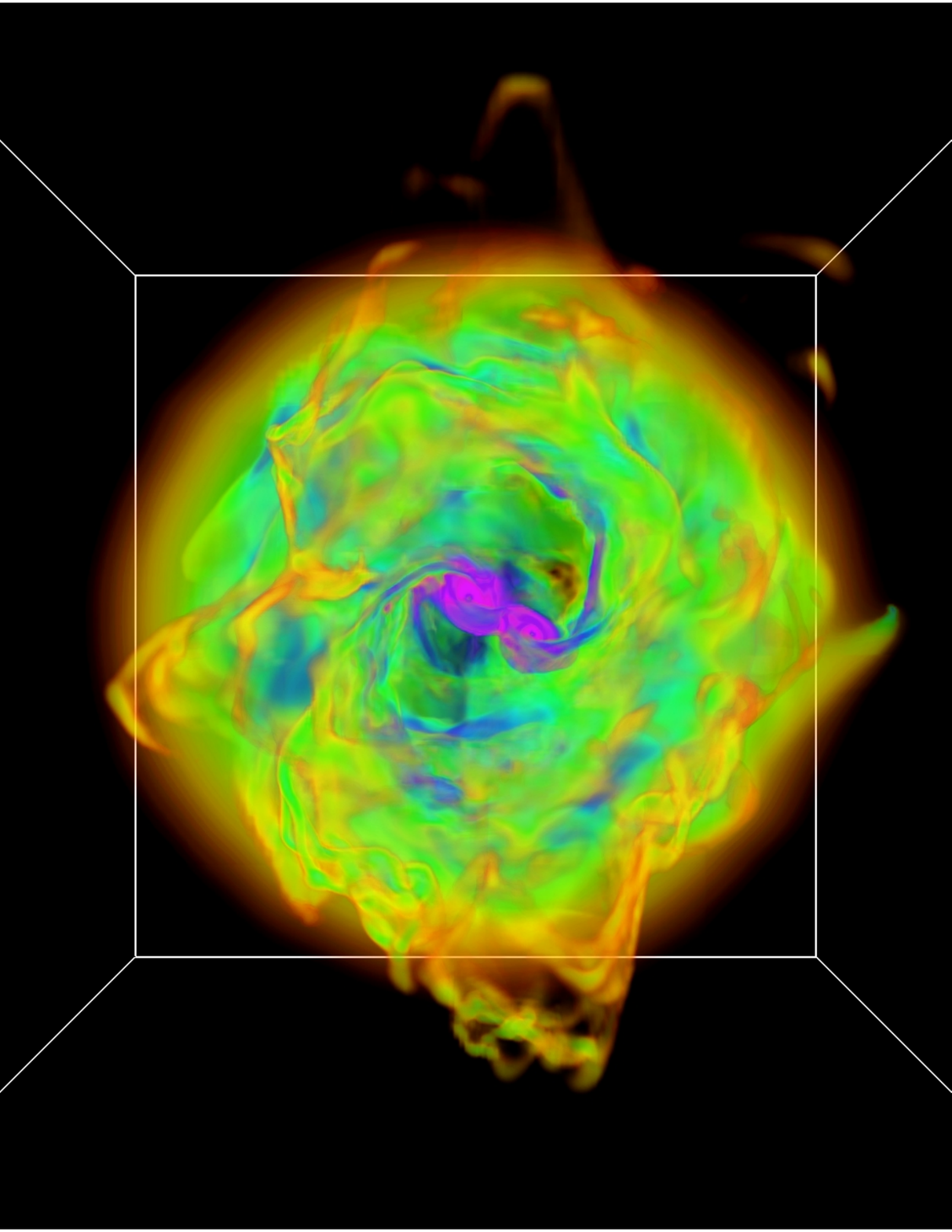}%
\vfill
}}}

\begin{document}

\AddToShipoutPicture*{\BackgroundPic}

\frontmatter


\thispagestyle{empty}


\maketitle 


\newpage
\begin{fullwidth}
~\vfill
\thispagestyle{empty}
\setlength{\parindent}{0pt}
\setlength{\parskip}{\baselineskip}
Original version: \the\year\ \thanklessauthor

\par\smallcaps{Published as part of the Open Astrophysics Bookshelf}

\par\smallcaps{\url{http://open-astrophysics-bookshelf.github.io/}}

\par Licensed under the Creative Commons 1.0 Universal License, \url{http://creativecommons.org/publicdomain/zero/1.0/}.\index{license}

\end{fullwidth}


\setcounter{tocdepth}{1}
\tableofcontents 


\listoffigures 




\cleardoublepage
~\vfill
\begin{doublespace}
\noindent\fontsize{18}{22}\selectfont\itshape
\nohyphenation
Dedicated to my family, Barbara, Alan, Ethan, and Rebecca, and with thanks to my students, who have contributed tremendously to the development of this book.
\end{doublespace}
\vfill
\vfill


\cleardoublepage
\chapter*{Introduction} 

This book is based on a series of lectures given by the author in his graduate class on star formation, taught from 2009 - 2016 at the University of California, Santa Cruz and Australian National University. It is intended for graduate students or advanced undergraduates in astronomy or physics, but does not presume detailed knowledge of particular areas of astrophysics (e.g., the interstellar medium or galactic structure). It is intended to provide a general overview of the field of star formation, at a level that would enable a student to begin independent research in the area.

This course covers the basics of star formation, ending at the transition to planet formation. The first two chapters, comprising part I, begin with a discussion of observational techniques, and the basic phenomenology they reveal. The goal is to familiarize students with the basic techniques that will be used throughout, and to provide a common vocabulary for the rest of the course. The next five chapters form part II, and provide a similar review of the basic physical processes that are important for star formation. Part III includes all the remaining chapters. These discuss star formation over a variety of scales, starting with the galactic scale and working our way down to the scales of individual stars and their disks, with slight deviations to discuss the particular problems of the formation of massive stars and of the first stars. The book concludes with the clearing of disks and the transition to planet formation.

The "texts" intended to go with these notes are the review articles "\href{http://adsabs.harvard.edu/abs/2014PhR...539...49K}{The Big Problems in Star Formation: the Star Formation Rate, Stellar Clustering, and the Initial Mass Function}", Krumholz, M. R., 2014, \textit{Physics Reports}, 539, 49, which provides a snapshot of the theoretical literature, and "\href{http://adsabs.harvard.edu/abs/2012ARA\%26A..50..531K}{Star Formation in the Milky Way and Nearby Galaxies}", Kennicutt, R.~C., \& Evans, N.~J., 2012, \textit{Annual Reviews of Astronomy \& Astrophysics}, 50, 531, which is more focused on observations. Another extremely useful reference is the series of review chapters from the \href{http://www.mpia.de/homes/ppvi/}{Protostars and Planets VI Conference}, which took place in July 2013. Suggested background readings to accompany most chapters are listed at the chapter beginning. In addition to these background materials, most chapters also include "suggested literature": papers from the recent literature whose content is relevant to the material covered in that chapter. These readings are included to help students engage with the active research literature, as well as the more general reviews.

In addition to the text and reading, this book contains five problem sets, which are interspersed with the chapters at appropriate locations. Solutions to the problems are included as an Appendix.


\setcounter{secnumdepth}{3}

\mainmatter


\part{Introduction and Phenomenology}

\chapter{Observing the Cold Interstellar Medium}
\label{ch:obscold}

\marginnote{\textbf{Suggested background reading:}
\begin{itemize}
\item \href{http://adsabs.harvard.edu/abs/2012ARA\%26A..50..531K}{Kennicutt, R.~C., \& Evans, N.~J. 2012, ARA\&A, 50, 531}, sections $1-2$ \nocite{kennicutt12a}
\end{itemize}
}

This first chapter focuses on observations of interstellar gas. Because the interstellar clouds that form stars are generally cold, most (but not all) of these techniques require in infrared, sub-millimeter, and radio observations. Interpretation of the results is often highly non-trivial. This will naturally lead us to review some of the important radiative transfer physics that we need to keep in mind to understand the observations. With this background complete, we will then discuss the phenomenology of interstellar gas derived from these observations.

\section{Observing Techniques}

\subsection{The Problem of H$_2$}

\begin{marginfigure}
\includegraphics[width=\linewidth]{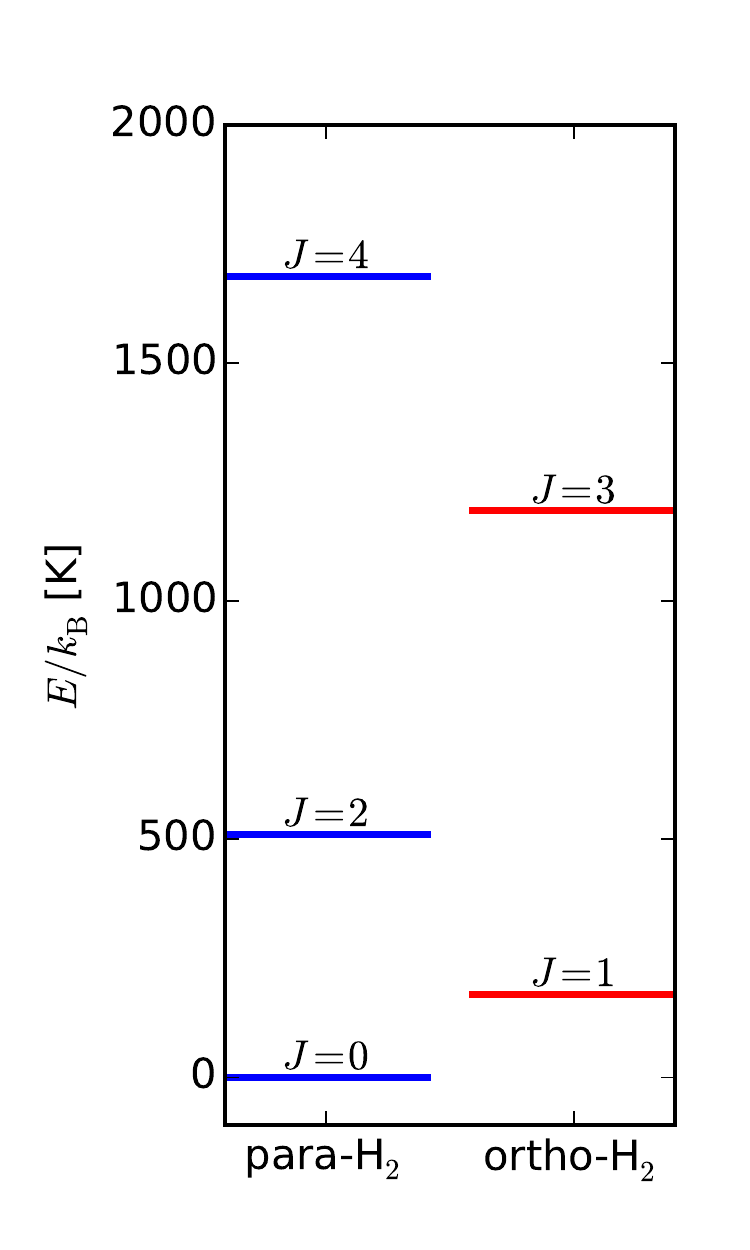}
\caption[H$_2$ level diagram]{
\label{fig:h2levels}
Level diagram for the rotational levels of para- and ortho-H$_2$, showing the energy of each level. Level data are taken from \url{http://www.gemini.edu/sciops/instruments/nir/wavecal/h2lines.dat}.
}
\end{marginfigure}

Before we dive into all the tricks we use to observe the dense interstellar medium (ISM), we have to start at the question of why it is necessary to be so clever. Hydrogen is the most abundant element, and when it is in the form of free atomic hydrogen, it is relatively easy to observe. Hydrogen atoms emit radio waves at a wavelength of 21 cm (1.4 GHz), associated with a hyperfine transition from a state in which the spin of the electron is parallel to that of the proton to a state where it is anti-parallel. The energy difference between these two states corresponds to a temperature $\ll 1$ K, so even in cold regions it can be excited. This line is seen in the Milky Way and in many nearby galaxies.
  
However, at the high densities where stars form, hydrogen tends to be molecular rather than atomic, and H$_2$ is extremely hard to observe directly. To understand why, we can look at an energy level diagram for rotational levels of H$_2$ (Figure \ref{fig:h2levels}). A diatomic molecule like H$_2$ has three types of excitation: electronic (corresponding to excitations of one or more of the electrons), vibrational (corresponding to vibrational motion of the two nuclei), and rotational (corresponding to rotation of the two nuclei about the center of mass). Generally electronic excitations are highest in energy scale, vibrational are next, and rotational are the lowest in energy. Thus the levels shown in Figure \ref{fig:h2levels} are the ones that lie closest to ground.

For H$_2$, the first thing to notice is that the first excited state, the $J=1$ rotational state, is $175$ K above the ground state. Since the dense ISM where molecules form is often also cold, $T\sim 10$ K (as we will see later), almost no molecules will be in this excited state. However, it gets even worse: H$_2$ is a homonuclear molecule, and for reasons of symmetry $\Delta J = 1$ radiative transitions are forbidden in homonuclear molecules. Indeed, there is no electronic process by which a hydrogen molecule with odd $J$ to turn into one with even $J$, and vice versa, because the allowed parity of $J$ is determined by the spins of the hydrogen nuclei. We refer to the even $J$ state as para-H$_2$, and the odd $J$ state as ortho-H$_2$.

The observational significance of this is that there is no $J=1\rightarrow 0$ emission. Instead, the lowest-lying transition is the $J=2\rightarrow 0$ quadrupole. This is very weak, because it's a quadrupole. More importantly, however, the $J=2$ state is 510 K above the ground state. This means that, for a population in equilibrium at a temperature of 10 K, the fraction of molecules in the $J=2$ state is $\sim e^{-510/10} \approx 10^{-22}$!\footnote{This oversimplifies things quite a bit, because in real molecular clouds there are usually shocked regions where the temperature is much greater than 10 K, and H$_2$ rotational emission is routinely observed from them. However, this emission tracers rare gas that is much hotter than the mean temperature in a cloud, not the bulk of the mass, which is cold.} In effect, in a molecular cloud there are simply no H$_2$ molecules in states capable of emitting. The reason such a high temperature is required to excite the H$_2$ molecule is its low mass: for a quantum oscillator or rotor, the level spacing varies with reduced mass as $m^{-1/2}$. Thus the levels of H$_2$ are much farther apart than the levels of other diatomic molecules (e.g., CO, O$_2$, N$_2$). It is the low mass of the hydrogen atom that creates our problems.
  
Given this result, we see that, for the most part, observations of the most abundant species can only be done by proxy. Only in very rare circumstances is it possible to observe H$_2$ directly -- usually when there is a bright background UV source that allows us to see it in UV absorption rather than in emission. Since these circumstances do not generally prevail, we are forced to consider alternatives.

\subsection{Dust Emission}

The most conceptually straightforward proxy technique we use to study star-forming clouds is thermal dust emission. Interstellar gas clouds are always mixed with dust, and the dust grains emit thermal radiation that we can observe. The gas, in contrast, does not emit thermal radiation because it is nowhere near dense enough to reach equilibrium with the radiation field. Instead, gas emission comes primarily in the form of lines, which we will discuss below.
  
\begin{marginfigure}
\includegraphics[width=\linewidth]{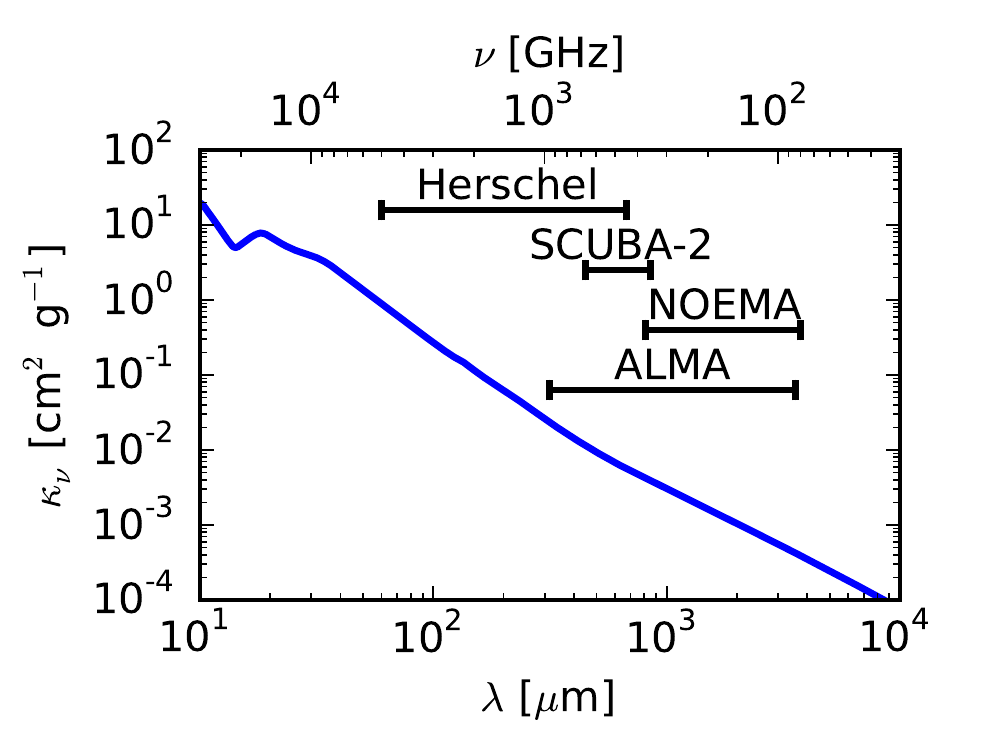}
\caption[Dust absorption opacity]{
\label{fig:draine03opacity}
Milky Way dust absorption opacities per unit gas mass as a function of wavelength $\lambda$ and frequency $\nu$ in the infrared and sub-mm range, together with wavelength coverage of selected observational facilities. Dust opacities are taken from the model of \citet{draine03a} for $R_V = 5.5$.
}
\end{marginfigure}

Consider a cloud of gas of mass density $\rho$ mixed with dust grains at a temperature $T$. The gas-dust mixture has an absorption opacity $\kappa_{\nu}$ to radiation at frequency $\nu$. Although the vast majority of the mass is in gas rather than dust, the opacity will be almost entirely due to the dust grains except for frequencies that happen to match the resonant absorption frequencies of atoms and molecules in the gas. Here we follow the standard astronomy convention that $\kappa_{\nu}$ is the opacity per gram of material, with units of cm$^2$ g$^{-1}$, i.e., we assign the gas an effective cross-sectional area that is blocked per gram of gas. For submillimeter observations, typical values of $\kappa_{\nu}$ are $\sim 0.01$ cm$^{2}$ g$^{-1}$. Figure \ref{fig:draine03opacity} shows a typical extinction curve for Milky Way dust.

Since essentially no interstellar cloud has a surface density $> 100$ g cm$^{-2}$, absorption of radiation from the back of the cloud by gas in front of it is completely negligible. Thus, we can compute the emitted intensity very easily. The emissivity for gas of opacity $\kappa_{\nu}$ is $j_{\nu} = \kappa_{\nu} \rho B_{\nu}(T)$, where $j_{\nu}$ has units of erg s$^{-1}$ cm$^{-3}$ sr$^{-1}$ Hz$^{-1}$, i.e.\ it describes (in cgs units) the number of ergs emitted in 1 second by 1 cm$^3$ of gas into a solid angle of 1 sr in a frequency range of 1 Hz. The quantity
\begin{equation}
B_{\nu}(T) = \frac{2 h\nu^3}{c^2} \frac{1}{e^{h\nu/\kb T}-1}
\end{equation}
is the Planck function.
  
Since none of this radiation is absorbed, we can compute the intensity transmitted along a given ray just by integrating the emission: 
  \begin{equation}
  I_{\nu} = \int j_{\nu} ds = \Sigma \kappa_{\nu} B_{\nu}(T) = \tau_{\nu} B_{\nu}(T)
  \end{equation}
where $\Sigma=\int \rho ds$ is the surface density of the cloud and $\tau_{\nu} = \Sigma \kappa_{\nu}$ is the optical depth of the cloud at frequency $\nu$. Thus if we observe the intensity of emission from dust grains in a cloud, we determine the product of the optical depth and the Planck function, which is determined solely by the observing frequency and the gas temperature. If we know the temperature and the properties of the dust grains, we can therefore determine the column density of the gas in the cloud in each telescope beam.

\begin{figure}
\includegraphics[width=\linewidth]{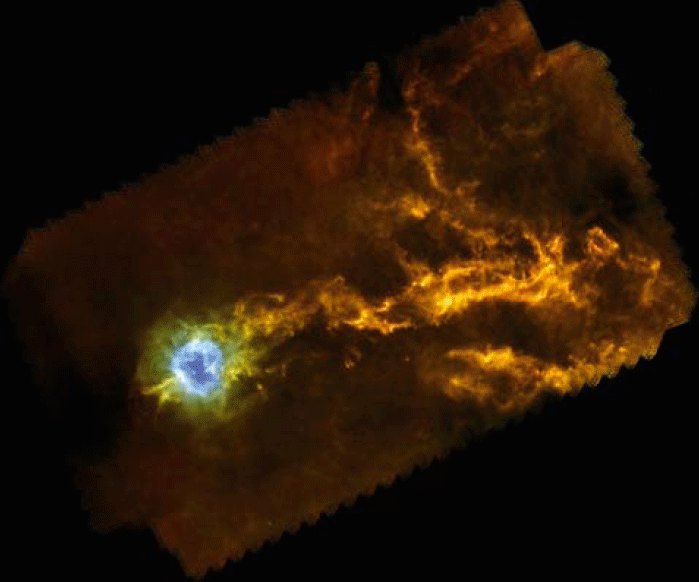}
\caption[\textit{Herschel} map of IC 5146]{
\label{fig:herschel_ic5146}
Three-color composite image of IC 5146 taken by the SPIRE and PACS instruments aboard \textit{Herschel}. Red is SPIRE 500 $\mu$m, green is SPIRE 250 $\mu$m plus PACS 160 $\mu$m, and blue is PACS 70 $\mu$m. Credit: \citeauthor{arzoumanian11a}, A\&A, 529, L6, 2011, reproduced with permission \copyright\,ESO.
}
\end{figure}

Figure \ref{fig:herschel_ic5146} show an example result using this technique. The advantage of this approach is that it is very straightforward. The major uncertainties are in the dust opacity, which we probably don't know better than a factor of few level, and in the gas temperature, which is also usually uncertain at the factor of $\sim 2$ level. The produces a corresponding uncertainty in the conversion between dust emission and gas column density. Both of these can be improved substantially by observations that cover a wide variety of wavelengths, since these allow one to simultaneously fit the column density, dust opacity curve, and dust temperature.

Before the \textit{Herschel} satellite (launched in 2009) such multi-wavelength observations were rare, because most of the dust emission was in at far-infrared wavelengths of several hundred $\mu$m that are inaccessible from the ground. \textit{Herschel} was specifically targeted at this wavelength range, and has greatly improved our knowledge of cloud properties from dust emission.

\subsection{Dust Absorption}

\begin{figure}
\includegraphics[width=\linewidth]{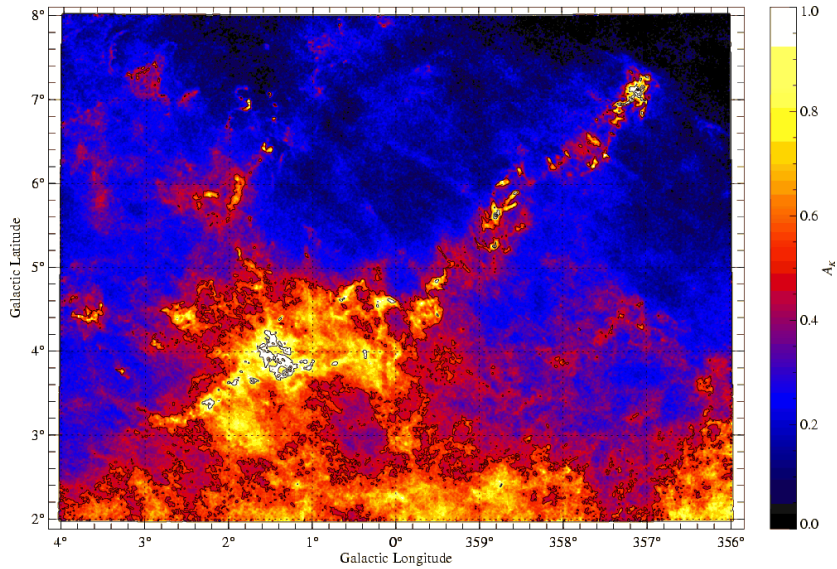}
\caption[Dust extinction map of the Pipe Nebula]{
\label{fig:pipe_lombardi06}
Extinction map of the Pipe Nebula. The color scale shows the extinction in K band. Credit: \citeauthor{lombardi06a}, A\&A, 454, 781, 2006, reproduced with permission \copyright\,ESO.
}
\end{figure}

A second related technique is, instead of looking at dust emission, looking at absorption of background starlight by dust, usually in the near infrared. In this case the calculation is even simpler. One measures the extinction of the background star and then simply divides by the gas opacity to get a column density. Probably the best example of this technique is the Pipe Nebula (Figure \ref{fig:pipe_lombardi06}).  

The advantages of this compared to dust thermal emission are threefold. First, since stars are bright compared to interstellar dust grains, and the observations are done in the near IR rather than the sub-mm, the available resolution is much, much higher. Second, since opacity doesn't depend on temperature, the uncertainty in converting what we see into a column density is reduced. Third, we know the dust opacity curve in the near infrared considerably better than we know it in the far-IR or sub-mm, further reducing the uncertainty. However, there are also drawbacks to this method. Due to the comparatively higher opacity in the infrared, it is only possible to use this technique for fairly diffuse regions; in denser regions the background stars are completely extincted. Moreover, one needs a good, clean field of background stars to get something like a map, and only a few clouds have such favorable geometry.

\subsection{Molecular Lines}
\label{ssec:molecular_lines}

Much of what we know about star forming gas comes from observations of line emission. These are usually the most complex measurements in terms of the modeling and theory required to understand them. However, they are also by far the richest in terms of the information they provide. They are also among the most sensitive, since the lines can be very bright compared to continuum emission. Indeed, the great majority of what we know about the ISM beyond the local group comes from studying emission in the rotational lines of the CO molecule, because these (plus the C~\textsc{ii} line found in atomic regions) are by far the easiest types of emission to detect from the cold ISM.

The simplest line-emitting system is an atom or molecule with exactly two energy states, but this example contains most of the concepts we will need. The generalization of these results to a multi-level system is given in Appendix \ref{app:multilevel_atoms}.

\paragraph{Einstein Coefficients and Collision Rates}

Consider an atom or molecule of species $X$ with two non-degenerate states that are separated by an energy $E$. Suppose we have a gas of such particles with number density $n_X$ at temperature $T$. The number density of atoms in the ground state is $n_0$ and the number density in the excited state is $n_1$. At first suppose that this system does not radiate. In this case collisions between the atoms will eventually bring the two energy levels into thermal equilibrium, and it is straightforward to compute $n_0$ and $n_1$. They just follow a Maxwellian distribution, so $n_1/n_0 = e^{-E/k_B T}$, and thus we have $n_0 = n_X /Z$ and $n_1 = n_X e^{-E/k_B T}/Z$, where $Z=1+e^{-E/k_B T}$ is the partition function.

Now let us consider radiative transitions between these states. There are three processes: spontaneous emission, stimulated emission, and absorption, which are described by the three Einstein coefficients. In studying star formation, we can often ignore stimulated emission and absorption, because the ambient radiation field is so weak that these processes occur at negligible rates. The exception to this is when lines become optically thick, so there are a lot of line photons bouncing around trapped inside a structure, or when the frequency of the transition in question is at very low energy, and interactions with CMB photons become significant. However, for simplicity we will begin by just focusing on spontaneous emission and ignoring absorption and stimulated emission. The full statistical mechanics problem including these processes is discussed in Appendix \ref{app:multilevel_atoms}.

An atom in the excited state can spontaneously emit a photon and decay to the ground state. The rate at which this happens is described by the Einstein coefficient $A_{10}$, which has units of s$^{-1}$. Its meaning is simply that a population of $n_1$ atoms in the excited state will decay to the ground state by spontaneous emission at a rate 
\begin{equation}
\left(\frac{dn_1}{dt}\right)_{\rm spon.~emis.} = -A_{10} n_1.
\end{equation}
In cgs units this quantity is measured in atoms per cm$^3$ per s, and this expression is equivalent to saying that the $e$-folding time for decay is $1/A_{10}$ seconds. For most of the molecules we will be considering in this book, decay times are typically at most a few centuries, which is short compared to pretty much any time scale associated with star formation. Thus if spontaneous emission were the only process at work, all molecules would quickly decay to the ground state and we wouldn't see any emission.

However, in the dense interstellar environments where stars form, collisions occur frequently enough to create a population of excited molecules. Of course collisions involving excited molecules can also cause de-excitation, with the excess energy going into recoil rather than into a photon. Since hydrogen molecules are almost always the most abundant species in the dense regions we're going to think about, with helium second, we can generally only consider collisions between our two-level atom and those partners. For the purposes of this exercise, we'll take an even simpler approach and ignore everything but H$_2$. Putting He back into the picture is easy, as it just requires adding extra collision terms that are completely analogous to the ones we will write down.

The rate at which collisions cause transitions between states is a horrible quantum mechanical problem. We cannot even confidently calculate the energy levels of single isolated molecules except in the simplest cases, let alone the interactions between two colliding ones at arbitrary velocities and relative orientations. Exact calculations of collision rates are generally impossible. Instead, we either make due with approximations (at worst), or we try to make laboratory measurements. Things are bad enough that, for example, we often assume that the rates for collisions with H$_2$ molecules and He atoms are related by a constant factor.

Fortunately, as astronomers we generally leave these problems to chemists, and instead do what we always do: hide our ignorance behind a parameter. We let the rate at which collisions between species $X$ and H$_2$ molecules induce transitions from the ground state to the excited state be
\begin{equation}
\left(\frac{dn_1}{dt}\right)_{\rm coll.~exc.} = k_{01} n_0 n,
\end{equation}
where $n$ is the number density of H$_2$ molecules and $k_{01}$ has units of cm$^3$ s$^{-1}$. In general $k_{01}$ will be a function of the gas kinetic temperature $T$, but not of $n$ (unless $n$ is so high that three-body processes start to become important, which is almost never the case in the ISM). 

The corresponding rate coefficient for collisional de-excitation is $k_{10}$, and the collisional de-excitation rate is
\begin{equation}
\left(\frac{dn_1}{dt}\right)_{\rm coll.~de-exc.} = -k_{10} n_1 n.
\end{equation}
A little thought should suffice to convince the reader that $k_{01}$ and $k_{10}$ must have a specific relationship. Consider an extremely dense region where $n$ is so large that collisional excitation and de-excitation both occur much, much more often than spontaneous emission, and we can therefore neglect the spontaneous emission term in comparison to the collisional ones. If the gas is in equilibrium then we have
\begin{eqnarray}
\frac{dn_1}{dt} = \left(\frac{dn_1}{dt}\right)_{\rm coll.~exc.} + \left(\frac{dn_1}{dt}\right)_{\rm coll.~de-exc.} & = & 0 \\
n (k_{01} n_0 - k_{10} n_1) & = & 0.
\end{eqnarray}
However, we also know that the equilibrium distribution is a Maxwellian, so $n_1/n_0 = e^{-E/k_B T}$. Thus we have
\begin{eqnarray}
n n_0 (k_{01} - k_{10} e^{-E/k_B T}) & = & 0 \\
k_{01} & = & k_{10} e^{-E/k_B T}.
\label{eq:detailed_balance}
\end{eqnarray}
This argument applies equally well between a pair of levels even for a complicated molecule with many levels instead of just 2. Thus, we only need to know the rate of collisional excitation or de-excitation between any two levels to know the reverse rate.

\paragraph{Critical Density and Density Inference}

We are now in a position to write down the full equations of statistical equilibrium for the two-level system. In so doing, we will see that we can immediately use line emission to learn a great deal about the density of gas. In equilibrium we have
\begin{eqnarray}
\frac{dn_1}{dt} & = & 0 \\
n_1 A_{10} + n n_1 k_{10} -n n_0 k_{01} & = & 0 \\
\frac{n_1}{n_0} \left(A_{10} + k_{10}n\right) - k_{01} n & = & 0\\
\frac{n_1}{n_0} & = & \frac{k_{01} n}{A_{10}+k_{10} n}\\
& = & e^{-E/k_B T} \frac{1}{1+A_{10}/(k_{10} n)}
\end{eqnarray}
This physical meaning of this expression is clear. If radiation is negligible compared to collisions, i.e., $A_{10} \ll k_{10} n$, then the ratio of level populations approaches the Maxwellian ratio $e^{-E/k_B T}$. As radiation becomes more important, i.e., $A_{10}/(k_{10} n)$ get larger, the fraction in the upper level drops -- the level population is sub-thermal. This is because radiative decays remove molecules from the upper state faster than collisions re-populate it.

Since the collision rate depends on density and the radiative decay rate does not, the balance between these two processes depends on density. This make it convenient to introduce a critical density $n_{\rm crit}$, defined by
\begin{equation}
\label{eq:ncrit}
n_{\rm crit} = \frac{A_{10}}{k_{10}},
\end{equation}
so that
\begin{equation}
\frac{n_1}{n_0} = e^{-E/k_B T} \frac{1}{1+n_{\rm crit}/n}.
\end{equation}
At densities much larger than $n_{\rm crit}$, we expect the level population to be close to the Maxwellian value, and at densities much smaller than $n_{\rm crit}$ we expect the upper state to be under-populated relative to Maxwellian; $n_{\rm crit}$ itself is simply the density at which radiative and collisional de-excitations out of the upper state occur at the same rate.

This process of thermalization has important consequences for the line emission we see from molecules. The energy emission rate per molecule from the line is 
\begin{eqnarray}
\frac{\mathcal{L}}{n_X} & = & \frac{E A_{10} n_1}{n_X} \\
& = & E A_{10} \frac{n_1}{n_0+n_1} \\
& = & E A_{10} \frac{n_1/n_0}{1+n_1/n_0} \\
& = & E A_{10} \frac{e^{-E/k_B T}}{1+e^{-E/k_B T}+n_{\rm crit}/n} \\
& = & E A_{10} \frac{e^{-E/k_B T}}{Z+n_{\rm crit}/n}
\end{eqnarray}
where again $Z$ is the partition function.

It is instructive to think about how this behaves in the limiting cases $n \ll n_{\rm crit}$ and $n\gg n_{\rm crit}$. In the limit $n\gg n_{\rm crit}$, the partition function $Z$ dominates the denominator, and we get $\mathcal{L}/n_X = E A_{10} e^{-E/k_B T}/Z$. This is just the energy per spontaneous emission, $E$, times the spontaneous emission rate, $A_{10}$, times the fraction of the population in the upper state when the gas is in statistical equilibrium, $e^{-E/k_B T}/Z$. This is density-independent, so this means that at high density the gas produces a fixed amount of emission per molecule of the emitting species. The total luminosity is just proportional to the number of emitting molecules.

For $n \ll n_{\rm crit}$, the second term dominates the denominator, and we get
\begin{equation}
\label{eq:cool_lowden}
\frac{\mathcal{L}}{n_X} \approx E A_{10} e^{-E/k_B T} \frac{n}{n_{\rm crit}}.
\end{equation}
Thus at low density each molecule contributes an amount of light that is proportional to the ratio of density to critical density. Note that this is the ratio of collision partners, i.e., of H$_2$, rather than the density of emitting molecules. The total luminosity varies as this ratio times the number of emitting molecules.

The practical effect of this is that different molecules tell us about different densities of gas in galaxies. Molecules with low critical densities reach the linear regime at low density, and since most of the mass tends to be at lower density, they probe this widespread, low-density component. Molecules with higher critical densities will have more of their emission contributed by higher density gas, and thus tell us about rarer, higher-density regions. This is all somewhat qualitative, since a transition between $\mathcal{L}/n_X \propto n$ and $\mathcal{L}/n_X \sim\mbox{constant}$ doesn't represent a particularly sharp change in behavior. Nonetheless, the luminosity ratios of lines with different critical densities are a very important diagnostic of the overall density distribution in the ISM.

As a caution, we should note that this is computed for optically thin emission. If the line is optically thick, we can no longer ignore stimulated emission and absorption processes, and not all emitted photons will escape from the cloud. In this case the effective critical density is reduced by a factor of the optical depth. CO, the most-commonly used tracer molecule, is usually optically thick.

\paragraph{Velocity and Temperature Inference}

We can also use molecular lines to infer the velocity and temperature structure of gas if the line in question is optically thin. For an optically thin line, the width of the line is determined primarily by the velocity distribution of the emitting molecules. The physics here is extremely simple. Suppose we have gas along our line of sight with a velocity distribution $\psi(v)$, i.e., the fraction of gas with velocities between $v$ and $v+dv$ is $\psi(v) dv$, and $\int_{-\infty}^{\infty} \psi(v) \, dv = 1$.

For an optically thin line, in the limit where natural and pressure-broadening of lines is negligible, which is almost always the case when observing the cold, dense, ISM, we can think of emission producing a delta function in frequency in the rest frame of the gas. There is a one-to-one mapping between velocity and frequency. Thus emission from gas moving at a frequency $v$ relative to us along our line of sight produces emission at a frequency $\nu \approx \nu_0 (1 - v/c)$, where $\nu_0$ is the central frequency of the line in the molecule's rest frame, and we assume $v/c \ll 1$. In this case the line profile is described trivially by $\phi(\nu)=\psi(c(1-\nu/\nu_0))$. 

We can measure $\phi(\nu)$ directly, and this immediately tells us the velocity distribution $\psi(v)$. In general the velocity distribution of the gas $\psi(v)$ is produced by a combination of thermal and non-thermal motions. Thermal motions arise from the Maxwellian velocity distribution of the gas, and produce a Maxwellian profile $\phi(\nu)\propto e^{-(\nu-\nu_{\rm cen})^2/2\sigma_\nu^2}$. Here $\nu_{\rm cen}$ is the central frequency of the line, which is $\nu_{\rm cen} = \nu_0 (1 - \bar{v}/c)$, where $\bar{v}$ is the mean velocity of the gas along our line of sight. The width is $\sigma_\nu = c^{-1}\sqrt{k_B T/\mu m_{\rm H}}$, where $T$ is the gas temperature and $\mu$ is the mean mass of the emitting molecule in units of hydrogen masses. This is just the 1D Maxwellian distribution.

\begin{marginfigure}
\includegraphics[width=\linewidth]{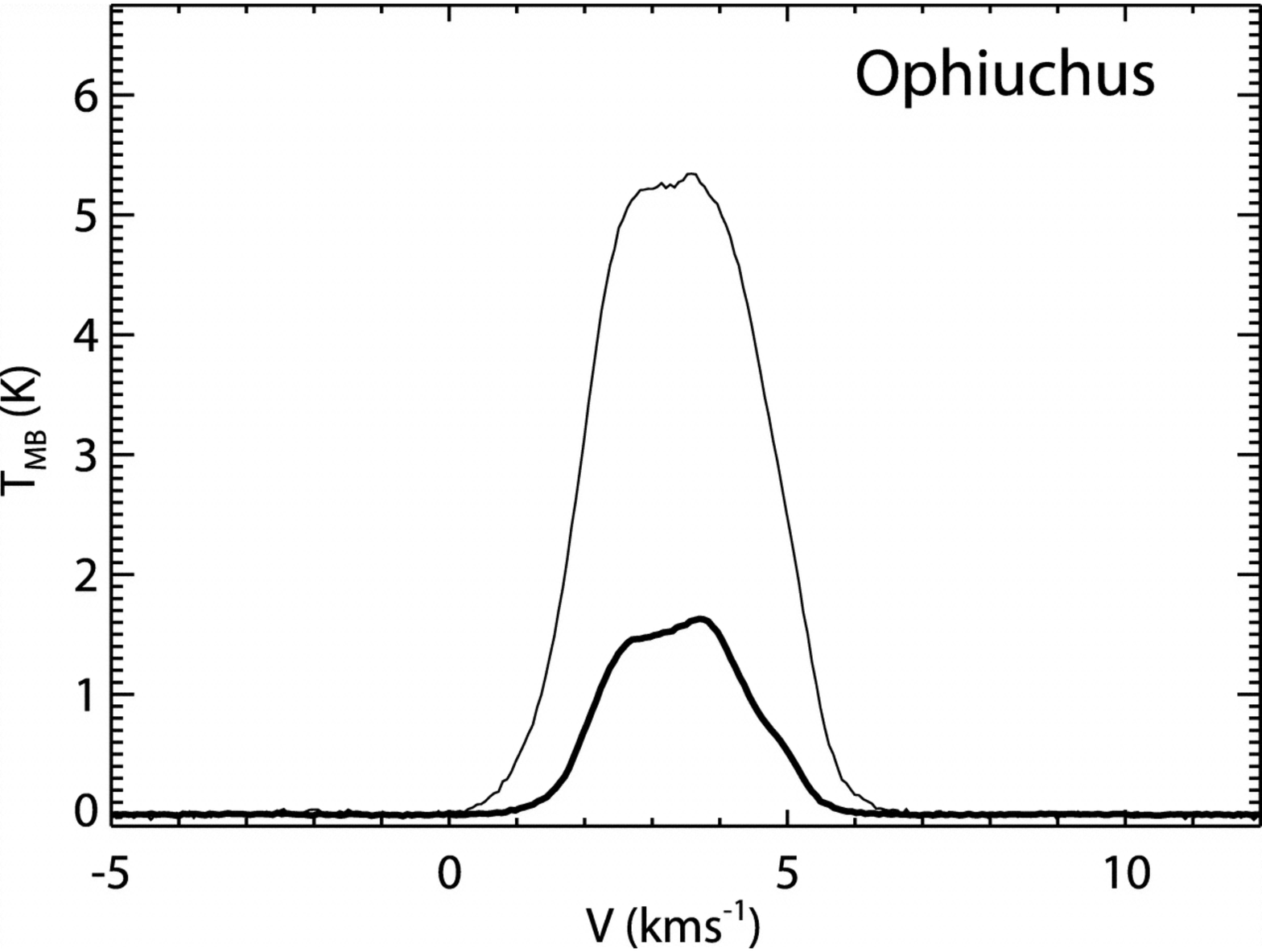}
\includegraphics[width=\linewidth]{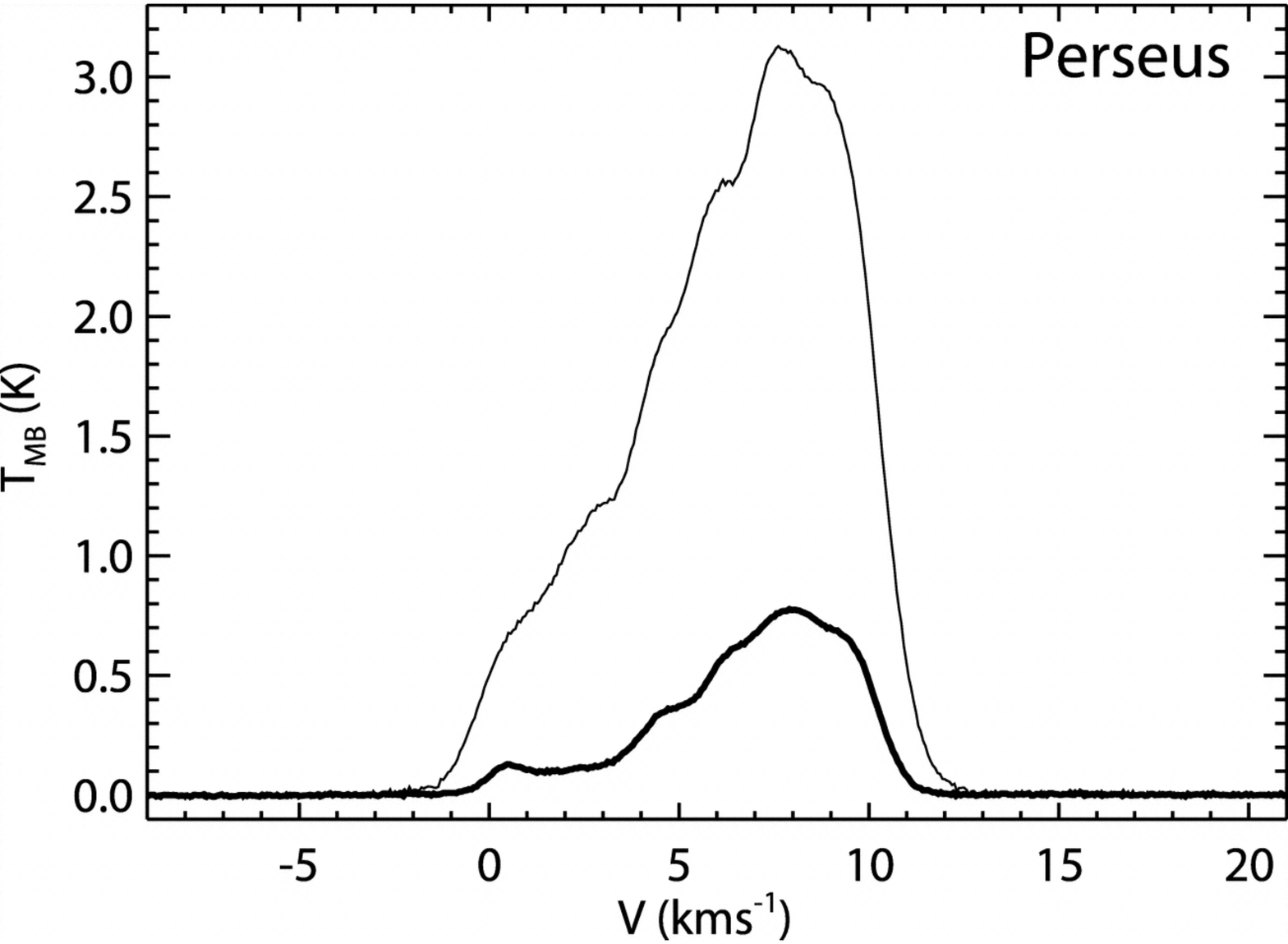}
\caption[COMPLETE spectra of Ophiuchus and Perseus]{
\label{fig:complete_ridge06}
Position-integrated velocity distributions of $^{12}$CO (\textit{thin lines}) and $^{13}$CO (\textit{thick lines}) for the Ophiuchus and Perseus clouds, measured the COMPLETE survey. The $y$ axis shows the beam temperature. Credit: \citet{ridge06a}, \copyright\, AAS. Reproduced with permission.
}
\end{marginfigure}

Non-thermal motions involve bulk flows of the gas, and can produce a variety of velocity distributions depending how the cloud is moving. Unfortunately even complicated motions often produce distributions that look something like Maxwellian distributions, just because of the central limit theorem: if you throw together a lot of random junk, the result is usually a Gaussian / Maxwellian distribution. Figure \ref{fig:complete_ridge06} shows an example of velocity distributions measured in two nearby star-forming clouds.

Determining whether a given line profile reflects predominantly thermal or non-thermal motion requires that we have a way of estimating the temperature independently. This can often be done by observing multiple lines of the same species. Our expression
\begin{equation}
\frac{\mathcal{L}}{n_X} = E A_{10} \frac{e^{-E/k_B T}}{Z + n_{\rm crit}/n}
\end{equation}
shows that the luminosity of a particular optically thin line is a function of the temperature $T$, the density $n$, and the number density of emitting molecules $n_X$. If we observe three transitions of the same molecule, then we have three equations in three unknowns and we can solve for $n$, $n_X$, and $T$ independently. Certain molecules, because of their level structures, make this technique particularly clean. The most famous example of this is ammonia, NH$_3$.

\paragraph{Complications}

Before moving on it is worth mentioning some complications that make it harder to interpret molecular line data. The first is optical depth: for many of the strongest lines and most abundant species, the line becomes optically thick. As a result observations in the line show only the surface a given cloud; emission from the back side of the cloud is absorbed by the front side. One can still obtain useful information from optically thick lines, but it requires a bit more thought. We will return to the topic of what we can learn from optically thick lines in Chapter \ref{ch:gmcs}.

The second complication is chemistry and abundances. The formation and destruction of molecules in the ISM is a complicated problem, and in general the abundance of any given species depends on the density, temperature, and radiation environment of the the gas. At the edges of clouds, certain molecules may not be present because they are dissociated by the interstellar UV field. At high densities and low temperatures, many species freeze out onto the surfaces of dust grains. This is true for example of CO. One often sees that peaks in density found in dust emission maps correspond to local minima of CO emission. This is because in the densest parts of clouds CO goes out of the gas phase and forms CO ice on the surfaces of dust grains.  Thus one must always be careful to investigate whether changes in molecular line emission are due to changes in gas bulk properties (e.g., density, temperature) or due to changes in the abundance of the emitting species.

\section{Observational Phenomenology}

\subsection{Giant Molecular Clouds}

As discussed above, we usually cannot observe H$_2$ directly, so we are forced to do so by proxy. The most common proxy is the rotational lines of CO. These are useful because CO is the single most abundance molecule in the ISM after H$_2$, it tends to be found in the same places as H$_2$ (for reasons that will become clear in Chapter \ref{ch:microphysics}, and the CO molecule has a number of transitions that can be excited at the low temperatures found in molecular clouds -- for example the CO $J=1$ state is only 5.5 K above the ground state. Indeed, the CO molecule is the primary coolant of molecular gas, so its excitation in effect sets the molecular gas temperature. 

In Chapter \ref{ch:gmcs} we will discuss how one infers the mass of an observed gas cloud from CO emission, and for the moment we will take it for granted that one can do so. By mass the Milky Way's ISM inside the solar circle is roughly 70\% H~\textsc{i} and 30\% H$_2$. The molecular fraction rises sharply toward the galactic center, reaching near unity in the molecular ring at $\sim 3$ kpc, then falling to $\sim 10\%$ our where we are. In other nearby galaxies the proportions vary from nearly all H~\textsc{i} to nearly all H$_2$.

\begin{figure}
\includegraphics[width=\linewidth]{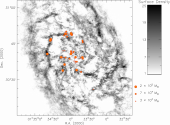}
\caption[Distribution of H~\textsc{i} and GMCs in M33]{
\label{fig:m33_imara}
Map of H~\textsc{i} in M33 (\textit{grayscale}), with giant molecular clouds detected in CO($1\rightarrow 0$) overlayed (\textit{circles}, sized by GMC mass). Credit: \citet{imara11b}, \copyright\, AAS. Reproduced with permission.
}
\end{figure}

\begin{figure}
\includegraphics[width=\linewidth]{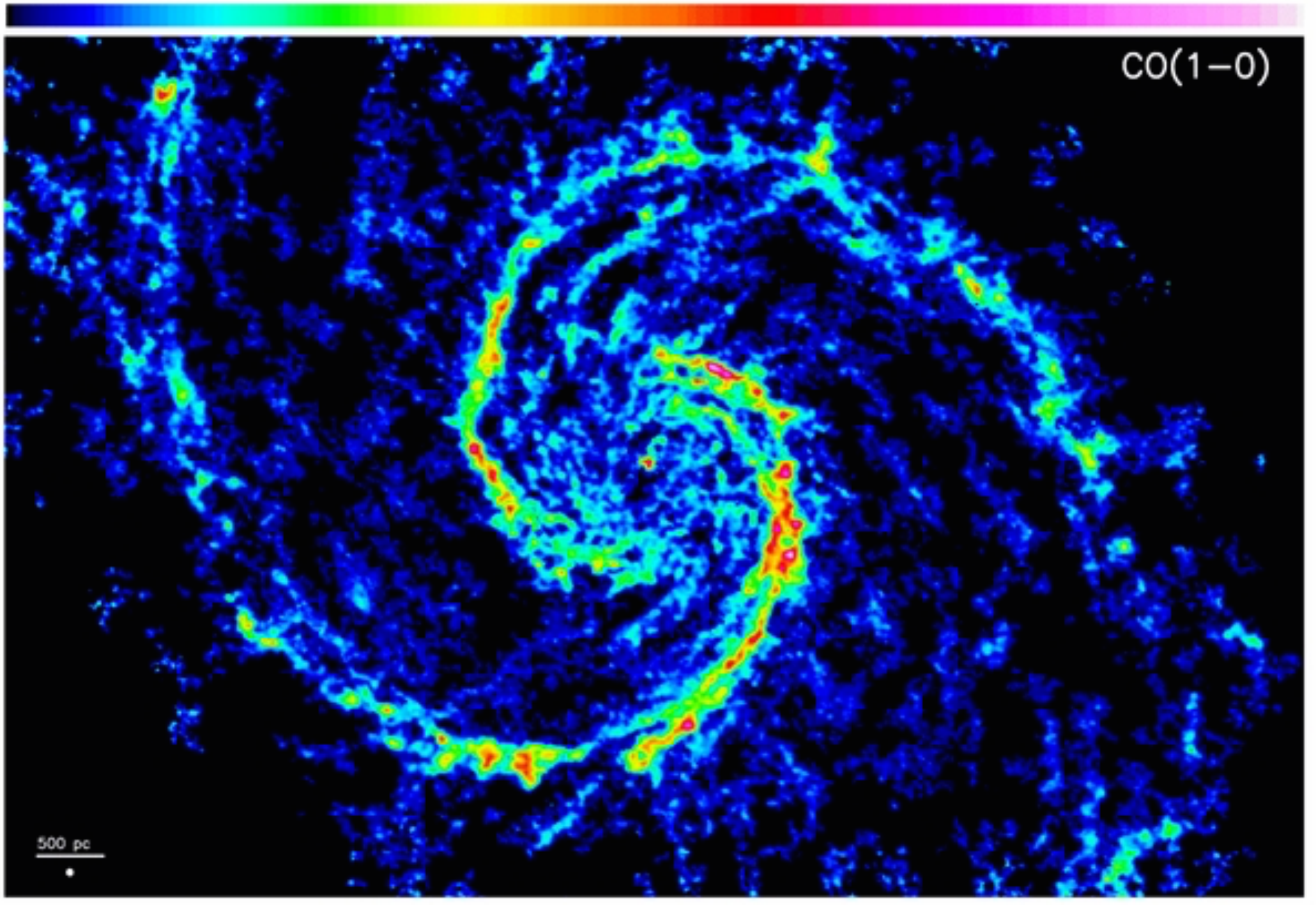}
\caption[Distribution of CO($1\rightarrow 0$) emission in M51]{
\label{fig:m51_schinnerer}
Map of CO($1\rightarrow 0$) emission in M51, as measured by the PdBI Arcsecond Whirlpool Survey (PAWS) project. Credit: \citet{schinnerer13a}, \copyright\, AAS. Reproduced with permission.
}
\end{figure}

In galaxies that are predominantly H~\textsc{i}, like ours, the atomic gas tends to show a filamentary structure, with small clouds of molecular gas sitting on top of peaks in the H~\textsc{i} distribution. In galaxies with large-scale spiral structure, the molecular gas closely tracks the optical spiral arms. Figures \ref{fig:m33_imara} and \ref{fig:m51_schinnerer} show examples of the former and the latter, respectively. The physical reasons for the associations between molecular gas and H~\textsc{i}, and between molecular clouds and spiral arms, are an interesting point that we will discuss in Chapter \ref{ch:microphysics}. 

As the images show, molecular gas in galaxies that are predominantly atomic tends to be organized into discreet clouds, called giant molecular clouds (GMCs). These can have a range of masses; in the Milky Way the most massive are a few million $\msun$, but there is a spectrum that seems to continue down to at least $10^4$ $\msun$. This organization into GMCs is clearest where the gas is predominantly atomic. In regions where molecules make up most of the mass, the clouds begin to run together and it is no longer possible to identify discrete clouds in a meaningful way.

\subsection{Internal structure of GMCs}

\begin{figure}
\includegraphics[width=\linewidth]{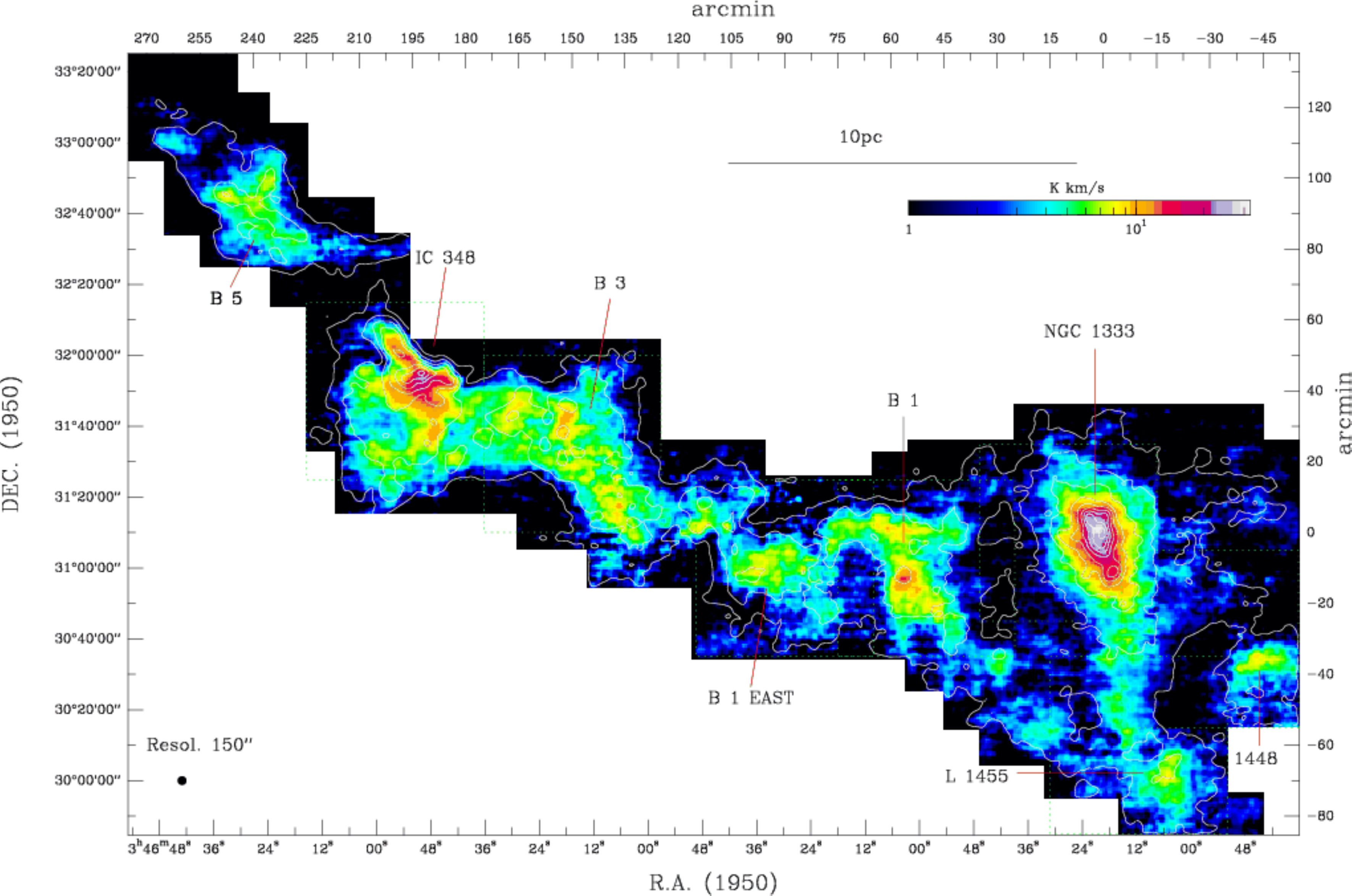}
\includegraphics[width=\linewidth]{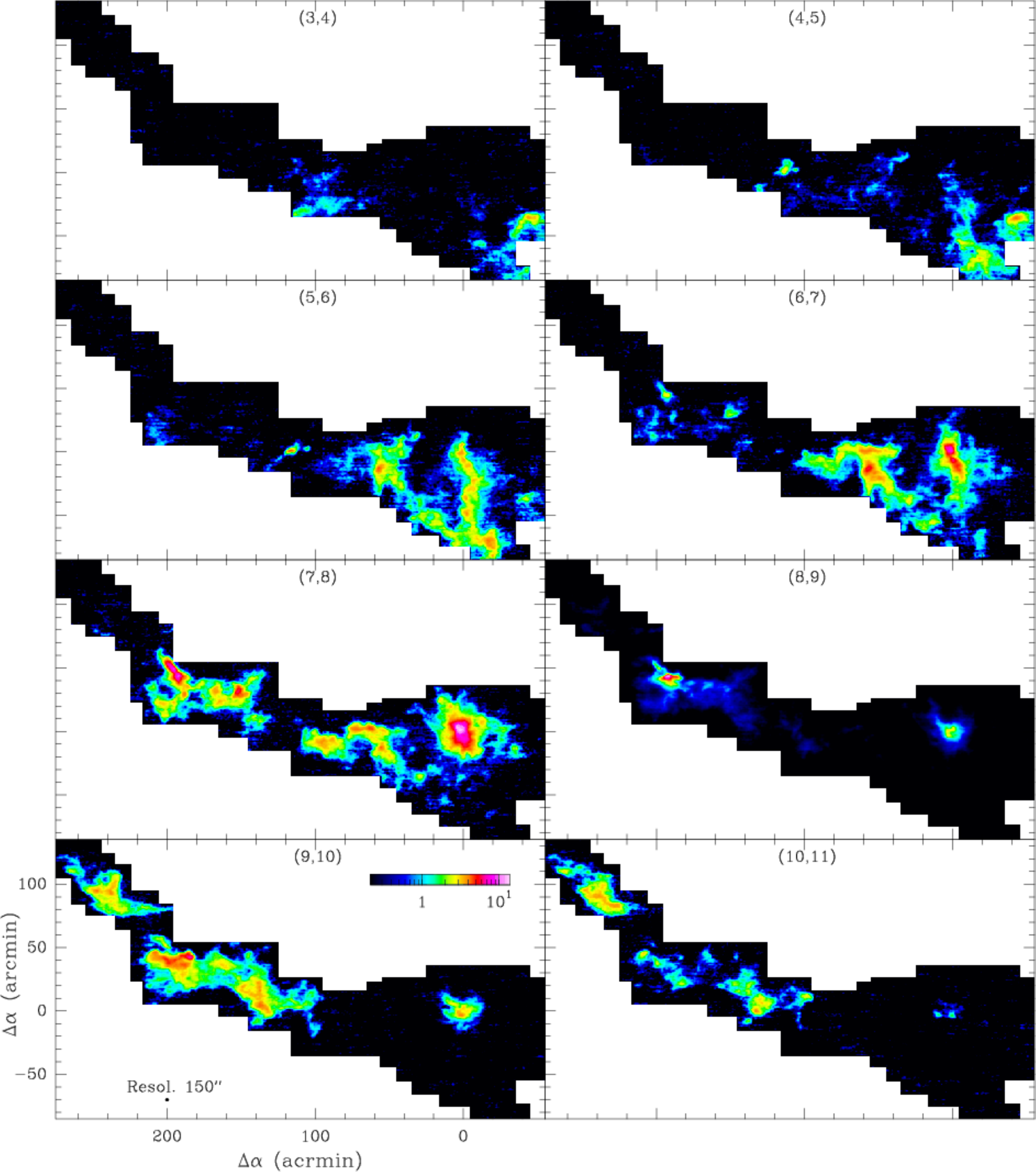}
\caption[$^{13}$CO($2\rightarrow 1$) maps of Perseus]{
\label{fig:perseus_sun06}
Map of the Perseus cloud in $^{13}$CO($2\rightarrow 1$). The top panel shows the emission integrated over all velocities, while the bottom panel shows maps integrated over different velocity channels. In each sub-panel in the bottom, the numbers at the top indicate the velocity range (in km s$^{-1}$) of the emission shown. Credit: \citeauthor{sun06a}, A\&A, 451, 539, 2006, reproduced with
permission \copyright ESO.
}
\end{figure}

Giant molecular clouds are not spheres. They have complex internal structures, as illustrated in Figure \ref{fig:perseus_sun06}. They tend to be highly filamentary and clumpy, with most of the mass in low density structures and only a little bit in very dense parts. However, if one computes a mean density by dividing the total mass by the rough volume occupied by the $^{12}$CO gas, the result is $\sim 100$ cm$^{-3}$. Typical size scales for GMCs are tens of pc -- the Perseus cloud shown in Figure \ref{fig:perseus_sun06} is a small one by Galactic standards, but the most massive ones are found predominantly in the molecular ring, so our high resolution images are all of nearby small ones.

This complex structure on the sky is matched by a complex velocity structure. GMCs typically have velocity spreads that are much larger than the thermal sound speed of $\sim 0.2$ km s$^{-1}$ appropriate to 10 K gas. One can use different tracers to explore the distributions of gas at different densities in position-position-velocity space -- at every position one obtains a spectrum that can be translated into a velocity distribution along that line of sight. The data can be slides into different velocities.

One can also get a sense of density and velocity structure by combining different molecular tracers. For example, the data set from COMPLETE (see Figure \ref{fig:complete_ridge06}) consists of three-dimensional cubes of $^{12}$CO and $^{13}$CO emission in position-position-velocity space, and from this one can draw isosurfaces. Generally the $^{12}$CO isosurfaces contain the $^{13}$CO ones, as expected since the $^{12}$CO traces less dense gas and the $^{13}$CO traces more dense gas. The density increases as one moves toward the cloud "center" in both position and velocity, but the morphology is not simple.   

\subsection{Cores}

As we zoom into yet smaller scales, the density rises to $10^5 - 10^7$ cm$^{-3}$ or more, while the mass decreases to a few $\msun$. These regions, called cores, tend to be strung out along filaments of lower density gas. Morphologically, cores tend to be closer to round than the lower-density material around them. These objects are thought to be the progenitors of single stars or star systems. Cores are distinguished not just by simple, roundish density structures, but by similarly simple velocity structures. Unlike in GMCs, where the velocity dispersion is highly supersonic, in cores it tends to be subsonic. This is indicated by a thermal broadening that is comparable to what one would expect from purely thermal motion.

\chapter{Observing Young Stars}
\label{ch:obsstars}

\marginnote{
\textbf{Suggested background reading:}
\begin{itemize}
\item \href{http://adsabs.harvard.edu/abs/2012ARA\%26A..50..531K}{Kennicutt, R.~C., \& Evans, N.~J. 2012, ARA\&A, 50, 531}, section 3 \nocite{kennicutt12a}
\item \href{http://adsabs.harvard.edu/abs/2014arXiv1402.0867K}{Krumholz, M.~R. 2014, Phys.~Rep., 539, 49}, section 2 \nocite{krumholz14c}
\end{itemize}
}

Having discussed how we observe interstellar gas that is forming stars, we now turn to the phenomenology of the young stars themselves. This chapter works form small to large scales, first discussing individual young stars, then resolved young stellar populations, and then ending with unresolved stellar populations in the Milky Way and nearby galaxies.

\section{Individual Stars}

Since we think star formation begins with a core that is purely gas, the first observable stage of star formation should be a cloud that is cold and lacks a central point source. Once a protostar forms, it will begin gradually heating up the cloud, while the gas in the cloud collapses onto the protostar, reducing the opacity. Eventually enough material accretes from the envelope to render it transparent in the near infrared and finally the optical, and we begin to be able to see the star directly for the first time. The star is left with an accretion disk, which gradually accretes and is then dispersed. Eventually the star contracts onto the main sequence.

This theoretical cartoon has been formalized into a system of classification of young stars based on observational diagnostics. At one end of this sequence lies purely gaseous sources where there is no evidence at all for the presence of a star, and at the other end lies ordinary main sequence stars. In between, objects are classified based on their emission in the infrared and sub-mm parts of the spectrum. These classifications probably give more of an impression of discrete evolutionary stages than is really warranted, but they nonetheless serve as a useful rough guide to the evolutionary state of a forming star.

Consider a core of mass $\sim 1$ $\msun$, seen in dust or molecular line emission. When a star first forms at its center, it will be very low mass and very low luminosity, and will heat up only the dust nearest to it, and only by a very small amount. Thus the total light output will still be dominated by the thermal emission of the dust at its equilibrium temperature. The spectral energy distribution of the source will therefore look just like that which prevailed before the star formed.

\begin{marginfigure}
\includegraphics[width=\linewidth]{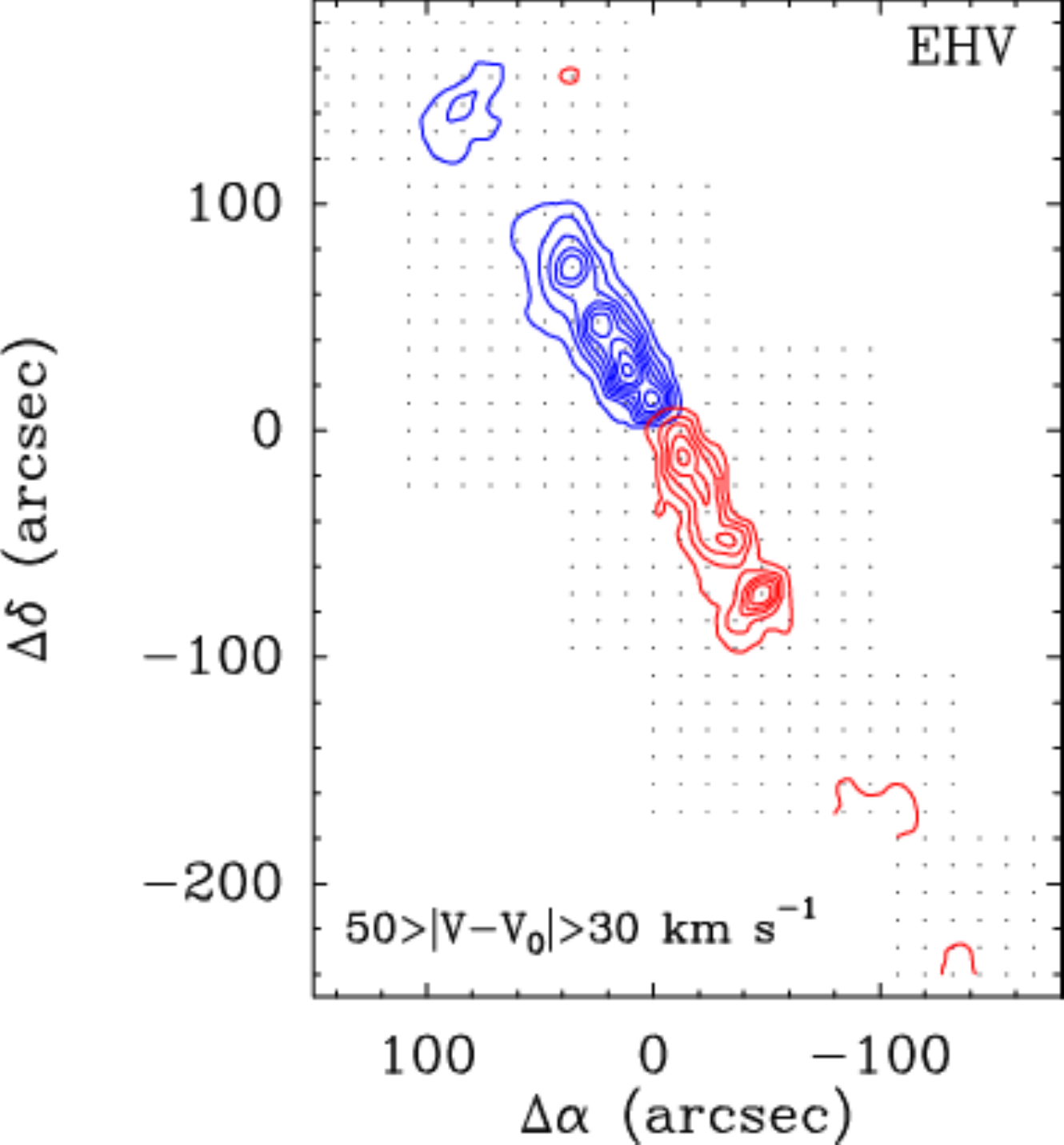}
\caption[Outflow in CO($2\rightarrow 1$)]{
\label{fig:outflow_tafalla04}
An integrated intensity map in CO($2\rightarrow 1$), showing material at velocities between $\pm 30-50$ km s$^{-1}$ (\textit{blue and red contours, respectively}) relative to the mean. Contours are spaced at intensities of 1 K km s$^{-1}$. The outflow shown is in the Taurus star-forming region. Credit: \citeauthor{tafalla04c}, A\&A,423, L21, 2004, reproduced with permission \copyright\, ESO.
}
\end{marginfigure}

However, there might be other indicators that a star has formed. For example, the density distribution might show a very sharp, unresolved peak. Another sign that a star has formed might be the presence of an outflow, which, as we discuss in Chapter \ref{ch:disks_obs}, all protostars seem to generate. Outflows coming from the center of a core can be detected in a few ways. Most directly, one can see bipolar, high velocity structures in molecular emission (Figure \ref{fig:outflow_tafalla04}). One can also detect indirect evidence of an outflow, from the presence of highly excited molecular line emission that is produced by shocks at hundreds of km s$^{-1}$. One example of such a line is SiO($2\rightarrow 1)$ line, which is generally seen in gas moving at several tens of km s$^{-1}$ with temperatures of several hundred K -- this is taken to be indication that emission in this line is produced in warm shocks. Since we know of no processes other than formation of a compact object with a $\gtrsim 100$ km s$^{-1}$ escape velocity that can accelerate gas in molecular clouds to such speeds, the presence of such an outflow is taken to indicate that a compact object has formed.

\begin{marginfigure}
\includegraphics[width=\linewidth]{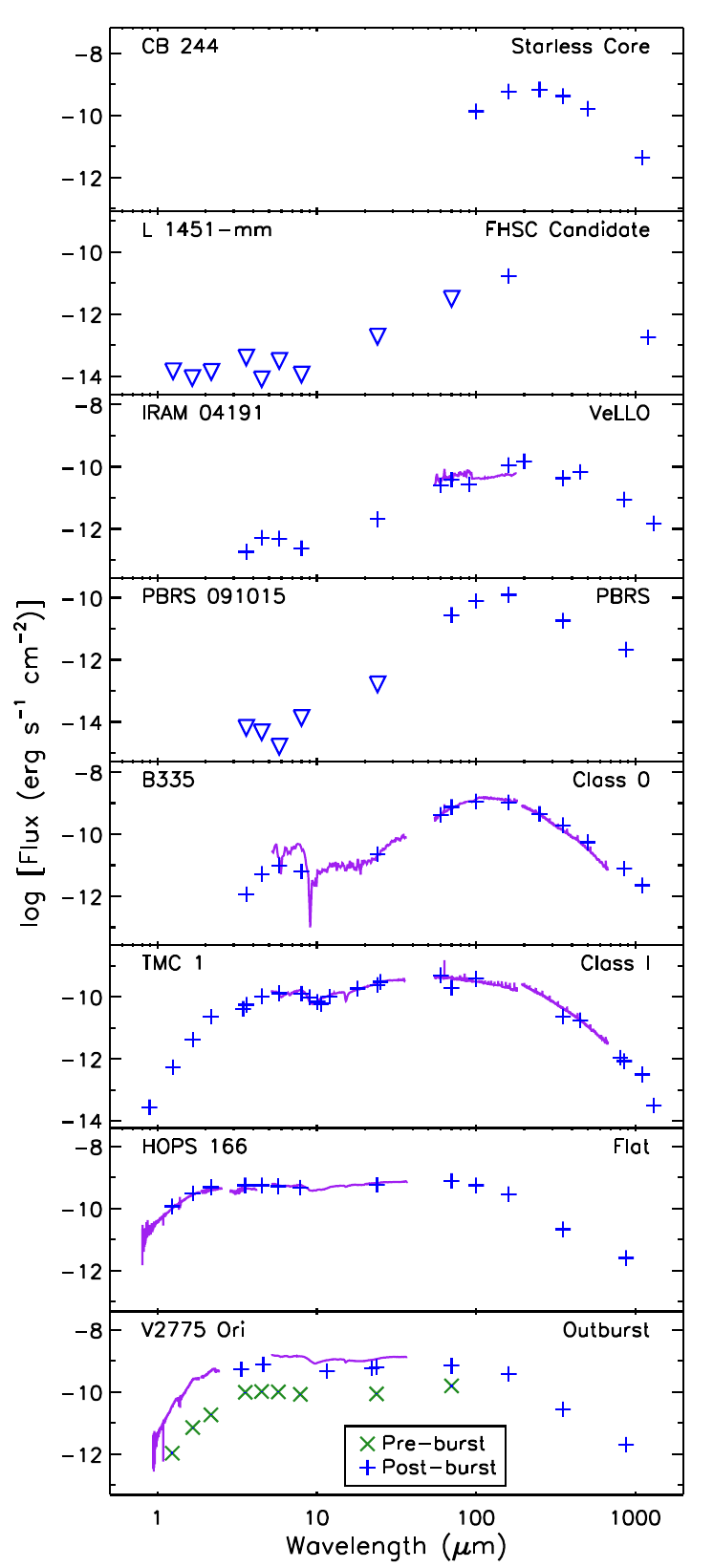}
\caption[Sample SEDs of protostellar cores]{
\label{fig:seds_dunham14}
Sample spectral energy distributions (SEDs) of protostellar cores, together with the assigned class, as collected by \citet{dunham14a}.
}
\end{marginfigure}

These are the earliest indications of star formation we have available to us. We call objects that show one of these signs, and do not fall into one of the other categories, class 0 sources. The dividing line between class 0 and class 1 is that the star begins to heat the dust around it to the point that there is non-trivial infrared emission. Before the advent of \textit{Spitzer} and \textit{Herschel}, the dividing line between class 0 and 1 was taken to be a non-detection in the IR, but as more sensitive IR telescopes became available, the detection limit went down, and it became necessary to specify a dividing line in terms of a luminosity cut. A source is said to be class 0 if more than 0.5\% of its total bolometric output emerges at wavelengths longer than $350$ $\mu$m, i.e., if $L_{\rm smm} / L_{\rm bol} > 0.5\%$, where $L_{\rm smm}$ is defined as the luminosity considering only wavelengths of 350 $\mu$m and longer (Figure \ref{fig:seds_dunham14}).

\begin{marginfigure}
\includegraphics[width=\linewidth]{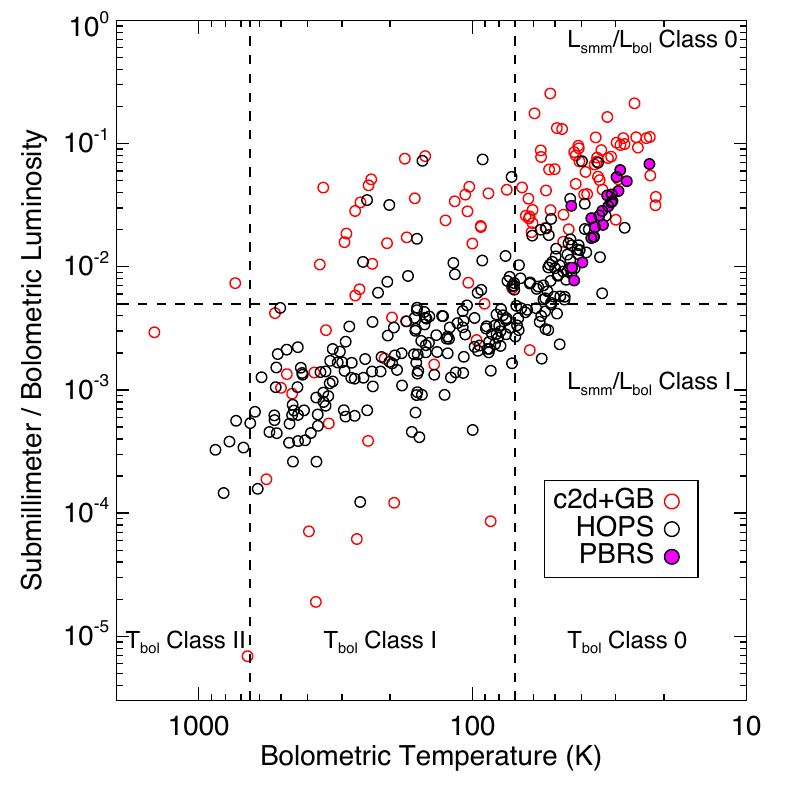}
\caption[Bolometric temperatures of protostellar cores]{
\label{fig:tbol_dunham14}
Bolometric temperatures of protostellar cores as compared to sub-mm to bolometric luminosity ratios \citep{dunham14a}. The samples shown are from three different surveys as indicated in the legend.
}
\end{marginfigure}

In practice, measuring $L_{\rm smm}$ can be tricky because it can be hard to get absolute luminosities (as opposed to relative ones) correct in the sub-mm, so it is also common to define the class 0-1 divide in terms of another quantity: the bolometric temperature $T_{\rm bol}$. This is defined as the temperature of a blackbody that has the same flux-weighted mean frequency as the observed spectral energy distribution (SED). That is, if $F_\nu$ is the flux as a function of frequency from the observed source, then we define $T_{\rm bol}$ by the implicit equation
\begin{equation}
\frac{\int \nu B_{\nu}(T_{\rm bol}) \, d\nu}{\int B_{\nu}(T_{\rm bol}) \, d\nu} = \frac{\int \nu F_{\nu}\, d\nu}{\int F_\nu \,d\nu}.
\end{equation}
The class 0-1 dividing line is also sometimes taken to be $T_{\rm bol} = 70$ K. In cases where $L_{\rm smm}$ is accurately measured, $T_{\rm bol}$ is observed to be a reasonably good proxy for $L_{\rm smm} / L_{\rm bol}$ (Figure \ref{fig:tbol_dunham14}).

Once protostars reach class I, their evolution into further classes is defined in terms of the infrared spectral energy distribution. The motivating cartoon is a follows. At early times, the envelope of dust around the protostar is very optically thick at visible and even near infrared wavelengths. As a result, we cannot directly observe the stellar photosphere. All the radiation is absorbed by the envelope. The dust is in thermal equilibrium, so it re-radiates that energy. Since the radius of the sphere of dust is much larger than that of the star, and the luminosity radiated by the dust must ultimately be equal to that of the star, this emission must be at lower temperature and thus longer wavelengths. Thus as the radiation propagates outward through the dust it is shifted to longer and longer wavelengths. However, at wavelengths longer than the characteristic sizes of the dust grains, the opacity decreases as roughly $\kappa_\lambda \propto \lambda^{-2}$. Thus eventually the radiation is shifted to wavelengths where the remaining dust is optically thin, and it escapes. What we observe is therefore not a stellar photosphere, but a "dust photosphere".

Given this picture, the greater the column density of the dust around the star, the further it will have to diffuse in wavelength in order to escape. Thus the wavelength at which the emission peaks, or, roughly equivalently, the slope of the spectrum at a fixed wavelength, is a good diagnostic for the amount of circumstellar dust. Objects whose SEDs peak closer to the visible are presumed to be more evolved, because they have lost more of their envelopes.

More formally, this classification scheme was based on fluxes as measured by the \textit{Infrared Astronomical Satellite (IRAS)}. We define
\begin{equation}
\alpha_{\rm IR} = \frac{d\log (\lambda F_{\lambda})}{d\log\lambda},
\end{equation}
as the infrared spectral index, and in practice we measure $\alpha_{\rm IR}$ using two points from the \textit{IRAS} SED: 2.2 $\mu$m and $10-25$ $\mu$m. More positive values of $\alpha_{\rm IR}$ indicate SEDs that peak at longer wavelengths, further into the IR, while more negative values indicate SEDs that peak closer to visible. We define sources with $\alpha_{\rm IR}\geq 0.0$, i.e., rising at longer wavelengths from 2 to 25 $\mu$m, as class I sources. Alternately, in terms of bolometric temperature, the class I to class II transition is generally taken to be at 650 K (Figure \ref{fig:seds_dunham14}).

As more of the envelope accretes, it eventually becomes optically thin at the peak emitting wavelengths of the stellar photosphere. In this case we see the stellar blackbody spectrum, but there is also excess infrared emission coming from the disk of warm, dusty gas that still surrounds the star. Thus the SED looks like a stellar blackbody plus some extra emission at near- or mid-infrared wavelengths. Stars in this class are also know as classical T Tauri stars, named for the first object of the class, although the observational definition of a T Tauri star is somewhat different than the IR classification scheme\footnote{T Tauri stars were first identified in the optical, long before the availability of infrared SEDs. They are defined by high levels of optical variability and the presence of strong chromospheric lines, indicating large amounts of circumstellar material. T Tauri stars are discussed further in Chapter \ref{ch:late_disk}.}, so the alignment may not be perfect. In terms of $\alpha_{\rm IR}$, these stars have indices in the range $-1.6 < \alpha_{\rm IR} < 0$.\footnote{Depending on the author, the breakpoint may  be placed at $-1.5$ instead of $-1.6$. Some authors also introduce an intermediate classification between 0 and I, called flat spectrum sources, which they take to be $-0.3 < \alpha_{\rm IR} < 0.3$.} A slope of around $-1.6$ is what we expect for a bare stellar photosphere without any excess infrared emission coming from circumstellar material. Since the class II phase is the last one during which there is a disk of any significant mass, this is also presumably the phase where planet formation must occur.

The final stages is class III, the category into which we places sources whose SEDs have $\alpha_{\rm IR} < -1.6$. Stars in this class correspond to weak line T Tauri stars. The SEDs of these stars look like bare stellar photospheres in the optical through the mid-infrared. If there is any IR excess at all, it is in the very far IR, indicating that the emitting circumstellar material is cool and located far from the star. The idea here is that the disk around them has begun to dissipate, and is either now optically thin at IR wavelengths or completely dissipated, so there is no strong IR excess. 

However, these stars are still not mature main sequence stars. First of all, their temperatures and luminosities do not correspond to those of main sequence stars. Instead, they are still puffed up to larger radii, so they tend to have either lower effective temperatures or higher bolometric luminosities (or both) than main sequence stars of the same mass. Second, they show extremely high levels of magnetic activity compared to main sequence stars, producing high levels of X-ray emission. Third, they show lithium absorption lines in their atmospheres. This is significant because lithium is easily destroyed by nuclear reactions at high temperatures, and no main sequence stars with convective photospheres show Li absorption. Young stars show it only because there has not yet been time for all the Li to burn.

\section{Statistics of Resolved Stellar Populations}

Young stars tend to be born in the presence of other stars, rather than by themselves. This is not surprising: the gas cores from which they form are very small fragments, $\sim 1$ $\msun$, inside much larger, $\sim 10^6$ $\msun$ clouds. It would be surprising if only one tiny fragment containing $\sim 10^{-6}$ of the total cloud mass were to collapse. We now pull back to somewhat larger scales to look at the formation of stars in groups.

\subsection{Multiplicity}

The smallest scale we can look at beyond a single star is multiple systems. When we do so, we find that a significant fraction of stars are members of multiple systems -- usually binaries, but also some triples, quadruples, and larger. The multiplicity is a strong function of stellar mass. The vast majority of B and earlier stars are multiples, while the majority of G, K, and M stars are singles. This means that most stars are single, but that most massive stars are multiples. The distribution of binary periods is extremely broad, ranging from hours to Myr. The origin of the distribution of periods, and of the mass-dependence of the multiplicity fraction, is a significant area of research in star formation theory, one to which we will return in Chapters \ref{ch:imf_obs}, \ref{ch:imf_th}, and \ref{ch:massivestar}.

\subsection{The Initial Mass Function}

If we observe a cluster of stars, the simplest thing to do is simply count up how many of them there are as a function of mass. The result is one of the most important objects in astrophysics, the initial mass function (IMF). This requires a bit of modeling, since of course what we can actually measure is a luminosity function, not a mass function. The problem of determining the IMF can be tackled in two ways: either by looking at stars in the solar neighborhood, or by looking at individual star clusters.

Looking at stars in the Solar neighborhood has the advantage that there are a lot of them compared to what you see in a clusters, so one gets a lot of statistical power. One also does not have to worry about two things that a major headache for studies of young clusters. First, young clusters usually have remaining bits of gas and dust around them, and this creates reddening that can vary with position and has to be modeled. Second, for clusters younger than $\sim 10$ Myr, the stars are not on the main sequence yet. Since young stars are brighter than main sequence stars of the same mass, this produces and age-mass degeneracy that you have to break by obtaining more information that just luminosities (usually temperatures or colors), and then making pre-main sequence evolutionary models.\footnote{Protostellar evolution is covered in Chapter \ref{ch:protostar_evol}.}

On the other hand, if we want to talk about the IMF of massive stars, we are largely stuck looking at young clusters. The same is also true for brown dwarfs. Since these fade with time, it is hard to find a large number of them outside of young clusters. An additional advantage of star clusters is that they are to good approximation chemically homogenous, so we need not worry about chemical variations masquerading as mass variations.

\begin{figure}
\includegraphics[width=\linewidth]{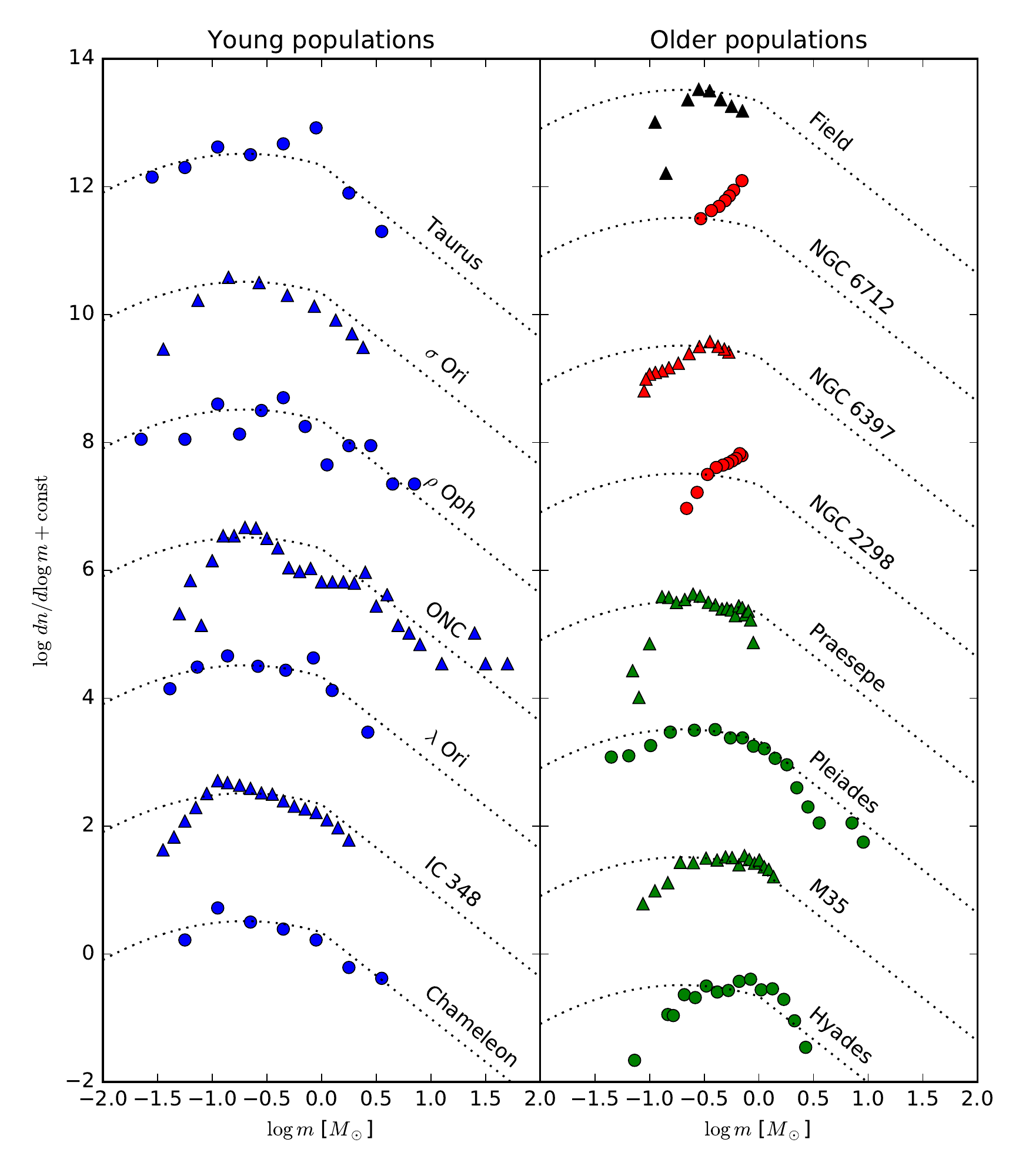}
\caption[Measured stellar IMFs in a variety of regions]{
\label{fig:imf_observed}
Stellar initial mass functions inferred for a wide variety of regions in the Milky Way (data compilation from \citealt{bastian10a}). The left panel shows young stellar populations (age under $\sim 5$ Myr), while the right panel shows older stellar populations in open clusters (green), globular clusters (red), and the Galactic field (black). The names of the regions are as indicated. The dotted black lines show the \citet{chabrier05a} fit to the IMF (equation \ref{eq:chabrier}). Note that vertical offsets in the plot are arbitrary, and the black dotted lines have been normalized to match the data at $m=0.5$ $M_\odot$. Finally, note that for many regions the data become incomplete below $\sim 0.2$ $M_\odot$.
}
\end{figure}

A big problem for either method is correction for unresolved binaries, particularly at the low mass end, where the companions of brighter stars are very hard to see. When one does all this, the result is the apparently universal or close-to-universal distribution illustrated in Figure \ref{fig:imf_observed}.\footnote{There have been recent claims of IMF variation from extragalactic observation, which we will discuss in Chapter \ref{ch:imf_obs}.} The basic features we see are a break peak centered around a few tenths of $\msun$, with a fairly steep fall off at higher masses that is well fit by a powerlaw function with a slope near $-2.3$. There is also a fall-off at lower masses, although some authors argue for a second peak in the brown dwarf regime. This is a difficult observational problem, both because brown dwarfs are hard to find, and because their evolutionary tracks are less secure than those for more massive stars.

The functional form shown in Figure \ref{fig:imf_observed} has been parameterized in a number of ways. Two of the most popular are from \citet{kroupa01a, kroupa02c} and \citet{chabrier03a, chabrier05a}.\footnote{See the reviews by \citet{bastian10a} and \citet{offner14a} for a thorough listing of alternate parameterizations.}  Both of these fit the field star data, and the data individual clusters, within the error bars. The functional form for Chabrier is
\begin{equation}
\label{eq:chabrier}
\frac{dn}{d\log m} \propto 
\left\{
\begin{array}{ll}
\exp\left[-\frac{(\log m-\log 0.22)^2}{2\times 0.57^2}\right], &
m<1 \\
\exp\left[-\frac{(-\log 0.22)^2}{2\times 0.57^2}\right] m^{-1.35}, &
m\ge 1
\end{array}
\right.,
\end{equation}
while the functional form for Kroupa is
\begin{equation}
\label{eq:kroupa}
\frac{dn}{d\log m} \propto
\left\{
\begin{array}{ll}
\left(\frac{m}{m_0}\right)^{-\alpha_0}, & m_0 < m < m_1\\
\left(\frac{m_1}{m_0}\right)^{-\alpha_0} \left(\frac{m}{m_1}\right)^{-\alpha_1}, &
m_1 < m < m_2 \\
\left[\prod_{i=1}{n} \left(\frac{m_i}{m_{i-1}}\right)^{-\alpha_i}\right] \left(\frac{m}{m_n}\right)^{-\alpha_n}, &
m_{n-1} < m < m_n
\end{array}
\right.,
\end{equation}
with
\begin{equation}
\begin{array}{ll}
\alpha_0 = -0.7\pm 0.7, & m_0 = 0.01 \\
\alpha_1 = 0.3\pm 0.5, & m_1 = 0.08 \\
\alpha_2 = 1.3\pm 0.3, & m_2 = 0.5 \\
\alpha_3 = 1.3\pm 0.7, & m_3 = 1 \\
\end{array}.
\end{equation}
In both of the above expressions, $m$ is understood to be in units of $M_\odot$.

\section{Unresolved Stellar Populations and Extragalactic Star Formation}

What about cases where we cannot resolve the stellar population, as is usually the case for extragalactic work? What can we learn about star formation in that case? The answer turns out to be that the thing we can most directly measure is the star formation rate, and that doing so yields some very interesting results.

\subsection{Measuring the Star Formation Rate: General Theory}

The most basic problem in working with unresolved stellar populations is how we distinguish young stars from main sequence ones. Except for the brightest stars in the nearest galaxies, we cannot obtain spectra, or even colors, for individual stars as we can in the Milky Way. Instead, the strategy we use to isolate young stars is to exploit the fact that massive stars have short lifetimes, so if we measure the total number of massive stars in a galaxy, or some patch of a galaxy, then we are effectively measuring we many such stars formed there over some relatively short period. We can formalize this theory a bit as follows.

Consider stars born with an initial mass function $dn/dm$. The mean stellar mass for this IMF is $\overline{m} = \int dm\, m(dn/dm)$. A time $t$ after a star is born, the star has a luminosity $L(m,t)$, where the luminosity can be bolometric, or integrated over some particular filter or wavelength range. First consider the simplest possible case, a population of stars all born at the same instant at time $0$. A time $t$ later, the luminosity of the stars is
\begin{equation}
L(t) = N_* \int_0^{\infty} dm\, L(m,t) \frac{dn}{dm},
\end{equation}
where $N_*$ is the total number of stars, and we have normalized the IMF so that $\int (dn/dm) \, dm = 1$. That is, we simply integrate the luminosity per star at time $t$ over the mass distribution of stars. Now consider a region, e.g., a galaxy, forming stars at a rate $\dot{M}_*(t)$; in terms of number, the star formation rate is $\dot{N}_*(t) = \dot{M}_*(t)/\overline{m}$. To find the luminosity of the stellar population that is present today, we simply take the expression we just derived and integrate over all the possible stellar ages. Thus we have
\begin{equation}
L =  \int_{0}^\infty dt\, \frac{\dot{M}_*(t)}{\overline{m}}  \int_0^{\infty} dm\, L(m,t) \frac{dn}{dm}.
\end{equation}

By itself this is of limited use, because the right hand side depends on the full star formation history $\dot{M}_*(t)$. However, let us assume that $\dot{M}_*$ is constant in time. The integral still converges as long as $L(m,t)$ reaches 0 after a finite time. In this case the integrals over $m$ and $t$ are separable, and we can rearrange them to
\begin{equation}
\label{eq:sfrtol}
L = \frac{\dot{M}_*}{\overline{m}} \int_0^{\infty} dm \, \frac{dn}{dm} \int_0^\infty dt \, L(m,t) \equiv \frac{\dot{M}_*}{\overline{m}} \int_0^{\infty} dm \, \frac{dn}{dm} \langle L t_{\rm life} \rangle_m 
\end{equation}
In the final step we defined a new quantity $\langle L t_{\rm life} \rangle_m$, which has a simple physical meaning: it is the total amount of radiant energy that a star of mass $m$ puts out over its lifetime.

Notice the expression on the right depends only on the constant star formation rate $\dot{M}_*$, the energy output $\langle L t_{\rm life}\rangle_m$, which we can generally calculate from stellar structure and evolution theory, and the IMF $dn/dm$. Thus if we measure $L$ and use the "known" values of  $\langle L t_{\rm life}\rangle_m$ and $dn/dm$, we can measure the star formation rate. The underlying physical assumption is that the stellar population being observed is in statistical equilibrium between new stars forming and old stars dying, so the total number of stars present and contributing to the light at any time is proportional to the rate at which they are forming. Thus a measurement of the light tells us about the star formation rate.

Is our assumption that $\dot{M}_*$ is constant reasonable? That depends on the system we are observing. For an entire galaxy that is forming stars quiescently and has not been externally perturbed, it is probably reasonable to assume that $\dot{M}_*$ cannot vary on timescales much shorter than the dynamical time of the galaxy, which is $\sim 200$ Myr for a galaxy like the Milky Way. If we choose to observe the luminosity at a wavelength where the light is coming mostly from stars with lifetimes shorter than this, so that $L(m,t)$ reaches 0 (at least to good approximation) at times much less than 200 Myr, then assuming constant $\dot{M}_*$ is quite reasonable.

\begin{marginfigure}
\includegraphics[width=\linewidth]{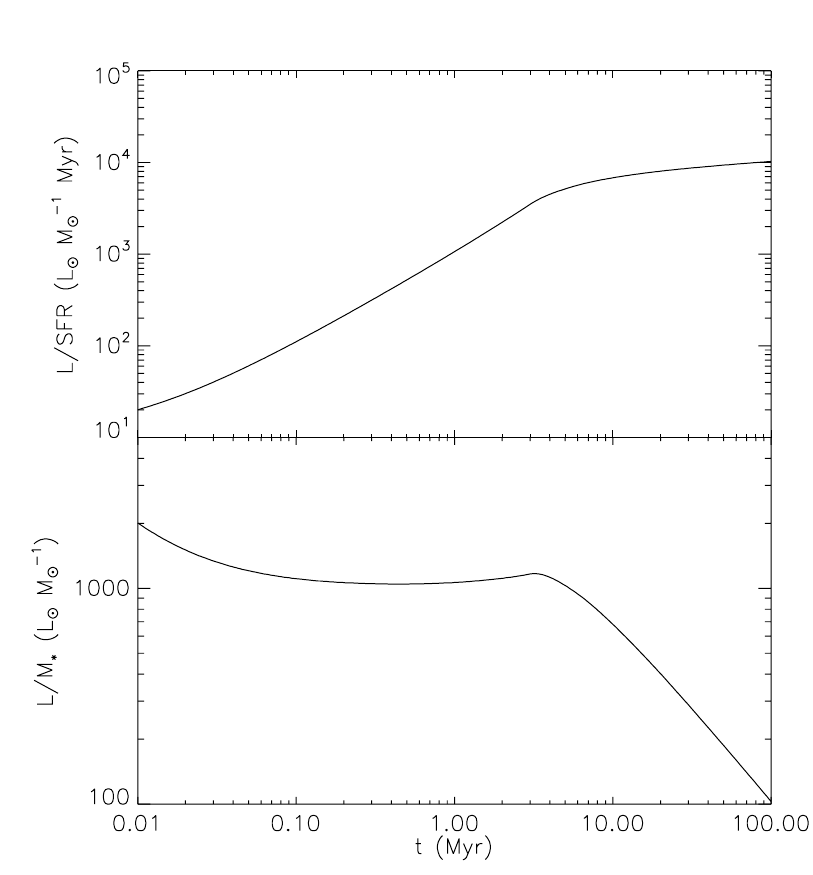}
\caption[Bolometric luminosity versus stellar population age]{
\label{fig:lvst_krumholz07}
Bolometric luminosity versus time for stellar populations as a function of population age. The top panel shows the luminosity normalized by the star formation rate, while the bottom shows the luminosity normalized by the total stellar mass. Credit: \citet{krumholz07e}, \copyright\,AAS. Reproduced with permission.
}
\end{marginfigure}

However, it is always important to keep this constraint in mind -- we can only measure the star formation rate as long as we believe it to be constant on timescales long compared to the lifetimes of the stars responsible for generating the luminosity we are measuring. One can actually see how the ratio of luminosity to star formation rate behaves in systems that do not satisfy the constraint by generating synthetic stellar populations. In the simple case of a system that begins with no stars and then forms star stars at a constant rate, the bolometric luminosity after the onset of star formation just increases linearly with time until the first stars star evolving off the main sequence, and only becomes constant after $\sim 4$ Myr (Figure \ref{fig:lvst_krumholz07}).

The need to satisfy this constraint generally drives us to look for luminosities that are dominated by very massive stars, because these have very short lifetimes. Thus we will begin by discussing what luminosities we can measure that are particularly good at picking out massive stars. This is far from an exhaustive list -- astronomers have invented many, many methods to infer star formation rates for galaxies at a range of redshifts. The accuracy of these techniques is highly variable, and in some cases amounts to little more than a purely empirical calibration. We focus here on the most reliable and widely used techniques that we can apply to relatively nearby galaxies.

\subsection{Recombination Lines}

\begin{figure}
\includegraphics[width=\linewidth]{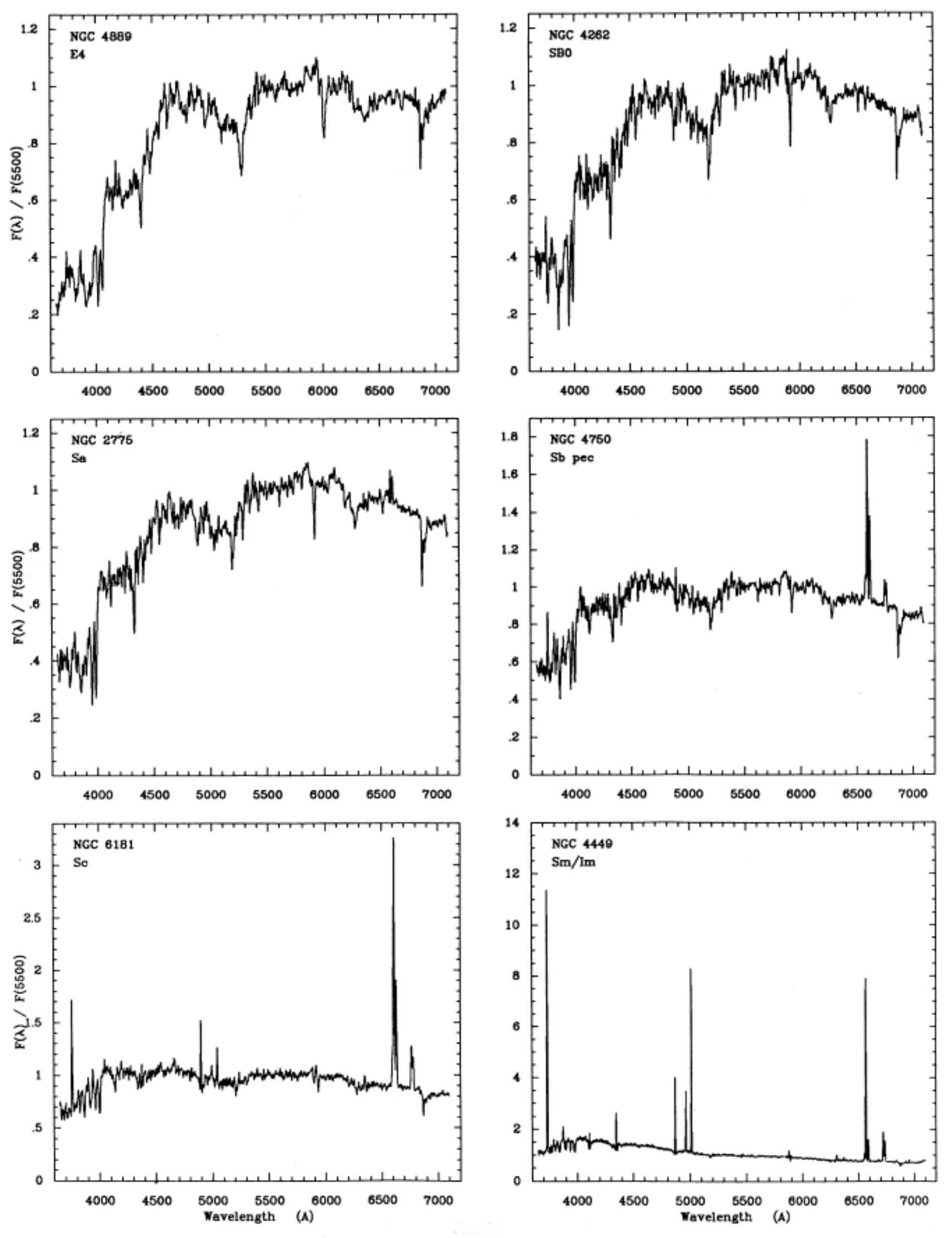}
\caption[Optical spectra of galaxies across the Hubble sequence]{
\label{fig:spectra_kennicutt92}
Example spectra of galaxies of varying Hubble type. In each panel, the galaxy name and Hubble type are listed. Credit: \citet{kennicutt92a}, \copyright\,AAS. Reproduced with permission.
}
\end{figure}

Probably the most common technique, and the only one that can be used from the ground for most galaxies, is hydrogen recombination lines. To illustrate why this is useful, it is helpful to look at some galaxy spectra (Figure \ref{fig:spectra_kennicutt92}). As we move from quiescent E4 and SB galaxies to actively star-forming Sc and Sm/Im galaxies, there is a striking different in the prominence of emission lines.

In the example optical spectra, the most prominent lines are the H$\alpha$ line at 6563 \AA\ and the H$\beta$ at 4861 \AA. These are lines produced by the $3\rightarrow 2$ and $4\rightarrow 2$, respectively, electronic transitions in hydrogen atoms. In the infrared (not shown in the figure) are the Paschen $\alpha$ and $\beta$ lines at 1.87 and 1.28 $\mu$m, and the Bracket $\alpha$ and $\gamma$ lines at 4.05 and 2.17 $\mu$m. These come from the $4\rightarrow 3$, $5\rightarrow 3$, $5\rightarrow 4$, and $7\rightarrow 4$ transitions.

Why are these related to star formation? The reason is that these lines come from H~\textsc{ii} regions: regions of ionized gas produced primarily by the ionizing radiation of young stars. Since only massive stars (lager than $10-20$ $\msun$) produce significant ionizing fluxes, these lines indicate the presence of young stars. Within these ionized regions, one gets hydrogen line emission because atoms sometimes recombine to excited states rather than to the ground state. These excited atoms then radiatively decay down to the ground state, producing line emission in the process.

Obtaining a numerical conversion between the observed luminosity in one of these lines and the star formation rate is a four-step process. First, one performs a quantum statistical mechanics calculation to compute the yield of photons in the various lines per recombination. This can be done very precisely from first principles. Second, one equates the total recombination rate to the total ionization rate, and uses this to determine the total rate of emission for the line in question per ionizing photon injected into the nebula. Third, one uses stellar models to compute $\langle L_{\rm ion} t_{\rm life}\rangle_m$, the total ionizing photon production by a star of mass $m$ over its lifetime. Four, one evaluates the integral over the IMF given by equation (\ref{eq:sfrtol}) to obtain the numerical conversion between star formation rate and luminosity. As of this writing, the most up-to-date resource for the results of such calculations is \citet{kennicutt12a}.

Note that there are significant uncertainties in these numbers, the dominant one of which is the IMF. The reason the IMF matters so much is that the light is completely dominated by the massive stars, while the mass is all in the low mass stars that are not observed directly. To give an example, for a Chabrier IMF at zero age, stars more massive than 15 $\msun$ contribute 99\% of the total ionizing flux for a stellar population, but constitute less than 0.3\% of the mass. Thus we are extrapolating by at least a factor of 30 in mass, and small changes in the IMF can produce large changes in the resulting ionizing luminosity to mass conversion.

Another complication is that some of the line emission is likely to be absorbed by dust grains within the source galaxy, and some of the ionizing photons are absorbed by dust grains rather than hydrogen atoms. Thus, one must make an extinction correction to the luminosities.

\subsection{Radio Free-Free}

A closely related method for measuring massive stars is to use free-free emission at radio wavelengths. An H~\textsc{ii} region emits optical lines from transitions between energy levels of hydrogen and other atoms, but it also emits free-free radiation in the radio. This is radiation produced by bremsstrahlung: free electrons scattering off ions, and emitting because accelerating charges emit. It is the opposite side of the coin from recombination line emission: the former occurs when free electrons and protons encounter one another and do not become bound, while the latter occurs when they do.

We will not treat bremsstrahlung or its application to H~\textsc{ii} regions here, but the relevant point for us is that the free-free luminosity of H~\textsc{ii} regions at radio wavelengths is proportional to $n_e n_i$, i.e., the product of the electron and ion densities. Since the recombination rate is also proportional to $n_e n_{\rm H^+}$, and due to chemical balance the recombination rate must equal the ionization rate, the free-free luminosity is directly proportional to the rate at which ionizing photons are injected into the H~\textsc{ii} region. Thus one can convert between free-free emission rate and ionization rate based on the physics of H~\textsc{ii} regions, and from then convert that into a star formation rate exactly as for optical recombination lines.

The free-free method has one major advantage, which is that radio emission is not obscured by dust, so one of the dust absorption corrections goes away. The correction for absorption of ionizing photons by dust grains within the H~\textsc{ii} region remains, but this is generally only a few tens of percent. Thus radio free-free measurements are more reliable that recombination line ones. Indeed, they are the only technique we can use for most H~\textsc{ii} regions in the Milky Way, since these tend to be located in the Galactic plane and thus suffer from heavy extinction at optical wavelengths. The downside is that the free-free emission is quite weak, and separating free-free from other sources of radio emission requires the ability to resolve individual H~\textsc{ii} regions. Thus at present this technique is useful primarily for the Milky Way and a few other nearby galaxies, since those are the only places where we can detect and resolve individual H~\textsc{ii} regions.

\subsection{Infrared}

The recombination line methods work well for galaxies that are like the Milky Way, but considerably less well for galaxies that are dustier and have higher star formation rates. This is because the dust extinction problem becomes severe, so that the vast majority of the Balmer line emission is absorbed. The Paschen and Bracket emission is much less sensitive to this, since those lines are in the IR, but even they can be extincted in very dusty galaxies, and they are also much harder to use than H$\alpha$ and H$\beta$ because they are $1-2$ orders of magnitude less bright intrinsically.

Instead, for dusty sources the tracer of choice is far infrared. The idea here is that, in a sufficiently dusty galaxy, essentially all stellar light will eventually be absorbed by dust grains. These grains will then re-emit the light in the infrared. As a result, the SED peaks in the IR. In this case one can simply use the total IR output of the galaxy as a sort of calorimeter, measuring the total bolometric power of the stars in that galaxy. In galaxies or regions of galaxies with high star formation rates, which tend to be where H$\alpha$ and other recombination line techniques fail, this bolometric power tends to be completely dominated by young stars. Since these stars die quickly, the total number present at any given time is simply proportional to the star formation rate.

The derivation of the conversion in this case is very straightforward -- the $L(m,t)$ that is required is just the total bolometeric output of the stars. Results are given in \citet{kennicutt12a}, and Problem Set 1 includes an example calculation. Of course IR emission has its its problems too. First of all, it misses all the optical and UV radiation from young stars that is not absorbed within the galaxy, which makes it a poor choice for dust-poor galaxies where a majority of the radiation from young stars escapes.

A second problem is that if the SFR is low, then old rather than young stars may dominate the bolometric output. In this case the IR indicator can give an artificially high SFR. A more common problem for the dusty galaxies where IR tends to be used most is contamination from an active galactic nucleus (AGN). If an AGN contributes significantly to the bolometric output of a galaxy, that can masquerade as star formation. This can be hard to detect in a very dusty galaxy where most of the AGN light, along with most of the starlight, is absorbed and reprocessed by dust.

\subsection{Ultraviolet}

Yet another way of measuring star formation rates is by the broadband ultraviolet (UV) flux at wavelengths that are longer than 912 \AA\ (corresponding to 13.6 eV, the energy required to ionized hydrogen) but shorter than where old stars put out most of their light. This range is roughly $1250-2500$ \AA. This light does not ionize hydrogen, so unlike shorter wavelengths it can get out of a galaxy.

For galaxies in the right redshift range this light gets redshifted into the visible, so we can see it from the ground. However, for local galaxies these wavelengths are only accessible from space (or least a balloon or rocket). For this reason this band was not used much until the launch of the \textit{Galaxy Evolution Explorer (GALEX)} satellite, which had detectors operating at 1300-1800 and 1800-2800 \AA (referred to as FUV and NUV, respectively). Emission in the FUV band is dominated by stars with masses $\sim 5$ $\msun$ and up, which have lifetimes of $\sim 50$ Myr, so the total FUV light measures the star formation rate integrated over this time scale. Sadly, \textit{GALEX} is no longer in operation, and there is no comparable mission on the immediate horizon, so this technique is largely of archival value for now.

FUV suffers from the same problems with dust extinction as H$\alpha$, and they are perhaps even more severe, since opacity increases as frequency does. On the other hand, FUV is less sensitive to the IMF than H$\alpha$, because ionizing photons come from hotter and thus more massive stars than FUV ones. For systems with low overall star formation rates, ionization-based star formation rate indicators can become quite noisy due to the rarity of the massive stars they trace. FUV has fewer problems in this regard. However, there is a corresponding disadvantage, in that the $\sim 50$ Myr lifetime of FUV-emitting stars is getting uncomfortably close to the typical orbital periods of galaxies, and so one can legitimately worry about whether the SFR has really be constant over the required timescale. This problem becomes even worse if one looks at small-subregions of galaxies, rather than galaxies as a whole. One also has to worry about stars moving from their birth locations over such long timescales.

\subsection{Combined Estimators}

As one might guess from the discussion thus far, none of the indicators by itself is particularly good. Recombination lines and UV get into trouble in dusty galaxies because they miss light from young stars that is obscured by dust, while IR gets into trouble because it misses light from young stars that is not dust-obscured. This suggests that the best way to proceed is to combine one or more estimators, and this is indeed the current state of the art. A number of combined indicators are suggested in \citet{kennicutt12a}.

\part{Physical Processes}

\chapter{Chemistry and Thermodynamics}
\label{ch:microphysics}

\marginnote{\textbf{Suggested background reading:}
\begin{itemize}
\item \href{http://adsabs.harvard.edu/abs/2014arXiv1402.0867K}{Krumholz, M.~R. 2014, Phys.~Rep., 539, 49}, sections $3.1-3.2$ \nocite{krumholz14c}
\end{itemize}
\textbf{Suggested literature:}
\begin{itemize}
\item \href{http://adsabs.harvard.edu/abs/2010MNRAS.404....2G}{Glover, S.~C.~O., Federrath, C., Mac Low, M.-M., \& Klessen, R.~S. 2010, MNRAS, 404, 2} \nocite{glover10a}
\end{itemize}
}

Having completed our whirlwind tour of the observational phenomenology, we now turn to the physical processes that govern the behavior of the star-forming ISM and its transformation into stars. The goal of this section is to develop physical intuition for how this gas behaves, and to develop some analytic tools for use through the remainder of the book. This chapter covers the microphysics of the cold ISM.

\section{Chemical Processes in the Cold ISM}

We will begin our discussion of the microphysics of the cold ISM with the goal of understanding something important that should be clear from the observational discussion: the parts of the ISM associated with star formation are overwhelmingly molecular gas. This is in contrast to the bulk of the ISM, at least in the Milky Way and similar galaxies, which is composed of atomic or ionized gas with few or no molecules. Our goal is to understand why the ISM in some places becomes predominantly molecular, and how this transition is related star formation. We will focus this discussion on the most important atoms in the ISM: hydrogen, carbon, and oxygen.

\subsection{Hydrogen Chemistry}
\label{ssec:Hchemistry}

Molecular hydrogen is a lower energy state than atomic hydrogen, so an isolated box of hydrogen left for an infinite amount of time will eventually become predominantly molecular. In interstellar space, though, the atomic versus molecular fraction in a gas is determined by a balance between formation and destruction processes.

Atomic hydrogen can turn into molecular hydrogen in the gas phase, but this process is extremely slow. This is ultimately due to the symmetry of the hydrogen molecule. To form an H$_2$ molecule, two H atoms must collide and then undergo a radiative transition that removes enough energy to leave the resulting pair of atoms in a bound state. However, two H atoms that are both in the ground state constitute a symmetric system, as does an H$_2$ molecule in its ground state. Because both the initial and final states are symmetric, the system has no dipole moment, and cannot emit dipole radiation. Thus transitions between the bound and unbound states are forbidden. Radiative transitions can in fact occur, but the rate is extremely small, and generally negligible under astrophysical circumstances.\footnote{One can be more precise about this based on an argument given by \citet{gould63a}. Consider the nuclei fixed, and examine the possible electronic state. The total electronic wave function must be anti-symmetric under particle exchange, so either the spatial part of the wave function must be symmetric and the spin part anti-symmetric, or vice versa. In turns out that the ground, bound state is spatially symmetric and spin anti-symmetric, so the two electrons have opposite spin and the total electronic spin is 0, making the state a singlet; we denote this state $\psi_{\uparrow\downarrow}$. The non-bound repulsive state is the opposite: spatially anti-symmetric, spin symmetric, so the total electronic spin is 1 and the state is a triplet; we denote this $\psi_{\uparrow\uparrow}$. The rate of electric dipole transitions between these two states, and thus the rate at which H$_2$ can form from atomic hydrogen in the gas phase, is proportional to the square of the matrix element $\langle\psi_{\uparrow\uparrow}|\mathbf{D}|\psi_{\uparrow\downarrow}\rangle$, where $\mathbf{D}$ is the electric dipole operator. However, $\mathbf{D}$ does not act on the spin parts of the wave functions, and since the spin parts of $\psi_{\uparrow\downarrow}$ and $\psi_{\uparrow\uparrow}$ are orthogonal, the matrix element should vanish. It does not do so exactly only because spin-spin and spin-orbit interactions slightly perturb the system so that the ground eigenstate is not exactly the pure singlet $\psi_{\uparrow\downarrow}$, but instead is a linear combination of mostly $\psi_{\uparrow\downarrow}$ with a small component of the triplet $\psi_{\uparrow\uparrow}$. The relative size of the triplet component is of order the ratio of the spin-orbit and spin-spin interaction energies to the electronic energies, which is $\sim \alpha^2$, where $\alpha \approx 1/137$ is the fine structure constant. Similarly, the repulsive state contains a singlet component of order $\alpha^2$ as well. Thus the bound and repulsive states are not purely orthogonal, but the matrix element is of order $\alpha^2$. Since the transition rate is proportional to the square of this matrix element, it is of order $\alpha^4 \sim 10^{-9}$ compared to allowed transitions.} One can circumvent this limitation by considering either starting or final states there are not symmetric (for example because one of the H atoms is in an excited state, or the final H$_2$ molecule is in an excited state), but this does not lead to a significant rate of gas phase H$_2$ formation either, because the lowest-lying energy states of the H$_2$ molecule are energetic enough that only a negligible fraction of collisions have enough energy to produce them. A third option for gas phase formation is to have three-way collisions, and we will return to this in Chapter \ref{ch:first_stars}. For now we simply remark that, since three-body collisions occur at a rate that depends on the cube of density, at typical interstellar densities they are generally negligible as well.

Due to this limitation, the dominant formation process is instead formation on the surfaces of dust grains. In this case the excess energy released by forming the molecule is transferred into vibrations in the dust grain lattice, and there is no need for forbidden photon emission. The rate of H$_2$ formation by surface catalysis is given by
\begin{equation}
\frac{1}{2} S(T,T_{\rm gr}) \eta(T_{\rm gr}) n_{\rm gr} n_{\rm H} \sigma_{\rm gr} v_{\rm H}.
\end{equation}
Here $S$ is the probability that a hydrogen molecule that hits a dust grain will stick, which is a function of both the gas temperature and the grain temperature. $\eta$ is the probability that a hydrogen atom that sticks will migrate across the grain surface and find another H atom before it is evaporated off the grain surface $n_{\rm gr}$ and $n_{\rm H}$ are the number densities of grains and hydrogen atoms, $\sigma_{\rm gr}$ is the mean cross section for a dust grain, and $v_{\rm H}$ is the thermal velocity of the hydrogen atoms.

The last three factors can be estimated reasonably well from observations of dust extinction and gas velocity dispersions, while the former two have to be determined by laboratory measurements and/or theoretical chemistry calculations. Rather than dive into this extensive literature, we will simply skip directly to the result: for conditions appropriate to cool atomic or molecular regions in the Milky Way, the formation rate is roughly
\begin{equation}
\mathcal{R} n n_{\rm H},
\end{equation}
where $n_{\rm H}$ and $n$ are the number densities of H atoms and H nuclei (in atomic or molecular form), respectively, and $\mathcal{R}\approx 3\times 10^{-17}$ cm$^3$ s$^{-1}$ is the rate coefficient. It may be a factor of a few lower in warmer regions where the sticking probability is reduced. This is for Milky Way dust content. If we go to a galaxy with less dust, the rate coefficient will be reduced proportionally.

The reverse process, destruction, is mostly due to photo-destruction. As with H$_2$ formation, things are somewhat complicated by the symmetry of the H$_2$ system. The binding energy of H$_2$ in the ground state is only 4.5 eV, but this doesn't mean that 4.5 eV photons can destroy it. A reaction of the form
\begin{equation}
{\rm H_2} + h\nu \rightarrow {\rm H}+{\rm H}
\end{equation}
is forbidden by symmetry for exactly the same reason as its inverse, and occurs at a negligibly small rate. Allowed transitions are possible if the H$_2$ molecule is in an excited state that thus asymmetric, or if one of the H atoms is left in an excited state. However, the former is almost never the case at the low temperatures found in molecular clouds, and the latter requires a photon energy of $14.5$ eV. Photons with an energy that high are not generally available, because they can ionize neutral hydrogen and thus have very short mean free paths through the interstellar medium.

Instead, the main H$_2$ destruction process proceeds in two stages. Hydrogen molecules have a series of excited electronic states with energies of $11.2-13.6$ eV (corresponding to $912-1100$ \AA) above the ground state, which produce absorption features known as the Lyman and Werner bands. Since these energies exceed the binding energy of the H$_2$ molecule (4.5 eV), absorptions into them undergo radiative decay to a ground electronic state that can be unbound. This happens roughly 10-15\% of the time, depending on exactly which excited state is decaying. Photons in the LW energy range are produced by hot stars, and the Galaxy is saturated with them, which is why most of the Galaxy's volume is filled with atomic or ionized rather than molecular gas. (There are some galaxies that are mostly molecular, for reasons we will see below.)

Consider a region where the number density of photons of frequency $\nu$ is given by $E^*_{\nu}$. The destruction rate of H$_2$ will then be
\begin{equation}
\int n_{{\rm H}_2} \sigma_{{\rm H}_2,\nu} c E^*_{\nu} f_{{\rm diss,}\nu}\, d\nu,
\end{equation}
where $n_{\rm H_2}$ is the molecular hydrogen number density, $\sigma_{{\rm H}_2,\nu}$ is the absorption cross-section at frequency $\nu$, and $f_{{\rm diss,}\nu}$ is the dissociation probability when a photon of frequency $\nu$ is absorbed. The expression inside the integral is just the number of hydrogen molecule targets times the cross section per target times the number of photons times the relative velocities of the photons and molecules ($=c$) times the probability of dissociation per collision. The integral in frequency goes over the entire LW band, from $912-1100$ \AA.

To understand the circumstances under which H$_2$ can become the dominant form of hydrogen, we can take a simple example. Suppose we have some cloud of gas, which we will treat as a uniform slab, which has a beam of UV radiation shining on its surface. The number density of hydrogen nuclei in the cloud is $n$, and the UV radiation field shining on the surface has a photon number density $E^*_0$. The photon flux is $F^* = c E^*_0$.

As a result of this radiation field, the outer parts of the cloud are atomic hydrogen. However, when a hydrogen molecule absorbs a photon and then re-emits that energy, the energy generally comes out in the form of multiple photons of lower energy, which are no longer able to excite resonant LW transitions. Thus photons are being absorbed as hydrogen forms, and the number of photons penetrating the cloud decreases as one moves further and further into it. Eventually the number of photons drops to near zero, and the gas becomes mostly molecular. This process is known as self-shielding.

We can get a rough estimate of when self-shielding is important by writing down two equations to describe this process. First, let us equate the rates of H$_2$ formation and destruction, i.e., assume the cloud is in chemical equilibrium. (This is generally true because the reaction rates go as $n^2$, so as long as turbulence produces high density regions, there will be places where the reaction occurs quite fast.) This gives
\begin{equation}
\label{eq:chembalance}
n_{\rm H} n \mathcal{R} = \int n_{{\rm H}_2} \sigma_{{\rm H}_2,\nu} c E^*_{\nu} f_{{\rm diss},\nu}\, d\nu
\approx f_{\rm diss}  \int n_{{\rm H}_2} \sigma_{{\rm H}_2,\nu} c E^*_{\nu}\, d\nu.
\end{equation}
In the second step we have made the approximation that $f_{\rm diss}$ is roughly frequency-independent, which is true, since it only varies by factors of less than order unity.

Second, let us write down the equation for photon conservation. This just says that the change in photon number density as we move into the cloud is given by the rate at which collisions with H$_2$ molecules remove photons:
\begin{equation}
\label{eq:dfdx}
\frac{dF^*_{\nu}}{dx}= c \frac{dE^*_{\nu}}{dx} = -n_{H_2} \sigma_{H_2,\nu} c E^*_{\nu}
\end{equation}
In principle there should be a creation term at lower frequencies, representing photons absorbed and re-emitted, but we are only interested in the higher LW frequencies, where there is only photon removal. The term on the right hand side is just the photon absorption rate we calculated above.

Now we can integrate the equation (\ref{eq:dfdx}) over frequency over the LW band. Doing so and dividing by a factor of $c$ gives
\begin{equation}
\frac{dE^*}{dx} = -\int n_{{\rm H}_2} \sigma_{{\rm H}_2,\nu} E^*_{\nu}\, d\nu,
\end{equation}
where $E^*$ is the frequency-integrated photon number density. If we combine this equation with the chemical balance equation (\ref{eq:chembalance}), we obtain
\begin{equation}
\frac{dE^*}{dx} = -\frac{n_{\rm H} n \mathcal{R}}{c f_{\rm diss}}
\end{equation}
This just says that the rate at which photons are taken out of the beam is equal to the recombination rate, increased by a factor of $1/f_{\rm diss}$ because only $\sim 1$ in 10 absorptions actually have to be balanced by a recombination.

If we make the further approximation that the transition from atomic to molecular hydrogen is sharp, so that $n_{\rm H}\approx n$ throughout the atomic layer, and we assume that $\mathcal{R}$ does not vary with position, then the equation is trivial to integrate. At any depth $x$ inside the slab,
\begin{equation}
E^*(x) = E^*_0 - \frac{n^2\mathcal{R}}{c f_{\rm diss}} x.
\end{equation}
The transition to molecular hydrogen occurs where $E^*$ reaches zero, which is at $x_{{\rm H}_2}= c f_{\rm diss} E^*_0 / (n^2 \mathcal{R})$. The total column of atomic hydrogen is
\begin{equation}
N_{\rm H} = nx_{{\rm H}_2} = \frac{c f_{\rm diss} E^*_0}{n\mathcal{R}}
\end{equation}

It is helpful at this point to put in some numbers. In the Milky Way, the observed interstellar UV field is $E^*_0=7.5\times 10^{-4}$ LW photons cm$^{-3}$, and we can take $n=100$ cm$^{-3}$ as a typical number density in a region where molecules might form. Plugging these in with $f_{\rm diss}=0.1$ and $\mathcal{R}=3\times 10^{-17}$ cm$^{-3}$ s$^{-1}$ gives $N_{\rm H} = 7.5\times 10^{20}$, or in terms of mass, a column of $\Sigma=8.4$ $\msun$ pc$^{-2}$. More precise calculations give numbers closer to $2\times 10^{20}$ cm$^{-2}$ for the depth of the shielding layer on one side of a GMC. (Of course a comparable column is required on the other side, too.) Every molecular cloud must be surrounded by an envelope of atomic gas with roughly this column density.

This has important implications. First, this means that molecular clouds with column densities of $100$ $\msun$ pc$^{-2}$ in molecules must have $\sim 10\%$ of their total mass in the form of an atomic shield around them. Second, it explains why most of the Milky Way's ISM in the Solar vicinity is not molecular. In the regions outside of molecular clouds, the mean column density is a bit under $10^{21}$ cm$^{-2}$, so the required shielding column is comparable to the mean column density of the entire atomic disk. Only when the gas clumps together can molecular regions form. This also explains why other galaxies which have higher column densities also have higher molecular fractions. To take an extreme example, the starburst galaxy Arp 220 has a surface density of a few $\times 10^4$ $\msun$ pc$^{-2}$ in its nucleus, and the molecular fraction there is at least 90\%, probably more.

\subsection{Carbon / Oxygen Chemistry}
\label{ssec:cochemistry}

H$_2$ is the dominant species in molecular regions, but it is very hard to observe directly for the reasons discussed in Chapter \ref{ch:obscold} -- the temperatures are too low for it to be excited. Moreover, as we will discuss shortly, H$_2$ is also not the dominant coolant for the same reason. Instead, that role falls to the CO molecule.

Why is CO so important? The main reason is abundances: the most abundant elements in the universe after H and He are O, C, and N, and CO is the simplest (and, under ISM conditions, most energetically favorable) molecule that can be made from them. Moreover, CO can be excited at very low temperatures because its mass is much greater than that of H$_2$, and its dipole moment is weak but non-zero. (A weak dipole moment lowers the energy of radiation emitted, which in turn lowers the temperature needed for excitation.)

Just as in the bulk of the ISM, hydrogen is mostly H, in the bulk of the ISM the oxygen is mostly O and the carbon is mostly C$^+$. It is C$^+$ rather than C because the ionization potential of carbon is less than that of hydrogen, and as a result it tends to be ionized by starlight. So how do we get from C$^+$ and O to CO?

The formation of CO is substantially different than that of H$_2$ in that it is dominated by gas-phase rather than grain-surface reactions. This is because there are no symmetric systems involved, and thus no symmetry barriers to radiation. However, since the temperatures in regions where CO is forming tend to be low, the key processes involve ion-neutral reactions. These are important because the rate at which they occur is to good approximation independent of temperature, while neutral-neutral reactions occur at a rate that declines with temperature as roughly $T^{1/2}$.\footnote{These dependencies are relatively easy to understand. For neutral-neutral reactions, there are no long-distance forces between particles, and thus the rate of collisions is proportional to the mean velocities of the particles involved, which scales as $T^{1/2}$. In contrast, for ion-neutral reactions the ion induces an electric dipole moment in the neutral and then attracts it via Coulomb forces. The slower the particles' relative velocities, the more important is this electric attraction, and this effect cancels out the lower overall rates of encounter caused by lower particle velocities.}

There are two main pathways to CO. One passes through the OH molecule, and involves a reaction chain that looks like
\begin{eqnarray}
{\rm H}_2 + {\rm CR} & \rightarrow & {\rm H}_2^+ + e^{-} + {\rm CR} \\
{\rm H}_2^+ + {\rm H}_2 & \rightarrow & {\rm H}_3^+ + {\rm H} \\
{\rm H}_3^+ + {\rm O} & \rightarrow & {\rm OH}^+ + {\rm H}_2 \\
{\rm OH}^+ + {\rm H}_2 & \rightarrow & {\rm OH}_2^+ + {\rm H} \\
{\rm OH}_2^+ + e^{-} & \rightarrow & {\rm OH} + {\rm H} \\
{\rm C}^+ + {\rm OH} & \rightarrow & {\rm CO}^+ + {\rm H} \\
{\rm CO}^+ + {\rm H}_2 & \rightarrow & {\rm HCO}^+ + {\rm H} \\
{\rm HCO}^+ + e^{-} & \rightarrow & {\rm CO} + {\rm H}.
\end{eqnarray}
Here CR indicates cosmic ray. There are also a number of possible variants (e.g., the OH$_2^+$ could form OH$_3^+$ before proceeding to OH). The second main route is through the CH molecule, where reaction chains tend to follow the general pattern
\begin{eqnarray}
{\rm C}^+ + {\rm H}_2 & \rightarrow & {\rm CH}_2^+ + h\nu \\
{\rm CH}_2^+ + e^{-} & \rightarrow & {\rm CH} + {\rm H} \\
{\rm CH} + {\rm O} & \rightarrow & {\rm CO} + {\rm H}.
\end{eqnarray}
The rate at which the first reaction chain manufactures CO is limited by the supply of cosmic rays that initiate the production of H$_2^+$, while the rate at which the second reaction chain proceeds is limited by the rate of the final neutral-neutral reaction. Which chain dominates depends on the cosmic ray ionization rate, density, temperature, and similar details. Note that both of these reaction chains require the presence of H$_2$. 

CO is destroyed via radiative excitation followed by dissociation in essentially the same manner as H$_2$. The shielding process for CO is slightly different however. As with H$_2$, photons that dissociate CO can be absorbed both by dust grains and by CO molecules. However, due to the much lower abundance of CO compared to H$_2$, the balance between these two processes is quite different than it is for hydrogen, with dust shielding generally the more important of the two. Moreover, there is non-trivial overlap between the resonance lines of CO and those of H$_2$, and thus there can be cross-shielding of CO by H$_2$.

At this point the problem is sufficiently complex that one generally resorts to numerical modeling. The net result is that clouds tend to have a layered structure. In poorly-shielded regions where the FUV has not yet been attenuated, H~\textsc{i} and C$^+$ dominate. Further in, where the FUV has been partly attenuated, H$_2$ and C$^+$ dominate. Finally a transition to H$_2$ and CO as the dominant chemical states occurs at the center. For typical Milky Way conditions, the final transition to a CO-dominated composition occurs once the V-band extinction $A_V$ exceeds $1-2$ mag. This corresponds to a column density of a few $\times 10^{21}$ cm$^{-2}$, or $\sim 20$ $\msun$ pc${-2}$, for Milky Way dust. In comparison, recall that typical GMC column densities are $\sim 10^{22}$ cm$^{-2}$, or $\sim 100$ $\msun$ pc$^{-2}$. This means that there is a layer of gas where the hydrogen is mostly H$_2$ and the carbon is still C$^+$, but it constitutes no more than a few tens of percent of the mass. However, in galaxies with lower dust to gas ratios, the layer where H$_2$ dominates but the carbon is not yet mostly CO can be much larger.

\section{Thermodynamics of Molecular Gas}

Having discussed the chemistry of molecular gas, we now turn to the problem of its thermodynamics. What controls the temperature of molecular gas? We have already seen that observations imply temperatures that are extremely low, $\sim 10$ K or even a bit less. How are such cold temperatures achieved? To answer this question, we must investigate what processes heat and cool the molecular ISM.

\subsection{Heating Processes}
\label{ssec:heatproc}

The dominant heating process in the atomic ISM is the grain photoelectric effect: photons from stars with energies of $\sim 8-13.6$ eV hit dust grains and eject fast electrons via the photoelectric effect. The fast electrons then thermalize and deposit their energy at heat in the gas. The rate per H nucleus at which this process deposits energy can be written approximately as\footnote{For a justification of this statement, and a much more complete description of the photoelectric heating process, see a general interstellar medium textbook such as \citet{tielens05a} or \citet{draine11a}.}
\begin{equation}
\Gamma_{\rm PE} \approx 4.0\times 10^{-26} \chi_{\rm FUV} Z_d' e^{-\tau_d}\mbox{ erg s}^{-1}
\end{equation}
where $\chi_{\rm FUV}$ is the intensity of the far ultraviolet radiation field scaled to its value in the Solar neighborhood, $Z'_d$ is the dust abundance scaled to the Solar neighborhood value, and $\tau_d$ is the dust optical depth to FUV photons. The result is, not surprisingly, proportional to the radiation field strength (and thus the number of photons available for heating), the dust abundance (and thus the number of targets for those photons), and the $e^{-\tau_d}$ factor by which the radiation field is attenuated.

At FUV wavelengths, typical dust opacities are $\kappa_d \approx 500$ cm$^2$ g$^{-1}$, so at a typical molecular cloud surface density $\Sigma\approx 50 - 100$ M$_\odot$ pc$^{-2}$, $\tau_d \approx 5-10$, and thus $e^{-\tau_d} \approx 10^{-3}$. Thus in the interiors of molecular clouds, photoelectric heating is strongly suppressed simply because the FUV photons cannot get in. Typical photoelectric heating rates are therefore of order a few $\times 10^{-29}$ erg s$^{-1}$ per H atom deep in cloud interiors, though they can obviously be much larger at cloud surfaces or in regions with stronger radiation fields.

We must therefore consider another heating process: cosmic rays. The great advantage of cosmic rays over FUV photons is that, because they are relativistic particles, they have much lower interaction cross sections, and thus are able to penetrate into regions where light cannot. The process of cosmic ray heating works as follows. The first step is the interaction of a cosmic ray with an electron, which knocks the electron off a molecule:
\begin{equation}
\mbox{CR}+\mbox{H}_2 \rightarrow \mbox{H}_2^+ +e^- + \mbox{CR}
\end{equation}
The free electron's energy depends only weakly on the CR's energy, and is typically $\sim 30$ eV.

The electron cannot easily transfer its energy to other particles in the gas directly, because its tiny mass guarantees that most collisions are elastic and transfer no energy to the impacted particle. However, the electron also has enough energy to ionize or dissociate other hydrogen molecules, which provides an inelastic reaction that can convert some of its 30 eV to heat. Secondary ionizations do indeed occur, but in this case almost all the energy goes into ionizing the molecule (15.4 eV), and the resulting electron has the same problem as the first one: it cannot effectively transfer energy to the much more massive protons.

Instead, there are a number of other channels that allow electrons to dump their energy into motion of protons, and the problem is deeply messy. The most up to date work on this is \citet{glassgold12a}, and we can very briefly summarize it here. A free electron can turn its energy into heat through three channels. The first is dissociation heating, in which the electron strikes an H$_2$ molecule and dissociates it:
\begin{equation}
e^- + {\rm H}_2 \rightarrow 2{\rm H} + e^{-}.
\end{equation}
In this reaction any excess energy in the electron beyond what is needed to dissociate the molecule (4.5 eV) goes into kinetic energy of the two recoiling hydrogen atoms, and the atoms, since they are massive, can then efficiently share that energy with the rest of the gas. A second pathway is that an electron can hit a hydrogen molecule and excite it without dissociating it. The hydrogen molecule then collides with another hydrogen molecule and collisionally de-excites, and the excess energy again goes into recoil, where it is efficiently shared. The reaction is
\begin{eqnarray}
e^- + {\rm H}_2 & \rightarrow & {\rm H}_2^* + e^{-} \\
{\rm H}_2^* + {\rm H}_2 & \rightarrow & 2 {\rm H}_2.
\end{eqnarray}
Finally, there is chemical heating, in which the H$_2^+$ ion that is created by the cosmic ray undergoes chemical reactions with other molecules that release heat. There are a large number of possible exothermic reaction chains, for example
\begin{eqnarray}
{\rm H}_2^+ + {\rm H}_2 & \rightarrow & {\rm H}_3^+ + {\rm H} \\
{\rm H}_3^+ + {\rm CO} & \rightarrow & {\rm HCO}^+ + {\rm H}_2 \\
{\rm HCO}^+ + e^- & \rightarrow & {\rm CO} + {\rm H}.
\end{eqnarray}
Each of these reactions produces heavy ions recoiling at high speed that can efficiently share their energy via collisions. Computing the total energy release requires summing over all these possible reaction chains, which is why the problem is ugly. The final results is that the energy yield per primary cosmic ray ionization is in the range $\sim 13$ eV under typical molecular cloud conditions, but that it can be several eV higher or lower depending on the local density, electron abundance, and similar variables.

Combining this with the primary ionization rate for cosmic rays in the Milky Way, which is observationally-estimated to be about  $\sim 10^{-16}$ s$^{-1}$ per H nucleus in molecular clouds, this gives a total heating rate per H nucleus
\begin{equation}
\Gamma_{\rm CR} \sim 2\times 10^{-27}\mbox{ erg s}^{-1}.
\end{equation}
The heating rate per unit volume is $\Gamma_{\rm CR} n$, where $n$ is the number density of H nuclei ($=2\times$ the density of H molecules). This is sufficient that, in the interiors of molecular clouds, it generally dominates over the photoelectric heating rate.

\subsection{Cooling Processes}

In molecular clouds there are two main cooling processes: molecular lines and dust radiation. Dust can cool the gas efficiently because dust grains are solids, so they are thermal emitters. However, dust is only able to cool the gas if collisions between dust grains and hydrogen molecules occur often enough to keep them thermally well-coupled. Otherwise the grains cool off, but the gas stays hot. The density at which grains and gas become well-coupled is around $10^4-10^5$ cm$^{-3}$, which is higher than the typical density in a GMC, so we will not consider dust cooling further at this point. We will return to it later in Chapter \ref{ch:protostar_form} when we discuss collapsing objects, where the densities do get high enough for dust cooling to be important.

The remaining cooling process is line emission, and by far the most important molecule for this purpose is CO, for the reasons stated earlier. The physics is fairly simple. CO molecules are excited by inelastic collisions with hydrogen molecules, and such collisions convert kinetic energy to potential energy within the molecule. If the molecule de-excites radiatively, and the resulting photon escapes the cloud, the cloud loses energy and cools.

Let us make a rough attempt to compute the cooling rate via this process. A diatomic molecule like CO can be excited rotationally, vibrationally, or electronically. At the low temperatures found in molecular clouds, usually only the rotational levels are important. These are characterized by an angular momentum quantum number $J$, and each level $J$ has a single allowed radiative transition to level $J-1$. Larger $\Delta J$ transitions are strongly suppressed because they require emission of multiple photons to conserve angular momentum.

Unfortunately the CO cooling rate is quite difficult to calculate, because the lower CO lines are all optically thick. A photon emitted from a CO molecule in the $J=1$ state is likely to be absorbed by another one in the $J=0$ state before it escapes the cloud, and if this happens that emission just moves energy around within the cloud and provides no net cooling. The cooling rate is therefore a complicated function of position within the cloud -- near the surface the photons are much more likely to escape, so the cooling rate is much higher than deep in the interior. The velocity dispersion of the cloud also plays a role, since large velocity dispersions Doppler shift the emission over a wider range of frequencies, reducing the probability that any given photon will be resonantly re-absorbed before escaping.

In practice this means that CO cooling rates usually have to be computed numerically, and will depend on the cloud geometry if we want accuracy to better than a factor of $\sim 2$. However, we can get a rough idea of the cooling rate from some general considerations. The high $J$ levels of CO are optically thin, since there are few CO molecules in the $J-1$ state capable of absorbing them, so photons they emit can escape from anywhere within the cloud. However, the temperatures required to excite these levels are generally high compared to those found in molecular clouds, so there are few molecules in them, and thus the line emission is weak. Moreover, the high $J$ levels also have high critical densities, so they tend to be sub-thermally populated, further weakening the emission.

On other hand, low $J$ levels of CO are the most highly populated, and thus have the highest optical depths. Molecules in these levels produce cooling only if they are within one optical depth the cloud surface. Since this restricts cooling to a small fraction of the cloud volume (typical CO optical depths are many tens for the $1\rightarrow 0$ line), this strongly suppresses cooling.

The net effect of combining the suppression of low $J$ transitions by optical depth effects and of high $J$ transitions by excitation effects is that cooling tends to be dominated a single line produced by the lowest $J$ level for which the line is not optically thick. This line is marginally optically thin, but is kept close to LTE by the interaction of lower levels with the radiation field. Which line this is depends on the column density and velocity dispersion of the cloud, but typical peak $J$ values in Milky Way-like galaxies range from $J=2\rightarrow 1$ to $J=5\rightarrow 4$.

For an optically thin transition of a quantum rotor where the population is in LTE, the rate of energy emission per H nucleus from transitions between angular momentum quantum numbers $J$ and $J-1$ is given by
\begin{eqnarray}
\label{eq:lambdaco}
\Lambda_{J,J-1} & = & x_{\rm em} \frac{(2J+1)e^{-E_J/k_B T}}{Z(T)} A_{J,J-1} (E_J - E_{J-1}) \\
E_J & = & h B J (J+1) \\
A_{J,J-1} & = & \frac{512\pi^4 B^3\mu^2}{3hc^3} \frac{J^4}{2J+1}.
\end{eqnarray}
Here $x_{\rm em}$ is the abundance of the emitting species per H nucleus, $T$ is the gas temperature, $Z(T)$ is the partition function, $A_{J,J-1}$ is the Einstein $A$ coefficient from transitions from state $J$ to state $J-1$, $E_J$ is the energy of state $J$, $B$ is the rotation constant for the emitting molecule, and $\mu$ is the electric dipole moment of the emitting molecule. The first equation is simply the statement that the energy loss rate is given by the abundance of emitters multiplied by the fraction of emitters in the $J$ state in question times the spontaneous emission rate for this state times the energy emitted per transition. Note that there is no explicit density dependence as a result of our assumption that the level with which we are concerned is in LTE. The latter two equations are general results for quantum rotors.

The CO molecule has $B=57$ GHz and $\mu=0.112$ Debye, and at Solar metallicity its abundance in regions where CO dominates the carbon budget is $x_{\rm CO} \approx 1.1\times 10^{-4}$. Plugging in these two values, and evaluating for $J$ in the range $2-5$, typical cooling rates are of order $10^{-27}-10^{-26}$ erg s$^{-3}$ when the temperature is $\sim 10$ K. This matches the heating rate we computed above, and this is why the equilibrium temperatures of molecular clouds are $\sim 10$ K.

\subsection{Implications}

The calculation we have just performed has two critical implications that strongly affect the dynamics of molecular clouds. First, the temperature will be relatively insensitive to variations in the local heating rate. The cosmic ray and photoelectric heating rates are to good approximation temperature-independent, but the cooling rate is extremely temperature sensitive because, for the dominant cooling lines of CO have level energies are large compared to $k_B T$. Equation (\ref{eq:lambdaco}) would in fact seem to suggest that the cooling rate is exponentially sensitive to temperature. In practice the sensitivity is not quite that great, because which $J$ dominates changes with temperature. Nonetheless, numerical calculations still show that $\Lambda_{\rm CO}$ varies with $T$ to a power of $p \sim 2-3$. This means that a factor $f$ increase in the local heating rate will only change the temperature by a factor $\sim f^{1/p}$. Thus we expect molecular clouds to be pretty close to isothermal, except near extremely strong local heating sources.

A second important point is the timescales involved. The gas thermal energy per H nucleus is\footnote{This equation is only approximate because this neglects quantum mechanical effects that are of order unity at these low temperatures. However, since the result we are after here is an order of magnitude one, we will not worry about this corrections.}
\begin{equation}
e \approx \frac{1}{2}\left(\frac{3}{2}k_B T\right) = 10^{-15} \left(\frac{T}{10\mbox{ K}}\right)\mbox{ erg}
\end{equation}
The factor of $1/2$ comes from 2 H nuclei per H$_2$ molecule. The characteristic cooling time is $t_{\rm cool} = e/\Lambda_{\rm CO}$. Suppose we have gas that is mildly out of equilibrium, say $T=20$ K instead of $T=10$ K. The heating and cooling are far out of balance, so we can ignore heating completely compared to cooling. At a cooling rate of $\Lambda_{\rm CO} \sim \mathrm{few} \times 10^{-26}$ erg s$^{-1}$ for 20 K gas (assuming the scaling $\Lambda_{\rm CO} \propto T^{2-3}$ as mentioned above), $t_{\rm cool} \sim 1$ kyr. In contrast, the crossing time for a molecular cloud is $t_{\rm cr} = L/\sigma \sim 10$ Myr for $L=30$ pc and $\sigma = 3$ km s$^{-1}$. The conclusion of this analysis is that radiative effects happen on time scales {\it much} shorter than mechanical ones. Gas that is driven out of thermal equilibrium by any hydrodynamic effect will return to its equilibrium temperature long before any mechanical motions can take place. For this reason, gas in molecular clouds is often approximated as isothermal.

\chapter{Gas Flows and Turbulence}
\label{ch:turbulence}

\marginnote{\textbf{Suggested background reading:}
\begin{itemize}
\item \href{http://adsabs.harvard.edu/abs/2014arXiv1402.0867K}{Krumholz, M.~R. 2014, Phys.~Rep., 539, 49}, section 3.3 \nocite{krumholz14c}
\end{itemize}
\textbf{Suggested literature:}
\begin{itemize}
\item \href{http://adsabs.harvard.edu/abs/2013MNRAS.436.1245F}{Federrath, C. 2013, MNRAS, 436, 1245} \nocite{federrath13b}
\end{itemize}
}

This chapter covers the physics of turbulence in the cold interstellar medium. This will be something of a whirlwind tour, since turbulence is an entire research discipline unto itself. Our goal is to understand the basic statistical techniques used to describe and model interstellar turbulence, so that we will be prepared to apply them in the context of star formation.

\section{Characteristic Numbers for Fluid Flow}

\subsection{The Conservation Equations}

To understand the origins of turbulence, both in the ISM and more generally, we start by examining the equations of fluid dynamics and the characteristic numbers that they define. Although the ISM is magnetized, we will first start with the simpler case of an unmagnetized fluid. Fluids are governed by a series of conservation laws. The most basic one is conservation of mass:
\begin{equation}
\frac{\partial}{\partial t} \rho = -\nabla \cdot (\rho\mathbf{v}).
\end{equation}
This equation asserts that the change in mass density at a fixed point is equal to minus the divergence of density times velocity at that point. Physically, this is very intuitive: density at a point changes at a rate that is simply equal to the rate at which mass flows into or out of an infinitesimal volume around that point.

We can write a similar equation for conservation of momentum:
\begin{equation}
\label{eq:momentum}
\frac{\partial}{\partial t}(\rho \mathbf{v}) = -\nabla \cdot(\rho \mathbf{v v}) - \nabla P + \rho \nu \nabla^2 \mathbf{v}.
\end{equation}
Note that the term $\mathbf{vv}$ here is a tensor product. This is perhaps more clear if we write things out in index notation:
\begin{equation}
\frac{\partial}{\partial t}(\rho v_i) = -\frac{\partial}{\partial x_j} (\rho v_i v_j) - \frac{\partial}{\partial x_i} P
+ \rho \nu \frac{\partial}{\partial x_j}\left(\frac{\partial}{\partial x_j} v_i\right)
\end{equation}
The intuitive meaning of this equation can be understood by examining the terms one by one. The term $\rho\mathbf{v}$ is the density of momentum at a point. The term $\nabla \cdot(\rho \mathbf{v v})$ is, in analogy to the equivalent term in the conservation of mass equation, the rate at which momentum is advected into or out of that point by the flow. The term $\nabla P$ is the rate at which pressure forces acting on the fluid change its momentum. Finally, the last term, $\rho \nu \nabla^2 \mathbf{v}$, is the rate at which viscosity redistributes momentum; the quantity $\nu$ is called the kinematic viscosity.

The last term, the viscosity one, requires a bit more discussion. All the other terms in the momentum equation are completely analogous to Newton's second law for single particles. The viscous term, on the other hand, is unique to fluids, and does not have an analog for single particles. It describes the change in fluid momentum due to the diffusion of momentum from adjacent fluid elements. We can understand this intuitively: a fluid is composed of particles moving with random velocities in addition to their overall coherent velocity. If we pick a particular fluid element to follow, we will notice that these random velocities cause some of the particles that make it up to diffuse across its boundary to the neighboring element, and some particles from the neighboring element to diffuse into the one we are following. The particles that wander across the boundaries of our fluid element carry momentum with them, and this changes the momentum of the element we are following. The result is that momentum diffuses across the fluid, and this momentum diffusion is called viscosity.

Viscosity is interesting and important because it's the only term in the equation that converts coherent, bulk motion into random, disordered motion. That is to say, the viscosity term is the only one that is dissipative, or that causes the fluid entropy to change. 

\subsection{The Reynolds Number and the Mach Number}
\label{ssec:reynolds_mach}

To understand the relative importance of terms in the momentum equation, it is helpful to make order of magnitude estimates of their sizes. Let us consider a system of characteristic size $L$ and characteristic velocity $V$; for a molecular cloud, we might have $L\sim 10$ pc and $V\sim 5$ km s$^{-1}$. The natural time scale for flows in the system is $L/V$, so we expect time derivative terms to be of order the thing being differentiated divided by $L/V$. Similarly, the natural length scale for spatial derivatives is $L$, so we expect spatial derivative terms to be order the quantity being differentiated divided by $L$. If we apply these scalings to the momentum equation, we expect the various terms to scale as follows:
\begin{equation}
\frac{\rho V^2}{L} \sim \frac{\rho V^2}{L} + \frac{\rho c_s^2}{L} + \rho \nu \frac{V}{L^2},
\end{equation}
where $c_s$ is the gas sound speed, and we have written the pressure as $P = \rho c_s^2$. Canceling the common factors, we get
\begin{equation}
1 \sim 1 + \frac{c_s^2}{V^2} + \frac{\nu}{VL}.
\end{equation}

From this exercise, we can derive two dimensionless numbers that are going to control the behavior of the equation. We define the Mach number and the Reynolds number as
\begin{eqnarray}
\mathcal{M} & \sim & \frac{V}{c_s} \\
\mathrm{Re} & \sim & \frac{LV}{\nu}.
\end{eqnarray}
The meanings of these dimensionless numbers are fairly clear from the equations. If $\mathcal{M} \ll 1$, then $c_s^2/V^2 \gg 1$, and this means that the pressure term is important in determining how the fluid evolves. In contrast, if $\mathcal{M} \gg 1$, then the pressure term is unimportant for the behavior of the fluid. In a molecular cloud,
\begin{equation}
c_s  =\sqrt{\frac{k_B T}{\mu m_H}} = 0.18 \left(\frac{T}{10\,\mathrm{K}}\right)^{1/2}\mbox{ km s}^{-1},
\end{equation}
where $\mu =2.33$ is the mean mass per particle in a gas composed of molecular hydrogen and helium in the usual cosmic abundance ratio of 1 He per 10 H atoms. Thus $\mathcal{M} = V/c_s \sim 20$, and we learn that pressure forces are unimportant.

The Reynolds number is a measure of how important viscous forces are. Viscous forces are significant for $\mathrm{Re} \sim 1$ or less, and are unimportant of $\mathrm{Re} \gg 1$. We can think of the Reynolds number as describing a characteristic length scale $L\sim \nu/V$ in the flow. This is the length scale on which diffusion causes the flow to dissipate energy. Larger scale motions are effectively dissipationless, while smaller scales ones are damped out by viscosity.

To estimate the Reynolds number in the molecular ISM, we must know its viscosity. For an ideal gas, the kinematic viscosity is $\nu=2\overline{u}\lambda$, where $\overline{u}$ is the RMS molecular speed (which is of order $c_s$) and $\lambda$ is the particle mean free-path. The mean free path is of order the inverse of cross-section times density, $\lambda \sim 1/(\sigma n) \sim [(1\mbox{ nm})^2 (100\mbox { cm}^{-3})]^{-1}\sim 10^{12}$ cm. Plugging this in then gives 
$\nu \sim 10^{16}$ cm$^2$ s$^{-1}$ and $\mathrm{Re} \sim 10^9$. Viscous forces are clearly unimportant in molecular clouds.

\begin{figure}
\includegraphics[width=\linewidth]{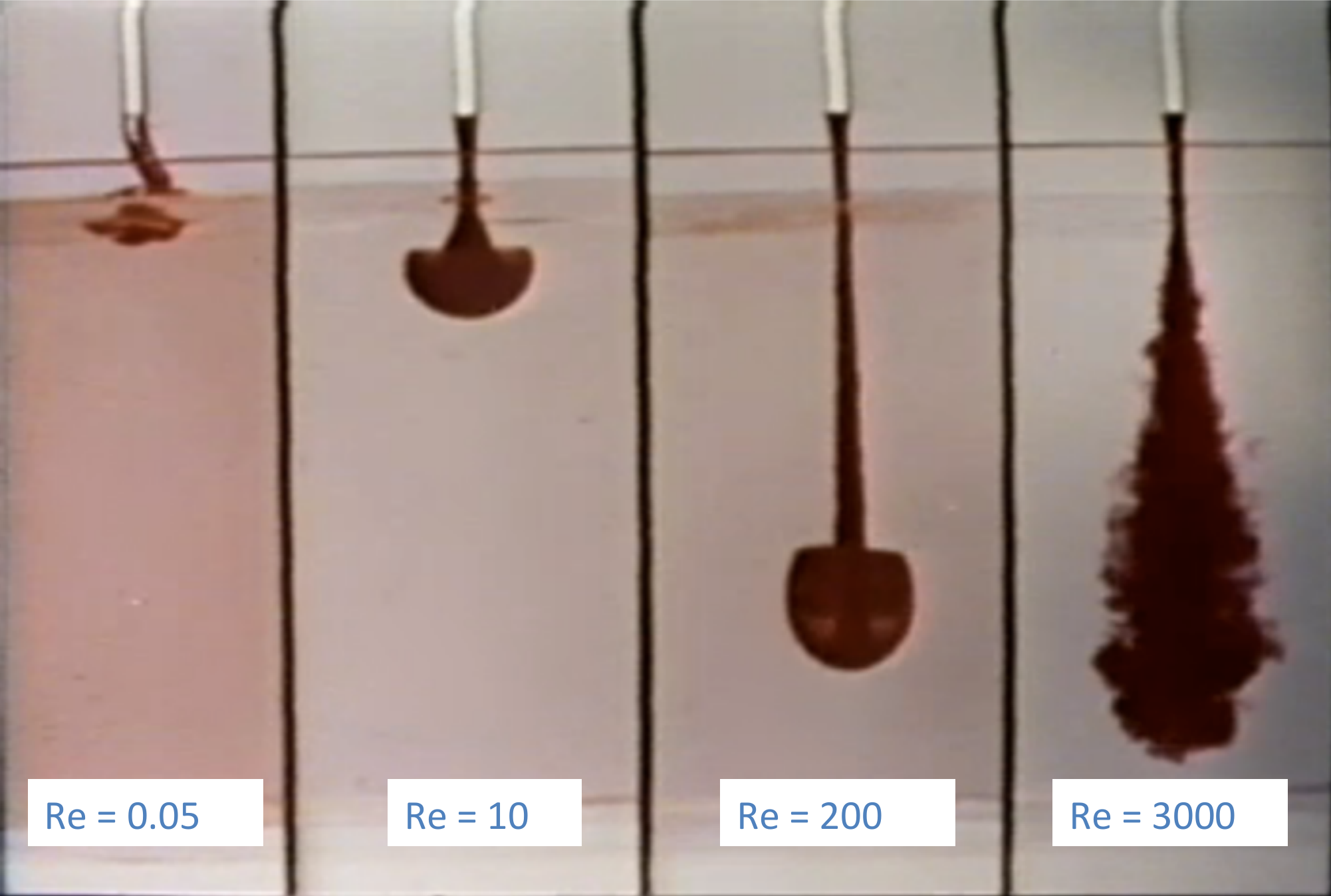}
\caption[Comparison of flows at varying Reynolds numbers]{
\label{fig:reynoldsnum_nsf}
Flows at varying Reynolds number Re. In each panel, a fluid that has been dyed red is injected from the top into the clear fluid on the bottom. The fluids are a glycerin-water mixture, for which the viscosity can be changed by altering the glycerin to water ratio. By changing the viscosity and the injection speed, it is possible to alter the Reynolds number of the injected flow. The frames show how the flow develops as the Reynolds number is varied. This image is a still from the \href{https://www.youtube.com/playlist?list=PL0EC6527BE871ABA3}{National Committee for Fluid Mechanics Film Series}  \citep{taylor64a}, which, once you get past the distinctly 1960s production values, are a wonderful resource for everything related to fluids.
}
\end{figure}

The extremely large value of the Reynolds number immediately yields a critical conclusion: molecular clouds must be highly turbulent, because flows with $\mathrm{Re}$ of more than $\sim 10^3-10^4$ invariable are. Figure \ref{fig:reynoldsnum_nsf} illustrates this graphically from laboratory experiments.

\section{Modeling Turbulence}

We have remarkably little understanding of how turbulence actually works. However, we have developed some simple models and tools to describe it, and we will next explore those.

\subsection{Velocity Statistics}

One quantity of interest in a turbulent medium is the structure of the velocity field. How does the velocity change from point to point?  In a turbulent medium velocity fluctuates in time and space, and so the best way to proceed is to study those fluctuations statistically. Many statistical tools exist to characterize turbulent motions, and many are used in astrophysics, but we will stick to a few of the simpler ones. We will also make two simplifying assumptions. First we assume that the turbulence is homogenous, in the sense that the turbulent motions vary randomly but not systematically, with position in the fluid. Second, we assume that it is isotropic, so that turbulent motions do not have a preferred directions. Neither of these are likely to be strictly true in a molecular cloud, particularly the second, since large-scale magnetic fields provide a preferred direction, but we will start with these assumptions and relax them later.

Let $\mathbf{v}(\mathbf{x})$ be the velocity at position $\mathbf{x}$ within some volume of interest $V$. To characterize how this varies with position, we define the autocorrelation function
\begin{equation}
A(\mathbf{r})\equiv \frac{1}{V} \int \mathbf{v}(\mathbf{x})\cdot \mathbf{v}(\mathbf{x}+\mathbf{r}) \, d\mathbf{x} \equiv \langle \vecv(\vecx) \cdot \vecv(\vecx+\vecr)\rangle,
\end{equation}
where the angle brackets indicate an average over all positions $\mathbf{x}$. Here, $A(0) = \langle|\vecv|^2\rangle$ is just the mean square velocity in the fluid. If the velocity field is isotropic, then clearly $A(\mathbf{r})$ cannot depend on the direction, and thus must depend only on $r=|\mathbf{r}|$. Thus $A(r)$ tells us how similar or different the velocities are at points separated by some distance $r$.

It is often more convenient to think about this in Fourier space than in real space, so rather than the autocorrelation function we often instead think about its Fourier transform. We define the Fourier transform of the velocity field in the usual way, i.e.,
\begin{equation}
\tilde{\vecv}(\veck) = \frac{1}{(2\pi)^{3/2}}\int \vecv(\vecx) e^{-i\veck\cdot\vecx} \,d\vecx.
\end{equation}
We then define the power spectrum
\begin{equation}
\Psi(\veck) \equiv | \tilde{\vecv}(\veck)|^2.
\end{equation}

Again, for isotropic turbulence, the power spectrum depends only on the magnitude of the wave number, $k=|\veck|$, not its direction, so it is more common to talk about the power per unit radius in $k$-space,
\begin{equation}
P(k) = 4\pi k^2 \Psi(k).
\end{equation}
This is just the total power integrated over some shell from $k$ to $k+dk$ in $k$-space. Note that Parseval's theorem tells us that
\begin{equation}
\int P(k)\, dk = \int | \tilde{\vecv}(\veck)|^2 \, d\veck = \int \vecv(\vecx)^2 \, d\vecx,
\end{equation}
i.e., the integral of the power spectral density over all wavenumbers is equal to the integral of the square of the velocity over all space, so for a flow with constant density (an incompressible flow) the integral of the power spectrum just tells us how much kinetic energy per unit mass there is in the flow. The Wiener-Khinchin theorem also tells us that $P(\veck)$ is just the Fourier transform of the autocorrelation function,
\begin{equation}
\Psi(\veck) = \frac{1}{(2\pi)^{3/2}} \int A(\vecr) e^{-i\veck\cdot\vecr} \,d\vecr.
\end{equation}

The power spectrum at a wavenumber $k$ then just tells us what fraction of the total power is in motions at that wavenumber, i.e., on that characteristic length scale. The power spectrum is another way of looking at the spatial scaling of turbulence. It tells us how much power there is in turbulent motions as a function of wavenumber $k=2\pi/\lambda$. A power spectrum that peaks at low $k$ means that most of the turbulent power is in large-scale motions, since small $k$ corresponds to large $\lambda$. Conversely, a power spectrum that peaks at high $k$ means that most of the power is in small-scale motions.

The power spectrum also tells us about the how the velocity dispersion will vary when it is measured over a region of some characteristic size. Suppose we consider a volume of size $\ell$, and measure the velocity dispersion $\sigma_v(\ell)$ within it. Further suppose that the power spectrum is described by a power law $P(k)\propto k^{-n}$. The total kinetic energy per unit mass within the region is, up to factors of order unity,
\begin{equation}
\mathrm{KE} \sim \sigma_v(\ell)^2,
\end{equation}
but we can also write the kinetic energy per unit mass in terms of the power spectrum, integrating over those modes that are small enough to fit within the volume under consideration:
\begin{equation}
\mathrm{KE} \sim \int_{2\pi/\ell}^\infty P(k) \, dk \propto \ell^{n-1}.
\end{equation}
It therefore follows immediately that
\begin{equation}
\label{eq:vdisp}
\sigma_v = c_s \left(\frac{\ell}{\ell_s}\right)^{(n-1)/2},
\end{equation}
where we have normalized the relationship by defining the sonic scale $\ell_s$ as the size of a region within which the velocity dispersion is equal to the thermal sound speed of the gas.

\subsection{The Kolmogorov Model and Turbulent Cascades}

The closest thing we have to a model of turbulence is in the case of subsonic, hydrodynamic turbulence; the basic theory for that goes back to \citet{kolmogorov41a}.\footnote{An English translation of \citet{kolmogorov41a} (which is in Russian) can be found in \citet{kolmogorov91a}.} Real interstellar clouds are neither subsonic nor hydrodynamic (since they are strongly magnetized, as we discuss in Chapter \ref{ch:magnetic})), but this theory is still useful for understanding how turbulence works. Kolmogorov's theory of turbulence begins with the realization that turbulence is a phenomenon that occurs when Re is large, so that there is a large range of scales where dissipation is unimportant. It is possible to show by Fourier transforming Equation (\ref{eq:momentum}) that for incompressible motion transfer of energy can only occur between adjacent wavenumbers. Energy at a length scale $k$ cannot be transferred directly to some scale $k' \ll k$. Instead, it must cascade through intermediate scales between $k$ and $k'$.

This gives a simple picture of how energy dissipates in fluids. Energy is injected into a system on some large scale that is dissipationless, and it cascades down to smaller scales until it reaches a small enough scale that $\mbox{Re}\sim 1$, at which point dissipation becomes significant. In this picture, if the turbulence is in statistical equilibrium, such that is neither getting stronger or weaker, the energy at some scale $k$ should depend only on $k$ and on the rate of injection or dissipation (which are equal) $\psi$.

This allows us to make the following clever dimensional argument: $k$ has units of $1/L$, i.e., one over length. The power spectrum $P(k)$ has units of energy per unit mass per unit radius in $k$-space. The energy per unit mass is like a velocity squared, so it has units $L^2/T^2$, and this is divided by $k$, so $P(k)$ has units of $L^3/T^2$. The injection and dissipation rates $\psi$ have units of energy per unit mass per unit time, which is a velocity squared divided by a time, or $L^2/T^3$.

Since $P(k)$ is a function only of $k$ and $\psi$, we can write $P(k) = C k^\alpha \psi^\beta$ for some dimensionless constant $C$. Then by dimensional analysis we have
\begin{eqnarray}
\frac{L^3}{T^2} & \sim & L^{-\alpha} \left(\frac{L^2}{T^3}\right)^\beta \\
L^3 & \sim & L^{-\alpha+2\beta} \\
T^{-2} & \sim & T^{-3\beta}\\
\beta & = & \frac{2}{3} \\
\alpha & = & 2\beta - 3 = -\frac{5}{3}
\end{eqnarray}

This immediately tell us three critical things. First, the power in the flow varies with energy injection rate to the $2/3$ power. Second, this power is distributed such that the power at a given wavenumber $k$ varies as $k^{-5/3}$. This means that most of the power is in the largest scale motions, since power diminishes as $k$ increases. Third, if we now take this spectral slope and use it to derive the scale-dependent velocity dispersion from equation (\ref{eq:vdisp}), we find that $\sigma_v \propto \ell^{1/3}$, i.e., velocity dispersion increase with size scale as size to the $1/3$ power. This is an example of what is known in observational astronomy as a linewidth-size relation -- linewidth because the observational diagnostic we use to characterize velocity dispersion is the width of a line. This relationship tells us that larger regions should have larger linewidths, with the linewidth scaling as the 1/3 power of size in the subsonic regime.

\begin{figure}
\includegraphics[width=\linewidth]{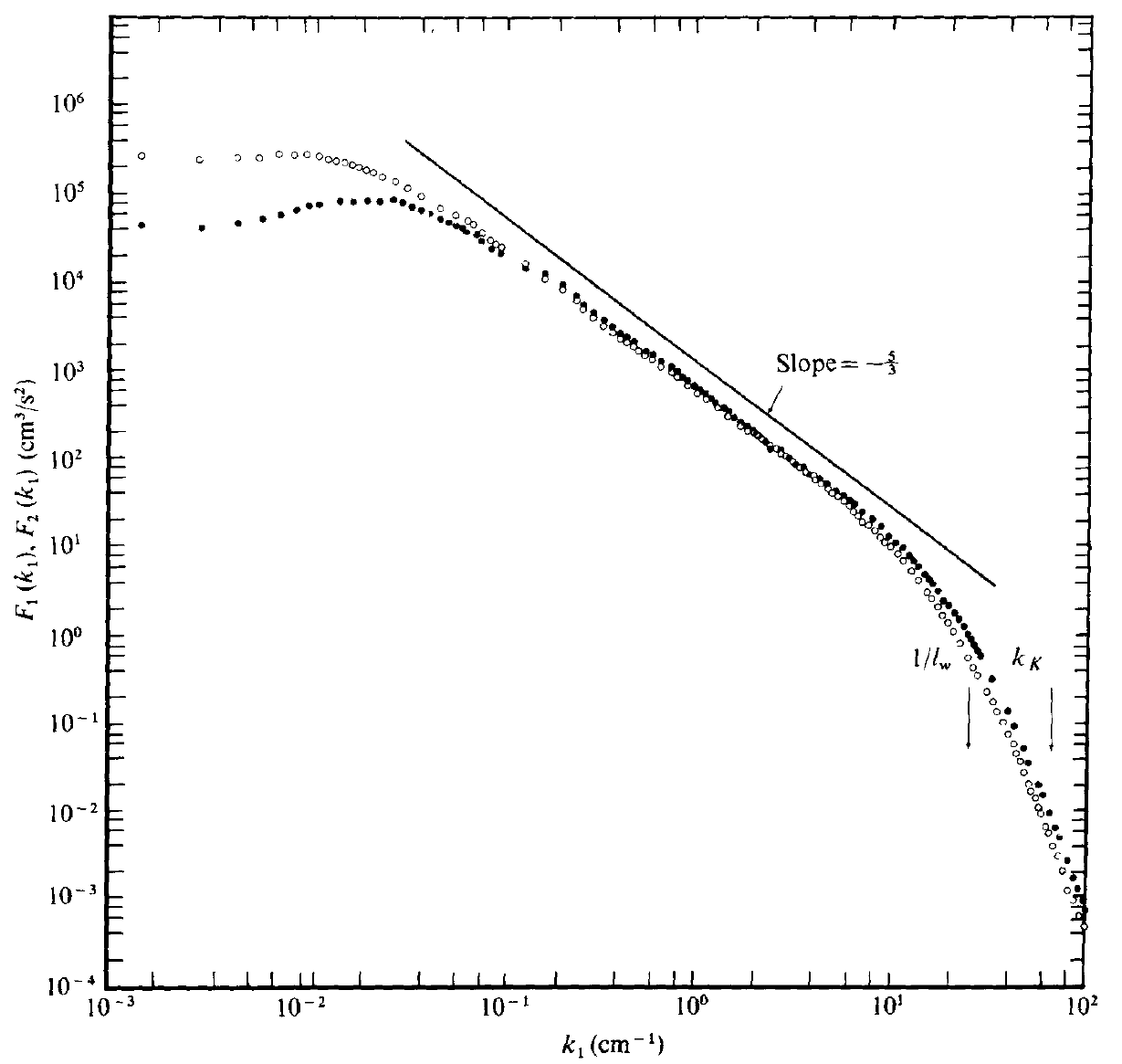}
\caption[Experimental power spectra for Kolmogorov turbulence]{
\label{fig:kolmogorov}
An experimentally-measured power spectrum for turbulence generated by an air jet. The $x$ axis is the wavenumber, and the open and filled points show the velocity power spectrum for the velocity components parallel and transverse to the stream, respectively. Credit: \citeauthor{champagne78a}, J.~Fluid.~Mech., 86, 67-108, 1978, reproduced with permission.
}
\end{figure}

The subsonic regime can be tested experimentally on Earth, and Kolmogorov's model provides an excellent fit to observations. Figure \ref{fig:kolmogorov} shows one example.

\section{Supersonic Turbulence}

\subsection{Velocity Statistics}

We have seen that real interstellar clouds not only have $\mathrm{Re} \gg 1$, they also have $\mathcal{M} \gg 1$, and so the flows within them are supersonic. This means that pressure is unimportant on size scales $L \gg \ell_s$. Since viscosity is also unimportant on large scales (since $\mbox{Re} \gg 1$), this means that gas tends to move ballistically on large scales. On small scales this will produce very sharp gradients in velocity, since fast-moving volumes of fluid will simply overtake slower ones. Since the viscosity term gets more important on smaller scales, the viscosity term will eventually stop the fluid from moving ballistically. In practice this means the formation of shocks -- regions where the flow velocity changes very rapidly, on a size scale determined by the viscosity.\footnote{In real interstellar clouds the relevant viscosity is the magnetic one, as we shall see in Chapter \ref{ch:magnetic}.}

\begin{marginfigure}
\includegraphics[width=\linewidth]{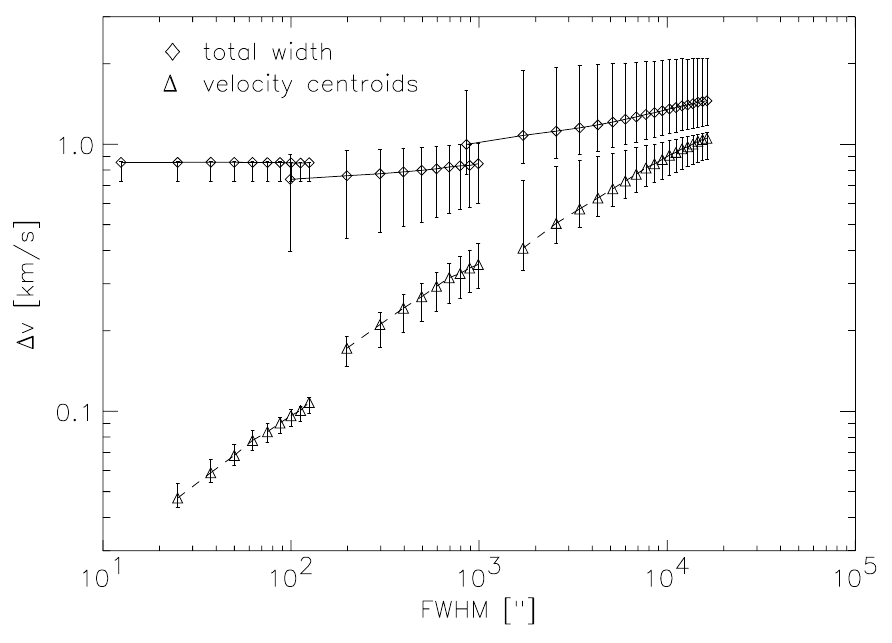}
\caption[Linewidth versus size in the Polaris Flare cloud]{
\label{fig:polaris}
Linewidth versus size in the Polaris Flare Cloud obtained from CO observations. Diamonds show the total measured velocity width within apertures of the size indicated on the $x$ axis, while triangles show the dispersion obtained by taking the centroid velocity in each pixel and measuring the dispersion of centroids. The three sets of points joined by lines represent measurements from three separate telescopes. Credit: \citeauthor{ossenkopf02a}, A\&A, 390, 307, 2002, reproduced by permission \copyright\,ESO.
}
\end{marginfigure}

We expect that the velocity field that results in this case will look like a series of step functions. The power spectrum of a step function is a power law $P(k)\propto k^{-2}$. One can establish this easily from direct calculation. Let us zoom in on the region around a shock, so that the change in velocity on either side of the shock is small. The Fourier transform of $v$ in 1D is
\begin{equation}
\tilde{v}(k) = \frac{1}{\sqrt{2\pi}} \int v(x) e^{-i kx}\, dx
\end{equation}
The integral of the periodic function $e^{-ikx}$ vanishes for all periods in the regions where $v$ is constant. It is non-zero only in the period that includes the shock. The amplitude of $\int v(x) e^{-i kx}\, dx$ during that period is simply proportional to the length of the period, i.e., to $1/k$. Thus, $\tilde{v}(k)\propto 1/k$. It then follows that $P(k)\propto k^{-2}$ for a single shock. An isotropic system of overlapping shocks should therefore also look approximately like a power law of similar slope. This gives a velocity dispersion versus size scale $\sigma_v\propto \ell^{1/2}$, and this is exactly what is observed. Figure \ref{fig:polaris} shows an example.

Note that, although the power spectrum is only slightly different than that of subsonic turbulence ($-2$ versus about $-5/3$), there is really an important fundamental difference between the two regimes. Most basically, in Kolmogorov turbulence decay of energy happens via a cascade from large to small scales, until a dissipative scale is reached. In the supersonic case, on the other hand, the decay of energy is via the formation of shocks, and as we have just seen a single shock generates a power spectrum $\propto k^{-2}$, i.e., it non-locally couples many scales. Thus, in supersonic turbulence there is no locality in $k$-space. All scales are coupled at shocks.

\subsection{Density Statistics}

In subsonic flows the pressure force is dominant. Thus if the gas is isothermal, then the density stays nearly constant -- any density inhomogeneities are ironed out immediately by the strong pressure forces. In supersonic turbulence, on the other hand, the flow is highly compressible. It is therefore of great interest to ask about the statistics of the density field.

Numerical experiments and empirical arguments (but not fully rigorous proofs) indicate that the density field for a supersonically-turbulent, isothermal medium is well-described by a lognormal distribution,
\begin{equation}
\label{eq:denpdf}
p(s) = \frac{1}{\sqrt{2\pi \sigma_s^2}} \exp\left[-\frac{(s-s_0)^2}{2\sigma_s^2}\right],
\end{equation}
where $s=\ln(\rho/\overline{\rho})$ is the log of the density normalized to the mean density $\overline{\rho}$. This distribution describes the probability that the density at a randomly chosen point will be such that $\ln(\rho/\overline{\rho})$ is in the range from $s$ to $s+ds$. The median of the distribution $s_0$ and the dispersion $\sigma_s$ must be related to one another, because we require that
\begin{equation}
\overline{\rho} = \int p(s) \rho \, ds.
\end{equation}
With a bit of algebra, one can show that this equation is satisfied if and only if
\begin{equation}
s_0 = -\sigma_s^2/2.
\end{equation}

\begin{marginfigure}
\includegraphics[width=\linewidth]{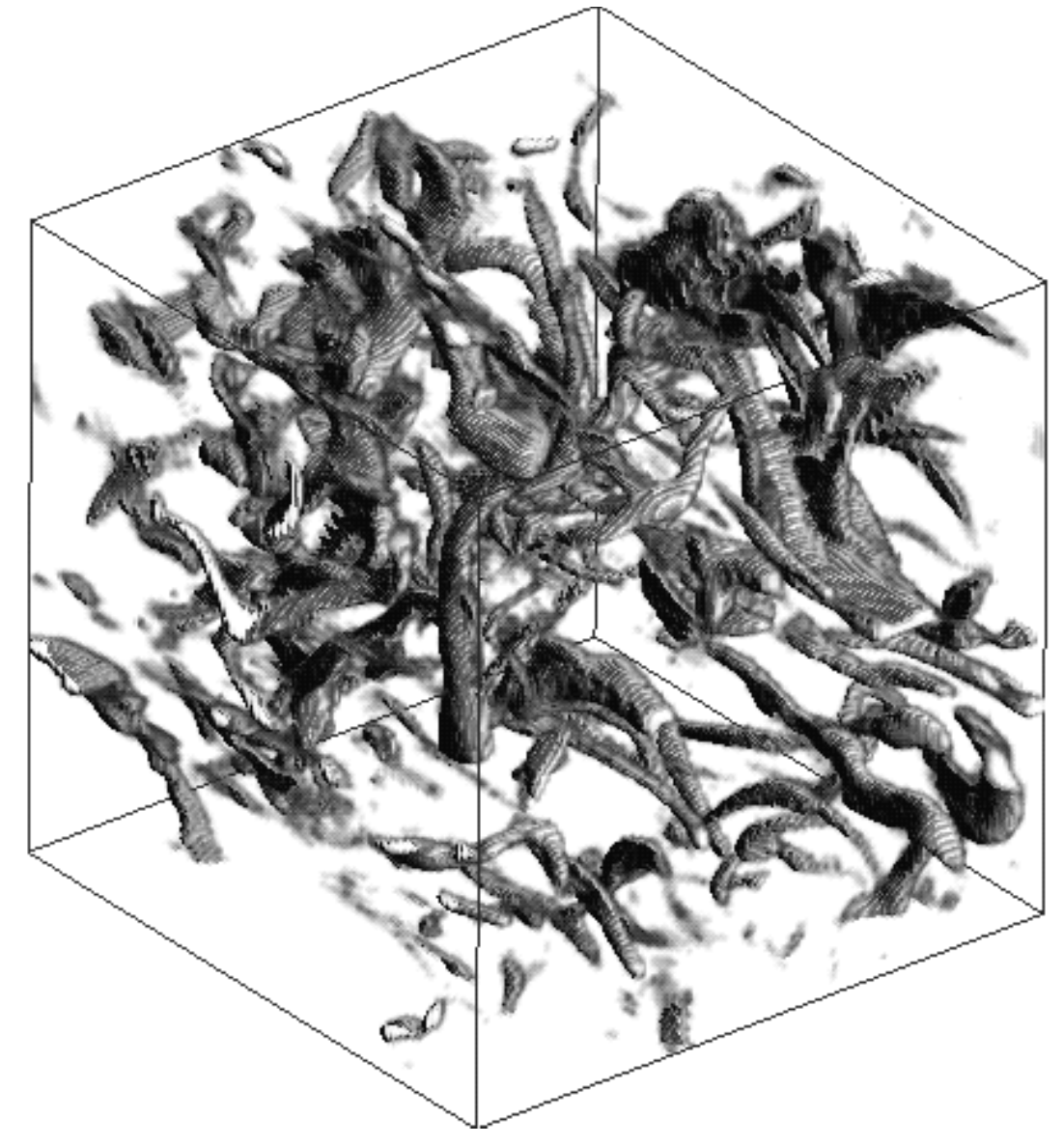}
\caption[Volume rendering of the density field for supersonic turbulence]{
\label{fig:turbrender}
Volume rendering of the density field in a simulation of supersonic turbulence. The surfaces shown are isosurfaces of density. Credit: \citet{padoan99a}, \copyright\,AAS. Reproduced with permission.
}
\end{marginfigure}

Instead of computing the probability that a randomly chosen point in space will have a particular density, we can also compute the probability that a randomly chosen mass element will have a particular density. This more or less amounts to a simple change of variables. Consider some volume of interest $V$, and examine all the material with density such that $\ln(\rho/\overline{\rho})$ is in the range from $s$ to $s+ds$. This material occupies a volume $dV = p(s) V$, and thus must have a mass
\begin{eqnarray}
dM & = & \rho p(s) \, dV \\
& = & \overline{\rho} e^s \cdot \frac{1}{\sqrt{2\pi \sigma_s^2}} \exp\left[-\frac{(s-s_0)^2}{2\sigma_s^2}\right]\,  dV \\
& = & \overline{\rho} \frac{1}{\sqrt{2\pi \sigma_s^2}} \exp\left[-\frac{(s+s_0)^2}{2\sigma_s^2}\right]\,  dV
\end{eqnarray}
It immediately follows that the mass PDF is simply
\begin{equation}
p_M(s) = \frac{1}{\sqrt{2\pi \sigma_s^2}} \exp\left[-\frac{(s+s_0)^2}{2\sigma_s^2}\right],
\end{equation}
i.e., exactly the same as the volume PDF but with the peak moved from $-s_0$ to $+s_0$. Physically, the meaning of these shifts is that the typical volume element in a supersonic turbulent field is at a density lower than the mean, because much of the mass is collected into shocks. The typical mass element lives in one of these shocked regions, and thus is at higher-than-average density. Figure \ref{fig:turbrender} shows an example of the density distribution produced in a numerical simulation of supersonic turbulence.

The lognormal functional form is not too surprising, given the central limit theorem. Supersonic turbulence consists of an alternating series of shocks, which cause the density to be multiplied by some factor, and supersonic rarefactions, which cause it to drop by some factor. The result of multiplying a lot of positive density increases by a lot of negative density drops at random tends to produce a normal distribution in the multiplicative factor, and thus a lognormal distribution in the density.

This argument does not, however, tell us about the dispersion of densities, which must be determined empirically from numerical simulations. The general result of these simulations \citep[e.g.,][]{federrath13b} is that
\begin{equation}
\sigma_s^2 \approx \ln \left(1 + b^2 \mathcal{M}^2 \frac{\beta_0}{\beta_0+1}\right),
\end{equation}
where the factor $b$ is a number in the range $1/3-1$ that depends on how compressive versus solenoidal the velocity field is, and $\beta_0$ is the ratio of thermal to magnetic pressure at the mean density and magnetic field strength, something we will discuss further in Chapter \ref{ch:magnetic}.

In addition to the density PDF, there are higher order statistics describing correlations of the density field from point to point. We will defer a discussion of these until we get to models of the IMF in Chapter \ref{ch:imf_th}, where they play a major role.

\problemset

\begin{enumerate}

\item \textbf{Molecular Tracers.}\\
Here we will derive a definition of the critical density, and use it to compute some critical densities for important molecular transitions. For the purposes of this problem, you will need to know some basic parameters (such as energy levels and Einstein coefficients) of common interstellar molecules. You can obtain these from the Leiden Atomic and Molecular Database (LAMDA, \url{http://www.strw.leidenuniv.nl/~moldata}). It is also worth taking a quick look through the associated paper \citep{schoier05a}\footnote{\href{http://adsabs.harvard.edu/abs/2005A\%26A...432..369S}{Sch\"{o}ier {\it et.~al}, 2005, A\&A, 432, 369}} so you get a feel for where these numbers come from.
\begin{enumerate}
\item Consider an excited state $i$ of some molecule, and let $A_{ij}$ and $k_{ij}$ be the Einstein $A$ coefficient and the collision rate, respectively, for transitions from state $i$ to state $j$. Write down expressions for the rates of spontaneous radiative and collisional de-excitations out of state $i$ in a gas where the number density of collision partners is $n$.
\item We define the critical density $n_{\rm crit}$ of a state as the density for which the spontaneous radiative and collisional de-excitation rates are equal.\footnote{There is some ambiguity in this definition. Some people define the critical density as the density for which the rate of radiative de-excitation equals the rate of {\it all} collisional transitions out of a state, not just the rate of collisional de-excitations out of it. In practice this usually makes little difference.} Using your answer to the previous part, derive an expression for $n_{\rm crit}$ in terms of the Einstein coefficient and collision rates for the state. 
\item When a state has a single downward transition that is far more common than any other one, as is the case for example for the rotational excitation levels of CO, it is common to refer to the critical density of the upper state of the transition as the critical density of the line. Compute critical densities for the following lines: CO $J=1\rightarrow 0$, CO $J=3\rightarrow 2$, CO $J=5\rightarrow 4$, and HCN $J=1\rightarrow 0$, using H$_2$ as a collision partner. Perform your calculation for the most common isotopes: $^{12}$C, $^{16}$O, and $^{14}$N. Assume the gas temperature is 10 K, the H$_2$ molecules are all para-H$_2$, and neglect hyperfine splitting.
\item Consider a molecular cloud in which the volume-averaged density is $n=100$ cm$^{-3}$. Assuming the cloud has a lognormal density distribution as given by equation (\ref{eq:denpdf}), with a dispersion $\sigma_s^2 = 5.0$, compute the fraction of the cloud mass that is denser than the critical density for each of these transitions. Which transitions are good tracers of the bulk of the mass in a cloud? Which are good tracers of the denser, and thus presumably more actively star-forming, parts of the cloud?
\end{enumerate} 

\vspace{0.2in}

\item \textbf{Infrared Luminosity as a Star Formation Rate Tracer.}\\
We use a variety of indirect indicators to measure the star formation rate in galaxies, and one of the most common is to measure the galaxy's infrared luminosity. The underlying assumptions behind this method are that (1) most of the total radiant output in the galaxy comes from young, recently formed stars, and (2) that in a sufficiently dusty galaxy most of the starlight will be absorbed by dust grains within the galaxy and then re-radiated in the infrared. We will explore how well this conversion works using the popular stellar population synthesis package Starburst99 \citep{leitherer99a, vazquez05a}, \url{http://www.stsci.edu/~science/starburst99}.
\begin{enumerate}
\item Once you have read enough of the papers to figure out what Starburst99 does, use it with the default parameters to compute the total luminosity of a stellar population in which star formation occurs continuously at a fixed rate $\dot{M}_*$. What is the ratio of $L_{\rm tot}/\dot{M}_*$ after 10 Myr? After 100 Myr? After 1 Gyr? Compare these ratios to the conversion factor between $L_{\rm TIR}$ and $\dot{M}_*$ given in Table 1 of \citet{kennicutt12a}\footnote{\href{http://adsabs.harvard.edu/abs/2012ARA\%26A..50..531K}{Kennicutt \& Evans, 2012, ARA\&A, 50, 531}}.
\item Plot $L_{\rm tot}/\dot{M}_*$ as a function of time for this population. Based on this plot, how old does a stellar population have to be before $L_{\rm TIR}$ becomes a good tracer of the total star formation rate?
\item Try making the IMF slightly top-heavy, by removing all stars below $0.5$ $\msun$. How much does the luminosity change for a fixed star formation rate? What do you infer from this about how sensitive this technique is to assumptions about the form of the IMF?
\end{enumerate}

\end{enumerate}

\chapter{Magnetic Fields and Magnetized Turbulence}
\label{ch:magnetic}

\marginnote{
\textbf{Suggested background reading:}
\begin{itemize} 
\item \href{http://adsabs.harvard.edu/abs/2012ARA\%26A..50...29C}{Crutcher, R.~M. 2012, ARA\&A, 50, 29} \nocite{crutcher12a}
\end{itemize}
\textbf{Suggested literature:}
\begin{itemize}
\item \href{http://adsabs.harvard.edu/abs/2008ApJ...684..380L}{Li, P.~S., McKee, C.~F., Klein, R.~I., \& Fisher, R.~T. 2008, ApJ, 684, 380} \nocite{li08a}
\end{itemize}
}

In our treatment of fluid flow and turbulence in Chapter \ref{ch:turbulence}, we concentrated on the hydrodynamic case. However, real star-forming clouds are highly magnetized. We therefore devote this chapter to the question of how magnetic fields change the nature of molecular cloud fluid flow.

\section{Observing Magnetic Fields}

\subsection{Zeeman Measurements}

We begin by reviewing the observational evidence for the existence and strength of magnetic fields in interstellar clouds. There are several methods that can be used to measure magnetic fields, but the most direct is the Zeeman effect. The Zeeman effect is a slight shift in the energy levels of an atom or molecule in the presence of a magnetic field. Ordinarily the energies of a level depend only the direction of the electron spin relative to its orbital angular momentum vector, not on the direction of the net angular momentum vector. However, in the presence of an external magnetic field, states with different orientations of the net angular momentum vector of the atom have slightly different energies due to the interaction of the electron magnetic moment with the external field. This causes levels that are normally degenerate to split apart slightly in energy. As a result, transitions into or out of these levels, which would normally produce a single spectral line, instead produce a series of separate lines at slightly different frequencies.

For the molecules with which we are concerned, the level is normally split into three sublevels -- one at slightly higher frequency than the unperturbed line, one at slightly lower frequency, and one at the same frequency. The strength of this splitting varies depending on the electronic configuration of the atom or molecule in question. For OH, for example, the splitting is $Z=0.98$ Hz/$\mu$G, where the parameter $Z$ is called the Zeeman sensitivity, and the shift is $\Delta \nu = BZ$, where $B$ is the magnetic field strength. Zeeman measurements target molecules where $Z$ is as large as possible, and these are generally molecules or atoms that have an unpaired electron in their outer shell. Examples include atomic hydrogen, OH, CN, CH, CCS, SO, and O$_2$.

\begin{marginfigure}
\includegraphics[width=\linewidth]{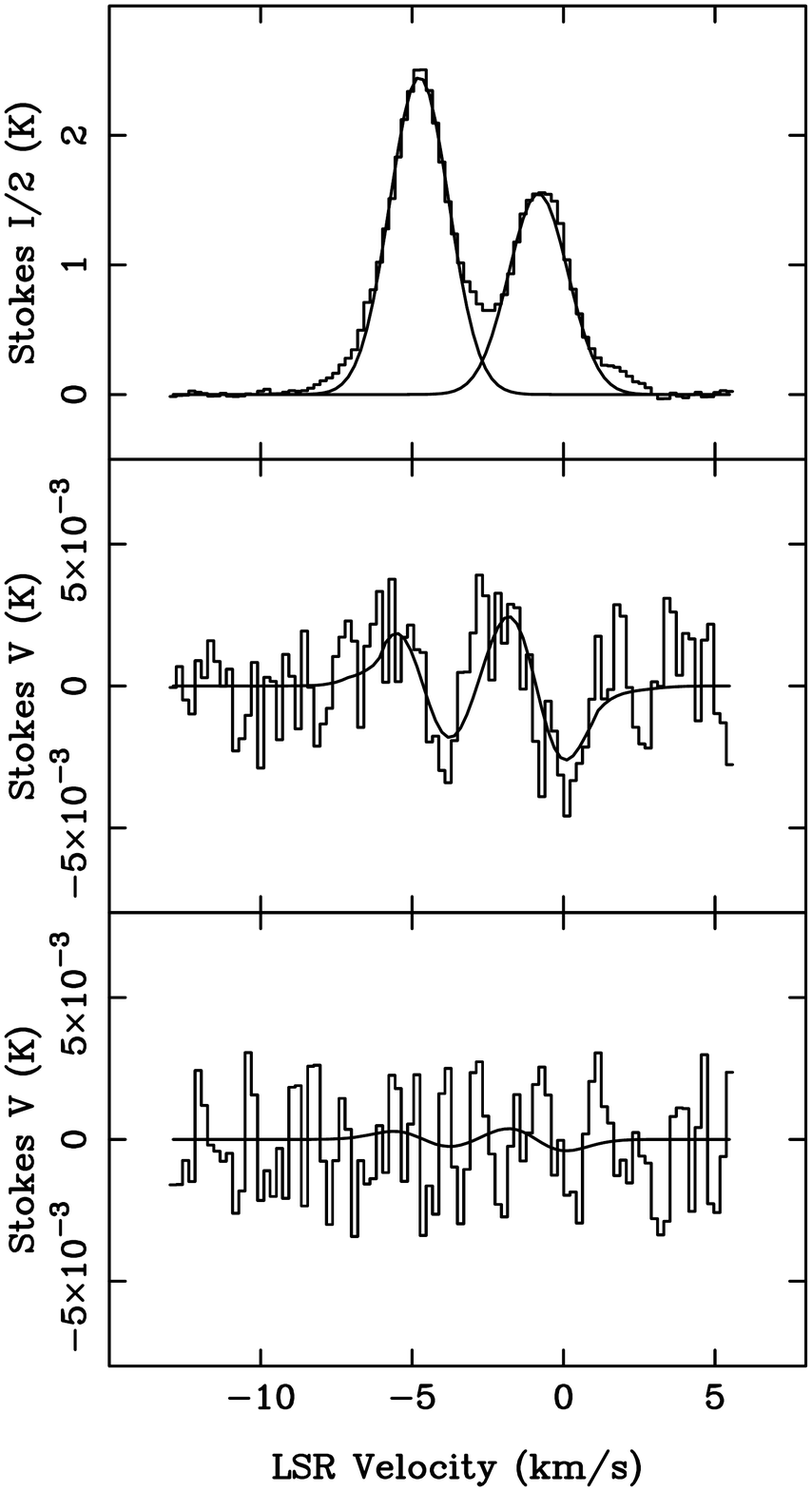}
\caption[Sample Zeeman detection of a magnetic field]{
\label{fig:zeeman}
Sample Zeeman detection of an interstellar magnetic field using the CN line in the region DR21(OH). The top panel shows the observed total intensity (Stokes $I$, red lines), which is well-fit by two different velocity components (blue lines). The CN molecule has 7 hyperfine components, of which 4 have a large Zeeman splitting and 3 have a small splitting. The middle panel shows the measured Stokes $V$ (circularly polarized emission) for the sum of the 4 strong splitting components, while the bottom panel shows the corresponding measurement for the 3 weak components. The smooth lines show the best fit, with the line of sight magnetic field as the fitting parameter. Credit: \citet{crutcher99b}, \copyright AAS. Reproduced with permission.
}
\end{marginfigure}
Zeeman splitting is not trivial to to measure due to Doppler broadening. To see why, consider the example of OH. The Doppler width of a line is $\sigma_{\nu} = \nu_0 (\sigma_v/c)$, where for the OH transition that is normally used for Zeeman measurements $\nu_0=1.667$ GHz. If the OH molecule has a velocity dispersion of order $0.1$ km s$^{-1}$, as expected even for the lowest observed velocity dispersions found on small scales in molecular clouds, then $(\sigma_v/c)\sim 10^{-6}$, so $\sigma_{\nu} \sim 1$ kHz. This means that, unless the field is considerably larger than $1000$ $\mu$G (1 mG), which it usually is not, the Zeeman splitting is smaller than the Doppler line width. Thus the split lines are highly blended, and cannot be seen directly in the spectrum.

However, there is a trick to avoid this problem: radiation from the different Zeeman sublevels has different polarization. If the magnetic field is along the direction of propagation of the radiation, emission from the higher frequency Zeeman sublevel is right circularly polarized, while radiation from the lower frequency level is left circularly polarized. The unperturbed level is unpolarized. Thus although one cannot see the line split if one looks at total intensity (as measured by the Stokes $I$ parameter), one can see that the different polarization components peak at slightly different frequencies, so that the circularly polarized spectrum (as measured by the Stokes $V$ parameter) looks different than the total intensity spectrum. One can deduce the magnetic field strength along the line of sight from the difference between the total and polarized signals. Figure \ref{fig:zeeman} shows a sample detection.

Applying this technique to line emission from molecular clouds indicates that they are threaded by magnetic fields whose strengths range from tens to thousands of $\mu$G, with higher density gas generally showing stronger fields. We can attempt to determine if this is dynamically important by a simple energy argument. For a low-density envelope of a GMC with $n\sim 100$ cm$^{-3}$ ($\rho \sim 10^{-22}$ g cm$^{-3}$), we might have $v$ of a few km s$^{-1}$, giving a kinetic energy density
\begin{equation}
e_K  = \frac{1}{2}\rho v^2 \sim 10^{-11}\mbox{ erg cm}^{-3}.
\end{equation}
For the magnetic field of $20$ $\mu$G, typical of molecular clouds on large scales, the energy density is
\begin{equation}
e_B = \frac{B^2}{8\pi} \sim 10^{-11} \mbox{ erg cm}^{-3}.
\end{equation}
Thus the magnetic energy density is comparable to the kinetic energy density, and is dynamically significant in the flow.

\subsection{The Chandrasekhar-Fermi Method}

While the Zeeman effect provides by far the most direct method of measuring magnetic field strengths, it is not the only method for making this measurement. Another commonly-used technique, which we will not discuss in any detail, is the Chandrasekhar-Fermi method \citep{chandrasekhar53a}. This method relies on the fact that interstellar dust grains are non-spherical, which has two important implications. First, a non-spherical grain acts like an antenna, in that it interacts differently with electromagnetic waves that are oriented parallel and perpendicular to its long axis. As a result, grains both absorb and emit light preferentially along their long axis. This would not matter if the orientations of grains in the interstellar medium were random. However, there is a second effect. Most grains are charged, and as a result they tend to become preferentially aligned with the local magnetic field. The combination of these two effects means that the dust in a particular region of the ISM characterized by a particular large scale magnetic field will produce a net linear polarization in both the light it emits and any light passing through it. The direction of the polarization then reveals the orientation of the magnetic field on the plane of the sky.

By itself this effect tells us nothing about the strength of the field -- in principle there should be some relationship between field strength and degree of dust polarization, but there are enough other compounding factors and uncertainties that we cannot with any confidence translate the observed degree of polarization into a field strength. However, if we have measurements of the field orientation as a function of position, then we can estimate the field strength from the morphology of the field. As we shall see below, the degree to which field lines are straight or bent is strongly correlated with the ratio of magnetic energy density to turbulent energy density, and so the degree of alignment becomes a diagnostic of this ratio. In fact, one can even attempt to make quantitative field strength estimates from this method, albeit with very large uncertainties.

\section{Equations and Characteristic Numbers for Magnetized Turbulence}

Now that we know that magnetic fields are present, we now turn to some basic theory for magnetized flow. To understand how magnetic fields affect the flows in molecular clouds, it is helpful to write down the fundamental evolution equation for the magnetic field in a plasma, known as the induction equation\footnote{One may find a derivation of this result in many sources. The notation we use here is taken from \citet{shu92a}.}
\begin{equation}
\label{eq:induction}
\frac{\partial \vecB}{\partial t} + \nabla \times (\vecB \times \vecv) = -\nabla \times (\eta \nabla \times \vecB)
\end{equation}
Here $\vecB$ is the magnetic field, $\vecv$ is the fluid velocity (understood to be the velocity of the ions, which carry all the mass, a distinction that will become important below), and $\eta$ is the electrical resistivity. If the resistivity is constant in space, we can use the fact that $\nabla \cdot \vecB=0$ to simplify this slightly to get
\begin{equation}
\label{eq:induction_eta}
\frac{\partial \vecB}{\partial t} + \nabla \times (\vecB \times \vecv) = \eta \nabla^2 \vecB.
\end{equation}
The last term here looks very much like the $\nu\nabla^2 \vecv$ term we had in the equation for conservation of momentum (equation \ref{eq:momentum}) to describe viscosity. That term described diffusion of momentum, while the one in this equation describes diffusion of the magnetic field.\footnote{We are simplifying quite a bit here. The real dissipation mechanism in molecular clouds is not simple resistivity, it is something more complex called ion-neutral drift, which is discussed in Section \ref{sec:non-ideal-mhd}. However, the qualitative analysis given in this section is unchanged by this complexity, and the algebra is significantly easier if we use a simple scalar resistivity.}

We can understand the implications of this equation using dimensional analysis much as we did for the momentum equation in Section \ref{ssec:reynolds_mach}. As we did there, we let $L$ be the characteristic size of the system and $V$ be the characteristic velocity, so $L/V$ is the characteristic timescale. Spatial derivatives have the scaling $1/L$, and time derivatives have the scaling $V/L$. We let $B$ be the characteristic magnetic field strength. Applying these scalings to equation (\ref{eq:induction_eta}), the various terms scale as
\begin{eqnarray}
\frac{BV}{L} + \frac{BV}{L} & \sim & \eta \frac{B}{L^2} \\
1 & \sim & \frac{\eta}{VL}
\end{eqnarray}
In analogy to the ordinary hydrodynamic Reynolds number, we define the magnetic Reynolds number by
\begin{equation}
\mbox{Rm} = \frac{LV}{\eta}.
\end{equation}
Magnetic diffusion is significant only if $\mathrm{Rm} \sim 1$ or smaller.

What is Rm for a typical molecular cloud? As in the hydrodynamic case, we can take $L$ to be a few tens of pc and $V$ to be a few km s$^{-1}$. The magnetic field $B$ is a few tens of $\mu$G. The electrical resistivity is a microphysical property of the plasma, which, for a weakly ionized plasma, depends on the ionization fraction in the gas and the ion-neutral collision rate. We will show in Section \ref{sec:non-ideal-mhd} that a typical value of the resistivity\footnote{Again keeping in mind that this is not a true resistivity, it is an order of magnitude effective resistivity associated with ion-neutral drift.} in molecular clouds is $\eta\sim 10^{22}-10^{23}$ cm$^2$ s$^{-1}$. If we consider a length scale $L\sim 10$ pc and a velocity scale $V\sim 3$ km s$^{-1}$, then $LV\sim 10^{25}$ cm$^2$ s$^{-1}$, then this implies that the Rm for molecular clouds is hundreds to thousands.

Again in analogy to hydrodynamics, this means that on large scales magnetic diffusion is unimportant for molecular clouds -- although it is important on smaller scales. The significance of a large value of Rm becomes clear if we write down the induction equation with $\eta=0$ exactly:
\begin{equation}
\frac{\partial \vecB}{\partial t} + \nabla \times (\vecB \times \vecv) = 0.
\end{equation}
To understand what this equation implies, it is useful consider the magnetic flux $\Phi$ threading some fluid element. We define this as
\begin{equation}
\Phi = \int_A  \vecB \cdot \nhat\, dA,
\end{equation}
where we integrate over some area $A$ that defines the fluid element. The time derivative of this is then
\begin{eqnarray}
\frac{d\Phi}{dt} & = & \int_A \frac{\partial\vecB}{\partial t}\cdot\nhat\, dA + \oint_{\partial A} \vecB\cdot \vecv \times d\mathbf{l} \\
& = & \int_A \frac{\partial\vecB}{\partial t}\cdot\nhat\, dA + \oint_{\partial A} \vecB\times \vecv \cdot d\mathbf{l}
\end{eqnarray}
where $\partial A$ is the boundary of $A$. Here the second term on the right comes from the fact that, if the fluid is moving at velocity $\vecv$, the area swept out by a vector $d\mathbf{l}$ per unit time is $\vecv\times d\mathbf{l}$, so the flux crossing this area is $\vecB\cdot \vecv\times d\mathbf{l}$. Then in the second step we used the fact that $\nabla\cdot \vecB = 0$ to exchange the dot and cross products.

If we now apply Stokes theorem again to the second term, we get
\begin{eqnarray}
\frac{d\Phi}{dt} 
& = & \int_A \frac{\partial\vecB}{\partial t}\cdot\nhat\, dA + \int_A \nabla\times (\vecB\times \vecv) \cdot \nhat\, dA \\
& = & \int_A \left[\frac{\partial\vecB}{\partial t} + \nabla\times (\vecB\times \vecv)\right]\cdot\nhat\,dA\\
& = & 0.
\end{eqnarray}
The meaning of this is that, when Rm is large, the magnetic flux through each fluid element is conserved. This is called flux-freezing, since we can envision it geometrically as saying that magnetic field lines are frozen into the fluid, and move along with it. Thus on large scales the magnetic field moves with the fluid. However, on smaller scales the magnetic Reynolds number is $\sim 1$, and the field lines are not tied to the gas. We will calculate this scale in Section \ref{sec:non-ideal-mhd}. Before doing so, however, it is helpful to calculate another important dimensionless number describing the MHD flows in molecular clouds.

The conservation of momentum equation including magnetic forces is
\begin{equation}
\frac{\partial}{\partial t}(\rho \mathbf{v}) = -\nabla \cdot(\rho \mathbf{v v}) - \nabla P + \rho \nu \nabla^2 \mathbf{v} + \frac{1}{4\pi} (\nabla\times\vecB)\times \vecB,
\end{equation}
and if we make order of magnitude estimates of the various terms in this, we have
\begin{eqnarray}
\frac{\rho V^2}{L} & \sim & -\frac{\rho V^2}{L} + \frac{\rho c_s^2}{L} + \frac{\rho \nu V}{L^2} + \frac{B^2}{L} \\
1 & \sim & 1 + \frac{c_s^2}{V^2} + \frac{\nu}{VL} + \frac{B^2}{\rho V^2}
\end{eqnarray}
The second and third terms on the right hand side we have already defined in terms of $\mathcal{M}= V/c_s$ and $\mathrm{Re} = LV/\nu$. We now define our fourth and final characteristic number,
\begin{equation}
\mathcal{M}_A \equiv \frac{V}{v_A},
\end{equation}
where
\begin{equation}
v_A = \frac{B}{\sqrt{4\pi \rho}}
\end{equation}
is the Alfv\'{e}n speed -- the speed of the wave that, in magnetohydrodynamics, plays a role somewhat analogous to the sound wave in hydrodynamics. In flows with $\mathcal{M}_A \gg 1$, which we refer to as super-Alfv\'{e}nic, the magnetic force term is unimportant, while in those with $\mathcal{M}_A \ll 1$, referred to as sub-Alfv\'{e}nic, it is dominant.

\begin{figure}
\includegraphics[width=\linewidth]{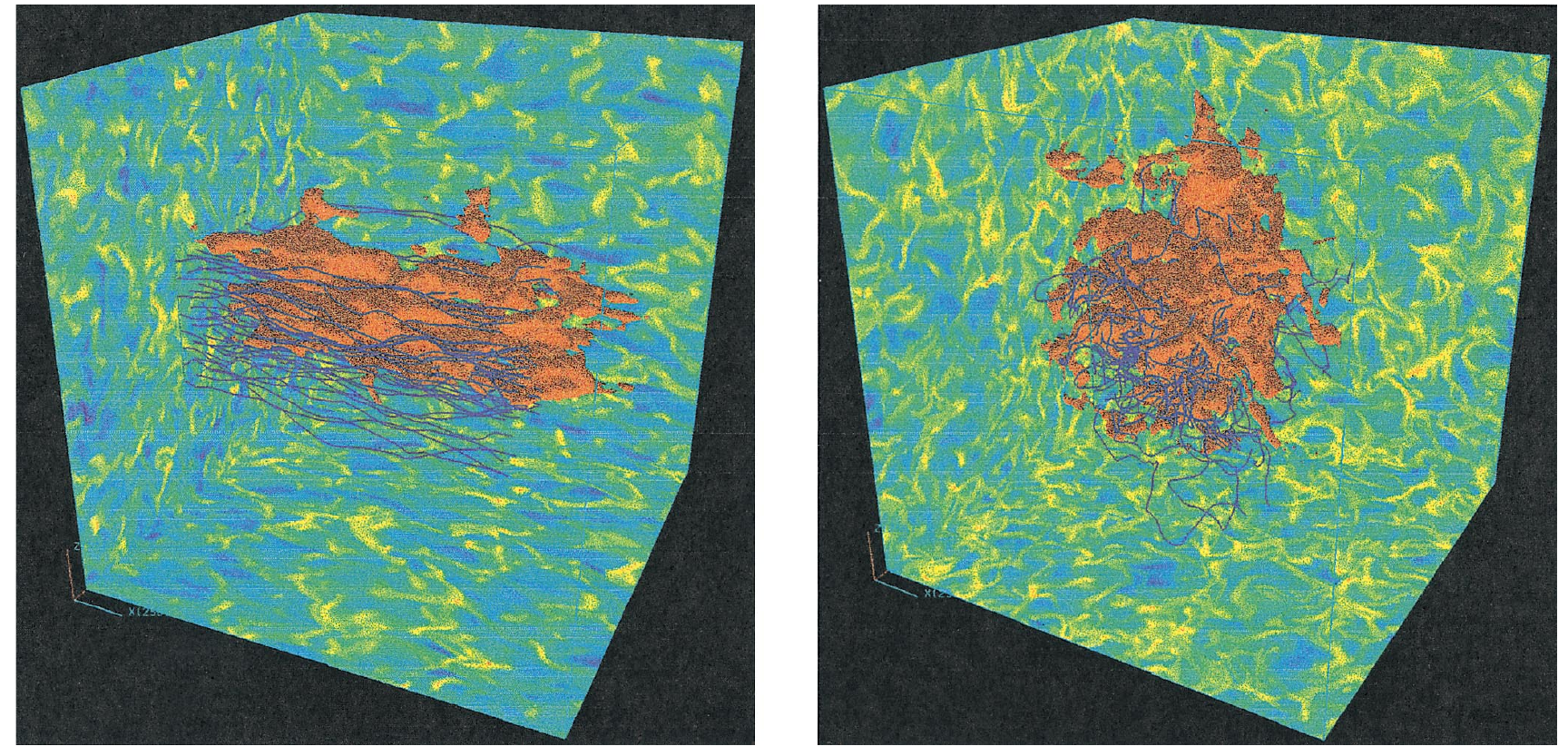}
\caption[Comparison of simulations of Alfv\'{e}nic and sub-Alfv\'{e}nic turbulence]{
\label{fig:alfvenmach}
Simulations of sub-Alfv\'{e}nic (left) and Alfv\'{e}nic (right) turbulence. Colors on the cube surface are slices of the logarithm of density, blue lines are magnetic field lines, and red surfaces are isodensity surfaces for a passive contaminant added to the flow. Credit: \citet{stone98a}, \copyright\,AAS. Reproduced with permission.
}
\end{figure}

For characteristic molecular cloud numbers $n\sim 100$ cm$^{-3}$, $B$ of a few tens of $\mu$G, and $V$ of a few km s$^{-1}$, we see that $v_A$ is of order a few km s$^{-1}$, about the same as the velocity. Thus the flows in molecular clouds are highly supersonic ($\mathcal{M}\gg 1$), but only trans-Alfv\'{e}nic ($\mathcal{M}_A \sim 1$), and magnetic forces have a significant influence. These forces make it much easier for gas to flow along field lines than across them, and result in a pattern of turbulence that is highly anisotropic (Figure \ref{fig:alfvenmach}).

\section{Non-Ideal Magnetohydrodynamics}
\label{sec:non-ideal-mhd}

We have just shown that the magnetic Reynolds number is a critical parameter for magnetized turbulence, and that this depends on the resistivity $\eta$. In the final part of this Chapter we will discuss in a bit more detail the physical origins of resistivity and related effects.

\subsection{Ion-Neutral Drift}

Molecular clouds are not very good plasmas. Most of the gas in a molecular cloud is neutral, not ionized. The ion fraction may be $10^{-6}$ or lower. Since only ions and electrons can feel the Lorentz force directly, this means that fields only exert forces on most of the particles in a molecular cloud indirectly. The indirect mechanism is that the magnetic field exerts forces on the ions and electrons (and mostly ions matter for this purpose), and these then collide with the neutrals, transmitting the magnetic force.

If the collisional coupling is sufficiently strong, then the gas acts like a perfect plasma. However, when the ion fraction is very low, the coupling is imperfect, and ions and neutrals do not move at exactly the same speed. The field follows the ions, since the are much less resistive, and flux freeizing for them is a very good approximation, but the neutrals are able to drift across field lines. This slippage between ions and neutrals is called ion-neutral drift, or ambipolar diffusion.

To estimate how this process works, we need to think about the forces acting on both ions and neutrals. The ions feel a Lorentz force
\begin{equation}
\vecf_L = \frac{1}{4\pi} (\nabla \times \vecB) \times \vecB.
\end{equation}
The other force in play is the drag force due to ion-neutral collisions, which is
\begin{equation}
\vecf_d = \gamma\rho_n \rho_i (\vecv_i-\vecv_n),
\end{equation}
where the subscript $i$ and $n$ refer to ions and neutrals, respectively, and $\gamma$ is the drag coefficient, which can be computed from the microphysics of the plasma. In a very weakly ionized fluid, the neutrals and ions very quickly reach terminal velocity with respect to one another, so the drag force and the Lorentz force must balance. Equating our expressions and solving for $\vecv_d=\vecv_i-\vecv_n$, the drift velocity, we get
\begin{equation}
\vecv_d  = \frac{1}{4\pi\gamma\rho_n\rho_i}(\nabla\times\vecB)\times\vecB
\end{equation}

To figure out how this affects the fluid, we write down the equation of magnetic field evolution under the assumption that the field is perfectly frozen into the ions:
\begin{equation}
\frac{\partial \vecB}{\partial t} + \nabla\times(\vecB\times \vecv_i) = 0.
\end{equation}
To figure out how the field behaves with respect to the neutrals, which constitute most of the mass, we simply use our expression for the drift speed $\vecv_d$ to eliminate $\vecv_i$. With a little algebra, the result is
\begin{equation}
\frac{\partial \vecB}{\partial t} + \nabla\times(\vecB\times \vecv_n) = 
\nabla \times \left\{ \frac{\vecB}{4\pi\gamma\rho_n\rho_i}\times\left[\vecB\times(\nabla\times\vecB)\right]\right\}.
\end{equation}
Referring back to the induction equation (\ref{eq:induction}), we can see that the resistivity produced by ion-neutral drift is not a scalar, and that it is non-linear, in the sense that it depends on $\vecB$ itself.

However, our scaling analysis still applies. The magnitude of the resistivity produced by ion-neutral drift is
\begin{equation}
\eta_{\rm AD} = \frac{B^2}{4\pi\rho_i\rho_n\gamma}.
\end{equation}
Thus, the magnetic Reynolds number is
\begin{equation}
\mbox{Rm} = \frac{LV}{\eta_{\rm AD}} = \frac{4\pi L V \rho_i\rho_n\gamma}{B^2} \approx \frac{4\pi L V \rho^2 x\gamma}{B^2},
\end{equation}
where $x=n_i/n_n$ is the ion fraction, which we've assumed is $\ll 1$ in the last step. Ion-neutral drift will allow the magnetic field lines to drift through the fluid on length scales $L$ such that $\mbox{Rm}\lesssim 1$. Thus, we can define a characteristic length scale for ambipolar diffusion by
\begin{equation}
L_{\rm AD} = \frac{B^2}{4\pi\rho^2 x \gamma V}
\end{equation}

In order to evaluate this numerically, we must calculate two things from microphysics: the ion-neutral drag coefficient $\gamma$ and the ionization fraction $x$. For $\gamma$, the dominant effect at low speeds is that ions induce a dipole moment in nearby neutrals, which allows them to undergo a Coulomb interaction. This greatly enhances the cross-section relative to the geometric value. We will not go into details of that calculation, and will simply adopt the results: $\gamma\approx 9.2 \times 10^{13}$ cm$^3$ s$^{-1}$ g$^{-1}$ (\citealt{smith97a}; note that \citealt{shu92a} gives a value that is lower by a factor of $\sim 3$, based on an earlier calculation).

The remaining thing we need to know to compute the drag force is the ion density. In a molecular cloud the gas is almost all neutral, and the high opacity excludes most stellar ionizing radiation. The main source of ions is cosmic rays, which can penetrate the cloud, although nearby strong x-ray sources can also contribute if present. We have already discussed cosmic rays in the context of cloud heating and chemistry, and here too they play a central role. Calculating the ionization fraction requires balancing this against the recombination rate, which is a nasty problem. That is because recombination is dominated by different processes at different densities, and recombinations are usually catalyzed by dust grains rather than occurring in the gas phase. We will not go into the details of these calculations here, and will instead simply quote a rough fit to their results given by \citet{tielens05a},
\begin{equation}
n_i \approx 2\times 10^{-3}\mbox{ cm}^{-3} \left(\frac{n_{\rm H}}{10^4\mbox{ cm}^{-3}}\right)^{1/2} \left(\frac{\zeta}{10^{-16}\mbox{ s}^{-1}}\right)^{1/2},
\end{equation}
where $n_{\rm H}$ is the number density and $\zeta$ is the cosmic ray primary\footnote{I.e., counting only ionizations caused by direct cosmic ray hits, as opposed to ionizations caused when primary electrons go on to ionize additional atoms or molecules.} ionization rate. Thus at a density $n_{\rm H} \sim 100$ cm$^{-3}$, we expect $x\approx 10^{-6}$.

Plugging this into our formulae, along with our characteristic numbers $L$ of a few tens of pc, $V\sim$ a few km s$^{-1}$, and $B\sim 10$ $\mu$G, we find
\begin{eqnarray}
\mbox{Rm} & \approx & 50 \\
L_{\rm AD}& \sim & 0.5\mbox{ pc}.
\end{eqnarray}
If we put in numbers for $L$ and $V$ more appropriate for cores than entire GMCs, we get $L_{\rm AD} \sim 0.05$ pc. Thus we expect the gas to act like a fully ionized gas on scales larger than this, but to switch over to behaving hydrodynamically on small scales.

\subsection{Turbulent Reconnection}

A final non-ideal MHD effect that may be important in molecular clouds, though this is still quite uncertain, is turbulent reconnection. The general idea of reconnection is that, in regions of non-zero resistivity where oppositely directed field lines are brought into close contact, the field lines can break and the field geometry can relax to a lower energy configuration. This may allow the field to slip out of the matter, and it always involves a reduction in magnetic pressure and energy density. The released energy is transformed into heat.

The simplest model of reconnection, the Sweet-Parker model, considers two regions of oppositely-directed field that meet at a plane. On that plane, a large current must flow in order to maintain the oppositely-directed fields on either side of it. Within this sheet reconnection can occur. As with ion-neutral drift, we can define a characteristic Reynolds-like number for this process, in this case called the Lundqvist number:
\begin{equation}
\mathcal{R}_L = \frac{LV}{\eta},
\end{equation}
where here $\eta$ is the true microphysical resistivity, as opposed to the term describing ion-neutral diffusion.

The rate at which reconnection can occur in the Sweet-Parker model is dictated by geometry. Matter is brought into the thin reconnection region, it reconnects, and then it must exit so that new reconnecting matter can be brought in. Matter can only exit the layer at the Alfv\'{e}n speed, and since the cross-section of the reconnection layer is small, this sets severe limits on the rate at which reconnection can occur. It turns out that one can show that the maximum speed at which matter can be brought into the reconnection region is of order $\mathcal{R}_L^{1/2} v_A$.

To figure out this speed, we need to know the resistivity, which is related to the electrical conductivity $\sigma$ by
\begin{equation}
\eta = \frac{c^2}{4\pi \sigma}.
\end{equation}
We will not provide a full calculation of plasma conductivity here\footnote{Again, \citet{shu92a} is a good source for those who wish to see a rigorous derivation.}, but we can sketch the basic outlines of the argument. The conductivity of a plasma is the proportionality constant between the applied electric field and the resulting current,
\begin{equation}
J = \sigma E.
\end{equation}
In a plasma the current is carried by motions of the electrons, which move much faster than the ions due to their lower mass, and the current is simply the electron charge times the electron number density times the mean electron speed: $J = e n_e v_e$. To compute the mean electron speed, one balances the electric force against the drag force exerted by collisions with neutrals (which dominate in a weakly ionized plasma), in precisely the same way we derived the mean ion-neutral drift speed by balancing the drag force against the Lorentz force. Not surprisingly $v_e$ ends up being proportional to $E$, and inversely proportional to the number density of the dominant non-electron species (H$_2$ in molecular clouds) and the cross section of this species for electron collisions. The final result of this procedure is
\begin{equation}
\sigma = \frac{n_e e^2}{m_e n_{\rm H_2} \langle\sigma v\rangle_{e-\rm H_2}} \approx 10^{17} x\mbox{ s}^{-1},
\end{equation}
where $\langle\sigma v\rangle_{e-\rm H_2}\approx 10^{-9}$ cm$^3$ s$^{-1}$ is the mean cross-section times velocity for electron-ion collisions. Plugging this into the resistivity gives
\begin{equation}
\eta \approx \frac{10^3\mbox{ cm}^2\mbox{ s}^{-1}}{x}
\end{equation}
Plugging in our typical value $x\sim 10^{-6}$ gives $\eta\sim 10^{9}$ cm$^2$ s$^{-1}$, and using $L\sim 10$ pc and $V$ of a few km s$^{-1}$, typical molecular cloud numbers, this implies
\begin{equation}
\mathcal{R}_L \sim 10^{16}.
\end{equation}

Of course this makes the reconnection speed truly tiny, of order $10^{-8}$ of $v_A$. So why is reconnection at all interesting? Why is it worth considering? The answer turns on the word turbulent. It turns out that the Sweet-Parker model underpredicts the observed reconnection rate in laboratory experiments or observed in Solar flares and the Earth's magnetosphere. Indeed, if Sweet-Parker were right, there would be no such things as Solar flares.

We currently lack a good understanding of reconnection, but a rough idea is that, in a turbulent medium, reconnection sheets are can be much wider due to turbulent broadening, and that this in turn removes the geometric constraint that makes the reconnection velocity much smaller than the Alfv\'{e}n speed. Exactly how and when this is important in molecular clouds is a subject of very active debate in the literature.

\chapter{Gravitational Instability and Collapse}
\label{ch:collapse}

\marginnote{
\textbf{Suggested background reading:}
\begin{itemize}
\item \href{http://adsabs.harvard.edu/abs/2014arXiv1402.0867K}{Krumholz, M.~R. 2014, Phys.~Rep., 539, 49}, section 3.4 \nocite{krumholz14c}
\end{itemize}
}

The previous two chapters provided a whirlwind tour of fluid dynamics and turbulence. However, in that discussion we completely omitted gravity, which is obviously critical to the process of star formation. We will now remedy that omission by bringing gravity back into the discussion.

\section{The Virial Theorem}

To open this topic, we will start by proving a powerful and general theorem about the behavior of fluids, known as the virial theorem.\footnote{Like the equations of motion, there is both an Eulerian form and a Lagrangian form of the virial theorem, depending on which version of the equations of motion we start with. We will derive the Eulerian form here, following the original proof by \citet{mckee92a}, but the derivation of the Lagrangian form proceeds in a similar manner, and can be found in many standard textbooks, for example \citet{shu92a}.} To derive the virial theorem, we begin with the MHD equations of motion, without either viscosity or resistivity (since neither of these are important for GMCs on large scales) but with gravity. We leave in the pressure forces, even though they are small, because they are also trivial to include. Thus we have
\begin{eqnarray}
\frac{\partial\rho}{\partial t} & = & -\nabla \cdot (\rho \vecv) \\
\frac{\partial}{\partial t}(\rho \vecv) & = & -\nabla\cdot(\rho\vecv\vecv) -\nabla P + \frac{1}{4\pi} (\nabla\times\vecB)\times\vecB - \rho \nabla \phi.
\end{eqnarray}
Here $\phi$ is the gravitational potential, so $-\rho \nabla \phi$ is the gravitational force per unit volume. These equations are the Eulerian equations written in conservative form.

Before we begin, life will be a bit easier if we re-write the entire second equation in a manifestly tensorial form -- this simplifies the analysis tremendously. To do so, we define two tensors: the fluid pressure tensor $\vecPi$ and the Maxwell stress tensor $\vecT_M$, as follows:
\begin{eqnarray}
\vecPi & \equiv & \rho \vecv\vecv + P\vecI \\
\vecT_M & \equiv & \frac{1}{4\pi} \left(\vecB\vecB - \frac{B^2}{2}\vecI\right)
\end{eqnarray}
Here $\vecI$ is the identity tensor. In tensor notation, these are
\begin{eqnarray}
(\vecPi)_{ij} & \equiv & \rho v_i v_j + P \delta_{ij} \\
(\vecT_M)_{ij} & \equiv & \frac{1}{4\pi} \left(B_i B_j - \frac{1}{2}B_k B_k \delta_{ij}\right).
\end{eqnarray}
With these definitions, the momentum equation just becomes
\begin{equation}
\frac{\partial}{\partial t}(\rho \vecv) = -\nabla\cdot(\vecPi-\vecT_M) - \rho \nabla\phi.
\end{equation}

The substitution for $\vecPi$ is obvious. The equivalence of $\nabla\cdot\vecT_M$ to $1/(4\pi) (\nabla\times\vecB)\times\vecB$ is easy to establish with a little vector manipulation, which is most easily done in tensor notation:
\begin{eqnarray}
(\nabla\times\vecB)\times\vecB & = & \epsilon_{ijk} \epsilon_{jmn} \left(\frac{\partial}{\partial x_m}B_n\right) B_k \\
& = & -\epsilon_{jik} \epsilon_{jmn} \left(\frac{\partial}{\partial x_m}B_n\right) B_k \\
& = & (\delta_{in}\delta_{km}-\delta_{im}\delta_{kn})\left(\frac{\partial}{\partial x_m}B_n\right) B_k \\
& = & B_k\frac{\partial}{\partial x_k} B_i - B_k\frac{\partial}{\partial x_i} B_k \\
& = & \left(B_k\frac{\partial}{\partial x_k} B_i + B_i \frac{\partial}{\partial x_k} B_k\right) - B_k\frac{\partial}{\partial x_i} B_k \\
& = & \frac{\partial}{\partial x_k}\left(B_i B_k\right) -\frac{1}{2} \frac{\partial}{\partial x_i} \left(B_k^2\right)\\
& = & \nabla\cdot \left(\vecB\vecB - \frac{B^2}{2}\right).
\end{eqnarray}

To derive the virial theorem, we begin by imagining a cloud of gas enclosed by some fixed volume $V$. The surface of this volume is $S$. We want to know how the overall distribution of mass changes within this volume, so we begin by writing down a quantity the represents the mass distribution. This is the moment of inertia,
\begin{equation}
I = \int_V \rho r^2\, dV.
\end{equation}

We want to know how this changes in time, so we take its time derivative:
\begin{eqnarray}
\dot{I} & = & \int_V \frac{\partial\rho}{\partial t} r^2 \,dV \\
& = & -\int_V \nabla \cdot (\rho \vecv) r^2\, dV \\
& = & -\int_V \nabla \cdot (\rho \vecv r^2)\, dV + 2\int_V \rho \vecv\cdot \vecr\, dV \\
& = & -\int_S (\rho \vecv r^2)\cdot d\vecS + 2\int_V \rho \vecv\cdot \vecr\, dV.
\end{eqnarray}
In the first step we used the fact that the volume $V$ does not vary in time to move the time derivative inside the integral. Then in the second step we used the equation of mass conservation to substitute. In the third step we brought the $r^2$ term inside the divergence. Finally in the fourth step we used the divergence theorem to replace the volume integral with a surface integral.

Now we take the time derivative again, and multiply by $1/2$ for future convenience:
\begin{eqnarray}
\frac{1}{2}\ddot{I} & = & -\frac{1}{2} \int_S r^2 \frac{\partial}{\partial t}(\rho\vecv)\cdot d\vecS +
\int_V \frac{\partial}{\partial t}(\rho\vecv)\cdot\vecr \, dV \\
& = & -\frac{1}{2} \frac{d}{dt} \int_S r^2 (\rho\vecv)\cdot d\vecS 
\nonumber \\
& & \quad {} -
\int_V \vecr \cdot \left[\nabla\cdot(\vecPi-\vecT_M)+ \rho\nabla \phi \right] \, dV.
\end{eqnarray}
The term involving the tensors is easy to simplify using a handy identity, which applies to an arbitrary tensor. This is a bit easier to follow in tensor notation:
\begin{eqnarray}
\int_V \vecr\cdot \nabla\cdot \vecT \, dV & = & \int_V x_i \frac{\partial}{\partial x_j} T_{ij}\,dV \\
& = & \int_V \frac{\partial}{\partial x_j}(x_i T_{ij})\,dV - \int_V T_{ij} \frac{\partial}{\partial x_j}x_i \, dV \\
& = & \int_S x_i T_{ij} \,dS_j - \int_V \delta_{ij} T_{ij} \, dV \\
& = & \int_S \vecr\cdot\vecT\cdot d\vecS - \int_V \mbox{Tr}\; \vecT \, dV,
\end{eqnarray}
where $\mbox{Tr}\;\vecT = T_{ii}$ is the trace of the tensor $\vecT$.

Applying this to our result our tensors, we note that
\begin{eqnarray}
\mbox{Tr}\; \vecPi & = & 3P + \rho v^2 \\
\mbox{Tr}\; \vecT_M & = & -\frac{B^2}{8\pi}
\end{eqnarray}
Inserting this result into our expression for $\ddot{I}$ gives the virial theorem, which we will write in a more suggestive form to make its physical interpretation clearer:
\begin{equation}
\frac{1}{2}\ddot{I} = 2(\mathcal{T} - \mathcal{T}_S) + \mathcal{B} + \mathcal{W} - \frac{1}{2}\frac{d}{dt} \int_S (\rho\vecv r^2)\cdot d\vecS,
\end{equation}
where
\begin{eqnarray}
\mathcal{T} & = & \int_V\left(\frac{1}{2}\rho v^2 + \frac{3}{2} P\right)\, dV \\
\mathcal{T}_S & = & \int_S \vecr \cdot \vecPi \cdot d\vecS\\
\mathcal{B} & = & \frac{1}{8\pi} \int_V B^2 \,dV + \int_S \vecr\cdot \vecT_M\cdot d\vecS\\
\mathcal{W} & = & -\int_V \rho \vecr\cdot\nabla\phi\,dV.
\end{eqnarray}

Written this way, we can give a clear interpretation to what these terms mean. $\mathcal{T}$ is just the total kinetic plus thermal energy of the cloud. $\mathcal{T}_S$ is the confining pressure on the cloud surface, including both the thermal pressure and the ram pressure of any gas flowing across the surface. $\mathcal{B}$ is the the difference between the magnetic pressure in the cloud interior, which tries to hold it up, and the magnetic pressure plus magnetic tension at the cloud surface, which try to crush it. $\mathcal{W}$ is the gravitational energy of the cloud. If there is no external gravitational field, and $\phi$ comes solely from self-gravity, then $\mathcal{W}$ is just the gravitational binding energy. The final integral represents the rate of change of the momentum flux across the cloud surface.

$\ddot{I}$ is the integrated form of the acceleration. For a cloud of fixed shape, it tells us the rate of change of the cloud's expansion of contraction. If it is negative, the terms that are trying to collapse the cloud (the surface pressure, magnetic pressure and tension at the surface, and gravity) are larger, and the cloud accelerates inward. If it is positive, the terms that favor expansion (thermal pressure, ram pressure, and magnetic pressure) are larger, and the cloud accelerates outward. If it is zero, the cloud neither accelerates nor decelerates.

We get a particularly simple form of the virial theorem if there is no gas crossing the cloud surface (so $\vecv=0$ at $S$) and if the magnetic field at the surface to be a uniform value $B_0$. In this case the virial theorem reduces to
\begin{equation}
\frac{1}{2}\ddot{I} = 2(\mathcal{T} - \mathcal{T}_S) + \mathcal{B} + \mathcal{W}
\end{equation}
with
\begin{eqnarray}
\mathcal{T}_S & = & \int_S rP \,dS\\
\mathcal{B} & = & \frac{1}{8\pi} \int_V (B^2-B_0^2) \,dV.
\end{eqnarray}
For this simplified physical setup, $\mathcal{T}_S$ just represents the mean radius times pressure at the virial surface, and $\mathcal{B}$ just represents the total magnetic energy of the cloud minus the magnetic energy of the background field over the same volume. Notice that, if a cloud is in equilibrium ($\ddot{I}=0$) and magnetic and surface forces are negligible, then we have $2\mathcal{T} = -\mathcal{W}$. Based on this result, we define the virial ratio
\begin{equation}
\label{eq:alpha_vir_th}
\alpha_{\rm vir} = \frac{2\mathcal{T}}{|\mathcal{W}|}.
\end{equation}
For an object for which magnetic and surface forces are negligible, and with no flow across the virial surface, a value of $\alpha_{\rm vir} > 1$ implies $\ddot{I} > 0$, and a value $\alpha_{\rm vir} < 1$ implies $\ddot{I}<0$. Thus $\alpha_{\rm vir} = 1$ roughly divides clouds that have enough internal pressure or turbulence to avoid collapse from those that do not.

\section{Stability Conditions}

Armed with the virial theorem, we are now in a position to understand, at least qualitatively, under what conditions a cloud of gas will be stable against gravitational contraction, and under what conditions it will not be. If we examine the terms on the right hand side of the virial theorem, we can group them into those that are generally or always positive, and thus oppose collapse, and those that are generally or always negative, and thus encourage it. The main terms opposing collapse are $\mathcal{T}$, which contains parts describing both thermal pressure and turbulent motion, and $\mathcal{B}$, which describes magnetic pressure and tension. The main terms favoring collapse are $\mathcal{W}$, representing self-gravity, and $\mathcal{T}_S$, representing surface pressure. The final term, the surface one, could be positive or negative depending on whether mass is flowing into our out of the virial volume. We will begin by examining the balance among these terms, and the forces they represent.

\subsection{Thermal Support and the Jeans Instability}

Gas pressure is perhaps the most basic force in opposing collapse. Unlike turbulent motions, which can compress in some places even as they provide overall support, gas pressure always tries to smooth out the gas. Similarly, self-gravity is the most reliable promoter of collapse. A full, formal analysis of the interaction between pressure and self-gravity was provided by James Jeans in 1902 \citep{jeans02a}, and we will go through it below. However, we can already see what the basic result will have to look like just from the virial theorem. We expect the dividing line between stability and instability to lie at $\alpha_{\rm vir} \approx 1$. For an isolated, isothermal cloud of mass $M$ and radius $R$ with only thermal pressure, we have
\begin{eqnarray}
\mathcal{T} & = & \frac{3}{2} M c_s^2 \\
\mathcal{W} & = & -a \frac{GM^2}{R},
\end{eqnarray}
where $a$ is a factor of order unity that depends on the internal density structure. Thus the condition $\alpha_{\rm vir} \gtrsim 1$ corresponds to
\begin{equation}
M c_s^2 \gtrsim \frac{GM^2}{R} \qquad\Longrightarrow\qquad R \gtrsim \frac{GM}{c_s^2},
\end{equation}
or, rewriting in terms of the mean density $\rho \sim M/R^3$,
\begin{equation}
R \gtrsim \frac{c_s}{\sqrt{G\rho}}.
\end{equation}

The formal analysis proceeds as follows. Consider a uniform, infinite, isothermal medium at rest. The density is $\rho_0$, the pressure is $P_0 = \rho_0 c_s^2$, and the velocity is $\vecv_0 = 0$. We will write down the equations of hydrodynamics and self-gravity for this gas:
\begin{eqnarray}
\frac{\partial}{\partial t}\rho + \nabla\cdot (\rho \vecv) & = & 0 \\
\frac{\partial}{\partial t}(\rho\vecv) + \nabla\cdot(\rho \vecv\vecv) & = & -\nabla P - \rho \nabla \phi \\
\nabla^2 \phi & = & 4\pi G\rho.
\end{eqnarray}

Here the first equation represents conservation of mass, the second represents conservation of momentum, and the third is the Poisson equation for the gravitational potential $\phi$. We take the background density $\rho_0$, velocity $\vecv_0 = 0$, pressure $P_0$, and potential $\phi_0$ to be an exact solution of these equations, so that all time derivatives are zero as long as the gas is not disturbed.

Note that this involves the "Jeans swindle": this assumption is actually not really consistent, because the Poisson equation cannot be solved for an infinite uniform medium unless $\rho_0 = 0$. In other words, there is no function $\phi_0$ such that $\nabla^2\phi_0$ is equal to a non-zero constant value on all space. That said, we will ignore this complication, since the approximation of a uniform infinite medium is a reasonable one for a very large but finite uniform medium. It is possible to construct the argument without the Jeans swindle, but doing so adds mathematical encumbrance without physical insight, so we will not do so.

That digression aside, now let us consider what happens if we perturb this system. We will write the density as $\rho = \rho_0 + \epsilon \rho_1$, where $\epsilon\ll 1$. Similarly, we write $\vecv=\epsilon \vecv_1$ and $\phi=\phi_0 + \epsilon \phi_1$. Since we can always use Fourier analysis to write an arbitrary perturbation as a sum of Fourier components, without loss of generality we will take the perturbation to be a single, simple Fourier mode. The reason to do this is that, as we will see, differential equations are trivial to solve when the functions in question are simple plane waves.

Thus we write $\rho_1 = \rho_a \exp[i(kx - \omega t)]$. Note that we implicitly understand that we use only the real part of this exponential. It is just easier to write things in terms of an $e^{i(kx-\omega t)}$ than it is to keep track of a bunch of sines and cosines. In writing this equation, we have chosen to orient our coordinate system so that the wave vector $\veck$ of the perturbation is in the $\vecx$ direction. Again, there is no loss of generality in doing so.

Given this density perturbation, what is the corresponding perturbation to the potential? From the Poisson equation, we have
\begin{equation}
\nabla^2 (\phi_0 + \epsilon \phi_1) = 4\pi G (\rho_0 + \epsilon \rho_1).
\end{equation}
Since by assumption $\rho_0$ and $\phi_0$ are exact solutions to the Poisson equation, we can cancel the $\phi_0$ and $\rho_0$ terms out of the equation, leaving
\begin{equation}
\nabla^2 \phi_1 = 4 \pi G \rho_1 = 4\pi G \rho_a e^{i(kx-\omega t)}.
\end{equation}
This equation is trivial to solve, since it is just of the form $y'' = a e^{bx}$. The solution is
\begin{equation}
\phi_1 = -\frac{4\pi G \rho_a}{k^2} e^{i(kx - \omega t)}.
\end{equation}
By analogy to what we did for $\rho_1$, we write this solution as $\phi_1 = \phi_a e^{i(kx-\omega t)}$, with
\begin{equation}
\phi_a = -\frac{4\pi G \rho_a}{k^2}.
\end{equation}

Now that we have found the perturbed potential, let us determine what motion this will induce in the fluid. To do so, we first take the equations of mass and momentum conservation and we linearize them. This means that we substitute in $\rho = \rho_0 + \epsilon \rho_1$, $\vecv=\epsilon \vecv_1$, $P=P_0+\epsilon P_1=c_s^2 (\rho_0 + \epsilon \rho_1)$, and $\phi=\phi_0+\epsilon \phi_1$. Note that $\vecv_0 = 0$. We then expand the equations in powers of $\epsilon$, and we drop all the terms that are of order $\epsilon^2$ or higher on the grounds that they become negligible in the limit of small $\epsilon$.

Linearizing the equation of mass conservation we get
\begin{eqnarray}
\frac{\partial}{\partial t} (\rho_0 + \epsilon \rho_1) + \nabla\cdot [(\rho_0 + \epsilon \rho_1)(\epsilon\vecv_1)] & = & 0 \\
\frac{\partial}{\partial t} \rho_0 + \epsilon \frac{\partial}{\partial t} \rho_1 + \epsilon \nabla \cdot (\rho_0\vecv_1) & = & 0 \\
\frac{\partial}{\partial t} \rho_1+ \nabla \cdot (\rho_0\vecv_1) & = & 0.
\label{eq:masscons_linear}
\end{eqnarray}
In the second step, we dropped a term of order $\epsilon^2$. In the third step we used the fact that $\rho_0$ is constant, i.e., that the background density has zero time derivative, to drop that term. Applying the same procedure to the momentum equation, we get
\begin{eqnarray}
\nonumber
\lefteqn{
\frac{\partial}{\partial t} [(\rho_0 + \epsilon \rho_1)(\epsilon\vecv_1)] + \nabla\cdot [(\rho_0 + \epsilon \rho_1)(\epsilon\vecv_1)(\epsilon\vecv_1)] 
}
\qquad\qquad\qquad
\\
& = & -c_s^2 \nabla (\rho_0 + \epsilon \rho_1) \nonumber \\
& & \qquad {} - (\rho_0 + \epsilon\rho_1)\nabla (\phi_0 + \epsilon \phi_1)
\\
\epsilon \frac{\partial}{\partial t} (\rho_0\vecv_1) & = & -c_s^2 \nabla \rho_0 - \rho_0 \nabla\phi_0
\nonumber \\
& & \qquad {}
- \epsilon \left(c_s^2 \nabla \rho_1 + \rho_1\nabla \phi_0 + \rho_0\nabla \phi_1\right) \\
\frac{\partial}{\partial t} (\rho_0\vecv_1) & = & -c_s^2 \nabla \rho_1 - \rho_0\nabla\phi_1.
\label{eq:momcons_linear}
\end{eqnarray}
In the second step we dropped terms of order $\epsilon^2$, and in the third step we used the fact that the background state is uniform to drop terms involving gradients of $\rho_0$ and $\phi_0$.

Now that we have our linearized equations, we're ready to find out what $\vecv_1$ must be. By analogy to what we did for $\rho_1$ and $\phi_1$, we take $\vecv_1$ to be a single Fourier mode, of the form
\begin{equation}
\vecv_1 = \vecv_a e^{i(kx-\omega t)}
\end{equation}
Substituting for $\rho_1$, $\phi_1$, and $\vecv_1$ into the linearized mass conservation equation (\ref{eq:masscons_linear}), we get
\begin{eqnarray}
\frac{\partial}{\partial t} \left(\rho_a e^{i(kx-\omega t)}\right) + \nabla\cdot (\rho_0 \vecv_a e^{i(kx-\omega t)}) & = & 0 \\
-i\omega \rho_a e^{i(kx-\omega t)} + i k \rho_0 v_{a,x} e^{i(kx-\omega t)} & = & 0 \\
-\omega\rho_a + k\rho_0 v_{a,x} & = & 0 \\
\frac{\omega\rho_a}{k\rho_0} & = & v_{a,x}
\end{eqnarray}
where $v_{a,x}$ is the $x$ component of $\vecv_a$.

We have now found the velocity perturbation in terms of $\rho_a$, $\omega$, and $k$. Similarly substituting into the linearized momentum equation (\ref{eq:momcons_linear}) gives
\begin{eqnarray}
\frac{\partial}{\partial t} \left(\rho_0 \vecv_a e^{i(kx-\omega t)}\right) & = & -c_s^2 \nabla(\rho_a e^{i(kx-\omega t)})
\nonumber \\
& & \qquad {}
 - \rho_0 \nabla (\phi_a e^{i(kx-\omega t)}) \\
-i\omega \rho_0 \vecv_a e^{i(kx-\omega t)} & = & -i k c_s^2 \rho_a \hat{\mathbf{x}} e^{i(kx-\omega t)} 
\nonumber \\
& & \qquad {}
- i k \rho_0 \phi_a e^{i(kx-\omega t)} \hat{\mathbf{x}}\\
\omega \rho_0 v_{a,x} & = & k \left(c_s^2 \rho_a + \rho_0 \phi_a\right).
\end{eqnarray}
Now let us take this equation and substitute in the values for $\phi_a$ and $v_{a,x}$ that we previously determined:
\begin{eqnarray}
\omega \rho_0 \left(\frac{\omega\rho_a}{k\rho_0}\right) & = & k c_s^2 \rho_a - k\rho_0 \left(\frac{4\pi G \rho_a}{k^2}\right) \\
\omega^2 & = & c_s^2 k^2 - 4\pi G \rho_0
\end{eqnarray}
This expression is known as a dispersion relation, because it describes the dispersion of the plane wave solution we have found, i.e., how that wave's spatial frequency $k$ relates to its temporal frequency $\omega$. 

To see what this implies, let us consider what happens when we put in a perturbation with a short wavelength or a large spatial frequency. In this case $k$ is large, and $c_s^2 k^2 - 4\pi G \rho_0>0$, so $\omega$ is a positive or negative real number. The density is $\rho=\rho_0 + \rho_a e^{i(kx-\omega t)}$, which represents a uniform background density with a small oscillation in space and time on top of it. Since $|e^{i(kx-\omega t)}| < 1$ at all times and places, the oscillation does not grow.

On the other hand, suppose that we impose a perturbation with a large spatial range, or a small spatial frequency. In this case $c_s^2 k^2 - 4\pi G \rho_0<0$, so $\omega$ is a positive or negative imaginary number. For an imaginary $\omega$, $|e^{-i\omega t}|$ either decays to zero or grows infinitely large in time, depending on whether we take the positive or negative imaginary root. Thus at least one solution for the perturbation will not remain small. It will grown in amplitude without limit.

This represents an instability: if we impose an arbitrarily small amplitude perturbation on the density at a sufficiently large wavelength, that perturbation will eventually grow to be large. Of course once $\rho_1$ becomes large enough, our linearization procedure of dropping terms proportional to $\epsilon^2$ becomes invalid, since these terms are no longer small. In this case we must follow the full non-linear behavior of the equations, usually with simulations.

We have, however, shown that there is a critical size scale beyond which perturbations that are stabilized only by pressure must grow to non-linear amplitude. The critical length scale is set by the value of $k$ for which $\omega=0$,
\begin{equation}
k_J = \sqrt{\frac{4\pi G\rho_0}{c_s^2}}.
\end{equation}
The corresponding wavelength is
\begin{equation}
\lambda_J = \frac{2\pi}{k_J} = \sqrt{\frac{\pi c_s^2}{G\rho_0}}.
\end{equation}
This is known as the Jeans length. One can also define a mass scale associated with this: the Jeans mass, $M_J=\rho\lambda_J^3/8$.\footnote{The definition of the Jeans mass is somewhat ambiguous, and multiple definitions can be found in the literature. The one we have chosen corresponds to considering the mass within a cube of half a Jeans length in size. Possible alternatives include choosing a cube one Jeans length in size or choosing a sphere one Jeans length in radius or diameter, to name just two possibilities. These definitions all scale with density and Jeans length in the same way, and differ only in their coefficients.}

If we plug in some typical numbers for a GMC, $c_s=0.2$ km s$^{-1}$ and $\rho_0 = 100 m_p$, we get $\lambda_J = 3.4$ pc. Since every GMC we have seen is larger than this size, and there are clearly always perturbations present, this means that molecular clouds cannot be stabilized by gas pressure against collapse. Of course one could have guessed this result just by evaluating terms in the virial theorem: the gas pressure term is very small compared to the gravitational one. Ultimately, the virial theorem and the Jeans instability analysis are just two different ways of extracting the same information from the equations of motion.

One nice thing about the Jeans analysis, however, is that it makes it obvious how fast we should expect gravitational instabilities to grow. Suppose we have a very unstable system, where $c_s^2 k^2 \ll 4 \pi G \rho_0$. This is the case for GMC, for example. There are perturbations on the size of the entire cloud, which might be 50 pc in size. This is a spatial frequency $k=2\pi/(50\mbox{ pc}) = 0.12$ pc$^{-1}$. Plugging this in with $c_s = 0.2$ km s$^{-1}$ and $\rho_0=100 m_p$ gives $c_s^2 k^2 / (4\pi G \rho) = 0.005$.
In this case we have
\begin{equation}
\omega \approx \pm i \sqrt{4\pi G\rho_0}.
\end{equation}

Taking the negative $i$ root, which corresponds to the growing mode, we find that
\begin{equation}
\rho_1 \propto \exp([4\pi G \rho_0]^{1/2} t).
\end{equation}
Thus the $e$-folding time for the disturbance to grow is $\sim 1/\sqrt{G\rho_0}$. We define the free-fall time as
\begin{equation}
t_{\rm ff} = \sqrt{\frac{3\pi}{32 G \rho_0}},
\end{equation}
where the numerical coefficient of $\sqrt{3\pi/32}$ comes from doing the closely related problem of the collapse of a pressureless sphere, which we will cover in Section \ref{sec:pressureless_collapse}. The free-fall time is the characteristic time scale required for a medium with negligible pressure support to collapse.

The Jeans analysis is of course only appropriate for a uniform medium, and it requires the Jeans swindle. Problem Set 2 contains a calculation of the maximum mass of a spherical cloud that can support itself against collapse by thermal pressure, called the Bonnor-Ebert mass \citep{ebert55a, bonnor56a}. Not surprisingly, the Bonnor-Ebert mass is simply $M_J$ times factors of order unity.

\subsection{Magnetic Support and the Magnetic Critical Mass}

We now examine another term that generally opposes collapse: the magnetic one. Let us consider a uniform spherical cloud of radius $R$ threaded by a magnetic field $\vecB$. We imagine that $\vecB$ is uniform inside the cloud, but that outside the cloud the field lines quickly spread out, so that the magnetic field drops down to some background strength $\vecB_0$, which is also uniform but has a magnitude much smaller than $\vecB$.

Here it is easiest to work directly with the virial theorem. The magnetic term in the virial theorem is
\begin{equation}
\mathcal{B} = \frac{1}{8\pi} \int_V B^2 \,dV + \int_S \vecr \cdot \vecT_M \cdot d\vecS
\end{equation}
where
\begin{equation}
\vecT_M = \frac{1}{4\pi} \left(\vecB\vecB - \frac{B^2}{2}\vecI\right).
\end{equation}

If the field inside the cloud is much larger than the field outside it, then the first term, representing the integral of the magnetic pressure within the cloud, is
\begin{equation}
\frac{1}{8\pi} \int_V B^2\, dV \approx \frac{B^2 R^3}{6}.
\end{equation}
Here we have dropped any contribution from the field outside the cloud. The second term, representing the surface magnetic pressure and tension, is
\begin{equation}
\int_S \vecx \cdot \vecT_M \cdot d\vecS = \int_S \frac{B_0^2}{8\pi} \vecx \cdot d\vecS
\approx \frac{B_0^2 R_0^3}{6}
\end{equation}

Since the field lines that pass through the cloud must also pass through the virial surface, it is convenient to rewrite everything in terms of the magnetic flux. The flux passing through the cloud is $\Phi_B = \pi B R^2$, and since these field lines must also pass through the virial surface, we must have $\Phi_B = \pi B_0 R_0^2$ as well. Thus, we can rewrite the magnetic term in the virial theorem as
\begin{equation}
\mathcal{B} \approx \frac{B^2 R^3}{6} - \frac{B_0^2 R_0^2}{6} = \frac{1}{6\pi^2} \left(\frac{\Phi_B^2}{R} - \frac{\Phi_B^2}{R_0}\right) \approx \frac{\Phi_B^2}{6\pi^2 R}.
\end{equation}

In the last step we used the fact that $R \ll R_0$ to drop the $1/R_0$ term. Now let us compare this to the gravitational term, which is
\begin{equation}
\mathcal{W} = -\frac{3}{5} \frac{GM^2}{R}
\end{equation}
for a uniform cloud of mass $M$. Comparing these two terms, we find that
\begin{equation}
\mathcal{B}+\mathcal{W} = \frac{\Phi_B^2}{6\pi^2 R} - \frac{3}{5} \frac{GM^2}{R} \equiv \frac{3}{5}\frac{G}{R} \left(M_{\Phi}^2-M^2\right)
\end{equation}
where
\begin{equation}
\label{eq:mphi}
M_{\Phi} \equiv \sqrt{\frac{5}{2}} \left(\frac{\Phi_B}{3 \pi G^{1/2}}\right)
\end{equation}
We call $M_{\Phi}$ the magnetic critical mass. Since $\Phi_B$ does not change as a cloud expands or contracts (due to flux-freezing), this magnetic critical mass does not change either.

The implication of this is that clouds that have $M>M_{\Phi}$ always have $\mathcal{B}+\mathcal{W} < 0$. The magnetic force is unable to halt collapse no matter what. Clouds that satisfy this condition are called magnetically supercritical, because they are above the magnetic critical mass $M_{\Phi}$. Conversely, if $M<M_{\Phi}$, then $\mathcal{B}+\mathcal{W} > 0$, and gravity is weaker than magnetism. Clouds satisfying this condition are called subcritical.

For a subcritical cloud, since $\mathcal{B}+\mathcal{W} \propto 1/R$, this term will get larger and larger as the cloud shrinks. In other words, not only is the magnetic force resisting collapse is stronger than gravity, it becomes larger and larger without limit as the cloud is compressed to a smaller radius. Unless the external pressure is also able to increase without limit, which is unphysical, then there is no way to make a magnetically subcritical cloud collapse. It will always stabilize at some finite radius. The only way to get around this is to change the magnetic critical mass, which requires changing the magnetic flux through the cloud. This is possible only via ion-neutral drift or some other non-ideal MHD effect that violates flux-freezing.

Of course our calculation is for a somewhat artificial configuration of a spherical cloud with a uniform magnetic field. In reality a magnetically-supported cloud will not be spherical, since the field only supports it in some directions, and the field will not be uniform, since gravity will always bend it some amount. Figuring out the magnetic critical mass in that case requires solving for the cloud structure numerically. A calculation of this effect by \citet{tomisaka98a} gives
\begin{equation}
M_{\Phi} = 0.12\frac{\Phi_B}{G^{1/2}}
\end{equation}
for clouds for which pressure support is negligible. The numerical coefficient we obtained for the uniform cloud case (equation \ref{eq:mphi}) is $0.17$, so this is obviously a small correction. It is also possible to derive a combined critical mass that incorporates both the flux and the sound speed, and which limits to the Bonnor-Ebert mass for negligible field and the magnetic critical mass for negligible pressure.

\begin{figure}
\includegraphics[width=\linewidth]{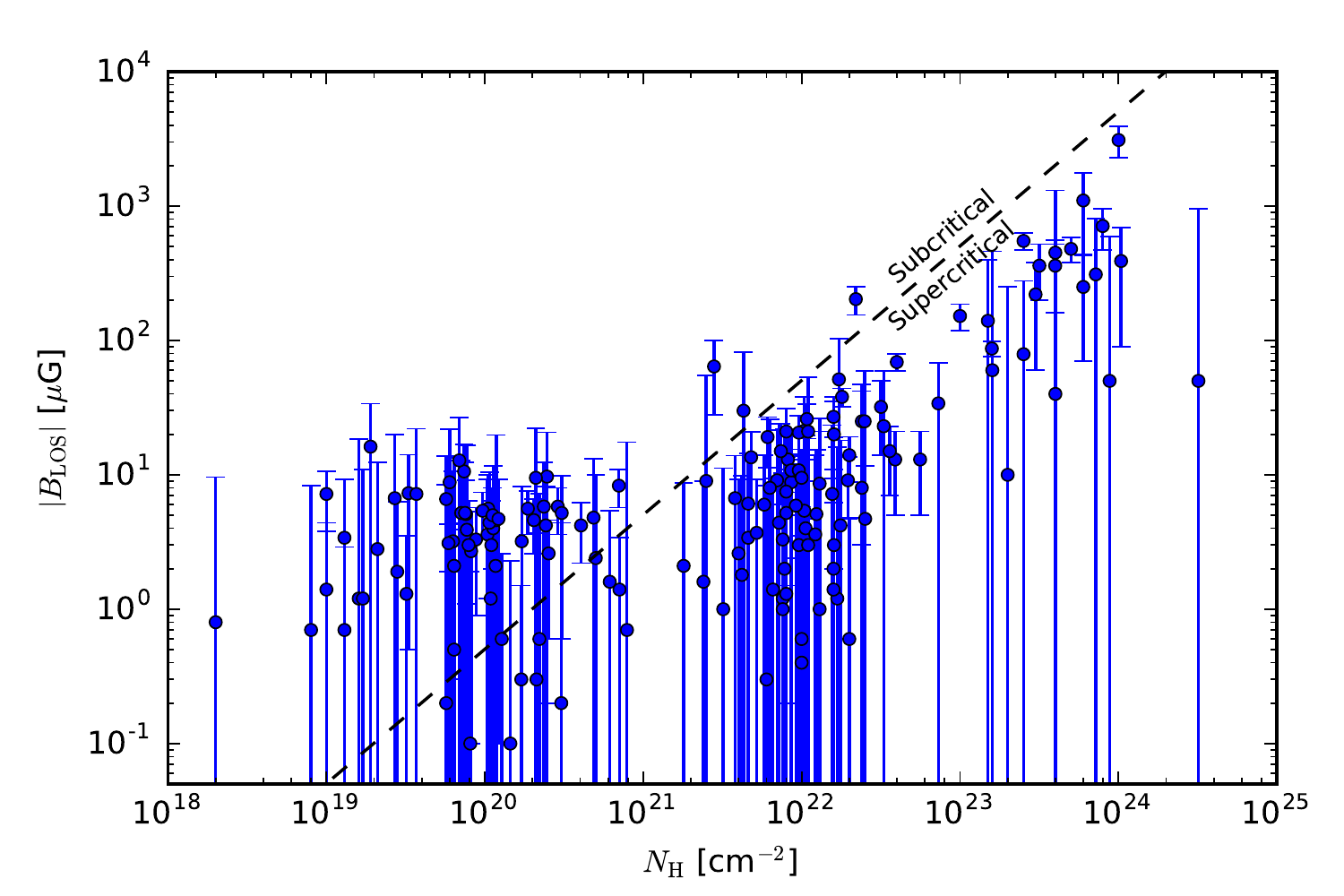}
\caption[Magnetic field strength measurements]{
\label{fig:bfields}
Measurements of the line of sight magnetic field strength from the Zeeman effect, versus total gas column density in H atoms cm$^{-2}$ (data from the compilation of \citealt{crutcher12a}). The three clumps of points represent, from left to right, measurements from the Zeeman splitting of H~\textsc{i}, OH, and CN. The dashed black line indicates the separation between field strengths that are large enough to render the gas subcritical, and those weak enough for it to be supercritical.
}
\end{figure}

It is not so easy to determine observationally whether the magnetic fields are strong enough to hold up molecular clouds.  The observations are somewhat complicated by the fact that, using the most common technique of Zeeman splitting, one can only measure the line of sight component of the field. This therefore gives only a lower limit on the magnetic critical mass. Nonetheless, for a large enough sample, one can estimate true magnetic field strengths statistically under the assumption of random orientations. When this analysis is performed, the current observational consensus is that magnetic fields in molecular clouds are not, by themselves, strong enough to prevent gravitation collapse. Figure \ref{fig:bfields} shows a summary of the current observations. Clearly atomic gas is magnetically subcritical, but molecular gas is supercritical.

\subsection{Turbulent Support}

There is one more positive term in the virial theorem, which is the turbulent component of $\mathcal{T}$. This one is not at all well understood, largely because we don't understand turbulence itself. This term almost certainly provides some support against collapse, but the amount is not well understood, and we will defer any further discussion of this effect until we get to our discussions of the star formation rate in Chapter \ref{ch:sflaw_th}.

\section{Pressureless Collapse}
\label{sec:pressureless_collapse}

As a final topic for this chapter, let us consider what we should expect to happen if gas does begin to collapse, in the simplest case of an initially-spherical cloud with an initial density distribution $\rho(r)$. We would like to know how the gas moves under the influence of gravity and thermal pressure, under the assumption of spherical symmetry. For convenience we define the enclosed mass
\begin{equation}
M_r =  \int_0^r 4\pi r'^2 \rho(r') \, dr'
\end{equation}
or equivalently
\begin{equation}
\frac{\partial M_r}{\partial r} = 4\pi r^2 \rho.
\end{equation}
The equation of mass conservation for the gas in spherical coordinates is
\begin{eqnarray}
\frac{\partial}{\partial t} \rho + \nabla\cdot (\rho \vecv) & = & 0 \\
\frac{\partial}{\partial t} \rho + \frac{1}{r^2}\frac{\partial}{\partial r}(r^2 \rho v) & = & 0,
\end{eqnarray}
where $v$ is the radial velocity of the gas. It is useful to write the equations in terms of $M_r$ rather than $\rho$, so we take the time derivative of $M_r$ to get
\begin{eqnarray}
\frac{\partial}{\partial t}M_r & = & 4\pi \int_{0}^{r} r'^2 \frac{\partial}{\partial t} \rho \,dr' \\
& = & -4\pi \int_{0}^{r} \frac{\partial}{\partial r'}(r'^2 \rho v)\, dr' \\
& = & -4\pi r^2 \rho v \\
& = & -v \frac{\partial}{\partial r}M_r.
\end{eqnarray}
In the second step we used the mass conservation equation to substitute for $\partial \rho/\partial t$, and in the final step we used the definition of $M_r$ to substitute for $\rho$.

To figure out how the gas moves, we write down the Lagrangean version of the momentum equation:
\begin{equation}
\rho \frac{Dv}{Dt} = -\frac{\partial}{\partial r}P - \mathbf{f}_g,
\end{equation}
where $\mathbf{f}_g$ is the gravitational force. For the momentum equation, we take advantage of the fact that the gas is isothermal to write $P=\rho c_s^2$. The gravitational force is $\mathbf{f}_g = -G M_r / r^2$. Thus we have
\begin{equation}
\frac{Dv}{Dt}= \frac{\partial}{\partial t}v + v\frac{\partial}{\partial r} v= -\frac{c_s^2}{\rho} \frac{\partial}{\partial r}{\rho} - \frac{G M_r}{r^2}.
\end{equation}

To go further, let us make one more simplifying assumption: that the sound speed $c_s$ is zero. This is not as bad an approximation as one might think. Consider the virial theorem: the thermal pressure term is just proportional to the mass, since the gas sound speed stays about constant. On the other hand, the gravitational term varies as $1/R$. Thus, even if pressure starts out competitive with gravity, as the core collapses the dominance of gravity will increase, and before too long the collapse will resemble a pressureless one.

In this case the momentum equation is trivial:
\begin{equation}
\frac{Dv}{Dt} = -\frac{GM_r}{r^2}.
\end{equation}
This just says that a shell's inward acceleration is equal to the gravitational force per unit mass exerted by all the mass interior to it, which is constant. We can then solve for the velocity as a function of position:
\begin{equation}
v = \dot{r} = -\sqrt{2GM_r}\left(\frac{1}{r_0}-\frac{1}{r}\right)^{1/2},
\end{equation}
where $r_0$ is the position at which a particular fluid element starts. 

The integral can be evaluated by the trigonometric substitution $r=r_0 \cos^2\xi$. The solution, first obtained by Hunter (1962), is
\begin{eqnarray}
-2 r_0 (\cos\xi \sin\xi) \dot{\xi} & = & -\sqrt{\frac{2GM_r}{r_0}} \left(\frac{1}{\cos^2\xi}-1\right)^{1/2} \\
2 (\cos\xi\sin\xi) \dot{\xi} & = & \sqrt{\frac{2GM_r}{r_0^3}}\tan\xi \\
2 \cos^2\xi\, d\xi & = & \sqrt{\frac{2GM_r}{r_0^3}} dt \\
\xi+\frac{1}{2}\sin 2\xi & = & t \sqrt{\frac{2GM_r}{r_0^3}}.
\end{eqnarray}
We are interested in the time at which a given fluid element reaches the origin, $r=0$. This corresponds to $\xi = \pi/2$, so this time is
\begin{equation}
t = \frac{\pi}{2}\sqrt{\frac{r_0^3}{2 G M_r}}.
\end{equation}

Suppose that the gas we started with was of uniform density $\rho$, so that $M_r = (4/3)\pi r_0^3 \rho$.
In this case we have
\begin{equation}
t = t_{\rm ff} = \sqrt{\frac{3\pi}{32 G \rho}},
\end{equation}
where we have defined the free-fall time $t_{\rm ff}$: it is the time required for a uniform sphere of pressureless gas to collapse to infinite density. This is of course just the characteristic growth time for the Jeans instability in the regime of negligible pressure, up to a factor of order unity.

For a uniform fluid this means that the collapse is synchronized -- all the mass reaches the origin at the exact same time. A more realistic case is for the initial state to have some level of central concentration, so that the initial density rises inward. Let us take the initial density profile to be $\rho = \rho_c (r/r_c)^{-\alpha}$, where $\alpha > 0$ so the density rises inward. The corresponding enclosed mass is
\begin{equation}
M_r = \frac{4}{3-\alpha}\pi \rho_c r_c^3 \left(\frac{r}{r_c}\right)^{3-\alpha} 
\end{equation}

Plugging this in, the collapse time is
\begin{equation}
t = \sqrt{\frac{(3-\alpha)\pi}{32 G \rho_c}} \left(\frac{r_0}{r_c}\right)^{\alpha/2}.
\end{equation}
Since $\alpha>0$, this means that the collapse time increases with initial radius $r_0$. This illustrates one of the most basic features of a collapse, which will continue to hold even in the case where the pressure is non-zero. Collapse of centrally concentrated objects occurs inside-out, meaning that the inner parts collapse before the outer parts.

Within the collapsing region near the star, the density profile also approaches a characteristic shape. If the radius of a given fluid element $r$ is much smaller than its initial radius $r_0$, then its velocity is roughly
\begin{equation}
v \approx v_{\rm ff}\equiv -\sqrt{\frac{2GM_r}{r}},
\end{equation}
where we have defined the free-fall velocity $v_{\rm ff}$ as the characteristic speed achieved by an object collapsing freely onto a mass $M_r$. The mass conservation equation is
\begin{equation}
\frac{\partial M_r}{\partial t} = -v\frac{\partial M_r}{\partial r}  = -4\pi r^2 v \rho
\end{equation}

If we are near the star so that $v\approx v_{\rm ff}$, then this implies that
\begin{equation}
\rho = \frac{(\partial M_r/\partial t) r^{-3/2}}{4\pi\sqrt{2 G M_r}}.
\end{equation}
To the extent that we look at a short interval of time, over which the accretion rate does not change much (so that $\partial M_r /\partial t$ is roughly constant), this implies that the density near the star varies as $\rho\propto r^{-3/2}$.

What sort of accretion rate do we expect from a collapse like this? For a core of mass $M_c = [4/(3-\alpha)]\pi \rho_c r_c^3$, the last mass element arrives at the center at a time
\begin{equation}
t_c = \sqrt{\frac{(3-\alpha)\pi}{32 G \rho_c}} = \sqrt{\frac{3-\alpha}{3}}t_{\rm ff}(\rho_c),
\end{equation}
so the time-averaged accretion rate is
\begin{equation}
\langle\dot{M}\rangle = \sqrt{\frac{3}{3-\alpha}} \frac{M_c}{t_{\rm ff}(\rho_c)}.
\end{equation}

In order to get a sense of the numerical value of this, let us suppose that our collapsing object is a marginally unstable Bonnor-Ebert sphere (see Problem Set 2). Such an object does not have negligible pressure, but the pressure will only change the collapse rate at order unity. Problem Set 2 includes a calculation of the structure of a maximum-mass Bonnor-Ebert sphere, so we will just quote the value. The maximum mass is
\begin{equation}
M_{\rm BE} = 1.18 \frac{c_s^4}{\sqrt{G^3 P_s}},
\end{equation}
where $P_s$ is the pressure at the surface of the sphere and $c_s$ is the thermal sound speed in the core.

Let us suppose that the surface of the core, at radius $r_c$, is in thermal pressure balance with its surroundings. Thus $P_s = \rho_c c_s^2$, so we may rewrite the Bonnor-Ebert mass as
\begin{equation}
M_{\rm BE} = 1.18 \frac{c_s^3}{\sqrt{G^3 \rho_c}}.
\end{equation}
A Bonnor-Ebert sphere does not have a powerlaw structure, but if we substitute into our equation for the accretion rate and say that the factor of $\sqrt{3/(3-\alpha)}$ is a number of order unity, we find that the accretion rate is
\begin{equation}
\langle\dot{M}\rangle \approx \frac{c_s^3/\sqrt{G^3\rho_c}}{1/\sqrt{G\rho_c}} = \frac{c_s^3}{G}.
\end{equation}

This is an extremely useful expression, because we know the sound speed $c_s$ from microphysics. Thus, we have calculated the rough accretion rate we expect to be associated with the collapse of any object that is marginally stable based on thermal pressure support. Plugging in $c_s=0.19$ km s$^{-1}$, we get $\dot{M} \approx 2\times 10^{-6}$ $\msun$ yr$^{-1}$ as the characteristic accretion rate for these objects. Since the typical stellar mass is a few tenths of $\msun$, based on the peak of the IMF, this means that the characteristic star formation time is of order $10^5-10^6$ yr. Of course this conclusion about the accretion rate only applies to collapsing objects that are supported mostly by thermal pressure. Other sources of support produce higher accretion rates, as we will see when we get to massive stars.

\chapter{Stellar Feedback}
\label{ch:feedback}

\marginnote{
\textbf{Suggested background reading:}
\begin{itemize}
\item \href{http://adsabs.harvard.edu/abs/2014prpl.conf..243K}{Krumholz, M.~R., et al. 2014, in ``Protostars and Planets VI", ed.~H.~Beuther et al., pp.~243-266} \nocite{krumholz14e}
\end{itemize}
\textbf{Suggested literature:}
\begin{itemize}
\item \href{http://adsabs.harvard.edu/abs/2010ApJ...709..191M}{Murray, N., Quataert, E., \& Thompson, T.~A. 2010, ApJ, 709, 191} \nocite{murray10a}
\item \href{http://adsabs.harvard.edu/abs/2014MNRAS.442..694D}{Dale, J.~E., Ngoumou, J., Ercolano, B., \& Bonnell, I.~A. 2014, MNRAS, 442, 694} \nocite{dale14a}
\end{itemize}
}

The final piece of physics we will cover before moving on to the star formation process itself is the interaction of stellar radiation, winds, and other forms of feedback with the interstellar medium. Our goal in this chapter is to develop a general formalism for describing the various forms of feedback that stars an exert on their environments, and to make an inventory of the most important processes.

\section{General Formalism}

\subsection{IMF-Averaged Yields}

In most cases when considering feedback, we will be averaging over many, many stars. Consequently, it makes sense to focus not on individual stars, but on the collective properties of stellar populations. For this reason, a very useful first step is to consider budgets of mass, momentum, and energy.

We have already encountered a formalism of this sort in our discussion of galactic star formation rate indicators in Chapter \ref{ch:obsstars}, and the idea is similar here. To begin, let us fix the IMF
\begin{equation}
\xi(m) \equiv \frac{dn}{d\ln m},
\end{equation}
with the normalization chosen so that $\int \xi(m) \, dm = 1$. Note that $\xi(m)$ is defined per unit logarithm mass rather than per unit mass, so that it describes the number of stars in a mass range from $\ln m$ to $\ln m + d\ln m$. However, this function also has a second interpretation, since $dn/d\ln m = m (dn/dm)$: this quantity is the total stellar mass found in stars with masses between $m$ and $m+dm$. Consequently, the mean stellar mass is
\begin{equation}
\overline{m} = \frac{\int_{-\infty}^\infty m \xi(m) \, d\ln m}{\int_{-\infty}^\infty \xi(m) \, d\ln m} = \frac{1}{\int_{-\infty}^\infty \xi(m) \, d\ln m},
\end{equation}
where the second step follows from our choice of normalization. The numerator here represents the total mass of the stars, and the denominator is the number of stars. Note that $\xi(m)$ is presumably zero outside some finite interval in mass -- we are writing the limits of integration as $-\infty$ to $\infty$ only for convenience.

We will further assume that, from stellar evolution, we know the rate $q$ at which stars produce some quantity $Q$ as a function of their starting mass and age, where $\dot{Q} = q$. For example if the quantity $Q$ we are concerned with is total radiant energy $E$, then $q$ is the bolometric luminosity $L(m,t)$ of a star of mass $m$ and age $t$. Now consider a population of stars that forms in a single burst at time 0. The instantaneous production rate for these stars is
\begin{equation}
q(t) = M \int_{-\infty}^{\infty} d\ln m \, \xi(m) q(m,t).
\end{equation}
We use this equation to define the IMF-averaged production rate,
\begin{equation}
\left\langle \frac{q}{M}\right\rangle = \int_{-\infty}^{\infty} d\ln m \, \xi(m) q(m,t).
\end{equation}
Note that this rate is a function of the age of the stellar population $t$. We can also define a lifetime-averaged yield. Integrating over all time, the total amount of the quantity produced is
\begin{equation}
Q = M \int_{-\infty}^{\infty} d\ln m \, \xi(m) \int_0^\infty dt\, q(M,t).
\end{equation}
We therefore define the IMF-averaged yield
\begin{equation}
\left\langle \frac{Q}{M} \right\rangle = \int_{-\infty}^{\infty} d\ln m \, \xi(m) \int_0^\infty dt\, q(M,t).
\end{equation}
The meaning of these quantities is that $\langle q/M\rangle$ is the instantaneous rate at which the stars are producing $Q$ per unit stellar mass, and $\langle Q/M\rangle$ is the total amount produced per unit mass of stars formed over the stars' entire lifetimes.

In practice we cannot really integrate to infinity for most quantities, since the lifetimes of some stars may be very, very long compared to what we are interested in. For example the luminous output of a stellar population will have a large contribution for $\sim 5$ Myr coming from massive stars, which is mostly what is of interest. However, if we integrate for $1000$ Gyr, we will find that the luminous output is dominated by the vast numbers of $\sim 0.2$ $\msun$ stars near the peak of the IMF that are fully convective and thus are able to burn all of their hydrogen to He. In reality, though, this is longer than the age of the Universe. In practice, therefore, we must define our lifetime averages as cutting off after some finite time.

It can also be useful to define a different IMF average. The quantities we have discussed thus far are yields per unit mass that goes into stars. Sometimes we are instead interested in the yield per unit mass that stays locked in stellar remnants for a long time, rather than the mass that goes into stars for $\sim 3-4$ Myr and then comes back out in supernovae. Let us define the mass of the remnant that a star of mass $m$ leaves as $w(m)$. If the star survives for a long time, $w(m) = m$. In this case, the mass that is ejected back into the ISM is
\begin{equation}
M_{\rm return} = M \int_{-\infty}^{\infty} d\ln m \, \xi(m) [m - w(m)] \equiv R M,
\end{equation}
where we define $R$ as the return fraction. The mass fraction that stays locked in remnants is $1-R$.

Of course ``long time" here is a vague term. By convention (defined by \citealt{tinsley80a}), we choose to take $w(m) = m$ for $m=1$ $\msun$. We take $w(m) = 0.7$ $\msun$ for $m=1-8$ $\msun$ and $w(m) = 1.4$ $\msun$ for $m>8$ $\msun$, i.e., we assume that stars from $1-8$ $\msun$ leave behind $0.7$ $\msun$ white dwarfs, and stars larger than that mass form $1.4$ $\msun$ neutron stars. If one puts this in for a \citet{chabrier05a} IMF, the result is $R=0.46$, meaning that these averages are larger by a factor of $1/0.54$.

Given this formalism, it is straightforward to use a set of stellar evolutionary tracks plus an IMF to compute $\langle q/M\rangle$ or $\langle Q/M\rangle$ for any quantity of interest. Indeed, this is effectively what starburst99 \citep{leitherer99a} and programs like it do. The quantities of greatest concern for massive star feedback are the bolometric output, ionizing photon output, wind momentum and energy output, and supernova output.

\subsection{Energy- versus Momentum-Driven Feedback}

Before discussing individual feedback mechanisms in detail, it is also helpful to lay out two general categories that can be used to understand them. Let us consider a population of stars surrounded by initially-uniform interstellar gas. Those stars eject both photons and baryons (in the form of stellar winds and supernova ejecta) into the surrounding gas, and these photons and baryons carry both momentum and energy. We want to characterize how the ISM will respond.

One important consideration is that, as we have already shown, it is very hard to raise the temperature of molecular gas (or even dense atomic gas) because it is able to radiate so efficiently. A factor of $\sim 10$ increase in the radiative heating rate might yield only a tens of percent increase in temperature. This is true as long as the gas is cold and dense, but at sufficiently high temperatures or if the gas is continuously illuminated then the cooling rate begins to drop off, and it is possible for gas to remain hot.

A critical distinction is therefore between mechanisms that are able to keep the gas hot for a time that is long enough to be significant (generally of order the crossing time of the cloud or longer), and those where the cooling time is much shorter. For the latter case, the energy delivered by the photons and baryons will not matter, only the momentum delivered will. The momentum cannot be radiated away. We refer to feedback mechanism where the energy is lost rapidly as momentum-driven feedback, and to the opposite case where the energy is retained for at least some time as energy-driven, or explosive, feedback.

To understand why the distinction between the two is important, let us consider two extreme limiting cases. We place a cluster of stars at the origin and surround it by a uniform region of gas with density $\rho$. At time $t=0$, the stars "turn on" and begin emitting energy and momentum, which is then absorbed by the surrounding gas. Let the momentum and energy injection rates be $\dot{p}_w$ and $\dot{E}_w$; it does not matter if the energy and momentum are carried by photons or baryons, so long as the mass swept up is significantly greater than the mass carried by the wind.

The wind runs into the surrounding gas and causes it to begin moving radially outward, which in turn piles up material that is further away, leading to an expanding shell of gas. Now let us compute the properties of that shell in the two extreme limits of all the energy being radiated away, and all the energy being kept. If all the energy is radiated away, then at any time the radial momentum of the shell must match the radial momentum injected up to that time, i.e.,
\begin{equation}
p_{\rm sh} = M_{\rm sh} v_{\rm sh} = \dot{p}_w t.
\end{equation}
The kinetic energy of the shell is
\begin{equation}
E = \frac{p_{\rm sh}^2}{2 M_{\rm sh}} = \frac{1}{2} v_{\rm sh} \dot{p}_w t.
\end{equation} 
For comparison, if none of the energy is radiated away, the energy is simply
\begin{equation}
E = \dot{E}_w t.
\end{equation}
Thus the energy in the energy-conserving case is larger by a factor of
\begin{equation}
\frac{1}{v_{\rm sh}} \cdot \frac{2\dot{E}_w}{\dot{p}_w}.
\end{equation}
If the energy injected by the stars is carried by a wind of baryons, then $2\dot{E}_w/\dot{p}_w$ is simply the speed of that wind, while if it is carried by photons, then $2\dot{E}_w/\dot{p}_w = 2 c$. Thus the energy in the energy-conserving case is larger by a factor of $2c/v_{\rm sh}$ for a photon wind, and $v_w/v_{\rm sh}$ for a baryon wind. These are not small factors: observed expanding shells typically have velocities of at most a few tens of km s$^{-1}$, while wind speeds from massive stars, for example, can be thousands of km s$^{-1}$. Thus it matters a great deal where a particular feedback mechanism lies between the energy- and momentum-conserving limits.

\section{Momentum-Driven Feedback Mechanisms}

We are now ready to consider individual mechanisms by which stars can deliver energy and momentum to the gas around them. Our goal is to understand what forms of feedback are significant and to estimate their relative budgets of momentum and energy.

\subsection{Radiation Pressure and Radiatively-Driven Winds}

The simplest form of feedback to consider is radiation pressure. Since the majority of the radiant energy deposited in the ISM will be re-radiated immediately, radiation pressure is (probably) a momentum-driven feedback. To evaluate the momentum it deposits, one need merely evaluate the integrals over the IMF we have written down using the bolometric luminosities of stars. \citet{murray10b} find
\begin{equation}
\left\langle \frac{L}{M}\right \rangle = 1140 \,\lsun\,\msun^{-1} = 2200\mbox{ erg g}^{-1},
\end{equation}
and the corresponding momentum injection rate is
\begin{equation}
\left\langle \frac{\dot{p}_{\rm rad}}{M}\right \rangle = \frac{1}{c} \left\langle \frac{L}{M}\right \rangle = 7.3\times 10^{-8}\mbox{ cm s}^{-2} = 23\mbox{ km s}^{-1}\mbox{ Myr}^{-1}
\end{equation}
The physical meaning of this expression is that for every gram of matter that goes into stars, those stars produce enough light over 1 Myr to accelerate another gram of matter to a speed of 23 km s$^{-1}$. For very massive stars, radiation pressure also accelerates winds off the star's surfaces; for such stars, the wind carries a bit under half the momentum of the radiation field. Including this factor raises the estimate by a few tens of percent.  However, these winds may also be energy conserving, a topic we will approach momentarily.

Integrated over the lifetimes of the stars, out 100 Myr the total energy production is
\begin{equation}
\left\langle \frac{E_{\rm rad}}{M}\right\rangle = 1.1\times 10^{51}\mbox{ erg}\,\msun^{-1}
\end{equation}
The majority of this energy is produced in the first $\sim 5$ Myr of a stellar population's life, when the massive stars live and die.

It is common to quote the energy budget in units of $c^2$, which gives a dimensionless efficiency with which stars convert mass into radiation. Doing so gives
\begin{equation}
\epsilon = \frac{1}{c^2} \left\langle \frac{E_{\rm rad}}{M}\right\rangle = 6.2\times 10^{-4}.
\end{equation}
The radiation momentum budget is simply this over $c$,
\begin{equation}
\left\langle \frac{p_{\rm rad,tot}}{M}\right\rangle = 190\mbox{ km s}^{-1}.
\end{equation}
This is an interesting number, since it is not all that different than the circular velocity of a spiral galaxy like the Milky Way. It is a suggestion that the radiant momentum output by stars may be interesting in pushing matter around in galaxies -- probably not by itself, but perhaps in conjunction with other effects.

\subsection{Protostellar Winds}

A second momentum-driven mechanism, that we will discuss in more detail in Chapters \ref{ch:disks_obs} and \ref{ch:disks_theory}, is protostellar jets. All accretion disks appear to produce some sort of wind that carries away some of the mass and angular momentum, and protostars are no exception. The winds from young stars carry a mass flux of order a few tens of percent of the mass coming into the stars, and eject it with a velocity of order the Keplerian speed at the stellar surface. Note that these winds are distinct from the radiatively-driven ones that come from main sequence O stars. They are very different in both their driving mechanism and physical characteristics.

Why do we expect protostellar winds to be a momentum-driven feedback mechanism instead of an energy-driven one? The key lies in their characteristic speeds. Consider a star of mass $M_*$ and radius $R_*$. Its wind will move at a speed of order
\begin{equation}
v_w \sim \sqrt{\frac{GM_*}{R_*}} = 250\mbox{ km s}^{-1}\left(\frac{M_*}{M_\odot}\right)^{1/2} \left(\frac{R_*}{3R_\odot}\right)^{-1/2},
\end{equation}
where the scalings are for typical protostellar masses and radii. The kinetic energy per unit mass carried by the wind is $v_w^2/2$, and when the wind hits the surrounding ISM it will shock and this kinetic energy will be converted to thermal energy. We can therefore find the post-shock temperature from energy conservation. The thermal energy per unit mass is
$(3/2) k_B T/\mu m_{\rm H}$, where $\mu$ is the mean particle mass in H masses. Thus the post-shock temperature will be
\begin{equation}
T = \frac{\mu m_{\rm H} v_w^2}{3 k_B} \sim 5 \times 10^6\mbox{ K}
\end{equation}
for the fiducial speed above, where we have used $\mu=0.61$ for fully ionized gas. This is low enough that gas at this temperature will be able to cool fairly rapidly, leaving us in the momentum-conserving limit.

So how much momentum can we extract? To answer that, we will use our formalism for IMF averaging. Let us consider stars forming over some timescale $t_{\rm form}$. This can be a function of mass if we wish. Similarly, let us assume for simplicity that the accretion rate during the formation stage is constant; again, this assumption actually makes no difference to the result, it just makes the calculation easier. Thus a star of mass $m$ accretes at a rate $\dot{m} = m/t_{\rm form}$ over a time $t_{\rm form}$, and during this time it produces a wind with a mass flux $f \dot{m}$ that is launched with a speed $v_K$. Thus IMF-averaged yield of wind momentum is
\begin{equation}
\left\langle\frac{p_w}{M}\right\rangle = \int_{-\infty}^{\infty} d\ln m \, \xi(m) \, \int_0^{t_{\rm form}} dt \, \frac{f m v_K}{t_{\rm form}}.
\end{equation}
In reality $v_K$, $f$, and the accretion rate probably vary over the formation time of a star, but to get a rough answer we can assume that they are constant, in which case the integral is trivial and evaluates to
\begin{equation}
\left\langle\frac{p_w}{M}\right\rangle =  f v_K \int_{-\infty}^{\infty} d\ln m \, \xi(m) m = f v_K
\end{equation}
where the second step follows from the normalization of the IMF. Thus we learn that winds supply momentum to the ISM at a rate of order $f v_K$. Depending on the exact choices of $f$ and $v_K$, this amounts to a momentum supply of a few tens of km s$^{-1}$ per unit mass of stars formed.

Thus in terms of momentum budget, protostellar winds carry over the full lifetimes of the stars that produce them about as much momentum as is carried by the radiation each Myr. Thus if one integrates over the full lifetime of even a very massive, short-lived star, it puts out much more momentum in the form of radiation than it does in the form of outflows. So why worry about outflows at all, in this case?

There are two reasons. First, because the radiative luminosities of stars increase steeply with stellar mass, the luminosity of a stellar population is dominated by its few most massive members. In small star-forming regions with few or no massive stars, the radiation pressure will be much less than our estimate, which is based on assuming full sampling of the IMF, suggests. On the other hand, protostellar winds produce about the same amount of momentum per unit mass accreted no matter what stars is doing the accreting -- this is just because $v_K$ is not a very strong function of stellar mass. (This is a bit of an oversimplification, but it is true enough for this purpose.) This means that winds will be significant even in regions that lack massive stars, because they can be produced by low-mass stars too.

Second, while outflows carry less momentum integrated over stars' lifetimes, when they are on they are much more powerful. Typical formation times, we shall see, are of order a few times $10^5$ yr, so the instantaneous production rate of outflow momentum is typically $\sim 100$ km s$^{-1}$ Myr$^{-1}$, a factor of several higher than radiation pressure. Thus winds can dominate over radiation pressure significantly during the short phase when they are on.

\section{(Partly) Energy-Driven Feedback Mechanisms}

\subsection{Ionizing Radiation}

Massive stars produce significant amounts of ionizing radiation. From \citet{murray10b}, the yield of ionizing photons from a zero-age population is
\begin{equation}
\left\langle\frac{S}{M}\right\rangle = 6.3\times 10^{46}\mbox{ photons s}^{-1}\,M_\odot^{-1}.
\end{equation}
The corresponding lifetime-averaged production of ionizing photons is
\begin{equation}
\left\langle \frac{S_{\rm tot}}{M}\right\rangle = 4.2\times 10^{60}\mbox{ photons}\,\msun^{-1}.
\end{equation}

\paragraph{H~\textsc{ii} Region Expansion}

We will not go into tremendous detail on how these photons interact with the ISM, but to summarize: photons capable of ionizing hydrogen will be absorbed with a very short mean free path, producing a bubble of fully ionized gas within which all the photons are absorbed. The size of this bubble can be found by equating the hydrogen recombination rate with the ionizing photon production rate, giving
\begin{equation}
S = \frac{4}{3} \pi r_i^3 n_e n_p \alphab,
\end{equation}
where $r_i$ is the radius of the ionized region, $n_e$ and $n_p$ are the number densities of electrons and protons, and $\alphab$ is the recombination rate coefficient for case B, and which has a value of roughly $3\times 10^{-13}$ cm$^3$ s$^{-1}$. Cases A and B, what they mean, and how this quantity is computed, are all topics discussed at length in standard ISM references such as \citet{osterbrock06a} and \citet{draine11a}, and here we will simply take $\alphab$ as a known constant.

The radius of the ionized bubble is known as the Str\"omgren radius after Bengt Str\"omgren, the person who first calculated it. If we let $\mu\approx 1.4$ be the mean mass per hydrogen nucleus in the gas in units of $m_{\rm H}$, and $\rho_0$ be the initial density before the photoionizing stars turn on, then $n_p = \rho_0/\mu m_{\rm H}$ and $n_e = 1.1 \rho_0/\mu m_{\rm H}$, with the factor of 1.1 coming from assuming that He is singly ionized (since its ionization potential is not that different from hydrogen's) and from a ratio of 10 He nuclei per H nucleus. Inserting these factors and solving for $r_i$, we obtain the Str\"omgren radius, the equilibrium radius of a sphere of gas ionized by a central source:
\begin{equation}
r_S = \left(\frac{3 S \mu^2 m_{\rm H}^2}{4(1.1) \pi \alphab \rho_0^2}\right)^{1/3} = 2.8 S_{49}^{1/3} n_2^{-2/3}\mbox{ pc},
\end{equation}
where $S_{49} = S/10^{49}$ s$^{-1}$, $n_2 = (\rho_0/\mu m_{\rm H})/100$ cm$^{-3}$, and we have used $\alphab = 3.46\times 10^{-13}$ cm$^3$ s$^{-1}$, the value for a gas at a temperature of $10^4$ K.

The photoionized gas will be heated to $\approx 10^4$ K by the energy deposited by the ionizing photons. The corresponding sound speed in the ionized gas will be
\begin{equation}
c_i = \sqrt{2.2 \frac{k_B T_i}{\mu m_{\rm H}}} = 11 T_{i,4}^{1/2}\mbox{ km s}^{-1},
\end{equation}
where $T_{i,4} = T_i/10^4$ K, and the factor of $2.2$ arises because there are 2.2 free particles per H nucleus (0.1 He per H, and 1.1 electrons per H). The pressure in the ionized region is $\rho_0 c_i^2$, which is generally much larger than the pressure $\rho_0 c_0^2$ outside the ionized region, where $c_0$ is the sound speed in the neutral gas. As a result, the ionized region is hugely over-pressured compared to the neutral gas around it. The gas in this region will therefore begin to expand dynamically.

The time to reach ionization balance is short compared to dynamical timescales, so we can assume that ionization balance is always maintained as the expansion occurs. Consequently, when the ionized region has reached a radius $r_i$, the density inside the ionized region must obey
\begin{equation}
\rho_i = \left[\frac{3 S \mu^2 m_{\rm H}^2}{4(1.1)\pi \alphab r_i^3}\right]^{1/2}.
\end{equation}
At the start of expansion $\rho_i = \rho_0$, but we see here that the density drops as $r_i^{-3/2}$ as expansion proceeds. Since the expansion is highly supersonic with respect to the external gas (as we will see shortly), there is no time for sound waves to propagate away from the ionization front and pre-accelerate the neutral gas. Instead, this gas must be swept up by the expanding H~\textsc{ii} region. However, since $\rho_i \ll \rho_0$, the mass that is swept up as the gas expands must reside not in the ionized region interior, but in a dense neutral shell at its edges. At late times, when $r_i \gg r_S$, we can neglect the mass in the shell interior in comparison to that in the shell, and simply set the shell mass equal to the total mass swept up. We therefore have a shell mass
\begin{equation}
M_{\rm sh} = \frac{4}{3} \pi \rho_0 r_i^3.
\end{equation}

We can write down the equation of motion for this shell. If we neglect the small ambient pressure, then the only force acting on the shell is the pressure $\rho_i c_i^2$ exerted by ionized gas in the H~\textsc{ii} region interior. Conservation of momentum therefore requires that
\begin{equation}
\frac{d}{dt} \left(M_{\rm sh} \dot{r}_i\right) = 4\pi r_i^2 \rho_i c_i^2.
\end{equation}
Rewriting everything in terms of $r_i$, we arrive at an ordinary differential equation for $r_i$:
\begin{equation}
\frac{d}{dt} \left(\frac{1}{3} r_i^3 \dot{r}_i\right) = c_i^2 r_i^2 \left(\frac{r_i}{r_S}\right)^{-3/2},
\end{equation}
where we have used the scaling $\rho_i = \rho_0 (r_i/r_S)^{-3/2}$.

This ODE is straightforward to solve numerically, but if we focus on late times when $r_i \gg r_S$, we can solve it analytically. For $r_i \gg r_S$, we can take $r_i \approx 0$ as $t\rightarrow 0$, and with this boundary condition the ODE cries out for a similarity solution. As a trial, consider $r_i = f r_S (t/t_S)^\eta$, where 
\begin{equation}
t_S = \frac{r_S}{c_i} = 240  S_{49}^{1/3} n_2^{-2/3} T_{i,4}^{-1/2}\mbox{ kyr}
\end{equation}
and $f$ is a dimensionless constant. Substituting this trial solution in, there are numerous cancellations, and in the end we obtain
\begin{equation}
\frac{1}{4} \eta (4\eta-1) f^4 \left(\frac{t}{t_S}\right)^{4\eta-2} = f^{1/2} \left(\frac{t}{t_S}\right)^{\eta/2}.
\end{equation}
Clearly we can obtain a solution only if $4\eta-2 = \eta/2$, which requires $\eta = 4/7$. Solving for $f$ gives $f=(49/12)^{2/7}$. We therefore have a solution
\begin{equation}
r_i = r_S \left(\frac{7 t}{2\sqrt{3} t_S}\right)^{4/7} = 9.4 S_{49}^{1/7} n_2^{-2/7} T_{i,4}^{2/7} t_6^{4/7}\mbox{ pc}
\end{equation}
at late times, where $t_6 = t/1$ Myr.

\paragraph{Feedback Effects of H~\textsc{ii} Regions}

Given this result, what can we say about the effects of an expanding H~\textsc{ii} region? There are several possible effects: ionization can eject mass, drive turbulent motions, and possibly even disrupt clouds entirely. First consider mass ejection. In our simple calculation, we have taken the ionized gas to be trapped inside a spherical H~\textsc{ii} region interior. In reality, though, once the H~\textsc{ii} region expands to the point where it encounters a low density region at a cloud edge, it will turn into a "blister" type region, and the ionized gas will freely escape into the low density medium.\footnote{This is a case where the less pleasant nomenclature has won out. Such flows are sometimes also called "champagne" flows, since the ionized gas bubbles out of the dense molecular cloud like champagne escaping from a bottle neck. However, the more common term in the literature these days appears to be blister. What this says about the preferences and priorities of the astronomical community is left as an exercise for the reader.} The mass flux carried in this ionized wind will be roughly
\begin{equation}
\dot{M} = 4\pi r_i^2 \rho_i c_i,
\end{equation}
i.e., the area from which the wind flows times the characteristic density of the gas at the base of the wind times the characteristic speed of the wind. Substituting in our similarity solution, we have
\begin{equation}
\dot{M} = 4\pi r_S^2 \rho_0 c_i \left(\frac{7t}{2\sqrt{3} t_S}\right)^{2/7} = 7.2\times 10^{-3} t_6^{2/7} S_{49}^{4/7} n_2^{-1/7} T_{i,4}^{1/7}\,\msun\mbox{ yr}^{-1}.
\end{equation}
We therefore see that, over the roughly $3-4$ Myr lifetime of an O star, it can eject $\sim 10^3 - 10^4$ $\msun$ of mass from its parent cloud, provided that cloud is at a relatively low density (i.e., $n_2$ is not too big). Thus massive stars can eject many times their own mass from a molecular cloud. In fact, some authors have used this effect to make an estimate of the star formation efficiency in GMCs \citep[e.g.,][]{matzner02a}.

We can also estimate the energy contained in the expanding shell. This is
\begin{eqnarray}
E_{\rm sh} & = & \frac{1}{2} M_{\rm sh} \dot{r}_i^2 = \frac{32}{147}\pi \rho_0 \frac{r_S^5}{t_S^2} \left(\frac{7t}{2\sqrt{3} t_S}\right)^{6/7}
\nonumber \\
& = & 8.1\times 10^{47} t_6^{6/7} S_{49}^{5/7} n_2^{-10/7} T_{i,4}^{10/7}\mbox{ erg}.
\end{eqnarray}
For comparison, the gravitational binding energy of a $10^5$ $\msun$ GMC with a surface density of $0.03$ g cm$^{-2}$ is $\sim 10^{50}$ erg. Thus a single O star's H~\textsc{ii} region provides considerably less energy than this. On the other hand, the collective effects of $\sim 10^2$ O stars, with a combined ionizing luminosity of $10^{51}$ s$^{-1}$ or so, can begin to produce H~\textsc{ii} regions whose energies rival the binding energies of individual GMCs. This means that H~\textsc{ii} region shells may sometimes be able to unbind GMCs entirely. Even if they cannot, they may be able to drive significant turbulent motions within GMCs.

We can also compute the momentum of the shell, for comparison to the other forms of feedback we discussed previously. This is
\begin{equation}
p_{\rm sh} = M_{\rm sh} \dot{r}_i = 1.1\times 10^5 n_2^{-1/7} T_{i,4}^{-8/7} S_{49}^{4/7} t_6^{9/7} \, M_\odot\mbox{ km s}^{-1}.
\end{equation}
Since this is non-linear in $S_{49}$ and in time, the effects of HII regions will depend on how the stars are clustered together, and how long they live. To get a rough estimate, though, we can take the typical cluster to have an ionizing luminosity around $10^{49}$ s$^{-1}$, since by number most clusters are small, and we can adopt an age of 4 Myr. This means that (also using $n_2 = 1$ and $T_{i,4} = 1$) the momentum injected per $10^{49}$ photons s$^{-1}$ of luminosity is $p = 3-5\times 10^5$ $M_\odot$ km s$^{-1}$. Recalling that we get $6.3\times 10^{46}$ photons s$^{-1}$ $M_\odot^{-1}$ for a zero-age population, this means that the momentum injected per unit stellar mass for HII regions is roughly
\begin{equation}
\left\langle\frac{p_{\rm HII}}{M}\right\rangle \sim 3\times 10^3\mbox{ km s}^{-1}.
\end{equation}
This is obviously a very rough calculation, and it can be done with much more sophistication, but this analysis suggests that H~\textsc{ii} regions are likely the dominant feedback mechanism compared to winds and H~\textsc{ii} regions.

There is one important caveat to make, though. Although in the similarity solution we formally have $v_i \rightarrow \infty$ as $r_i \rightarrow 0$, in reality the ionized region cannot expand faster than roughly the ionized gas sound speed: one cannot drive a 100 km s$^{-1}$ expansion using gas with a sound speed of 10 km s$^{-1}$. As a result, all of these effects will not work in any cluster for which the escape speed or the virial velocity exceeds $\sim 10$ km s$^{-1}$. This is not a trivial limitation, since for very massive star clusters the escape speed can exceed this value. An example is the R136 cluster in the LMC, which has a present-day stellar mass of $5.5\times 10^4$ $\msun$ inside a radius of $1$ pc \citep{hunter96a}. The escape speed from the stars alone is roughly 20 km s$^{-1}$. Assuming there was gas in the past when the cluster formed, the escape speed must have been even higher. For a region like this, H~\textsc{ii} regions cannot be important.

\subsection{Hot Stellar Winds}

Next let us consider the effects of stellar winds. As we alluded to earlier, O stars launch winds with velocities of $v_w \sim 1000-2500$ km s$^{-1}$ and mass fluxes of $\dot{M}_w \sim 10^{-7}$ $\msun$ yr$^{-1}$. We have already seen that the momentum carried by these winds is fairly unimportant in comparison to the momentum of the protostellar outflows or the radiation field, let alone the momentum provided by H~\textsc{ii} regions. However, because of the high wind velocities, repeating the analysis we performed for protostellar jets yields a characteristic post-shock temperature that is closer to $10^8$ K than $10^6$ K. Gas at such high temperatures has a very long cooling time, so we might end up with an energy-driven feedback. We therefore consider that case next.

Since the winds are radiatively driven, they tend to carry momenta comparable to that carried by the stellar radiation field. The observed correlation between stellar luminosity and wind momentum \citep[e.g.,][]{repolust04a} is that
\begin{equation}
\dot{M}_w v_w \approx 0.5 \frac{L_*}{c},
\end{equation}
where $L_*$ is the stellar luminosity. This implies that the mechanical luminosity of the wind is
\begin{equation}
L_w = \frac{1}{2} \dot{M}_w v_w^2 = \frac{L_*^2}{8 \dot{M}_w c^2} = 850 L_{*,5}^2 \dot{M}_{w,-7}^{-1} \, \lsun.
\end{equation}
This is not much compared to the star's radiant luminosity, but that radiation will mostly not go into pushing the ISM around. The wind, on the other hand might. Also notice that over the integrated power output is
\begin{equation}
E_w = L_w t = 1.0\times 10^{50}  L_{*,5}^2 \dot{M}_{w,-7}^{-1} t_6\mbox{ erg},
\end{equation}
so over the $\sim 4$ Myr lifetime of a very massive star, one with $L_{*,5}\sim 3$, the total mechanical power in the wind is not much less than the amount of energy released when the star goes supernova.

If energy is conserved, and we assume that about half the available energy goes into the kinetic energy of the shell and half is in the hot gas left in the shell interior,\footnote{This assumption is not quite right. See \citet{castor75a} and \citet{weaver77a} for a better similarity solution. However, for an order of magnitude estimate, which is of interest to us, this simple assumption suffices.} conservation of energy then requires that
\begin{equation}
\frac{d}{dt} \left(\frac{2}{3}\pi \rho_0 r_b^3 \dot{r}_b^2\right) \approx \frac{1}{2} L_w.
\end{equation}
As with the H~\textsc{ii} region case, this cries out for similarity solution. Letting $r_b = A t^\eta$, we have
\begin{equation}
\frac{4}{3} \pi \eta^2 (5\eta-2) \rho_0 A^5 t^{5\eta-3} \approx L_w.
\end{equation}
Clearly we must have $\eta=3/5$ and $A=[25 L_w/(12\pi \rho_0)]^{1/5}$. Putting in some numbers,
\begin{equation}
r_b = 16 L_{*,5}^{2/5} \dot{M}_{w,-7}^{-1/5} n_2^{-1/5} t_6^{3/5}\mbox{ pc}.
\end{equation}
Note that this is greater than the radius of the comparable H~\textsc{ii} region, so the wind will initially move faster and drive the H~\textsc{ii} region into a thin ionized layer between the hot wind gas and the outer cool shell -- {\it if the energy-driven limit is correct}. A corollary of this is that the wind would be even more effective than the ionized gas at ejecting mass from the cloud.

However, this may not be correct, because this solution assumes that the energy carried by the wind will stay confined within a closed shell. This may not be the case: the hot gas may instead break out and escape, imparting relatively little momentum. Whether this happens or not is difficult to determine theoretically, but can be address by observations. In particular, if the shocked wind gas is trapped inside the shell, it should produce observable X-ray emission. We can quantify how much X-ray emission we should see with a straightforward argument. It is easiest to phrase this argument in terms of the pressure of the X-ray emitting gas, which is essentially what an X-ray observation measures.

Consider an expanding shell of matter that began its expansion a time $t$ ago. In the energy-driven case, the total energy within that shell is, up to factors of order unity, $E_w = L_w t$. The pressure is simply $2/3$ of the energy density (since the gas is monatomic at these temperatures). Thus,
\begin{equation}
P_X = \frac{2 E_w}{3 [(4/3)\pi r^3]} = \frac{L_*^2 t}{16\pi \dot{M}_w c^2 r^3}.
\end{equation}
It is useful to compute the ratio of this to the pressure exerted by the radiation, which is simply twice that exerted by the wind in the momentum-driven limit. This is
\begin{equation}
P_{\rm rad} = \frac{L_*}{4\pi r^2 c}.
\end{equation}
We define this ratio as the trapping factor:
\begin{equation}
f_{\rm trap} = \frac{P_X}{P_{\rm rad}} = \frac{L_* t}{4\dot{M}_w c r} \approx \frac{L_*}{4\dot{M}_w c v},
\end{equation}
where in the last step we used $v \approx r/t$, where $v$ is the expansion velocity of the shell. If we now use the relation $\dot{M}_w v_w \approx (1/2)L_*/c$, we finally arrive at
\begin{equation}
f_{\rm trap} \approx \frac{v_w}{2v}.
\end{equation}
Thus if shells expand in the energy-driven limit due to winds, the pressure of the hot gas within them should exceed the direct radiation pressure by a factor of roughly $v_w/v$, where $V$ is the shell expansion velocity and $v_w$ is the wind launch velocity. In contrast, the momentum driven limit gives $P_X / P_{\rm rad} \sim 1/2$, since the hot gas exerts a force that is determined by the wind momentum, which is roughly has the momentum carried by the stellar radiation field.

\citet{lopez11a} observed the 30 Doradus H~\textsc{ii} region, which is observed to be expanding with $v\approx 20$ km s$^{-1}$, giving a predicted $f_{\rm trap} = 20$ for a conservative $v_w = 1000$ km s$^{-1}$. They then measured the hot gas pressure from the X-rays and the direct radiation pressure from the stars optical emission. The result is that $f_{\rm trap}$ is much closer to 0.5 than 20 for 30 Doradus, indicating that the momentum-driven solution is closer to reality there. \citet{harper-clark09a} reached a similar conclusion about the Carina Nebula.

\subsection{Supernovae}

We can think of the energy and momentum budget from supernovae as simply representing a special case of the lifetime budgets we've computed. In this case, we can simply think of $q(M,t)$ as being a $\delta$ function: all the energy and momentum of the supernova is released in a single burst at a time $t=t_l(m)$, where $t_l(m)$ is the lifetime of the star in question. We normally assume that the energy yield per star is $10^{51}$ erg, and have to make some estimate of the minimum mass at which a SN will occur, which is roughly 8 $\msun$. We can also, if we want, imagine mass ranges where other things happen, for example direct collapse to black hole, pair instability supernovae that produc more energy, or something more exotic. These choices usually do not make much difference, though, because they affect very massive stars, and since the supernova energy yield (unlike the luminosity) is not a sharp function of mass, the relative rarity of massive stars means they make a small contribution. Thus it usually safe to ignore these effects.

Given this preamble, we can write the approximate supernova energy yield per unit mass as
\begin{equation}
\left\langle \frac{E_{\rm SN}}{M}\right\rangle = E_{\rm SN} \int_{m_{\rm min}}^\infty d\ln m\, \xi(m) \equiv E_{\rm SN} \left\langle \frac{N_{\rm SN}}{M}\right\rangle,
\end{equation}
where $E_{\rm SN} = 10^{51}$ erg is the constant energy per SN, and $m_{\rm min} = 8$ $\msun$ is the minimum mass to have a supernova. Note that the integral, which we have named $\langle N_{\rm SN}/M\rangle$, is simply the number of stars above $m_{\rm min}$ per unit mass in stars total, which is just the expected number of supernovae per unit mass of stars. For a Chabrier IMF from $0.01-120$ $\msun$, we have
\begin{equation}
\left\langle \frac{N_{\rm SN}}{M}\right\rangle = 0.011\, \msun^{-1}
\quad
\left\langle \frac{E_{\rm SN}}{M}\right\rangle = 1.1\times 10^{49}\mbox{ erg}\,\msun^{-1}=6.1\times 10^{-6} c^2.
\end{equation}
A more detailed calculation from starburst99 agrees very well with this crude estimate. Note that this, plus the Milky Way's SFR of $\sim 1$ $\msun$ yr$^{-1}$, is the basis of the oft-quoted result that we expect $\sim 1$ supernova per century in the Milky Way.

The momentum yield from SNe can be computed in the same way. This is slightly more uncertain, because it is easier to measure the SN energy than its momentum -- the latter requires the ability to measure the velocity or mass of the ejecta before they are mixed with significant amounts of ISM. However, roughly speaking the ejection velocity is $v_{\rm ej} \approx 10^9$ cm s$^{-1}$, which means that the momentum is $p_{\rm SN} = 2 E_{\rm SN}/v_{\rm ej}$. Adopting this value, we have
\begin{equation}
\left\langle \frac{p_{\rm SN}}{M}\right\rangle = \frac{2}{v_{\rm ej}} \left\langle \frac{E_{\rm SN}}{M}\right\rangle = 55 v_{\rm ej,9}^{-1} \mbox{ km s}^{-1}.
\end{equation}
Physically, this means that every $\msun$ of matter than goes into stars provides enough momentum to raise another $\msun$ of matter to a speed of 55 km s$^{-1}$. This is not very much compared to other feedbacks, but of course supernovae, like stellar winds, may have an energy-conserving phase where their momentum deposition grows. We will discuss the question of supernova momentum deposition more in Chapter \ref{ch:sflaw_th} in the context of models for regulation of the star formation rate.

\part{Star Formation Processes and Problems}

\chapter{Giant Molecular Clouds}
\label{ch:gmcs}

\marginnote{
\textbf{Suggested background reading:}
\begin{itemize}
\item \href{http://adsabs.harvard.edu/abs/2014prpl.conf....3D}{Dobbs, C.~L., et al. 2014, in ``Protostars and Planets VI", ed.~H.~Beuther et al., pp.~3-26} \nocite{dobbs14a}
\end{itemize}
\textbf{Suggested literature:}
\begin{itemize}
\item \href{http://adsabs.harvard.edu/abs/2014ApJ...784....3C}{Colombo, D., et al. 2014, ApJ, 784, 3} \nocite{colombo14a}
\end{itemize}
}

We now begin our top-down study of star formation, from large to small scales. This chapter focuses on observations of the bulk properties of giant molecular clouds (GMCs), primarily in the Milky Way and in nearby galaxies where we can resolve individual GMCs. The advantage of looking at the Milky Way is of course higher resolution. The advantage of looking at other galaxies is that, unlike in the Milky Way, we can get an unbiased view of all the GMCs, with much smaller distance uncertainties and many fewer confusion problems. This allows us to make statistical inferences that are often impossible to check with confidence locally. This study will be a preparation for the next two chapters, which discuss the correlation of molecular clouds with star formation and the problem of the star formation rate.

\section{Molecular Cloud Masses}

\subsection{Mass Measurement}

The most basic quantity we can measure for a molecular cloud is its mass. However, this also turns out to be one of the trickiest quantities to measure. The most commonly used method for inferring masses is based on molecular line emission, because lines are bright and easy to see even in external galaxies. The three most commonly-used species on the galactic scale are $^{12}$CO, $^{13}$CO, and, more recently, HCN.

\paragraph{Optically Thin Lines}

Conceptually, $^{13}$CO is the simplest, because its lines are generally optically thin. For emitting molecules in LTE at temperature $T$, it is easy to show from the radiative transfer equation that the intensity emitted by a cloud of optical depth $\tau_{\nu}$ at frequency $\nu$ is simply
\begin{equation}
I_{\nu} = \left(1-e^{-\tau_{\nu}}\right) B_{\nu}(T),
\end{equation}
where $B_{\nu}(T)$ is the Planck function evaluated at frequency $\nu$ and temperature $T$.

Although we will not derive this equation here\footnote{See any standard radiative transfer reference, for example \citet{rybicki86a} or \citet{shu91a}.}, it behaves exactly as one would expect intuitively. In the limit of a very optically thick cloud, $\tau_{\nu}\gg 1$, the exponential factor becomes zero, and the intensity simply approaches the Planck function, which is the intensity emitted by a black body. In the limit of a very optically thin cloud, $\tau_{\nu}\ll 1$, the exponential factor just becomes $1-\tau_{\nu}$, so the intensity approaches that of a black body multiplied by the (small) optical depth. Thus the intensity is simply proportional to the optical depth, which is proportional to the number of atoms along the line of sight.

These equations allow the following simple method of deducing the column density from an observation of the $^{13}$CO and $^{12}$CO $J=1\rightarrow 0$ lines (or any similar pair of $J$ lines) from a molecular cloud. If we assume that the $^{12}$CO line is optically thick, as is almost always the case, then we can approximate $1-e^{-\tau_{\nu}}\approx 1$ at line center, so $I_{\nu} \approx B_{\nu}(T)$. If we measure $I_{\nu}$, we can therefore immediately deduce the temperature $T$. We then assume that the $^{13}$CO molecules are at the same temperature, so that $B_{\nu}(T)$ is the same for $^{12}$CO and $^{13}$CO except for the slight shift in frequency. Then if we measure $I_\nu$ for the center of the $^{13}$CO line, we can solve the equation
\begin{equation}
I_{\nu} = \left(1-e^{-\tau_{\nu}}\right) B_{\nu}(T),
\end{equation}
for $\tau_{\nu}$, the optical depth of the $^{13}$CO line. If $N_{\rm ^{13}CO}$ is the column density of $^{13}$CO atoms, then for gas in LTE the column densities of atoms in the level 0 and 1 states are
\begin{eqnarray*}
N_0 & = & \frac{N_{^{13}CO}}{Z} \\
N_1 & = & e^{-T/T_1}\frac{N_{^{13}CO}}{Z}
\end{eqnarray*}
where $Z$ is the partition function, which is a known function of $T$, and $T_1=5.3$ K is the temperature corresponding to the first excited state.

The opacity to line absorption at frequency $\nu$ is
\begin{equation}
\kappa_{\nu} = \frac{h\nu}{4\pi} (n_0 B_{01} - n_1 B_{10}) \phi(\nu),
\end{equation}
where $B_{01}$ and $B_{10}$ are the Einstein coefficients for spontaneous absorption and stimulated emission, defined by 
\begin{eqnarray}
B_{10} & = & \frac{c^2}{2 h \nu^3} A_{10} \\
B_{01} & = & \frac{g_1}{g_0} B_{10}.
\end{eqnarray}
The quantity $\phi(\nu)$ is the line shape function (see Chapter \ref{ch:obscold}). The corresponding optical depth at line center is
\begin{equation}
\tau_{\nu} = \frac{h\nu}{4\pi} (N_0 B_{01} - N_1 B_{10}) \phi(\nu).
\end{equation}
Since we know $\tau_{\nu}$ from the line intensity, we can measure $\phi(\nu)$ just by measuring the shape of the line, and $N_0$ and $N_1$ depend only on $N_{\rm ^{13}CO}$ and the (known) temperature, we can solve for $N_{\rm ^{13}CO}$.  In practice we generally do this in a slightly more sophisticated way, by fitting the optical depth and line shape as a function of frequency simultaneously, but the idea is the same. We can then convert to an H$_2$ column density by assuming a ratio of $^{12}$CO to H$_2$, and of $^{13}$CO to $^{12}$CO.

This method also has some significant drawbacks that are worth mentioning. The need to assume ratios of $^{13}$CO to $^{12}$CO and $^{12}$CO to H$_2$ are obvious ones. The former is particularly tricky, because there is strong observational evidence that the $^{13}$C to $^{12}$C ratio varies with galactocentric radius. We also need to assume that the $^{12}$CO and $^{13}$CO molecules are at the same temperature, which may not be true because the $^{12}$CO emission comes mostly from the cloud surface and the $^{13}$CO comes from the entire cloud. Since the cloud surface is usually warmer than its deep interior, this will tend to make us overestimate the excitation temperature of the $^{13}$CO molecules, and thus underestimate the true column density. This problem can be even worse because the lower abundance of $^{13}$CO means that it cannot self-shield against dissociation by interstellar UV light as effectively at $^{12}$CO. As a result, it may simply not be present in the outer parts of clouds at all, leading us to miss their mass and underestimate the true column density.

Another serious worry is the assumption that the $^{13}$CO molecules are in LTE. As shown in Problem Set 1, the $^{12}$CO $J=1$ state has a critical density of a few thousand cm$^{-3}$, which is somewhat above the mean density in a GMC even when we take into account the effects of turbulence driving mass to high density. The critical density for the $^{13}$CO $J=1$ state is similar. For the $^{12}$CO $J=1$ state, the effective critical density is lowered by optical depth effects, which thermalize the low-lying states. Since $^{13}$CO is optically thin, however, there is no corresponding thermalization for it, so in reality the excitation of the gas tends to be sub-LTE. The result is that the emission is less than we would expect based on an LTE assumption, and so we tend to underestimate the true $^{13}$CO column density, and thus the mass, using this method.

A final point to mention about this method is that, since the $^{13}$CO line is optically thin, it is simply not as bright as an optically thick line would be. Consequently, this method is generally only used within the Galaxy, not for external galaxies.

\paragraph{Optically Thick Lines}

Optically thick lines are nice and bright, so we can see them in distant galaxies. The challenge for an optically thick line is how to infer a mass, given that we are really only seeing the surface of a cloud. Our standard approach here is to define an "X factor": a scaling between the observed frequency-integrated intensity along a given line of sight and the column density of gas along that line of sight. For example, if we see a frequency-integrated CO $J=1\rightarrow 0$ intensity $I_{\rm CO}$ along a given line of sight, we define $X_{\rm CO}=N/I_{\rm CO}$, where $N$ is the true column density (in H$_2$ molecules per cm$^2$) of the cloud. Note that radio astronomers work in horrible units, so the X factor is defined in terms of a velocity-integrated brightness temperature, rather than a frequency-integrated intensity -- specifically, the usual units for $X_{\rm CO}$ are $\mbox{cm}^{-2} / (\mbox{K km s}^{-1})$. The brightness temperature corresponding to a given intensity at frequency $\nu$ is just defined as the temperature of a blackbody that produces that intensity at that frequency. Integrating over velocity just means that we integrate over frequency, but that we measure the frequency in terms of the Doppler shift in velocity it corresponds to.

The immediate question that occurs to us after defining the X factor is: why should such a scaling exist at all? Given that the cloud is optically thick, why should there be a relation between column density and intensity at all? On the face of it, this is a bit like claiming that the brightness of the thermal emission from a wall in a building is somehow related to the thickness of that wall. The reason this works is that a spectral line, even an optically thick one, contains much more information than continuum emission. Consider optically thick line emission from a cloud of mass $M$ and radius $R$ at temperature $T$. The mean column density is $N= M/(\mu m_{\rm H} \pi R^2)$, where $\mu=2.3$ is the mass per H$_2$ molecule in units of $m_{\rm H}$. The total integrated intensity we expect to see from the line is
\begin{equation}
\int I_{\nu}\, d\nu = \int (1 - e^{-\tau_{\nu}}) B_{\nu}(T)\,d\nu.
\end{equation}

Suppose this cloud is in virial balance between kinetic energy and gravity, i.e., $\mathcal{T}=\mathcal{W}/2$ so that $\ddot{I}=0$, neglecting magnetic and surface terms. The gravitational-self energy is $\mathcal{W}=a GM^2/R$, where $a$ is a constant of order unity that depends on the cloud's geometry and internal mass distribution. For a uniform sphere $a=3/5$. The kinetic energy is $\mathcal{T}=(3/2)M\sigma_{\rm 1D}^2$, where $\sigma_{\rm 1D}$ is the one dimensional velocity dispersion, including both thermal and non-thermal components.

We define the observed virial ratio as
\begin{equation}
\label{eq:alpha_vir_obs}
\avir =  \frac{5\sigma_{\rm 1D}^2 R}{GM}.
\end{equation}
For a uniform sphere, which has $a=3/5$, this definition reduces to $\avir=2\mathcal{T}/\mathcal{W}$, which is the virial ratio we defined previously based on the virial theorem (equation \ref{eq:alpha_vir_th}). Thus $\avir=1$ corresponds to the ratio of kinetic to gravitational energy in a uniform sphere of gas in virial equilibrium between internal motions and gravity. In general we expect that $\avir\approx 1$ in any object supported primarily by internal turbulent motion, even if its mass distribution is not uniform.

Re-arranging this definition, we have
\begin{equation}
\sigma_{\rm 1D} = \sqrt{\left(\frac{\avir}{5}\right)\frac{GM}{R}}.
\end{equation}
To see why this is relevant for the line emission, consider the total frequency-integrated intensity that the cloud will emit in the line of interest. We have as before
\begin{equation}
I_{\nu} = \left(1-e^{-\tau_{\nu}}\right) B_{\nu}(T),
\end{equation}
so integrating over frequency we get
\begin{equation}
\int I_{\nu}\,d\nu = \int \left(1-e^{-\tau_{\nu}}\right) B_{\nu}(T)\,d\nu.
\end{equation}
The optical depth at line center is $\tau_{\nu_0}\gg 1$, and for a Gaussian line profile the optical depth at frequency $\nu$ is
\begin{equation}
\tau_{\nu} =\tau_{\nu_0} \exp\left[-\frac{(\nu-\nu_0)^2}{2 (\nu_0 \sigma_{\rm 1D}/c)^2}\right],
\end{equation}
where $\nu_0$ is the frequency of line center. Since the integrated intensity depends on the integral of $\tau_{\nu}$ over frequency, and the frequency-dependence of $\tau_{\nu}$ is determined by $\sigma_{\rm 1D}$, we therefore expect that the integrated intensity will depend on $\sigma_{\rm 1D}$.  

To get a sense of how this dependence will work, let us adopt a very simplified yet schematically correct form for $\tau_{\nu}$. We will take the opacity to be a step function, which is infinite near line center and drops sharply to 0 far from line center. The frequency at which this transition happens will be set by the condition $\tau_{\nu}=1$, which gives
\begin{equation}
\Delta \nu = |\nu-\nu_0| = \nu_0 \sqrt{2\ln \tau_{\nu_0}} \frac{\sigma_{\rm 1D}}{c}.
\end{equation}
The corresponding range in Doppler shift is
\begin{equation}
\Delta v = \sqrt{2\ln \tau_{\nu_0}} \sigma_{\rm 1D}.
\end{equation}
For this step-function form of $\tau_{\nu}$, the emitted brightness temperature is trivial to compute. At velocity $v$, the brightness temperature is
\begin{equation}
T_{B,v} = \left\{
\begin{array}{ll}
T, \qquad & |v-v_0| < \Delta v \\
0, &  |v-v_0| > \Delta v
\end{array}
\right.,
\end{equation}
where $v_0$ is the cloud's mean velocity.

If we integrate this over all velocities of emitting molecules, we get
\begin{equation}
I_{\rm CO} = \int T_{B,\nu}\, dv = 2 T_B \Delta v = \sqrt{8\ln\tau_{\nu_0}} \sigma_{\rm 1D} T.
\end{equation}
Thus, the velocity-integrated brightness temperature is simply proportional to $\sigma_{\rm 1D}$. The dependence on the line-center optical depth is generally negligible, since that quantity enters only as the square root of the log. We therefore have
\begin{eqnarray*}
X\mbox{ [cm}^{-2}\mbox{ (K km s}^{-1}\mbox{)}^{-1}\mbox{]} & = & \frac{M/(\mu \pi R^2)}{I_{\rm CO}} \\
& = & 10^5 \frac{(8\ln\tau_{\nu_0})^{-1/2}}{T\mu\pi} \frac{M}{\sigma_{\rm 1D} R^2} \\
& = & 10^5 \frac{(\mu m_{\rm H} \ln\tau_{\nu_0})^{-1/2}}{T} \sqrt{\frac{5n}{6\pi \avir G}},
\end{eqnarray*}
where $n= 3M/(4\pi \mu m_{\rm H} R^3)$ is the number density of the cloud, and the factor of $10^5$ comes from the fact that we are measuring $I_{\rm CO}$ in km s$^{-1}$ rather than cm s$^{-1}$. To the extent that all molecular clouds have comparable volume densities on large scales and are virialized, this suggests that there should be a roughly constant CO X factor. If we plug in $T=10$ K, $n=100$ cm$^{-3}$, $\avir=1$, and $\tau_{\nu_0}=100$, this gives $X_{\rm CO}=5\times 10^{19}$ cm$^{-2}$ (K km s$^{-1}$)$^{-1}$.

This argument is a simplified version of a more general technique of converting between molecular line luminosity and mass called the large velocity gradient approximation, introduced by \citet{goldreich74a}. The basic idea of all these techniques is the same: for an optically thick line, the total luminosity that escapes will be determined not directly by the amount of gas, but instead by the range in velocity or frequency that the cloud occupies, multiplied by the gas temperature.

Of course this calculation has a few problems -- we have to assume a volume density, and there are various fudge factors like $a$ floating around. Moreover, we had to assume virial balance between gravity and internal motions. This implicitly assumes that both surface pressure and magnetic fields are negligible, which they may not be. Making this assumption would necessarily make it impossible to independently check whether molecular clouds are in fact in virial balance between gravity and turbulent motions.

In practice, the way we get around these problems is by determining X factors by empirical calibration. We generally do this by attempting to measure the total gas column density by some tracer that measures all the gas along the line of sight, and then subtracting off the observed atomic gas column -- the rest is assumed to be molecular.

One way of doing this is measuring $\gamma$ rays emitted by cosmic rays interacting with the ISM. The $\gamma$ ray emissivity is simply proportional to the number density of hydrogen atoms independent of whether they are in atoms or molecules (since the cosmic ray energy is very large compared to any molecular energy scales). Once produced, the $\gamma$ rays travel to Earth without significant attenuation, so the $\gamma$ ray intensity along a line of sight is simply proportional to the total hydrogen column. Using this method, \citet{strong96a} obtained $X \approx 2\times 10^{20}$ cm$^{-2}$ (K km s$^{-1}$)$^{-1}$, and more recent work from Fermi \citep{abdo10b} gives about the same value.

Another way is to measure the infrared emission from dust grains along the line of sight, which gives the total dust column. This is then converted to a mass column using a dust to gas ratio. Based on this technique, \citet{dame01a} obtained $X\approx 2 \times 10^{20}$ cm$^{-2}$ (K km s$^{-1}$)$^{-1}$; more recently, \citet{draine07a} got about twice this, $X\approx 4 \times 10^{20}$ cm$^{-2}$ (K km s$^{-1}$)$^{-1}$. However, all of these techniques give numbers that agree to within a factor of two in the Milky Way, so we can be fairly confident that the X factor works to that level. It is important to emphasize, however, that this is only under Milky Way conditions. We will see shortly that there is good evidence that it does not work under very different conditions.

Note that we can turn the argument around. These other calibration methods, which make no assumptions about virialization, give conversions that are in quite good agreement with what we get by assuming virialization between gravity and turbulence. This suggests that molecular clouds cannot be too far from virial balance between gravity and turbulence. Neither magnetic fields nor surface pressure can be completely dominant in setting their structures, nor can clouds have large values of $\ddot{I}$.

It is also worth mentioning some caveats with this method. The most serious one is that it assumes that CO will be found wherever H$_2$ is, so that the mass traced by CO will match the mass traced by H$_2$. This seems to be a pretty good assumption in the Milky Way, but it may begin to break down in lower metallicity galaxies due to the differences in how H$_2$ and CO are shielded against dissociation by the interstellar UV field.

\subsection{Mass Distribution}

Armed with these techniques for measuring molecular cloud masses, what do we actually see? The answer is that in both the Milky Way and in a collection of nearby galaxies, the molecular cloud mass distribution in the cloud seems to be well-fit by a truncated powerlaw,
\begin{equation}
\frac{d\mathcal{N}}{d M} =
\left\{
\begin{array}{ll}
\mathcal{N}_u \left(\frac{M_u}{M}\right)^{\gamma}, \qquad & M \leq M_u \\
0, & M > M_u
\end{array}
\right..
\end{equation}
Here $M_u$ represents an upper mass limit for GMCs -- there are no clouds in a galaxy larger than that mass. The number of clouds with masses near the upper mass limit is $N_u$. Below $M_u$, the mass distribution follows a powerlaw of slope $\gamma$. Note that, since we have given this as the number per unit log mass rather than the number per unit mass, we can think of this index as telling us the total mass of clouds per decade in mass.

\begin{marginfigure}
\hspace{0.06\linewidth}\includegraphics[width=0.94\linewidth]{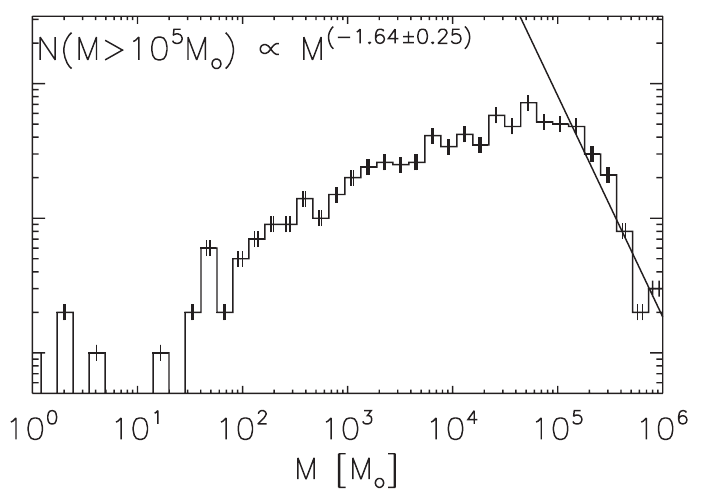}
\includegraphics[width=\linewidth]{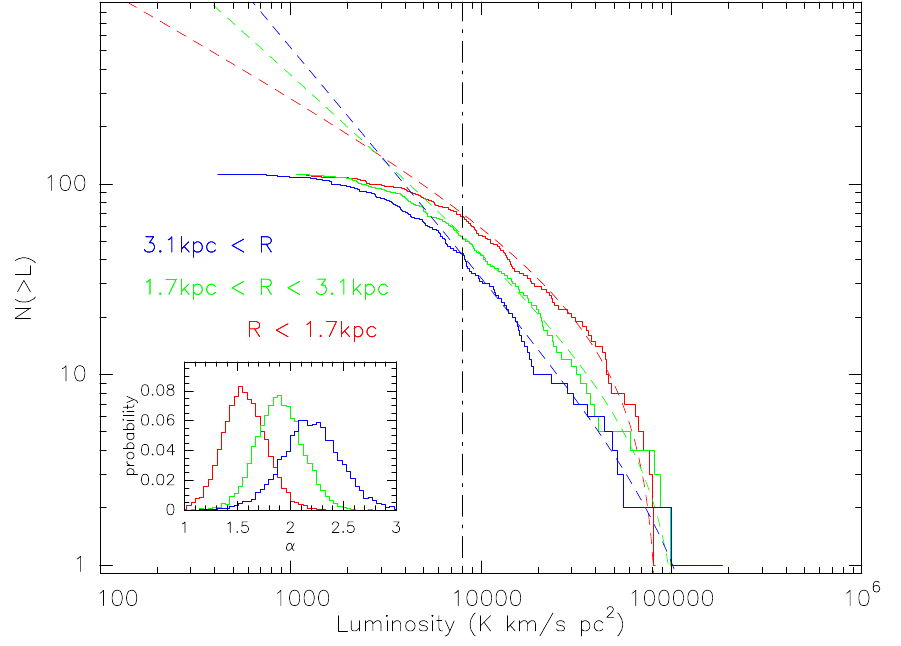}
\caption[GMC mass spectra]{
\label{fig:gmcmass}
Two measurements of the GMC mass spectrum. The top panel shows the mass spectrum for the inner Milky Way determined from $^{13}$CO measurements; the sample is complete at masses above $\sim 10^5$ $M_\odot$. The bottom panel shows the mass spectrum in M33 using $^{12}$CO. Note that these are cumulative distributions in luminosity, whereas the top panel shows a differential distribution in mass. The three colors show three different galactocentric regions: the inner galaxy (red), the mid-disk (green), and the outer galaxy (blue).
Credit: top panel: \citet{roman-duval10a}, \copyright\,AAS, reproduced with permission; bottom panel: \citeauthor{gratier12a}, A\&A, 542, A108, 2012, reproduced with permission \copyright\,ESO.
}
\end{marginfigure}

So what are $N_u$, $M_u$, and $\gamma$? It depends on where we look, as illustrated in Figure \ref{fig:gmcmass}. In the inner, H$_2$-rich parts of galaxies, the slope is typically $\gamma \sim -2$ to $-1.5$. In the outer, molecule-poor regions of galaxies, and in dwarf galaxies, it is $-2$ to $-2.5$. These measurements imply that, since the bulk of the molecular mass is found in regions with $\gamma > -2$, most of the molecular mass is in large clouds rather than small ones. This is just because the mass in some mass range is proportional to $\int (d\mathcal{N}/dM) M \, dM \propto M^{2+\gamma}$. A value of $\gamma=-2$ therefore represents a critical line separating distributions that are dominated by large and small masses.

\section{Scaling Relations}

Once we have measured molecular cloud masses, the next thing to investigate is their other large-scale properties, and how they scale with mass. Observations of GMCs in the Milky Way and in nearby galaxies yield three basic results, which are known as Larson's Laws, since they were first pointed out by \citet{larson81a}. The physical significance of these observational correlations is still debated today.

\begin{marginfigure}
\hspace{0.12\linewidth}\includegraphics[width=0.88\linewidth]{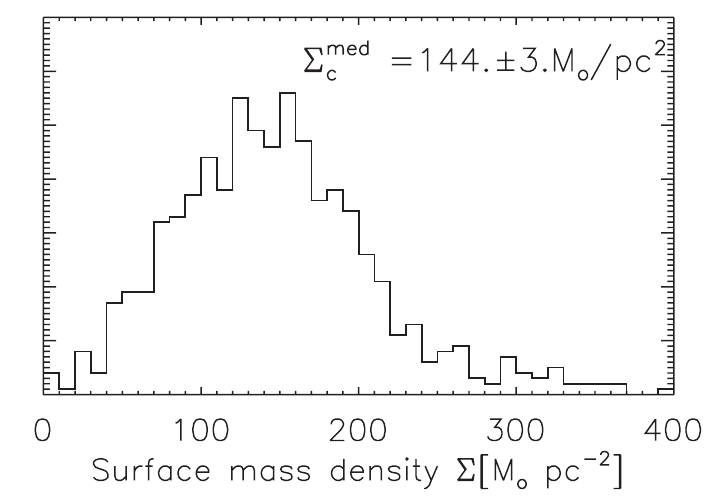}
\includegraphics[width=\linewidth]{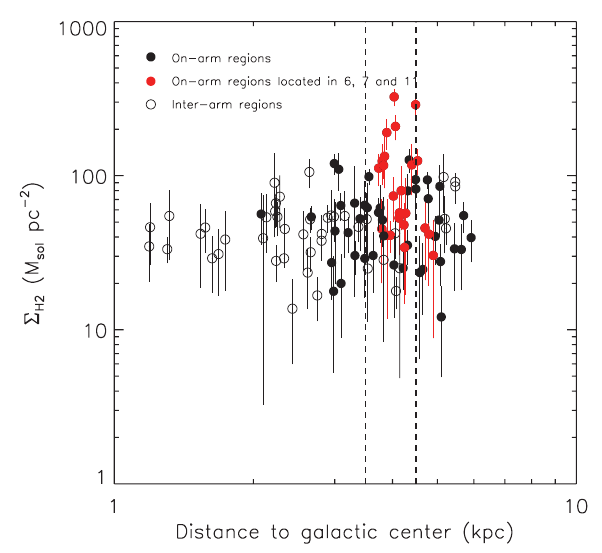}
\caption[GMC surface densities]{
\label{fig:gmcsigma}
Two measurements of GMC surface densities. The top panel shows the distribution of surface densities for the inner Milky Way determined from $^{13}$CO measurements. The bottom panel shows GMC surface density versus galactocentric radius in NGC 6946, measured from both $^{12}$CO and $^{13}$CO. Credit: top panel: \citet{roman-duval10a}; bottom panel: \citet{rebolledo12a}. \copyright AAS. Reproduced with permission.
}
\end{marginfigure}

The first is the molecular clouds have characteristic surface densities of $\sim 100$ $M_\odot$ pc$^{-2}$ (Figure \ref{fig:gmcsigma}). This appears to be true in the Milky Way and in all nearby galaxies where we can resolve individual clouds. There may be some residual weak dependence on the galactic environment -- $\sim 50$ $M_\odot$ pc$^{-2}$ in low surface density, low metallicity galaxies like the Large Magellanic Cloud (LMC), up $\sim 200$ $M_\odot$ pc$^{-2}$ in molecule- and metal-rich galaxies like M51, but generally around that value.

Note that the universal column density combined with the GMC mass spectrum implies are characteristic volume density for GMCs
\begin{equation}
n = \frac{3M}{4\pi R^3\mu m_{\rm H}} = \left(\frac{3\pi^{1/2}}{4\mu m_{\rm H}}\right) \sqrt{\frac{\Sigma^3}{M}} = 23 \Sigma_2^{3/2} M_6^{-1/2}\mbox{ cm}^{-2},
\end{equation}
where $\Sigma_2 = \Sigma/(100\,\msun\mbox{ pc}^{-2})$ and $M_6=M/10^6$ $\msun$. This is the number density of H$_2$ molecules, using a mean mass per molecule $\mu=2.3$. There is an important possible caveat to this, however, which is sensitivity bias: GMCs with surface densities much lower than this value may be hard to detect in CO surveys. However, there is no reason that higher surface density regions should not be detectable, so it seems fairly likely that this is a physical and not just observational result (though that point is disputed).

\begin{marginfigure}
\includegraphics[width=\linewidth]{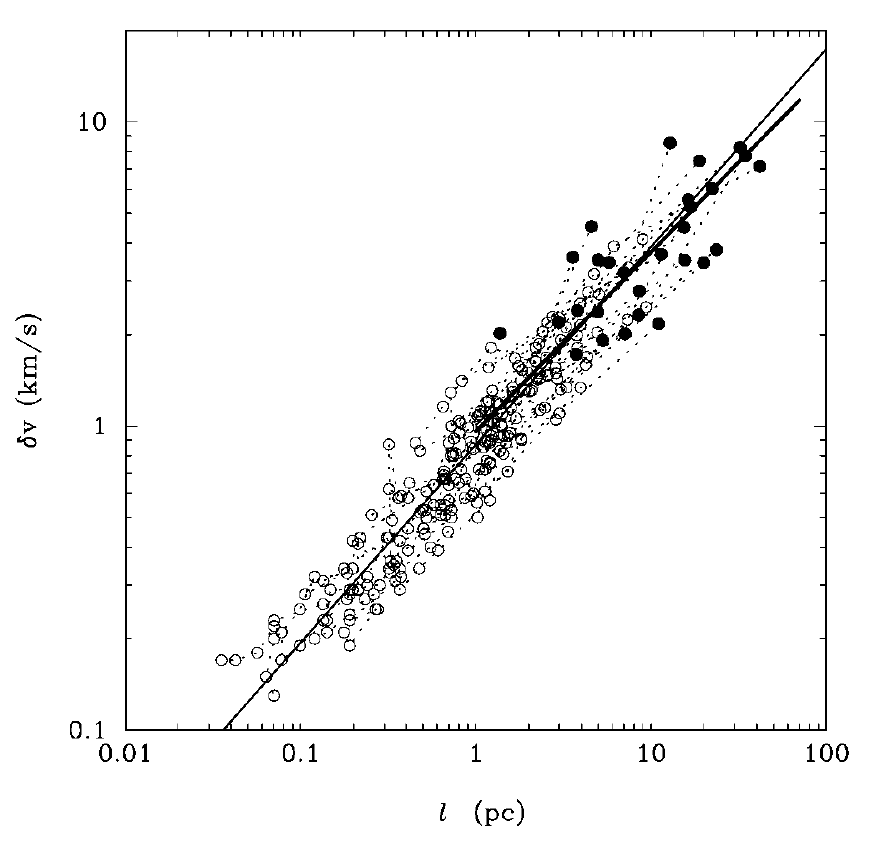}
\caption[GMC linewidth-size relation]{
\label{fig:gmclws}
Measured correlation between GMC linewidth $\delta v$ and size scale $\ell$ for Milky Way clouds. Credit: \citet{heyer04a}, \copyright\,AAS. Reproduced with permission.
}
\end{marginfigure}

The second of Larson's Laws is that GMCs obey a linewidth-size relation. The velocity dispersion of a given cloud depends on its radius. \citet{solomon87a} find $\sigma = (0.72\pm 0.07)R_{\rm pc}^{0.5\pm 0.05}$ km s$^{-1}$ in the Milky Way, where $R_{\rm pc}$ is the cloud radius in units of pc. For a sample of a number of external galaxies, \citet{bolatto08a} find $\sigma = 0.44^{+0.18}_{-0.13} R_{\rm pc}^{0.60\pm 0.10}$ km s$^{-1}$. Within individual molecular clouds in the Milky Way, \citet{heyer04a} find $\sigma=0.9 L_{\rm pc}^{0.56\pm 0.02}$ km s$^{-1}$ (Figure \ref{fig:gmclws}).

One interesting thing to notice here is that the exponent of the observed linewidth-size relation within a single cloud is quite close to the scaling $\sigma\propto \ell^{0.5}$ that we expect from supersonic turbulence. However, turbulence alone does not explain why all molecular clouds follow the {\it same} linewidth-size relation, in the sense that not only is the exponent the same, but the normalization is the same. It would be fully consistent with supersonic turbulence for different GMCs to have very different levels of turbulence, so that two clouds of equal size could have very different velocity dispersions. Thus the fact that turbulence in GMCs is universal is an important observation.

Larson's final law is that GMCs have $\avir\approx 1$, i.e., they are in rough virial balance between gravity and internal turbulence. We have already noted the good agreement between the value of X that we derived from a trivial virial assumption and the value derived by $\gamma$ ray and dust observations, which suggest exactly this result. In practice, the way we compute the virial ratio is to measure a mass using an X factor calibrated by $\gamma$ rays or dust, compute a radius from the observed size of the cloud on the sky and its estimated distance, and measure the velocity dispersion from the width of the line in frequency. Using the method, \citet{solomon97a} get $\avir=1.1$ as their mean within the Galaxy, and \citet{bolatto08a} get a similar result for external galaxies. This result only appears to hold for sufficiently massive clouds. Clouds with masses below $\sim 10^4$ $\msun$ have virial ratios $\avir \gg 1$. The interpretation is that these objects are confined by external pressure rather than gravity.

\begin{marginfigure}
\includegraphics[width=\linewidth]{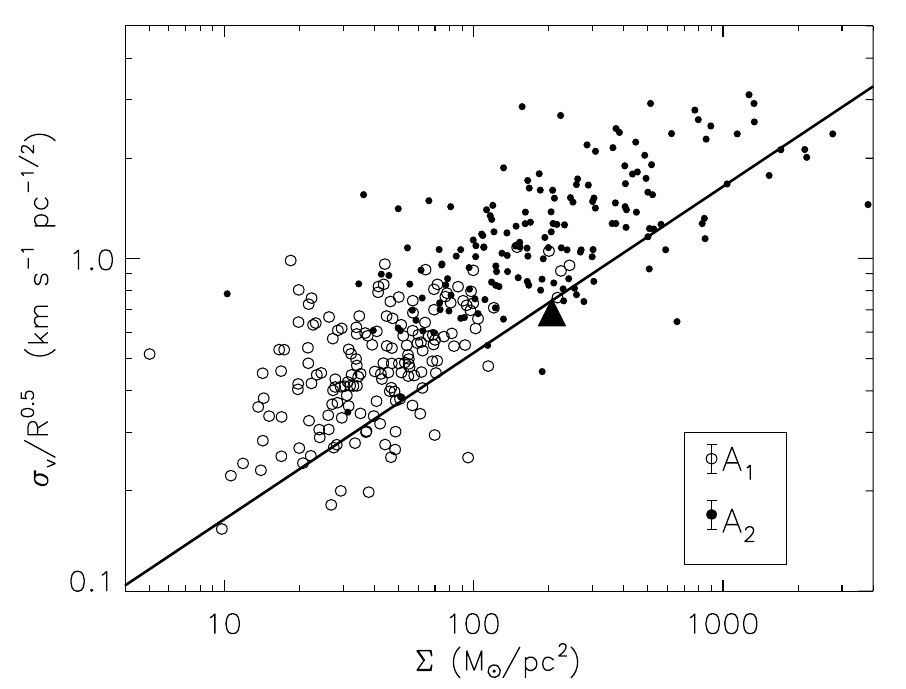}
\caption[GMC virial ratios]{
\label{fig:gmcalpha}
Correlation between GMC surface density $\Sigma$ and the combination $\sigma_v/R^{1/2}$, where $\sigma_v$ is the velocity dispersion and $R$ is the radius. The solid line represents the relationship that has $\alpha_{\mathrm{vir}} = 1$. Open circles indicate values derived with the lowest detectable contour, while closed ones indicate values derived using the half maximum CO isophote. Credit: \citet{heyer09a}, \copyright\,AAS. Reproduced with permission.
}
\end{marginfigure}

It is important to realize that Larson's three laws are not independent. If we write the linewidth-size relation as $\sigma=\sigma_{\rm pc} R_{\rm pc}^{1/2}$, then
\begin{equation}
\avir = \frac{5\sigma^2 R}{G M} = \left(\frac{5}{\pi \mbox{ pc}}\right) \frac{\sigma_{\rm pc}^2}{G\Sigma} = 3.7 \left(\frac{\sigma_{\rm pc}}{1\mbox{ km s}^{-1}}\right)^2 \left(\frac{100\,\msun\mbox{ pc}^{-2}}{\Sigma}\right).
\end{equation}
This shows that the universality of the linewidth-size relation is equivalent to the universality of the molecular cloud surface density, and vice-versa. The normalization of the linewidth-size relation is equivalent to the statement that $\avir=1$, and vice-versa. This is indeed what is observed (Figure \ref{fig:gmcalpha}).

It is also instructive to compute the pressure in GMCs that these relations imply. The kinetic pressure is $P = \overline{\rho} \sigma^2 = 3\Sigma\sigma_{\rm pc}^2/(4\mbox{ pc})$ Plugging in the observed linewidth-size relation, this gives $P/k_{\rm B}\approx 3\times 10^5$ K cm$^{-3}$. This is much larger than the mean pressure in the disk of the Milky Way or similar galaxies, which is typically closer to $10^4$ K cm$^{-3}$.

\section{Molecular Cloud Timescales}

Perhaps the most difficult thing to observe about GMCs are the timescales associated with their behavior. These are always long compared to any reasonable observation time, so we must instead infer timescales indirectly. In order to help understand the physical implications of GMC timescales, it is helpful to compare these to the characteristic timescales implied by Larson's Laws.

One of these is the crossing time,
\begin{equation}
t_{\rm cr} \equiv \frac{R}{\sigma} = \frac{0.95}{\sqrt{\avir G}} \left(\frac{M}{\Sigma^3}\right)^{1/4} = 14\, \avir^{-1/2} M_6^{1/4} \Sigma_2^{-3/4}\mbox{ Myr}.
\end{equation}
This is the characteristic time that it will take a signal to cross a cloud. The other is the free-fall time,
\begin{equation}
t_{\rm ff} \equiv \sqrt{\frac{3\pi}{32 G \rho}} = \frac{\pi^{1/4}}{\sqrt{8G}}\left(\frac{M}{\Sigma^3}\right)^{1/4} = 
7.0\, M_6^{1/4} \Sigma_2^{-3/4}\mbox{ Myr}
\end{equation}
For a virialized cloud, $\avir=1$, the free-fall time is half the crossing time, and both timescales are $\sim 10$ Myr. Thus, when discussing GMCs, we will compare our timescales to 10 Myr.

\subsection{Depletion Time}

The first timescale to think about is the one defined by the rate at which GMCs form stars. We call this the depletion time -- the time required to turn all the gas into stars. Formally, $t_{\rm dep} = M_{\rm gas}/\dot{M}_*$ for a cloud, or, if we're talking about an extra-Galactic observation where we measure quantities over surface areas of a galactic disk, $t_{\rm dep} = \Sigma_{\rm gas}/\dot{\Sigma}_*$. This is sometimes also referred to gas the gas consumption timescale.

This is difficult to determine for individual GMCs, in large part because stars destroy their parent clouds after they form. This means that we do not know how much gas mass a cloud started with, just how much gas is left at the time when we observe it. If the GMC is young we might see a lot of gas and few stars, and if it is old we might see many stars and little gas, but the depletion time might be the same.

We can get around this problem by studying a galactic population of GMCs. This should contain a fair sample of GMCs in all evolutionary stages, and tell us what the value of the star formation rate is when averaged over all these clouds. \citet{zuckerman74a}, pointed out that for the Milky Way the depletion time is remarkably long. Inside the Solar circle the Milky Way contains $\sim 10^9$ $\msun$ of molecular gas, and the star formation rate in the Milky Way is $\sim 1$ $\msun$ yr$^{-1}$, so $t_{\rm dep}\approx 1$ Gyr. This is roughly 100 times the free-fall time or crossing time of $\sim 10$ Myr. \citet{krumholz05c} pointed out that this ratio is a critical observational constraint for theories of star formation, and defined the dimensionless star formation rate per free-fall time as $\epsilon_{\rm ff} = t_{\rm ff} / t_{\rm dep}$. This is the fraction of a GMC's mass that it converts into stars per free-fall time.

\begin{figure}
\includegraphics[width=\linewidth]{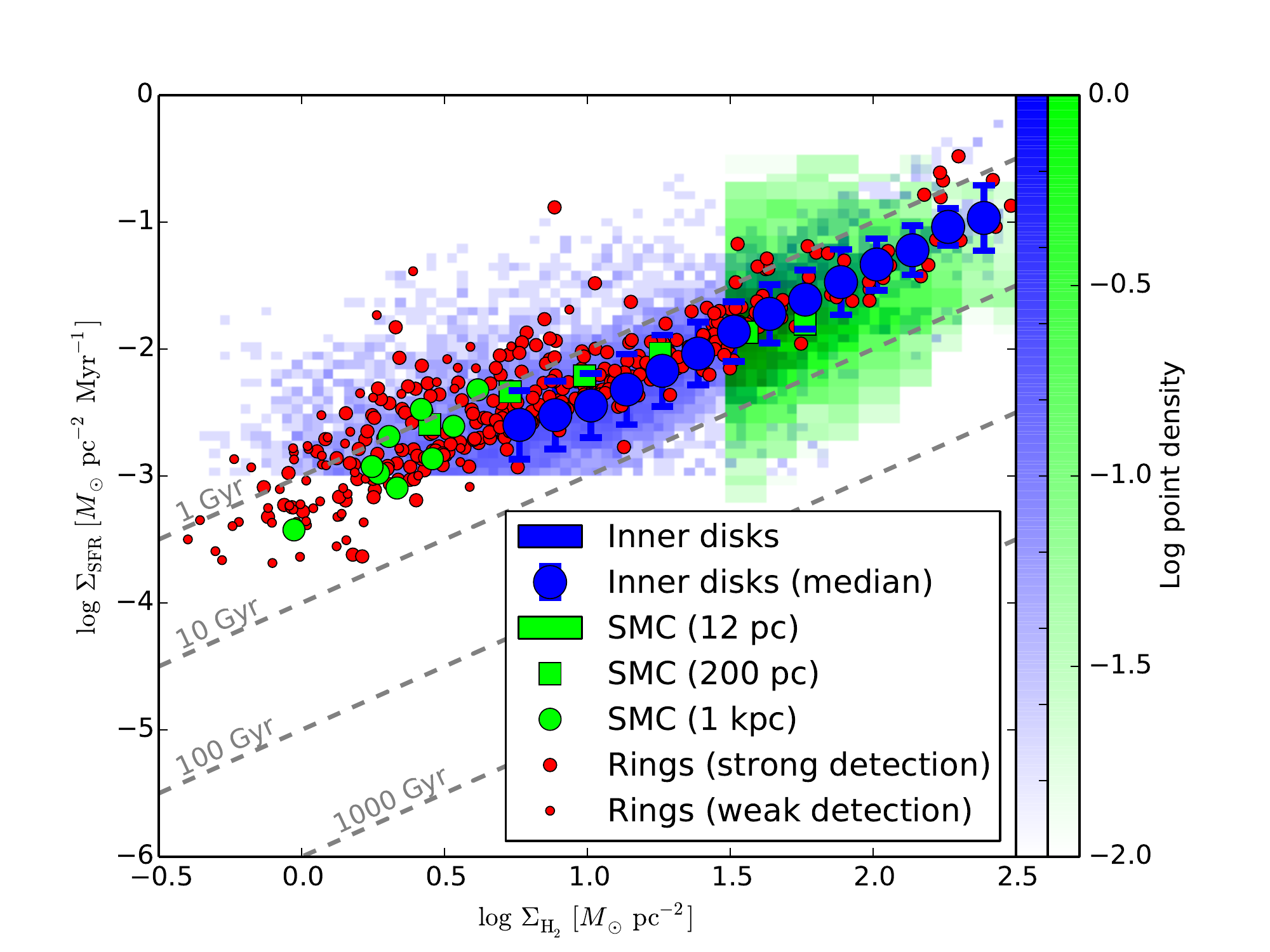}
\caption[Surface densities of gas and star formation]{
\label{fig:sfh2_krumholz14}
Surface density of star formation versus surface density of gas. Blue pixels show the distribution of pixels in the inner parts of nearby galaxies, resolved at $\sim 750$ pc scales \citet{leroy13a}, while green pixels show the SMC resolved at 12 pc scales \citet{bolatto11a}; other green and blue points show various averages of the pixels. Red points show azimuthal rings in outer galaxies \citet{schruba11a}, in which CO emission can be detected only by stacking all the pixels in a ring. Gray lines show lines of constant depletion time $t_{\mathrm{dep}}$. Reprinted from Phys. Rep., 539, \citeauthor{krumholz14c}, "The big problems in star formation: The star formation rate, stellar clustering, and the initial mass function", 49-134, 2014, with permission from Elsevier.
}
\end{figure}

Since 1974 these calculations have gotten more sophisticated and have been done for a number of nearby galaxies. Probably the cleanest, largest sample of nearby galaxies comes from the recent HERACLES survey \citep{leroy13a}. Surveys of local galaxies consistently find a typical depletion time $t_{\rm dep}=2$ Gyr for the molecular gas over. A wider by lower resolution survey, COLD GASS \citep{saintonge11a, saintonge11b}, found a non-constant depletion time over a wider range of galaxies, but still relatively little variation.\footnote{It is unclear what accounts for the difference between HERACLES and COLD GASS. The samples are quite different, in that HERACLES looks at individual patches within nearby well-resolved galaxies, while COLD GASS only has one data point per galaxy, and the observations are unresolved. On the other hand, COLD GASS has a much broader range of galaxy morphologies and properties. It possible that some of the COLD GASS galaxies are in a weak starburst, while there are no starbursts present in HERACLES.} Figure \ref{fig:sfh2_krumholz14} summarizes the current observations for galaxies close enough to be resolved.

\begin{figure}
\includegraphics[width=\linewidth]{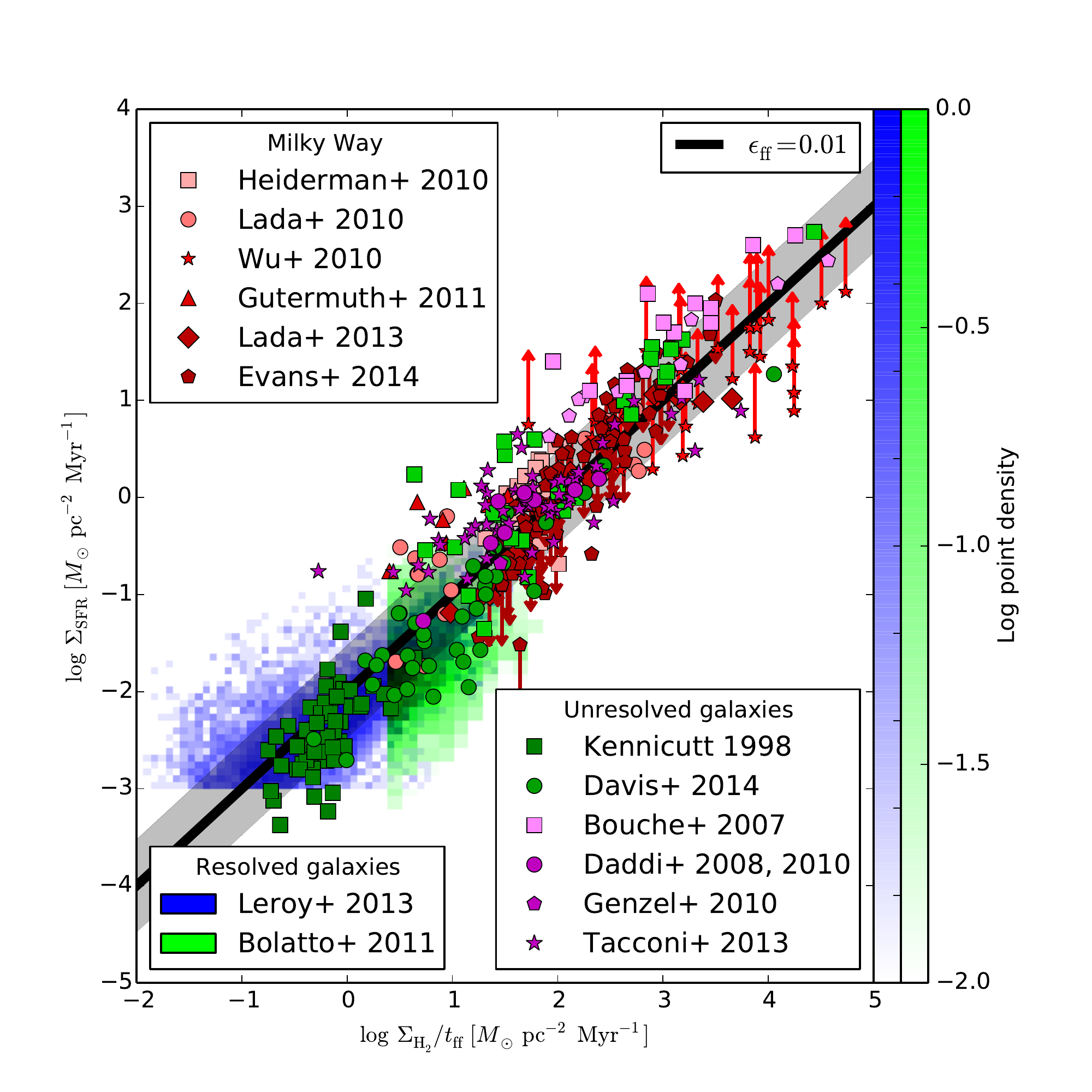}
\caption[Surface density of star formation versus surface density of gas normalized by free-fall time]{
\label{fig:eff_krumholz14}
Surface density of star formation versus surface density of gas normalized by free-fall time. Blue and green pixels are the same as in Figure \ref{fig:sfh2_krumholz14}, while points represent measurements of marginally-resolved galaxies ($\sim 1$ beam per galaxy). Points are color-coded: green indicates local galaxies, purple indicates high-$z$ galaxies, and red indicates individual Milky Way clouds. The thick black line represents $\epsilon_{\mathrm{ff}} = 0.01$, while the gray band shows a factor of 3 scatter about it. Reprinted from Phys. Rep., 539, \citeauthor{krumholz14c}, "The big problems in star formation: The star formation rate, stellar clustering, and the initial mass function", 49-134, 2014, with permission from Elsevier.
}
\end{figure}

\citet{krumholz07e} and \citet{krumholz12a} performed this analysis for a variety of tracers of mass other than CO and for a variety of galaxies, and for individual clouds within the Milky Way, and found that $\epsilon_{\rm ff}\sim 0.01$ for essentially all of them (Figure \ref{fig:eff_krumholz14}).

\subsection{Lifetime}

The second quantity of interest observationally is how long an individual GMC survives. This is a difficult problem in part because clouds are filled with structures on all scales, and authors are not always consistent regarding the scales on which a lifetime is being measured. When clouds have complex, hierarchical structures, things can depend tremendously on whether we say that a region consists of a single, sub-structured, big cloud or of many small ones. This makes it particularly difficult to compare Galactic and extra-galactic data. In extragalactic observations where resolution is limited, we tend to label things as large clouds with smaller densities and thus longer free-fall and crossing timescales. The same cloud placed within the Milky Way might be broken up and assigned much shorter timescales. The moral of this story is that, in estimating cloud lifetimes, it is important to be consistent in defining the sample and the methods used to estimate its lifetime. There are many examples in the literature of people being less than careful in this regard.

\begin{figure}
\includegraphics[width=\linewidth]{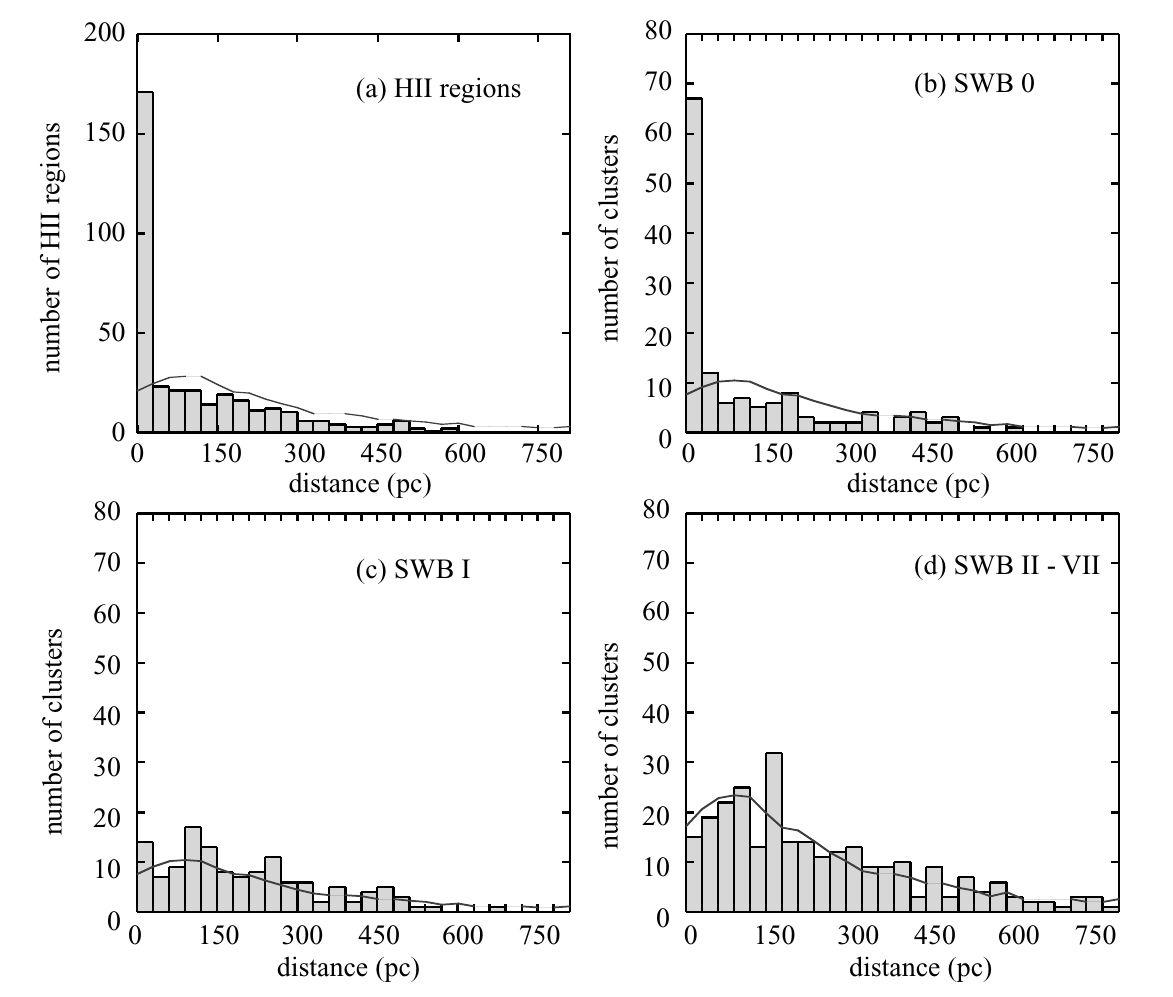}
\caption[Histogram of distances to nearest GMC]{
\label{fig:gmcdist_kawamura}
Histogram of projected distances to the nearest GMC in the LMC for H~\textsc{ii} regions, star clusters $<10$ Myr old (SWB 0), star clusters $10-30$ Myr old (SWB I), and star clusters older than 30 Myr (SWB II-IV), as indicated. In each panel, the lines show the frequency distribution that results from random placement of each category of object relative to the GMCs. Credit: \citet{kawamura09a}, \copyright\,AAS. Reproduced with permission.
}
\end{figure}

Probably the best determination of GMC lifetimes comes from extragalactic studies, where many biases and confusions can be eliminated. In the LMC, the NANTEN group catalogued the positions of all the molecular clouds \citet{fukui08a}, all the H~\textsc{ii} regions, and all the star clusters down to a reasonable completeness limit ($\sim 10^{4.5}$ $\msun$ for the GMCs). Star clusters' ages can be estimated from their colors, and thus the clusters can be broken into different age bins. They then compute the minimum projected distance between each cluster or H~\textsc{ii} region and the nearest GMC, and compare the distribution to what one would expect if the spatial distribution were random \citet[Figure \ref{fig:gmcdist_kawamura}]{kawamura09a}. There is clearly an excess of H~\textsc{ii} regions and clusters in the class SWB0, which are those with ages $\leq 10$ Myr, at small separations from GMCs. This represents a physical association between GMCs and these objects -- the clusters or H~\textsc{ii} regions are near their parent GMCs. There is no comparable excess for the older clusters. 

This allows us to estimate the GMC lifetime as follows. First, we note that roughly 60\% of the SWB 0 clusters are in the excess spike at small separations. This implies that, on average, 60\% of their $\sim 10$ Myr lifetime must be spent near their parent GMC, i.e., the phase of a GMC's evolution when it has a visible nearby cluster is 6 Myr. To estimate the total GMC lifetime, we note that only a minority of GMCs have visible nearby clusters. \citet{kawamura09a} find 39 GMCs are associated with nearby star clusters. In contrast, 88 are associated with H~\textsc{ii} regions but not star clusters, and 44 are associated with neither. If we assume that we are seeing these clouds are random stages in their lifetimes, then the fraction associated with star clusters must represent the fraction of the total GMC lifetime for which this association lasts. Thus the lifetime of each phase is just proportional to the fraction of clouds in that phase, i.e.,
\begin{equation}
t_{\rm HII} = \frac{N_{\rm HII}}{N_{\rm cluster}} t_{\rm cluster}
\end{equation}
and similarly for $t_{\rm quiescent}$.
Plugging in the numbers of clouds, and given that $t_{\rm cluster}=6$ Myr, we obtain $t_{\rm quiescent} = 7$ Myr, $t_{\rm HII} = 14$ Myr, and $t_{\rm life} = t_{\rm starless} + t_{\rm HII} + t_{\rm cluster} = 27$ Myr. This is $\sim 2-3$ crossing times, or $4-6$ free-fall times.

Notice that for this argument to work is it {\it not} necessary that the different phases be arranged in any particular sequence. \citeauthor{kawamura09a}~suggest that there is in fact a sequence, with GMCs without clusters or HII regions forming the earliest phase, GMCs with HII regions but not clusters forming the second phase, and GMCs with both HII regions and optically visible clusters forming the third phase. However, recent theoretical work by \citet{goldbaum11a} suggests that this is not necessarily the case.

\begin{marginfigure}
\includegraphics[width=\linewidth]{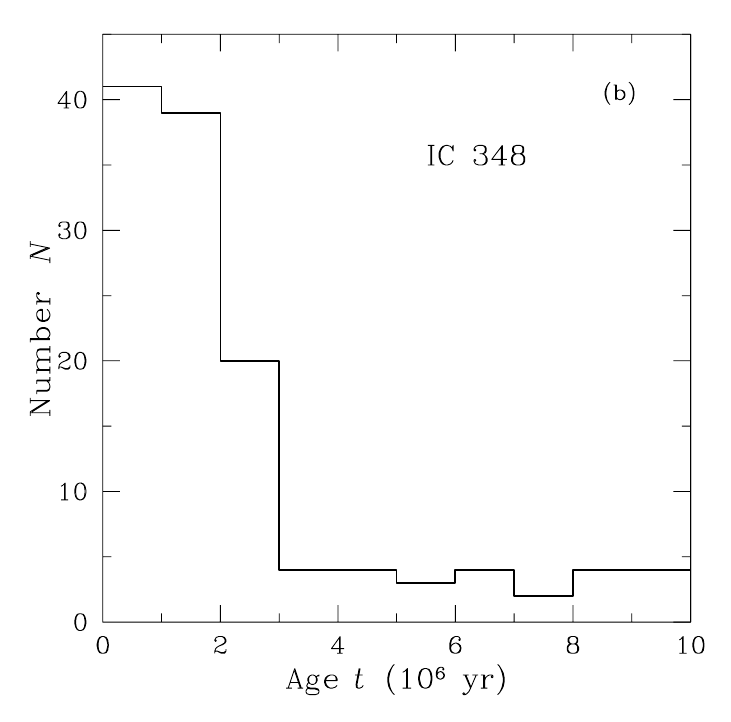}
\caption[Histogram of stellar ages in IC 348]{
\label{fig:ic348_palla}
Histogram of inferred stellar ages in the cluster IC 348. Credit: \citet{palla00a}, \copyright\,AAS. Reproduced with permission.
}
\end{marginfigure}

Within the galaxy and on smaller scales exercises like this get vastly trickier. If we look at individual star clusters, which we can age-date using pre-main sequence Hertzsprung-Russell diagrams (see Chapter \ref{ch:protostar_evol}) we find that they usually cease to be embedded in gaseous envelopes by the time the stellar population is $2-3$ Myr old (Figure \ref{fig:ic348_palla}). Interpreting this as a true cluster formation age is tricky due to numerous observational biases, e.g., variable extinction masquerading as age spread (which tends to raise the age estimate) and a bias against finding older stars because they are dimmer (which tends to reduce the age estimate). There are also uncertainties in the theoretical models themselves used to estimate the ages.

However, an individual GMC generally makes many cluster. The typical star clusters is only a few hundred $\msun$, compared to GMC masses of $10^5-10^6$ $\msun$, and we see associations made up of many clusters with age spreads of $10-15$ Myr. This suggests that the smaller pieces of a GMC (like the lumps we see in Perseus, shown in Figure \ref{fig:perseus_sun06}) clear away their gas relatively quickly, but that their larger-scale GMCs are not completely destroyed by this process.  The small regions therefore have lifetimes of a few Myr, but they also are much denser and thus have shorter crossing / free-fall times. For example, if the Orion Nebula cluster were smeared out into gas, its current stellar mass ($\sim 4000$ $\msun$) and surface density ($\Sigma \sim 0.1$ g cm$^{-2}$) suggest a crossing time of $0.7$ Myr. Given that the cluster has almost certainly lost some mass and spread out to somewhat lower surface density since it dispersed its gas, the true crossing time of the parent cloud was almost certainly shorter. This suggests an age of several crossing times for the ONC, but given the uncertainties in the true age spread of several crossing times. However, this is an extremely uncertain and controversial subject, and other authors have argued for shorter lifetimes on these smaller scales.

\subsection{Star Formation Lag Time}

A third important observable timescale is the time between GMC formation and the onset of star formation, defined as the lag time. We can estimate the lag time either statistically or geometrically. Statistically, we can do this using a technique much like what we did for the total lifetime in the LMC: compare the number of starless GMCs to the number with stars.

For the LMC, if we accept the \citet{kawamura09a} age sequence, the quiescent phase is 7 Myr. However, there may be star formation for some time before H~\textsc{ii} regions detectable at extragalactic distances begin to appear, or there may be clouds where H~\textsc{ii} regions appear and then go off, leading a cloud without a visible cluster or H~\textsc{ii} region, but still actively star-forming. This is what \citet{goldbaum11a} suggest.

In the solar neighborhood, within 1 kpc of the Sun, the ratio of clouds with star formation to clouds without is between $7:1$ and $14:1$, depending on the level of evidence on demands for star formation activity. If we take the time associated with star formation for these clouds to be $\sim 2-3$ Myr, this suggests a lag time less than a few tenths of a Myr in these high-density knots. Since this is comparable to or smaller than the crossing time, this suggests that these regions must begin forming stars while the are still in the process of forming.

\begin{figure}
\includegraphics[width=\linewidth]{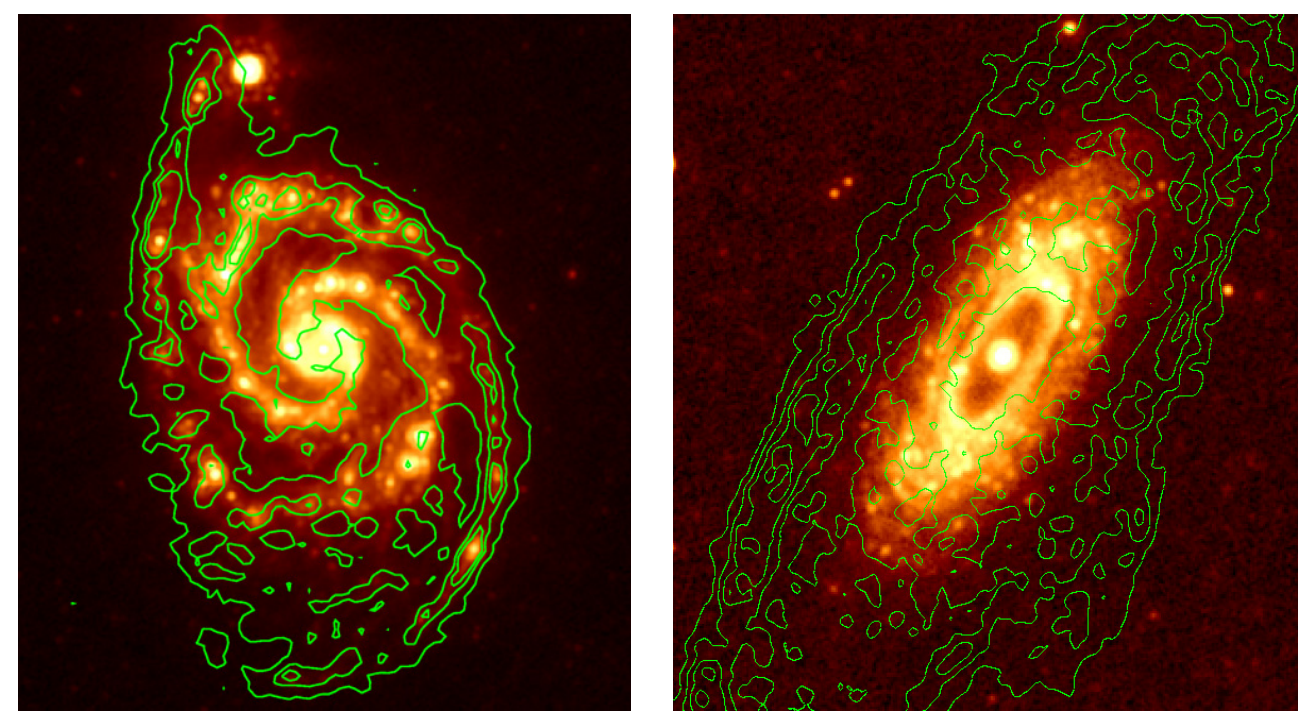}
\caption[H~\textsc{i} and 24 $\mu$m maps of NGC 5194 and 2841]{
\label{fig:sflag_tamburro08}
The galaxies NGC 5194 (left) and NGC 2841 (right), imaged in H~\textsc{i} from the THINGS survey with the VLA, and 24 $\mu$m from \textit{Spitzer}. Credit: \citet{tamburro08a}, \copyright AAS. Reproduced with permission.
}
\end{figure}

Geometric arguments provide similar conclusions. The way geometric arguments work is to look at a spiral galaxy and locate the spiral shock in H~\textsc{i} or CO. Generally some tracer of star formation, e.g., H$\alpha$ emission or 24 $\mu$m IR emission, will appear at some distance behind the spiral arm. If one can measure the pattern speed of the spiral arm, then the physical distance between the spiral shock and the onset of star formation, as indicated by the tracer of choice, can be identified with a timescale. This technique is illustrated in Figure \ref{fig:sflag_tamburro08}. In effect, one wants to take this image and measure by what angle the green contours (tracing H~\textsc{i}) should be rotated so that those arms peak at the same place as the 24 $\mu$m map, and then associate a time with that.

Performing this exercise with 24$\mu$m emission indicates lag timescales of $1-3$ Myr \citet{tamburro08a}. Performing it with H$\alpha$ as the tracer gives $t_{\rm lag}\sim 5$ Myr \citet{egusa04a}.The difference is probably because the H$\alpha$ better traces the bulk of the star formation, what 24 $\mu$m traces the earliest phase, when the stars are still embedded in their parent clouds. The latter is therefore probably a better estimate of the lag time. Since this is again comparable to or smaller than the molecular cloud crossing / free-fall timescale, we again conclude the GMCs must start forming stars while they are still being assembled.

\problemset

\begin{enumerate}

\item \textbf{The Bonnor-Ebert Sphere.}\\
Here we will investigate the properties of hydrostatic spheres of gas supported by thermal pressure. These are reasonable models for thermally-supported molecular cloud cores. Consider an isothermal, spherically-symmetric cloud of gas with mass $M$ and sound speed $c_s$, confined by some external pressure $P_{\mathrm{s}}$ on its surface.
\begin{enumerate}
\item For the moment, assume that the gas density inside the sphere is uniform. Use the virial theorem to derive a relationship between $P_{\mathrm{s}}$ and the cloud radius $R$. Show that there is a maximum surface pressure $P_{\mathrm{s,max}}$ for which virial equilibrium is possible, and derive its value.
\item Now we will compute the true density structure. Consider first the equation of hydrostatic balance,
\begin{displaymath}
-\frac{1}{\rho}\frac{d}{dr} P = \frac{d}{dr} \phi,
\end{displaymath}
where $P = \rho c_s^2$ is the pressure and $\phi$ is the gravitational potential. Let $\rho_c$ be the density at $r=0$, and choose a gauge such that $\phi = 0$ at $r=0$. Integrate the equation of hydrostatic balance to obtain an expression relating $\rho$, $\rho_c$, and $\phi$.
\item Now consider the Poisson equation for the potential,
\begin{displaymath}
\frac{1}{r^2}\frac{d}{dr}\left(r^2 \frac{d\phi}{dr}\right) = 4 \pi G \rho.
\end{displaymath}
Use your result from the previous part to eliminate $\rho$, and define $\psi \equiv \phi/c_s^2$. Show that the resulting equation can be non-dimensionalized to give the isothermal Lane-Emden equation:
\begin{displaymath}
\frac{1}{\xi^2}\frac{d}{d\xi}\left(\xi^2 \frac{d\psi}{d\xi}\right) = e^{-\psi}.
\end{displaymath}
where $\xi = r/r_0$. What value of $r_0$ is required to obtain this equation?
\item Numerically integrate the isothermal Lane-Emden equation subject to the boundary conditions $\psi=d\psi/d\xi = 0$ at $\xi=0$; the first of these conditions follows from the definition of $\psi$, and the second is required for the solution to be non-singular. From your numerical solution, plot both $\psi$ and the density contrast $\rho/\rho_c = e^{-\psi}$ versus $\xi$.
\item The total mass enclosed out to a radius $R$ is
\begin{displaymath}
M = 4\pi \int_0^R \rho r^2 \, dr.
\end{displaymath}
Show that this is equivalent to
\begin{displaymath}
M =\frac{c_s^4}{\sqrt{4\pi G^3 P_s}} \left(e^{-\psi/2}\xi^2 \frac{d\psi}{d\xi}\right)_{\xi_s},
\end{displaymath}
where
\begin{eqnarray*}
\xi_s & \equiv & \frac{R}{r_0} \\
P_s & \equiv & \rho_s c_s^2.
\end{eqnarray*}
Hint: to evaluate the integral, it is helpful to use the isothermal Lane-Emden equation to substitute.
\item Plot the dimensionless mass $m = M/(c_s^4/\sqrt{G^3 P_s})$ versus the dimensionless density contrast $\rho_c/\rho_s=e^{-\psi_s}$, where $\psi_s$ is the value of $\psi$ at $\xi=\xi_s$. You will see that $m$ reaches a finite maximum value $m_{\mathrm{max}}$ at a particular value of $\rho_c/\rho_s$. Numerically determine $m_{\mathrm{max}}$, along with the density contrast $\rho_c/\rho_s$ at which it occurs.
\item The existence of a finite maximum $m$ implies that, for a given dimensional mass $M$, there is a maximum surface pressure $P_s$ at which a cloud of that mass can be in hydrostatic equilibrium. Solve for this maximum, and compare your result to the result you obtained in part (a).
\item Conversely, for a given surface pressure $P_s$ and sound speed $c_s$ there exists a maximum mass at which the cloud can be in hydrostatic equilibrium, called the Bonnor-Ebert mass $M_{\mathrm{BE}}$. Obtain an expression for $M_{\mathrm{BE}}$ in terms of $P_s$ and $c_s$. In a typical low-mass star-forming region, the surface pressure on a core might be $P_{\mathrm{s}}/k_{\rm B} = 3\times 10^5$ K cm$^{-3}$. Compute this mass for a core with a temperature of 10 K, assuming the standard mean molecular weight $\mu=2.3$.\\
\end{enumerate}

\item \textbf{Driving Turbulence with Protostellar Outflows.}\\
Consider a collapsing protostellar core that delivers mass to an accretion disk at its center at a constant rate $\dot{M}_d$. A fraction $f$ of the mass that reaches the disk is ejected into an outflow, and the remainder goes onto a protostar at the center of the disk. The material ejected into the outflow is launched at a velocity equal to the escape speed from the stellar surface. The protostar has a constant radius $R_*$ as it grows.
\begin{enumerate}
\item Compute the momentum per unit stellar mass ejected by the outflow in the process of forming a star of final mass $M_*$. Evaluate this numerically for $f=0.1$, $M_* = 0.5$ $\msun$. and $R_* = 3$ $\rsun$.
\item The material ejected into the outflow will shock and radiate energy as it interacts with the surrounding gas, so on large scales the outflow will conserve momentum rather than energy. The terminal velocity of the outflow material will be roughly the turbulent velocity dispersion $\sigma$ in the ambient cloud. If this cloud is forming a cluster of stars, all of mass $M_*$, with a constant star formation rate $\dot{M}_{\rm cluster}$, compute the rate at which outflows inject kinetic energy into the cloud.
\item Suppose the cloud obeys Larson's relations, so its velocity dispersion, mass $M$, and size $L$ are related by $\sigma = \sigma_1 (L/\mbox{pc})^{0.5}$ and $M=M_1 (L/\mbox{pc})^2$, where $\sigma_1 \approx 1$ km s$^{-1}$ and $M_1\approx 100$ $\msun$ are the velocity dispersion and mass of a 1 pc-sized cloud. Assuming the turbulence in the cloud decays exponentially on a timescale $t_{\rm cr}=L/\sigma$, what star formation rate is required for energy injected by outflows to balance the energy lost via the decay of turbulence? Evaluate this numerically for $L = 1, 10$ and $100$ pc.
\item If stars form at the rate required to maintain the turbulence, what fraction of the cloud mass must be converted into stars per cloud free-fall time? Assume the cloud density is $\rho=M/L^3$. Again, evaluate numerically for $L = 1,10$ and $100$ pc. Are these numbers reasonable? Conversely, for what size clouds, if any, is it reasonable to neglect the energy injected by protostellar outflows?\\
\end{enumerate}

\item \textbf{Magnetic Support of Clouds.}\\
Consider a spherical cloud of gas of initial mass $M$, radius $R$, and velocity dispersion $\sigma$, threaded by a magnetic field of strength $B$. In Chapter \ref{ch:collapse} we showed that there exists a critical magnetic flux $M_\Phi$ such that, if the cloud's mass $M<M_\Phi$, the cloud is unable to collapse.
\begin{enumerate}
\item Show that the the cloud's Alfv\'en Mach number $\mathcal{M}_A$ depends only on its virial ratio $\alpha_{\rm vir}$ and on $\mu_\Phi \equiv M/M_\Phi$ alone. Do not worry about constants of order unity. 
\item Your result from the previous part should demonstrate that, if any two of the dimensionless quantities $\mu_\Phi$, $\alpha_{\rm vir}$, and $\mathcal{M}_A$ are of order unity, then the third quantity must be as well. Give an intuitive explanation of this result in terms of the ratios of energies (or energy densities) in the cloud.
\item Magnetized turbulence naturally produces Alfv\'en Mach numbers $\mathcal{M}_A \sim 1$. Using this fact plus your responses to the previous parts, explain why this makes it difficult to determine observationally whether clouds are supported by turbulence or magnetic fields.
\end{enumerate}

\end{enumerate}

\chapter{The Star Formation Rate at Galactic Scales: Observations}
\label{ch:sflaw_obs}

\marginnote{
\textbf{Suggested background reading:}
\begin{itemize}
\item \href{http://adsabs.harvard.edu/abs/2012ARA\%26A..50..531K}{Kennicutt, R.~C., \& Evans, N.~J. 2012, ARA\&A, 50, 531}, sections $5-6$ \nocite{kennicutt12a}
\end{itemize}
\textbf{Suggested literature:}
\begin{itemize}
\item \href{http://adsabs.harvard.edu/abs/2010AJ....140.1194B}{Bigiel et al., 2010, ApJ, 709, 191} \nocite{bigiel10a}
\item \href{http://adsabs.harvard.edu/abs/2013AJ....146...19L}{Leroy et al., 2013, ApJ, 146, 19} \nocite{leroy13a}
\end{itemize}
}

In the previous chapter we discussed observations of the bulk properties of giant molecular clouds. Now we will discuss the correlation of gas with star formation, a topic known loosely as star formation "laws". This chapter will focus on the observational situation, and the following one will focus on theoretical models that attempt to make sense of the observations. This is an extremely active area of research, and much of the available data is only a few years old. Most of the models are of similarly recent vintage. The central questions with which all of these models and data are concerned are: what determines the rate at which a galaxy transforms its gas content into stars? What determines where in the galaxy, both in terms of location and in terms of the physical state of the ISM, this transformation will take place? What physical mechanisms regulate this transformation? 

\section{The Star Formation Rate Integrated Over Whole Galaxies}

\subsection{Methodology}

Research into the star formation "law" was really kicked off by the work of Robert Kennicutt, who wrote a groundbreaking paper in 1998 \citep{kennicutt98a} collecting data on the gas content and star formation of a large number of disk and starburst galaxies in the local Universe. Today this is one of the most cited papers in astrophysics, and the relationship that Kennicutt discovered is often called the Kennicutt Law in his honor. (It is also sometimes referred to as the Schmidt Law, after the paper \citet{schmidt59a}, which introduced the conjecture that there is a scaling between gas density and star formation rate.) Before diving into this, though, we must first discuss how the measurements are made.

We are interested in the correlation between neutral gas and star formation averaged over an entire galaxy. To obtain information about the gas content, we need a method of tracing the molecular gas and the neutral hydrogen. For neutral hydrogen, the standard technique is to measure the flux in the 21 cm line, which can be translated more or less directly into a hydrogen mass under the assumption that the line is optically thin. There are a few caveats with this conversion, mostly involving the possibility of the line becoming optically thick in cold regions of high column density, but these are unlikely to make more then a tens of percent difference when we consider measurements over entire galaxies. The main problem is that the line is both weak and at a very low frequency, so in practice it can only be observed in the local Universe. There are at present no detections of 21 cm emission at high redshift.

The molecular content requires a proxy, and in large surveys this is almost always the $J=1\rightarrow 0$ or $J=2\rightarrow 1$ line of CO. This is then converted to a total mass using the X factor that discussed in Chapter \ref{ch:gmcs}. This is subject to non-trivial uncertainties. As discussed in that Chapter, the X factor depends on the volume density, temperature, and virial ratio of the molecular gas, albeit not tremendously strongly. In the Milky Way and in some nearby galaxies we have cross-checks against other methods like gamma rays and dust emission, and we are starting to get dust cross-checks at high redshift, but there is still significant uncertainty.

The star formation rate also requires a proxy. Depending on the survey, this can be one of several things: H$\alpha$ emission for nearby galaxies with relatively modest levels of dust obscuration, FUV continuum for either nearby or high redshift galaxies with fairly modest dust obscuration, and infrared emission for very dusty galaxies. The best cases combine multiple proxies for star formation to capture both the light that is and is not reprocessed by dust.

A fourth ingredient sometimes included in these studies is a measurement of the rotation rate of the galaxy. This can be obtained from a map in H~\textsc{i} or CO that is even modestly resolved, since the difference in Doppler shift of the line across the galaxy provides a direct measurement. One must of course choose a point at which to measure the rotation rate and, what is usually the more interesting parameter, the galactic rotation period, and there is some uncertainty in this choice. The convention is to use the "outer edge of the star-forming disk."\footnote{If that definition sounds nebulous and author-dependent, it is. There is no standard convention for where exactly in a galaxy the rotation period should be measured.}

\subsection{Nearby Galaxies}

\begin{marginfigure}
\includegraphics[width=\linewidth]{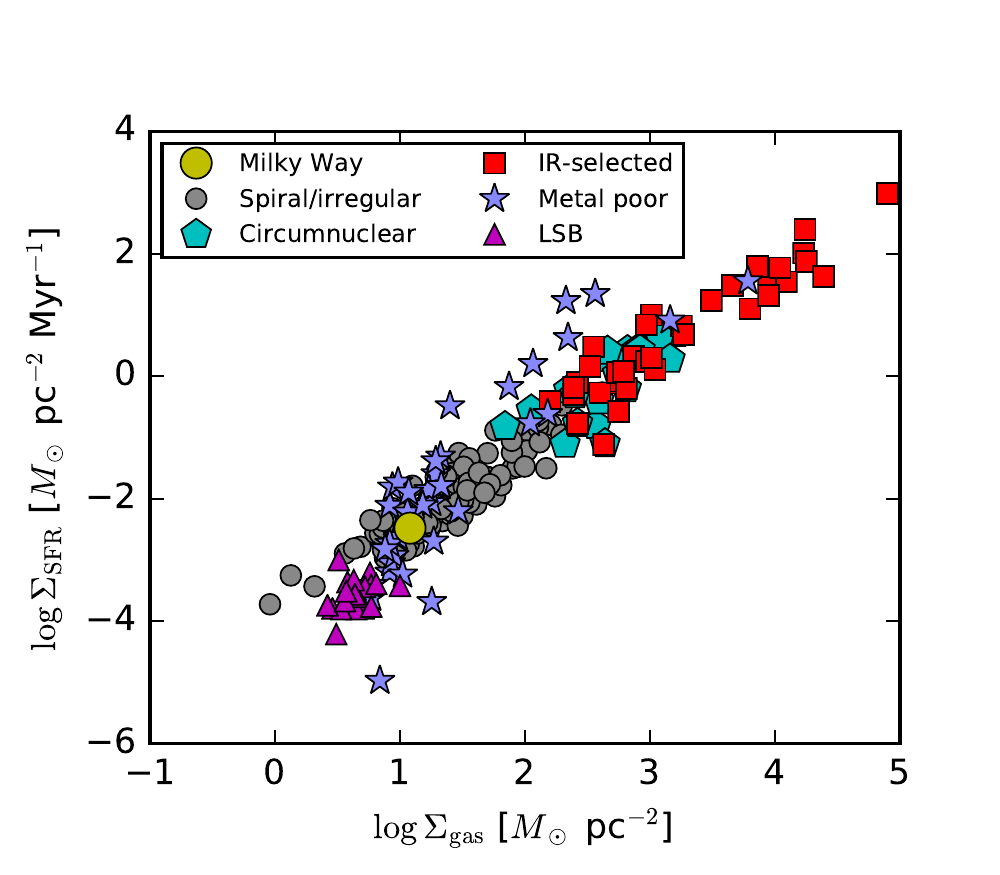}
\caption[Whole-galaxy Kennicutt-Schmidt relation]{
\label{fig:ksrelation}
The observed collection between gas surface density $\Sigma_{\mathrm{gas}}$ and star formation surface density $\Sigma_{\mathrm{SFR}}$, integrating over whole galaxies. Galaxy classes are as indicated in the legend; circumnuclear indicates circumnuclear starburst, IR-selected is galaxies selected based on their high-infrared luminosity, metal-poor is galaxies with substantially sub-solar metallicity, and LSB is low surface-brightness galaxies. Data from \citet{kennicutt12a}.
}
\end{marginfigure}

So what is the outcome of these studies? Not surprisingly, if one simply plots something like star formation rate against total gas mass, there is a strong correlation. This is mostly a matter of "the bigger they are, the bigger they are": galaxies that are larger overall tend to have more star formation and more gas content. Somewhat more interesting is the case where the galaxy is at least marginally resolved, and thus we can normalize out the projected area. In this case we can measure the relationship between gas mass per unit area, $\Sigma_{\rm gas}$, and star formation rate per unit area, $\Sigma_{\rm SFR}$. \citet{kennicutt98a} was the first to assemble a large sample of such measurements, and he found that there was a strong correlation over a wide range in gas surface density. Figure \ref{fig:ksrelation} shows this correlation using a modern data set of local galaxies. The data are reasonably well fit by a correlation

\begin{equation}
\Sigma_{\rm SFR} \propto \Sigma_{\rm gas}^{1.4}.
\end{equation}

There are a few caveats to this. This fit uses the same value of $X_{\rm CO}$ for all galaxies, but there is excellent evidence that $X_{\rm CO}$ is lower for starbursts and higher for metal-poor galaxies. Correcting for this effect would tend to move the metal-poor galaxies that lie above the relation back toward it (by increasing their inferred $\Sigma_{\rm gas}$), while steepening the relation overall (by moving the galaxies with the highest star formation rates systematically to lower $\Sigma_{\rm gas}$). Correcting for this effect increases the slope from $\sim 1.4$ to something more like $\sim 1.7-1.8$ \citep[e.g.,][]{narayanan12a}, but with a significantly larger uncertainty. For extreme but not utterly implausible scalings of $X_{\rm CO}$ with star formation rate or gas content, one can get slopes as steep as $\sim 2$.

\begin{marginfigure}
\includegraphics[width=\linewidth]{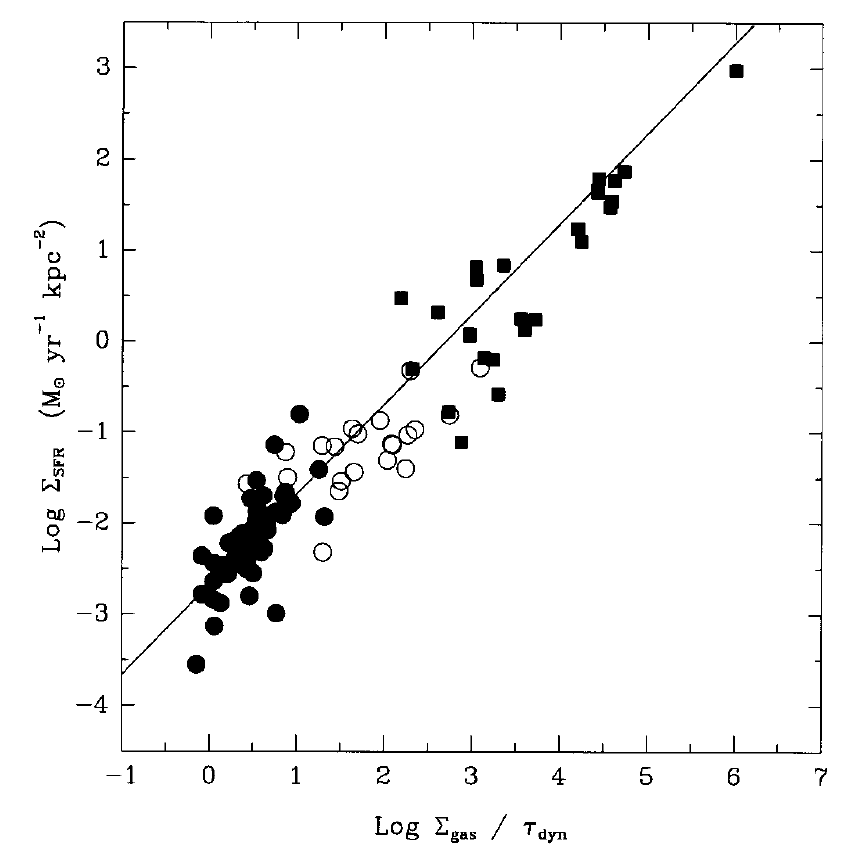}
\caption[Whole-galaxy Kennicutt-Schmidt relation, including orbital time]{
\label{fig:ksrelation_torb}
The observed collection between gas surface density divided by galaxy orbital period $\Sigma_{\mathrm{gas}}/\tau_{\mathrm{dyn}}$ and star formation surface density $\Sigma_{\mathrm{SFR}}$, integrating over whole galaxies. Filed circles are normal disk galaxies, open circles are circumnuclear starbursts, and filled squares are starburst galaxies. Credit: \citet{kennicutt98a}, \copyright AAS. Reproduced with permission.
}
\end{marginfigure}

While this is one way of plotting the data, another way is to make use of the galactic rotation curve. The star formation rate per unit area has units of mass per unit time per unit area, so it is natural to compare this to the gas mass per unit area divided by the galactic orbital period $t_{\rm orb}$, which has the same units. Physically, this relationship describes what fraction of the gas mass is transformed into stars per orbital period. Making this plot yields a relationship that actually fits the data every bit as well as the $\Sigma_{\rm gas} - \Sigma_{\rm SFR}$ plot (Figure \ref{fig:ksrelation_torb}), and with a slope of unity, i.e., $\Sigma_{\rm SFR} \propto \Sigma_{\rm gas}/t_{\rm orb}$.

\subsection{High-Redshift Galaxies}

\begin{marginfigure}
\includegraphics[width=\linewidth]{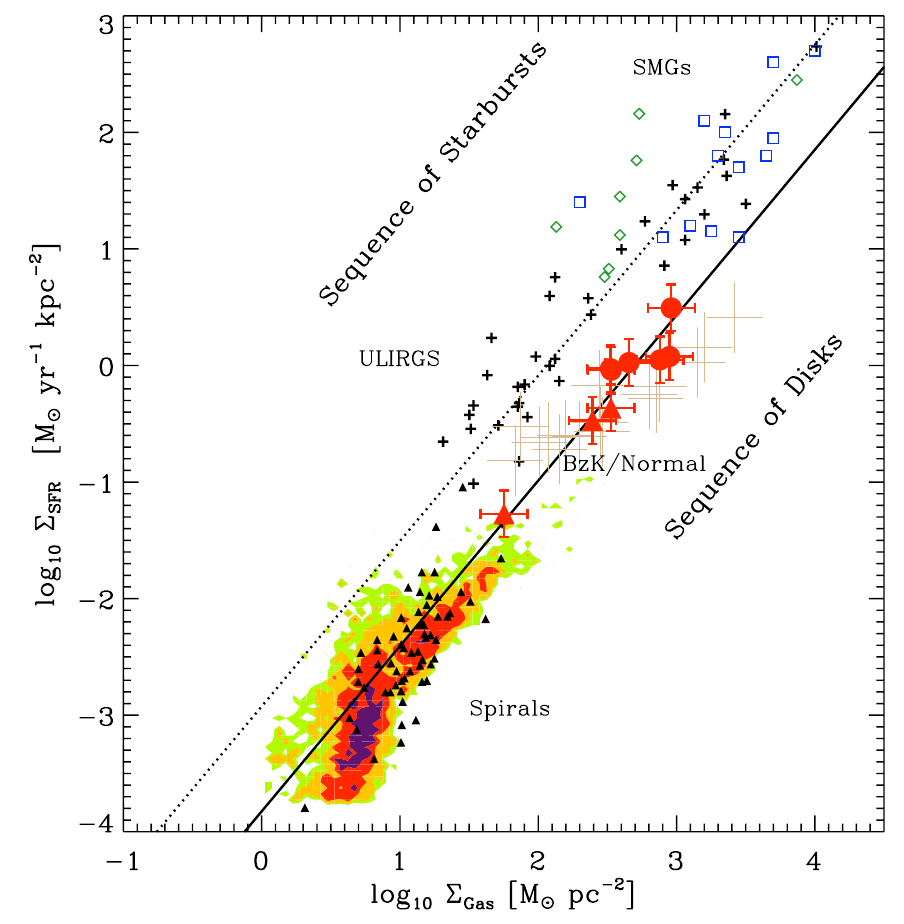}
\caption[Kennicutt-Schmidt relation, with additional high-redshift data]{
\label{fig:sfsequences}
Kennicutt-Schmidt relation including an expanded high-redshift sample, with two proposed sequences (``disks" and ``starbursts") indicated. Points are integrated-galaxy measurements, while contours are spatially-resolved regions (see below). Credit: \citet{daddi10a}, \copyright AAS. Reproduced with permission.
}
\end{marginfigure}

Since Kennicutt's initial collection, a number of other authors have added much more data to this plot, principally but not exclusively from the high redshift Universe. The expanded data set suggests that there isn't a single relationship between $\Sigma_{\rm gas}$ and $\Sigma_{\rm SFR}$, but that instead "normal galaxies" and "starbursts" occupy different loci on the $\Sigma_{\rm gas} - \Sigma_{\rm SFR}$ plane (Figure \ref{fig:sfsequences}).

This result should be taken with a considerable grain of salt. In part, the bimodality is exaggerated by the use of different $X_{\rm CO}$ factors for the two sequences, which spreads them further apart. If one uses a single $X_{\rm CO}$, the bimodality is far less clear. As mentioned above, there are excellent reasons to think that $X_{\rm CO}$ is not in fact constant, but conversely there are no good reasons to think that it is bimodal as opposed to changing continuously. A second issue is one of selection: the samples that occupy the two loci are selected in different ways, and this may well lead to an artificial bimodality that is not present in the real galaxy population.

\begin{marginfigure}
\includegraphics[width=\linewidth]{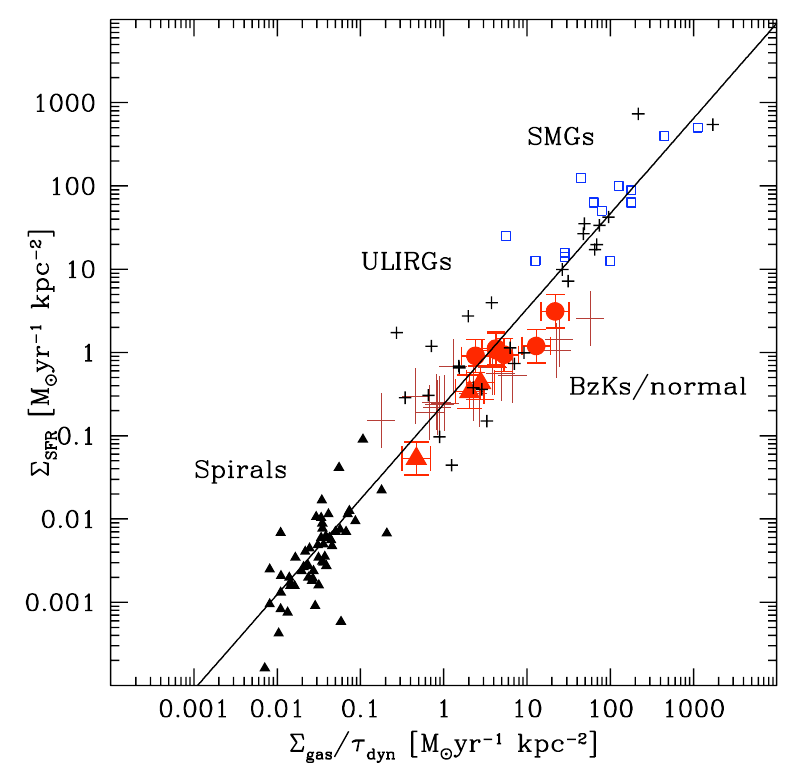}
\caption[Kennicutt-Schmidt relation, orbital time version, with additional high-redshift data]{
\label{fig:sftorb}
Kennicutt-Schmidt relation in its $\Sigma_{\rm SFR}-\Sigma_{\rm gas}/t_{\rm orb}$ form, including an expanded high-redshift sample. Points are the same as in Figure \ref{fig:sfsequences}, except that points for which the orbital time are unavailable have been omitted. Credit: \citet{daddi10a}, \copyright AAS. Reproduced with permission.
}
\end{marginfigure}

Nonetheless, the point remains that it is far from clear that there is a single, uniform relationship between $\Sigma_{\rm gas}$ and $\Sigma_{\rm SFR}$. On the other hand, the $\Sigma_{\rm gas}/t_{\rm orb}$ versus $\Sigma_{\rm SFR}$ relationship appears to persist even in the expanded data set (Figure \ref{fig:sftorb}).

\subsection{Dwarfs and low surface brightness galaxies}

A second area in which Kennicutt's original sample has been greatly expanded is in the study of dwarf galaxies. There were a few dwarfs in Kennicutt's original sample, but not that many, due to the difficulty of measuring star formation rates in low luminosity systems. Kennicutt's original sample used star formation rates primarily based on H$\alpha$ and infrared, but these are difficult to use on dwarfs: the H$\alpha$ is faint and hard to pick out above the sky background due to the low overall star formation rate, and the IR is faint because dwarfs tend to have little dust and thus reprocess little of their starlight into the IR. The situation improved greatly with the launch of \textit{GALEX} in 2003, which allowed the study of dwarfs in the FUV. The FUV has the advantage that, from space, the background is nearly zero, and thus much lower levels of star formation activity can be detected much more easily.

Another problem that does remain for dwarfs is that the CO to H$_2$ conversion factor is almost certainly different than in spirals, and the CO is often so faint as to be undetectable. This makes it impossible to measure the molecular gas content of many dwarfs without using a better proxy like dust. Only with the launch of \textit{Herschel} has this been possible with even a modest sample of dwarfs; prior to that, with the exception of the Small Magellanic Cloud (which could be mapped in dust with \textit{IRAS} due to its large size on the sky). Nonetheless, the H~\textsc{i} can certainly be measured, and since the H~\textsc{i} almost certainly dominates the total gas, the relationship between total gas content and star formation could also be measured.

\begin{marginfigure}
\includegraphics[width=\linewidth]{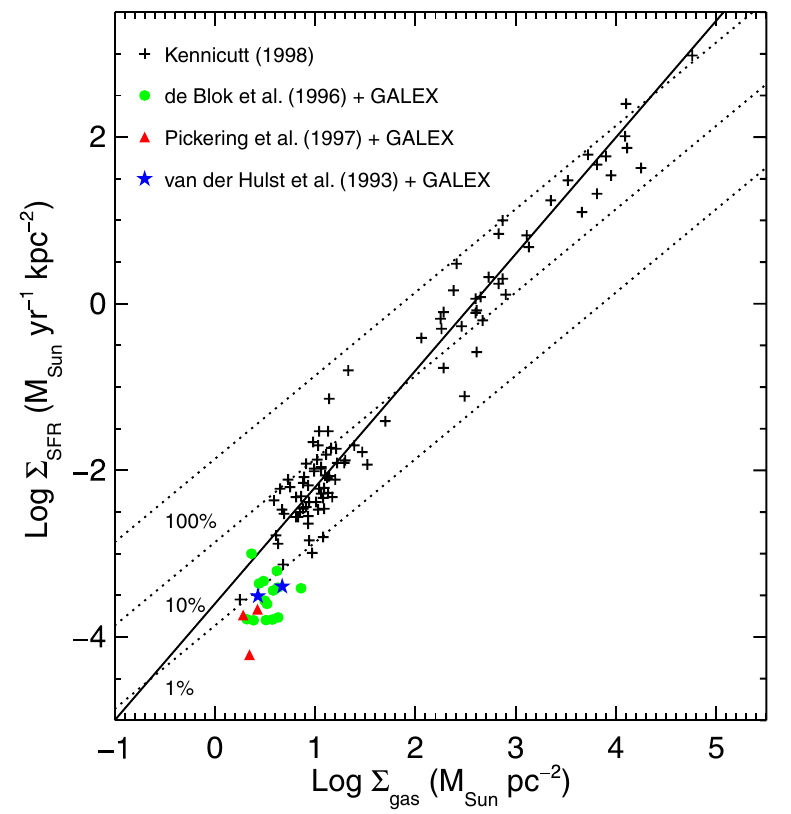}
\caption[Kennicutt-Schmidt relation, with additional low surface brightness sample]{
\label{fig:ks_lsb}
Kennicutt-Schmidt relation including an expanded sample of low surface brightness galaxies. The black points are the original \citet{kennicutt98a} sample, while the colored points are the low surface brightness sample. Credit: \citet{wyder09a}, \copyright AAS. Reproduced with permission.
}
\end{marginfigure}

When the data are plotted, the result is that dwarfs generally lie below the linear extrapolation of the Kennicutt relationship when one considers their total gas content (Figure \ref{fig:ks_lsb}).

\section{The Spatially-Resolved Star Formation Rate}

The previous section summarizes the observational state of play as far as single points per galaxy goes, but what about if we start to resolve galaxies? Starting around 2006-7, instrumentation reached the point where it became possible to make spatially resolved maps of the gas and star formation in galaxies. For gas, the key development was the advent of heterodyne receiver arrays, which greatly increased mapping speed and made it possible to produce maps of the CO in nearby galaxies at resolutions of $\sim 1$ kpc or better in reasonable amounts of observing time. For star formation, the key was the development of space-based infrared telescopes, first \textit{Spitzer} and then \textit{Herschel}, that could make images of the dust-reprocessed light from a galactic disk. Armed with these new technologies, a number of groups began to make maps of the relationship between gas and star formation within the disks of nearby galaxies, starting at $\sim 1$ kpc or better scales and eventually going in some cases to $\sim 10$ pc scales.

\subsection{Relationship to Molecular Gas}

\begin{figure}
\includegraphics[width=\linewidth]{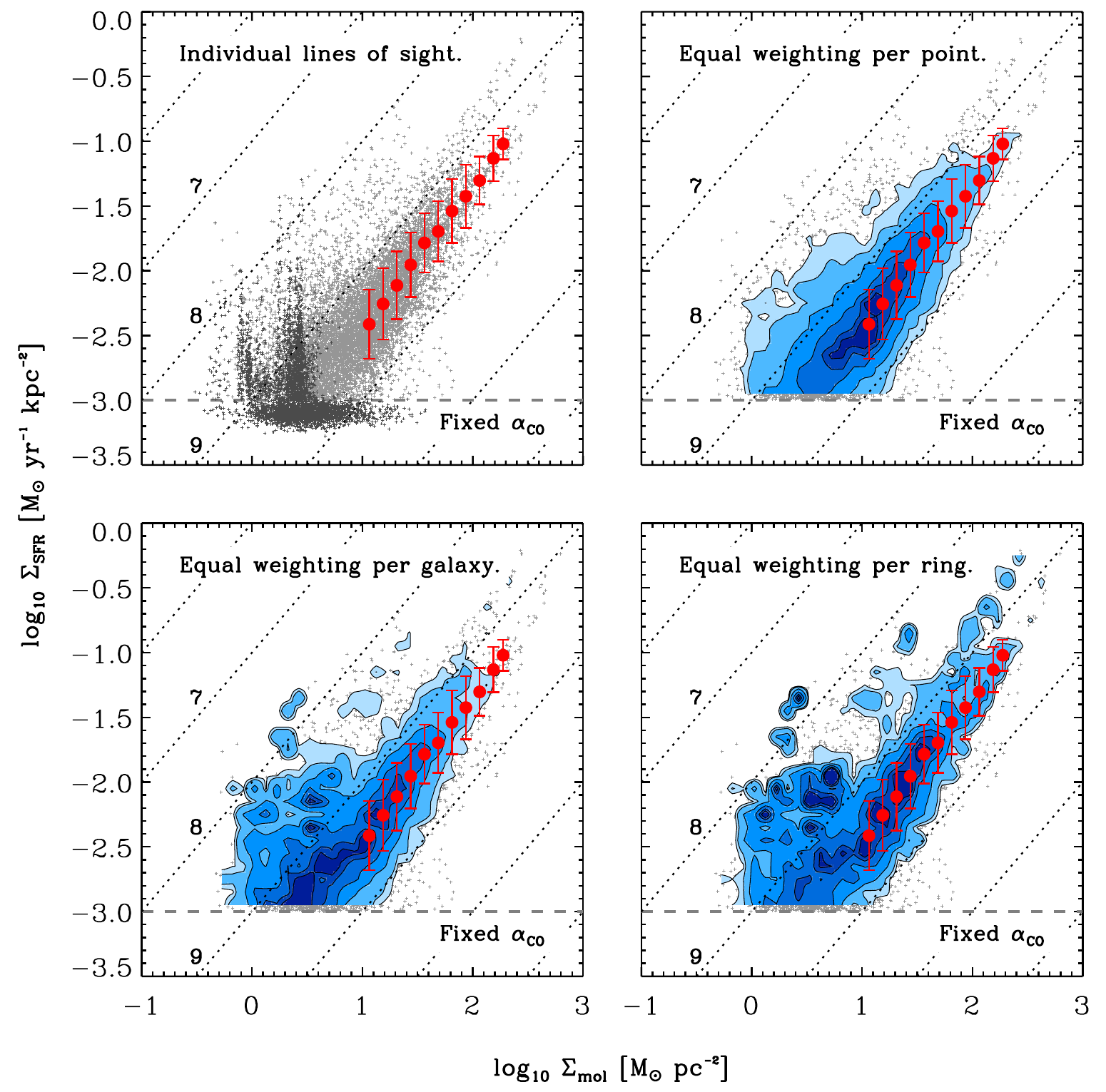}
\caption[Kennicutt-Schmidt relation for galaxies resolved at $\sim$kpc scales]{
\label{fig:tdeph2_leroy13}
Kennicutt-Schmidt relation for $\sim$kpc-sized lines of sight through a sample of nearby galaxies, computed with fixed CO-H$_2$ conversion factor. The four panels show points individual lines of sight (top left), contours with equal weighting per line of sight (top right), contours with equal weighting per galaxy (bottom left) and contours with equal weighting per azimuthal ring (bottom right). Dotted lines of slope unity are lines of constant $t_{\mathrm{dep}} = \Sigma_{\mathrm{SFR}} / \Sigma_{\mathrm{mol}}$, with the number indicating the log of the depletion time in yr. Gray horizontal dashed lines mark the star formation rate sensitivity limit. Credit: \citet{leroy13a}, \copyright AAS. Reproduced with permission.
}
\end{figure}

One of the first results to emerge from these studies was the strikingly-good correlation between molecular gas and star formation when both are measured at $\sim 0.5-1$ kpc scales (Figure \ref{fig:tdeph2_leroy13}). The correlation between molecular gas and star formation is noticeably tighter than the galaxy-averaged correlation first explored by Kennicutt. In nearby galaxies, at least in the inner disks where CO is bright enough to be detectable, there appears to be a roughly constant depletion time $t_{\rm dep} = \Sigma_{\rm H_2}/\Sigma_{\rm SFR} \approx 2$ Gyr. There is considerable debate about whether the depletion time is actually constant, or whether it increases or decreases slightly with $\Sigma_{\rm H_2}$. This debate mostly turns on technical questions of how to handle background subtraction and correct for contamination, and on how to properly fit a very noisy data set. Thus indices within a few tenths of $1.0$ for $\Sigma_{\rm SFR}$ versus $\Sigma_{\rm H_2}$ cannot be ruled out. Nonetheless, the correlation is clear and striking.

Also striking is the extent to which this depletion time is \textit{in}sensitive to any other properties of the galaxy.  Varying the stellar surface density or the local orbital timescale, or the dust to gas ratio (once a dust to gas-dependent $X_{\rm CO}$ factor has been used) appears to have no significant effect on the star formation rate per unit molecular gas mass. Note that the lack of dependence on the orbital time scale is in striking contrast to the results for whole galaxy star formation rates, where plotting things in terms of surface density does not yield a single, simple sequence, but plotting in terms of surface density normalized by orbital time does.

\begin{figure}
\includegraphics[width=\linewidth]{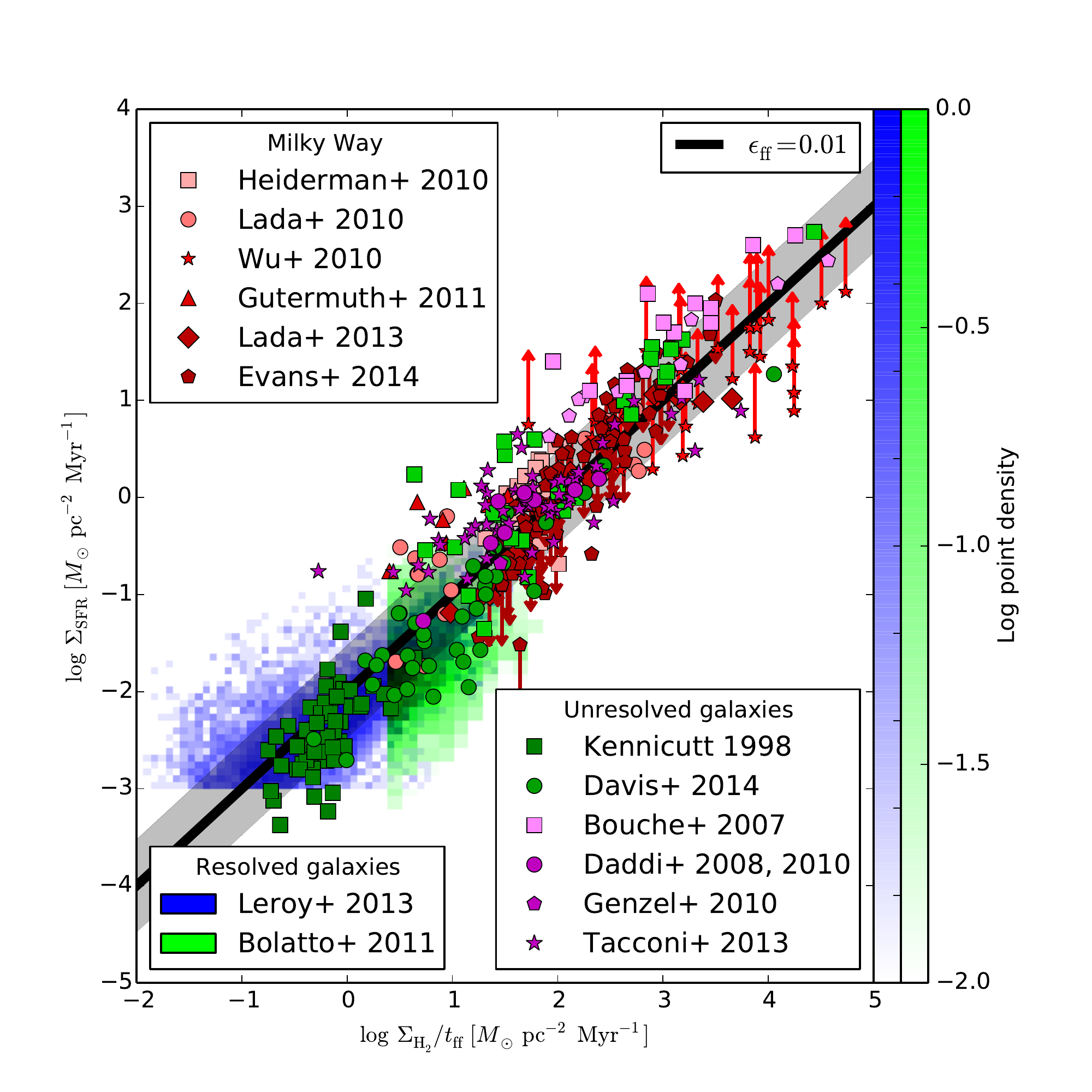}
\caption[Kennicutt-Schmidt relation normalized by the free-fall time]{
\label{fig:ks_krumholz14}
Kennicutt-Schmidt relation normalized by the estimated free-fall time. Points plotted include resolved pixels in nearby galaxies (blue and green rasters), unresolved galaxies at low (green) and high (purple) redshift, and individual clouds within the Milky Way (red). Reprinted from Phys. Rep., 539, \citeauthor{krumholz14c}, "The big problems in star formation: The star formation rate, stellar clustering, and the initial mass function", 49-134, 2014, with permission from Elsevier.
}
\end{figure}

How does this compare to the free-fall time in these clouds, which is the natural times scale on which they evolve? We have no direct access to the volume densities in these clouds, so we cannot answer the question directly. However, \citet{krumholz12a} suggest a simple \textit{ansatz} to estimate free-fall times. The idea was to exploit the fact that observed GMCs seem to have surface densities of $\sim 100$ $M_\odot$ pc$^{-2}$ in normal galaxies. They also have characteristic masses comparable to the galactic Toomre mass,
\begin{equation}
M_{\rm GMC} = \frac{\sigma^4}{G^2 \Sigma_{\rm tot}},
\end{equation}
where $\sigma$ is the galactic velocity dispersion and $\Sigma_{\rm tot}$ is the total gas surface density. From a total mass and a surface density, one can compute a mean density and a corresponding free-fall time:
\begin{equation}
\rho_{\rm GMC} = \frac{3\sqrt{\pi}}{4} \frac{G \sqrt{\Sigma_{\rm GMC}^3 \Sigma_{\rm tot}}}{\sigma^2}.
\end{equation}
This must break down once the mean density at the mid-plane of the galaxy rises too high, as it must in some galaxies where the total gas surface density is $\gg 100$ $M_\odot$ pc$^{-2}$. To be precise, the mid-plane pressure in a galactic disk can be written
\begin{equation}
P = \rho \sigma^2 = \frac{\pi}{2} \phi_P G \Sigma_{\rm tot}^2,
\end{equation}
where $\phi_P$ is a constant of order unity that depends on the ratio of gas to stellar mass. For a pure gas disk, in the diffuse matter class we have shown that $\phi_P = 1$, but realistic values in actual galaxy disks are $\sim 3$. Combining these statements, we obtain
\begin{equation}
\rho_{\rm mp} = \frac{\pi \phi_P G \Sigma_{\rm tot}^2}{2\sigma^2}.
\end{equation}
The simple approximation suggested by \citeauthor{krumholz12a} is just to use the larger of $\rho_{\rm GMC}$ and $\rho_{\rm mp}$. If one does so, then it becomes possible to estimate $t_{\rm ff}$ from observable quantities. The result of this exercise is that the observed depletion times seen in external galaxies are generally consistent with $\epsilon_{\rm ff} \approx 0.01$, with a scatter of about a factor of 3. The same is true if we put the whole-galaxy points on the plot, although for them the uncertainties are considerably greater (Figure \ref{fig:ks_krumholz14}).

Finally, some important caveats are in order. First, this result is limited to the inner parts of galaxies where there is significant CO emission. In outer disks where there is little molecular gas and CO is faint, molecular emission can be detected only by stacking entire rings or focusing on local patches of strong emission, so the sort of pixel-by-pixel unbiased analysis done for inner galaxies is not yet possible. Second, this sample covers a very limited range of galaxy properties, certainly compared to the high-$z$ data. The pixel by pixel analysis can only be done for a large sample of local galaxies, within $\sim 20$ Mpc, and this volume does not contain any of the starbursts that form the upper part of the sequences seen in the local Kennicutt or high-$z$ samples.

\begin{marginfigure}
\includegraphics[width=\linewidth]{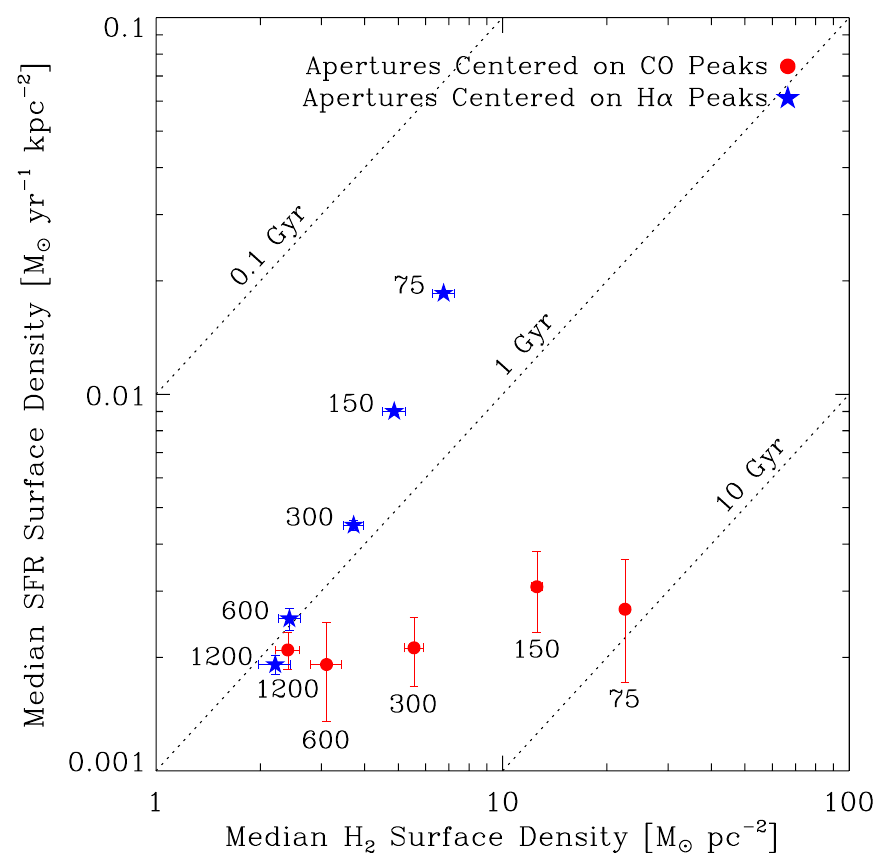}
\caption[Kennicutt-Schmidt relation averaged on different size scales in M33]{
\label{fig:ksscale_schruba10}
Kennicutt-Schmidt relation on different size scales. The points show the median surface densities of gas and star formation, using apertures of $75-1200$ pc in size, centered in CO peaks (red) and H$\alpha$ peaks (blue). Dotted lines of slope unity are lines of constant $t_{\mathrm{dep}} = \Sigma_{\mathrm{SFR}} / \Sigma_{\mathrm{mol}}$, with the number indicating the depletion time. Credit: \citet{schruba10a}, \copyright AAS. Reproduced with permission.
}
\end{marginfigure}

A third and final caveat has to do with scale-dependence. Depending on the scales over which one averages, the correlation between molecular gas and star formation can be better or worse. Generally speaking, as one goes to smaller and smaller scales, the scatter in the $\Sigma_{\rm H_2} - \Sigma_{\rm SFR}$ correlation increases, and systematic biases start to appear. If one focuses on peaks of the H$_2$ distribution, one obtains systematically longer depletion times than for similar apertures centered on peaks of the inferred star formation rate distribution. Figure \ref{fig:ksscale_schruba10} illustrates the observational situation.

The most likely explanation for this is that, on sufficiently small scales, the central assumption that we are looking at an "average" piece of a galaxy begins to break down. If we focus on peaks of the H$_2$ distribution, we are looking at places where molecular gas is just now accumulating and there has not yet been time for much star formation to take place. In terms of the classification scheme discussed in Chapter \ref{ch:gmcs}, these represent class I clouds. If we focus on peaks in the H$\alpha$ distribution (the usual proxy for star formation rate in this sort of study), we are looking at H~\textsc{ii} regions where a molecular cloud once was, and which has since mostly been dispersed. In terms of the molecular cloud types discussed in Chapter \ref{ch:gmcs}, these are class III clouds.

In this case our proxies are misleading -- the CO tells us about the instantaneous amount of molecular gas present, while the H$\alpha$ tells us about the average number of stars formed over the last $\sim 5$ Myr, and those are not exactly what we want to compare. We want either to compare the instantaneous molecular mass and star formation rate, or the averages of both molecular mass and star formation rate over similar timescales. If we average over a large enough piece of the galaxy, our beam encompasses clouds in all stages of evolution, so we get a representative average, but that ceases to be true as we go to smaller and smaller scales. Indeed, the characteristic scale at which that ceases to be true can be used as something of a proxy for characteristic molecular cloud lifetime, a point made recently by \citet{kruijssen14c}.

\subsection{Relationship to Atomic Gas and All Neutral Gas}

\begin{marginfigure}
\includegraphics[width=\linewidth]{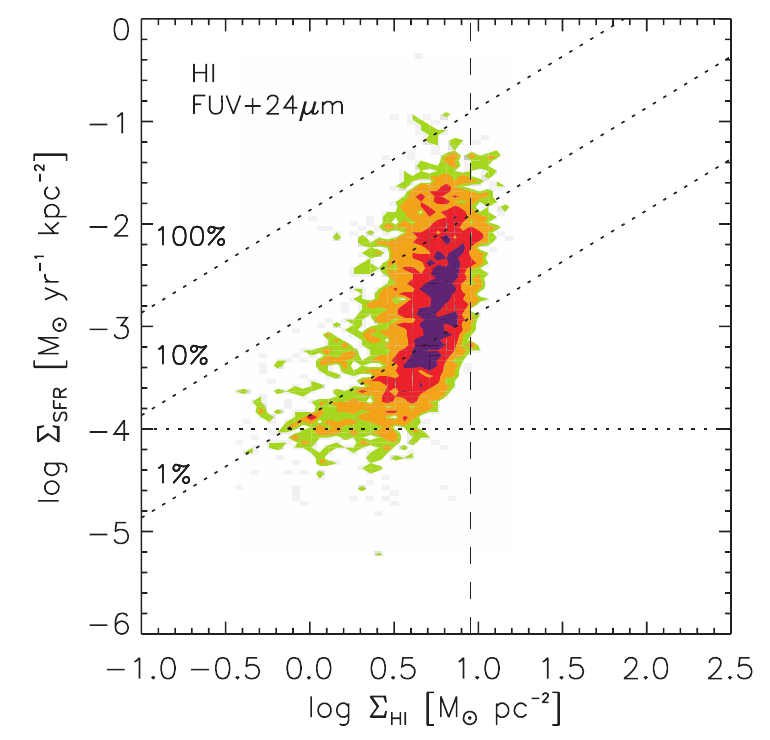}
\caption[Kennicutt-Schmidt relation for H~\textsc{i} gas in inner galaxies]{
\label{fig:kshi_bigiel08}
Kennicutt-Schmidt relation for H~\textsc{i} gas in inner galaxies, averaged on $\sim 750$ pc scales. Contours indicate the density of points. Credit: \citet{bigiel08a}, \copyright AAS. Reproduced with permission.
}
\end{marginfigure}

The results for molecular gas are in striking contrast to the results for total gas or just atomic gas. If one considers only atomic gas, one finds that the H~\textsc{i} surface density reaches a maximum value which it does not exceed, and that the star formation rate is essentially uncorrelated with the H~\textsc{i} surface density when it is at this maximum (Figure \ref{fig:kshi_bigiel08}). In the inner parts of galaxies, star formation does not appear to care about H~\textsc{i}.

\begin{figure}
\includegraphics[width=\linewidth]{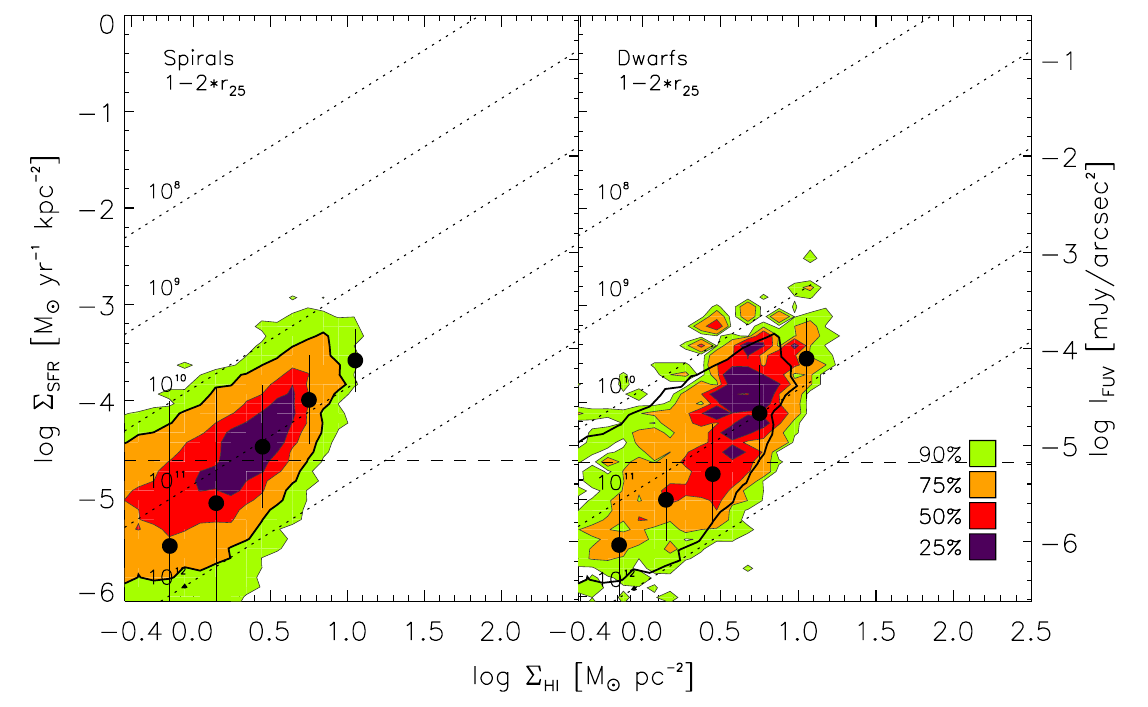}
\caption[Kennicutt-Schmidt relation for H~\textsc{i} gas in outer galaxies]{
\label{fig:kshi_bigiel10}
Kennicutt-Schmidt relation for H~\textsc{i} gas in outer galaxies, averaged on $\sim 750$ pc scales. Contours indicate the density of points, and the two panels are for spirals and dwarfs, respectively. Black points with error bars indicate the mean and dispersion in bins of $\Sigma_{\mathrm{HI}}$. Credit: \citet{bigiel10a}, \copyright AAS. Reproduced with permission.
}
\end{figure}

On the other hand, if one considers the outer parts of galaxies, there is a correlation between H~\textsc{i} content and star formation, albeit with a very, very large scatter (Figure \ref{fig:kshi_bigiel10}). While there is a correlation, the depletion time is extremely long -- typically $\sim 100$ Gyr. It is important to point out that, while it is not generally possible to detect CO emission over broad areas in these outer disks, when one stacks the data, the result is that the depletion time in molecular gas is still $\sim 2$ Gyr. Thus these very long depletion times appear to be a reflection of a very low H$_2$ to H~\textsc{i} ratio, but one that does not go all the way to zero, and instead stops at a floor of $\sim 1-2\%$.

\begin{figure}
\includegraphics[width=\linewidth]{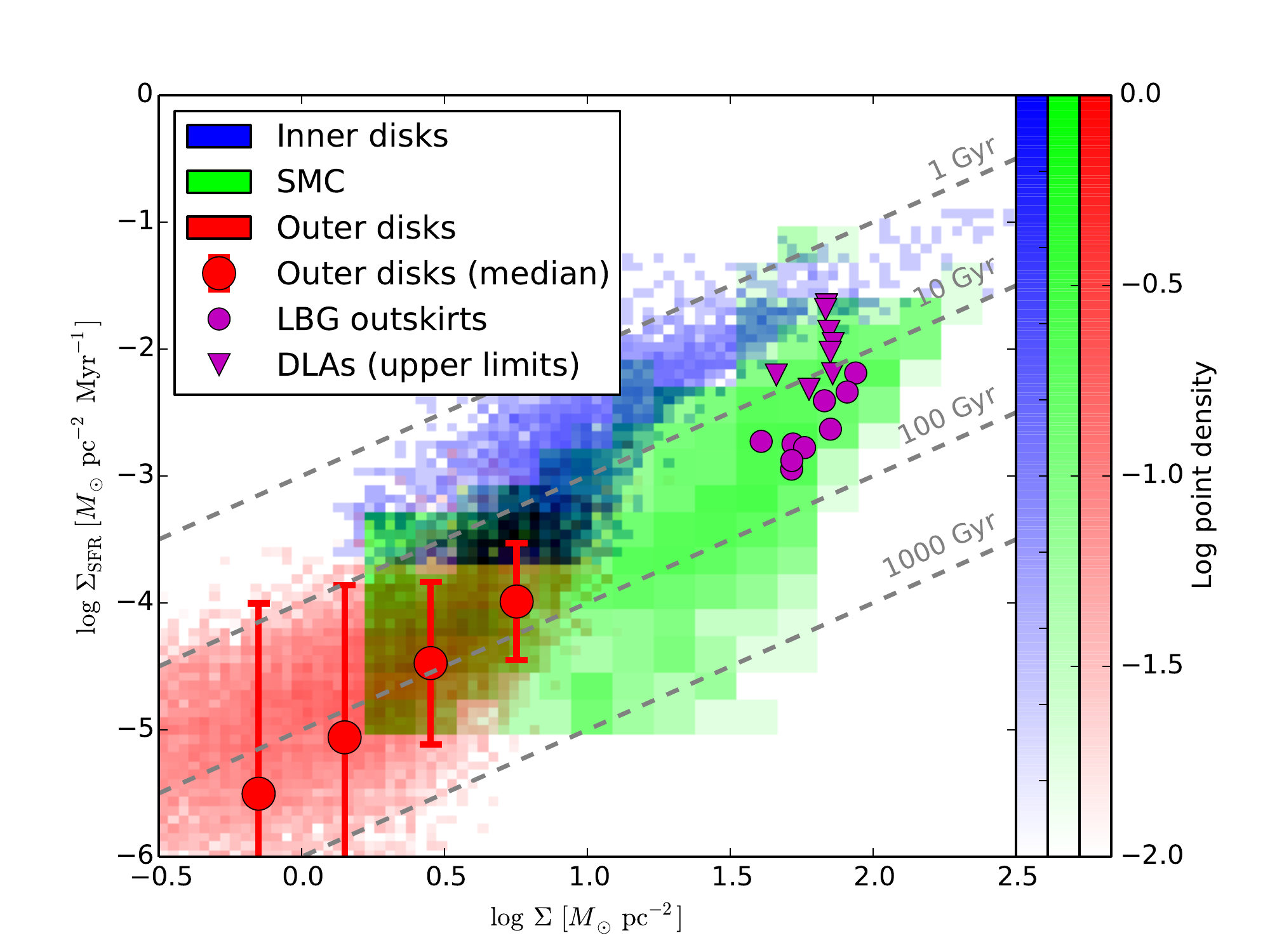}
\caption[Kennicutt-Schmidt relation for total gas in resolved galaxies]{
\label{fig:kstot_krumholz14}
Kennicutt-Schmidt relation for all gas, atomic plus molecular. The rasters show lines of sight through inner and outer galaxies and through the Small Magellanic Cloud, as indicated. Purple points indicate individual lines of sight through high-redshift systems, where H~\textsc{i} columns are measured in Ly$\alpha$ absorption. Reprinted from Phys. Rep., 539, \citeauthor{krumholz14c}, "The big problems in star formation: The star formation rate, stellar clustering, and the initial mass function", 49-134, 2014, with permission from Elsevier.
}
\end{figure}

If instead of plotting just atomic or molecular gas on the $x$-axis, one plots total gas, then a clear relationship emerges. At high gas surface density, the ISM is mostly H$_2$, and this gas forms stars with a constant depletion time of $\sim 2$ Gyr. In this regime, the H~\textsc{i} surface density saturates at $\sim 10$ $M_\odot$ pc$^{-2}$, and has no relationship to the star formation rate. This constant depletion time begins to change at a total surface density of $\sim 10$ $M_\odot$ pc$^{-2}$, at which point the ISM begins to transition from H$_2$-dominated to H~\textsc{i}-dominated. Below this critical surface density, the star formation rate drops precipitously, and the depletion time increases by a factor of $\sim 50$ over a very small range in total gas surface density. Finally, below $\sim 10$ $M_\odot$ pc$^{-2}$, the star formation rate does correlate with both the total and H~\textsc{i} surface densities (which are roughly the same), but the depletion time is extremely long, and there is an extremely large amount of scatter. Figure \ref{fig:kstot_krumholz14} summarizes the data.

\subsection{Additional Parameters}

In the H$_2$-dominated regime, as we have seen nothing seems to affect the star formation rate per unit molecular mass. However, that is not the case in the H~\textsc{i}-dominated regime, where the scatter is large and "second parameters" seem to have an effect. This regime is not understood very well, and the data are still incomplete, but two striking correlations are apparent in the data. First, in the H~\textsc{i}-dominated regime, the metallicity of the gas seems to matter. This is obvious is we compare the Small Magellanic Cloud, at metallicity 20\% of Solar, damped Lyman $\alpha$ systems (which have $\sim 10\%$ of Solar metallicity), and other low-metallicity dwarf galaxies (Figure \ref{fig:kstot_krumholz14}) to the bulk of the sample, which has near Solar metallicity. Indeed, the main effect of a low metallicity seems to be that the characteristic value of $\sim 10$ $M_\odot$ pc$^{-2}$ at which the gas goes from H~\textsc{i}- to H$_2$-dominated is shifted to higher surface densities.

Another parameter that appears to matter is the stellar surface density. Higher stellar surface densities appear to yield higher H$_2$ fractions and higher star formation rates at fixed gas surface density in the H~\textsc{i}-dominated regime. This correlation appears to be on top of the correlation with metallicity. Similarly, galactocentric radius seems to matter. Since all of these quantities are correlated with one another, it is hard to know what the driving factor(s) are.

\section{Star Formation in Dense Gas}

\subsection{Alternatives to CO}

The far all the observations we have discussed have used CO (or, in a few cases, dust emission) as the proxy of choice for H$_2$. This is by far the largest and richest data set available right now. However, it is of great interest to consider other tracers as well, in particular tracers of gas at higher densities. Doing so makes it possible, in principle, to map out the density distribution within the gas in another galaxy, and thereby to gain insight into how gas at different densities is correlated with star formation.

Moving past H$_2$, the next-brightest molecular line (not counting isotopologues of CO, which are generally found under the same conditions) in most galaxies is HCN. Other bright molecules are HCO$^+$, CS, and HNC, but we will focus on HCN as a synecdoche for all of these tracers. Like CO, the HCN molecule has rotational transitions that can be excited at low temperatures, and is abundant because it combines some of the most abundant elements. Thus the data set for correlations of HCN with star formation is the second-largest after CO. However, it is important to realize that this data set is still quite limited, and biased toward starburst galaxies where the HCN/CO ratio is highest. In normal galaxies HCN is $\sim 10$ times dimmer than CO, leading to $\sim 100\times$ larger mapping times in order to reach the same signal to noise. As a result, we are with HCN today roughly where we were with CO back in the time of \citet{kennicutt98a}, though that is starting to change.

Before diving into the data, let us pause for a moment to compare CO and HCN. The first few excited rotational states of CO lie 5.5, 16.6, 33.3, and 55.4 K above ground; the corresponding figures for HCN are 4.3, 12.8, 25.6, and 42.7 K. Thus the temperature ranges probed are quite similar, and all lines are relatively easy to excite at the temperatures typically found in molecular clouds. For CO, the collisional de-excitation rate coefficient for the $1-0$ transition is $k_{10} = 3.3\times 10^{-11}$ cm$^3$ s$^{-1}$ (at 10 K, for pure para-H$_2$ for simplicity), and the Einstein $A$ for the same transition is $A_{10} = 7.2\times 10^{-8}$ s$^{-1}$, giving a critical density
\begin{equation}
n_{\rm crit} = \frac{A_{10}}{k_{10}} = 2200\mbox{ cm}^{-3}.
\end{equation}
As discussed in Chapter \ref{ch:obscold}, radiative transfer effects lower the effective critical density significantly. In contrast, the collisional de-excitation rate coefficient and Einstein $A$ for HCN $1-0$ are $k_{10} = 2.4\times 10^{-11}$ cm$^3$ s$^{-1}$ and $A_{10}=2.4\times 10^{-5}$ s$^{-1}$, giving $n_{\rm crit} = 1.0\times 10^6$ cm$^{-3}$. Again, this is lowered somewhat by optical depth effects, but there is nonetheless a large contrast with CO. In effect, CO emission switches from rising quadratically with density to rising linearly with density at much lower volume density than does HCN, and thus HCN emission is considerably more weighted to denser gas. For this reason, HCN is often thought of as a tracer of the "dense" gas in galaxies.

\subsection{Correlations}

The first large survey of HCN emission from galaxies was undertaken by \citet{gao04b, gao04a}. This study had no spatial resolution -- it was simply one beam per galaxy. They found that, while CO luminosity measured in the same one-beam-per-galaxy fashion was correlated non-linearly with infrared emission (which was the proxy for star formation used in this study), HCN emission in contrast correlated almost linearly with IR emission. \citet{wu05a} showed that individual star-forming clumps in the Milky Way fell on the same linear correlation as the extragalactic observations.

\begin{figure}
\includegraphics[width=\linewidth]{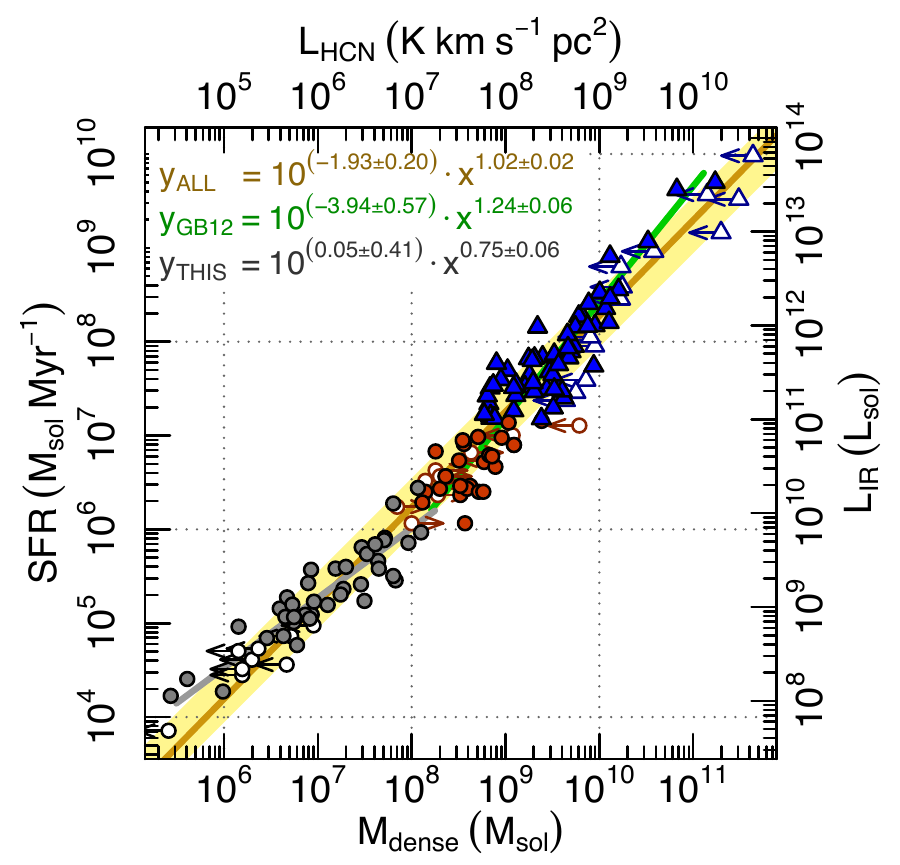}
\caption[Infrared-HCN luminosity correlation]{
\label{fig:hcn-ir}
Observed correlation between HCN luminosity (converted to a mass of "dense" gas using an $X$ factor) and infrared luminosity (converted to a star formation rate). Gray points show resolved observations within galaxy disks, while red and blue points show unresolved observations of entire galaxy disks. Open points indicate upper limits. Credit: \citet{usero15a}, \copyright AAS. Reproduced with permission.
}
\end{figure}

This linear correlation was at first taken to be a sign that the HCN-emitting gas was the "dense" gas that was actively star-forming. In this picture, the high rates of star formation found in starburst galaxies are associated with the fact that they have high "dense" gas fractions, as diagnosed by high HCN to CO ratios. More recent studies, however, have shown that the correlation is not as linear as the initial studies suggested. Partly this is a matter of technical corrections to the existing data (e.g., observations that covered more of the disk of a galaxy), partly a matter of obtaining more spatially-resolved data (as opposed to one beam per galaxy), and partly a matter of expanded samples. Figure \ref{fig:hcn-ir} shows a recent compilation from \citet{usero15a}.

Despite these revisions, it is clear that there is a generic trend that HCN and other tracers that have higher critical densities have a correlation with star formation that is flatter than for lower critical density tracers -- that is, a power law fit of the form
\begin{equation}
L_{\rm IR} \propto L_{\rm line}^p
\end{equation}
will recover an index $p$ that is closer to unity for higher critical density lines and further from unity for lower critical density ones. This has now been seen not just between HCN and CO, but also with higher $J$ lines of CO, and with HCO$^+$, another fairly bright line for which large enough data sets exist to make correlations.

\subsection{Physical Interpretation and Depletion Times}

To go beyond the sheer correlations, one must attempt to convert the observed quantities into physical ones. For infrared emission, this is straightforward: the galaxies where we have dense gas tracers are almost exclusively ones with high star formation rates, gas surface densities, and dust content. For them it is safe to assume that the great majority of the light from young stars is reprocessed into infrared emission, and, conversely, that IR emission is driven primarily by newly-formed stars.

To convert the HCN emission into a mass, we require an HCN X factor analogous to $X_{\rm CO}$. Since the HCN $J=1-0$ line, and other low $J$ lines, are generally optically thick, such a conversion factor can be derived from theoretical arguments much like the ones we used to estimate $X_{\rm CO}$. The conversion factor does not depend on the HCN abundance, which is good, because that is not tremendously well known. However, the resulting conversion is still significantly more uncertain that for CO, because, unlike the case for CO, it has not been calibrated against independent tracers of the mass like dust or $\gamma$-rays.

There is also a real physical ambiguity worth noting. For CO, we are essentially looking at all the gas where CO is present, because the critical density is low enough (once radiative transfer effects are accounted for) that we can assume that most of the gas is in the regime where emission is linear in number of emitting molecules. For HCN, on the other hand, some gas is in the high-density linear regime and some in the low-density quadratic regime, and thus it is not entirely clear what mass we are measuring. It will be a complicated, density-weighted average, which will tell us something about the mass of gas denser than the mean, but how much is not quite certain.

If one ignores all these complications and converts an observed HCN luminosity into a mass and an observed IR luminosity into a star formation rate, one can then derive a depletion time for the HCN-emitting gas. Typical depletion times are $\sim 10-100$ Myr, much smaller than for CO. On the other hand, we are also looking at much denser gas. If one makes a reasonable guess at the density, one can make a corresponding estimate of the free-fall time. At a density of $10^5$ cm$^{-3}$ (probably about the right density once one takes radiative transfer effects into account), the free-fall time is $t_{\rm ff} = 100$ kyr, so a star formation timescale of $t_{\rm dep} = 10$ Myr corresponds to
\begin{equation}
\epsilon_{\rm ff} = \frac{t_{\rm ff}}{t_{\rm dep}} \sim 10^{-3} - 10^{-2},
\end{equation}
with a fairly large uncertainty. However, $\epsilon_{\mathrm{ff}} \sim 1$ is clearly ruled out.

\chapter{The Star Formation Rate at Galactic Scales: Theory}
\label{ch:sflaw_th}

\marginnote{
\textbf{Suggested background reading:}
\begin{itemize}
\item \href{http://adsabs.harvard.edu/abs/2014arXiv1402.0867K}{Krumholz, M.~R. 2014, Phys.~Rep., 539, 49}, section 4 \nocite{krumholz14c}
\end{itemize}
\textbf{Suggested literature:}
\begin{itemize}
\item \href{http://adsabs.harvard.edu/abs/2011ApJ...731...41O}{Ostriker, E.~C., \& Shetty, R. 2011, ApJ, 731, 41} \nocite{ostriker11a}
\item \href{http://adsabs.harvard.edu/abs/2013MNRAS.436.2747K}{Krumholz, M.~R. 2013, MNRAS, 436, 2747} \nocite{krumholz13c}
\end{itemize}
}

Chapter \ref{ch:sflaw_obs} was a brief review of the current state of the observations describing the correlation between star formation and gas in galaxies. This chapter will focus on theoretical models that attempt to unify and make sense of these observations. To recap, there are a few broad observational results we would like any successful model to reproduce:
\begin{itemize}
\item Star formation appears to be a very slow or inefficient process, measured on both the galactic scale and the scale of individual molecular clouds (at least for local clouds). The depletion time is $\sim 100$ times larger than the free-fall time.
\item In unresolved observations, the rate of star formation appears to rise non-linearly with the total gas content.
\item In the central disks of galaxies, where most star formation takes place, star formation appears to correlate strongly with the molecular phase of the ISM, and poorly or not at all with the atomic phase.
\item The depletion time in molecular gas is nearly constant in nearby, "normal" galaxies, though a weak dependence on total gas surface density cannot be ruled out given the observational uncertainties. In more actively star-forming galaxies with higher gas surface densities than any found within $\sim 20$ Mpc of the Milky Way, the depletion time does appear to be smaller.
\item  A correlation between star formation and atomic gas appears only in regions where the ISM is completely dominated by atomic gas, but with a very large scatter, and with a depletion time in the atomic gas is $\sim 2$ order of magnitude larger than that in molecular gas. In such regions, "second parameters" such as the metallicity or the stellar mass density appear to affect the star formation rate in ways that they do not in inner disks.
\item If one uses tracers of higher density gas such as HCN, the depletion time is shorter than for the bulk of the molecular gas, but still remains much longer than any plausible estimate of the free-fall time in the emitting gas.
\end{itemize}
As we shall see, there is at present no theory that is capable of fully, self-consistently explaining all the observations. However, there are a number of approaches that appear to successfully explain at least some of the observations, and may serve as the nucleus for a fuller theory in the future.

\section{The Top-Down Approach}

Theoretical attempts to explain the correlation between gas and star formation in galaxies can be roughly divided into two categories: those that focus on regulation by galactic scale processes, and those that focus on regulation within individual molecular clouds. We will generically refer to the former as "top-down" models, and the latter as "bottom-up" models.

\subsection{Hydrodynamics Plus Gravity}

\begin{marginfigure}
\includegraphics[width=\linewidth]{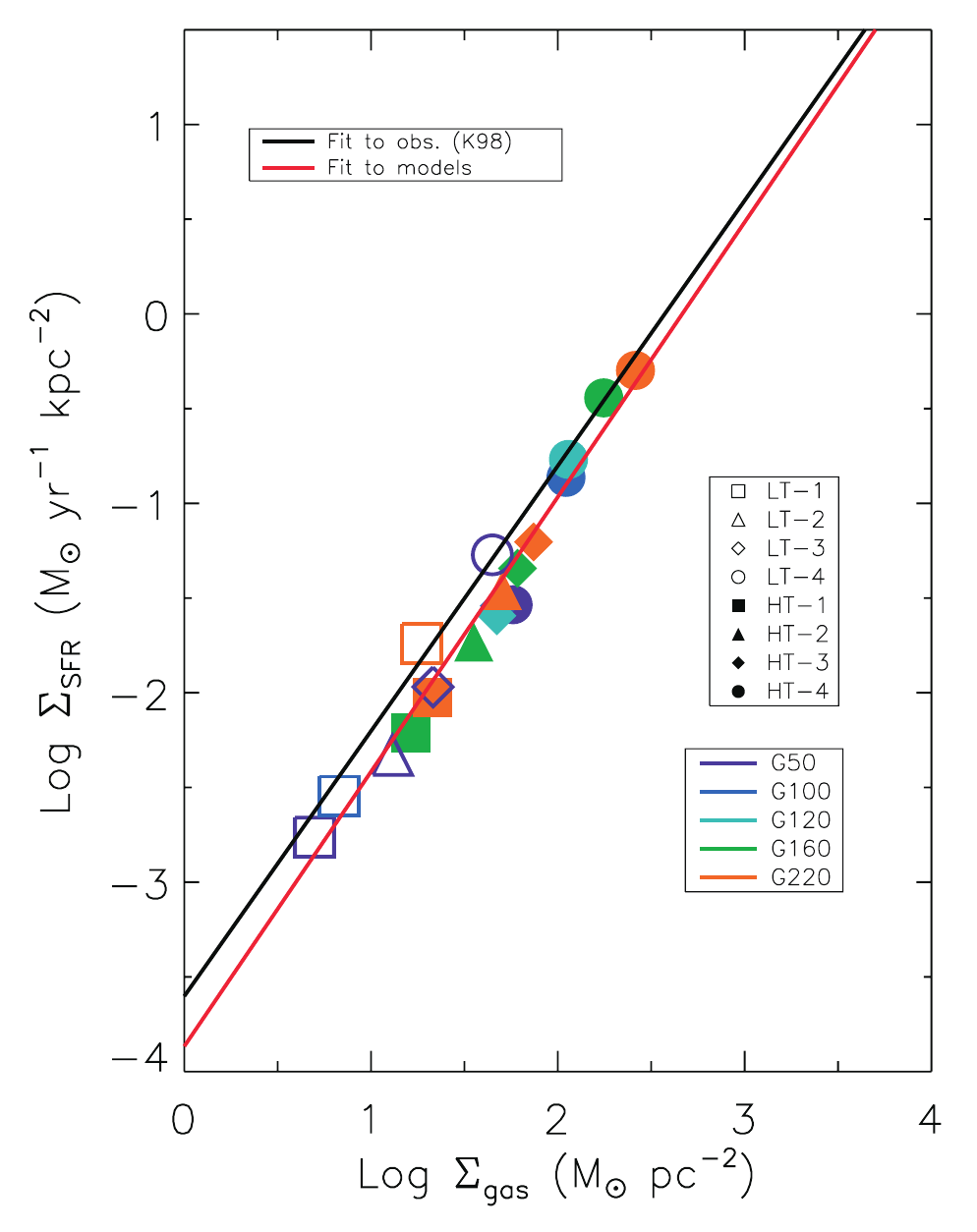}
\caption[Kennicutt-Schmidt relation from simulations with only gravity and hydrodynamics]{
\label{fig:kslaw_sim_li05}
Relationship between gas surface density $\Sigma_{\mathrm{gas}}$ and star formation surface density $\Sigma_{\mathrm{SFR}}$, measured from a series of simulations using no physics except hydrodynamics and gravity. Credit: \citet{li05a}, \copyright AAS. Reproduced with permission.
}
\end{marginfigure}

The simplest approach to the problem of the star formation rate is to consider no physics beyond hydrodynamics and gravity, including no stellar feedback. Models with only these ingredients form a useful baseline against which more sophisticated models may be compared. A key question for such models is the extent to which large-scale gravitational instability is expected. This is parameterized by the \citet{toomre64a} $Q$ parameter, where
\begin{equation}
Q = \frac{\Omega \sigma}{\pi G \Sigma}.
\end{equation}
Here $\Omega$ is the angular velocity of the disk rotation, $\sigma$ is the gas velocity dispersion, and $\Sigma$ is the gas surface density. Problem set 3 contains a calculation showing that systems with $Q < 1$ are unstable to axisymmetric perturbations, while those with $Q>1$ are stable. Observed galactic disks appear to have $Q \approx 1$ over most of their disks, rising to $Q > 1$ at the edges of the disk.

The fact that galactic disks tend to reach $Q \sim 1$ suggests that gravitational instability occurring on galactic scales might be an important driver of star formation. If this is in fact the case, then the rate of star formation is likely to be non-linearly sensitive to the value of $Q$, and thus to the gas surface density, potentially giving rise to the non-linear correlation between gas surface density and star formation rate seen in the unresolved observations. Some simulations are able to reproduce exactly this effect (Figure \ref{fig:kslaw_sim_li05}).

On the other hand, the fact that star formation rate does not go to zero in outer disks suggests that star formation can still occur even in regions of galaxies where $Q > 1$, if at a lower rate. We might expect this because in a multi-phase ISM there will still be places where the gas becomes locally cold and has $\sigma$ much less than the disk average, producing a local value of $Q$ that is lower than the mean. Such regions of dense, cold gas are expected to appear wherever the density is driven up by spiral arms or similar global structures. In this case we might expect that the star formation timescale should be proportional to the frequency with which spiral arms pass through the disk, and thus we should have
\begin{equation}
\Sigma_{\rm SFR} \propto \frac{\Sigma}{t_{\rm orb}},
\end{equation}
consistent with one of the observed parameterizations for galaxy-averaged star formation rates. In this approach, the key physics driving star formation is not the self-gravity of a galactic disk, but instead the ability of the gas to cool to low temperatures behind spiral shocks.

As a theory for the star formation rate, these models are mainly useful for target practice. (In fairness, they are often intended to study things other than the star formation rate, and thus make only minimal efforts to get this rate right.) We can identify a few obvious failings by comparing to our observational checklist. First of all, in these models, once gravitationally bound clouds form, there is nothing to stop them from collapsing on a timescale comparable to $t_{\rm ff}$. As a result, the star formation rate in molecular gas that these models predicts tends to be $\epsilon_{\rm ff} \sim 1$, rather than $\sim 0.01$, unless the models introduce an artificial means to lower the star formation rate -- in other words, these models produce rapid, efficient star formation rather than slow, inefficient star formation as required by the data.

A second problem is that these models do not naturally predict any metallicity-dependence. Gravitational instability and large-scale spiral waves do not obviously care about the metallicity of the gas, but observations strongly suggest that metallicity does matter. It is possible that the interaction of spiral arms with cooling might give rise to a metallicity dependence in the star formation rate, but this has not be been explored.

\subsection{Feedback-Regulated Models}

\paragraph{Derivation}

The usual response to the failures of gravity plus hydro-only models has been to invoke "feedback". The central idea for these models can be understood analytically quite simply. The argument we give here is taken mostly from \citet{ostriker11a}. We begin by considering the gas momentum equation, ignoring viscosity but including magnetic fields:
\begin{equation}
\frac{\partial}{\partial t} (\rho\vecv) = - \nabla\cdot (\rho\vecv\vecv) - \nabla P + \frac{1}{4\pi} \nabla \cdot \left(\vecB\vecB - \frac{B^2}{2} \vecI\right) + \rho \vecg,
\end{equation}
where $\vecg$ is the gravitational force per unit mass, and the pressure $P$ includes all sources of pressure -- thermal pressure plus radiation pressure plus cosmic ray pressure. Let us align our coordinate system so that the galactic disk lies in the $xy$ plane. The $z$ component of this equation, corresponding to the vertical component, is simply
\begin{equation}
\frac{\partial}{\partial t} (\rho v_z) = - \nabla\cdot (\rho\vecv v_z) - \frac{dP}{dz} + \frac{1}{4\pi} \nabla \cdot \left(\vecB B_z\right) - \frac{1}{8\pi} \frac{d}{dz} B^2 + \rho g_z.
\end{equation}

Now let us consider some area $A$ at constant height $z$, and let us average the above equation over this area. The equation becomes
\begin{eqnarray}
\frac{\partial}{\partial t} \langle \rho v_z\rangle & = & -\frac{1}{A} \int_A \nabla\cdot (\rho\vecv v_z) \, dA - \frac{d\langle P\rangle}{dz} +
\frac{1}{4\pi A} \int_A \nabla \cdot \left(\vecB B_z\right)\, dA 
\nonumber \\
& & {} - \frac{1}{8\pi} \frac{d}{dz} \langle B^2 \rangle + \left\langle\rho g_z\right\rangle,
\end{eqnarray}
where for any quantity $Q$ we have defined
\begin{equation}
\langle Q\rangle \equiv \frac{1}{A}\int_A Q\, dA.
\end{equation}
We can simplify this a bit by separating the $x$ and $y$ components from the $z$ components of the divergences and making use of the divergence theorem:
\begin{eqnarray}
\frac{\partial}{\partial t} \langle \rho v_z\rangle & = & - \frac{d\langle P\rangle}{dz} - \frac{1}{8\pi} \frac{d}{dz} \langle B^2 \rangle + \left\langle\rho g_z\right\rangle  - \frac{d}{dz} \langle \rho v_z^2\rangle + \frac{1}{4\pi} \frac{d}{dz} \langle B_z^2\rangle \nonumber \\
& & {}- \frac{1}{A} \int_A \nabla_{xy}\cdot (\rho\vecv v_z) \, dA
\nonumber \\
& & {}  +
\frac{1}{4\pi A} \int_A \nabla_{xy} \cdot \left(\vecB B_z\right)\, dA \\
& = & - \frac{d\langle P\rangle}{dz} - \frac{1}{8\pi} \frac{d}{dz} \langle B^2 \rangle + \left\langle\rho g_z\right\rangle  - \frac{d}{dz} \langle \rho v_z^2\rangle + \frac{1}{4\pi} \frac{d}{dz} \langle B_z^2\rangle \nonumber \\
& & {}-\frac{1}{A} \int_{\partial A} v_z \rho \vecv \cdot \nhat \, d\ell + \frac{1}{4\pi A} \int_{\partial A} B_z \vecB \cdot \nhat \, d\ell.
\end{eqnarray}
where $\partial A$ is the boundary of the area $A$, and $\nhat$ is a unit vector normal to this boundary, which always lies in the $xy$ plane.

Now let us examine the last two terms, representing integrals around the edge of the area. The first of these integrals represents the advection of $z$ momentum $\rho v_z$ across the edge of the area. If we consider a portion of a galactic disk that has no net flow of material within the plane of the galaxy, then this must, on average, be zero. Similarly, the second integral is the rate at which $z$ momentum is transmitted across the boundary of the region by magnetic stresses. Again, if we are looking at a galactic disk in steady state with no net flows or advection in the plane, this must be zero as well. Thus the last two integrals are generally zero and can be dropped.

If we further assume that the galactic disk is approximately time steady, the time derivative is also obviously zero. We therefore arrive at an equation of hydrostatic balance for a galactic disk,
\begin{equation}
\frac{d}{dz} \left\langle P + \rho v_z^2 + \frac{B^2}{8\pi}\right \rangle  - \frac{d}{dz} \left\langle\frac{B_z^2}{4\pi}\right\rangle - \left\langle \rho g_z\right\rangle = 0
\end{equation}
The first term represents the upward force due to gradients in the total pressure, including the turbulent pressure $\rho v_z^2$ and the magnetic pressure $B^2/8\pi$. The third term represents the downward force due to gravity. The middle term represents forces due to magnetic tension, and is usually sub-dominant because it requires a special geometry to exert significant forces -- the field would need to be curved upward (think of a hammock) or downward (think of an arch) over most of the area of interest. Thus we are left with balancing the first and last terms.

The quantities in angle brackets can be thought of as forces, but they can equivalently be thought of as momentum fluxes. Each one represents the rate per unit area at which momentum is transported upward or downward through the disk, and in hydrostatic equilibrium these transport rates must match. The central \textit{ansatz} in the feedback-regulated model is to equate the rate of momentum transport represented by the first term with the rate of momentum injection by feedback. To be precise, one approximates that
\begin{equation}
\left \langle P + \rho v_z^2 + \frac{B^2}{8\pi}\right \rangle \sim \left\langle\frac{p}{M}\right\rangle \Sigma_{\rm SFR}
\end{equation}
where $\langle p/M\rangle$ is the momentum yield per unit mass of stars formed, due to whatever feedback processes we think are important. The quantity on the right hand side is the rate of momentum injection per unit area by star formation.

What follows from this assumption? To answer that, we have to examine the gravity term. For an infinite thin slab of material of surface density $\Sigma$, the gravitational force per unit mass above the slab is
\begin{equation}
g_z = 2\pi G \Sigma.
\end{equation}
Note that $\Sigma$ here should be the total mass per unit area within roughly 1 gas scale height of the mid-plane, including the contribution from both gas and stars. If we plug this in, then we get
\begin{equation}
\frac{d}{dz} \left(\left\langle \frac{p}{M}\right\rangle \Sigma_{\rm SFR}\right) \sim 2\pi G \Sigma \rho
\quad \Longrightarrow \quad
\Sigma_{\rm SFR} \sim 2\pi G \left\langle \frac{p}{M}\right\rangle^{-1} \Sigma \Sigma_{\rm gas} ,
\end{equation}
where we have taken the vertical derivative $d/dz$ to be of order $1/h$, where $h$ is the gas scale height, and we have taken $\rho h \sim \Sigma_{\rm gas}$.

Thus in a feedback-regulated model, we expect a star formation rate that scales as the product of the gas surface density and the total surface density. In regions where gas dominates the gravity, so that $\Sigma \sim \Sigma_{\rm gas}$, we will have a star formation law
\begin{equation}
\Sigma_{\rm SFR} \propto \Sigma_{\rm gas}^2,
\end{equation}
while in regions where stars dominate we will instead have
\begin{equation}
\Sigma_{\rm SFR} \propto \Sigma_{\rm gas} \Sigma_*.
\end{equation}

To the extent that we think we know the momentum yield from star formation, $\langle p/M\rangle$, we can make the calculation quantitative and predict the actual rate of star formation, not just the proportionality. For example, estimates for the total momentum yield for supernovae give $\langle p/M\rangle \sim 3000$ km s$^{-1}$ by the end of the energy-conserving phase. Using this number, we obtain (this is equation 13 of \citealt{ostriker11a})
\begin{equation}
\Sigma_{\rm SFR} \sim 0.09 M_\odot\mbox{ pc}^{-2}\mbox{ Myr}^{-1} \left(\frac{\Sigma}{100\,M_\odot\,\mathrm{pc}^{-2}}\right)^2,
\end{equation}
which is in the right ballpark for the observed star formation rate at that gas surface density.

\begin{figure}
\includegraphics[width=\linewidth]{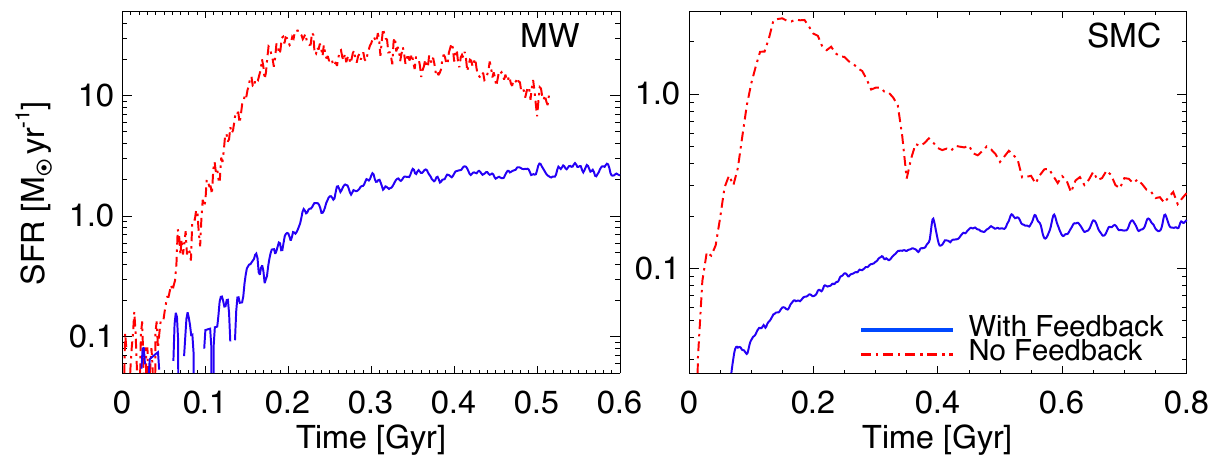}
\caption[Star formation rates in galaxy simulations with and without stellar feedback]{
\label{fig:sffeedback_hopkins11}
Star formation rates versus time measured in simulations of isolated galaxies performed with (blue) and without (red) a subgrid model for stellar feedback. One simulation shown is for a galaxy with properties chosen to be similar to the Milky Way (left), and one is for a galaxy chosen to resemble the Small Magellanic Cloud (right). Credit: \citeauthor{hopkins11a}, 2011, MNRAS, 417, 950, reproduced by
permission of Oxford University Press on behalf of the RAS.
}
\end{figure}

A number of simulations of models of this type have been conducted, and they seem to show that one can indeed produce star formation rates that are in rough agreement with observation, for plausible choices of $\langle p/M\rangle$ and/or plausible implementations of stellar feedback. Figure \ref{fig:sffeedback_hopkins11} shows an example.

\paragraph{Successes and Failures of Feedback-Regulated Models}

Models of this sort have a number of appealing features. They are physically-motivated and allow quantitative calculation of the star formation rate, both in simulations and analytically. Another virtue of these models is that they allow one to calculate the star formation rate independent of any knowledge of how star formation operates within individual molecular clouds. Only the mean momentum balance of the ISM matters. This is particularly nice from the standpoint of galactic and cosmological simulations, because these almost always lack the resolution to follow the formation of stars directly. Instead, they must rely in subgrid recipes for star formation. If the star formation rate is controlled only by feedback, however, then the choice of this recipe does not matter much to the results. A final virtue of this approach is that the linear scaling between $\Sigma_{\rm SFR}$ and $\Sigma_{\rm gas}$ expected in the regime where stars dominate the matter, and the scaling with stellar surface density, agree pretty well with what we see in outer galaxies.

However, there are also significant problems and omissions with this sort of model. First of all, the quantitative prediction is only as good as one's estimate of $\langle p/M\rangle$. As we saw in Chapter \ref{ch:feedback}, there are significant uncertainties in this quantity. Second, in models of this sort we expect to have $\Sigma_{\rm SFR} \propto \Sigma_{\rm gas}^2$ in the gas-dominated regime found in starburst galaxies. This is noticeably steeper than the observed scaling between $\Sigma_{\rm SFR}$ and $\Sigma_{\rm gas}$, which, as we have seen, has an index $\sim 1.5$. One can plausibly get this close to 2 if one adopts a non-constant value of $X_{\rm CO}$ that depends on star formation rate in the right way, but the validity of such a scaling has not been demonstrated. Third, while this model goes a reasonable job of explaining how things might work in outer disks and why stars matter there, its predictions about the impact of metallicity appear to be in strong tension with the observations. Nothing in the argument we just made has anything to do with metallicity, and it is not at all clear how one could possibly shoehorn metallicity into this model. Thus the natural prediction of the feedback-regulated model is that metallicity does not matter. In contrast, as we discussed, the available evidence suggests quite the opposite.

A related issue is that it is not clear how the chemical state of the gas (i.e., whether it is atomic or molecular) fits into this story. All that matters in this model is the weight of the ISM, which is unaffected by the chemical state of the gas, One could plausibly say that molecular gas simply forms wherever there is gas collapsing to stars, but then it is not clear why the depletion time in the molecular gas should be so much longer than the free-fall time -- if molecular gas is formed \textit{en passant} as atomic gas collapses to stars, why is it not depleted on a free-fall time scale?

\begin{marginfigure}
\includegraphics[width=\linewidth]{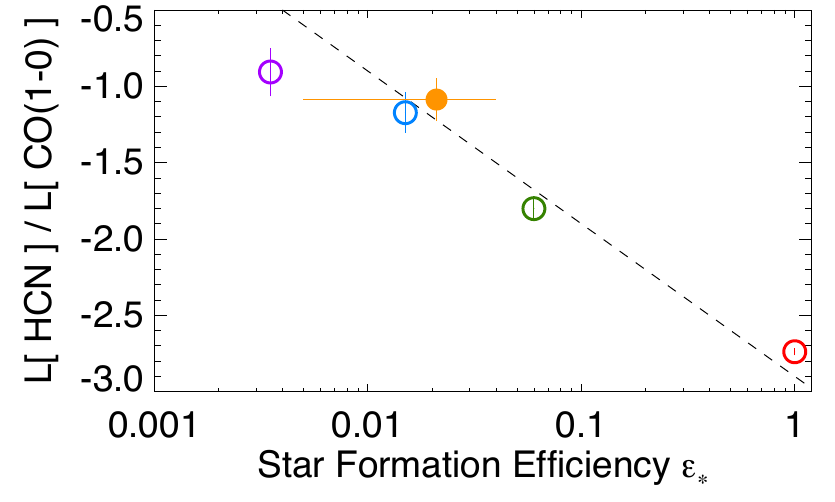}
\caption[Ratio of HCN to CO luminosity as a function of subgrid star formation recipe]{
\label{fig:hcn-co_hopkins13}
Ratio of HCN to CO luminosity computed from simulations of galaxies that are identical except for their subgrid model for the star formation rate in dense gas, parameterized by $\epsilon_*$. Credit: \citeauthor{hopkins13c}, 2013, MNRAS, 433, 69, reproduced by permission of Oxford University Press on behalf of the RAS.
}
\end{marginfigure}

A fourth and final issue is that the independence of the predicted star formation rate on the local star formation law, which we praised as a virtue above, is also a defect. Observations appear to require that star formation be about as slow and inefficient within individual molecular clouds and dense regions as it is within galaxies as a whole. There are two independent lines of evidence to this effect: the low star formation rates measured in Solar neighborhood clouds, and the correlation between infrared and HCN luminosity. However, in a feedback-regulated model there is no reason why this should be the case. Indeed, one can check this explicitly using simulations by changing the small scale star formation law used in the simulations (Figure \ref{fig:hcn-co_hopkins13}). If one changes the parameter describing how gas turns to stars within individual clouds, the star formation rate in the galaxy as a whole is unchanged, but the star formation rate within individual clouds, and the correlation between HCN emission and IR luminosity, changes dramatically.

One can sharpen the problem even more: in this story, the star formation rate is regulated primarily by feedback from massive stars. However, the star formation rate is observed to be low even in Solar neighborhood clouds where there are no stars larger than a few $M_\odot$, and where there probably will never be any because the stellar population is too small and low mass to be likely to produce any more massive stars. Why then is the star formation rate low in these clouds?

\section{The Bottom-Up Approach}

The alternate approach to the problem of the star formation rate has been to focus first on what happens inside individual clouds, and then to try to build up the galactic star formation law as simply the result of adding up many independent, small star-forming regions. The argument proceeds in two steps: first one attempts to determine which parts of the galaxy's ISM are "eligible" to form stars, which under Milky Way-like conditions more or less reduces to the question how the ISM will be partitioned between a star-forming molecular phase and an inert atomic phase. The second step is to ask about the star formation rate within individual molecular clouds.

\subsection{Which Gas is Star-Forming?}

Observationally, stars form primarily or exclusively in molecular gas, and so it is natural to identify the star-forming part of the ISM with the molecular part. However, we would like to have a physical explanation for this correlation. The first explanation one might think of is that the formation of H$_2$ and CO lead to rapid cooling of the gas, allowing it to collapse. However, while CO is a very good coolant, it turns out that it is not much better than C$^+$, the main coolant in the cool atomic ISM. Moreover, in galaxies where there is a significant amount of H$_2$ that is not traced by CO, such as the Small Magellanic Cloud, star formation appears to correlate with the presence of H$_2$, not the presence of CO. This also suggests that CO cooling is not important.

Instead, the explanation that appears to have become accepted over the past few years is that H$_2$ is associated with star formation because of the importance of shielding. Let us recall from Section \ref{ssec:heatproc} the processes that heat the dense ISM. In a region without significant heating due to photoionization, the main heating processes are the grain photoelectric effect and cosmic ray heating. We can write the summed heating rate per H nucleus from both of these as
\begin{equation}
\Gamma = \left(4\times 10^{-26} \chi_{\rm FUV} Z'_d e^{-\tau_d} + 2\times 10^{-27} \zeta'\right) \mbox{ erg s}^{-1},
\end{equation}
where $\chi_{\rm FUV}$, $Z'_d$, and $\zeta'$ are the local FUV radiation field, dust metallicity, and cosmic ray ionization rate, all normalized to the Solar neighborhood value, and $\tau_d$ is the dust optical depth.

If we are in a region where the carbon has not yet formed CO, the main coolant will emission in the C$^+$ fine structure line at 91 K. This is fairly easy to compute. Assuming the gas is optically thin and well below the critical density (both reasonable assumptions), then the cooling rate is simply equal to the collisional excitation rate multiplied by the energy of the level, since every collisional excitation will lead to a radiative de-excitation that will remove energy. Thus we have a cooling rate per H nucleus
\begin{equation}
\Lambda_{\rm CII} = k_{\rm CII-H} \delta_C E_{\rm CII} n_{\rm H},
\end{equation}
where $k_{\rm CII-H} \approx 8\times 10^{-10} e^{-T_{\rm CII}/T}$ cm$^3$ s$^{-1}$ is the excitation rate coefficient, $T_{\rm CII} = 91$ K is the energy of the excited state measured in K, $\delta_C\approx 1.1\times 10^{-4} Z'_d$ is the carbon abundance relative to hydrogen, $E_{\rm CII} = k_B T_{\rm CII}$ is the energy of the level, and $n_{\rm H}$ is the hydrogen number density.

We can obtain the equilibrium temperature by setting the heating and cooling rates equal and solving. The result is
\begin{equation}
T = -\frac{T_{\rm CII}}{\ln \left( 0.36 \chi_{\rm FUV} e^{-\tau_d} + 0.018 \zeta'/Z'_d\right) - \ln n_{\rm H,2}},
\end{equation}
where $n_{\rm H,2} = n_{\rm H}/100$ cm$^{-3}$. Clearly there will be two possible behaviors of this solution, depending on whether the term $0.36 \chi_{\rm FUV} e^{-\tau_d}$ is larger or smaller than the term $0.018\zeta'/Z_d'$. If the first, FUV heating term, dominates, then we have
\begin{equation}
T \approx \frac{91\mbox{ K}}{1.0 + \tau_d - \ln \chi_{\rm FUV} + \ln n_{\rm H,2}}
\end{equation}
while if the second, cosmic ray term dominates, we have
\begin{equation}
T \approx \frac{91\mbox{ K}}{4.0 - \ln \zeta'/Z'_d + \ln n_{\rm H,2}}.
\end{equation}
The transition between the two regimes occurs when $\tau_d \sim 3$.

In the cosmic ray-dominated regime, for $\zeta'/Z'_d = 1$ we get $T = 23$ K. Thus the gas can cool down to almost as low a temperature as we would get in a CO-dominated region (which will be closer to 10 K). On the other hand, if the cosmic ray heating rate is negligible compared to the FUV heating rate, and the optical depth is small, will have a temperature that is an order of magnitude higher than what we normally expect in molecular clouds. The corresponding Jeans mass,
\begin{equation}
M_J = \rho \lambda_J^3 = \rho \left(\frac{\pi c_s^2}{G \rho}\right)^{3/2} = 4.8\times 10^3\,M_\odot n_{\rm H,2}^{-1/2} T_2^{3/2}
\end{equation}
where $T_2 = T/100$ K, will differ between the two cases by a factor of $\sim (91/23)^{1.5} \approx 8$. Thus the presence of a high optical depth that suppresses FUV heating lowers the mass that can be supported against collapse by roughly an order of magnitude (or possibly more, if the local FUV radiation field is more intense than in the Solar neighborhood, as we would expect closer to a galactic center).

\begin{marginfigure}
\includegraphics[width=\linewidth]{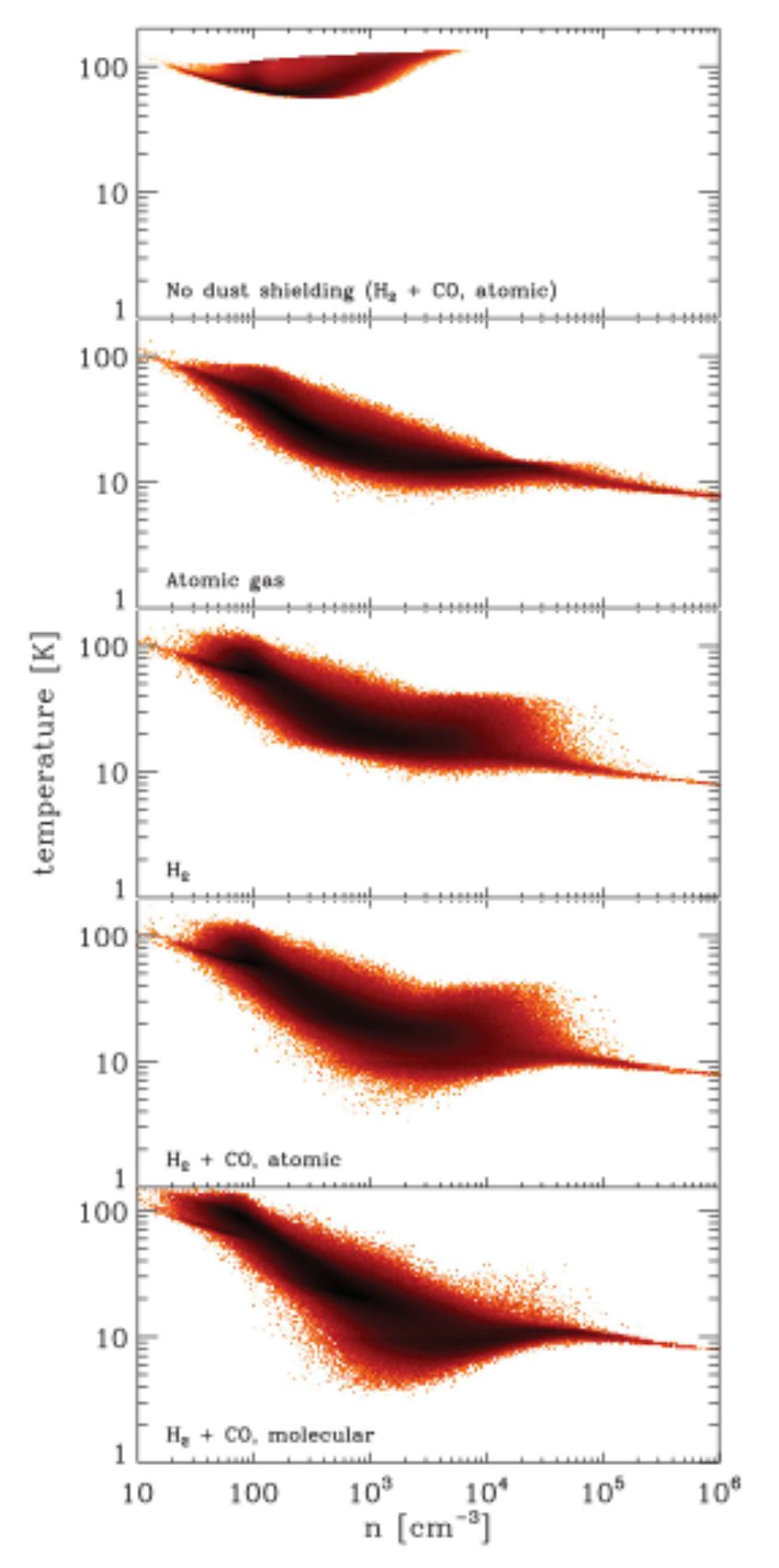}
\caption[Density-temperature distribution for different cooling models]{
\label{fig:shielding_gc12}
Density-temperature distributions measured in simulations with different treatments of ISM thermodynamics and chemistry. All simulations use identical initial conditions, but vary in how the gas heating and cooling rates are calculated. The top panel ignores dust shielding, but includes full chemistry and heating and cooling. The bottom panel includes all chemistry and cooling. The middle three panels turn off, respectively, H$_2$ formation, CO formation, and CO cooling. The tail of material proceeding to high density in some simulations is indicative of star formation. Credit: \citeauthor{glover12a}, 2012, MNRAS, 421, 9, reproduced by permission of Oxford University Press on behalf of the RAS.
}
\end{marginfigure}

The central \textit{ansatz} in bottom-up models is that this dramatic change in Jeans mass has important implications for the regulation of star formation: in regions where the temperature is warm, the gas will be too thermally supported to collapse to form stars, while in regions where it gets cold star formation will proceed efficiently. There is some evidence for this from simulations (Figure \ref{fig:shielding_gc12}).

So what does all of this have to do with H$_2$? To answer that, recall that the transition to H$_2$ also depends critically upon shielding. We saw in Section \ref{ssec:Hchemistry} that the shielding column of atomic hydrogen that has to be present before a transition to H$_2$ occurs is
\begin{equation}
N_{\rm H} = \frac{c f_{\rm diss} E_0^*}{n \mathcal{R}} \approx 7.5\times 10^{20} \chi_{\rm FUV} n_{\rm H,2}^{-1} (Z'_d)^{-1}\mbox{ cm}^{-2},
\end{equation}
or, in terms of mass surface density,
\begin{equation}
\Sigma = N_{\rm H} \mu m_{\rm H} = 8.4 \chi_{\rm FUV} n_{\rm H,2}^{-1} (Z'_d)^{-1}\,M_\odot\mbox{ pc}^{-2}.
\end{equation}

It is even more illuminating to write this in terms of the dust optical depth $\tau_d$. For FUV photons, the dust cross section per H nucleus is $\sigma_d \approx 10^{-21} Z'_d$ cm$^{-2}$, and so the dust optical depth one expects for the typical H~\textsc{i} shielding column is
\begin{equation}
\tau_d = N_{\rm H} \sigma_d = 7.5 \chi_{\rm FUV} n_{\rm H,2}^{-1}
\end{equation}
Thus the optical depth at which the gas becomes molecular is more or less the same optical depth at which the gas transitions from the FUV heating-dominated regime to the cosmic ray-dominated one. Moreover, Krumholz et al.~(2009) pointed out that the quantity $\chi_{\rm FUV} n_{\rm H,2}^{-1}$ appearing in these equations is not actually a free parameter -- in the main disks of galaxies where the atomic ISM forms a two-phase equilibrium, the cold phase will change its characteristic density in response to the local FUV radiation field, so that $\chi_{\rm FUV} n_{\rm H,2}^{-1}$ will always have about the same value (which turns out to be a few tenths).

Thus those models provide a natural, physical explanation for why star formation should be correlated with molecular gas, and why there is a turn-down in the relationship between $\Sigma_{\rm gas}$ and $\Sigma_{\rm SFR}$ at $\sim 10$ $M_\odot$ pc$^{-2}$. Gas that is cold enough to form stars is also generally shielded enough to be molecular, and vice versa. Gas that is not shielding enough to be molecular will also be too warm to form stars. The physical reason behind this is simple: the photons that dissociate H$_2$ are the same ones that are responsible for photoelectric heating, so shielding against one implies shielding against the other as well. Detailed models reproduce this qualitative conclusion (e.g., \citealt{krumholz11b}).

\begin{figure}
\includegraphics[width=\linewidth]{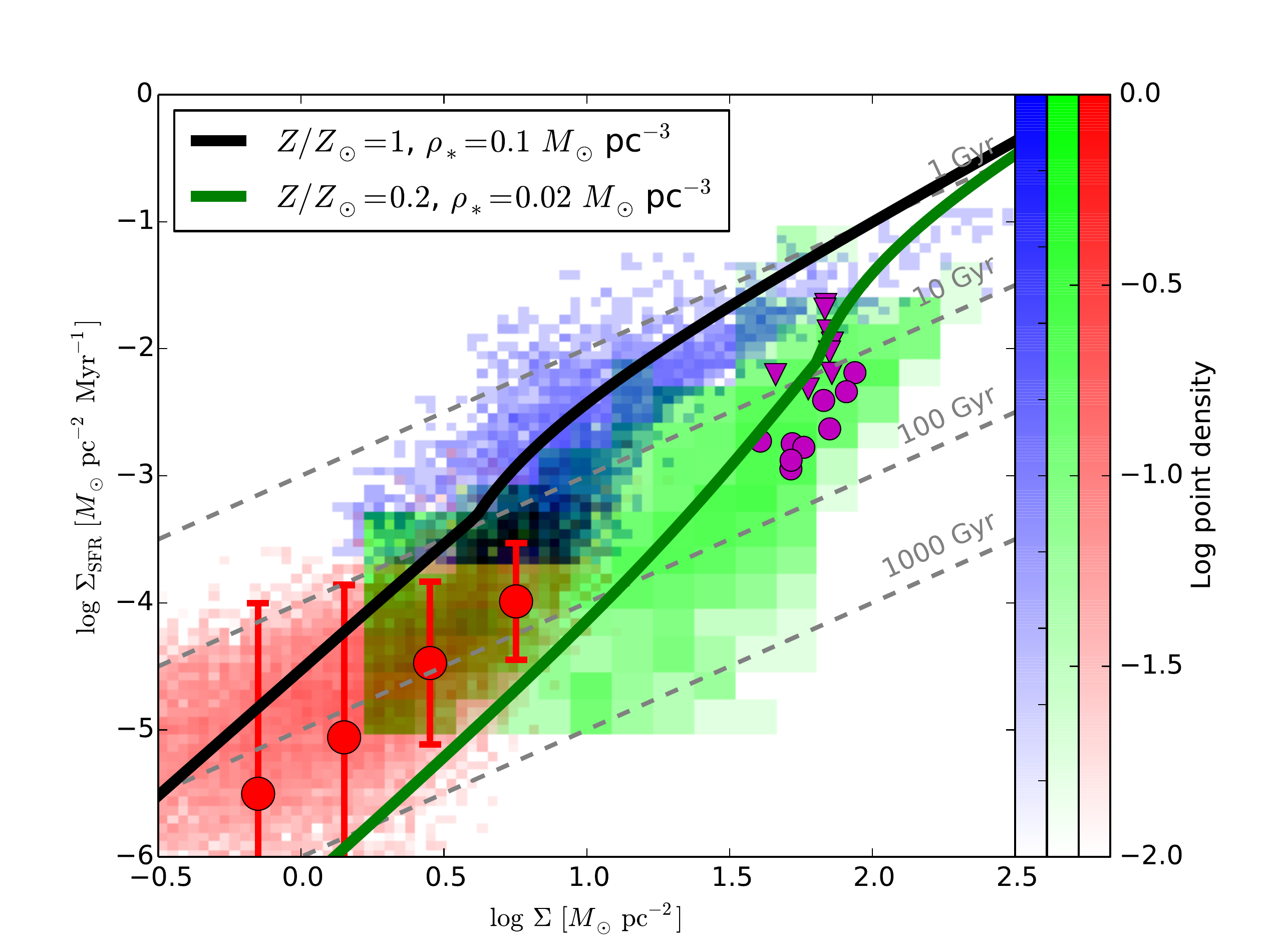}
\caption[Theoretical model for metallicity-dependence of the star formation rate]{
\label{fig:kshi_krumholz13}
Relationship between star formation rate surface density $\Sigma_{\mathrm{SFR}}$ and total gas surface density $\Sigma$, from \citet{krumholz13c}. Pixels and points show observations, and are the same as in Figure \ref{fig:kstot_krumholz14}. Solid black and green lines are theoretical models for two different combinations of metallicity normalized to Solar, $Z/Z_\odot$, and mid-plane stellar density, $\rho_*$, as indicated in the legend.
}
\end{figure}

This model also naturally explains the observed metallicity-dependence of both the H~\textsc{i} / H$_2$ transition and the star formation. With a bit more work, it can also explain the linear dependence of $\Sigma_{\rm SFR}$ and $\Sigma_{\rm gas}$ in the H~\textsc{i}-dominated regime -- in essence, once one gets to the regime of very low star formation and weak FUV fields, the quantity $\chi_{\rm FUV} n_{\rm H,2}^{-1}$ cannot stay constant any more, because $n_{\rm H,2}$ cannot fall below the minimum required to maintain hydrostatic balance. This puts a floor on the fraction of the ISM that is dense and shielded enough to form stars, which is linearly proportional to $\Sigma_{\rm gas}$. Figure \ref{fig:kshi_krumholz13} shows the result.

\subsection{The Star Formation Rate in Star-Forming Clouds}
\label{ssec:eff_th}

Thus far the model we have outlined explains the metallicity-dependence and the overall shape of the relationship between total gas and star formation, but it does not say anything about the overall rate of star formation in molecular regions. Why is the star formation rate in molecular gas so low?

One potential explanation focuses on the role of turbulent support. This model was first developed quantitatively by \citet{krumholz05c}, and has subsequently been refined and improved by a large number of authors (e.g., \citealt{hennebelle11b, padoan11a, padoan14a, federrath12a}). The argument is fairly simple, and it relies on the statistical properties of the turbulence discussed in Chapter \ref{ch:turbulence}. Consider a turbulent medium with a linewidth-size relation
\begin{equation}
\sigma(l)=c_s\left(\frac{l}{\lambda_s}\right)^{1/2},
\end{equation}
where $c_s$ is the sound speed and $\lambda_s$ is the sonic length. We want to know what parts of this flow will go Jeans-unstable and begin to collapse. The maximum mass that can held up against turbulence is the Bonnor-Ebert mass, the computation of which is included in Problem Set 2:
\begin{equation}
M_{\rm BE} = 1.18 \frac{c_s^3}{\sqrt{G^3\rho}} = \frac{1.18}{\pi^{3/2}}\rho \lambda_J^3,
\end{equation}
where $c_s$ is the isothermal sound speed and $\rho$ is the {\it local} gas density, i.e., the density at the surface of the Bonnor-Ebert sphere. The corresponding radius is
\begin{equation}
R_{\rm BE} = 0.37\lambda_J
\end{equation}
Let us evaluate the various terms in the virial theorem for this object. The gravitational energy is
\begin{equation}
\mathcal{W} = -a \frac{G M_{\rm BE}^2}{R_{\rm BE}} = -1.06 \frac{c_s^5}{G^{3/2} \rho^{1/2}},
\end{equation}
where $a$ is a geometric factor that depends on the density distribution, and for the numerical evaluation we used $a=0.73$, the numerical value for a maximum-mass Bonnor-Ebert sphere. The corresponding thermal energy is
\begin{equation}
\mathcal{T}_{\rm th} = \frac{3}{2}M_{\rm BE} c_s^2 = 1.14 |\mathcal{W}|.
\end{equation}
Finally, to estimate the turbulent energy we will use the linewidth-size relation, and assume that the velocity dispersion is given by $\sigma(2 R_{\rm BE})$, i.e., by the linewidth-size relation evaluated at a length scale equal to diameter of the sphere. This gives
\begin{equation}
\mathcal{T}_{\rm turb} = \frac{3}{2} M_{\rm BE} \sigma(2R_{\rm BE})^2 = 0.89 \left(\frac{\lambda_J}{\lambda_s}\right) |\mathcal{W}|.
\end{equation}
More sophisticated treatments, as given in some of the papers cited above, include magnetic support as well. For simplicity, though, we will omit that here.

Now let us turn this around and hypothesize that the collapsing parts of the flow are those for which the density is unusually high, such that potential energy is comparable to or larger than the turbulent energy. Based on what we just calculated, for this condition to be true it must be the case that the local Jeans length $\lambda_J$ is comparable to or smaller than the sonic length $\lambda_s$. As an ansatz, we therefore say that collapse will occur in any region where $\lambda_J \lesssim \lambda_s$. It is convenient to write this in terms of the Jeans length at the mean density
\begin{equation}
\lambda_{J0}=\sqrt{\frac{\pi c_s^2}{G\overline{\rho}}},
\end{equation}
where $\overline{\rho}$ is the mean density. If we let $s = \ln \rho/\overline{\rho}$, then $\lambda_J = \lambda_{J0}/e^{s/2}$. The condition that $\lambda_J \lesssim \lambda_s$ therefore requires that the overdensity $s$ satisfy
\begin{equation}
s > s_{\rm crit} \equiv 2 \ln \left(\phi_s \frac{\lambda_{J0}}{\lambda_s}\right),
\end{equation}
where we have replaced the $\lesssim$ simply with a firm inequality, and introduced $\phi_s$, a dimensionless number of order unity.

The nice thing is that we can now determine what fraction of the mass satisfies this condition simply from knowing the density PDF. Specifically,
\begin{eqnarray}
f & = & \int_{s_{\rm crit}}^\infty p_M(s) \,ds \\
& = & \frac{1}{\sqrt{2\pi \sigma_s^2}}
\int_{s_{\rm crit}}^\infty \exp\left[-\frac{(s -\sigma_s^2/2)^2}{2\sigma_s^2}\right] \, ds \\
& = & \frac{1}{2} \left[1+\mbox{erf}\left(\frac{-2 s_{\rm crit}+\sigma_s^2}{2^{3/2}\sigma_s}\right)\right],
\end{eqnarray}
where $\sigma_s \approx [\ln(1+3\mathcal{M}^2/4)]^{1/2}$ and $\mathcal{M}$ is the 1D Mach number. If we then hypothesize that a fraction $\sim f$ of the cloud will collapse every cloud free-fall time, the total star formation rate per free-fall time in the simulations should follows
\begin{equation}
\epsilon_{\rm ff} = \frac{1}{2 \phi_t} \left[1+\mbox{erf}\left(\frac{-2 s_{\rm crit}+\sigma_{\rho}^2}{2^{3/2}\sigma_s}\right)\right],
\end{equation}
where $\phi_t$ is another fudge factor of order unity.

Another assumption here is that the collapse time is given by the global free-fall time, as opposed to a density-dependent local free-fall time. Again, this is an area where subsequent work by \citet{hennebelle11b} and \citet{federrath12a} have improved on the original model. It turns out to be a better assumption that the collapse happens on a local free-fall timescale instead, in which case we instead have a star formation rate
\begin{equation}
\epsilon_{\rm ff} = \frac{1}{\phi_t} \int_{s_{\rm crit}}^\infty p(s) e^{s/2} \, ds
= \frac{1}{2\phi_t} \left[1+\mbox{erf}\left(\frac{-s_{\rm crit}+\sigma_s^2}{2^{1/2}\sigma_s}\right)\right] \exp\left(\frac{3}{8}\sigma_s^2\right),
\end{equation}
where the extra factor of $e^{s/2}$ inside the integral comes from the fact that $t_{\rm ff} \propto \rho^{-1/2}$, so higher density regions get weighted more because they collapse faster.

We can write the critical ratio $\lambda_{J0}/\lambda_s$ in terms of quantities that we can determine by observations. If we have a region for which the virial ratio is
\begin{equation}
\avir = \frac{5\sigma^2 R}{GM},
\end{equation}
with $\sigma$ here representing the velocity dispersion over the entire region, then the linewidth-size relation is
\begin{equation}
\sigma(l) = \sigma_{\rm 2R} \left(\frac{l}{2R}\right)^{1/2}.
\end{equation}
We therefore have
\begin{equation}
\lambda_s = 2R \left(\frac{c_s}{\sigma_{2R}}\right)^2.
\end{equation}
Similar, we can re-write the mean-density Jeans length as
\begin{equation}
\lambda_{J0} = \sqrt{\frac{\pi c_s^2}{G\overline{\rho}}} = 2\pi c_s \sqrt{\frac{R^3}{3 GM}}.
\end{equation}
Putting this together, we get
\begin{equation}
s_{\rm crit} = \left(\phi_s \frac{\lambda_{J0}}{\lambda_s}\right)^2 = \frac{\pi^2 \phi_s^2}{15} \avir \mathcal{M}^2 \approx \avir \mathcal{M}^2,
\end{equation}
Since $\epsilon_{\rm ff}$ is a function only of $s_{\rm crit}$ and $\sigma_s$, and these are now both known in terms of $\avir$ and $\mathcal{M}$, we have now written the star formation rate in terms of $\avir$ and $\mathcal{M}$. Numerical evaluation is straightforward, and full 3D simulations show that the theory works reasonably well \citep{federrath12a}. 

For $\alpha_{\rm vir} \sim 2$, comparable to observed values, and using the value of $\phi_t$ and $\phi_s$ that agree best with simulations, the numerical value of $\epsilon_{\rm ff}$ typically comes out a bit too high compared to observations, closer to $\sim 0.1$ than $\sim 0.01$. However, localized sources of feedback like protostellar outflows are likely able to reduce that further. Such feedback would be required in any event, since without it the turbulence would decay and the star formation rate would rise.

\subsection{Strengths and Weaknesses of Bottom-Up Models}

Comparing the bottom-up models to our observational constraints, we see that they do better than the top-down ones on some metrics, not quite as well on others. As already mentioned, the bottom up models naturally reproduce the observed dependence of star formation on the phase of the ISM, and on the metallicity. This is a major difference from the top-down models, which struggle on these points.

A second strength of the bottom-up model, at least one in which turbulence is assumed to regulate the star formation rate within molecular clouds, is that it automatically reproduces the observation that $\epsilon_{\rm ff} \sim 0.01$ on all scales, from individual clouds to small dense regions to entire galaxies; indeed, the central assumption of the model is that the galactic star formation law is simply a sum of local cloud ones. Thus the local-global connection is made naturally.

However, the model has two major weaknesses as well. First, the explanation why $\epsilon_{\rm ff} \sim 0.01$, as opposed to $\sim 0.1$, is still somewhat hazy, and relies on generalized appeals to local feedback processes that are not tremendously well understood. The central problem is that of dynamic range. If local feedback processes like H~\textsc{ii} regions or protostellar outflows are what drives the low rate of star formation within clouds, not larger-scale things like supernovae, then the problem is much harder to solve numerically due to the far larger dynamic range involved. No one has ever successfully simulated an entire galaxy, following the self-consistent formation and evolution of molecular clouds, with enough resolution to capture the turbulence and all the local feedback processes that drive it within individual molecular clouds.

A second weakness is that the appeal to the thermodynamics of the gas as an explanation for the origin of the low star formation rate does not address the question of global regulation of the ISM and its hydrostatic balance. To put it another way, in principle one could have a region of the ISM where the gas surface density is low enough that almost all the gas is atomic and there is very little star formation. In that case, however, what maintains vertical hydrostatic balance? If thermal pressure alone is not enough to do so, where does the required turbulent pressure come from in the absence of star formation? This is an unsolved problem in the local model, one that is avoided in the global model simply by adopting hydrostatic balance as a starting assumption.

\chapter{Stellar Clustering}
\label{ch:clustering}

\marginnote{
\textbf{Suggested background reading:}
\begin{itemize}
\item \href{http://adsabs.harvard.edu/abs/2014arXiv1402.0867K}{Krumholz, M.~R. 2014, Phys.~Rep., 539, 49}, section 5 \nocite{krumholz14c}
\end{itemize}
\textbf{Suggested literature:}
\begin{itemize}
\item \href{http://adsabs.harvard.edu/abs/2012MNRAS.426.3008K}{Kruijssen, J.~D.~M. 2012, MNRAS, 426, 3008} \nocite{kruijssen12a}
\end{itemize}
}

The previous two chapters focused on star formation at the scale of galaxies, with attention to what determines the overall rate at which stars form. In this chapter we will now zoom in a bit, and ask how star formation is arranged in space and time within a single molecular cloud, and how these arrangements evolve over time as star formation proceeds and eventually ceases. The central goal of this analysis will be to understand a striking observational feature of star formation: sometimes, but not often, it produces gravitationally-bound clusters of stars.

\section{Observations of Clustering}

We will start our discussion with a review of the observational situation, focusing first on young stars and gas and then moving on to older populations of stars that have become gas-free.

\subsection{Spatial and Kinematic Distributions of Gas and Young Stars}

\begin{figure}
\includegraphics[width=\linewidth]{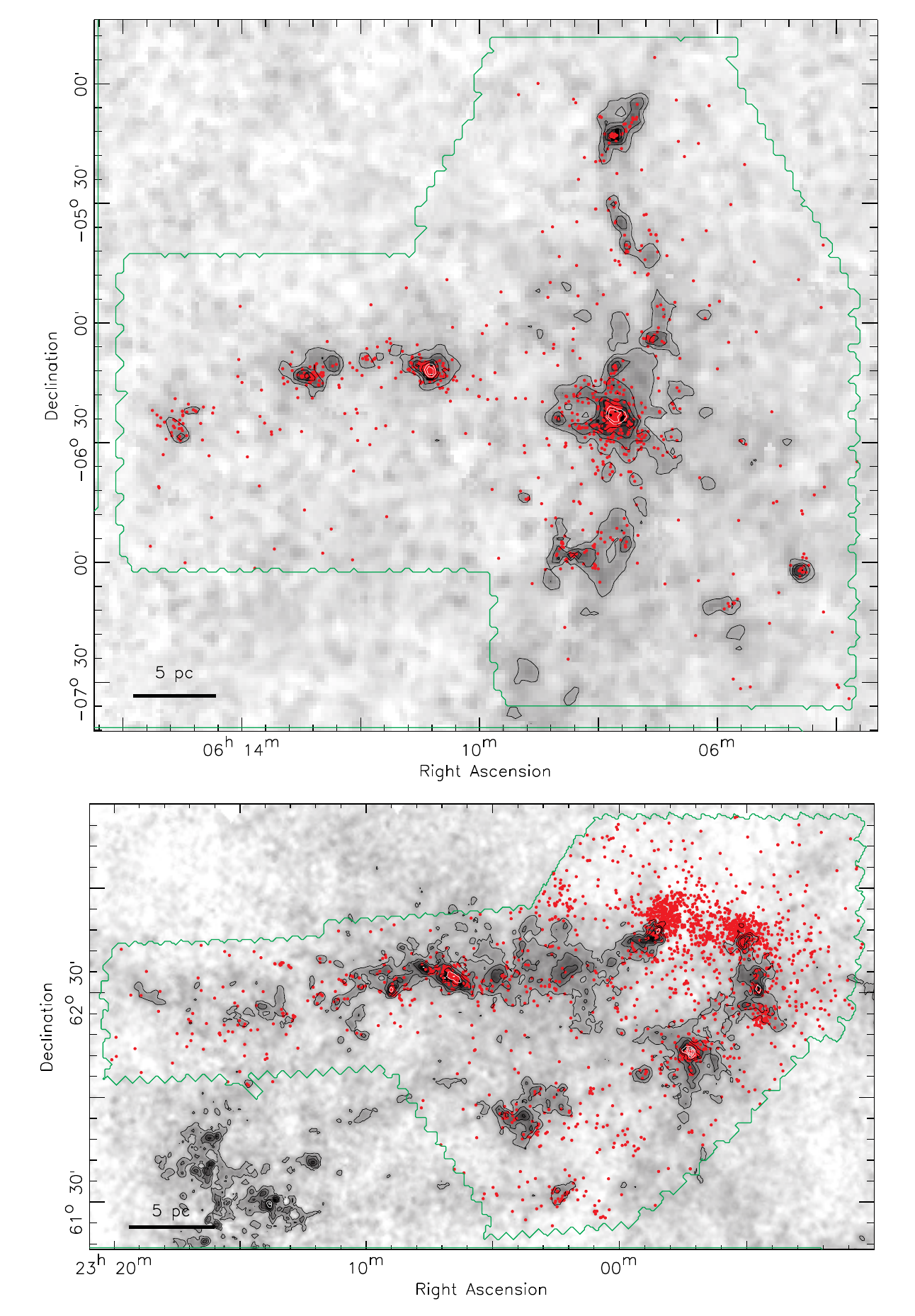}
\caption[Maps of gas and young stars in two clouds]{
\label{fig:maps_gutermuth11}
Maps showing the distribution of gas (grayscale) and young stellar objects (YSOs, red points) in the MonR2 (top) and CepOB3 (bottom) clouds. The grayscale gas maps are measured from dust extinction, which is plotted using a linear stretch from extinction $A_V = -1$ to 10 mag; contours start at $A_V=3$ mag, and are at 2 mag intervals. Credit: \citet{gutermuth11a}, \copyright AAS. Reproduced with permission.
}
\end{figure}

Newborn stars are, like the gas in molecular clouds, distributed in a highly structured and inhomogeneous fashion. The gas is arranged in filaments, and young stars are largely arranged along those filaments, at least in the youngest regions. In somewhat older regions we start to see clusters of stars where the is no gas and the filamentary structure has dissolved, but with gas morphologies highly suggestive that it is being blown away by the young stars. Figure \ref{fig:maps_gutermuth11} shows examples of both a younger, filament-dominated region and an older one where substantial gas clearing has taken place.

Such an inhomogeneous structure calls for a statistical description, and a number of statistical techniques have been used to describe gas and star arrangements. For gas we have already encountered some of these, in the form of power spectra. Power spectra can be computed for velocity structure, but they can also be computed for density structure. They can be computed for both 2D projected images of density as well as true 3D data.

Stars, on the other hand, are point objects, and so one cannot compute a power spectrum for them as one would for a continuous field like the density. However, one can compute a closely-related quantity, the two-point correlation function. Recall that, for a continuous vector field (say the velocity), we defined the autocorrelation function as
\begin{equation}
A_{\vecv}(\vecr) = \frac{1}{V} \int \vecv(\vecx) \cdot \vecv(\vecx + \vecr)\, d\vecx.
\end{equation}
For a scalar field, say density $\rho$, we can just replace the dot product with a simple multiplication. It is also common to slightly modify the definition by subtracting off the mean density so that we get a quantity that depends only on the shape of the density distribution, not its mean value. This quantity is
\begin{equation}
\label{eq:autocorr}
\xi(\vecr) = \frac{1}{V} \int [\rho(\vecx)-\overline{\rho}] [\rho(\vecx + \vecr)-\overline{\rho}]\, d\vecx,
\end{equation}
where $\overline{\rho} = (1/V) \int \rho(\vecx)\, d\vecx$. If the function is isotropic, then the autocorrelation function depends only on $r=|\vecr|$.

This is still defined for a continuous field, but we can extend the definition to the positions of a collection of point particles by imagining that the point particles represent samples drawn from an underlying probability density function. That is, we imagine that there is a continuous probability $(dP(\vecr)/dV)\, dV$ of finding a star in a volume of size $dV$ centered at some position $\vecr$, and that the actual stars present represent a random draw from this distribution.

In this case one can show that the autocorrelation function can be defined by the following procedure. Imagine drawing stellar positions from the PDF until the mean number density is $n$, and then imagine choosing a random star from this sample. Now consider a volume $dV$ that is displaced by a distance $\vecr$ from the chosen star. If $dP(\vecr)/dV$ were uniform, i.e., if there were no correlation, then the probability of finding another star at that point would simply be $n\, dV$. The two-point correlation function is then the \textit{excess} probability of finding a star over and above this value. That is, if the actual probability of finding a star is $dP_2(\vecr)/dV$, we define the two-point correlation function by
\begin{equation}
\frac{dP_2}{dV}(\vecr) = n \left[1 + \xi(\vecr)\right].
\end{equation}
Defined this way, the quantity $\xi(\vecr)$ is known as the two-point correlation function. It is possible to show that, with this definition, $\xi(\vecr)$ is equivalent to that given by equation (\ref{eq:autocorr}) applied to the underlying probability density function. As above, if the distribution is isotropic, then $\xi$ depends only on $r$, not $\vecr$. Also note that this is a 3D distribution, but if one only has 2D data on positions (the usual situation in practice), one can also define a 2D version of this where the volume is simply interpreted as representing annuli on the sky rather than shells in 3D space.

How does one go about estimating $\xi(r)$ in practice? There are a few ways. The most sophisticated is to take the measured positions and randomize them to create a random catalog, measure the numbers of object pairs in bins of separation, and use the difference between the random and true catalogs as an estimate of $\xi(r)$. This is the normal procedure in the galaxy community where surveys have well-defined areas and selection functions. In the star formation community, things are a bit more primitive, and the usual procedure is just to count the mean surface density of neighbors as a function of distance around a star, that is, to estimate that
\begin{equation}
\Sigma(r) = n \left[1 + \xi(r)\right]
\end{equation}
where $r$ is taken to be the projected separation. This is quite rough, and is vulnerable to considerable biases arising from things like edge effects (formally the correlation function is only defined over an infinite volume, but in reality of course surveys are finite in size), but it is what the star formation community generally uses.

\begin{figure}
\includegraphics[height=5.2cm]{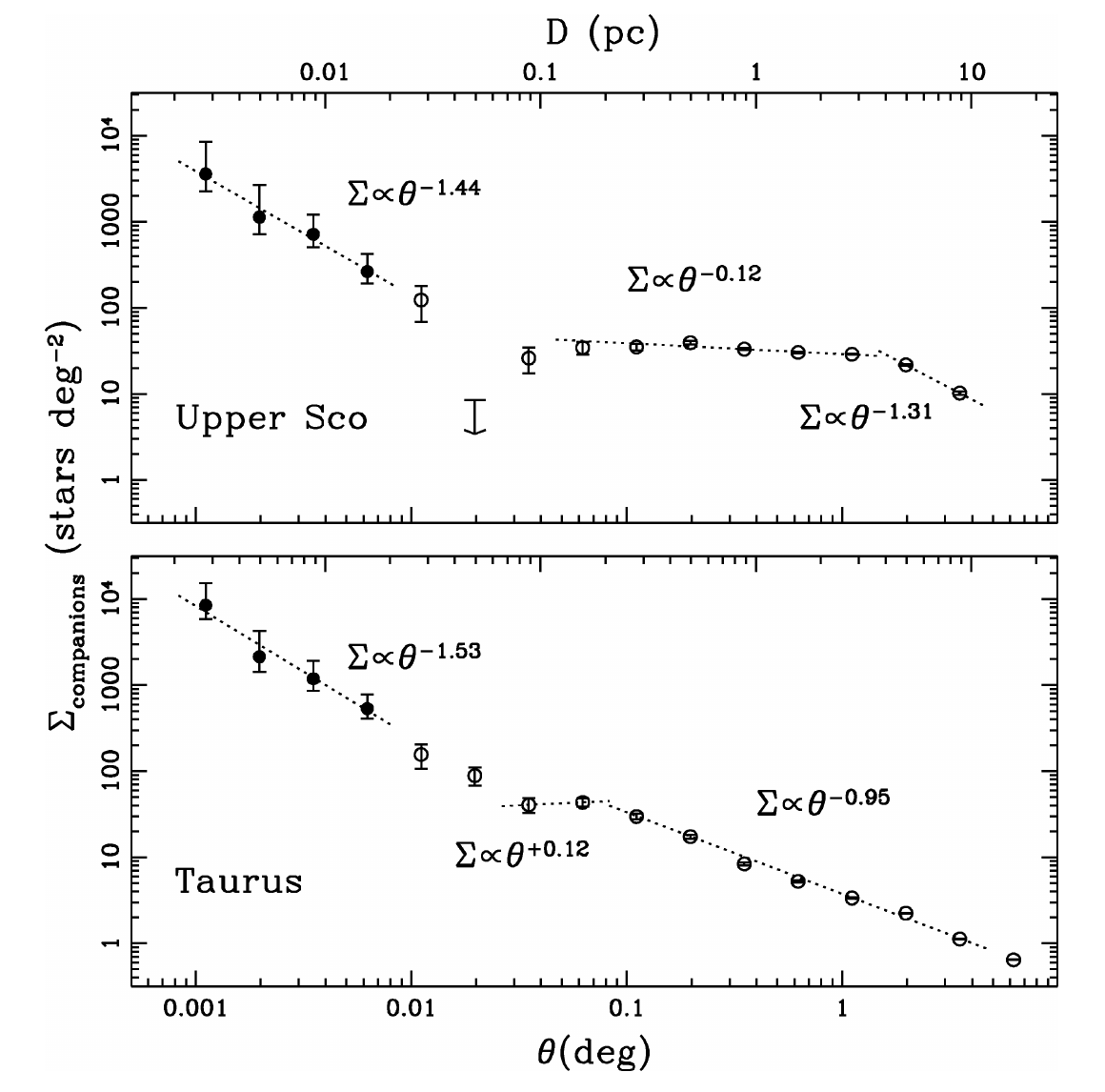}\includegraphics[height=5cm]{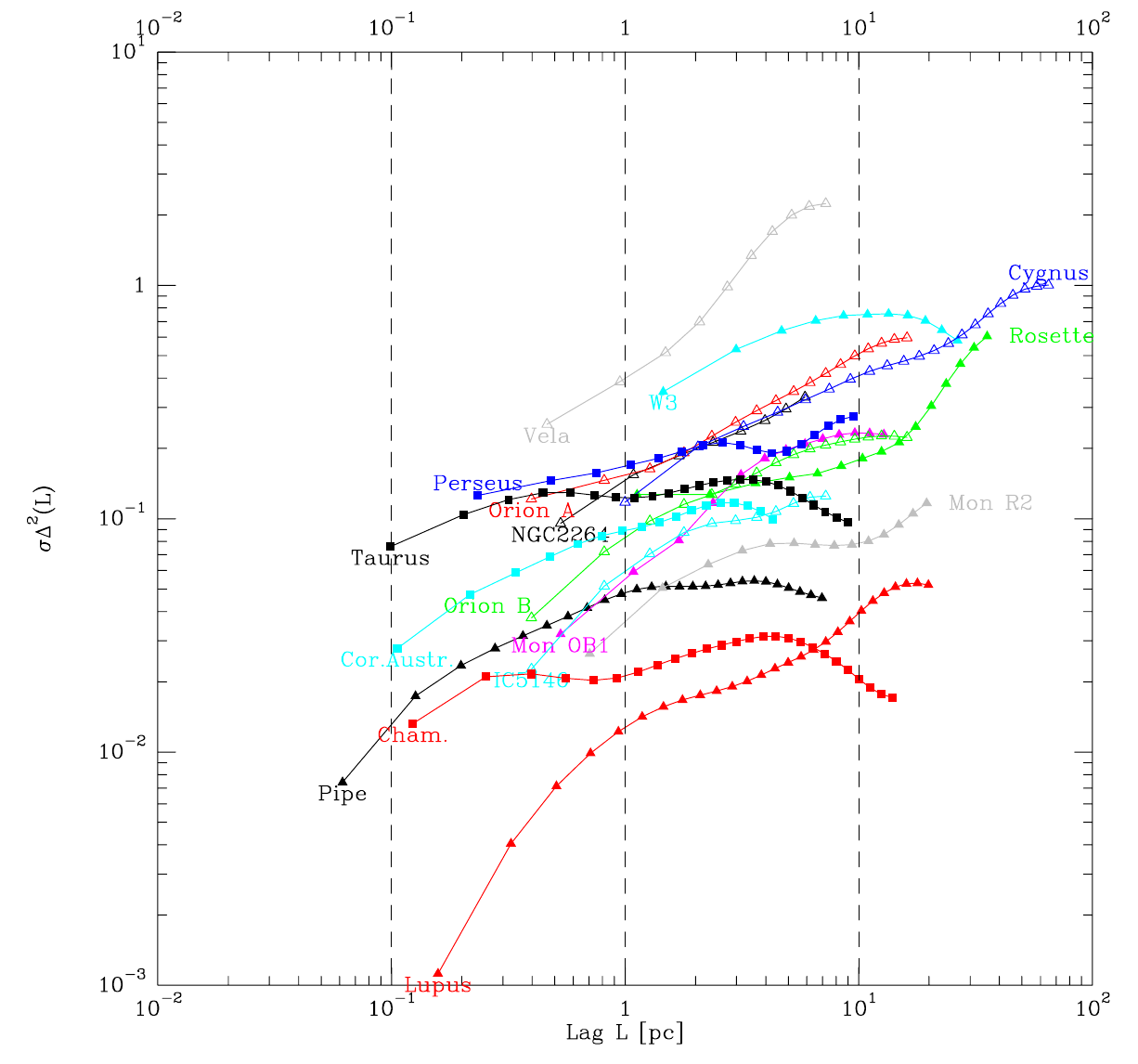}
\caption[Correlation functions for gas and stars]{
\label{fig:correlation_stargas}
Measurements of the stellar and gas correlations in nearby star-forming regions. The two figures on the left show the surface density of neighbors around each star $\Sigma(r)$ in Upper Sco and Taurus. The right panel shows measurements, for a large number of nearby molecular clouds, of a statistic called the $\Delta$ variance, $\sigma \Delta^2$, which is related to the correlation function. Credit: left panel: \citet{kraus08a}, \copyright\, AAS, reproduced with permission; right panel: \citeauthor{schneider11a}, A\&A, 529, A1, 2011, reproduced with permission, \copyright\, ESO.
}
\end{figure}

With that formal throat-clearing out of the way, we are now in a position to look at actual data, and, since we have these clean definitions, we can talk about gas and stars on essentially equal footing. So what do autocorrelation functions of gas and star look like? Figure \ref{fig:correlation_stargas} shows some example measurements.

In the stellar distributions, we can identify a few features. At small separations, we see one powerlaw distribution. This is naturally identified as representing wide binaries. This falls off fairly steeply, until it breaks to a shallower falloff at larger separations, which can be interpreted as describing the distribution of stars within the cluster. This is also a powerlaw, covering several orders of magnitude in separation. That fact that the distribution is well fit by a powerlaw indicates that the stars follow a self-similar, scale-free structure. One can interpret such a structure as a fractal, and the index of the powerlaw is related to the dimensionality of the fractal; typical values for the dimensionality are $\sim 1$, consistent with a highly filamentary structure.

\begin{marginfigure}
\includegraphics[width=\linewidth]{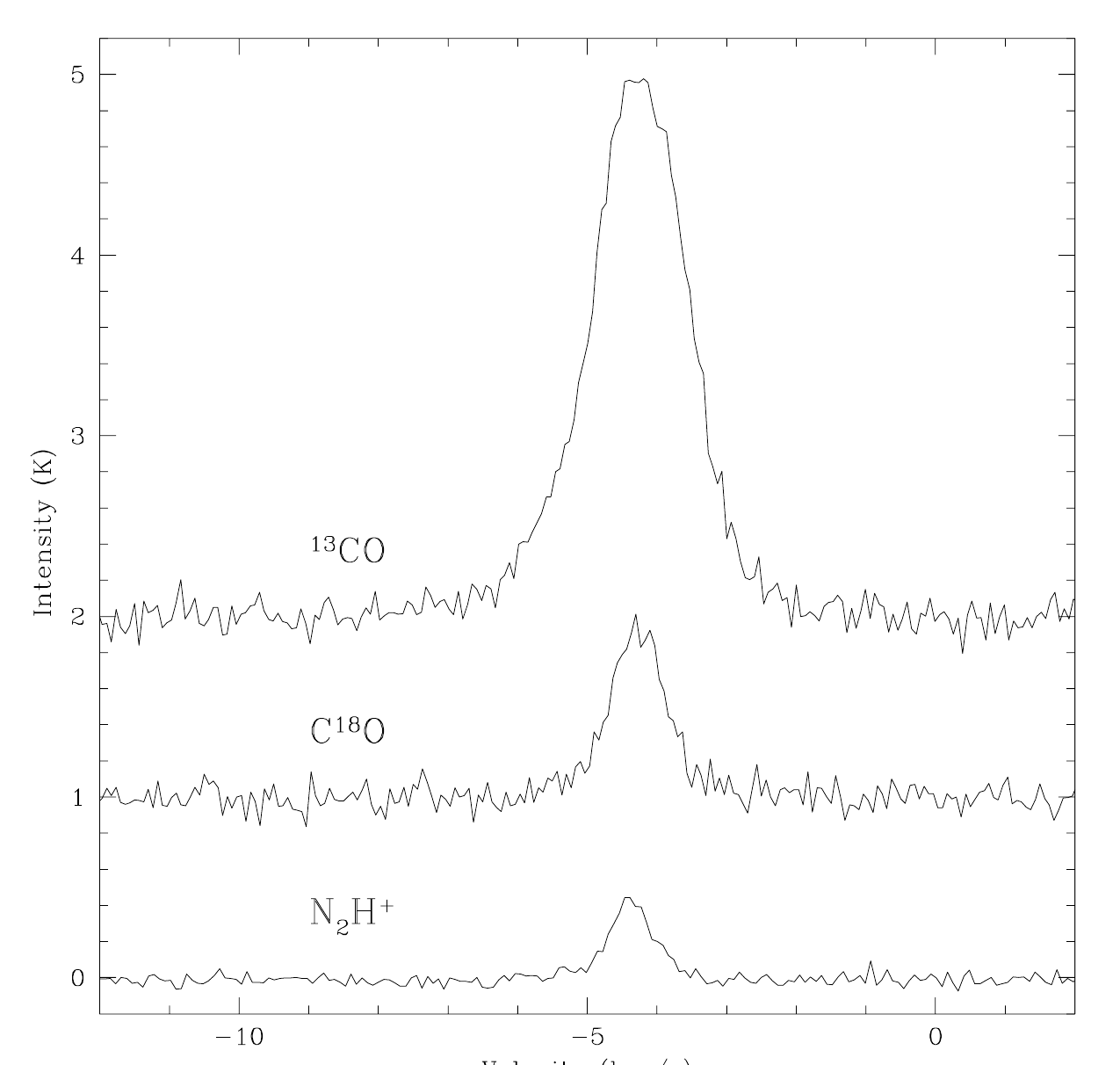}
\caption[Velocity distributions in varying molecular lines]{
\label{fig:denvel_walsh04}
Velocity distributions measured toward a nearby protostellar core using three different molecular line tracers, as indicated. The transitions $^{13}$CO, C$^{18}$O, and N$_2$H$^+$ should be roughly ordered from lowest to highest in terms of the density of gas that produces them. Creidt: \citet{walsh04a}, \copyright AAS. Reproduced with permission.
}
\end{marginfigure}

For the gas, one tends to obtain a similar powerlaw structure over a broad range of scales, with possible breaks at the high and low end. Thus the basic conclusion is that the stars and gas are in highly structured, fractal-like distributions. At young ages, the gas and stellar distributions are highly correlated with one another, which is not surprising. For older stellar populations, the correlation begins to break down.

\begin{marginfigure}
\includegraphics[width=\linewidth]{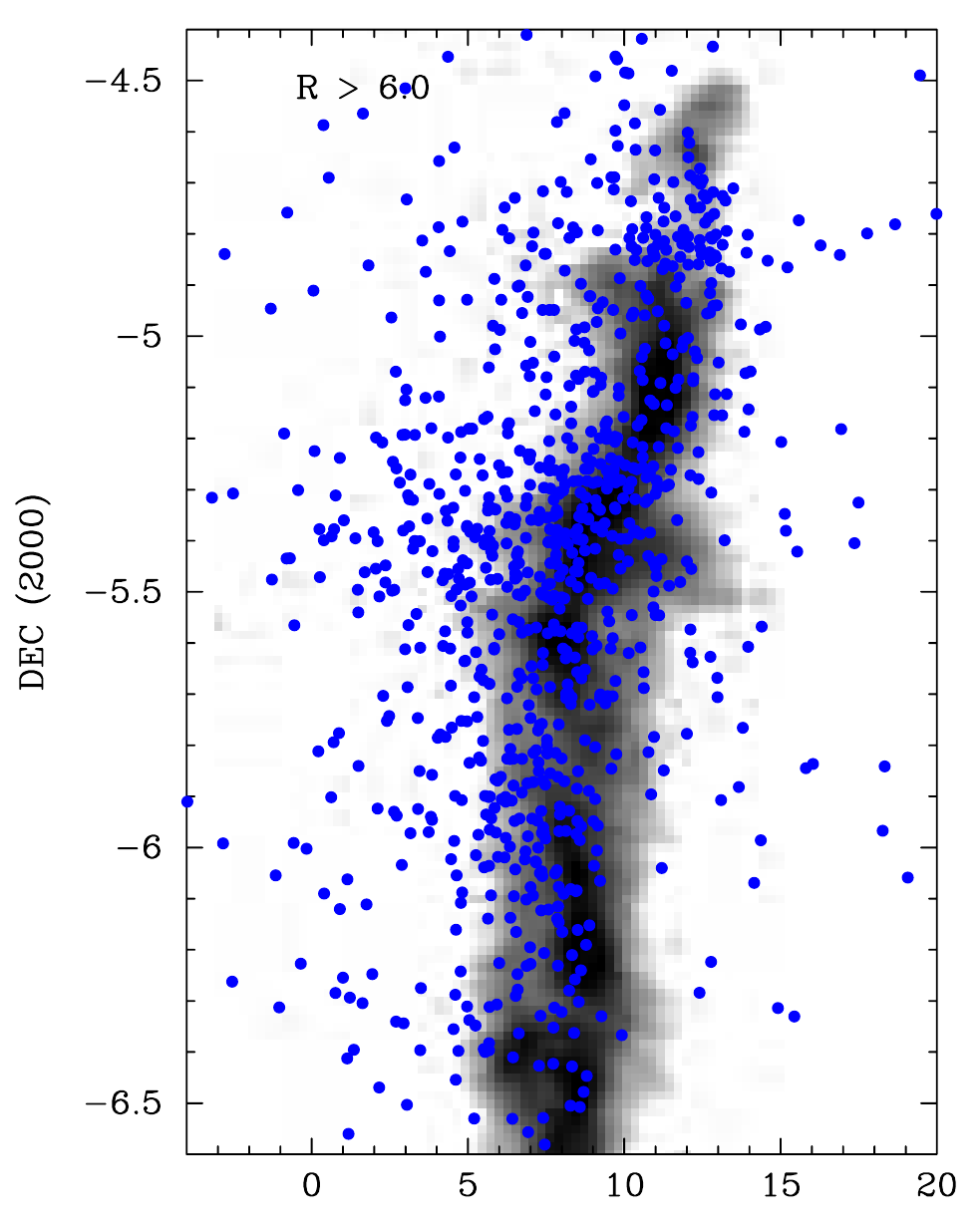}
\caption[Spatial and velocity distributions of gas and stars]{
\label{fig:gasstar_tobin09}
Measured distributions of $^{13}$CO (grayscale) and young stellar objects (blue points) in velocity ($x$ axis) and position on the sky in one dimension ($y$ axis) for the Orion Nebula Cluster. Credit: \citet{tobin09a}, \copyright AAS. Reproduced with permission.
}
\end{marginfigure}

In addition to the spatial distribution of stars and gas, one can also ask about their kinematics. Stellar kinematics can be determined by spectroscopy, and gas kinematics by molecular line observations. Depending on the choice of line, one learns about the kinematics of either lower or higher density regions of gas. These studies show that both the dense gas and the stars have much lower velocity dispersions than the less dense gas (Figures \ref{fig:denvel_walsh04} and \ref{fig:gasstar_tobin09}), but that the mean velocities are quite well correlated. The lower velocity dispersion will prove important below.

\subsection{Time Evolution of the Stellar Distribution}

As discussed in Chapter \ref{ch:gmcs}, stars stay associated with the gas from which they form for only a relatively short period. One can see this transition directly by comparing older and younger stellar populations. The younger the stellar population, the better the star-gas correlation. By stellar ages of $\sim 5-10$ Myr, there is usually no associated gas at all. However, it is still interesting to investigate how the stars evolve, because this contains important clues about how the formed.

The typical star-forming environment is vastly denser than the mean of the ISM, and as a result the stars that form are also vastly denser, in terms of either mass or number of stars per unit volume, than the mean density of gas in the ISM or stars near the Galactic midplane. More than 90\% of star formation observed within 2 kpc of the Sun takes place in regions where the stellar mass density exceeds $1$ $M_\odot$ pc$^{-3}$, corresponding to a number density $n > 30$ cm$^{-3}$ \citet{lada03a}. In comparison, the stellar mass density in the Solar neighborhood is $\sim 0.01$ $M_\odot$ pc$^{-3}$ \citep{holmberg00a}), and the mean density of gas in the ISM is $\sim 1$ cm$^{-3} \approx 0.03$ $M_\odot$ pc$^{-3}$.

\begin{marginfigure}
\includegraphics[width=\linewidth]{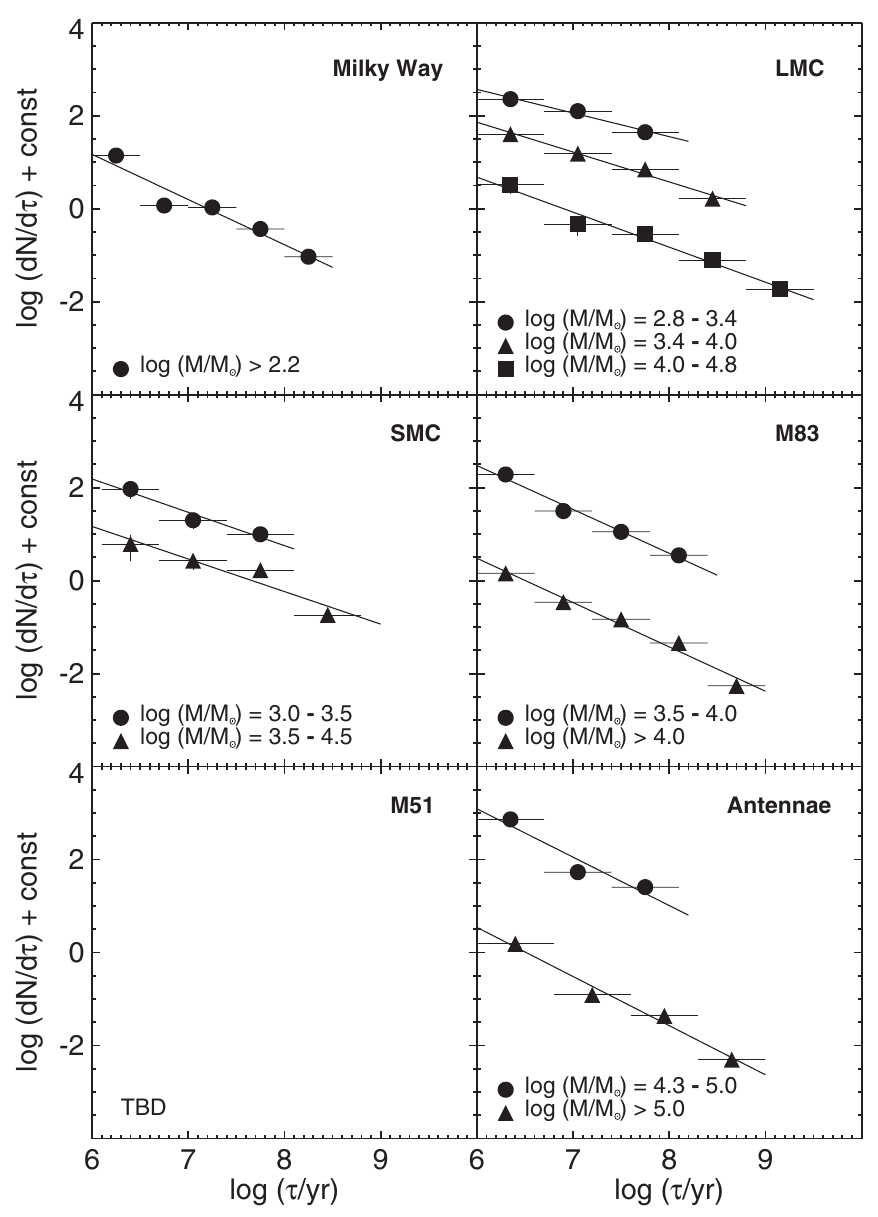}
\caption[Star cluster age distributions]{
\label{fig:clusterage_fall12}
Measured distributions of star cluster ages in several galaxies. Clusters have been binned in mass, and different symbols show different mass bins, as indicated. Credit: \citet{fall12a}, \copyright AAS. Reproduced with permission.
}
\end{marginfigure}

However, these high densities do not last. If one examines stars at an age of $\sim 100$ Myr, the ratio is flipped -- only $\sim 10\%$ are in star clusters with a density identifiably higher than that of the stellar field, while $\sim 90\%$ have dispersed and can no longer be identified as members of discrete clusters (Figure \ref{fig:clusterage_fall12}). (They can, however, still be grouped by their kinematics, which take much longer to be randomized than their positions. Collections of stars that are now at low density and no longer show up as clusters based on their positions, but that remain very close together in velocity space, are called moving groups.)

The exact functional form of this decline in number of star clusters, and whether the fraction that remain in clusters after some period of time varies with the large-scale properties of the galaxy, are both uncertain. The answers seems to depend at least in part on how one chooses to define "cluster" at very young ages when the stars are still in a fractal, non-relaxed distribution. Nonetheless, the fact that the stars disperse tells us something very important, which is that they must have formed via a mechanism that leaves the resulting stellar system gravitationally unbound. Only in very rare cases does a bound stellar system remain after the gas is removed. This is an important constraint for theories of star formation.

\begin{marginfigure}
\includegraphics[width=\linewidth]{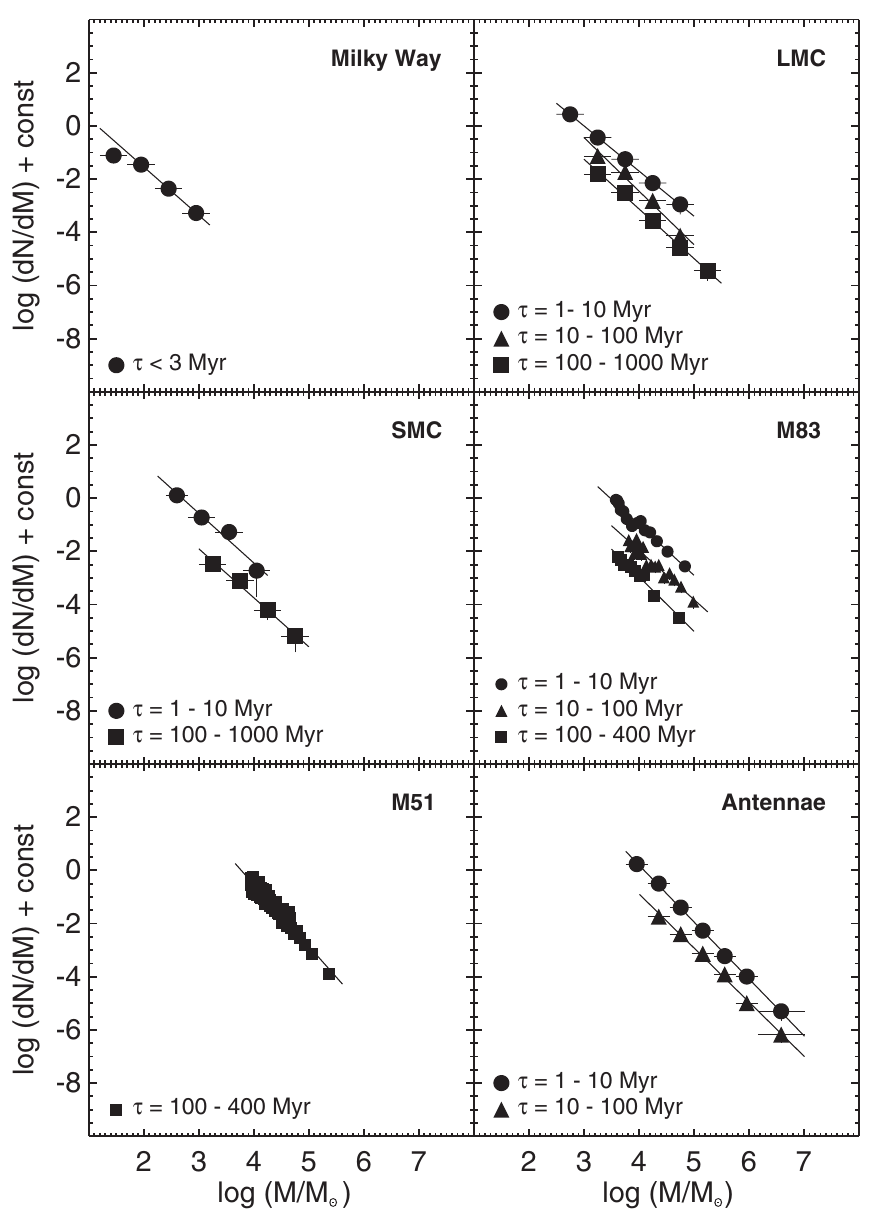}
\caption[Star cluster mass distributions]{
\label{fig:clustermass_fall12}
Measured distributions of star cluster mass in several galaxies. Clusters have been binned in age, and different symbols show different age bins, as indicated. Credit: \citet{fall12a}, \copyright AAS. Reproduced with permission.
}
\end{marginfigure}

A second important observational constraint is that the star clusters that do remain always show mass distribution that is close to a powerlaw of the form $dN/dM \propto M^{-2}$ (Figure \ref{fig:clustermass_fall12}), meaning equal mass per logarithmic bin in cluster mass. This mass function is recovered in essentially all galaxies that have been examined, and does not appear to vary with large-scale galaxy properties. The origin of this distribution is also currently debated.

\section{Theory of Stellar Clustering}

Having discussed the observational situation, we now turn to theoretical models for the origin of stellar clustering. The models here are somewhat less developed than for either the star formation rate or the IMF, but the problem is no less important and interesting.

\subsection{Origin of the Gas and Stellar Distributions}

The origin of the spatial and kinematic distributions of gas and stars, and the correlation between them, ultimately seems to lie in very general behaviors of cold gas. The characteristic timescale for gravitational collapse is the free-fall time, which varies with density as $t_{\rm ff} \propto \rho^{-1/2}$. As a result, the densest regions tend to run away and form stars first, leading a a highly structured distribution in which stars are concentrated in the densest regions of gas. Quantitatively, simulations of turbulent flows are able to reproduce the powerlaw-like two point correlation functions that are observed \citep{hansen12a}.

\begin{marginfigure}
\includegraphics[width=\linewidth]{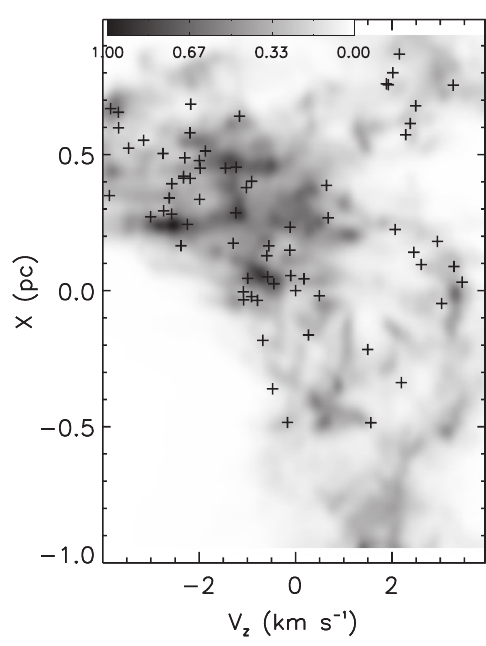}
\caption[Spatial and velocity distributions of gas and stars in a simulation]{
\label{fig:gasstar_offner09}
Distributions of $^{13}$CO (grayscale) and young stellar objects (black crosses) in velocity ($x$ axis) and position on the sky in one dimension ($y$ axis) in a simulation of the Orion Nebula Cluster. Credit:\citet{offner09b}, \copyright AAS. Reproduced with permission.
}
\end{marginfigure}

The kinematics also arise from the properties of cold, turbulent gas. One general feature of such flows is a density-velocity anti-correlation. The densest regions of gas are produced by strong converging shocks, and immediately after the passage of such a shock the velocity is small because of the cancellation of opposing fluid velocities. The stars form from these dense, shocked regions, and so they inherit the low velocities of the dense gas out of which they form -- in some sense the stars are simply the tip of the density distribution. Again, simulations can qualitatively and quantitatively reproduce the observed kinematics. Figure \ref{fig:gasstar_offner09} shows an example.

\subsection{Gas Removal and the Transition to Gas-Free Evolution}

It seems that the spatial and kinematic arrangements of young stars are understood reasonably well. This is mainly because the physics that is responsible for them -- gravity plus hydrodynamics -- is well understood and easy to simulate.  Where we start to run into trouble is when we try to follow the transition from gas-dominated to gas-free evolution, where stellar feedback almost certainly plays a role.

First of all, as a baseline, let us consider what happens if we do not include any feedback. We have already seen that this creates dimensionless star formation rates $\epsilon_{\rm ff}$ that are $\sim 2$ orders of magnitude too high. However, omitting feedback also leads to problems when it comes to stellar clustering, because if one omits feedback, then most of the available gas is transformed into stars. The result is that, if the gas cloud from which the stars formed was bound to begin with, the resulting stellar system is also bound, and thus all star formation occurs in bound clusters.

In fact, the situation is even worse than that: even if one starts with an \textit{unbound} gas cloud, then if star formation feedback does not prevent consumption of most of the gas, the result is still that most of the stars are members of bound clusters. This happens because most of the kinetic energy is on large scales, so that, even if the entire cloud is unbound, there are plenty of sub-regions within it that become bound as the turbulence dissipates \citep{clark04a}. The result is that unbound clouds wind up fragmenting into a few clusters that are unbound from one another, but that are at internally bound. Thus explaining the observed fact that most stars are not members of bound clusters requires some mechanism to truncate star formation well before the majority of the mass is transformed to stars. 

\paragraph{Rapid Versus Adiabatic Mass Loss}

To see what fraction of the gas mass must be lost to render the system unbound, we can begin with a simple argument. Let us consider a system of gas and stars with total mass $M$ in virial equilibrium, and with negligible support from magnetic fields. In this case, we have
\begin{equation}
2\mathcal{T} + \mathcal{W} = 0,
\end{equation}
where $\mathcal{T}$ is the total thermal plus kinetic energy, and $\mathcal{W}$ is the gravitational potential energy. Now let us consider what happens if we remove mass from the system, reducing the mass from $M$ to $\epsilon M$. We can envision that this is because a fraction $\epsilon$ of the starting gas mass has been turned into stars, while the remaining fraction $1-\epsilon$ is in the form of gas that is removed by some form of stellar feedback.

First suppose the removal is very rapid, on a timescale much shorter than the crossing time or free-fall time. In this case there will be no time for the system to adjust, and all the particles that remain will keep the same velocity and temperature. Thus the new kinetic energy is
\begin{equation}
\mathcal{T}' = \epsilon \mathcal{T}.
\end{equation}
Similarly, the positions of all particles that remain will be unchanged, so if the mass removal is uniform (i.e., we remove mass by randomly removing a certain fraction of the particles, without regard for their location) then the new potential energy will be
\begin{equation}
\mathcal{W}' = \epsilon^2 \mathcal{W}.
\end{equation}
The total energy of the system after mass removal is
\begin{equation}
E' = \mathcal{T}' + \mathcal{W'} = \epsilon \mathcal{T} + \epsilon^2 \mathcal{W} = \epsilon(1-2\epsilon) \mathcal{T} = \epsilon\left(\epsilon-\frac{1}{2}\right) \mathcal{W}
\end{equation}
Since $\mathcal{T} > 0$ and $\mathcal{W} < 0$, it immediately follows that the total energy of the system after mass removal is negative if and only if $\epsilon > 1/2$. Thus the system remains bound only if we remove less than $1/2$ the mass, and becomes unbound if we remove more than $1/2$ the mass. 

If the system remains bound, it will eventually re-virialize at a new, larger radius. We can solve for this radius from the equations we have already written down. The total energy of a system in virial equilibrium is
\begin{equation}
E = \frac{\mathcal{W}}{2} = -a\frac{GM^2}{2R},
\end{equation}
so if the system re-virializes the new radius $R'$ must obey
\begin{equation}
E' = -a\frac{G(\epsilon M)^2}{2R'}.
\end{equation}
However, we also know that
\begin{equation}
E' = \epsilon\left(\epsilon-\frac{1}{2}\right) \mathcal{W} = -\epsilon\left(\epsilon-\frac{1}{2}\right)a\frac{GM^2}{R},
\end{equation}
and combining these two statements we find that the new radius is
\begin{equation}
R' = \frac{\epsilon}{2\epsilon-1} R.
\end{equation}

Now consider the opposite limit, where mass is removed very slowly compared to the crossing time. To see what happens in this case, it is helpful to imagine the mass loss as occurring in very small increments, and after each increment of mass loss waiting for the system to re-establish virial equilibrium before removing any more mass. Such mass loss if referred to as adiabatic. If we change the mass by an amount $dM$ (with the sign convention that $dM < 0$ indicates mass loss), we can find the change in radius $dR$ by Taylor expanding the equation we just derived for the new radius, recalling that $\epsilon = 1 + dM/M$:
\begin{equation}
R' = \frac{\epsilon}{2\epsilon-1} R = \frac{1 + dM/M}{1 + 2 dM/M} R = \left[1 - \frac{dM}{M} + O\left(\frac{dM^2}{M^2}\right)\right] R
\end{equation}
Thus
\begin{equation}
\frac{dR}{R} = \frac{R' - R}{R} = -\frac{dM}{M},
\end{equation}
and if we integrate both sides then we obtain
\begin{equation}
\ln R = -\ln M + \mbox{const} \qquad\Longrightarrow\qquad R' \propto \frac{1}{M}.
\end{equation}
Thus if we reduce the mass from $M$ to $\epsilon M$ but do it adiabatically, the radius changes from $R$ to $R/\epsilon$. The system remains bound at all times, and just expands smoothly.

These simple arguments would suggest that mass loss should produce a bound cluster if the star formation efficiency is $>1/2$ and or the mass removal is slow, and an unbound set of stars if the efficiency is $<1/2$ and the mass removal is fast. In reality, life is more complicated than these simple arguments suggest, for a few reasons.

First, even if gas removal is rapid, some stars will still become unbound even if $\epsilon > 1/2$, and some will still remain bound even if $\epsilon < 1/2$. This is because the energy is not perfectly shared among the stars. Instead, at any given instant, some stars are moving faster than their average speed, and some are moving slower. Those that are moving rapidly at the instant when mass is removed will simply sail on out of the much-reduced potential well without sharing their energy, and thus can be lost even if $\epsilon > 1/2$. Conversely, the slowest-moving stars will not escape even if there is a very large reduction in the potential well, because they will not have time to acquire energy from the faster stars that escape. Thus for rapid mass loss, there is not a sharp boundary at $\epsilon = 1/2$. Instead, there is more of a smooth transition from no stars becoming unbound at $\epsilon \sim 1$ to no stars remaining at $\epsilon \sim 0$.

Second, we have done our calculations in a vacuum, but in reality star clusters exist inside a galactic potential, and this creates a tidal gravitational field. If a star wanders too far from the cluster, the tidal field of the galaxy will pull it off. Thus our conclusion that, in the adiabatic case, the cluster always remains bound and simply expands, must fail once the expansion proceeds too far. The outermost parts of the cluster will start to be stripped if they expand too far, and if the expansion proceeds so far that the mean density of the cluster becomes too low, it will be pulled apart entirely.

Third, the calculation we have just gone through assumes that the system starts in virial equilibrium, with the stars moving at the speed expected for virial balance. However, as we discussed before, this is not a good assumption: the stars have a much lower velocity dispersion than the gas when the cluster is young, and thus are much harder to unbind than the above argument suggests. If star formation continues for more than a single crossing time, this should become less and less of a problem as time passes and the stars are able to relax and dynamically heat up in the potential well of the gas. However, if star formation is ended very rapidly, in a crossing time, then the efficiency will have to be even lower than the value we have just estimated to be able to unbind the stars, since they are starting from much lower kinetic energies than they would have in virial balance.

\paragraph{The Cluster Formation Efficiency}

Given the theoretical modeling we have just performed, what can we say about what fraction of star formation will result in bound stellar clusters that will survive the initial gas expulsion? To address this question, we must be able to calculate the star formation efficiency, which is of course a very difficult problem, quite analogous to the problem of understanding what limits the rate of star formation overall. The answer almost certainly involves some sort of stellar feedback, so we can study a simple model for how that might work, drawn from \citet{fall10a}.

Let us consider a spherical gas cloud of initial mass $M$ and radius $R$, which begins forming stars. Star formation ends when the stars are able to inject momentum into the remaining gas at a rate high enough to raise that gas to a speed of order the escape speed in a time comparable to the crossing time. The requisite speed is
\begin{equation}
v_e \sim \sqrt{\frac{G M}{R}}.
\end{equation}
If the stellar mass at any given time is $\epsilon M$, then the momentum injection rate is
\begin{equation}
\dot{p} = \left\langle\frac{\dot{p}}{M_*}\right\rangle \epsilon M,
\end{equation}
where the quantity in angle brackets is momentum per unit time per unit stellar mass provided by a zero age population of stars. Thus our condition is that star formation ceases when
\begin{equation}
M v_e \sim \dot{p} t_{\rm cr} \sim \left\langle\frac{\dot{p}}{M_*}\right\rangle \epsilon M \frac{R}{v_e},
\end{equation}
where, since we are dropping factors of order unity, we have simply taken $t_{\rm cr} \sim R/v_e$.

Re-arranging, we conclude that star formation should cease and gas should be expelled when
\begin{equation}
\epsilon \sim \left\langle\frac{\dot{p}}{M_*}\right\rangle^{-1} \frac{v_e^2}{R} \sim \left\langle\frac{\dot{p}}{M_*}\right\rangle^{-1} G\Sigma,
\end{equation}
where $\Sigma \sim M/R^2$ is the surface density of the cloud.

Thus we expect to achieve a star formation efficiency of $\epsilon \sim 0.5$ when
\begin{equation}
\left\langle\frac{\dot{p}}{M_*}\right\rangle\sim G \Sigma.
\end{equation}
Just to give a sense of what this implies, we showed in \autoref{ch:feedback} that $\left\langle\dot{p}/M_*\right\rangle$ for stellar radiation is 23 km s$^{-1}$ Myr$^{-1}$, and plugging this in we obtain $\Sigma \sim 1$ g cm$^{-2}$. Thus regions with surface densities of $\sim 1$ g cm$^{-2}$ should be able to form bound clusters, while those with lower surface densities should not. This might plausibly explain why most regions do not form bound clusters.

However, this is an extremely crude calculation, and it assumes that one can define a well-defined "cloud" with a well-defined surface density. Real clouds, of course, have complex fractal structures. The suggested literature reading for this chapter, \citet{kruijssen12a}, is an attempt to develop a theory somewhat like this for a more realistic model of the structure of a cloud.

\paragraph{The Cluster Mass Function}

As a final topic for this chapter, what are the implications of this sort of analysis for the cluster mass function? Again, we will proceed with a spherical cow style of analysis. Consider a collection of star-forming gas clouds with an observed mass spectrum $dN_{\rm obs}/dM_g$. Each such cloud lives for a time $t_\ell(M_g)$ before forming its stars and dispersing, so the cluster formation rate is
\begin{equation}
\frac{dN_{\rm form}}{dM_g} \propto \frac{1}{t_{\ell}(M_g)} \frac{dN_{\rm obs}}{dM_g}.
\end{equation}

Now let $\epsilon$ be the final star formation efficiency for a cloud of mass $M_g$, and let $f_{\rm cl}(\epsilon)$ be the fraction of the stars that remain bound following gas removal. Thus the final mass of the star cluster formed will be
\begin{equation}
M_c = f_{\rm cl} \epsilon M_g.
\end{equation}
From this we can calculate the formation rate for star clusters of mass $M_c$:
\begin{eqnarray}
\frac{dN_{\rm form}}{dM_c} & = & \left(\frac{dM_c}{dM_g}\right)^{-1} \frac{dN_{\rm form}}{dM_g} \\
& \propto & \left[\epsilon f_{\rm cl} + \left(f_{\rm cl}+ \frac{df_{\rm cl}}{d\ln \epsilon}\right) \frac{d\epsilon}{d\ln M_g}\right]^{-1} 
\nonumber \\
& & {} \cdot
\frac{1}{t_{\ell}(M_g)} \frac{dN_{\rm obs}}{dM_g}.
\end{eqnarray}

Let us unpack this result a bit. It tells us how to translate the observed cloud mass function into a formation rate for star clusters of different masses. This relationship depends on several factors. The factor $1/t_\ell(M_g)$ simply accounts for the fact that our observed catalog of clouds oversamples the clouds that stick around the longest. The factor $\epsilon f_{\rm cl}$ just translates from gas cloud mass to cluster mass. The remaining factor, $(f_{\rm cl} + df_{\rm cl}/d\ln\epsilon)(d\epsilon/d\ln M_g)$, compensates for the way the gas cloud mass function gets compressed or expanded due to any non-linear mapping between gas cloud mass and final star cluster mass. The mapping will be non-linear if the star formation efficiency is not constant with gas cloud mass, i.e., if $d\epsilon/d\ln M_g$ is non-zero.

Since observed gas cloud mass functions are not too far from the $dN/dM \propto M^{-2}$ observed for the final star cluster mass function, this implies that the terms in square brackets cannot be extremely strong functions of $M_g$. This is interesting, because it implies that the star formation efficiency $\epsilon$ cannot be a very strong function of gas cloud mass.

\chapter{The Initial Mass Function: Observations}
\label{ch:imf_obs}

\marginnote{
\textbf{Suggested background reading:}
\begin{itemize}
\item \href{http://adsabs.harvard.edu/abs/2014prpl.conf...53O}{Offner, S.~S.~R., et al. 2014, in ``Protostars and Planets VI", ed.~H.~Beuther et al., pp.~53-75} \nocite{offner14a}
\end{itemize}
\textbf{Suggested literature:}
\begin{itemize}
\item \href{http://adsabs.harvard.edu/abs/2010Natur.468..940V}{van Dokkum, P.~G., \& Conroy, C. 2010, Nature, 468, 940} \nocite{van-dokkum10a}
\item \href{http://adsabs.harvard.edu/abs/2012ApJ...748...14D}{da Rio, N., et al. 2012, ApJ, 748, 14} \nocite{da-rio12a}
\end{itemize}
}

As we continue to march downward in size scale, we now turn from the way gas clouds break up into clusters to the way clusters break up into individual stars. This is the subject of the initial mass function (IMF), the distribution of stellar masses at formation. The IMF is perhaps the single most important distribution in stellar and galactic astrophysics. Almost all inferences that go from light to physical properties for unresolved stellar populations rely on an assumed form of the IMF, as do almost all models of galaxy formation and the ISM.

\section{Resolved Stellar Populations}

There are two major strategies for determining the IMF from observations. One is to use direct star counts in regions where we can resolve individual stars. The other is to use integrated light from more distant regions where we cannot.

\subsection{Field Stars}

The first attempts to measure the IMF were by \citet{salpeter55a},\footnote{This has to be one of the most cited papers in all of astrophysics -- nearly 5,000 citations as of this writing.} using stars in the Solar neighborhood, and the use of Solar neighborhood stars remains one of the main strategies for measuring the IMF today. Suppose that we want to measure the IMF of the field stars within some volume or angular region around the Sun. What steps must we carry out? 

\paragraph{Constructing the Luminosity Function}

The first step is to construct a luminosity function for the stars in our survey volume in one or more photometric bands. This by itself is a non-trivial task, because we require absolute luminosities, which means we require distances. If we are carrying out a volume-limited instead of a flux-limited survey, we also require distances to determine if the target stars are within our survey volume.

The most accurate distances available are from parallax, but this presents a challenge. To measure the IMF, we require a sample of stars that extends down to the lowest masses we wish to measure. As one proceeds to lower masses, the stars very rapidly become dimmer, and as they become dimmer it becomes harder and harder to obtain accurate parallax distances. For $\sim 0.1$ $M_\odot$ stars, typical absolute V band magnitudes are $M_V \sim 14$, and parallax catalogs at such magnitudes are only complete out to $\sim 5-10$ pc. A survey of this volume only contains $\sim 200-300$ stars and brown dwarfs, and this sample size presents a fundamental limit on how well the IMF can be measured. If one reduces the mass range being studied, parallax catalogs can go out somewhat further, but then one is trading off sample size against the mass range that the study can probe. Hopefully \textit{Gaia} will improve this situation significantly.

\begin{marginfigure}
\includegraphics[width=\linewidth]{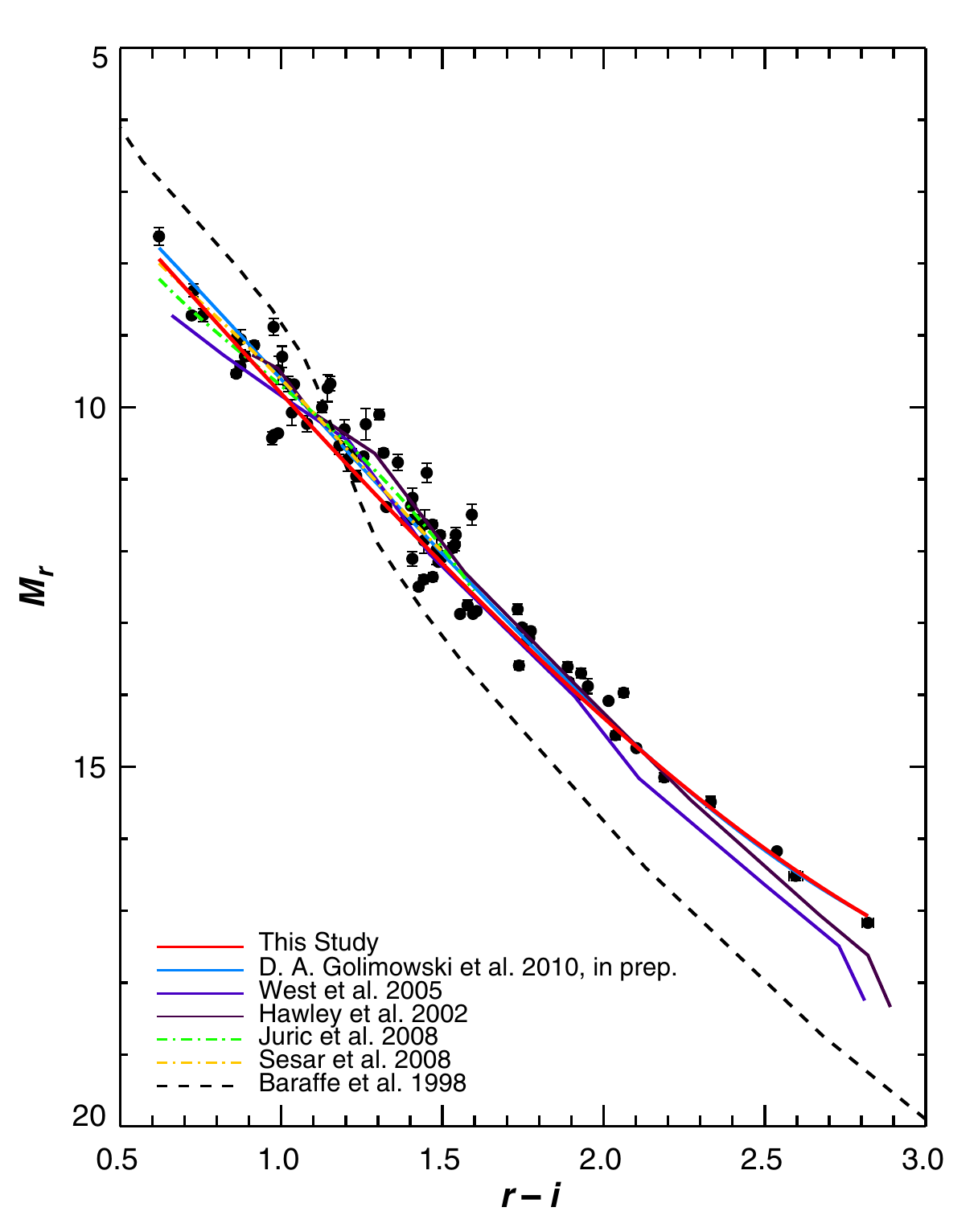}
\caption[Color-magnitude diagram of nearby stars]{
\label{fig:cmd_bochanski10}
Color-magnitude diagram for stars with well-measured parallax distances. The filters used are the SDSS $r$ and $i$. Credit: \citet{bochanski10a}, \copyright AAS. Reproduced with permission.
}
\end{marginfigure}

For these reasons, more recent studies have tended to rely on less accurate spectroscopic or photometric distances. These introduce significant uncertainties in the luminosity function, but they are more than compensated for by the vastly larger number of stars available, which in the most recent studies can be $>10^6$. The general procedure for photometric distances is to construct color-magnitude (CMD) diagrams in one or more colors for Solar neighborhood stars using the limited sample of stars with measured parallax distances, perhaps aided by theoretical models. Figure \ref{fig:cmd_bochanski10} shows an example of such a CMD. Each observed star with an unknown distance is then assigned an absolute magnitude based on its color and the CMD. The absolute magnitude plus the observed magnitude also gives a distance. The spectroscopic parallax method is analogous, except that one uses spectral type - magnitude diagrams (STMD) in place of color-magnitude ones to assign absolute magnitudes. This can be more accurate, but requires at least low resolution spectroscopy instead of simply photometry.

\paragraph{Bias Correction}

Once that procedure is done, one has in hand an absolute luminosity function, either over a defined volume or (more-commonly) a defined absolute magnitude limit. The next step is to correct it for a series of biases. We will not go into the technical details of how the corrections are made, but it is worth going through the list just to understand the issues, and why this is not a trivial task.

\textit{Metallicity bias:} the reference CMDs or STMDs used to assign absolute magnitudes are constructed from samples very close to the Sun with parallax distances. However, there is a known negative metallicity gradient with height above the galactic plane, so a survey going out to larger distances will have a lower average metallicity than the reference sample. This matters because stars with lower metallicity have higher effective temperature and earlier spectral type than stars of the same mass with lower metallicity. (They have slightly higher absolute luminosity as well, but this is a smaller effect.) As a result, if the CMD or spectral type-magnitude diagram used to assign absolute magnitudes is constructed for Solar metallicity stars, but the star being observed is sub-Solar, then we will tend to assign too high an absolute luminosity based on the color, and, when comparing with the observed luminosity, too large a distance. We can correct for this bias if we know the vertical metallicity gradient of the galaxy.

\textit{Extinction bias:} the reference CMDs / STMDs are constructed for nearby stars, which are systematically less extincted than more distant stars because their light travels through less of the dusty Galactic disk. Dust extinction reddens starlight, which causes the more distant stars to be assigned artificially red colors, and thus artificially low magnitudes. This in turn causes their absolute magnitudes and distances to be underestimated, moving stars from their true luminosities to lower values. These effects can be mitigated with knowledge of the shape of the dust extinction curve and estimates of how much extinction there is likely to be as a function of distance.

\textit{Malmquist bias:} there is some scatter in the magnitudes of stars at fixed color, both due to the intrinsic physical width of the main sequence (e.g., due to varying metallicity, age, stellar rotation) and due to measurement error. Thus at fixed color, magnitudes can scatter up or down. Consider how this affects stars that are near the distance of magnitude limit for the survey: stars whose true magnitude should place them just outside the survey volume or flux limit will be artificially scatter into the survey if they scatter up but not if they scatter down, and those whose true magnitude should place them within the survey will be removed if they scatter to lower magnitude. This asymmetry means that, for stars near the distance or magnitude cutoff of the survey, the errors are not symmetric; they are much more likely to be in the direction of positive than negative flux. This effect is known as Malmquist bias. It can be corrected to the extent that one has a good idea of the size of the scatter in magnitude and understands the survey selection.

\textit{Binarity:} many stars are members of binary systems, and all but the most distant of these will be unresolved in the observations and will be mistaken for a single star. This has a number of subtle effects, which we can think of in two limiting cases. If the binary is far from equal mass, say $q = M_2/M_1 \sim 0.3$ or less, then the secondary star contributes little light, and the system colors and absolute magnitude will not be that different from those of an isolated primary of the same mass. Thus the main effect is that we correctly include the primary in our survey, but we miss the secondary entirely, and therefore undercount the number of low luminosity stars. On the other hand, if the mass ratio $q\sim 1$, then the main effect is that the color stays about the same, but using our CMD we assign the luminosity of a single star when the true luminosity is actually twice that. We therefore underestimate the distance, and artificially scatter things into the survey (if it is volume limited) or out of the survey (if it is luminosity-limited). At intermediate mass ratios, we get a little of both effects.

The main means of correcting for this is, if we have a reasonable estimate of the binary fraction and mass ratio distribution, to guess a true luminosity function, determine which stars are binaries, add them together as they would be added in the observation, filter the resulting catalog through the survey selection, and compare to the observed luminosity function. This procedure is then repeated, adjusting the guessed luminosity function, until the simulated observed luminosity function matches the actually observed one. 

\begin{marginfigure}
\includegraphics[width=\linewidth]{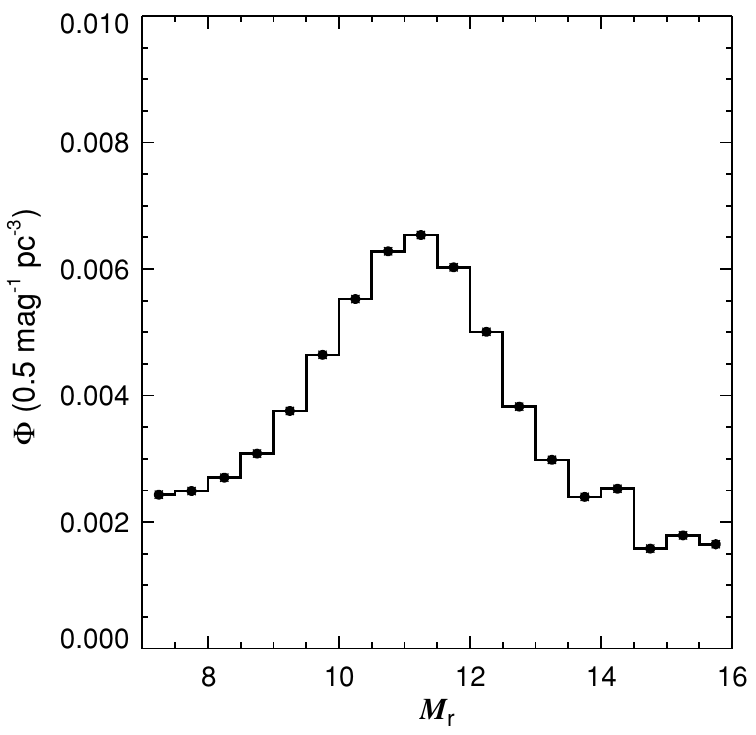}
\includegraphics[width=\linewidth]{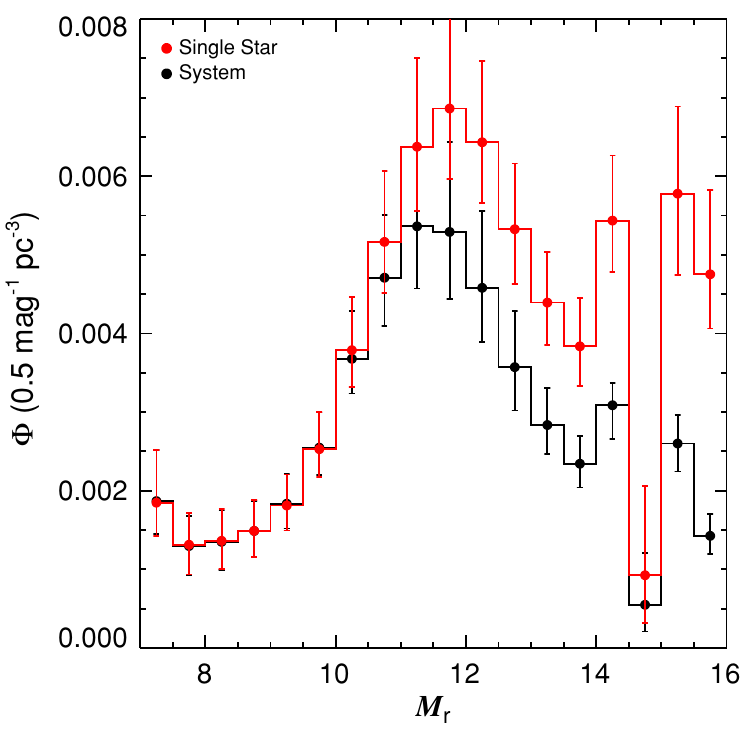}
\caption[Color-magnitude diagram of nearby stars]{
\label{fig:lumfunction_bochanski10}
Luminosity function for Milky Way stars before (top) and after (bottom) bias correction. Credit: \citet{bochanski10a}, \copyright AAS. Reproduced with permission.
}
\end{marginfigure}

Once all these bias corrections are made, the result is a corrected luminosity function that (should) faithfully reproduce the actual luminosity function in the survey volume. Figure \ref{fig:lumfunction_bochanski10} shows an example of raw and corrected luminosity functions.

\paragraph{The Mass-Magnitude Relation}

\begin{marginfigure}
\includegraphics[width=\linewidth]{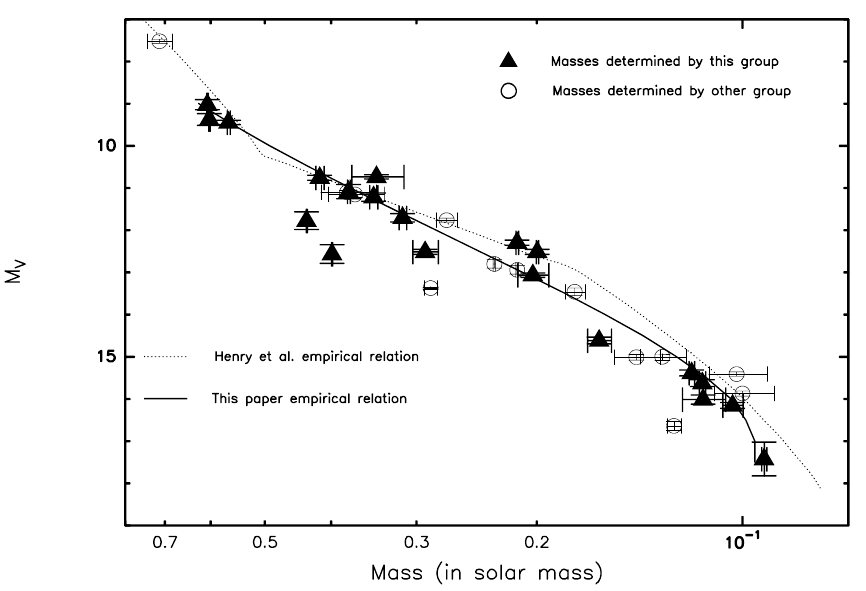}
\caption[Mass-magnitude relationship]{
\label{fig:mmr_delfosse00}
Empirically-measured mass-magnitude relationship in $V$ band. Credit: \citeauthor{delfosse00a}, A\&A, 364, 217, 2000, reproduced with permission \copyright\, ESO.
}
\end{marginfigure}

The next step is to convert the luminosity function into a mass function, which requires knowledge of the mass-magnitude relation (MMR) in whatever photometric band we have used for our luminosity function. This must be determined by either theoretical modelling, empirical calibration, or both. Particularly at the low mass end, the theoretical models tend to have significant uncertainties arising from complex atmospheric chemistry that affects the optical and even near-infrared colors. 
For empirical calibrations, the data are only as good as the empirical mass determinations, which must come from orbit modelling. This requires the usual schemes for measuring stellar masses from orbits, e.g., binaries that are both spectroscopic and eclipsing and thus have known inclinations, or visual binaries with measured radial velocities. Figure \ref{fig:mmr_delfosse00} shows an example empirical MMR.

As with the luminosity function, there are a number of possible biases, because the stars are not uniform in either age or metallicity, and as a result there is no true single MMR. This would only introduce a random error if the age and metallicity distribution of the sample used to construct the MMR were the same as that in the IMF survey. However, there is no reason to believe that this is actually the case. The selection function used to determine the empirical mass-magnitude sample is complex and poorly characterized, but it is certainly biased towards systems closer to the Sun, for example. Strategies to mitigate this are similar to those used to mitigate the corresponding biases in the luminosity function.

Once the mass-magnitude relationship and any bias corrections have been applied, the result is a measure of the field IMF. The results appear to be well-fit by a lognormal distribution or a broken powerlaw, along the lines of the \citet{chabrier05a} and \citet{kroupa02c} IMFs introduced in Chapter \ref{ch:obsstars}.

\paragraph{Age Correction}

The strategy we have just described works fine for stars up to $\sim 0.7$ $M_\odot$ in mass. However, it fails with higher mass stars, for one obvious reason: stars with masses larger than this can evolve off the main sequence on timescales comparable to the mean stellar age in the Solar neighborhood. Thus the quantity we measure from this procedure is the present-day mass function (PDMF), not the IMF. Even that is somewhat complicated because stars' luminosities start to evolve non-negligibly even before they leave the main sequence, so there are potential errors in assigning masses based on a MMR calibrated from younger stars.

One option in this case is simply to give up and not say anything about the IMF at higher masses. However, there is another option, which is to try to correct for the bias introduced by stellar evolution. Suppose that we think we know both the star formation history of the region we are sampling, $\dot{M}_*(t)$, and the initial mass-dependent main-sequence stellar lifetime, $t_{\rm MS}(m)$. Let $dn/dm$ be the IMF. In this case, the total number of stars formed of the full lifetime of the galaxy in a mass bin from $m$ to $m+dm$ is
\begin{equation}
\frac{dn_{\rm form}}{dm} =  \frac{dn}{dm} \int_{-\infty}^0 dt \, \dot{M}_*(t)
\end{equation}
where $t=0$ represents the present. In contrast, the number of stars per unit mass still on the main sequence is
\begin{equation}
\frac{dn_{\rm MS}}{dm} = \frac{dn}{dm} \int_{-t_{\rm MS}(m)}^0 dt \, \dot{M}_*(t)
\end{equation}
Thus if we measure the main sequence mass distribution $dn_{\rm MS}/dm$, we can correct it to the IMF just by multiplying:
\begin{equation}
\frac{dn}{dm} \propto \frac{dn_{\rm MS}}{dm} \frac{\int_{-t_{\rm MS}(m)}^0 dt \, \dot{M}_*(t)}{\int_{-\infty}^0 dt \, \dot{M}_*(t)}.
\end{equation}
This simply reduces to scaling the number of observed stars by the fraction of stars in that mass bin that are still alive today.

Obviously this correction is only as good as our knowledge of the star formation history, and it becomes increasingly uncertain as the correction factor becomes larger. Thus attempts to measure the IMF from the Galactic field even with age correction are generally limited to masses of no more than a few $M_\odot$.

\subsection{Young Clusters}

To measure the IMF for more massive stars requires a different technique: surveys of young star clusters. The overall outline of the technique is essentially the same as for the field: construct a luminosity function, correct for biases, then use a mass-magnitude relation to convert to a mass function. However, compared to the field, studying a single cluster offers numerous advantages:
\begin{itemize}
\item If the population is young enough, then even the most massive stars will remain on the main sequence, so there is no need to worry about correcting from the PDMF to the IMF. Even for somewhat older clusters, one can probe to higher masses than would be possible with the $\sim 5-10$ Gyr old field population.
\item The stellar population is generally uniform in metallicity or very close to it, so there are no metallicity biases.
\item The entire stellar population is at roughly the same distance, so there are no Malmquist or extinction biases. Moreover, in some cases the distance to the cluster is known to better than 10\% from radio parallax -- some young stars flare in the radio, and with radio interferometry it is possible to obtain parallax measurements at much larger distances than would be possible for the same stars in the optical. 
\item Low-mass stars and brown dwarfs are significantly more luminous at young ages, and so the same magnitude limit will correspond to a much lower mass limit, making it much easier to probe into the brown dwarf regime.
\end{itemize}

These advantages also come with some significant costs.
\begin{itemize}
\item The statistics are generally much worse than for the field. The most populous young cluster that is close enough for us to resolve individual stars down to the hydrogen burning limit is the Orion Nebula Cluster, and it contains only $\sim 10^3 - 10^4$ stars, as compared to $\sim 10^6$ for the largest field surveys.
\item The MMR that is required to convert an observed magnitude into a mass is much more complex in a young cluster, because a significant fraction of the stars may be pre-main sequence. For such stars, the magnitude is a function not just of the mass but also the age, and one must fit both simultaneously, and with significant theoretical uncertainty. We will discuss this issue further in Chapter \ref{ch:protostar_evol}. How much of a problem this is depends on the cluster age -- for a 100 Myr-old cluster like the Pleiades, all the stars have reached the main sequence, while for a $\sim 1-2$ Myr-old cluster like Orion, almost none have. However, there is an obvious tradeoff here: in a Pleiades-aged cluster, the correction for stars leaving the main sequence is significant, while for an Orion-aged cluster it is negligible.
\item For the youngest clusters, there is usually significant dust in the vicinity of the stars, which introduces extinction and reddening that is not the same from star to star. This introduces scatter, and also potentially bias because the extinction may vary with position, and there is a systematic correlation between position and mass (see next point).
\item Mass segregation can be a problem. In young clusters, the most massive stars are generally found closer to the center -- whether this is a result of primordial mass segregation (the stars formed there) or dynamical mass segregation (they formed elsewhere but sank to the center), the result is the same. Conversely, low mass stars are preferentially on the cluster outskirts. This means that studies must be extremely careful to measure the IMF over the full cluster, not just its outskirts or core; this can be hard in the cluster center due to problems with crowding. Moreover, if the extinction is not spatially uniform, more massive stars toward the cluster center are likely to suffer systematically more extinction that low-mass ones.
\item Dynamical effects can also be a problem. A non-trivial fraction of O and B stars are observed to be moving with very high spatial velocities, above $\sim 50$ km s$^{-1}$. These are known as runaways. They are likely created by close encounters between massive stars in the core of a newly-formed cluster that lead to some stars being ejected at speeds comparable to the orbital velocities in the encounter. Regardless of the cause, the fact that this happens means that, depending on its age and how many ejections occurred, the cluster may be missing some of its massive stars. Conversely, because low-mass stars are further from the center, if there is any tidal stripping, that will preferentially remove low-mass stars.
\item Binary correction is harder for young stars because the binary fraction as a function of mass is much less well known for young clusters than it is for field stars.
\end{itemize}

Probably the best case for studying a very young cluster is the Orion Nebula Cluster, which is 415 pc from the Sun. Its distance is known to a few percent from radio interferometry \citep{sandstrom07a, menten07a, kim08a}. It contains several thousand stars, providing relatively good statistics, and it is young enough that all the stars are still on the main sequence. It is close enough that we can resolve all the stars down to the brown dwarf limit, and even beyond. However, the ONC's most massive star is only 38 $M_\odot$, so to study the IMF at even higher masses requires the use of more distant clusters within which we cannot resolve down to low masses. 

For somewhat older clusters, the best case is almost certainly the Pleiades, which has an age of about 120 Myr. It obviously has no very massive stars left, but there are still $\sim 10$ $M_\odot$ stars present, and it is also close and very well-studied. The IMF inferred for the Pleiades appears to be consistent with that measured in the ONC.

\subsection{Globular Clusters}

A final method for studying the IMF is to look at globular clusters. Compared to young clusters, globular cluster lack the massive stars because they are old, and suffer somewhat more from confusion problems due to their larger distances. Otherwise they are quite similar in terms of methodological advantages and disadvantages.

The main reason for investigating globular clusters is that they provide us with the ability to measure the IMF in an environment as different as possible from that of young clusters forming in the disk of the Milky Way today. The stars in globular clusters are ancient and metal poor, and they provide the only means of accessing that population without resorting to integrated light measurements. They are therefore a crucial bridge to the integrated light methods we will discuss shortly.

The major challenge for globular clusters is that all the dynamical effects are much worse, due to the longer time that the clusters have had to evolve. Over long times, globular clusters systematically lose low-mass stars due to tidal shocking and a phenomenon known as two-body evaporation, whereby the cluster attempts to relax to a Maxwellian velocity distribution, but, due to the fact that the cluster is sitting in a tidal potential, the tail of that distribution keeps escaping. This alters the IMF. There can also be stellar collisions, which obviously move low mass stars into higher mass bins.

Accounting for all these effects is a major challenge, and the usual method is to adopt a proposed IMF and then try to simulate the effects of dynamical evolution over the past $\sim 13$ Gyr in order to predict the PDMF that would result. This is then compared to the observed PDMF, and the underlying IMF is iteratively adjusted until they match. This is obviously subject to considerable uncertainties.

\subsection{General Results}

The general result of these studies is that the IMF appears to be fairly universal. There are claims for variation in the literature, but they are generally based on statistical analyses that ignore (or underestimate) systematic errors, which are pervasive. This is not to say that the IMF certainly is universal, just that there is as yet no strongly convincing evidence for its variation. One possible exception is in the nuclear star cluster of the Milky Way, where \citet{lu13a} report an IMF that is somewhat flatter than usual at the high mass end. It is unclear if this is a true IMF effect resulting from the very strange formation environment, or a dynamical effect.

\section{Unresolved Stellar Populations}

The main limitation of studying the IMF using resolved stars is that it limits our studies to the Milky Way and, if we are willing to forgo observing below $\sim 1$ $M_\odot$, the Magellanic Clouds. This leaves us with a very limited range of star-forming environments to study, at least compared with the diversity of galaxies that has existed over cosmological time, or even that exist in the present-day Universe. To measure the IMF in more distant systems, we must resort to techniques that rely on integrated light from unresolved stars.

\subsection{Stellar Population Synthesis Methods}

One method for working with integrated light is stellar population synthesis: one starts with a proposed IMF, and then generates a prediction for the stellar light from it. In the case of star clusters or other mono-age populations, the predicted frequency-dependent luminosity from a stellar population of mass $M_*$ is
\begin{equation}
L_\nu = M_* \int_0^\infty dm \,\frac{dn}{dm} L_\nu(m,t),
\end{equation}
where $L_\nu(m,t)$ is the predicted specific luminosity of a star of mass $m$ and age $t$. For a population with a specified star formation history (usually constant), one must further integrate over the star formation history
\begin{equation}
L_\nu = \int_0^{\infty} dt\, \dot{M}_*(t) \int_0^\infty dm\, \frac{dn}{dm} L_\nu(m,t)
\end{equation}
where $\dot{M}_*(t)$ is the star formation rate a time $t$ in the past. The predicted spectrum can then be compared to observations to test whether the proposed IMF is consistent with them.

\paragraph{The Upper IMF} In practice when using this method to study the IMF, one selects combinations of photometric filters or particular spectral features that are particularly sensitive to certain regions of the IMF. One prominent example of this is the ratio of H$\alpha$ emission to emission in other bands (or to inferred total mass). This probes the IMF because H$\alpha$ emission is produced by recombinations, and thus the H$\alpha$ emission rate is proportional to the ionizing luminosity. This in turn is dominated by $\sim 50$ $M_\odot$ and larger stars. In contrast, other bands are more sensitive to lower masses -- how low depends on the choice of band, but even for the bluest non-ionizing colors (e.g., \textit{GALEX} FUV), at most $\sim 20$ $M_\odot$. Thus the ratio of H$\alpha$ to other types of emission serves as a diagnostic of the number of very massive stars per unit total mass or per unit lower mass stars, and thus of the shape of the upper end of the IMF.

When comparing models to observations using this technique, one must be careful to account for stochastic effects. Because very massive stars are rare, approximating the IMF using the integrals we have written down will produce the right averages, but the dispersion about this average may be very large and asymmetric. In this case Monte Carlo sampling of the IMF is required. Once one does that, the result is a predicted probability distribution of the ratio of H$\alpha$ to other tracers, or to total mass. One can then compare this to the observed distribution of luminosity ratio in a sample of star clusters in order to study whether those clusters' light is consistent with a proposed IMF. One can also use the same technique on entire galaxies (which are assumed to have constant star formation rates) in order to check if the integrated light from the galaxy is consistent with the proposed IMF.

This technique has been deployed in a range of nearby spirals and dwarfs, and the results are that, when the stochastic correction is properly included, the IMF is consistent with the same high end slope of roughly $dn/dm\propto m^{-2.3}$ seen in resolved star counts.

\paragraph{The Low-Mass IMF in Ellipticals} A second technique has been to target two spectral features that are sensitive to the low mass end of the IMF: the Na~\textsc{i} doublet and the Wing-Ford molecular FeH band. Both of these regions are useful because they are produced by absorption by species found only in M type stars, but they are also gravity-sensitive, so they are \textit{not} found in the spectra of M giants. They therefore filter out a contribution from red giants, and only include red dwarfs. The strength of these two features therefore measures the ratio of M dwarfs to K dwarfs, which is effectively the ratio of $\sim 0.1-0.3$ $M_\odot$ stars to $\sim 0.3-0.5$ $M_\odot$ stars.

\citet{van-dokkum10a} used this technique on stacked spectra of ellipticals in the Coma and Virgo clusters, and found that the spectral features there were \textit{not} consistent with the IMF seen in the Galactic field and in young clusters. Instead, they found that the spectrum required an IMF that continues to rise down to $\sim 0.1$ $M_\odot$ rather than having a turnover. This result was, and continues to be, highly controversial due to concerns about unforeseen systematics hiding in the stellar population synthesis modelling.

\begin{figure}
\includegraphics[width=\linewidth]{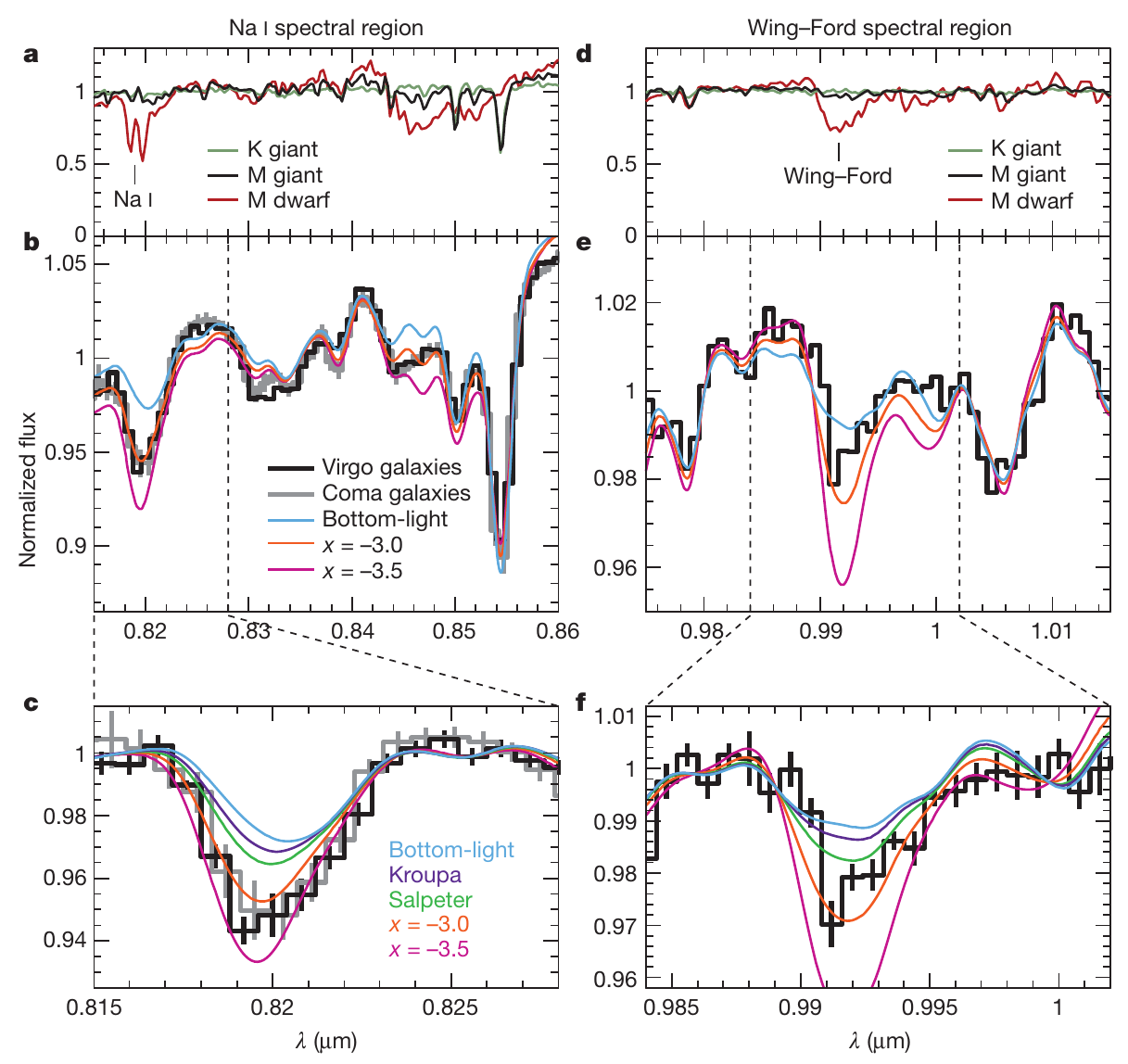}
\caption[Elliptical galaxy spectra in the Na~\textsc{i} and Wing-Ford regions]{
\label{fig:spectra_vdk10}
Top panels: sample spectra of K and M giants and M dwarfs in the Na~\textsc{i} and Wing-Ford spectral regions. Middle panels: averaged spectra for Virgo cluster (black) and Coma cluster (gray) ellipticals, overlayed with predicted model spectra for four possible IMFs, ranging from "bottom light" (few dwarfs) to powerlaws of increasing steepness (more dwarfs). Bottom panels: zoom-ins on the Na~\textsc{i} and Wing-Ford regions in the previous panels. Reprinted by permission from Macmillan Publishers Ltd:
Nature, 468, 940, \citeauthor{van-dokkum10a}, \copyright 2010.
}
\end{figure}

\subsection{Mass to Light Ratio Methods}

A second method of probing the IMF in unresolved stellar populations relies in measuring the mass independently of the starlight and thereby inferring a mass to light ratio that can be compared to models. As with the Na~\textsc{i} and Wing-Ford methods, this is most easily applied to old, gas-free stellar populations with no gas to complicate the modelling. One can obtain an independent measurement of the mass in two ways: from lensing of background objects, or from dynamical modelling in systems where the stellar velocity distribution as a function of position has been determined using an integrated field unit (IFU) or similar technique to get a spectrum at each position. Once it is obtained, the mass map is divided by the light map to form a mass to light ratio.

The main complication in comparing the mass to light ratio to theoretical predictions from stellar population synthesis is that one must account for dark matter, which can raise the ratio compared to that of a pure stellar system. This requires some modelling, and is probably the most uncertain part of the procedure. Of course if one allows a completely arbitrary distribution of dark matter, then one can produce any light to mass ratio that is heaver than one produced by the stars alone. However, this might require extremely implausible dark matter distributions. Thus the general procedure is to consider a set of "reasonable" dark matter distributions and infer limits on the stellar mass to light ratio from the extreme limiting cases.

A number of authors have used this technique \citep[e.g.,][]{cappellari12a} and tentatively found results consistent with those of \citet{van-dokkum10a}, i.e., that in giant elliptical galaxies the mass to light ratio is such that one must have an IMF that produces less light per unit mass than the Milky Way IMF.

\section{Binaries}

While this chapter is mostly about the IMF, the IMF is inextricably bound up with the properties of binary star systems. This is partly for observational reasons -- the need to correct observed luminosity functions for binarity -- and partly for theoretical reasons, which we will cover in Chapter \ref{ch:imf_th}. We will therefore close this chapter with a discussion of the observational status of the properties of binary stars, or stellar multiples more generally.

\subsection{Finding Binaries}

Before diving in, we will briefly review how we find stellar binaries. The history of this is interesting, because binary stars are one of the first examples of successful use of statistical inference in astronomy. Of course there are many stars that appear close together on the sky, but it is non-trivial to determine which are true companions and which are chance alignments. In the 1700s, it was not known if there were any true binary stars. However, in 1767 the British astronomer John Michell performed a statistical analysis of the locations of stars on the sky, and showed that there were more close pairs than would be expected from random placement. Here therefore concluded that there must be true binaries. What is particularly impressive is that this work predates a general understanding of Poisson distributions, which were not fully understood until Poisson's work in 1838.

Today binaries can be identified in several ways.
\begin{itemize}
\item {\it Spectroscopic binaries}: these are systems where the spectral lines of a star show periodic radial velocity variations that are consistent with the star moving in a Keplerian orbit. Single-lined spectroscopic binaries are those where only one star's moving lines are seen, and double-lined ones are systems where two sets of lines moving in opposite senses are seen. Spectroscopic detection is generally limited to binaries that are quite close, both for reasons of velocity sensitivity and for reasons of timescale -- wide orbits take too long to produce a noticeable change in radial velocity.
\item {\it Eclipsing binaries}: these are systems that show periodic light curve variation consistent with one stars occulting the disk of another star. As with spectroscopic binaries, this technique is mostly sensitive to very close systems, because the probability of occultation and the fraction of the a stellar disk blocked (and thus the strength of the photometric variation) are higher for closer systems.
\item {\it Visual binaries}: these are systems where the stars are far enough apart to be resolved by a telescope, perhaps aided by adaptive optics or similar techniques to improve contrast and angular resolution. This technique is obviously sensitive primarily to binaries with relatively wide orbits. Of course seeing two stars close together does not prove they are related, and so this category breaks into sub-categories depending on how binarity is confirmed. 
\begin{itemize}
\item One way of confirming the stars are related is measuring their proper motions and showing that they have the same space velocity. Systems of this sort are called {\it common proper motions binaries}. 
\item Even better, if the stars have a short enough orbital period one may be able to see the stars complete all or part of an orbit around one another. Stars in this category are called {\it astrometric binaries}.
\item Finally, if the stars are close enough in the sky, one may simply argue on probabilistic grounds that a chance alignment at that small a separation is very unlikely, and therefore argue that the stars are likely a binary on statistical grounds.
\end{itemize}
\end{itemize}

Given these techniques, it is important to note that the hardest binaries to find are usually those at intermediate separations -- too close to be visually resolved, but too distant to produce detectable radial velocity variation, and too distant for eclipses to be likely. The problem is exacerbated for more distant stars, since the minimum physical separation for which it is possible to resolve a binary visually is obviously inversely proportional to distance. Massive stars, which are rare and therefore tend to be distant, are the worst example of this. For example very little is known about companions to O stars at $\sim 100$ AU separations and mass ratios not near unity.

\subsection{Binary Properties}

Having reviewed the observational techniques, we now consider what the observations reveal. There are a few basic facts about binaries that any successful theory should be able to reproduce (but none really do very well).

\begin{marginfigure}
\includegraphics[width=\linewidth]{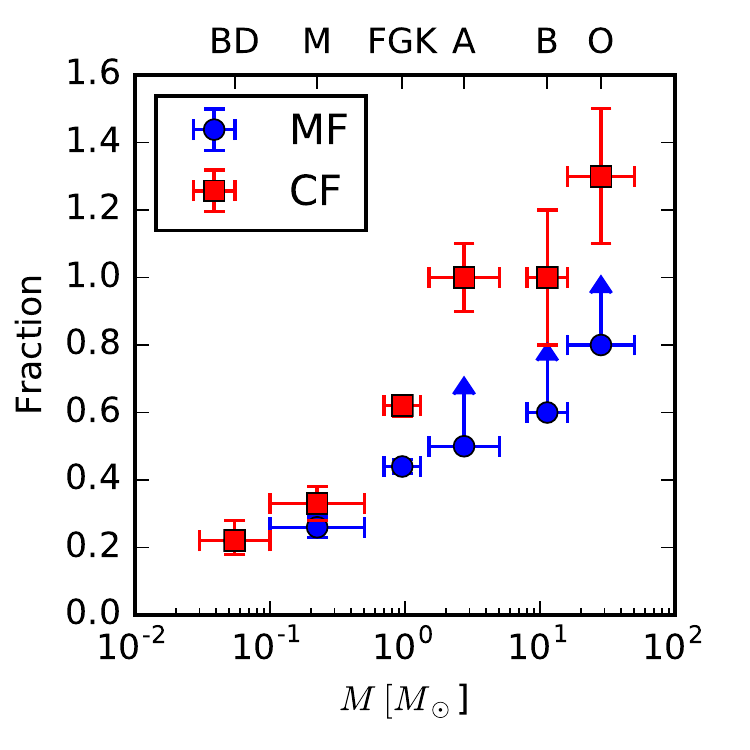}
\caption[Multiple system fraction versus stellar mass]{
\label{fig:multiplicity}
Multiple system fraction (blue) and companion fraction (red) versus primary star mass for field stars. Horizontal error bars show ranges of mass, and the upper axis shows the spectral type corresponding to that mass, with BD short for brown dwarf. Vertical error bars and limits indicate observational uncertainties. The multiple system fraction is the fraction of stars of that mass that are in multiple systems, while the companion fraction is the mean number of companions per star. The data plotted are taken from Table 1 of \citep{duchene13a}.
}
\end{marginfigure}

First, the binary fraction is a strong function of the mass of the primary star. For O stars it approaches 100\%, while for M and earlier stars it is closer to 20\%. Since the IMF is heavily weighted toward low mass stars (by number), the majority of stars are single -- \citet{lada06a} estimates the single star fraction in the disk today as $60-70\%$. Thus the binary formation mechanism must be strongly mass-dependent. Figure \ref{fig:multiplicity} summarizes this dependence.

Second, the binary period (or separation) distribution is extremely broad and lacks many obvious features \citep{duquennoy91a}. Depending on the stellar mass and the range of periods to which the data are sensitive, this may be fit either by a lognormal in period, or by a flat distribution in $\log P$. The latter is known as \"{O}pik's Law, and it states that there are equal numbers of binaries per logarithmic bin in period (or in semi-major axis). That seems to break down at very large and very small separations, but there is a broad plateau that is close to flat.

For massive stars, there is some evidence for an excess at small separations, indicating an excess of close binaries above what a flat distribution would produce \citep{sana11a}. However it is not entirely clear how much weight to put on this result, since it requires combining data sets gathered in highly different ways (i.e., putting spectroscopic and visual binaries together), and because the selection biases for both data sets are highly complex.

Third, close stellar companions do not appear to be drawn randomly from the IMF. Instead, they are far more likely that a random drawing from the IMF would predict to have masses close to the mass of the primary. In contrast, long-period binaries are consistent with random drawing from the IMF.  We define the mass ratio of a binary consisting of two stars $M_1$ and $M_2$ as $q=M_2/M_1$, where by convention $M_1 > M_2$, so $q$ runs from 0 to 1. Since that the IMF peaks near $0.2$ $\msun$, we would expect random drawing from a sample with primary masses well above $0.2$ $\msun$ to produce many more binaries with small $q$ than large $q$. This is exactly what is seen for distant binaries ($>1000$ day periods), but the opposite trend is seen for closer binaries \citep{mazeh92a, sana11a}.

\problemset

\begin{enumerate}

\item \textbf{Toomre Instability.}\\
Chapter \ref{ch:sflaw_th} discusses the Toomre instability as a potentially important factor in driving star formation. It may also be relevant to determining the maximum masses of molecular clouds. In this problem we will calculate the stability condition and related quantities. Consider a uniform, infinitely thin disk of surface density $\Sigma$ occupying the $z=0$ plane. The disk has a flat rotation curve with velocity $v_R$, so the angular velocity is $\veco=\Omega \ehat_z$, with $\Omega = v_R/r$ at a distance $r$ from the disk center. The velocity of the fluid in the $z=0$ plane is $\vecv$ and its vertically-integrated pressure is $\Pi=\int_{-\infty}^{\infty} P \, dz = \Sigma c_s^2$. 
\begin{enumerate}
\item Consider a coordinate system co-rotating with the disk, centered at a distance $R$ from the disk center, oriented so that the $x$ direction is radially outward and the $y$ direction is in the direction of rotation. In this frame, the vertically-integrated equations of motion and the Poisson equation are
\begin{eqnarray*}
\frac{\partial \Sigma}{\partial t} + \nabla \cdot (\Sigma \vecv) & = & 0 \\
\frac{\partial \vecv}{\partial t} + (\vecv\cdot\nabla)\vecv & = & -\frac{\nabla \Pi}{\Sigma} - \nabla \phi - 2\veco \times \vecv + \Omega^2 (x \ehat_x + y \ehat_y) \\
\nabla^2 \phi & = & 4 \pi G \Sigma \delta(z).
\end{eqnarray*}
The last two terms in the second equation are the Coriolis and centrifugal force terms.
We wish to perform a stability analysis of these equations. Consider a solution $(\Sigma_0, \phi_0)$ to these equations in which the gas is in equilibrium (i.e., $\vecv=0$), and add a small perturbation: $\Sigma=\Sigma_0 + \epsilon \Sigma_1$, $\vecv = \vecv_0 + \epsilon \vecv_1$, $\phi=\phi_0 + \epsilon \phi_1$, where $\epsilon \ll 1$. Derive the perturbed equations by substituting these values of $\Sigma$, $\vecv$, and $\phi$ into the equations of motion and keeping all the terms that are linear in $\epsilon$.
\item The perturbed equations can be solved by Fourier analysis. Consider a trial value of $\Sigma_1$ described by a single Fourier mode $\Sigma_1 = \Sigma_a \exp[i(kx - \omega t)]$, where we choose to orient our coordinate system so that the wave vector $\mathbf{k}$ for this mode is in the $x$ direction. As an {\it ansatz} for $\phi_1$, we will look for a solution of the form $\phi_1 = \phi_a \exp[i(kx - \omega t) - |k z|]$. (One can show that the solution must take this form, but we will not do so here.) Derive the relationship between $\phi_a$ and $\Sigma_a$.
\item Now try a similar single-Fourier mode form for the perturbed velocity: $\vecv_1 = (v_{ax} \ehat_x + v_{ay} \ehat_y) \exp[i(kx - \omega t)]$. Derive three equations relating the unknowns $\Sigma_a$, $v_{ax}$, and $v_{ay}$. You will find it useful to expand $\Omega$ in a Taylor series around the origin of your coordinate system, i.e., write $\Omega = \Omega_0 + (d\Omega/dx)_0 x$, where $\Omega_0 = v_R/R$ and $(d\Omega/dx)_{0} = -\Omega_0/R$.
\item Show that these equations have non-trivial solutions only if
\begin{displaymath}
\omega^2 = 2 \Omega_0^2 - 2 \pi G \Sigma_0 |k| + k^2 c_s^2.
\end{displaymath}
This is the dispersion relation for our rotating thin disk.
\item Solutions with $\omega^2 > 0$ correspond to oscillations, while those with $\omega^2 < 0$ correspond to pairs of modes, one of which decays with time and one of which grows. We refer to the growing modes as unstable, since in the linear regime they become arbitrarily large. Show that an unstable mode exists if $Q<1$, where
\begin{displaymath}
Q = \frac{\sqrt{2} \Omega_0 c_s}{\pi G \Sigma_0}.
\end{displaymath}
is called the Toomre parameter. Note that this stability condition refers only to axisymmetric modes in infinitely thin disks; non-axisymmetric instabilities in finite thickness disks usually appear around $Q\approx 1.5$.
\item When an unstable mode exists, we define the Toomre wave number $k_T$ as the wave number that corresponds to mode for which the instability grows fastest. Calculate $k_T$ and the corresponding Toomre wavelength, $\lambda_T = 2\pi / k_T$.
\item The Toomre mass, defined as $M_T =  \lambda_T^2 \Sigma_0$, is the characteristic mass of an unstable fragment produced by Toomre instability. Compute $M_T$, and evaluate it for $Q=1$, $\Sigma_0=12$ $\msun$ pc$^{-2}$ and $c_s = 6$ km s$^{-1}$, typical values for the atomic ISM in the solar neighborhood. Compare the mass you find to the maximum molecular cloud mass observed in the Milky Way as reported by \href{http://adsabs.harvard.edu/abs/2005PASP..117.1403R}{Rosolowsky (2005, {\it PASP}, 117, 1403)}. \nocite{rosolowsky05b}\\
\end{enumerate}

\item \textbf{The Origin of Brown Dwarfs.}\\
For the purposes of this problem, we will define a brown dwarf as any object whose mass is below $M_{\rm BD} = 0.075$ $\msun$, the hydrogen burning limit. We would like to know if these could plausibly be produced via turbulent fragmentation, as appears to be the case for stars.
\begin{enumerate}
\item For a \citet{chabrier05a} IMF (see Chapter \ref{ch:obsstars}, equation \ref{eq:chabrier}), compute the fraction $f_{\rm BD}$ of the total mass of stars produced that are brown dwarfs.
\item In order to collapse the brown dwarf must exceed the Bonnor-Ebert mass. Consider a molecular cloud of temperature 10 K. Compute the minimum ambient density $n_{\rm min}$ that a region of the cloud must have in order for the thermal pressure to be such that the Bonnor-Ebert mass is less than the brown dwarf mass.
\item Assume the cloud has a lognormal density distribution; the mean density is $\overline{n}$ and the Mach number is $\mathcal{M}$. Plot a curve in the $(\overline{n}$, $\mathcal{M})$ plane along which the fraction of the mass at densities above $n_{\rm min}$ is equal to $f_{\rm BD}$. Does the gas cloud that formed the cluster IC 348 ($\overline{n} \approx 5\times 10^4$ cm$^{-3}$, $\mathcal{M}\approx 7$) fall into the part of the plot where the mass fraction is below or above $f_{\rm BD}$?
\end{enumerate}

\end{enumerate}

\chapter{The Initial Mass Function: Theory}
\label{ch:imf_th}

\marginnote{
\textbf{Suggested background reading:}
\begin{itemize}
\item \href{http://adsabs.harvard.edu/abs/2014arXiv1402.0867K}{Krumholz, M.~R. 2014, Phys.~Rep., 539, 49}, section 6 \nocite{krumholz14c}
\end{itemize}
\textbf{Suggested literature:}
\begin{itemize}
\item \href{http://adsabs.harvard.edu/abs/2012MNRAS.423.2037H}{Hopkins, 2012, MNRAS, 423, 2037} \nocite{hopkins12d}
\item \href{http://adsabs.harvard.edu/abs/2012ApJ...754...71K}{Krumholz et al., 2012, ApJ, 754, 71} \nocite{krumholz12b}
\end{itemize}
}

The previous chapter discussed observations of the initial mass function, both how they are made and what they tell us. We now turn to theoretical attempts to explain the IMF. As with theoretical models of the star formation rate, there is at present no completely satisfactory theory for the origin of the IMF, just different ideas that do better or worse at various aspects of the problem. To recall, the things we would really like to explain most are (1) the slope of the powerlaw at high masses, and (2) the location of the peak mass. We would also like to explain the little-to-zero variation in these quantities with galactic environment. Furthermore, we would like to explain the origin of the distribution of binary properties.

\section{The Powerlaw Tail}

We begin by considering the powerlaw tail at high masses, $dn/dm \propto m^{-\Gamma}$ with $\Gamma \approx 2.3$. There are two main classes of theories for how this powerlaw tail is set: competitive accretion and turbulence. Both are scale-free processes that could plausibly produce a powerlaw distribution of masses comparable to what is observed.

\subsection{Competitive Accretion}

One hypothesis for how to produce a powerlaw mass distribution is to consider what will happen in a region where a collection of small "seed" stars form, and then begin to accrete at a rate that is a function of their current mass. Quantitatively, and for simplicity, suppose that every star accretes at a rate proportional to some power of its current mass, i.e.,
\begin{equation}
\frac{dm}{dt} \propto m^\eta.
\end{equation}
If we start with a mass $m_0$ and accretion rate $\dot{m}_0$ at time $t_0$, this ODE is easy to solve for the mass at later times. We get
\begin{equation}
m(t) = m_0 \left\{
\begin{array}{ll}
[1 - (\eta-1)\tau]^{1/(1-\eta)}, & \eta \neq 1 \\
\exp(\tau), & \eta = 1
\end{array}
\right.,
\end{equation}
where $\tau = t / (m_0/\dot{m}_0)$ is the time measured in units of the initial mass-doubling time. The case for $\eta = 1$ is the usual exponential growth, and the case for $\eta > 1$ is even faster, running away to infinite mass in a finite amount of time $\tau = 1/(\eta-1)$.

Now suppose that we start with a collection of stars that all begin at mass $m_0$, but have slightly different values of $\tau$ at which they stop growing, corresponding either to growth stopping at different physical times from one star to another, to stars stopping at the same time but having slightly different initial accretion rates $\dot{m}_0$, or some combination of both. What will the mass distribution of the resulting population be? If $dn/d\tau$ is the distribution of stopping times, then we will have
\begin{equation}
\frac{dn}{dm} \propto \frac{dn/d\tau}{dm/d\tau} \propto m(\tau)^{-\eta} \frac{dn}{d\tau}.
\end{equation}
Thus the final distribution of masses will be a powerlaw in mass, with index $-\eta$, going from $m(\tau_{\rm min})$ to $m(\tau_{\rm max})$; a powerlaw distribution naturally results.

The index of this powerlaw will depend on the index of the accretion law, $\eta$. What should this be? In the case of a point mass accreting from a uniform, infinite medium at rest, the accretion rate onto a point mass was worked out by \citet{hoyle46a} and \citet{bondi52a}, and the problem is known as Bondi-Hoyle accretion. The accretion rate scales as $\dot{m} \propto m^2$, so if this process describes how stars form, then the expected mass distribution should follow $dn/dm\propto m^{-2}$, not so far from the actual slope of $-2.3$ that we observe. A number of authors have argued that this difference can be made up by considering the effects of a crowded environment, where the feeding regions of smaller stars get tidally truncated, and thus the growth law winds up begin somewhat steeper than $\dot{m}\propto m^2$.

\begin{marginfigure}
\includegraphics[width=\linewidth]{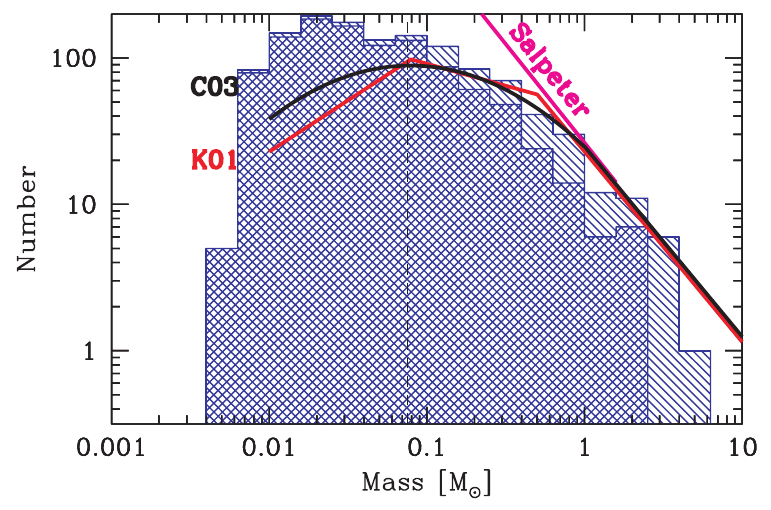}
\caption[IMF in a competitive accretion simulation]{
\label{fig:imf_bate09}
The IMF measured in a simulation of the collapse of a 500 $M_\odot$ initially uniform density cloud. The single-hatched histogram shows all objects in the simulation, while the double-hatched one shows objects that have stopped accreting. Credit: \citeauthor{bate09b}, 2009, MNRAS, 392, 590, reproduced by permission of Oxford University Press on behalf of the RAS.
}
\end{marginfigure}
This is an extremely simple model, requiring no physics but hydrodynamics and gravity, and thus it is easy to simulate. Simulations done based on this model do sometimes return a mass distribution that looks much like the IMF, as illustrated in Figure \ref{fig:imf_bate09}. However, this appears to depend on the choice of initial conditions. Generally speaking, one gets about the right IMF if one stars with something with a viral ratio $\alpha_{\rm vir} \sim 1$ and no initial density structure, just velocities. Simulations that start with either supervirial or sub-virial initial conditions, or that begin with turbulent density structures, do not appear to grow as predicted by competitive accretion (e.g., \citealt{clark08a}).

Another potential problem with this model is that it only seems to work in environments where there is no substantial feedback to drive the turbulence or eject the gas. In simulations where this is not true, there appears to be no competitive accretion. The key issue is that competitive accretion seems to require a global collapse where all the stars fall together into a region where they can compete, and this is hard to accomplish in the presence of feedback.

Yet a third potential issue is that this model has trouble making sense of the IMF peak, as we will discuss in Section \ref{sec:imfpeak}.

\subsection{Turbulent Fragmentation}

A second class of models for the origin of the powerlaw slope is based on the physics of turbulence. The first of these models was proposed by \citet{padoan97a}, and there have been numerous refinements since \citep[e.g.,][]{padoan02a, padoan07a, hennebelle08b, hennebelle09a, hopkins12e, hopkins12d}. The basic assumption in the turbulence models is that the process of shocks repeatedly passing through an isothermal medium leads to a broad range of density distributions, and that stars form wherever a local region happens to be pushed to the point where it becomes self-gravitating. We then proceed as follows. Suppose we consider the density field smoothed on some size scale $\ell$. The mass of an object of density $\rho$ in this smoothed field is
\begin{equation}
m \sim \rho \ell^3,
\end{equation}
and the total mass of objects with characteristic density between $\rho$ and $\rho+d\rho$ is
\begin{equation}
dM_{\rm tot} \sim \rho p(\rho) \,d\rho,
\end{equation}
where $p(\rho)$ is the density PDF. Then the total number of objects in the mass range from $m$ to $m+dm$ on size scale $\ell$ can be obtained just by dividing the total mass of objects at a given density by the mass per object, and integrating over the density PDF on that size scale
\begin{equation}
\frac{dn_\ell}{dm} = \frac{dM_{\rm tot}}{m} \sim  \ell^{-3} \int p(\rho) \,d\rho.
\end{equation}

Not all of these structures will be bound. To filter out the ones that are, we impose a density threshold, analogous to the one we used in computing the star formation rate in Section \ref{ssec:eff_th}.\footnote{Indeed, several of the models discussed there allow simultaneous computation of the star formation rate and the IMF.} We assert that an object will be bound only if its gravitational energy exceeds its kinetic energy, that is, only if the density exceeds a critical value given by
\begin{equation}
\frac{Gm^2}{\ell} \sim m \sigma(\ell)^2
\qquad\Longrightarrow\qquad
\rho_{\rm crit} \sim \frac{\sigma(\ell)^2}{G \ell^2},
\end{equation}
where $\sigma(\ell)$ is the velocity dispersion on size scale $\ell$, which we take from the linewidth-size relation, $\sigma(\ell) = c_s (\ell/\ell_s)^{1/2}$. Thus we have a critical density
\begin{equation}
\rho_{\rm crit} \sim \frac{c_s^2}{G \ell_s \ell},
\end{equation}
and this forms a lower limit on the integral.

There are two more steps in the argument. One is simple: just integrate over all length scales to get the total number of objects. That is,
\begin{equation}
\frac{dn}{dm} \propto \int \frac{dn_\ell}{dm} \, d\ell.
\end{equation}
The second is that we must know the functional form $p(\rho)$ for the smoothed density PDF. One can estimate this in a variety of ways, but to date no one has performed a fully rigorous calculation. For example, \citet{hopkins12e} assumes that the PDF is lognormal no matter what scale it is smoothed on, and all that changes as one alters the smoothing scale is the dispersion. He obtains this by setting the dispersion on some scale $\sigma_\ell$ equal to an integral over the dispersions on all smaller scales. In contrast, \citet{hennebelle08b, hennebelle09a} assume that the density power spectrum is a powerlaw, and derive the density PDF from that. These two assumptions yield similar but not identical results.

\begin{marginfigure}
\includegraphics[width=\linewidth]{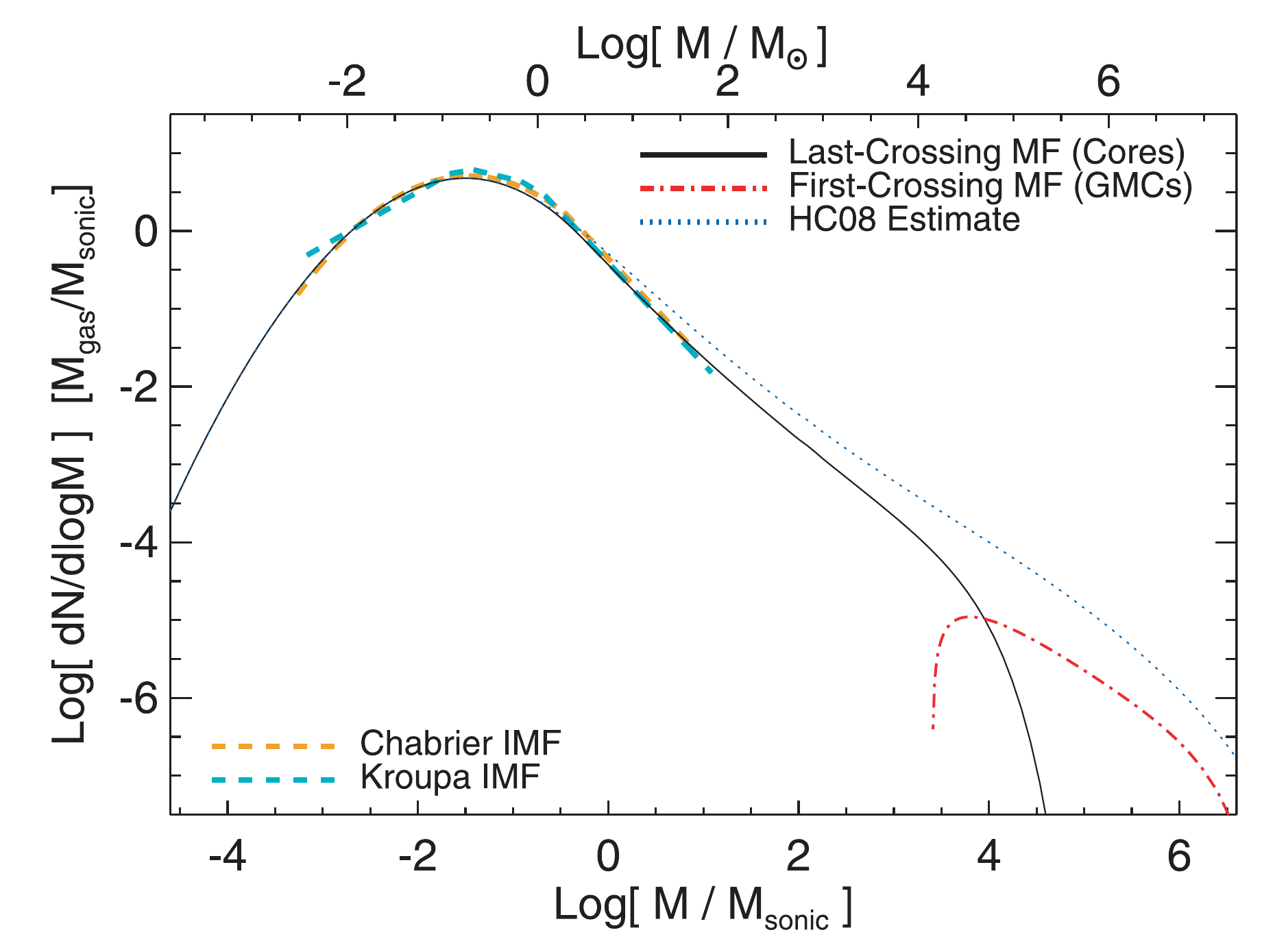}
\caption[IMF from an analytic model of turbulent fragmentation]{
\label{fig:imf_hopkins12}
The IMF predicted by an analytic model of turbulent fragmentation. Credit: \citeauthor{hopkins12d}, MNRAS, 423, 2037, reproduced by permission of Oxford University Press on behalf of the RAS.
}
\end{marginfigure}
At this point we will cease following the mathematical formalism in detail; interested readers can find it worked out in the papers referenced above. We will simply assert that one can at this point evaluate all the integrals to get an IMF. The result clearly depends only on two dimensional quantities: the sound speed $c_s$ and the sonic length $\ell_s$. However, at masses much greater than the sonic mass $M_s \approx c_s^2 \ell_s/G$, the result is close to a powerlaw with approximately the right index. Figure \ref{fig:imf_hopkins12} shows an example prediction.

As with the competitive accretion model, this hypothesis encounters certain difficulties. First, there is the technical problem that the choice of smoothed density PDF estimate is not at all rigorous, and there are noticeable differences in the result depending on how the choice is made. Second, the dependence on the sonic length is potentially problematic, because real molecular clouds do not have constant sonic lengths. Regions of massive star formation are observed to be systematically more turbulent. Third, the theory does not address the question of why gravitationally-bound regions do not sub-fragment as they collapse. Indeed, \citet{guszejnov15a} and \citet{guszejnov16a} argue that, when this effect is taken into account, the IMF (as opposed to the core mass distribution) becomes a pure powerlaw. As a result, the model has trouble explaining the IMF peak.

\section{The Peak of the IMF}
\label{sec:imfpeak}

\subsection{Basic Theoretical Considerations}

A powerlaw is scale-free, but the peak has a definite mass scale. This mass scale is one basic observable that any theory of star formation must be able to predict. Moreover, the presence of a characteristic mass scale immediately tells us something about the physical processes that must be involved in producing the IMF. We have thus far thought of molecular clouds as consisting mostly of isothermal, turbulent, magnetized, self-gravitating gas. However, we can show that there \textit{must} be additional processes beyond these at work in setting a peak mass.

We can see this in a few ways. First we can demonstrate it in a more intuitive but not rigorous manner, and then we can demonstrate it rigorously. The intuitive argument is as follows. In the system we have described, there are four energies in the problem: thermal energy, bulk kinetic energy, magnetic energy, and gravitational potential energy. From these energies we can define three dimensionless ratios, and the behavior of the system will be determined by these three ratios. As an example, we might define the Mach number, plasma beta, and Jeans number via
\begin{equation}
\mathcal{M} = \frac{\sigma}{c_s} \qquad \beta = \frac{8\pi \rho c_s^2}{B^2}
\qquad n_J = \frac{\rho L^2}{c_s^3/\sqrt{G^3\rho}}.
\end{equation}
The ratios describe the ratio of kinetic to thermal energy, the ratio of thermal to magnetic energy, and the ratio of thermal to gravitational energy. All other dimensionless numbers we normally use can be derived from these, e.g., the Alfv\'enic Mach number $\mathcal{M}_A = \mathcal{M}\sqrt{\beta/2}$ is simply the ratio of kinetic to magnetic energy.

Now notice the scalings of these numbers with density $\rho$, velocity dispersion $\sigma$, magnetic field strength $B$, and length scale $L$:
\begin{equation}
\mathcal{M} \propto \sigma
\qquad
\beta \propto \rho B^{-2}
\qquad
n_J \propto \rho^{3/2} L^3.
\end{equation}
If we scale the problem by $\rho\rightarrow x \rho$, $L\rightarrow x^{-1/2} L$, $B\rightarrow x^{1/2} B$, all of these dimensionless numbers remain fixed. Thus the behavior of two systems, one with density a factor of $x$ times larger than the other one, length a factor of $x^{-1/2}$ smaller, and magnetic field a factor of $x^{1/2}$ stronger, are simply rescaled versions of one another. If the first system fragments to make a star out of a certain part of its gas, the second system will too. Notice, however, that the {\it masses} of those stars will not be the same! The first star will have a mass that scales as $\rho L^3$, while the second will have a mass that scales as $(x\rho) (x^{-1/2} L)^3 = x^{-1/2} \rho L^3$. We learn from this an important lesson: isothermal gas is scale-free. If we have a model involving only isothermal gas with turbulence, gravity, and magnetic fields, and this model produces stars of a given mass $m_*$, then we can rescale the system to obtain an arbitrarily different mass.

Now that we understand the basic idea, we can show this a bit more formally. Consider the equations describing this system. For simplicity we will omit both viscosity and resistivity. These are
\begin{eqnarray}
\frac{\partial \rho}{\partial t} & = & -\nabla \cdot (\rho \vecv) \\
\frac{\partial}{\partial t}(\rho \vecv) & = & -\nabla \cdot (\rho\vecv\vecv) - c_s^2 \nabla \rho
\nonumber \\
& & {} + \frac{1}{4\pi} (\nabla \times \vecB) \times \vecB - \rho \nabla \phi \\
\frac{\partial\vecB}{\partial t} & = & -\nabla \times (\vecB\times\vecv) \\
\nabla^2 \phi & = & 4\pi G \rho
\end{eqnarray}
One can non-dimensionalize these equations by choosing a characteristic length scale $L$, velocity scale $V$, density scale $\rho_0$, and magnetic field scale $B_0$, and making a change of variables $\mathbf{x} = \mathbf{x}'L$, $t = t' L/V$, $\rho = r \rho_0$, $\mathbf{B} = \mathbf{b} B_0$, $\mathbf{v} = \mathbf{u} V$, and $\phi = \psi G\rho_0 L^2$. With fairly minimal algebra, the equations then reduce to
\begin{eqnarray}
\frac{\partial r}{\partial t'} & = & -\nabla'\cdot (r\mathbf{u}) \\
\frac{\partial}{\partial t'}(r \mathbf{u}) & = & -\nabla' \cdot \left(r\mathbf{uu}\right) - \frac{1}{\mathcal{M}^2}\nabla' r \nonumber \\
& & {} + \frac{1}{\mathcal{M}_A^2} (\nabla'\times\mathbf{b})\times\mathbf{b} - \frac{1}{\alpha_{\rm vir}} \nabla' \psi \\
\frac{\partial\mathbf{b}}{\partial t'} & = & -\nabla'\times\left(\mathbf{b}\times\mathbf{u}\right) \\
\nabla'^2 \psi & = & 4\pi r,
\end{eqnarray}
where $\nabla'$ indicates differentiation with respect to $x'$. The dimensionless ratios appearing in these equations are
\begin{eqnarray}
\mathcal{M} & = & \frac{V}{c_s} \\
\mathcal{M}_A & = & \frac{V}{V_A} = V\frac{\sqrt{4\pi \rho_0}}{B_0} \\
\alpha_{\rm vir} & = & \frac{V^2}{G \rho_0 L^2},
\end{eqnarray}
which are simply the Mach number, Alfv\'{e}n Mach number, and virial ratio for the system. These equations are fully identical to the original ones, so any solution to them is also a valid solution to the original equations. In particular, suppose we have a system of size scale $L$, density scale $\rho_0$, magnetic field scale $B_0$, velocity scale $V$, and sound speed $c_s$, and that the evolution of this system leads to a star-like object with a mass
\begin{equation}
m \propto \rho_0 L^3.
\end{equation}
One can immediately see that system with length scale $L' = y L$, density scale $\rho_0' = \rho_0 / y^2$, magnetic field scale $B_0' = B_0/y$, and velocity scale $V' = V$ has exactly the same values of $\mathcal{M}$, $\mathcal{M}_A$, and $\alpha_{\rm vir}$ as the original system, and therefore has exactly the same evolution. However, in this case the star-like object will instead have a mass
\begin{equation}
m' \propto \rho' L'^3 = y m
\end{equation}
Thus we can make objects of arbitrary mass just by rescaling the system.

This analysis implies that explaining the IMF peak requires appealing to some physics beyond that of isothermal, magnetized turbulence plus self-gravity. This immediately shows that the competitive accretion and turbulence theories we outlined to explain the powerlaw tail of the IMF cannot be adequate to explaining the IMF peak, at least not by themselves. Something must be added, and models for the origin of the IMF peak can be broadly classified based on what extra physics they choose to add.

\subsection{The IMF From Galactic Properties}

One option is hypothesize that the IMF is set at the outer scale of the turbulence, where the molecular clouds join to the atomic ISM (in a galaxy like the Milky Way), or on sizes of the galactic scale-height (for a molecule-dominated galaxy). Something in this outer scale picks out the characteristic mass of stars at the IMF peak.

This hypothesis comes in two flavors. The simplest is that characteristic mass is simply set by the Jeans mass at the mean density $\overline{\rho}$ of the cloud, so that
\begin{equation}
m_{\rm peak} \propto \frac{c_s^3}{\sqrt{G^3 \overline{\rho}}}
\end{equation}
While simple, this hypothesis immediately encounters problems. Molecular clouds have about the same temperature everywhere, but they do not all have the same density -- indeed, based on our result that the surface density is about constant, the density should vary with cloud mass as $M^{1/2}$. Thus at face value this hypothesis would seem to predict a factor of $\sim 3$ difference in characteristic peak mass between $10^4$ and $10^6$ $M_\odot$ clouds in the Milky Way. This is pretty hard to reconcile with observations. The problem is even worse if we think about other galaxies, where the range of density variation is much greater and thus the predicted IMF variation is too. One can hope for a convenient cancellation, whereby an increase in the density is balanced by an increase in temperature, but this seems to require a coincidence.

A somewhat more refined hypothesis, which is adopted by all the turbulence models, is that the IMF peak is set by the sound speed and the normalization of the linewidth-size relation. As discussed above, in the turbulence models the only dimensional free parameters are $c_s$ and $\ell_s$, and from them one can derive a mass in only one way:
\begin{equation}
m_{\rm peak} \sim \frac{c_s^2 \ell_s}{G}.
\end{equation}
\citet{hopkins12d} calls this quantity the sonic mass, but it's the same thing as the characteristic masses in the other models.

This value can be expressed in a few ways. Suppose that we have a cloud of characteristic mass $M$ and radius $R$. We can write the velocity dispersion in terms of the virial parameter:
\begin{equation}
\alpha_{\rm vir} \sim \frac{\sigma^2 R}{G M}
\qquad\Longrightarrow\qquad
\sigma \sim \sqrt{\alpha_{\rm vir} \frac{G M}{R}}.
\end{equation}
This is the velocity dispersion on the outer scale of the cloud, so we can also define the Mach number on this scale as
\begin{equation}
\mathcal{M} = \frac{\sigma}{c_s} \sim \sqrt{\alpha_{\rm vir} \frac{G M}{R c_s^2}}
\end{equation}
The sonic length is just the length scale at which $\mathcal{M} \sim 1$, so if the velocity dispersion scales with $\ell^{1/2}$, then we have
\begin{equation}
\ell_s \sim \frac{R}{\mathcal{M}^2} \sim \frac{c_s^2}{\alpha_{\rm vir} G \Sigma},
\end{equation}
where $\Sigma\sim M/R^2$ is the surface density. Substituting this in, we have
\begin{equation}
m_{\rm peak} \sim \frac{c_s^4}{\alpha_{\rm vir} G^2 \Sigma},
\end{equation}
and thus the peak mass simply depends on the surface density of the cloud. We can obtain another equivalent expression by noticing that
\begin{equation}
\frac{M_J}{\mathcal{M}} \sim \frac{c_s^3}{\sqrt{G^3 \overline{\rho}}} \sqrt{\frac{R c_s^2}{\alpha_{\rm vir} G M}}
\sim \frac{c_s^4}{\alpha_{\rm vir} G^2 \Sigma} \sim m_{\rm peak}
\end{equation}
Thus, up to a factor of order unity, this hypothesis is also equivalent to assuming that the characteristic mass is simply the Jeans mass divided by the Mach number.

An appealing aspect of this argument is that it naturally explains why molecular clouds in the Milky Way all make stars at about the same mass. A less appealing result is that it would seem to predict that the masses could be quite different in regions of different surface density, and we observe that there are star-forming regions where $\Sigma$ is indeed much higher than the mean of the Milky Way GMCs. This is doubly-true if we extend our range to extragalactic environments. One can hope that this will cancel because the temperature will be higher and thus $c_s$ will increase, but this again seems to depend on a lucky cancellation, and there is no \textit{a priori} reason why it should.

\subsection{Non-Isothermal Fragmentation}

The alternative to breaking the isothermality at the outer scale of the turbulence is to relax the assumption that the gas is isothermal on small scales. This has the advantage that it avoids any ambiguity about what constitutes the surface density or linewidth-size relation normalization for a "cloud".

\paragraph{Fixed equation of state models.}

\begin{marginfigure}
\includegraphics[width=\linewidth]{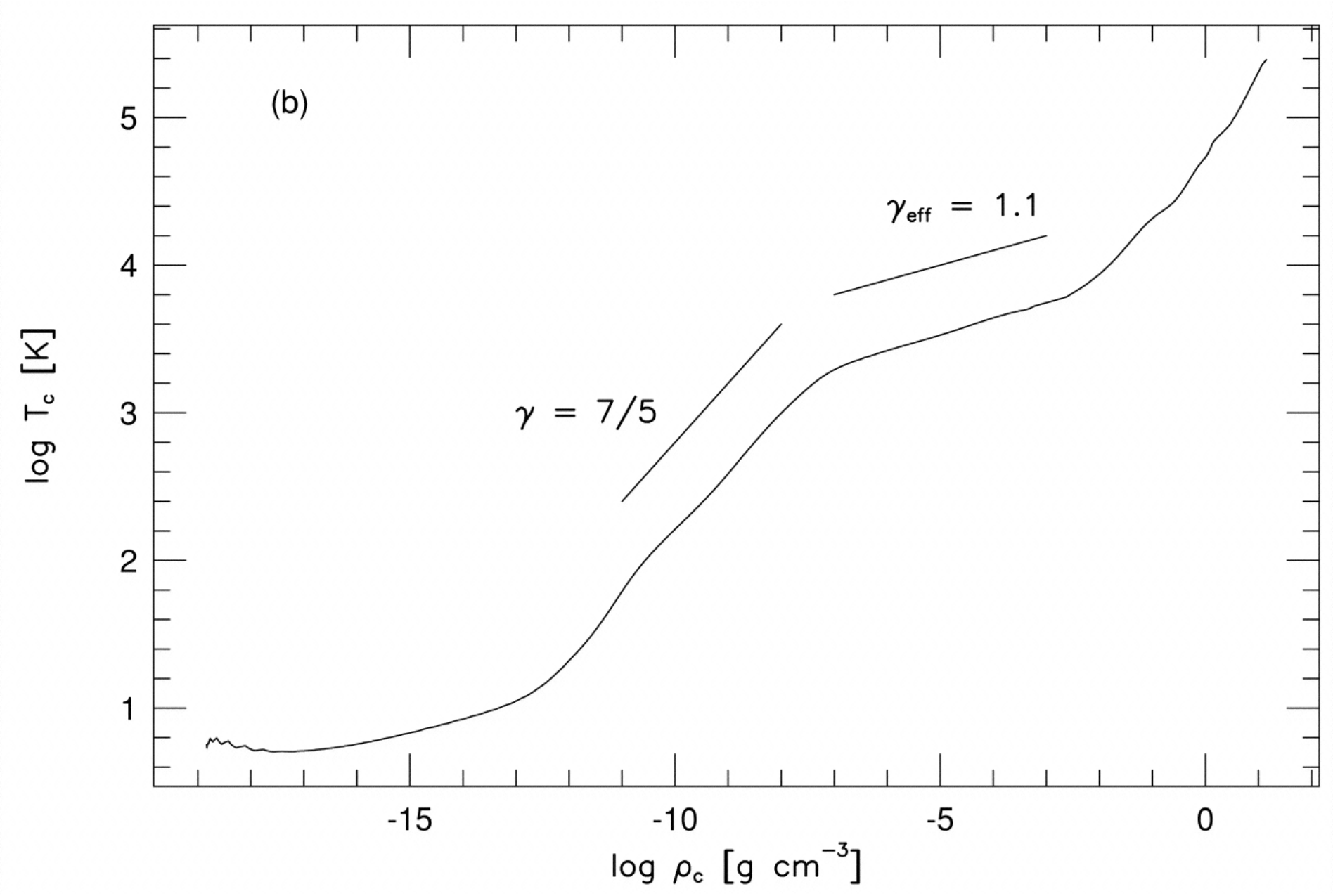}
\caption[Temperature versus density in a collapsing core]{
\label{fig:rhot_masunaga00}
Temperature versus density found in a one-dimensional calculation of the collapse of a $1$ $M_\odot$ gas cloud, at the moment immediately before a central protostar forms. Credit: \citet{masunaga00a}, \copyright AAS. Reproduced with permission.
}
\end{marginfigure}
The earliest versions of these models were proposed by \citet{larson05a}, and followed up by \citet{jappsen05a}. The basic idea of these models is that the gas in star-forming clouds is only approximately isothermal. Instead, there are small deviations from isothermality, which can pick out preferred mass scales. We will discuss these in more detail in Chapters \ref{ch:first_stars} and \ref{ch:protostar_form}, but for now we assert that there are two places where significant deviations from isothermality are expected (Figure \ref{fig:rhot_masunaga00}).

At low density the main heating source is cosmic rays and UV photons, both of which produce a constant heating rate per nucleus if attenuation is not significant. This is because the flux of CRs and UV photons is about constant, and the rate of energy deposition is just proportional to the number of target atoms or dust grains for them to interact with. Cooling is primarily by lines, either of CO once the gas is mostly molecular, or of C$^+$ or O where it is significantly atomic.

In both cases, at low density the gas is slightly below the critical density of the line, so the cooling rate per nucleus or per molecule is an increasing function of density. Since heating per nucleus is constant but cooling per nucleus increases, the equilibrium temperature decreases with density. As one goes to higher density and passes the CO critical density this effect ceases. At that point one generally starts to reach densities such that shielding against UV photons is significant, so the heating rate goes down and thus the temperature continues to drop with density.

This begins to change at a density of around $10^{-18}$ g cm$^{-3}$, $n\sim 10^5 - 10^6$ cm$^{-3}$. By this point the gas and dust have been thermally well-coupled by collisions, and the molecular lines are extremely optically thick, so dust is the main thermostat. As long as the gas is optically thin to thermal dust emission, which it is at these densities, the dust cooling rate per molecule is fixed, since the cooling rate just depends on the number of dust grains. Heating at these densities comes primarily from compression as the gas collapses, i.e., it is just $P\, dV$ work. If the compression were at a constant rate, the heating rate per molecule would be constant. However, the free-fall time decreases with density, so the collapse rate and thus the heating rate per molecule increase with density. The combination of fixed cooling rate and increasing heating rate causes the temperature to begin rising with density. At still higher densities, $\sim 10^{-13}$ g cm$^{-3}$, the gas becomes optically thick to dust thermal emission. At this point the gas simply acts adiabatically, with all the $P\,dV$ work being retained, so the heating rate with density rises again.

\citet{larson05a} pointed out that deviations from isothermality are particularly significant for filamentary structures, which dominate in turbulent flows.  It is possible to show that a filament cannot go into runaway collapse if $T$ varies with $\rho$ to a positive number, while it can collapse if $T$ varies as $\rho$ to a negative number. This suggests that filaments will collapse indefinitely in the low-density regime, but that their collapse will then halt around $10^{-18}$ g cm$^{-3}$, forcing them to break up into spheres in order to collapse further. The upshot of all these arguments is that the Jeans or Bonnor-Ebert mass one should be using to estimate the peak of the stellar mass spectrum is the one corresponding to the point where there is a changeover from sub-isothermal to super-isothermal.

In other words, the $\rho$ and $T$ that should be used to evaluate $M_J$ or $M_{\rm BE}$ are the values at that transition point. Larson proposes an approximate equation of state to represent the first break in the EOS:
Combining all these effects, \citet{larson05a} proposed a single simple equation of state
\begin{equation}
T = \left\{
\begin{array}{ll}
4.4 \,\rho_{18}^{-0.27}\mbox{ K}, \qquad & \rho_{18} < 1 \\
4.4 \,\rho_{18}^{0.07}\mbox{ K}, & \rho_{18} \ge 1
\end{array}
\right.
\end{equation}
where $\rho_{18}=\rho/(10^{-18}\mbox{ g cm}^{-3})$. Conveniently enough, the Bonnor-Ebert mass at the minimum temperature here is $M_{\rm BE} = 0.067$ $\msun$, which is not too far off from the observed peak of the IMF at $M=0.2$ $\msun$. (The mass at the second break is a bit less promising. At $\rho = 10^{-13}$ g cm$^{-3}$ and $T=10$ K, we have $M_{\rm BE} = 7\times 10^{-4}$ $\msun$.)

\begin{figure}
\includegraphics[width=\linewidth]{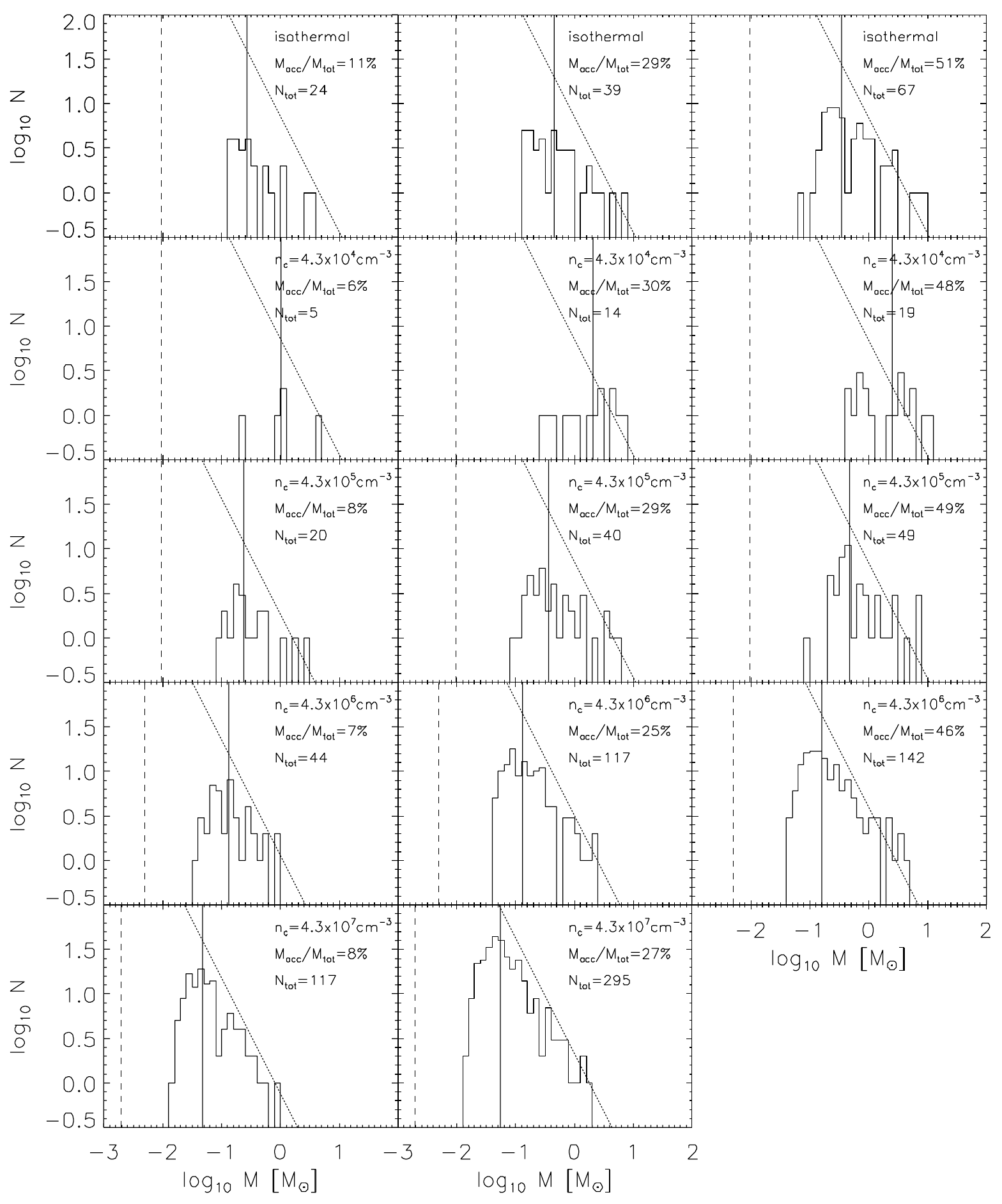}
\caption[IMF from simulations of non-isothermal fragmentation]{
\label{fig:imf_jappsen05}
Measured stellar mass distributions in a series of simulations of turbulent fragmentation using non-isothermal equations of state. Each row shows a single simulation, measured at a series of times, characterized by a particular mass in stars as indicated in each panel. Different rows use different equations of state, with the vertical line in each panel indicating the Jeans mass evaluated at the temperature minimum of the equation of state. Histograms show the mass distributions measured for the stars. Credit: \citeauthor{jappsen05a}, A\&A, 435, 611, 2005, reproduced with permission \copyright\, ESO.
}
\end{figure}

Simulations done adopting this proposed equation of state seem to verify the conjecture that the characteristic fragment mass does depend critically on the break on the EOS (Figure \ref{fig:imf_jappsen05}).

\paragraph{Radiative models.}

While this is a very interesting result, there are two problems. First, the proposed break in the EOS occurs at $n=4\times 10^5$ cm$^{-3}$. This is a fairly high density in a low mass star-forming region, but it is actually quite a low density in more typical, massive star-forming regions. For example, the Orion Nebula cluster now consists of $\approx 2000$ $\msun$ of stars in a radius of $0.8$ pc, giving a mean density $n\approx 2\times 10^4$ cm$^{-3}$. Since the star formation efficiency was less than unity and the cluster is probably expanding due to mass loss, the mean density was almost certainly higher while the stars were still forming. Moreover, recall that, in a turbulent medium, the bulk of the mass is at densities above the volumetric mean density. The upshot of all this is that almost all the gas in Orion was probably over \citet{larson05a}'s break density while the stars were forming. Since Orion managed to form a normal IMF, it is not clear how the break temperature could be relevant.

\begin{figure}
\includegraphics[width=\linewidth]{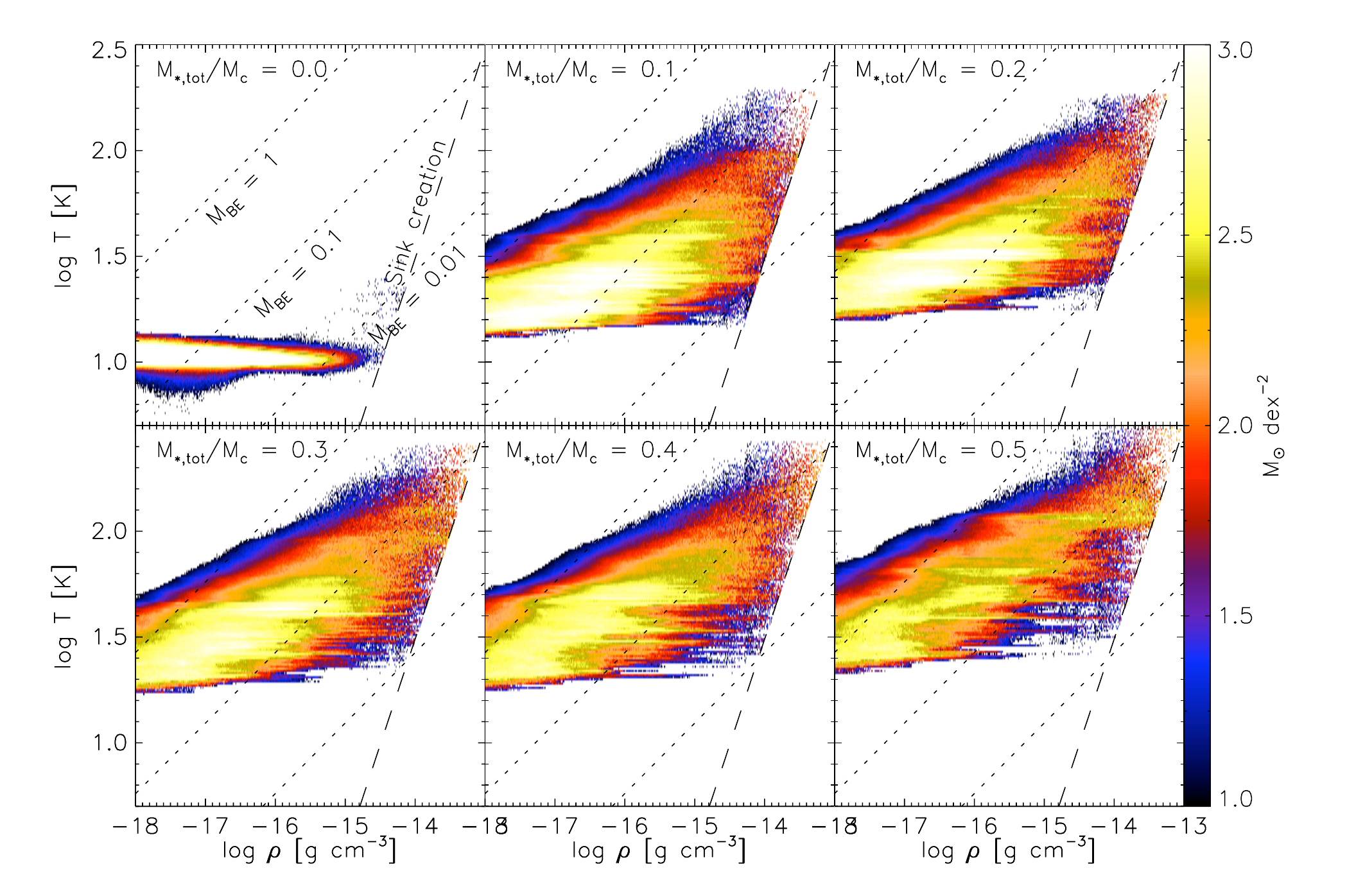}
\caption[Density-temperature distribution from a simulation of the formation of the ONC]{
\label{fig:rhot_krumholz11}
Density-temperature distributions measured from a simulation of the formation of an ONC-like star cluster, including radiative transfer and stellar feedback. The panels show the distribution at different times in the simulation, characterized by the fraction of mass that has been turned into stars. Doted lines show lines of constant Bonnor-Ebert mass (in $M_\odot$), while dashed lines show the threshold for sink particle formation in the simulation. Credit: \citet{krumholz11c}, \copyright AAS. Reproduced with permission.
}
\end{figure}

A second problem is that, in dense regions like the ONC, the simple model proposed by \citet{larson05a} is a very bad representation of the true temperature structure, because it ignores the effects of radiative feedback from stars. In dense regions the stars that form will heat the gas around them, raising the temperature. Figure \ref{fig:rhot_krumholz11} shows the density-temperature distribution of gas in simulations that include radiative transfer, and that have conditions chosen to be similar to those of the ONC.

These two observations suggest that one can build a model for the IMF around radiative feedback. There are a few numerical and analytic papers that attempt to do so, including \citet{bate09a, bate12a}, \citet{krumholz11e}, \citet{krumholz12b}, and \citet{guszejnov16a}. The central idea for these models is that radiative feedback shuts off fragmentation at a characteristic mass scale that sets the peak of the IMF.

Suppose that we form a first, small protostellar that radiates at a rate $L$. The temperature of the material at a distance $R$ from it, assuming the gas is optically thick, will be roughly
\begin{equation}
L \approx 4 \pi \sigma_{\rm SB} R^2 T^4,
\end{equation}
where $\sigma_{\rm SB}$ is the Stefan-Boltzmann constant. Now let us compute the Bonnor-Ebert mass using the temperature $T$:
\begin{equation}
M_{\rm BE} \approx \frac{c_s^3}{\sqrt{G^3\rho}} = \sqrt{\left(\frac{k_B T}{\mu m_{\rm H} G}\right)^3 \frac{1}{\rho}},
\end{equation}
where $\mu=2.33$ is the mean particle mass, and we are omitting the factor of $1.18$ for simplicity. Note that $M_{\rm BE}$ here is a function of $R$. At small $R$, the temperature is large and thus $M_{\rm BE}$ is large, while at larger distances the gas is cooler and $M_{\rm BE}$ falls.

Now let us compare this mass to the mass enclosed within the radius $R$, which is $M=(4/3)\pi R^3 \rho$. At small radii, $M_{\rm BE}$ greatly exceeds the enclosed mass, while at large radii $M_{\rm BE}$ is much less than the enclosed mass. A reasonable hypothesis is that fragmentation will be suppressed out to the point where $M \approx M_{\rm BE}$. If we solve for the radius $R$ and mass $M$ at which this condition is met, we obtain
\begin{equation}
M  \approx \left(\frac{1}{36\pi}\right)^{1/10} \left(\frac{k_B}{G \mu m_{\rm H}}\right)^{6/5} \left(\frac{L}{\sigma_{\rm SB}}\right)^{3/10} \rho^{-1/5}.
\end{equation}

To go further, we need to know the luminosity $L$. The good news is that, for reasons to be discussed in Chapter \ref{ch:protostar_evol}, the luminosity is dominated by accretion, and the energy produced by accretion is simply the accretion rate multiplied by a roughly fixed energy yield per unit mass. In other words, we can write
\begin{equation}
L \approx \psi \dot{M},
\end{equation}
where $\psi \approx 10^{14}$ erg g$^{-1}$, and can in fact be written in terms of fundamental constants. Taking this on faith for now, if we further assume that stars form over a time of order a free-fall time, then
\begin{equation}
\dot{M} \approx M \sqrt{G\rho},
\end{equation}
and substituting this into the equation for $M$ above and solving gives
\begin{eqnarray}
M & \approx & \left(\frac{1}{36\pi}\right)^{1/7} \left(\frac{k_B}{G \mu m_{\rm H}}\right)^{12/7} \left(\frac{\psi}{\sigma_{\rm SB}}\right)^{3/7} \rho^{-1/14} \\
& = & 0.3 \left(\frac{n}{100\mbox{ cm}^{-3}}\right)^{-1/14} M_\odot,
\end{eqnarray}
where $n = \rho/(\mu m_{\rm H})$. Thus we get a characteristic mass that is a good match to the IMF peak, and that depends only very, very weaky on the ambient density.

Simulations including radiation seem to support the idea that this effect can pick out a characteristic peak ISM mass. The main downside to this hypothesis is that it has little to say by itself about the powerlaw tail of the IMF. This is not so much a problem with the model as an omission, and a promising area of research seems to be joining a non-isothermal model such as this onto a turbulent fragmentation or competitive accretion model to explain the full IMF.

\chapter{Protostellar Disks and Outflows: Observations}
\label{ch:disks_obs}

\marginnote{
\textbf{Suggested background reading:}
\begin{itemize}
\item \href{http://adsabs.harvard.edu/abs/2014prpl.conf..173L}{Li, Z.-Y., et al. 2014, in "Protostars and Planets VI", ed.~H.~Beuther et al., pp.~173-194}, sections 1-2 \nocite{li14a}
\end{itemize}
\textbf{Suggested literature:}
\begin{itemize}
\item \href{http://adsabs.harvard.edu/abs/2012Natur.492...83T}{Tobin et al., 2012, Nature, 492, 83} \nocite{tobin12a}
\end{itemize}
}

We now zoom in even further on the star formation process, and examine the dominant circumstellar structures found around young stars: accretion disks. We will spend two chapters on this subject. In the first we will discuss the observational phenomenology of disks, including the outflows they generate. There are a wide range of observational techniques for studying the properties of disks around young stars, and we will certainly not exhaust the list here. We will focus on a few of the most widely used methods, and develop an understanding of how they work and what we can learn from them.

\section{Observing Disks}

\subsection{Dust at Optical Wavelengths}

\begin{marginfigure}
\includegraphics[width=\linewidth]{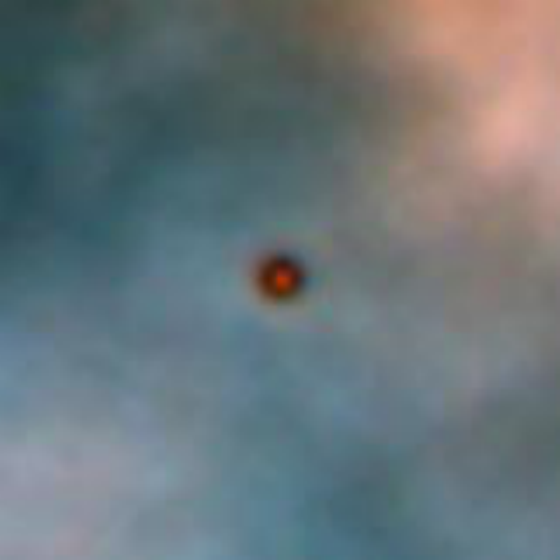}
\includegraphics[width=\linewidth]{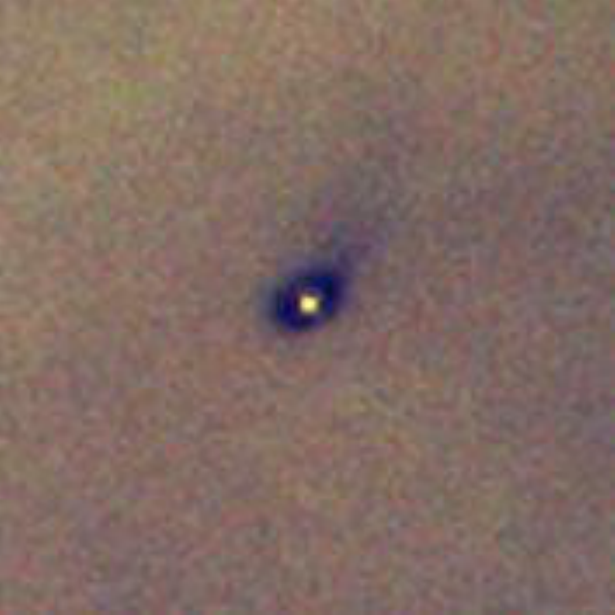}
\caption[Protostellar disks in absorption in the ONC]{
\label{fig:proplyds}
Two disks in the Orion Nebula seen in absorption against the nebula using the \textit{Hubble Space Telescope}. Taken from \url{http://hubblesite.org/newscenter/archive/releases/1995/45/image/g/}.
}
\end{marginfigure}

The first idea that might occur to an astronomer who wants to study disks would be to work in the optical. The main challenge to that is that for the most part disks do not emit optical light, because they are too cool. This leaves only a couple of options in the optical. One is that we can detect the disk in scattered starlight. This is very hard, because the light is very faint, and the geometry has to be just right. Polarization can help in this case, since the scattered light will be polarized. There are a few examples of this.

The other possibility for the optical is to work in absorption. This requires a bright, extended background source against which the disk can be detected in silhouette. Fortunately, massive young stars produce H~\textsc{ii} regions, which are bright diffuse sources, and can provide a nice backlight for absorption work. The most spectacular examples of this technique are in the Orion Nebula, as illustrated in Figure \ref{fig:proplyds}.

In this case, since we are working in optical, we get excellent spatial resolution. The disks we see in this case are typically hundreds of AU in size. In such images we can also see very clearly that protostellar jets are launched perpendicular to the disks, confirming the central role of disks in producing them, as we will discuss in Chapter \ref{ch:disks_theory}.

While the optical offers spectacular pictures, its restriction to the cases where we have favorable geometry, a nice backlight, or some combination of the two limits its usefulness as a general tool for studying disks. A further complication is that optical only lets us study disks once their parent cores, which are opaque at optical wavelengths, have dissipated. This limits optical techniques to studying the later stages of disk evolution.

\subsection{Dust Emission in the Infrared and Sub-mm}

A much more broadly used technique is to detect the dust in a disk in the infrared or sub-mm. As discussed in Chapter \ref{ch:obsstars}, young stars often show significantly more IR and sub-mm emission that would be expected from a bare stellar photosphere. The natural candidate for producing this emission is warm dust grains near the star. The fact that we see the stellar photosphere at all, and that it is not hugely reddened, implies that the grains cannot be in any sort of shell or spherical distribution. A disk is the natural candidate geometry.

\begin{marginfigure}
\includegraphics[width=\linewidth]{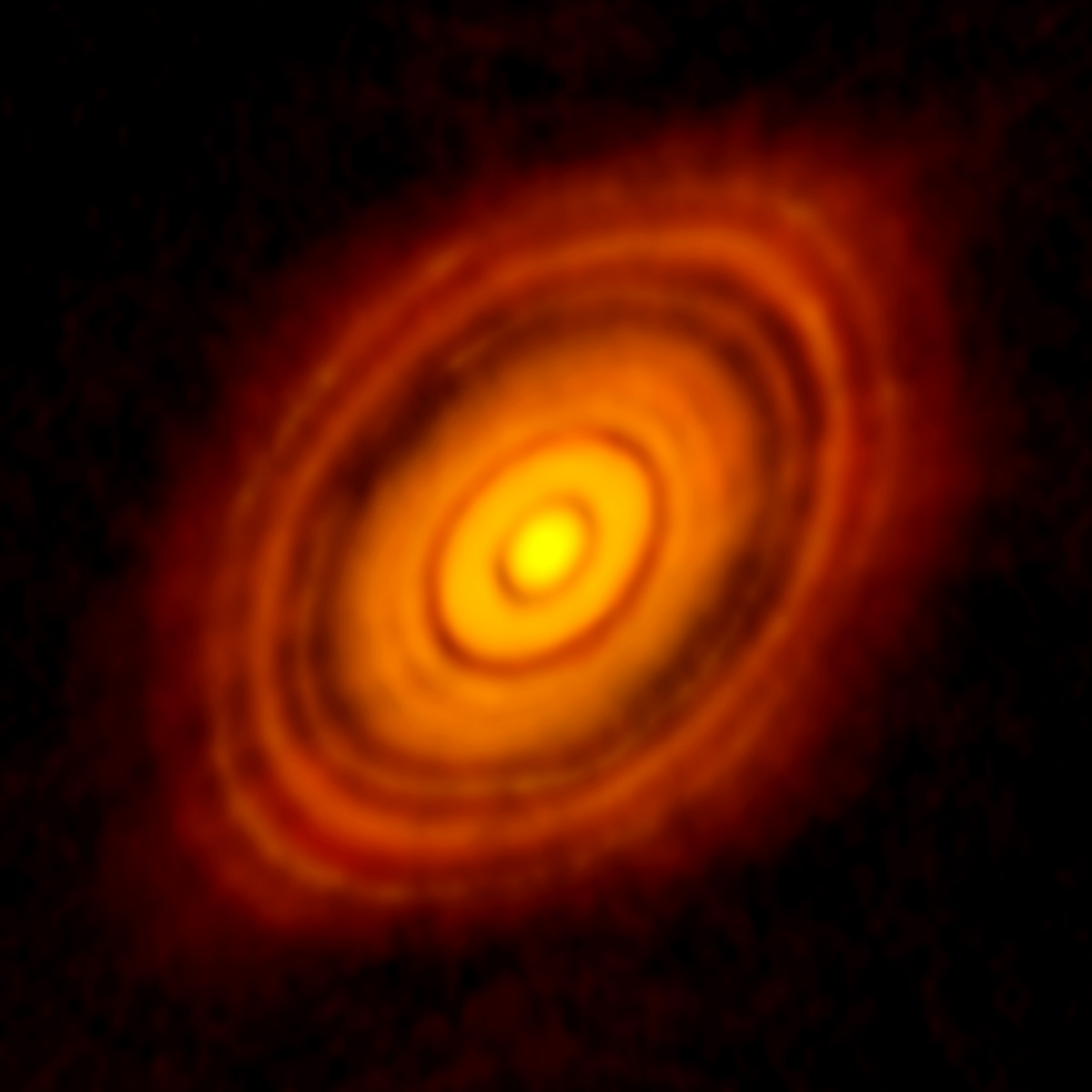}
\caption[ALMA image of the disk around HL Tau]{
\label{fig:hltau_nrao}
An image of the disk around the young star HL Tau made by the Atacama Large Millimeter Array (ALMA). The image shows dust continuum emission. Image from \url{https://public.nrao.edu/static/pr/planet-formation-alma.html}.
}
\end{marginfigure}

In some cases we can spatially resolve a disk in IR or sub-mm observations (Figure \ref{fig:hltau_nrao} shows a spectacular example), and in some cases the disks are unresolved. In either case, in order to interpret these images, we need to think a bit about which parts of the disk we expect to see at which wavelengths. Consider a geometrically thin disk of material of surface density $\Sigma(\varpi)$ and temperature $T(\varpi)$ beginning at a radius $\varpi_0$ around the star and extending out to radius $\varpi_1$. The dust has opacity $\kappa_\lambda$ at wavelength $\lambda$. The entire disk is inclined relative to our line of sight at angle $\theta$. The flux we receive from the disk at wavelength $\lambda$ is
\begin{equation}
F_{\lambda} = \int I_{\lambda}\, d\Omega,
\end{equation}
where $I_{\lambda}$ is the intensity emitted by a portion of the disk at wavelength $\lambda$, and the integral goes over the solid angle $\Omega$ occupied by the disk.

To evaluate the flux, note the ring of material at radius $\varpi$ has an area $2\pi \varpi \,d\varpi$. It is inclined relative to the line of sight by $\theta$, however, so its projected area is $2\pi \varpi \cos\theta\, d\varpi$. The case $\theta=0$ corresponds to the ring being seen perfectly face-on, and the case $\theta=1$ corresponds to perfectly edge-on, and gives 0 in the limit of an infinitely thin disk. This is the projected area, and to covert this to a projected solid angle we divide by $D^2$, where $D$ is the distance to the disk. Thus the flux is
\begin{equation}
F_{\lambda} = \frac{2\pi \cos\theta}{D^2} \int_{\varpi_0}^{\varpi_1} I_\lambda(\varpi) \varpi\, d\varpi.
\end{equation}

To make further progress we must specify the intensity, which is a function of $\Sigma$ and $T$. The optical depth of the disk will be
\begin{equation}
\tau_{\lambda} = \frac{\kappa_{\lambda}\Sigma}{\cos\theta}
\end{equation}
Note that the inclination factor $\cos\theta$ appears on the bottom here, as it should: for $\theta=0$, face-on, we just get the ordinary surface density, but that gets boosted as we incline the disk. The intensity produced by a slab of material of uniform temperature $T$ and optical depth $\tau_{\lambda}$ is
\begin{equation}
I_{\lambda} = B_{\lambda}(T) \left(1-e^{-\tau_{\lambda}}\right),
\end{equation}
where $B_{\lambda}(T)$ is the Planck function. Plugging this in, we have
\begin{equation}
\label{eq:diskflux}
F_{\lambda} = \frac{2\pi \cos\theta}{D^2} \int_{\varpi_0}^{\varpi_1} B_{\lambda}(T) \left[1-\exp\left(-\frac{\kappa_{\lambda}\Sigma}{\cos\theta}\right)\right] \varpi\, d\varpi.
\end{equation}

For a given disk model it is clearly straightforward to evaluate this integral to obtain the emitted flux. However, the model is underspecified, in the sense that we are fitting only one function, $F_{\lambda}$, and we have two free functions to use: $T(\varpi)$ and $\Sigma(\varpi)$. This is even if we assume that the opacity is known, which we will see is not a great assumption.

In order to deduce things like $\Sigma$ and $T$ we need to have a physical model of how the disk behaves, and to deduce either $\Sigma$, $T$, or a relationship between them in order to obtain strong constraints on either one from an observed SED. In general such models can be quite complicated, because the disk's temperature distribution depends on both internal heating via viscous dissipation, and external illumination due to the star. Problem Set 4 contains a problem in which such a model is developed. Even without such a sophisticated model, however, it is possible to learn very interesting things simply from the behavior of the flux in certain limits.

\paragraph{The optically thick limit.}

First, suppose that the disk is optically thick at some wavelength, i.e., $\tau_{\lambda} = \kappa_{\lambda}\Sigma/\cos\theta \gg 1$. This is likely to be true for shorter (e.g., near-IR) wavelengths, where the opacity is high, and where most emission is coming form close to the star where the surface density is highest. In this case it is reasonable to set the exponential factor in equation (\ref{eq:diskflux}) to 0, and we are simply left with the integral of the Planck function of the disk temperature over radius.

Note that in this limit $\Sigma$ drops out, which makes intuitive sense: if the disk is optically thick then we only get to see its surface, and adding or removing material beneath this surface will not change the amount of light we see. Substituting in the Planck function, in the optically thick case we now have
\begin{equation}
F_{\lambda} = \frac{4\pi\cos\theta}{D^2} \frac{hc^2}{\lambda^5} \int_{\varpi_0}^{\varpi_1} \frac{\varpi}{\exp[hc/(k_B T)]-1}\,d\varpi.
\end{equation}

If we further assume that the temperature varies with radius as a powerlaw, $T=T_0(\varpi/\varpi_0)^{-q}$, then we can evaluate the integral via the substitution
\begin{equation}
\label{eq:xsubstitute}
x=\left(\frac{hc}{\lambda k_B T_0}\right)^{1/q} \frac{\varpi}{\varpi_0},
\end{equation}
which gives
\begin{eqnarray}
F_{\lambda} & = & 
\frac{4\pi \cos\theta}{D^2} \frac{hc^2}{\lambda^5} \left(\frac{\varpi_0}{x_0}\right)^2 \int_{x_0}^{x_1} \frac{x}{\exp(x^q)-1}\,dx \\
& = & 
\frac{4\pi \cos\theta}{D^2} \frac{hc^2}{\lambda^5} \left(\frac{hc}{\lambda k_B \varpi_0^q T_0}\right)^{-2/q} \int_{x_0}^{x_1} \frac{x}{\exp(x^q)-1}\,dx,
\end{eqnarray}
and $x_0$ and $x_1$ are obtained by plugging $\varpi_0$ and $\varpi_1$ into equation (\ref{eq:xsubstitute}).

If we look at the part of the spectral energy distribution (SED) where emission is dominated neither by the inner edge of the disk nor the outer optically thin parts, which will be the case over most of the IR, then we can set $x_0\approx 0$ and $x_1\approx \infty$ in the integral. In this case the integral is simply a numerical function of $q$ alone. Since the integral then does not depend on the wavelength, our expression for $F_{\lambda}$ immediately tells us the wavelength-dependence of the emission:
\begin{equation}
\lambda F_{\lambda} \propto \lambda^{(2-4q)/q}.
\end{equation}

Conversely, this means that if we observe the SED of the disk at relatively short wavelengths, for example near-IR, we can invert the wavelength dependence to deduce how the temperature falls with radius. If we also know the distance $D$ and the inclination $\theta$, we can also clearly deduce the combination of variables $\varpi_0^q T_0$ from the observed value of $F_\lambda$.

\paragraph{The optically thin limit.}

Now let us consider the opposite limit, of an optically thin disk. This limit is likely to hold at long wavelengths, such as far-IR and sub-mm, where the dust opacity is low, and where most emission comes from the outer disk where the surface density is low. In the optically thin limit, we can take
\begin{equation}
1-\exp\left(-\frac{\kappa_{\lambda}\Sigma}{\cos\theta}\right) \approx \frac{\kappa_{\lambda}\Sigma}{\cos\theta},
\end{equation}
and substituting this into equation (\ref{eq:diskflux}) for the flux gives
\begin{equation}
F_{\lambda} = \frac{2\pi}{D^2} \int_{\varpi_0}^{\varpi_1} B_{\lambda}(T) \kappa_{\lambda}\Sigma \varpi\,d\varpi.
\end{equation}
Note that in this case the inclination factor $\cos\theta$ drops out, which makes sense: if the disk is optically thin we see all the material in it, so how it is oriented on the sky does not matter.

Even more simplification is possible if we concentrate on emission at wavelengths sufficiently long that we are on the Rayleigh-Jeans tail of the Planck function. This will be true for most sub-mm work, for example: at 1 mm, $hc/(k_B\lambda) = 14$ K, and even the cool outer parts of the disk will be warm enough for emission at this wavelength to fall into the low-energy powerlaw part of the Planck function. In the Rayleigh-Jeans limit
\begin{equation}
B_{\lambda}(T) \approx \frac{2c k_B T}{\lambda^4},
\end{equation}
and substituting this in gives
\begin{equation}
F_{\lambda} = \frac{4\pi c k_B \kappa_{\lambda}}{D^2\lambda^4} \int_{\varpi_0}^{\varpi_1} \Sigma T \varpi \,d\varpi.
\end{equation}

Note that, again, all the wavelength-dependent terms are now outside the integral, and we therefore again expect to be able to predict the wavelength-dependence of the emission without knowing anything about the disk's density or temperature structure. If the dust opacity varies as $\kappa_\lambda\propto \lambda^{-\beta}$, then we have
\begin{equation}
\lambda F_\lambda \propto \lambda^{-3-\beta}.
\end{equation}

This is a particularly important result because it means that we can use the sub-mm SED of a protostellar disk to measure the wavelength-dependence of the dust opacity. In the ISM, $\beta$ is generally observed to be $2$ in diffuse regions, going down to $\sim 1$ as we go to dense regions. The powerlaw index describing how $\kappa_{\lambda}$ varies with $\lambda$ is determined primarily by the size distribution of the dust grains, with larger grains giving smaller $\beta$. This means that reductions in $\beta$ indicate grain growth, an important prelude to planet formation.

One big caveat here is that this only applies in the optically thin limit, and at shorter wavelengths one is probing closer to the star, where the gas is closer to optically thick. This can fool us into thinking we are seeing grain growth. To see why, note that for $\beta=1-2$, the typical values for non-disk interstellar grains, we expect $\lambda F_{\lambda}$ to vary as a powerlaw with index between $-4$ and $-5$ in the optically thin limit. Smaller $\beta$, which we expect to occur when grains grow, would make this value shallower.

However, recall that in the optically thick limit $\lambda F_\lambda \propto \lambda^{(2-4q)/q}$, where $q$ is the powerlaw index describing how the temperature varies with radius. The value of $q$ depends on the thermal structure of the disk, but for observed optically thick sources values of $q$ in the range $0.5-1$ are commonly inferred. In this case $\lambda F_{\lambda}$ varies as a powerlaw with index between $0$ and $-2$. In other words, a transition from optically thin to optically thick {\it also} causes the SED to flatten. Thus, when we see a flattening, we have to be very careful to be sure that it is due to changes in the grain population and not in the optical depth.

The best way to get around this is with spatially resolved observations, which let us look at a single radius in the disk, thereby getting rid of the effects of radial temperature variation.

\paragraph{Mass estimates.}

By combining the optically thin and optically thick parts of these curves, it is also possible to obtain estimates of the mass of disks, provided that we think we understand the properties of the dust. The general procedure is to observe the disk in the IR, where it is assumed to be optically thick. As discussed earlier, this lets us figure out $q$ and $\varpi_0^q T_0$. This means that the temperature distribution $T(\varpi)$ can be considered known. If one plugs this into the equation for the optically thin flux,
\begin{equation}
F_{\lambda} = \frac{4\pi c k_B \kappa_{\lambda}}{D^2\lambda^4} \int_{\varpi_0}^{\varpi_1} \Sigma T \varpi \,d\varpi,
\end{equation}
then the only remaining unknowns are $\kappa_{\lambda}$ and $\Sigma$. If one assumes a known $\kappa_\lambda$ (a questionable assumption), then $\Sigma$ is the only unknown.

The problem in this case is no longer underdetermined. The flux $F_{\lambda}$ is one known function, and it determined the unknown function $\Sigma(\varpi)$ uniquely through an integral equation. This can be solved numerically to obtain $\Sigma(\varpi)$, which in turn gives the disk mass. Typical T Tauri disk masses determined via this technique range from $10^{-3}-10^{-1}$ $\msun$, although with an obviously large uncertainty coming from the unknown grain properties, and from the need to convert a dust mass into a total mass.

\subsection{Disks in Molecular Lines}

The optical and IR / sub-mm continuum techniques both target the dust, but they do not directly tell us about the gas in disks, which dominates the mass. To observe the gas we must detect line emission. The lines detected can be in the infrared, which will mostly tell us about the warm portions of the disk very close to the star, or in the radio / sub-mm, which can tell us about the cool material far from the star.

\begin{figure}
\includegraphics[width=\linewidth]{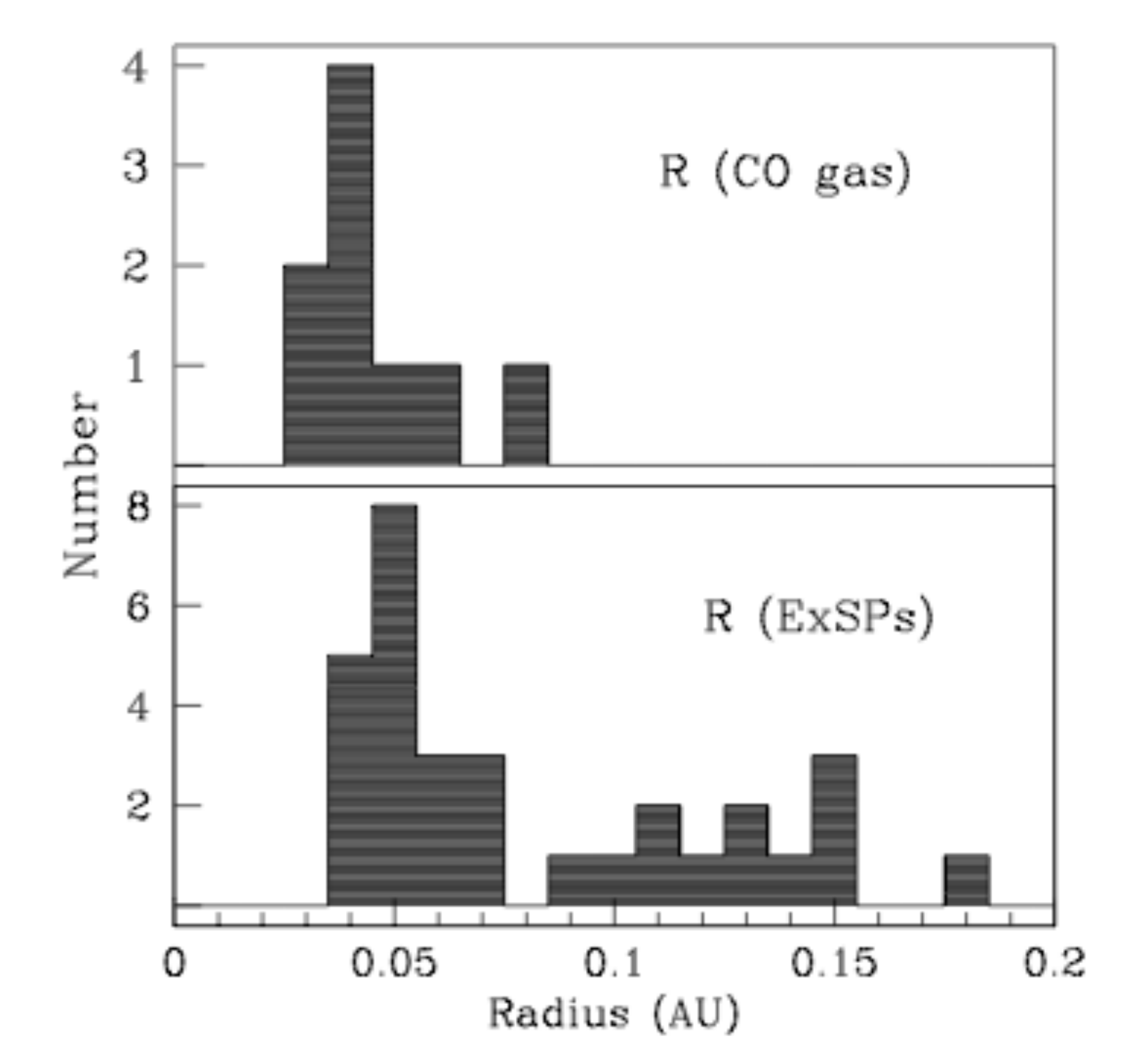}
\caption[Inner disk radii from CO line emission]{
\label{fig:diskradius_najita07}
The top panel shows inner truncation radii of disks as inferred from the maximum velocity of CO vibrational emission. For comparison, the bottom panel shows the radial distribution of the hot Jupiter expolanets known at the time \citep{najita07b}.
}
\end{figure}

For the former case, lines that have been detected include the vibrational and ro-vibrational transitions of CO, OH, water, and molecular hydrogen. These generally probe regions within a few tenths of an AU of the star, simply because of the high temperatures required for the upper levels to be significantly populated. One particularly important use of these techniques is to infer the inner radii at which disks become truncated. Since this is line emission, we determine the velocity of the gas. If we assume that rotation near the star is Keplerian, and we can measure the stellar mass and inclination by other means, then the maximum measured rotation velocity directly tells us innermost radius at which there is a dense disk. Using this technique suggests that disks are truncated at inner radii of $\sim 0.04$ AU (Figure \ref{fig:diskradius_najita07}).

\begin{figure}
\includegraphics[width=\linewidth]{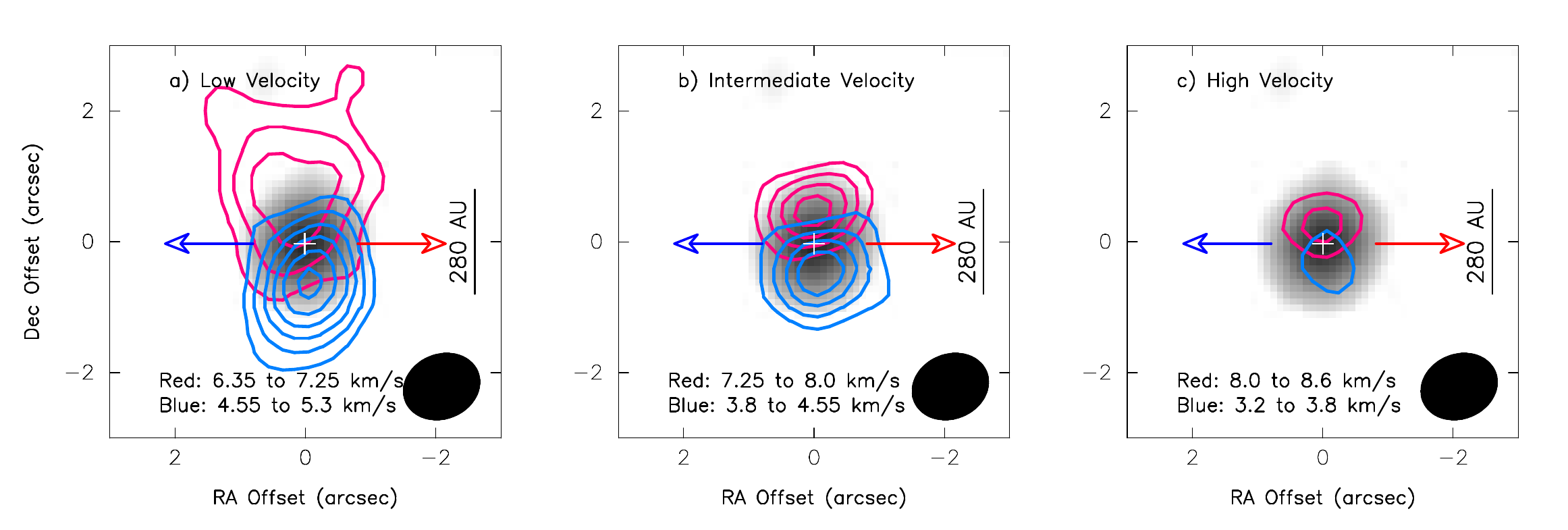}
\caption[$^{13}$CO channel maps of the disk in L1527]{
\label{fig:diskvel_tobin12}
Observed $^{13}$CO line emission from the disk in the core L1527. In each panel, grayscale shows dust continuum emission, white plus signs mark the location of the star, and the red and blue contours show the emission observed in the indicated velocity range. Black ovals show the observational beam, and red and blue arrows show the axis defined by an observed molecular outflow. Reprinted by permission from Macmillan Publishers Ltd: Nature, 492, 83, \citeauthor{tobin12a}, \copyright 2012.
}
\end{figure}

For the sub-mm and radio, detections have mostly involved CO and its isotopologues. The main advantage of thes data, as opposed to the dust continuum, is that we obtain kinematic information. This can then be used to determine whether the (usually) poorly-resolved objects we see in the continuum have a velocity structure consistent with Keplerian rotation.  Figure \ref{fig:diskvel_tobin12} shows an example. Note that higher velocity emission tends to come from closer to the star, exactly as would be expected for a Keplerian disk. Indeed, when one fits the data to Keplerian rotation curves, they are entirely consistent. The number of sources for which this analysis has been done is not large, but it is growing.

\section{Observations of Outflows}

\subsection{Outflows in the Optical}

In addition to observing the disks themselves, we can observe the outflows that they drive. Outflows were first noticed in the 1950s based on optical observations by Herbig and Haro, working independently. The class of objects they discovered are known as Herbig-Haro, or HH, objects in their honor. HH objects were first seen as small patches of optical emission containing both continuum and a number of lines, most prominently H$\alpha$. The H$\alpha$ indicates the presence of ionization, but, unlike the large ionized regions generated by massive stars, where all species are highly ionized, HH objects also show signs of emission from neutral or weakly ionized species such as O~\textsc{i} and N~\textsc{ii}.

The standard interpretation of this sort of ionization structure is that we are seeing a fast shock. The shocked material is ionized, producing H$\alpha$ emission as it recombines. Both upstream and downstream of the shock itself, however, there is neutral material that is warm, either because it has had a chance to recombine but not to cool (for the downstream gas) or because it has been pre-heated by radiation from the shock (for the upstream gas). This produces the neutral or weakly ionized emission lines.

\begin{figure}
\includegraphics[width=\linewidth]{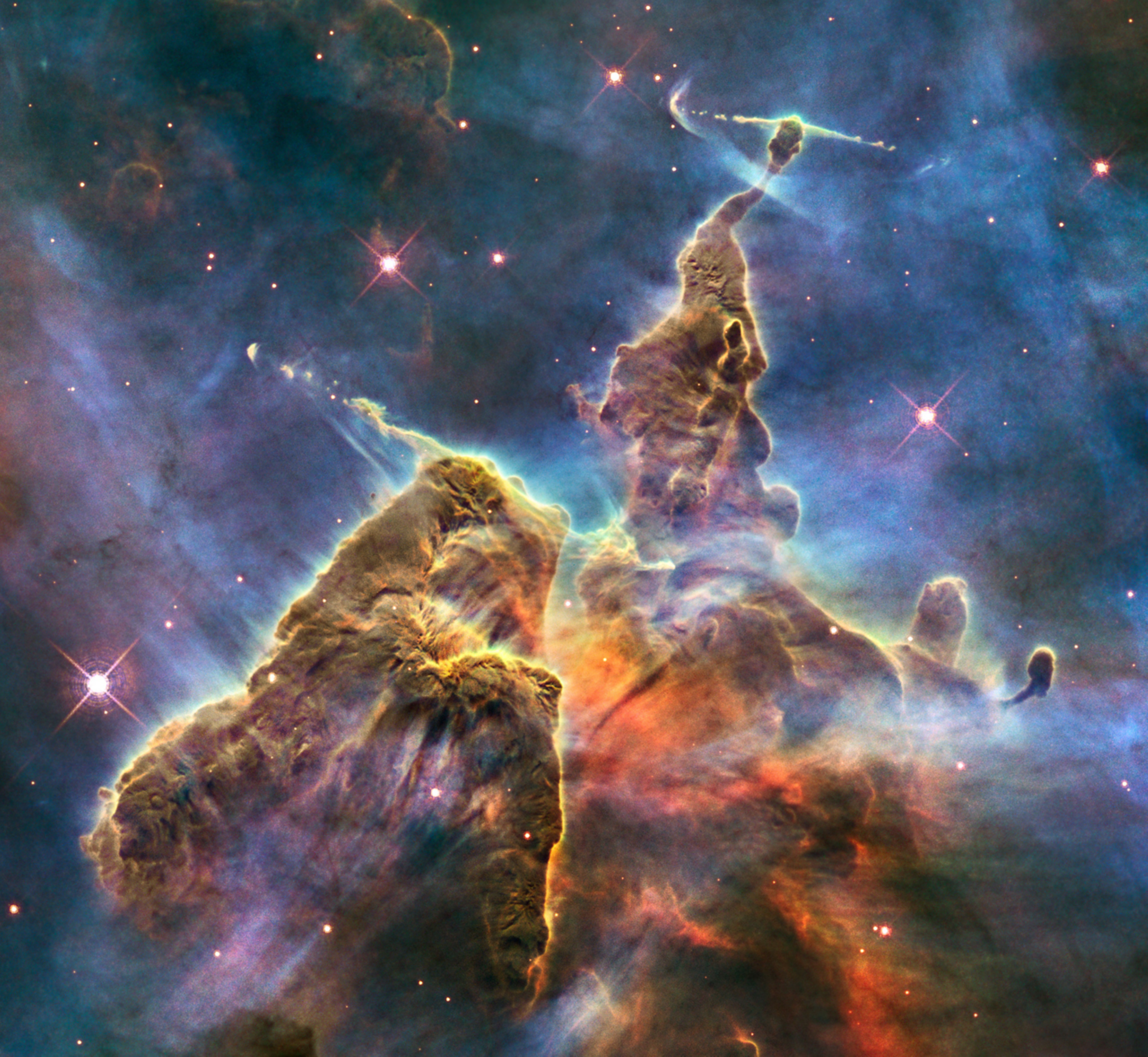}
\caption[Herbig-Haro jets from \textit{HST}]{
\label{fig:hhjet}
Herbig-Haro jets imaged with the \textit{Hubble Space Telescope}. Two jets are visible; one is at the tip of the ``pillar" near the top of the image, and another is near the edge of the structure in the middle-left part of the image. Bow shocks from the jets are clearly visible. Taken from \url{http://hubblesite.org/newscenter/archive/releases/2010/13/image/a/}.
}
\end{figure}

More sensitive measurements in the 1970s revealed that the bright emission knots Herbig and Haro saw are in fact connected by linear structures that also emit in optical, just with lower surface brightness. We can also see bow shocks at the heads of jets, where the plough into dense molecular gas (Figure \ref{fig:hhjet}).

Today our interpretation of the HH knots is that they are locations where the jet has either encountered a dense region of interstellar material, producing a strong shock and bright emission, or where some variation in the velocity or mass flux being launched into the jet has caused an internal shock. The weaker emission in between the knots is caused by the interaction of the jet with a lower density environment. One can also detect this component in radio free-free emission, produced by electrons in the jet plasma.

The HH objects move fast enough to produce noticeable shifts in the positions of the bright knots over spans of $\sim 10$ years. The inferred velocities are typically hundreds of km s$^{-1}$. These velocities are also consistent with what we infer from Doppler shifts in cases where the jets is partly oriented toward us.

An important point is that these HH jets are usually bipolar, meaning that there is a clear driving star at the base of two HH objects propagating in opposite directions. Sometimes the knots of emission are even mirror symmetric, suggesting that they are produced by variations in the outflow velocity or mass flux originating at the point where the jet is launched, rather than any property of the environment.

Estimates of the density of the outflowing material based on models of the shocks suggest mass fluxes that range from $10^{-6}$ $\msun$ yr$^{-1}$ in class 0 sources, dropping to $10^{-8}-10^{-7}$ $\msun$ yr$^{-1}$ for classical T Tauri stars / class I sources. The inferred momentum flux is therefore of order $10^{-6}-10^{-3}$ $\msun$ km s$^{-1}$ yr$^{-1}$. These estimates are quite uncertain however, since they are based on shock diagnostics, and tell us relatively little about the material in between the bright HH knots.

\subsection{Outflows in the Radio}

Optical and near-IR emission traces the regions where strong shocks heat the gas enough to excite transitions at these wavelengths. However, the jets of fast moving material only show the tip of the iceberg as far as the outflow is concerned. Observations in molecular lines reveal that narrow optical HH jets are accompanied by a much wider-angle, slower-moving, and more massive molecular outflow. 

Molecular observations show large masses of molecular gas moving at $\sim 10$ km s$^{-1}$ -- the velocity is based on the Doppler shifts of the molecular lines used to observe the outflow. Again, we generally see a bipolar morphology. Depending on the outflow direction, this can consist of two lobes pointing in opposite directions on the sky, two lobes in the same position on the sky but with distinct red- and blue-shifted components, or some combination of the two.

Despite their lower velocities, these wind components actually contain the bulk of the outflow momentum, typically $10^{-4}-10^{-1}$ $\msun$ km s$^{-1}$ yr$^{-1}$, depending on the luminosity of the driving source. The molecular outflows are thought to consist primarily not of material ejected directly by the launching mechanism, but of ambient gas that has been swept up by this gas as it flows outward. The interaction is via shocks that can radiate, so energy is not conserved, but momentum is. This entrainment explains why the velocities of this material are so low compared to the material in the jets.

\chapter{Protostellar Disks and Outflows: Theory}
\label{ch:disks_theory}

\marginnote{
\textbf{Suggested background reading:}
\begin{itemize}
\item \href{http://adsabs.harvard.edu/abs/2014prpl.conf..173L}{Li, Z.-Y., et al. 2014, in ``Protostars and Planets VI", ed.~H.~Beuther et al., pp.~173-194}, sections 3-6 \nocite{li14a}
\end{itemize}
\textbf{Suggested literature:}
\begin{itemize}
\item \href{http://adsabs.harvard.edu/abs/2013MNRAS.432.3320S}{Seifried, D., et al., 2013, MNRAS, 432, 3320} \nocite{seifried13a}
\end{itemize}
}

The previous chapter introduced observations of protostellar disks and their outflows. This companion chapter reviews theoretical models of such disks, with particular attention to how they form, why they accrete, and why they launch outflows.

\section{Disk Formation}

\subsection{The Angular Momentum of Protostellar Cores}

To understand why disks form, we must start with the question of the angular momentum content of the gas that will eventually form the star. To determine this observationally, one maps a core in an optically thin tracer and measures the mean velocity on every line of sight through the core. If there is a systematic gradient in the mean velocity, that is indicative of some net rotation. Doing this for a sample of cores yields a distribution of rotation rates.

It is most convenient to express the resulting distribution dimensionlessly, in terms of the ratio of kinetic energy in rotation to gravitational binding energy. If the angular velocity of the rotation is $\Omega$ and the moment of inertia of the core is $I$, this is
\begin{equation}
\beta = \frac{(1/2)I\Omega^2}{a G M^2/R},
\end{equation}
where $a$ is our usual numerical factor that depends on the mass distribution. For a sphere of uniform density $\rho$, we get
\begin{equation}
\beta = \frac{1}{4\pi G \rho}\Omega^2 = \frac{\Omega^2 R^3}{3 G M}
\end{equation}
Thus we can estimate $\beta$ simply given the density of a core and its measured velocity gradient. Observed values of $\beta$ typically a few percent \citep[e.g.,][]{goodman93a}.

This implies that cores are not primarily supported by rotation. In fact, we can understand the observed rotation rates as being a property of the turbulence. Although cores are primarily thermally-supported, they do still have some turbulent motions present at transsonic or subsonic levels. Since most of the power in this turbulence is on large scales, there is likely to be a net gradient. When one performs the experiment of generating random turbulent velocity fields with a variety of power spectra and analyzing them as an observer would, the result is a $\beta$ distribution that agrees very well with the observed one \citep{burkert00a}.

\subsection{Rotating Collapse: the Hydrodynamic Case}

Given this small amount of rotation, how can we expect it to affect the collapse? Let us take the simplest case of a cloud in solid body rotation at a rate $\Omega$. Consider a fluid element that is initially at some distance $\varpi_0$ from the axis of rotation. We will consider it to be in the equatorial plane, since fluid elements at equal radius above the plane have less angular momentum, and thus will fall into smaller radii.

Its initial angular momentum in the direction along the rotation axis is $j=\varpi_0^2\Omega$. If pressure forces are insignificant for this fluid element, it will travel ballistically, and its specific angular momentum and energy will remain constant as it travels. At its closest approach to the central star plus disk, its radius is $\varpi_{\rm min}$ and by conservation of energy its velocity is $v_{\rm max} = \sqrt{2 G m_*/\varpi_{\rm min}}$, where $m_*$ is the mass of the star plus the disk material interior to this fluid element's position.
Conservation of angular momentum them implies that $j=\varpi_{\rm min} v_{\rm max}$.

Combining these two equations for the two unknowns $\varpi_{\rm min}$ and $v_{\rm max}$, we have
\begin{equation}
\varpi_{\rm min} = \frac{\varpi_0^4 \Omega^2}{G m_*} = \frac{4\pi \rho \beta \varpi_0^4}{m_*},
\end{equation}
where we have substituted in for $\Omega^2$ in terms of $\beta$. This tells us the radius at which infalling material must go into a disk because conservation of angular momentum and energy will not let it get any closer.

We can equate the stellar mass $m_*$ with the mass that started off interior to the position of the fluid element we are considering. This amounts to assuming that the collapse is perfectly inside-out, and that the mass that collapses before the fluid element under consideration all makes it onto the star. If we make this approximation, then $m_*=(4/3)\pi \rho r_0^3$, and we have
\begin{equation}
\varpi_{\rm min} = 3 \beta \varpi_0,
\end{equation}
i.e., the radius at which the fluid element settles into a disk is simply proportional to $\beta$ times a numerical factor of order unity.

We should not take the factor too seriously, since of course real clouds are not uniform spheres in solid body rotation, but the result that rotation starts to influence collapse and force disk formation at a radius that is a few percent of the core radius is interesting. It implies that for cores that are $\sim 0.1$ pc in size and have $\beta$ values typical of what is observed, they should start to become rotationally flattened at radii of several hundred AU. This should be the typical size scale of protostellar disks in the hydrodynamic regime.

\subsection{Rotating Collapse: the Magnetohydrodynamic Case}

Magnetic fields can greatly complicate this picture, due to magnetic braking. As a core contracts, conservation of angular momentum causes it to spin faster, but this in turn twists up the magnetic field. This creates a tension force that opposes the rotation rate, and tries to keep the core rotating as a solid body.

To analyze this effect, let us work in cylindrical coordinates $(\varpi, \phi, z)$. Consider a fluid element in a disk at a distance $\varpi$ from the star, whose dimensions are $d\varpi$, $d\phi$, $dz$ in the $\varpi$, $\phi$, and $z$ directions. The fluid element is rotating around the star with a velocity $v_{\phi}$ in the $\phi$ direction. The fluid element is threaded by a magnetic field $\vecB=(B_\varpi, B_\phi, B_z)$. For future convenience we define the poloidal component of the field to be
\begin{equation}
\vecB_p = (B_\varpi, B_z),
\end{equation}
i.e., it is the component of the field not associated with wrapping around the rotation axis. The $\phi$ component of the field is called the toroidal component, since it represents the part of the field that is wrapped in the rotation direction. To help visualize this, imagine drawing a two-dimensional plot of the system in the $(\varpi, z)$-plane. In this plot, the poloidal component is the one on the page, and the toroidal component is the one going into or out of the page.

We will assume that both the fluid and the magnetic field are axisymmetric, so that they do not vary with $\phi$, although the field does have a $\phi$ component. The magnetic field exerts a Lorentz force per unit volume on the fluid element given by
\begin{eqnarray}
\mathbf{f} & = & \frac{1}{4\pi} \left[(\nabla\times\vecB)\times \vecB\right] \\
& = & \frac{1}{4\pi} \left[\frac{B_\varpi}{\varpi} \frac{\partial(\varpi B_\phi)}{\partial \varpi} + B_z  \frac{\partial B_\phi}{\partial z}\right] \hat{\phi}\\
& = & \frac{1}{4\pi\varpi} \vecB_p \cdot \nabla_p (\varpi B_{\phi}) \hat{\phi},
\end{eqnarray}
where all the components except the $\phi$ one vanish by symmetry, and in the final step we have defined the poloidal gradient as $\nabla_p = (\partial/\partial\varpi, \partial/\partial z)$, i.e., it is just the components of the gradient in the $\varpi$ and $z$ directions.

The rate of change of the momentum associated with the Lorentz force alone is
\begin{equation}
\frac{\partial}{\partial t}(\rho \vecv) = \mathbf{f},
\end{equation}
so writing down the $\phi$ component of this equation and multiplying on both sides by $\varpi$, we have
\begin{equation}
\frac{\partial}{\partial t} (\rho v_{\phi} \varpi) = \frac{1}{4\pi} \vecB_p\cdot\nabla_p(\varpi B_\phi)
\end{equation}
The left hand side of this equation just represents the time rate of change of the angular momentum per unit volume $\rho v_{\phi} \varpi$, while the right hand side represents the torque per unit volume exerted by the field.

Given this equation, how quickly can a magnetic field stop rotation? We can define a magnetic braking time by
\begin{equation}
t_{\rm br} = \frac{\rho v_{\phi} \varpi}{\frac{\partial}{\partial t} (\rho v_{\phi} \varpi)}
= \frac{4 \pi \rho v_{\phi} \varpi}{\vecB_p \cdot\nabla_p (\varpi B_\phi)}
\end{equation}
To evaluate this timescale, consider the case of a fluid element that is part of a collapsing cloud, and is trying to rotate at a velocity $v_{\phi}$ equal to the Keplerian velocity, i.e.,
\begin{equation}
v_{\phi} = \sqrt{\frac{GM}{\varpi}},
\end{equation}
where $M$ is the mass interior to the fluid element.

If we started with a uniform cloud of density $\rho$, the mass interior to our element is $M\approx (4\pi/3) \rho \varpi^3$, so $v_\phi \approx \sqrt{(4\pi/3) G \rho \varpi^2}$. Plugging this into the timescale, we have
\begin{equation}
t_{\rm br} \approx \frac{(4\pi \rho)^{3/2} G^{1/2} \varpi^2}{\vecB_p \cdot\nabla_p (\varpi B_\phi)}.
\end{equation}

To make an order of magnitude estimate of what this implies, let us suppose that the poloidal and toroidal components of the field are comparable, and that the characteristic length scale on which the field varies is $\varpi$, so the field is fairly smooth on all scales smaller than the size of the region that is currently collapsing. In this case $\vecB_p \cdot \nabla_p (\varpi B_\phi) \sim B^2$, so the time scale becomes
\begin{eqnarray}
t_{\rm br} & \sim & \frac{G^{1/2}\rho^{3/2} \varpi^2}{B^2} \\
& \sim & \frac{(G\rho)^{1/2} \varpi^2}{v_A^2} \\
& \sim & \frac{t_{\rm cr}^2}{t_{\rm ff}},
\end{eqnarray}
where we are dropping constants of order unity. Note that in the second step we wrote $B$ in terms of the Alfven speed $v_A = B/\sqrt{4\pi \rho}$, and in the final step we wrote $t_{\rm cr} = \varpi/v_A$, where $t_{\rm cr}$ is the Alfven crossing time of the cloud.

\begin{marginfigure}
\includegraphics[width=\linewidth]{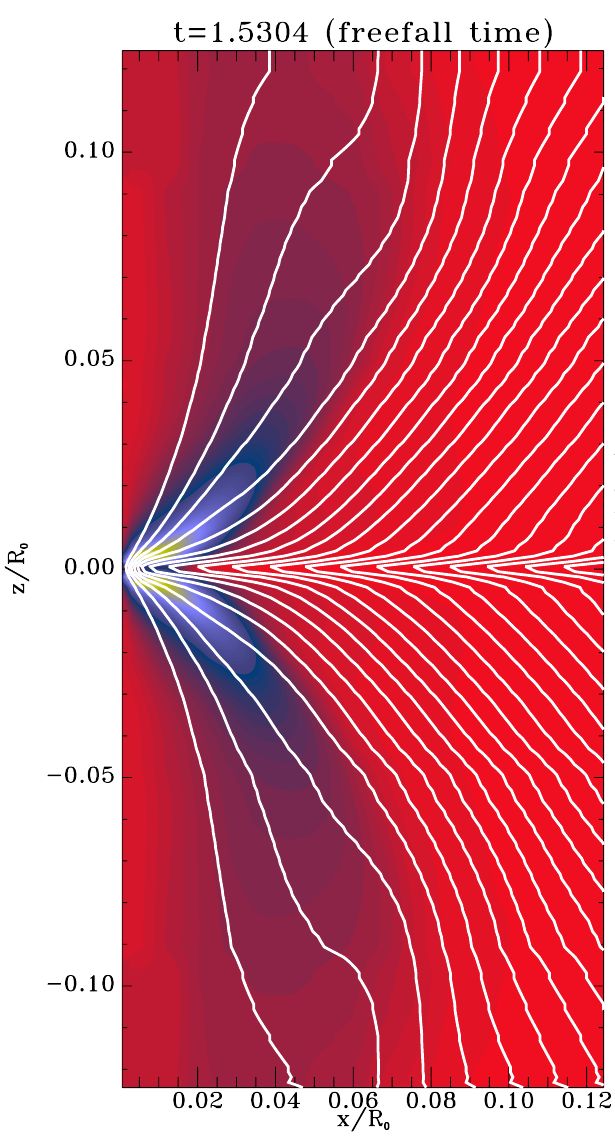}
\includegraphics[width=\linewidth]{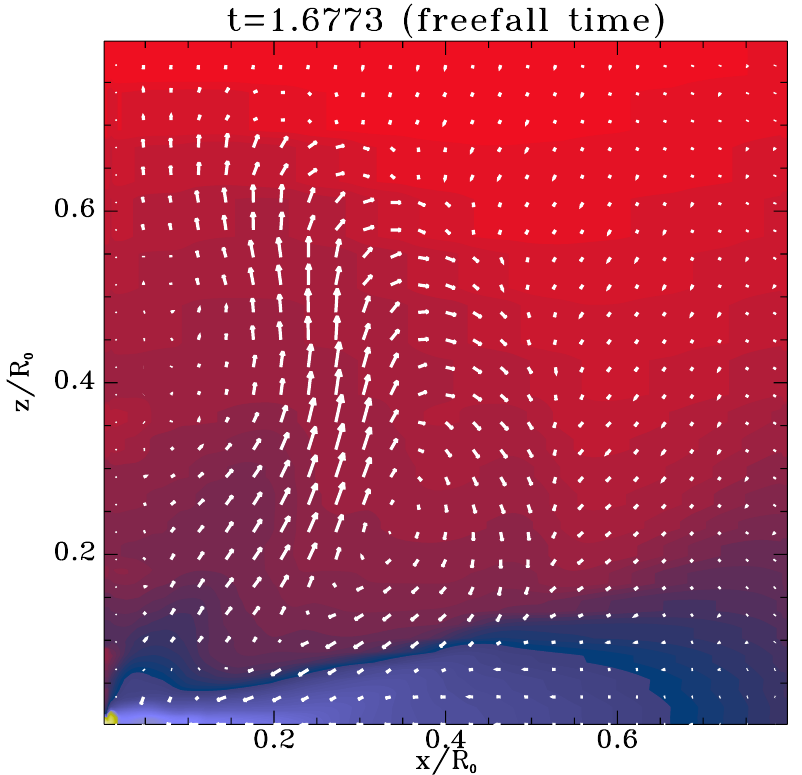}
\caption[Simulations of magnetized rotating collapse]{
\label{fig:magdisk_hennebelle}
Results from a simulation of magnetized rotating collapse. The top panel shows the magnetic field structure; solid lines are poloidal magnetic field lines, while color indicates the azimuthally-averaged total magnetic field strength, on a scale from $0-3.5$ mG. The bottom panel shows the density (color) and velocity (arrows) structure at a slightly later time in the simulation. The structure in the mid-plane is a non-rotating pseudo-disk. Credit:  \citeauthor{hennebelle08c}, A\&A, 477, 9, 2008, reproduced with
permission \copyright\,ESO.
}
\end{marginfigure}

If a cloud starts out with a magnetic field near equipartition with gravity and thermal energies, we expect $t_{\rm ff} \sim t_{\rm cr}$, so this means that $t_{\rm br}\sim t_{\rm cr}$. This is an order of magnitude calculation, but its implication is clear: if we have a field that is even marginally wound up, such that the poloidal and toroidal components become comparable, this field is capable of stopping Keplerian rotation in a time scale comparable to the collapse or crossing time. This can effectively prevent formation of a Keplerian disk at all if the magnetic field is strong enough. Indeed, this is what simulations seem to show happening (Figure \ref{fig:magdisk_hennebelle}).

\subsection{The Magnetic Braking Problem and Possible Solutions}

The calculation of magnetic braking we have just performed presents us with a fundamental problem: it naively seems like magnetic fields should prevent disks from forming at all, but we observe that they do. We even observe disks present in class 0 sources, where the majority of the gas is still in the envelope. So how can we get out of this? This is not a completely solved problem, but we can make a few observations about what a solution might look like.

\begin{marginfigure}
\includegraphics[width=\linewidth]{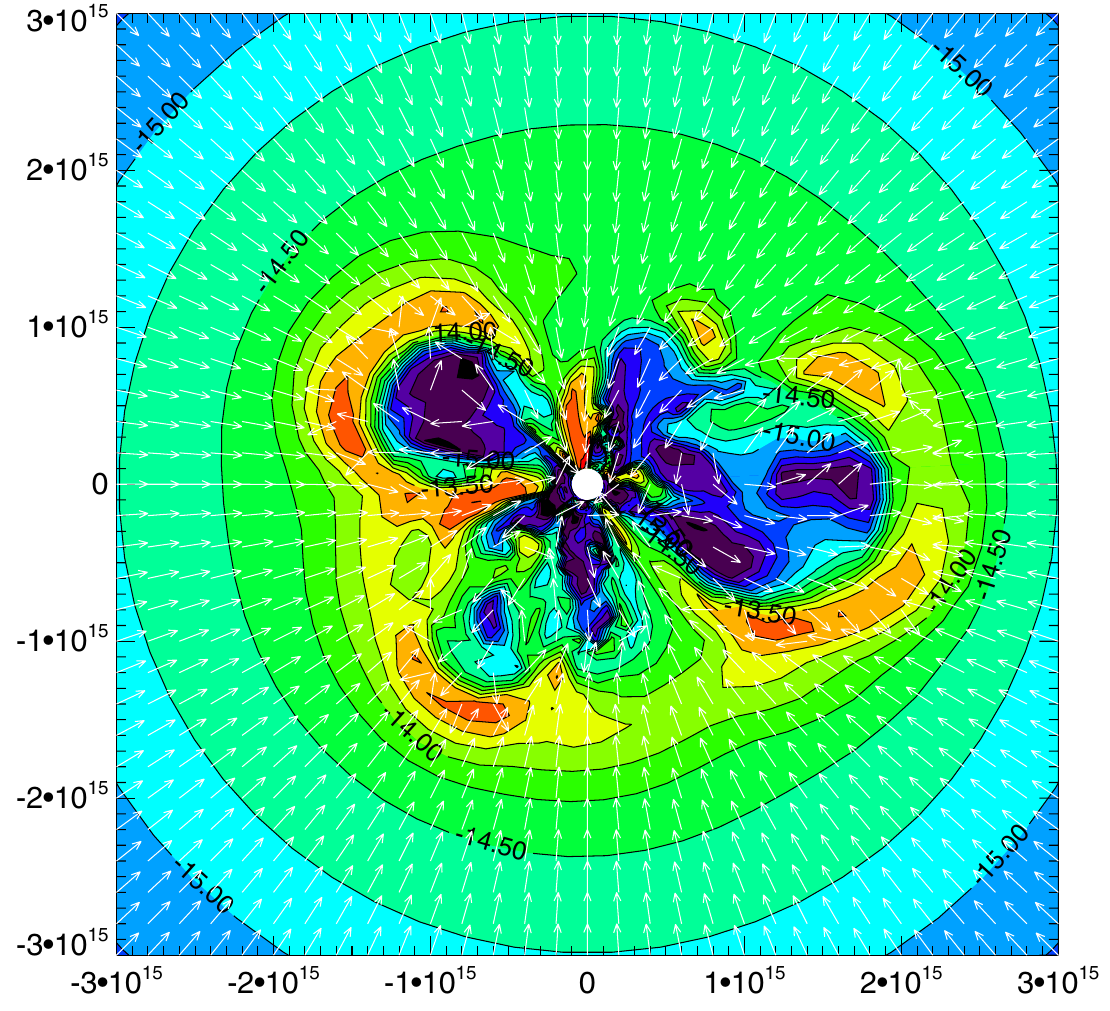}
\caption[Simulations of magnetized rotating collapse with non-ideal MHD]{
\label{fig:magdisk_krasnopolsky12}
Results from a simulation of magnetized rotating collapse including the effects of ion-neutral drift and Ohmic dissipation. Lengths on the axes are in units of cm. Colors and contours show the density in the equatorial plane, on a logarithmic scale from $10^{-16.5}$ to $10^{-12.5}$ g cm$^{-3}$. Arrows show velocity vectors. Credit: \citet{krasnopolsky12a}, \copyright AAS. Reproduced with permission.
}
\end{marginfigure}

We can first ask whether ion-neutral drift might offer a way out. Recall in Chapter \ref{ch:magnetic}, we showed that, at the densities and velocities typical of protostellar cores, ion-neutral drift should allow gas to decouple from the magnetic field on scales below $L_{\mathrm{AD}} \sim 0.05$ pc. One might expect that this would make it possible to form disks below the decoupling scale. However, simulations suggest that this solution does not work. The flux that is released from the gas by ion-neutral drift does not disappear. Instead, it builds up flux tubes near the star with relatively little mass on them, and these flux tubes prevent a disk from forming (Figure \ref{fig:magdisk_krasnopolsky12}).

A more promising solution appears to be misalignment between the rotation axis of the gas and the magnetic field, or, more generally, the presence of turbulence in the collapsing gas. In simulations where the gas is turbulent, the magnetic field lines tend to be bent or misaligned relative to the disk, and this greatly reduces the efficiency of magnetic braking. However, the problem of how disks form is still not fully solved.

\section{Disk Evolution}

Given that disks do exist, in the real universe if not in our models, we wish to understand how they evolve, and how they accrete onto their parent stars. We therefore sketch here a basic theory for how disks behave.

\subsection{Steady Thin Disks}

\paragraph{Evolution equations.}

Consider a thin disk of surface density $\Sigma$ orbiting at an angular velocity $\Omega$. We take the disk to be cylindrically symmetric, so that $\Sigma$ and $\Omega$ are functions of the radius $\varpi$ only. We assume it is very thin in the vertical direction, so we only need to solve the equations in the plane $z=0$. In addition to its orbital velocity $v_{\phi}=r\Omega$, the gas has a radial velocity $v_\varpi$, which we assume is much less than $v_{\phi}$. This allows the gas to accrete onto a central object.

For this system, the general equation of mass conservation is
\begin{equation}
\frac{\partial}{\partial t}\rho + \nabla\cdot(\rho \vecv) = \frac{\partial}{\partial t}\rho + \frac{1}{\varpi} \frac{\partial}{\partial \varpi}(\varpi \rho v_\varpi) = 0,
\end{equation}
where we have written out the divergence for cylindrical coordinates, and we have used the cylindrical symmetry of the problem to drop the components of the divergence in the $z$ and $\phi$ directions. Since we have a thin disk, the volume density is $\Sigma \delta(z)$, i.e., it is zero off the plane, infinite in the plane, and integrates to $\Sigma$. Integrating the mass conservation equation over $z$ then immediately gives
\begin{equation}
\frac{\partial}{\partial t}\Sigma + \frac{1}{\varpi} \frac{\partial}{\partial \varpi}(\varpi \Sigma v_\varpi) = 0.
\end{equation}
This equation just says that the change in the surface density at some point is equal to the net rate of radial mass flow into or out of it. It is convenient at this point to introduce the mass accretion rate $\dot{M}=-2\pi \varpi \Sigma v_\varpi$, which represents the rate of inward mass flux across the cylinder at radius $\varpi$. With this definition, the mass conservation equation becomes
\begin{equation}
\label{eq:continuity_disk}
\frac{\partial}{\partial t}\Sigma - \frac{1}{2\pi \varpi} \frac{\partial}{\partial \varpi}\dot{M} = 0.
\end{equation}

Next we can write down the Navier-Stokes equation for the fluid,
\begin{equation}
\rho\left(\frac{\partial}{\partial t}\vecv + \vecv\cdot\nabla\vecv\right) = -\nabla p - \rho\nabla\psi + \nabla\cdot \vecT,
\end{equation}
where $p$ is the pressure, $\psi$ is the gravitational potential, and $\vecT$ is the viscous stress tensor. We choose to write the equation in this form, rather than in the conservative formulation we have used elsewhere in this book, because it makes the dependence on the viscous stress tensor particularly explicit, which will become useful below. Integrating this equation over $z$ gives
\begin{equation}
\Sigma \left(\frac{\partial}{\partial t}\vecv + \vecv\cdot\nabla\vecv\right) = -\nabla P - \Sigma \nabla\psi + \int \nabla\cdot \vecT\,dz,
\end{equation}
where $P$ is the vertically-integrated pressure.

Now consider the $\phi$ component of this equation. This is particularly simple, because all $\phi$ derivatives vanish due to symmetry, and the pressure and gravitational forces therefore drop out. This gives\footnote{Note there is some subtlety here in writing out the gradient of a tensor in cylindrical coordinates. \citet{shu92a} has a useful appendix for vector and tensor operations in non-Cartesian coordinate systems.}
\begin{equation}
\Sigma \left[\frac{\partial}{\partial t} v_\phi + \frac{v_\varpi}{\varpi} \frac{\partial}{\partial \varpi}(\varpi v_\phi)\right] = \int \frac{1}{\varpi^2} \frac{\partial}{\partial \varpi}(\varpi^ 2 T_{\varpi\phi})\,dz.
\end{equation}
If we multiply through by $2\pi \varpi^2$, we obtain
\begin{equation}
2\pi \varpi \Sigma \left(\frac{\partial}{\partial t} j + v_\varpi \frac{\partial}{\partial \varpi}j\right) = \int \frac{\partial}{\partial \varpi}(2\pi \varpi^2 T_{\varpi\phi})\,dz = \frac{\partial}{\partial \varpi} \mathcal{T}
\end{equation}
where we have defined $j=\varpi v_\phi$ as the angular momentum per unit mass of the material and
\begin{equation}
\mathcal{T} = 2\pi \varpi \int \varpi T_{\varpi\phi}\, dz.
\end{equation}
Thus we see that this represents an evolution equation for the angular momentum of the gas. The factor $2\pi \varpi \Sigma$ is just the mass per unit radius in a thin ring, so $2\pi \varpi\Sigma j$ is the angular momentum in the ring.

The quantity $\mathcal{T}$ represents the torque exerted on the ring due to viscosity. This is clear if we examine its components. The viscous stress tensor component $T_{\varpi\phi}$ represents the force per unit area created by viscosity. This is multiplied by $\varpi$, so we have $\varpi T_{\varpi\phi}$, which is just the torque per unit area, since it is a force times a lever arm. Finally, this is multiplied by $2\pi \varpi$ and integrated over $z$, which is just the area of the cylindrical surface over which this torque is applied. Thus, $\mathcal{T}$ is the total torque. We take its derivative with respect to $\varpi$ to obtain the difference in torque between the ring immediately interior to the one we are considering and the ring immediately exterior to it. 

Suppose we look for solutions of this equation in which the angular momentum per unit mass at a given location stays constant, i.e., $\partial j/\partial t=0$. This will be the case, for example, of a disk where the azimuthal motion is purely Keplerian at all times, or more generally for any disk orbiting in a fixed potential. In this case the evolution equation just becomes
\begin{equation}
\label{eq:angmom_disk}
-\dot{M} \frac{\partial j}{\partial \varpi} = \frac{\partial \mathcal{T}}{\partial \varpi}.
\end{equation}
This equation describes a relationship between the accretion rate and the viscous torques in a disk. Its physical meaning is that the accretion rate $\dot{M}$ is controlled by the rate at which viscous torques remove angular momentum from material closer to the star and give it to material further out. 

To make further progress, let us write down the viscous stress $T_{\varpi\phi}$ a little more specifically. We will assume that the gas in the disk is Newtonian, meaning that the viscous stress is proportional to the rate of strain in the fluid. We want to know $T_{\varpi\phi}$, meaning the force per unit area in the $\phi$ direction, exerted on the radial face of a fluid element. Consider an observer in a frame comoving with the orbiting fluid at some particular distance $\varpi$ from the star, and consider a fluid element that is initially on the same radial ray as the observer, but a distance $d\varpi$ further from the star. If the rotation is solid body, then the fluid element and the observer will always lie on the same radial ray, so there is no strain, and there will be no viscous stress. On the other hand, if there is differential rotation, such that the fluid further the star has a longer orbital period (as we expect for Keplerian motion), the fluid element will gradually fall behind the observer. This represents a strain in the fluid.

How quickly does the element fall behind? The difference in angular velocity between the observer and the fluid element is $d\Omega = (d\Omega/d\varpi) d\varpi$, and so the difference in spatial velocity is $\varpi (d\Omega/d\varpi) d\varpi$. The rate of strain is defined as one over the time it takes the fluid element to be displaced a distance $d\varpi$ downstream from the observer, i.e., the time it takes for the differential rotation to stretch the fluid in between by an amount of order unity. Thus, the rate of strain is $\varpi (d\Omega/d\varpi) d\varpi/d\varpi = \varpi (d\Omega/d\varpi)$.

The viscous stress is equal to this rate of strain times the dynamic viscosity $\mu$, so
\begin{equation}
T_{\varpi\phi} = \mu \varpi \frac{d\Omega}{d\varpi} = \rho \nu \varpi \frac{d\Omega}{d\varpi},
\end{equation}
where $\nu=\mu/\rho$ is the kinematic viscosity, which will be more convenient to work with, because it is an intensive quantity that depends on the properties of the fluid but not directly on its density. If we plug this into our definition of the viscous torque, we obtain
\begin{equation}
\label{eq:torque}
\mathcal{T} = 2\pi \varpi \int \varpi T_{\varpi\phi}\, dz = 2\pi \varpi^3 \Sigma \nu \frac{d\Omega}{d\varpi}.
\end{equation}
This combined with the equation giving the relationship between $\dot{M}$ and $d\mathcal{T}/d\varpi$ immediately give us the accretion rate for any steady disk of known surface density and angular momentum profile $j(\varpi)$.

The most interesting case in star formation is for $j$ and $\Omega$ corresponding to Keplerian rotation, $j=\sqrt{GM_*\varpi}$ and $\Omega=\sqrt{GM_*/\varpi^3}$, where $M_*$ is the mass of the central star.\footnote{If we were interested in galactic disks, we might instead have considered a flat rotation curve, $j\propto \varpi$.} If we now take the continuity equation (\ref{eq:continuity_disk}) and the angular momentum equation (\ref{eq:angmom_disk}), express the torque in terms of $\nu$ using equation (\ref{eq:torque}), and plug in the Keplerian values of $j$ and $\Omega$, a little algebra shows that the resulting equation is
\begin{equation}
\label{eq:disk_keplerian}
\frac{\partial \Sigma}{\partial t} = \frac{3}{\varpi} \frac{\partial}{\partial \varpi} \left[\varpi^{1/2} \frac{\partial}{\partial \varpi}\left(\nu \Sigma \varpi^{1/2}\right)\right],
\end{equation}
and the corresponding equation for the radial drift velocity is
\begin{equation}
v_\varpi = -\frac{3}{\Sigma \varpi^{1/2}} \frac{\partial}{\partial \varpi}(\nu \Sigma \varpi^{1/2}).
\end{equation}

\begin{marginfigure}
\includegraphics[width=\linewidth]{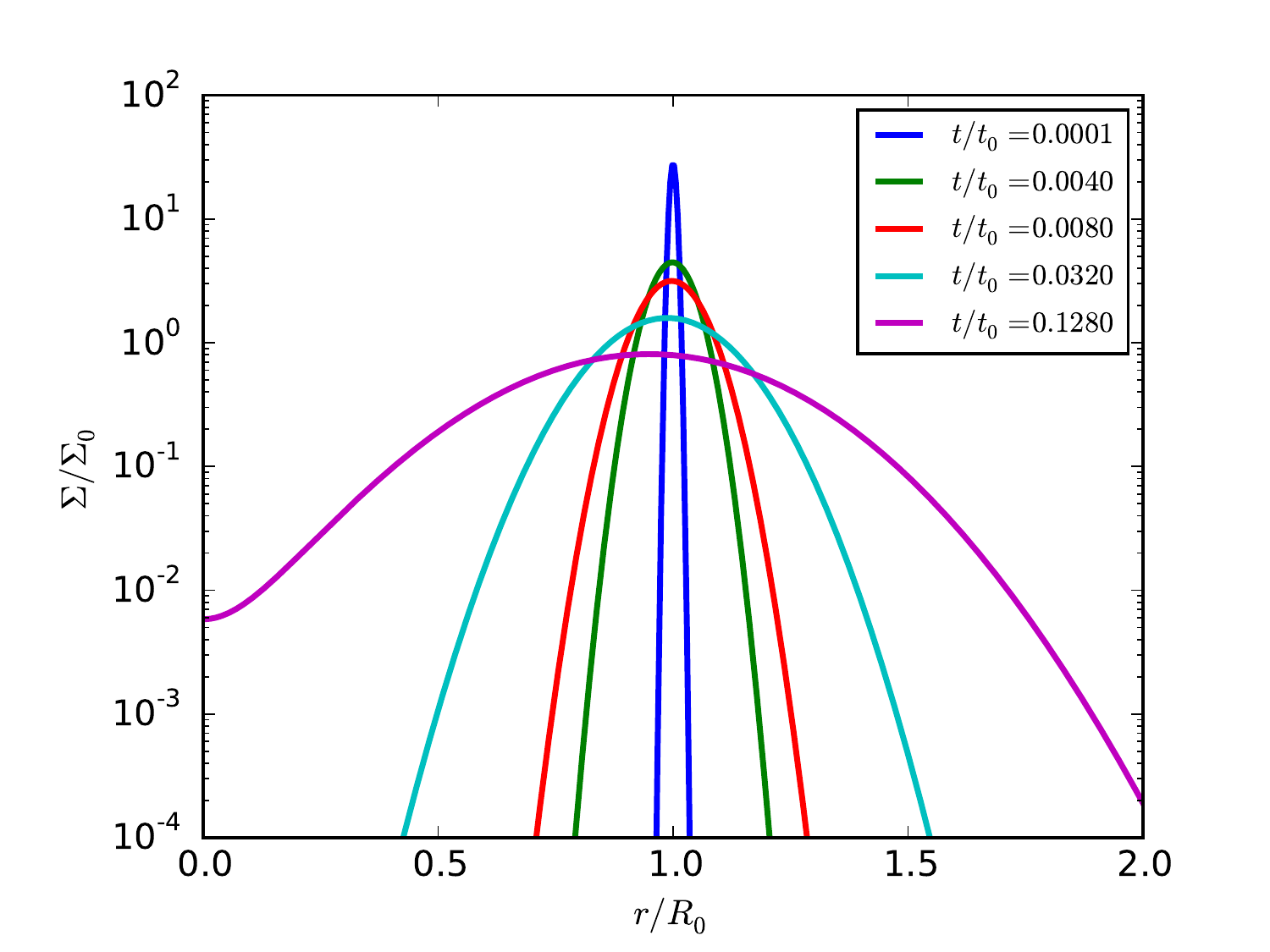}
\caption[Viscous ring evolution]{
\label{fig:viscring}
Analytic solution for the viscous ring of material with constant kinematic viscosity $\nu$. At time $t=0$, the column density distribution is $\Sigma = \Sigma_0 \delta(r-R_0)$. Colored lines show the surface density distribution at later times, as indicated in the legend. Times are normalized to the characteristic viscous diffusion time $t_0 = R_0^2/12\nu$. The analytic solution shown is that of \citet{pringle81a}.
}
\end{marginfigure}

These equations can be solved numerically for a given value of $\nu$, and they can be solved analytically in special cases, but it is useful to examine their general behavior first. First, note that equation (\ref{eq:disk_keplerian}) for the evolution of the surface density $\Sigma$ involves a partial time derivative on the left hand side and a second spatial derivative on the right hand side. This is the form of a diffusion equation; because of the extra factor of $\varpi$ and $\nu$ inside the spatial derivatives, it is a non-linear diffusion equation, meaning that the diffusivity is not constant in space. However, the behavior is qualitatively unchanged by the non-linearity, and the system still behaves diffusively. Thus if we start with a sharply peaked $\Sigma$, say a surface density that looks like a ring, it will spread it out. In fact, the case in which $\nu$ is constant and $\Sigma\propto \delta(\varpi-R_0)$ at time 0 can be solved analytically. The analytic solution is shown in Figure \ref{fig:viscring}.

How quickly to do rings spread, and does mass move inward? To answer that, we must evaluate $\partial \mathcal{T}/\partial \varpi$ under the assumption that $\Sigma$ and $\nu$ are relatively constant with radius, so we can take them out of the derivative, and that the disk is in steady state. We again assume Keplerian background rotation to evaluate $\Omega$. Under these assumptions
\begin{equation}
\frac{d\mathcal{T}}{d\varpi} = 2\pi \Sigma \nu \frac{d}{d\varpi}\left(\varpi^3 \frac{d\Omega}{d\varpi}\right) = -3\pi\Sigma \nu \frac{d}{d\varpi}(\varpi^2\Omega) = -3\pi \Sigma \nu \frac{dj}{d\varpi},
\end{equation}
and the angular momentum evolution equation (\ref{eq:angmom_disk}) trivially reduces to
\begin{equation}
\dot{M} = 3\pi\Sigma\nu.
\end{equation}
Thus the accretion rate is just proportional to the viscosity and the disk surface density. The radial velocity of the material under these assumptions is
\begin{equation}
v_\varpi = -\frac{3}{2} \frac{\nu}{\varpi}.
\end{equation}
The time required for a given fluid element to reach the star, therefore, is $t_{\rm acc}\sim \varpi/v_\varpi \sim \varpi^2/\nu$.

\paragraph{The $\alpha$ model.}

Before delving into the physical origin of the viscosity, it is helpful to non-dimensionalize the problem. We will write down the viscosity in terms of a dimensionless number called $\alpha$, following the original model first described by \citet{shakura73a}. The model is fairly straightforward. The viscous stress  $T_{\varpi\phi}$ has units of a pressure, so let us normalize it to the disk pressure. This is not as arbitrary as it sounds. If, for example, the mechanism responsible for producing fast angular momentum transport is fluctuating magnetic fields, then we would expect the strength of this effect to scale with the energy density in the magnetic field, which is in turn proportional to the magnetic pressure. Similar arguments can be made for other plausible mechanisms.

The $\alpha$-disk ansatz is simply to set
\begin{equation}
T_{\varpi\phi} = -\alpha p \frac{\Omega/\varpi}{d\Omega/d\varpi},
\end{equation}
where the dimensionless factor $(\Omega/\varpi)/(d\Omega/d\varpi)$ is inserted purely for convenience. This is equivalent to setting
\begin{equation}
\nu = \frac{T_{\varpi\phi}}{\rho \varpi (d\Omega/d\varpi)} = \frac{\alpha c_s^2}{\Omega} = \alpha c_s H,
\end{equation}
where $c_s$ is the sound speed in the disk and $H = c_s/\Omega$ is the disk scale height. Note that $c_s$ and $H$ include both thermal pressure and magnetic pressure. If we now substitute this into our simplified expression for $\dot{M}$, we get an accretion rate
\begin{equation}
\label{eq:mdot_steady}
\dot{M} = 3\pi \Sigma \alpha c_s H = 3\pi \Sigma \alpha \frac{c_s^2}{\Omega}
\end{equation}
This if we know the disk thermal structure, i.e.\ we know $c_s$ and $H$, and we know its surface density $\Sigma$, then $\alpha$ tells us its accretion rate.

The physical meaning of this result becomes a bit clearer if we put in an order of magnitude estimate that $\Sigma \approx M_d / R_d^2$, where $M_d$ and $R_d$ are the disk mass and radius. Putting this in we have
\begin{equation}
\dot{M} \approx \alpha \frac{M_d}{R_d^2} \frac{c_s^2}{\Omega}
\end{equation}
If we define $t_{\rm acc} = M_d/\dot{M}$ as the accretion timescale (the characteristic time to accrete the entire disk), $t_{\rm cross} = R_d/c_s$ as the sound crossing time of the disk, and $t_{\rm orb} = 2\pi/\Omega$ as the disk orbital period, then with a little algebra it is easy to show that this expression reduces to
\begin{equation}
t_{\rm acc} = \frac{1}{\alpha} \left(\frac{t_{\rm cross}}{t_{\rm orb}}\right)^2 t_{\rm orb}.
\end{equation}
Thus the time required to drain the disk is of order $(1/\alpha)(t_{\rm cross}/t_{\rm orb})^2$ orbits. Note that $t_{\rm orb} \ll t_{\rm cross}$, because orbital speeds are highly supersonic (as they must be for a thin disk to form). In a disk with $\alpha = 1$, the number of orbits required to drain the disk is this ratio squared, and with $\alpha < 1$ it takes longer, with the number of orbits required scaling as $\alpha^{-1}$. Based on observations of the accretion rates in disks and these properties, \citet{hartmann98a} estimate that $\alpha \sim 10^{-2}$ in nearby T Tauri star disks. It is probably larger at earlier phases in the star formation process.

\subsection{Physical Origins of Disk Viscosity}

We have established that there must be a viscous mechanism to transport angular momentum and mass through accretion disks, and we have even estimated its strength from observations, but we have not yet specified what that mechanism is.

\paragraph{Ordinary fluid viscosity.}

The obvious place to start is to examine the ordinary hydrodynamic viscosity we expect all fluids to have. The kinematic viscosity of a diffuse gas is $\nu = 2\overline{u}\lambda$, where $\overline{u}$ is the RMS particle speed and $\lambda$ is the mean free path. Let us consider a protostellar accretion disk with the typical properties density $n=10^{12}$ cm$^{-3}$ and temperature 100 K. In this case the velocity $\overline{u} = 0.6$ km s$^{-1}$, and assuming a particle-particle cross section of $\sigma=(1\mbox{ nm})^2$, the mean free path is $\lambda \sim 1/(n\sigma)=100$ cm, and $\nu \sim 10^8$ cm$^{-2}$ s. We can put this in terms of $\alpha$ if we remember that $\nu = \alpha (c_s^2/\Omega)$. If the material under consideration is orbiting 100 AU from a 1 $\msun$ star, then $\Omega = 6.3\times 10^{-3}$ yr$^{-1}$, and we have $\alpha \approx 6\times 10^{-12}$.

This obviously a problem. Suppose the gas starts out $\sim 100$ AU from the star. The time required for the gas to accrete is then $t_{\rm acc}\sim \varpi^2/\nu \sim (100\mbox{ AU})^2/\nu \sim 10^{22}$ s, or just shy of $10^{15}$ yr. In other words, longer than the age of the universe. The obvious conclusion from this is that ordinary hydrodynamic viscosity is completely ineffective at producing accretion. If that were the only source of angular momentum transport in a disk, then stars would never form. Something else must be at work.

\paragraph{Turbulent hydrodynamic viscosity.}

One possible explanation solution to this problem is turbulent hydrodynamic viscosity. If there are large-scale radial motions within a disk, then the effective value of $\overline{u}\lambda$ could be significantly larger than the microphysical one we calculated. In effect, these motions will mix material from different radii within the disk, exchanging angular momentum between inner and outer parts of the disk. This would require the existence of an instability capable of generating and sustaining large radial turbulent motions. Although several such mechanisms have been proposed, it is essentially impossible to determine the amount of angular momentum transport that will be produced based on purely analytic calculations. That is because the transport will depend on the non-linear saturation amplitude of any instability, which is not something that one can generally determine analytically.

Numerical simulations have been attempted, and seem to find that hydrodynamic mechanisms do not produce significant angular momentum transport, but they may be compromised by limited resolution. In a numerical simulation, the maximum possible Reynolds number is set by the ratio of the size of the computational domain to the size of a grid cell, since flows are always smoothed on the grid scale. Even for the largest calculations ever performed this is at most a few thousand, whereas we have seen that the Reynolds numbers in real astrophysical systems are typically $\sim 10^9$. Thus, if the saturation were Reynolds number-dependent, numerical simulations would not get it right.

The question of whether hydrodynamic mechanisms could be responsible for angular momentum transport is sufficiently complex and interesting that the latest frontier is laboratory experiment. Researchers construct counter-rotating cylinders filled with a fluid, and set the cylinders rotating to produce a Keplerian-like rotation profile. They then measure the force exerted on the inner and outer cylinders to measure the rate of angular momentum transport. The laboratory experiments can reach Reynolds numbers of $\sim 10^6$, and seem to find negligible transport, $\alpha < 10^{-6}$ \citep{ji06a}. Given these results, most researchers are convinced that purely hydrodynamic mechanisms cannot explain the observed lifetimes and rates of angular momentum transport in disks. Instead, some other mechanism is required.

\paragraph{Magneto-rotational instability.}

Magnetic fields offer one opportunity for angular momentum transport. We have already mentioned magnetic braking as a possibility, but that requires that the matter be well-coupled to the field, and that the field be dragged inward into the disk so that there is a large net flux. This may not happen due to non-ideal MHD effects, however.

Another mechanism is possible that does not require a large net flux, and that allows weaker (although not zero) coupling. This is the magneto-rotational instability (MRI), first discovered mathematically by \citet{chandrasekhar61a}, and later re-discovered and applied to astrophysical systems by \citet{balbus91a}. The full theory of MRI has been explored extensively both analytically and numerically.

The basic idea is that magnetic field lines threading the disk connect annuli at different radii. As the disk rotates and the annuli shear, this stretches the magnetic field line connecting them. This causes an opposing magnetic tension, which attempts to force the two points to stay close together, and thus to force them into co-rotation. This speeds up the outermost fluid element, which is falling behind, and slows down the innermost one, and thus it moves angular momentum outward. However, when one removes angular momentum from a fluid element it tends to fall toward the center, so the innermost fluid element falls even closer to the star. Similarly, the outermost fluid element gains angular momentum, and so it wants to move outward. This increases the tension even more, and the system goes unstable due to this positive feedback loop.

Simulators are still working to try to come up with a general result about the value of $\alpha$ produced by the MRI, but in at least some cases $\alpha$ as high as $0.1$ seems to be possible. This would nicely explain the observed accretion rates and lifetimes of T Tauri star disks. MRI is not the end of the story, however. The problem with MRI is that it only operates as long as matter is sufficiently coupled to the magnetic field, which in turn depends on its ionization state. MRI will only operate if the mechanism we have described is able to generate turbulent fluctuations in the magnetic field to transport angular momentum. In turn, this requires that no non-ideal mechanism, of which there are several possibilities, be able to smooth out the field over the scale of the accretion disk.

\begin{marginfigure}
\includegraphics[width=\linewidth]{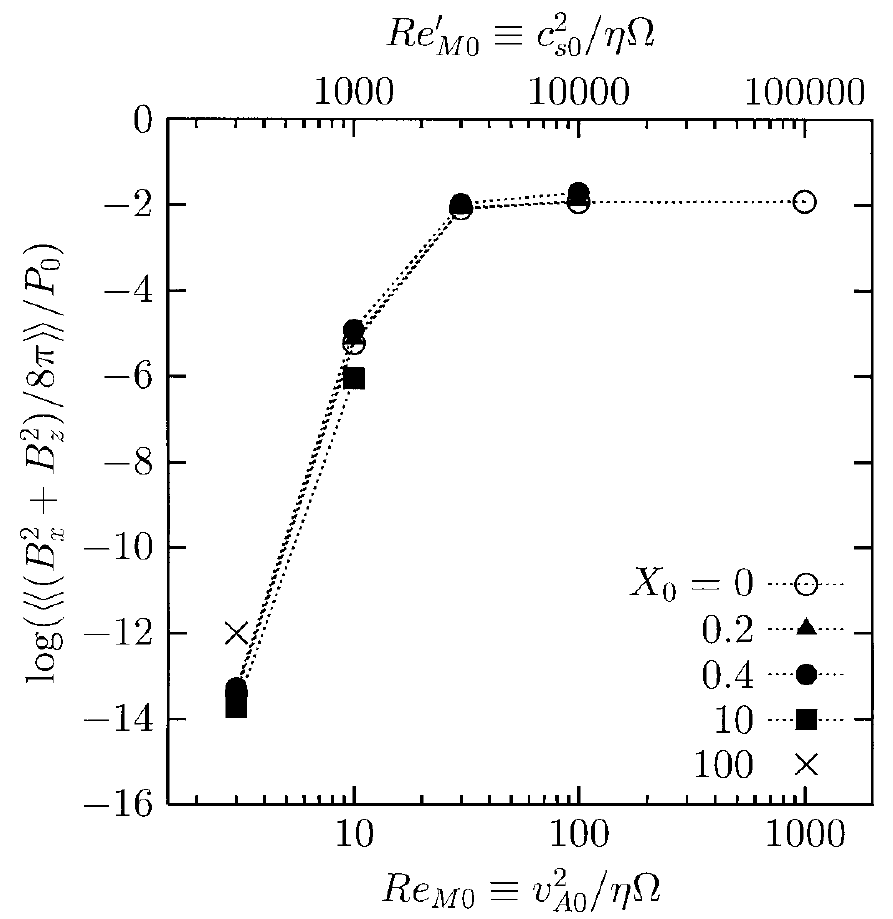}
\caption[Maxwell stress in non-ideal MHD simulations of the MRI]{
\label{fig:mri_sano02}
Results from a series of simulation of magneto-rotational instability with non-ideal MHD. The $y$ axis shows the mean Maxwell stress measured in the simulation once it reaches statistical steady state, normalized by the gas pressure. This is roughly the same as $\alpha$. Simulations are shown at a range of magnetic Reynolds numbers $\mbox{Re}_{\mathrm{M}}$. Different values of the parameter $X_0$ correspond to different strengths of Hall diffusivity. Credit: \citet{sano02a}, \copyright AAS. Reproduced with permission.
}
\end{marginfigure}

The question of how well coupled the gas and the field are turns on details of the ionization structure of the disk. Turbulence requires large values of the magnetic Reynolds number $\mbox{Re}_{\rm M} = v_A^2 /(\eta\Omega)$, where $v_A$ is the Alfven speed and $\eta$ is the magnetic diffusivity. Numerical experiments suggest that MRI shuts of when $\mbox{Re}_{\rm M} \lesssim 3000$ (Figure \ref{fig:mri_sano02}). The diffusivity, in turn, depends on the electron fraction: $\eta = c^2/(4\pi \sigma_e)$, where $\sigma_e = n_e e^2/(m_e \nu_c)$ is the conductivity and $\nu_c$ is the frequency of electron-neutral collisions. If there are few electrons, $\sigma_e$ is small and $\eta$ is large, making $\mbox{Re}_{\rm M}$ small. This means that, in order to know where and whether MRI will operate, we need to know the ionization fraction in the disk.

This is an incredibly complex problem, because the ionization is generally non-thermal, non-LTE, and it only takes a tiny number of electrons to make MRI operate: an electron fraction $\sim 10^{-9}$ is sufficient. In the very inner disk near the star, where the temperature is $\sim 2000$ K, thermal ionization of alkali metals will provide the necessary electrons. In regions where the disk column density is $\lesssim 100$ g cm$^{-2}$, X-rays from the central star and cosmic rays can penetrate the disk, providing free electrons. However, for comparison the estimated column density of the minimum mass Solar nebula (the minimum mass required to make all the planets -- to be discussed further in Chapter \ref{ch:late_disk}) is 1700 g cm$^{-2}$ at 1 AU, and the equilibrium temperature is $\ll 2000$ K.

In such high column density, cool regions, the electron fraction depends on such complex questions as the mean size of dust grains (since these can absorb free electrons) and the rate of vertical transport of electrons from the surface layers down to the disk midplane. One possible result of this is that MRI would operate only at the surface of disk, leaving the midplane a "dead zone". Another possibility is that there may be radial dead zones with no MRI. Material would move inward to such regions, but then get stuck there, potentially making accretion bursty.

\paragraph{Gravitational transport mechanisms.}

Magnetic fields provide on potential source of transport, but, as we have seen, they may fail if the gas is not sufficiently ionized. If the accretion rate onto the disk is large enough, it is also possible that MRI may operate, but not may not provide sufficiently rapid angular momentum transport to stop gas from building up the disk -- this is likely to occur particularly for massive stars. In this case, mass can build up in the disk, leading to gravitational instability.

We can understand when gravitational instability is likely to set in using our theory of disks. In steady state we showed that the accretion rate depends on $\Sigma$, $\alpha$, $c_s$, and $\Omega$ (equation \ref{eq:mdot_steady}). Since we are interested in gravitational stability, let us introduce the Toomre $Q$ parameter for our disk,
\begin{equation}
Q = \frac{\Omega c_s}{\pi G \Sigma}
\end{equation}
where we have $\Omega$ rather than $\sqrt{2}\Omega$ because the rotation curve is Keplerian rather than flat as in a galactic disk, and where we are assuming the non-thermal velocity dispersion in the disk is subsonic. Solving for $\Sigma$ in terms of $Q$ and inserting the result into equation (\ref{eq:mdot_steady}), we obtain
\begin{equation}
\dot{M} = \frac{3\alpha}{Q} \frac{c_s^3}{G}.
\end{equation}

Thus we see that, if $Q > 1$, so the disk is gravitationally stable, and $\alpha < 1$, as we expect for MRI or almost any other local transport mechanism, the maximum rate at which the disk can move matter inward is roughly $c_s^3/G$. This is also the characteristic rate at which matter falls onto the disk from a thermally-supported core, provided that we use the sound speed in the core rather than in the disk.

Normally disks are somewhat warmer than the cores around them, both because the star shines on the disk and because viscous dissipation in the disk releases heat. However, what this result shows is that, in any regions where the disk is not significantly warmer than the core that is feeding it, for example the outer parts of the disk where stellar and viscous heating are small, the disk cannot transport matter inward as quickly as it is fed. The result will be that the surface density will rise and $Q$ will decrease, giving rise to gravitational instability.

This can in turn generate transport of angular momentum via gravitational torques. Transport of this sort comes in two flavors: local and global. Local instability happens when the disk clumps up on small scales due to its own gravity. This depends on the Toomre $Q$ of the disk. If $Q\sim 1$ it will begin to clump up, and these clumps can transport angular momentum by interacting with one another gravitationally, sending mass inward and angular momentum outward. However, this clumping will heat the gas via the release of gravitational potential energy, which in turn tends to drive $Q$ back to higher values.

What happens then depends on how the gas radiates away the excess energy. If the radiation rate is too low, the gas will heat up until it smoothes out, and the Toomre $Q$ will be pushed up. If it is too high, the gas will fragment entirely and collapse into bound objects in the disk -- something like what happens in a galactic disk. If it is in between, the disk can enter a state of sustained gravitationally-driven turbulence in which there is no fragmentation but the rate of heating by compression balances the rate of radiation, and there is a net transport of mass and angular momentum.

The global variety of gravitational instability occurs when the disk clumps up on scales comparable to the entire disk. This occurs when the disk mass becomes comparable to the mass of the star it is orbiting; instability sets in at disk masses of $30-50\%$ of the total system mass. This generally manifests as the appearance of spiral arms. These instabilities transport angular momentum because the disk is no longer axisymmetric, and instead has a significant moment arm that can exert torques, or on which torques can be exerted. Transport of angular momentum can then occur in several ways. If there is an envelope outside the disk, the disk can spin up the envelope, sending angular momentum outward in that way.

The disk can also transfer angular momentum to the star by forcing the star to move away from the center of mass. In this configuration the disk develops a one-armed spiral, and the star in effect goes into a binary orbit with the overdensity in the disk. In this case angular momentum is transported inward rather than out, with the excess angular momentum going into orbital motion of the star. This phenomenon is known as the Sling instability \citet{shu90a}.

\section{Outflow Launching}

The final topic for this chapter is how and why disks launch the ubiquitous jets and winds that observations reveal. The topic of jets is not limited to the star formation context, of course, and much of the theory for it was originally developed in the context of active galactic nuclei and compact objects. We will only scratch the surface of this theory here. Our goal is just to get a general understanding of how and why we expect winds to be launched from disks, and what general properties we expect them to have.

\subsection{Mechanisms}

We begin by considering what mechanisms could be responsible for launching winds. We can start by discarding the two mechanisms that we usually invoke to explain the winds of main sequence stars. All main sequence stars, including the Sun, produce winds. For low mass stars like the Sun, the driving mechanism is thermal. MHD waves propagating into the low-density solar corona heat the gas to temperatures of up to $\sim 10^6$ K. The high pressure in the hot region drives flows of gas outward; for the Sun, the mass loss rate is roughly $10^{-14}$ $\msun$ yr$^{-1}$, and the mechanical luminosity is $\sim 10^{-4}$ $\lsun$. In contrast, the mechanical power input required to explain the observed outflows from young stars is closer to $\sim 0.1$ $\lsun$. This is far greater than the thermal energy available in the hot X-ray corona of the star -- a corona capable of providing this much power would exceed the total stellar bolometric output.

For massive main sequence stars, the main driving mechanism is the pressure exerted by stellar photons on the gas. The problem with this mechanism is the momentum budget. Observed outflow momentum is generally $1-2$ orders of magnitude larger than $L/c$, the amount of momentum available in the stellar radiation field. In contrast, for the winds of main sequence stars the outflow momentum flux is always $\lesssim L/c$. Thus the stellar photon field does not have enough momentum to drive the observed outflows of young stars. Moreover, neither the thermal winds of low mass main sequence stars, nor the radiatively-driven winds of massive ones, show highly collimated features like the HH jets.

Having discarded these two mechanisms, we must seek an alternative source of energy. The most natural one is the gravitational potential energy being liberated by the accretion flow, which, combined with magnetic fields, can produce highly collimated outflows. The question then becomes exactly how the combination of gravitational power and magnetic fields produces the observed outflows.

\subsection{Stability Analysis for Magnetocentrifugal Winds}

There are a range of theoretical models for the exact mechanism by which winds are launched. However, the general picture of all of these mechanisms is to combine centrifugal force with magnetic fields. Consider a disk of material in Keplerian orbit, and consider an open field line passing through the disk; here by "open" we mean that the field line does not loop back into the disk, but instead goes out, formally to infinity. We write the field in the vicinity of the disk as the sum of a poloidal and a toroidal component,
\begin{equation}
\vecB = \vecB_p + B_\phi \hat{e}_\phi
\end{equation}
The field exerts negligible forces within the disk, but for the much lower-density region above the disk (the corona), magnetic forces are non-negligible.

Let us consider a test fluid element that is, for whatever reason, lofted slightly above the disk, into the corona. We will assume ideal MHD, so the fluid element is constrained to move only along the field line. We can think of the test fluid element as a bead stuck on a wire. We will further assume that the density of material above the disk is very small, so that magnetic forces dominate and the field simply rotates as a rigid body. Now let us consider how this fluid element will evolve in time.

In a frame co-rotating with the launch point of the fluid element, there are two potentials to worry about: the gravitational potential of the central star, and the centrifugal potential that arises from the fact that we have chosen to work in a rotating reference frame. The former is simply the usual
\begin{equation}
\psi_{g} = -\frac{GM_*}{\sqrt{\varpi^2 + z^2}},
\end{equation}
where $M_*$ is the star's mass, and we are working in cylindrical coordinates. For the latter, we are working in a frame rotating at an angular velocity equal to the Keplerian value at the fluid element's launch point $\varpi_0$, which is $\Omega = \sqrt{GM_*/\varpi_0^3}$. The centrifugal potential is therefore
\begin{equation}
\psi_c = -\frac{1}{2}\Omega^2 \varpi^2 = -\frac{1}{2} G M_* \frac{\varpi^2}{\varpi_0^3}.
\end{equation}
Thus the total potential is
\begin{equation}
\psi = -\frac{GM_*}{\varpi_0} \left[\frac{1}{2}\left(\frac{\varpi}{\varpi_0}\right)^2 + \frac{\varpi_0}{\sqrt{\varpi^2+z^2}} \right].
\end{equation}

To determine the evolution of the test fluid element, we must consider the forces associated with this potential. The force per unit mass  is simply minus the gradient of the potential, and thus we have
\begin{equation}
\mathbf{f} = -\nabla \psi = -GM_* \left\{ \varpi \left[\frac{1}{(\varpi^2+z^2)^{3/2}} - \frac{1}{\varpi_0^3}\right]\hat{e}_\varpi + \frac{z}{(\varpi^2+z^2)^{3/2}} \hat{e}_z \right\}
\end{equation}
If we plug in our starting point, $\varpi = \varpi_0$ and $z=0$, we see that the gradient is exactly zero, which is what we expect: the starting point is, by assumption, in equilibrium between centrifugal and gravitational forces.

Now let us consider our perturbed fluid element. It has been moved a distance $ds$ from its starting point, and it is moving along the field line, which has radial and vertical components $B_\varpi$ and $B_z$. For convenience, let us define the angle of the field line relative to the horizontal by
\begin{equation}
\cos \theta = \frac{B_\varpi}{\sqrt{B_\varpi^2 + B_z^2}}.
\end{equation}
An angle $\theta=90^\circ$ corresponds to a field line that has zero radial component, and $\theta=0^\circ$ corresponds to one that has zero vertical component. The coordinates of the displaced fluid elements are $\varpi = \varpi_0 + \cos\theta \, ds$ and $z = \sin\theta \, ds$. To determine the force experienced by the fluid element, we simply plug these coordinates into $\mathbf{f}$ and expand to first order:
\begin{equation}
d\mathbf{f} = \frac{GM_*}{\varpi_0^3} \left(3 \cos\theta\,\hat{e}_\varpi - \sin\theta\,\hat{e}_z\right) ds.
\end{equation}

We are interested in the component of this force parallel to the field line, since the fluid element is constrained to move along the field line. That is, we are interested in
\begin{equation}
df_\parallel = d\mathbf{f} \cdot (\cos\theta\,\hat{e}_\varpi +  \sin \theta\,\hat{e}_z) = \frac{GM_*}{\varpi_0^3} \left(3\cos^2 \theta - \sin^2 \theta\right) ds.
\end{equation}
The force is therefore positive, indicating that it is pushing the fluid element further away from the launch point, if
\begin{equation}
3 \cos^2 \theta - \sin^2 \theta > 0
\qquad\Longrightarrow\qquad
\theta < 60^\circ.
\end{equation}
We have therefore derived a condition under which a disk threaded by open field lines will be unstable to the formation of a wind. If the field lines make an angle of $<60^\circ$ off the plane, then any fluid element that is lofted infinitesimally above the disk will be forced further down the field line by the centrifugal force, forming a wind.

\subsection{Properties of the Wind}

If the disk is unstable to wind formation, the next question is what properties that wind will have. We can answer this question at least approximately from the following elementary consideration. We have thus far assumed that the field lines above and below the disk are perfectly rigid, but of course that cannot be strictly true out to infinite radius. If we choose a large enough radius, then maintaining perfect solid body rotation would require a velocity larger than the speed of light, which is obviously forbidden by relativity. However, there is an even more restrictive limit: the field line can remain rigid only as long as the matter attached to it has negligible inertia. If the inertia of the material is significant, it will slow down the field lines, causing them to deviate from rigid rotation.

Recalling our dimensional analysis of the MHD equations in Chapter \ref{ch:turbulence}, the relative importance of the terms describing inertia and magnetic force is determined by the Alfv\'en Mach number,
\begin{equation}
\mathcal{M}_A \sim \frac{v}{v_A},
\end{equation}
where $v_A = B/\sqrt{4\pi \rho}$ is the Alfv\'{e}n speed. The material starts at zero velocity, and accelerates as it moves outward, so that $\mathcal{M}_A$ increases along any given field line. We expect that the field lines will cease to be rigid once the material along them is accelerated to a velocity such that $\mathcal{M}_A \sim 1$. This transition between sub- and super-Alfv\'{e}nic motion will occur at a critical radius $\varpi_A$ (which is not necessarily the same along every field line), called the Alfv\'{e}n radius.

Once the field line starts to unwind at $\varpi_A$, it will no longer be able to impart significant angular momentum or energy to the fluid parcels that travel along it. Returning to our bead on a wire analogy, it is as if the rigid wires that are accelerating the beads are beginning to bend. We therefore expect the terminal velocity of the wind to be of order the wind speed at $\varpi_A$, which is
\begin{equation}
v_\infty \sim \Omega_0 \varpi_A \sim v_{K,0} \frac{\varpi_A}{\varpi_0},
\end{equation}
where $\Omega_0$ is the angular velocity at the launch point, and $v_{K,0}$ is the Keplerian velocity at that point. Thus the wind speed is comparable to the Keplerian speed times a factor of order the ratio of the Alfv\'{e}n radius to the launch radius.

The specific angular momentum of the material ejected in the wind will be
\begin{equation}
j_w \sim \varpi_A v_\infty \sim v_{K,0} \frac{\varpi_A^2}{\varpi_0}.
\end{equation}
For comparison, the specific angular momentum of the material that remains in the disk is
\begin{equation}
j_d = v_{K,0} \varpi_0.
\end{equation}
Thus the specific angular momentum of wind material exceeds that of disk material by a factor of
\begin{equation}
\frac{j_w}{j_d} \sim \left(\frac{\varpi_A}{\varpi_0}\right)^2.
\end{equation}
One factor of $\varpi_A/\varpi_0$ comes from the greater level arm of the material being launched into the wind, and the second factor comes from the greater velocity.

If the wind is predominantly responsible for removing the angular momentum of the disk and allowing accretion, this implies that the rates of mass accretion $\dot{M}$ and wind launch $\dot{M}_w$ must be related by
\begin{equation}
\dot{M} \sim \left(\frac{\varpi_A}{\varpi_0}\right)^2 \dot{M}_w.
\end{equation}
Thus wind launching provides an efficient means to allow accretion, since for even a relatively modest Alfv\'{e}n radius, say $\varpi_A/\varpi_0 \sim 3$, it will enable accretion to occur using only $\sim 1/10$ of the available accreting mass. Of course we have not self-consistently calculated $\varpi_A$, and we will not do so here. Schematically, one must do so by taking a wind mass launching rate (called the mass loading) as a function of radius, and then self-consistently solving for the structure of the magnetic field and the velocity above the disk. The Alfv\'{e}n radius then appears as a critical point of the solution along each streamline / magnetic field line. The first such self-consistent calculation was provided by \citet{blandford82a}, although this calculation still had to leave the mass loading as a free parameter.

\chapter{Protostar Formation}
\label{ch:protostar_form}

\marginnote{
\textbf{Suggested background reading:}
\begin{itemize}
\item \href{http://adsabs.harvard.edu/abs/2014prpl.conf..173L}{Dunham, M.~M., et al. 2014, in "Protostars and Planets VI", ed.~H.~Beuther et al., pp.~195-218}, sections 1-4 \nocite{dunham14a}
\end{itemize}
\textbf{Suggested literature:}
\begin{itemize}
\item \href{http://adsabs.harvard.edu/abs/2013ApJ...763....6T}{Tomida, K., et al., 2013, ApJ, 763, 6} \nocite{tomida13a}
\end{itemize}
}

The next two chapters focus on the structure and evolution of protostars. Our goal will be to understand when and why collapse stops, leading to formation of a pressure-supported object, and how those objects subsequently evolve into main sequence stars. This chapter focuses on the dynamics and thermal behavior of the material at the center of a collapsing core as it settles into something we can describe as a star, and on the structure of the envelope around this protostar. Chapter \ref{ch:protostar_evol} is focused on the evolution of this object, both internally and in its appearance on the HR diagram. 

\section{Thermodynamics of a Collapsing Core}

We will begin by considering what happens at the center of a collapsing core where the density in is rising rapidly material collapses. 

\subsection{The Isothermal-Adiabatic Transition}
\label{ssec:iso_adiabat}

Thus far we have treated the gas in star-forming regions as approximately isothermal, but this assumption must break down at some point. At low density there are minor deviations from isothermality that result from the density dependence of various heating and cooling processes, but these are fairly minor, in the sense that they are unable to significantly impede collapse. For example, the proposed \citet{larson05a} EOS discussed in chapter \ref{ch:imf_th} only gets as stiff as $T\propto \rho^{0.07}$ at high density, corresponding to a polytrope $P\propto \rho^{\gamma}$ with $\gamma=1.07$. Spherical objects can only be stable if $\gamma>4/3$, so gas $\gamma=1.07$ is still in the unstable regime. In contrast, if the gas is not able to radiate at all, it will behave adiabatically. This means it will approach a polytrope with $\gamma=7/5$ or $5/3$, depending on whether the gas temperature is high enough to excite the rotational and vibrational levels of H$_2$ or not.\footnote{In actuality the value of $\gamma$ for H$_2$ is more complicated than that, but this detail is unimportant for our purposes.} Either of these values is $>4/3$, and thus sufficient to halt collapse.

Let us make some estimates of when deviations from isothermality that are significant enough to slow collapse will occur. Since we are dealing with the collapse of the first region to fall in, we can probably safely assume that this material has very low angular momentum and treat the collapse as spherical -- higher angular momentum material will only fall in later, since removal of angular momentum by the disk takes a while. The behavior of this material has been studied by a number of authors, going all the way back to \citet{larson69a}, but the treatment here follows that of \citet{masunaga98a} and \citet{masunaga00a}.

At high densities inside a core immediately before a central star forms and begins to radiate, the dominant source of energy is adiabatic compression of the gas. Let $e$ be the thermal energy per unit mass of a particular gas parcel, and let $\Gamma$ and $\Lambda$ be the rates of change in $e$ due to heating and cooling processes, i.e.,
\begin{equation}
\frac{de}{dt} = \Gamma - \Lambda.
\end{equation}
As the gas collapses it will heat up due to adiabatic compression. The first law of thermodynamics tells us that the heating rate due to this process is
\begin{equation}
\label{eq:gamma_ad}
\Gamma = -p \frac{d}{dt}\left(\frac{1}{\rho}\right),
\end{equation}
where $\rho$ and $p=\rho c_s^2$ are the gas density and pressure, and $c_s$ is the isothermal sound speed. Since $1/\rho$ is the specific volume, meaning the volume per unit mass occupied by the gas, this term is just $p\,dV$, the work done on the gas in compressing it. If the gas is collapsing in free-fall, the compression time scale is about the free-fall timescale $t_{\rm ff} = \sqrt{3\pi/32G\rho}$, so we expect
\begin{equation}
\Gamma = C_1 c_s^2 \sqrt{4\pi G\rho},
\end{equation}
where $C_1$ is a constant of order unity that will depend on the exact collapse solution, and the factor of $\sqrt{4\pi}$ has been inserted for future convenience.

The main cooling source is thermal emission by dust grains, which at the high densities with which we are concerned are thermally very well coupled to the gas. Let us first consider the case where the gas is optically thin to this thermal radiation, so the cooling rate per unit mass is simply given by the rate of thermal emission,
\begin{equation}
\Lambda_{\rm thin} = 4\kappa_{\rm P} \sigma_{\rm SB} T^4.
\end{equation}
Here $\sigma_{\rm SB}$ is the Stefan-Boltzmann constant and $\kappa_{\rm P}$ is the Planck mean specific opacity of the gas-dust mixture. As long as $\Lambda \gtrsim \Gamma$, the gas will remain isothermal. (Strictly speaking if $\Lambda > \Gamma$ the gas will cool, but that is because we have left out other sources of heating, such as cosmic rays and the fact that the gas and dust are bathed in a background IR radiation field from other stars.) If we equate the heating and cooling rates, using for $T$ the temperature in the isothermal gas, we therefore will obtain a characteristic density beyond which the gas can no longer remain isothermal. Doing so gives
\begin{eqnarray}
\rho_{\rm thin} & = & \frac{4}{\pi} \frac{\kappa_{\rm P}^2 \sigma_{\rm SB}^2 \mu^2 m_{\rm H}^2 T^6}{C_1^2 G k_B^2} \\
& = & 5\times 10^{-15} \mbox{ g cm}^{-3}\, C_1^{-2} \kappa_{\rm P,-2}^2 T_1^6
\end{eqnarray}
where $\mu$ is the mean mass per particle in units of $m_{\rm H}$ ($\mu=2.3$ for fully molecular gas), and we have set $c_s = \sqrt{k_B T/\mu m_{\rm H}}$. In the second line, $T_1 = T/10$ K and $\kappa_{\rm P,-2} = \kappa_{\rm P}/0.01$ cm$^2$ g$^{-1}$, a typical value for thermal radiation at a temperature of $\approx 10$ K and Milky Way dust grains. Thus we find that compressional heating and optically thin cooling to balance at about $10^{-14}$ g cm$^{-3}$.

A second important density is the one at which the gas starts to become optically thick to its own re-emitted infrared radiation. Suppose that the optically thick region at the center of our core has some mean density $\rho$ and radius $R$. The condition that the optical depth across it be unity then reduces to
\begin{equation}
\label{eq:thick_thin}
2 \kappa_{\rm P} \rho R \approx 1.
\end{equation}
If this central region corresponds to the size of the region that is no longer in free-fall collapse and is instead thermally supported, then its size must be comparable to the Jeans length at its lowest temperature, i.e., $R\sim \lambda_J = \sqrt{\pi c_s^2/(G\rho)}$. Thus we set
\begin{equation}
R = C_2 \frac{2\pi c_s}{\sqrt{4\pi G \rho}},
\end{equation}
where $C_2$ is again a constant of order unity, and Masunaga et al.\ find based on numerical results that $C_2\approx 0.75$. Plugging this value of $R$ into equation (\ref{eq:thick_thin}), we obtain the characteristic density at which the gas transitions from optically thin to optically thick,
\begin{eqnarray}
\rho_{\tau\sim 1} & = & \frac{1}{4\pi} C_2^{-2} \frac{\mu m_{\rm H} G}{\kappa_{\rm P}^2 k_B T} \\
& = & 
1.5\times 10^{-13}\mbox{ g cm}^{-3}\,  C_2^{-2} \kappa_{\rm P}^{-2} T_1^{-1}
\end{eqnarray}
This is not very different from the value for $\rho_{\rm thin}$, so in general for reasonable collapse conditions we expect that cores transition from isothermal to close to adiabatic at a density of $\sim 10^{-13}-10^{-14}$ g cm$^{-3}$.

It is worth noting that ratio of $\rho_{\rm thin}$ to $\rho_{\tau\sim 1}$ depends extremely strongly on both $\kappa_{\rm P}$ (to the 4th power) and $T$ (to the 7th), so any small change in either can render them very different. For example, if the metallicity is super-solar then $\kappa_{\rm P}$ will be larger, which will increase $\rho_{\rm thin}$ and decrease $\rho_{\tau\sim 1}$.
Similarly, if the region is somewhat warmer, for example due to the presence of nearby massive stars, then $\rho_{\rm thin}$ will increase and $\rho_{\tau\sim 1}$ will decrease.

If $\rho_{\tau\sim 1} < \rho_{\rm thin}$, the collapsing gas will become optically thick before heating becomes faster than optically thin cooling. In this case we must compare the heating rate due to compression with the cooling rate due to optically thick cooling instead of optically thin cooling. The cooling rate for an optically thick region is determined how quickly radiation can diffuse out. If we have a central region of optical depth $\tau\gg 1$, the effective speed of the radiation moving through it is $c/\tau$, so the time required for the radiation to diffuse out is
\begin{equation}
t_{\rm diff} = \frac{l\tau}{c} = \frac{\kappa_{\rm P} \rho l^2}{c}
\end{equation}
where $l$ is the characteristic size of the core. Inside the optically thick region matter and radiation are in thermal balance, so the radiation energy density approaches the blackbody value $a_R T^4$. The radiation energy per unit mass is therefore $a_R T^4/\rho$. Putting all this together, and taking $l=2 R$ as we did before in computing $\rho_{\tau\sim 1}$, the optically thick cooling rate per unit mass is
\begin{equation}
\Lambda_{\rm thick} = \frac{a_R T^4/\rho}{t_{\rm diff}} = \frac{\sigma_{\rm SB} T^4}{\kappa_{\rm P} \rho^2 R^2},
\end{equation}
where $\sigma_{\rm SB} =ca_R /4$. If we equate $\Lambda_{\rm thick}$ and $\Gamma$, we get the characteristic density where the gas becomes non-isothermal in the optically thick regime
\begin{eqnarray}
\rho_{\rm thick} & = & \left(\frac{C_1^2 G \sigma_{\rm SB}^2 \mu^4 m_{\rm H}^4 T^4}{4\pi^3 C_2^4 k_B^4 \kappa_{\rm P}^2}\right)^{1/3} \\
& = & 5\times 10^{-14}\mbox{ g cm}^{-3}\, \frac{C_1^{2/3}}{C_2^{4/3}} \kappa_{\rm P,-2}^{-2/3} T_1^{4/3}.
\end{eqnarray}
This is much more weakly dependent on $\kappa_{\rm P}$ and $T$, so we can now make the somewhat more general statement that, even for supersolar metallicity or warmer regions, we expect a transition from isothermal to adiabatic behavior somewhere in the vicinity of $10^{-14}-10^{-13}$ g cm$^{-3}$.

\subsection{The First Core}

The transition to an adiabatic equation of state, with $\gamma>4/3$, means that the collapse must at least temporarily halt. The result will be a hydrostatic object that is supported by its own internal pressure. This object is known as the first core, or sometimes a Larson's first core, after Richard Larson, who first predicted this phenomenon.

We can model the first core reasonably well as a simple polytrope, with index $n$ defined by $n=1/(\gamma-1)$. When the temperature in the first core is low, $\gamma\approx 5/3$ and $n\approx 3/2$, and for a more massive, warmer core $\gamma \approx 7/5$ ($n\approx 5/2$). The theory of polytropes can be found in many standard stellar structure textbooks \citep[e.g.,][]{chandrasekhar39a, kippenhahn94a}, and so we will not rehearse the topic here, and will simply quote the result. For a polytrope of central density $\rho_c$, the radius and mass are
\begin{eqnarray}
R & = & a \xi_1\\
M & = & -4\pi a^3 \rho_c \left(\xi^2\frac{d\theta}{d\xi}\right)_1,
\end{eqnarray}
where $\xi=r/a$ is the dimensionless radius, $\theta = (\rho/\rho_c)^{1/n}$ is the dimensionless density, the subscript $1$ refers to the value at the edge of the sphere (where $\theta=0$), the factors $\xi_1$ and $(\xi\, d\theta/d\xi)_1$ can be determined by integrating the Lane-Emden equation, and the scale factor $a$ is defined by
\begin{equation}
a^2 = \frac{(n+1) K}{4\pi G} \rho_c^{\frac{1-n}{n}}.
\end{equation}
The factor $K = p/\rho^{\gamma}$ is the polytropic constant, which is determined by the specific entropy of the gas.

For our first core, the specific entropy will just be determined by the density at which the gas transitions from isothermal to adiabatic. If we let $\rho_{\rm ad}$ be the density at which the gas becomes adiabatic, then the pressure at this density is $p = \rho_{\rm ad} c_{s0}^2$, where $c_{s0}$ is the sound speed in the isothermal phase, and $K = c_{s0}^2 \rho_{\rm ad}^{1-\gamma}$. For $\gamma=5/3$ ($n=1.5$) we have $\xi_1=3.65$ and $(\xi^2\, d\theta/d\xi)_1=-2.71$, and plugging in we get
\begin{eqnarray}
R & = & 2.2\mbox{ AU} \; T_1^{1/2} \rho_{c,-10}^{1/6} \rho_{\rm ad,-13}^{-1/3} \\
M & = & 0.059\,\msun\; T_1^{1/2} \rho_{c,-10}^{7/6} \rho_{\rm ad,-13}^{-1/3},
\end{eqnarray}
where $\rho_{c,-10} = \rho_c / 10^{-10}$ g cm$^{-3}$ and $\rho_{\rm ad,-13} = \rho_{\rm ad}/10^{-13}$ g cm$^{-3}$. Our decision to scale $\rho_c$ to $10^{-10}$ g cm$^{-3}$ will be justified in a moment. Repeating the exercise for $\gamma=7/5$ ($n=2.5$) gives almost identical results, with slightly different leading constants. We therefore conclude that the first core is an object a few AU in size, with a mass of a few hundredths of a Solar mass.

\subsection{Second Collapse}

The first core is a very short-lived phase in the evolution of the protostar. To see why, let us estimate its temperature. The temperature inside the sphere rises as $T\propto \rho^{\gamma-1}$, so the central temperature is
\begin{equation}
T_c = T_0 \left(\frac{\rho_c}{\rho_{\rm ad}}\right)^{\gamma-1},
\end{equation}
where $T_0$ is the temperature in the isothermal phase. Thus the central temperature will be higher than the boundary temperature by a factor that is determined by how high the central density has risen, which in turn will be determined by the amount of mass that has accumulated on the core.

In general we have $M\propto \rho_c^{(3+n)/(2n)}$, or $M\propto \rho_c^{(3\gamma-2)/2}$. We also have $T_c\propto \rho_c^{\gamma-1}$. Combining these results, we have
\begin{equation}
T_c \propto M^{(2\gamma-2)/(3\gamma-2)}.
\end{equation}
The exponent is $0.44$ for $\gamma=5/3$ and $0.36$ for $\gamma=7/5$. Plugging in some numbers, $M = 0.06\msun$, $\rho_{\rm ad}=10^{-13}$ g cm$^{-3}$, and $\gamma=5/3$ gives $\rho_c = 10^{-10}$ g cm$^{-3}$ and $T_c=1000$ K. Thus we see that by the time anything like $0.1$ $\msun$ of material has accumulated on the first core, compression will have caused its central temperature to rise to $1000$ K or more.

This causes yet another change in the thermodynamics of the gas, because all the hydrogen is still molecular, and molecular hydrogen has a binding energy of $4.5$ eV. In comparison, the kinetic energy per molecule for molecular hydrogen at a temperature $T$ is $(3/2) k_B T = 0.13 T_3$ eV, where $T_3=T/(1000\mbox{ K})$. At 1000 K this means that the mean molecule still has only a few percent of the kinetic energy that would be required to dissociate it. However, there is a non-negligible tail of the Maxwellian distribution that is moving fast enough for collisions to produce dissociation. Each of these dissociative collisions removes $4.5$ eV from the kinetic energy budget of the gas and puts it into chemical energy instead. Since dissociations are occurring on the tail of the Maxwellian, any slight increase in the temperature dramatically increases the dissociation rate, moving even more kinetic energy into chemical energy.

This effectively acts as a thermostat for the gas, in much the same way that a boiling pot of water stays near the boiling temperature of water even when energy is added, because all the extra energy that is provided goes into changing the chemical phase of the water rather than raising its temperature. Detailed numerical calculations of this effect show that at temperatures above $1000-2000$ K, the equation of state becomes closer to $T\propto \rho^{0.1}$, or $\gamma=1.1$. This is again below the critical value of $\gamma=4/3$ required to have a hydrostatic object, and as a result the center of the first core again goes into something like free-fall collapse.

This is called the second collapse. The time required for it is set by the free-fall time at the central density of the first core, which is only a few years. This collapse continues until all the hydrogen dissociates. The hydrogen also ionizes during this collapse, since the ionization potential of $13.6$ eV is not very different from the dissociation potential of $4.5$ eV. Only once all the hydrogen is dissociated and ionized can a new hydrostatic object form. At this point the gas is warmer than $\sim 10^4$ K, is fully ionized, and the new hydrostatic object is a true protostar. It is supported by degeneracy pressure at first when its mass is low, and then as more mass arrives it heats up and becomes supported by thermal pressure. 

An important point to make here is that this discussion implies that brown dwarfs, at least those of sufficiently low mass, do not undergo a prompt second collapse. Instead, their first cores never accumulate enough mass to dissociate the molecules at their center. This is not to say that dissociation never happens in them, and that second collapse never occurs. A brown dwarf-mass first core will still radiate from its surface and, lacking any internal energy source, this energy loss will have to be balanced by compression. As the gas compresses the temperature and entropy will rise, and, if the object does not become supported by degeneracy pressure first, the central temperature will eventually rise enough to produce second collapse. The difference for a brown dwarf is that this will only occur once slow radiative losses cause a temperature rise, which may take a very long time compared to formation. For stars, in contrast, there is enough mass to reach the critical temperature by compression during formation.

\section{The Protostellar Envelope}

Once a protostar is born at the center of a collapsing cloud, we can ask both about the structure immediately around it and about its internal structure. We defer the latter to Chapter \ref{ch:protostar_evol}, and focus here on the envelope around the newborn protostar.

\subsection{Accretion Luminosity}

The temperature of the gas around the newborn protostar is determined by the radiation that the central star emits. At early times the star has not reached the main sequence or ignited any nuclear burning, so gravity is the only important energy source in the problem. Even if nuclear burning does start, we will see that it is negligible for low mass stars. The protostar is a hydrostatic object, although it undergoes secular contraction, so that gas striking its surface comes to a halt in an accretion shock. In this shock its kinetic energy is converted to heat, which is then radiated away.

The detailed structure of the accretion shock was first worked out by \citet{stahler80a, stahler80b}. The summary is that the energy radiated away at the shock is roughly
\begin{equation}
L_{\rm acc} = \frac{G M_* \dot{M}_*}{R_*},
\end{equation}
where $M_*$, $\dot{M}_*$, and $R_*$ are the mass, accretion rate, and radius for the protostar. We will see in Chapter \ref{ch:protostar_evol} that $R_*$ is typically a few $\rsun$ (and indeed this is consistent with the observed radii of T Tauri stars). We have previously calculated typical accretion rates of $\dot{M}_*\sim 10^{-5}$ $\msun$ yr$^{-1}$ for low mass stars.
Plugging in these numbers, we find
\begin{equation}
L_{\rm acc} = 30\lsun\, \dot{M}_{*,-5} M_{*,0} R_{*,1}^{-1},
\end{equation}
where $\dot{M}_{*,-5}=\dot{M}_*/(10^{-5}\msun\mbox{ yr}^{-1})$, $M_{*,0}=M_*/\msun$, and $R_{*,1}=R_*/(10\rsun)$.
Thus a typical low mass protostar can easily put out many tens of $\lsun$ in accretion power, far greater than what it would produce from nuclear burning on the main sequence. 

We can also estimate the effective temperature of the stellar surface due to accretion. The infalling gas arrives in free-fall at a velocity 
\begin{equation}
v_{\rm ff}=\sqrt{\frac{2GM_*}{R_*}} = 200\mbox{ km s}^{-1} \, M_{*,0}^{1/2} R_{*,1}^{-1/2}.
\end{equation}
The vastly exceeds the sound speed of a few km s$^{-1}$ in gas at a temperature of $\sim 10^3-10^4$ K, so the gas must decelerate in a strong shock with a Mach number of order 100. For a strong shock, one where the Mach number $\mathcal{M} \gg 1$, the Rankine-Hugoniot jump conditions tell us that the post-shock temperature is
\begin{equation}
T_2 = \frac{2\gamma(\gamma-1)}{(\gamma+1)^2} \mathcal{M}^2 T_1
= \frac{2\gamma(\gamma-1)}{(\gamma+1)^2} \frac{v_{\rm shock}^2}{c_1^2} T_1
= \frac{2(\gamma-1)}{(\gamma+1)^2} \frac{\mu m_{\rm H}}{k_B} v_{\rm shock}^2,
\end{equation}
where $c_1$ is the adiabatic sound speed in the pre-shock gas.

Taking $v_{\rm shock}=v_{\rm ff}$, $\gamma=5/3$ for a monatomic gas, and $\mu=1.4$ for the pre-shock gas (assuming it to be neutral hydrogen), and plugging in we get
\begin{equation}
T_2 = 1.2\times 10^6 \, M_{*,0} R_{*,1}^{-1} \mbox{ K}.
\end{equation}
In other words, the post-shock gas is heated to temperatures such that it emits in UV and x-rays. The incoming gas will be extremely opaque to this radiation due to the opacity provided by both free electrons and numerous lines of multiply ionized metal atoms such as iron. As a result all the radiation emitted by the post-shock gas will be absorbed in a small region immediately outside the shock and reprocessed until it becomes blackbody emission. The stellar surface therefore emits as a blackbody, whose temperature we can calculate in the standard way:
\begin{eqnarray}
L_{\rm acc} & = & 4\pi R_*^2 \sigma_{\rm SB} T_*^4 \\
T_* & = & 4300 \dot{M}_{*,-5}^{1/4} M_{*,0}^{1/4} R_{*,1}^{-3/4} \mbox{ K}.
\end{eqnarray}
Thus the star is effectively a blackbody at a surface temperature comparable to that of a main sequence star.

\subsection{The Dust Destruction Front}

Now let us consider the effect of this luminosity on the gas around the protostar. Consider a spherical black dust grain of radius $a$ some distance $r$ from the star. It absorbs radiation at a rate
\begin{equation}
\Gamma = \frac{L_{\rm acc}}{4\pi r^2} \pi a^2 = \pi a^2 \sigma_{\rm SB} T_*^4 \left(\frac{R_*}{r}\right)^2
\end{equation}
and radiates it at a rate\footnote{This expression is only valid if the wavelengths characteristic of the peak of the blackbody curve at temperature $T_d$ are small compared to the circumference of the grain. For the dust temperature of $\approx 1000$ K we will insert below, this implies that the result is valid for grains with characteristic sizes $\gtrsim 1$ $\mu$m. Smaller grains will have lower values of $\Lambda$ and thus higher equilibrium temperatures.}
\begin{equation}
\Lambda = 4\pi a^2 \sigma_{\rm SB} T_d^4,
\end{equation}
where $T_d$ is the dust grain's temperature. Equating these two, the temperature of the grain is
\begin{equation}
T_d = \left(\frac{R_*}{2r}\right)^{1/2} T_*
\end{equation}

Even the most refractory materials out of which interstellar dust is made, such as graphite and silicate, will vaporize at temperatures larger than $\sim 1000-1500$ K. The exact temperature depends on the chemical composition of the grains. Thus when $r/R_*$ is too small grains cannot survive. They are vaporized. We therefore expect the protostar to be surrounded by dust-free region.

Since the ionizing radiation produced at the shock at the stellar surface all gets absorbed close to the shock, and the star is shining into this dust-free region as a blackbody at a temperature of only a few thousand K, the gas in this region is primarily neutral. Neutral atomic gas with no dust in it is essentially transparent to visible radiation, so in this region the opacity is tiny, and stellar radiation is able to free-stream outward. The dust-free neutral region is called the opacity gap.

As one moves away from the star the equilibrium grain temperature drops, and eventually one reaches a surface where dust grains can exist. This is called the dust destruction radius, since incoming gas that reaches this radius has its grains destroyed. If we plug the grain destruction temperature into our equation for $T_d$, we can solve for the dust destruction radius:
\begin{equation}
r_d = \frac{R_*}{2} \left(\frac{T_*}{T_d}\right)^2 = 0.4 \,  T_{d,3}^{-2} \dot{M}_{*,-5}^{1/2} M_{*,0}^{1/2} R_{*,1}^{-1/2} \mbox{ AU},
\end{equation}
where $T_{d,3}=T_d/(1000\mbox{ K})$ is the dust destruction temperature in units of 1000 K. Thus the dust-free region extends to $\sim 1$ AU around an accreting protostar.

\subsection{Temperature Structure and Observable Properties}

Now let us consider the material beyond the dust destruction front. At the front the gas density is given roughly by the condition
\begin{eqnarray}
\dot{M}_* & = & 4\pi r_d^2 \rho v_{\rm ff} \\
\rho & = & \frac{\dot{M}_*}{\sqrt{8 \pi^2 G M_* r_d^3}} \\
& = & 4\times 10^{-13} \, \dot{M}_{*,-5}^{1/4} M_{*,0}^{-7/4} R_{*,1}^{3/4} T_{d,3}^3\mbox{ g cm}^{-3}.
\end{eqnarray}
Just inside the front, the stellar spectrum is nearly a blackbody at a temperature of a few thousand Kelvin, so the peak wavelength is
\begin{equation}
\lambda \approx \frac{hc}{4k_BT} = 440 \, \dot{M}_{*,-5}^{-1/4} M_{*,0}^{-1/4} R_{*,1}^{3/4}\mbox{ nm},
\end{equation}
placing it in the visible.

The opacity of gas with Milky Way dust composition at 440 nm is roughly $\kappa=8000$ cm$^2$ g$^{-1}$, so the mean free-path of a stellar photon moving through the dust destruction front is $(\kappa\rho)^{-1} \approx 3\times 10^8$ cm. This is a tiny length scale compared to any other scale in the problem, such as the size of the core, the size of the opacity gap, or even the radius of the protostar. Thus all the starlight that strikes the dust destruction front will immediately be absorbed by the dust grains. They will re-emit it as thermal radiation with a peak wavelength determined by their blackbody temperature, which will be a factor of $\sim 4$ lower than the stellar surface temperature. At around 1.8 $\mu$m, a factor of 4 longer wavelength than the 440 nm we started with, the opacity is drops to around 1000 cm$^2$ g$^{-1}$, so the mean free path is a factor of 8 larger. Nonetheless, this is still tiny, so all the re-emitted radiation will also be absorbed.

Since we are in a situation where all the radiation is absorbed and re-emitted many times, it is reasonable to treat this as a diffusion problem. Protostellar radiation free-streams from the surface, through the opacity gap, and is absorbed and thermalized at the dust destruction front. Then it must diffuse out through the dust envelope. This is essentially the same calculation that is made for radiation diffusing outward through a star, and the equation describing it is the same:
\begin{equation}
F = -\frac{c}{3\rho \kappa_R} \nabla E,
\end{equation}
where $F$ is the radiation flux, $E$ is the radiation energy density, and $\kappa_R$ here is the Rosseland mean opacity, meaning the mean of the frequency-dependent opacity using a weighting function that is equal to the temperature derivative of the Planck function. Note here that $\kappa_R$ is a function of $T$.

The repeated absorption and re-emission of radiation forces it into thermal equilibrium with the gas, so $E$ is simple the energy density of a thermal radiation field at the gas temperature: $E=a_R T^4$. Since no energy is added or removed from the radiation field as it diffuses outward through the envelope, $F=L_{\rm acc}/(4\pi r^2)$. Putting this together, we have
\begin{equation}
L_{\rm acc}=-\frac{16\pi ca_R r^2}{3\rho\kappa_R} T^3 \frac{dT}{dr}
\end{equation}

For a given density structure and a model of dust grains that specifies $\kappa_R(T)$, this equation allows us to estimate the temperature structure in the protostellar envelope. For reasonable grain models we expect $\kappa_R\propto T^\alpha$ with $\alpha\approx 0.8$ in the temperature range of a few hundred K. Let us suppose that the density distribution in the envelope looks something like a powerlaw, so $\rho\propto r^{-k_{\rho}}$. Finally, let us also suppose that the temperature also behaves like a powerlaw in radius, $T\propto r^{-k_T}$. The left hand side of the equation is a constant, and we have now worked out how the right hand side varies with $r$. Plugging in all the radial dependences on the RHS, and knowing that they must sum to zero since the LHS is a constant, we get
\begin{equation}
k_T = \frac{k_{\rho} + 1}{4-\alpha}.
\end{equation}

Thus in the freely-falling part of the envelope, where $k_{\rho}\approx 3/2$, we have $k_T\approx 0.8$. In our fiducial example, where the temperature is 1000 K at 0.4 AU, we would expect the temperature to drop to 300 K at around 2 AU, to 100 K at around 8 AU, and back to the background temperature of 10 K at around 150 AU. In the outer part of the envelope the falloff in temperature can be either steeper or shallower depending on how the density falls off -- sharper density falloffs (larger $k_{\rho}$) lead to sharper temperature falls (larger $k_T$) as well.

Of course this approximation only applies as long as the radiation is trapped by the dust, and the dust opacity is highest for high frequency radiation. Once the dust temperature falls off to less than $\sim 100$ K, depending on the size of the core, the radiation is free to escape instead. Even further in, where the dust temperature is higher, long wavelength radiation can escape freely.

As a result the spectrum inside the core is never truly a blackbody, since radiation at long wavelengths never reaches thermal equilibrium. The emitted spectrum is also complicated by this behavior. We can think of this as follows: for a star, there is something close to a single well-defined photosphere at all frequencies because the density drops off sharply. For a dust cloud, on the other hand, the density drop is not sharp, and so the photosphere, the surface of optical depth unity (or 2/3 if you prefer) is in different places at different frequencies. At high frequencies it is near the core surface because the opacity is high, and at low frequencies the low opacity allows it to be much farther in. For this reason, centrally-heated cores do not emit as blackbodies.

In order to truly determine the temperature distribution within a core it is necessary to either use a more sophisticated analytic treatment (for example one is given in \citealt{chakrabarti05a}) or to proceed numerically. If one wants a more sophisticated density structure that is not spherical, numerical methods are also required. Of course all of this only applies as long as a great deal of mass remains in the envelope, so that it is optically thick to both the star's direct radiation and to the re-radiated thermal radiation from the dust destruction front. In terms of our evolutionary classes, all of this applies to class 0 and class I sources.

\chapter{Protostellar Evolution}
\label{ch:protostar_evol}

\marginnote{
\textbf{Suggested background reading:}
\begin{itemize}
\item \href{http://adsabs.harvard.edu/abs/2014prpl.conf..173L}{Dunham, M.~M., et al. 2014, in "Protostars and Planets VI", ed.~H.~Beuther et al., pp.~195-218}, sections 5-9 \nocite{dunham14a}
\end{itemize}
\textbf{Suggested literature:}
\begin{itemize}
\item \href{http://adsabs.harvard.edu/abs/2011ApJ...738..140H}{Hosokawa, T., Offner, S.~S.~R., \& Krumholz, M,~R., 2011, ApJ, 738, 140} \nocite{hosokawa11a}
\end{itemize}
}

This chapter considers the behavior of the stellar objects that form at the centers of collapsing clouds. Our goal is to understand how the usual theory of stellar structure can be adapted to the case of protostars that are not yet on the main sequence. Since stellar structure is a vast topic by itself, and there are numerous textbooks covering it, we will not attempt to re-derive it in its entirety here. Instead, we will focus on how the theory must be modified for protostars, and to follow the implications of this modification.

\section{Fundamental Theory}

\subsection{Time Scales}

The fundamental reason that stars can be hydrostatic objects is that the time they require to reach a mechanical equilibrium where the inward force of gravity is balanced by outward pressure is small compared to the time required for their energies, and thus pressures, to change. This is in contrast to molecular clouds, which cannot be hydrostatic because they are able to radiate away energy faster than they can reach force balance. We therefore begin our discussion by verifying that the mechanical and thermal equilibration timescales for protostars are, like those timescale for main sequence stars, very well separated.

The time required for a star to reach mechanical equilibrium is the sound crossing time, $t_s\sim R/c_s$, where $R$ is the stellar radius and $c_s$ is the sound speed. The virial theorem tells use that the sound speed inside the star (potentially including the contribution from radiation pressure) must be of order $\sqrt{GM/R}$, where $M$ is the stellar mass. Thus the mechanical equilibration timescale is
\begin{equation}
t_s \sim \sqrt{\frac{R^3}{GM}} = 35\, M_0^{-1/2} R_1^{3/2}\mbox{ hours},
\end{equation}
where $M_0 = M / \msun$ and $R_{1} = R/(10\rsun)$. Here the radius to which we have scaled is a typical one for protostars, as we will see below.

In contrast, the time required to reach thermal equilibrium is the Kelvin-Helmholtz (KH) time, which is defined as roughly the time required for the star to radiate away its own binding energy,
\begin{equation}
t_{\rm KH} = \frac{GM^2}{RL} = 3\times 10^5\, M_0^2 R_1^{-1} L_1^{-1}\mbox{ yr},
\end{equation}
where $L_1 = L/(10 \lsun)$; again this is a typical protostellar value, as we show below. Thus the star reaches mechanical equilibrium essentially instantaneously compared to the time required to reach thermal equilibrium. It is therefore reasonable to assume that at all times the star is in hydrostatic balance, and then to describe its subsequent evolution movement from one hydrostatic state to another, with the change in state dictated by the evolution of the energy and entropy of the gas. 

For future reference, it is also useful to think about how long accretion will last. At an accretion rate of $10^{-5}$ $\msun$ yr$^{-1}$, the formation of a 1 $\msun$ star takes $10^5$ yr. Thus, the accretion time is generally shorter than the KH time, so that stars will cease accreting before they reach thermal equilibrium. Note that this is true only for low mass stars, not high mass ones. We will discuss the case of high mass stars further in Chapter \ref{ch:massivestar}.

\subsection{Evolution Equations}

Now that we have shown that we can treat protostars as hydrostatic equilibrium objects, let us proceed to write down the evolution equations that govern the protostar. These should be familiar from stellar structure. An important caveat is that what we cover here represents an extremely simple approach to stellar structure, and that all of the complications that arise in real stellar structure calculations (e.g., rotation, convective overshooting, real stellar atmospheres, etc.) apply equally well to protostellar evolution. The goal here is simply to sketch the basic theory, so that we can understand how it changes for protostars as opposed to main sequence stars.

As in other stellar structure calculations, it is most convenient to work in Lagrangian coordinates, where we let $M_r$ be the mass interior to radius $r$, so that $M_r$ runs from 0 to M. We then solve for stellar properties as a function of $M_r$. The first equation is the standard definition of mass in terms of density and radius:
\begin{equation}
\label{mass}
\frac{\partial r}{\partial M_r} = \frac{1}{4\pi r^2 \rho}.
\end{equation}
The second equation is the equation of hydrostatic balance. In Eulerian coordinates it is
\begin{equation}
\label{eq:hydrobalance}
\frac{\partial P}{\partial r} = -\frac{G M_r \rho}{r^2},
\end{equation}
and converting to Lagrangian coordinates by dividing by the relationship between $r$ and $M_r$ gives
\begin{equation}
\frac{\partial P}{\partial M_r} = -\frac{G M_r}{4\pi r^4}.
\end{equation}

The third equation is the equation of radiation diffusion:
\begin{equation}
F = \frac{L}{4\pi r^2} = -\frac{c}{3\rho\kappa_{\rm R}} \frac{\partial E}{\partial r},
\end{equation}
where $F$ is the radiation flux, $L$ is the luminosity passing through radius $r$, and $\kappa_{\rm R}$ is the Rosseland mean opacity of the gas. Writing $E=a_R T^4 = \sigma_{\rm SB} T^4/(4c)$ and again converting to Lagrangian coordinates by dividing by $\partial r/\partial M_r$ gives
\begin{equation}
T^3 \frac{\partial T}{\partial M_r} = - \frac{3\kappa_{\rm R} L}{256 \pi^2 \sigma_{\rm SB} r^4}.
\end{equation}
This applies only as long as the protostar is stable against convection, $\partial s/\partial M_r > 0$, i.e., the entropy increases outward. If it is unstable to convection, we instead have
\begin{equation}
\frac{\partial s}{\partial M_r} = 0,
\end{equation}
or some more sophisticated treatment of convection based on mixing-length theory.

Finally, the last equation describes how the internal energy of the fluid evolves:\footnote{This evolution equation is simply a form of the fundamental thermodynamic relationship $dU = T\, dS - P\, dV$, where $dU$ is the change in internal energy, $dS$ is the change in total entropy, and $dV$ is the change in volume. The term on the left hand side is $T \, dS$, and the term on the right hand side is the change in internal energy due to nuclear reactions ($\rho \epsilon$) plus the change due to radiative transfer ($-\nabla \mathbf{F} = (1/r^2) (\partial/\partial r)(r^2 F)$ in spherical symmetry). Since the system is hydrostatic, the change in volume $dV$ is zero.}
\begin{equation}
\rho T\frac{\partial s}{\partial t} = \rho \epsilon - \frac{1}{r^2}\frac{\partial}{\partial r}(r^2 F),
\end{equation}
where $s$ is the entropy per unit mass of the gas and $\epsilon$ is the rate of nuclear energy generation per unit mass. Substituting $L = 4 \pi r^2 F$ gives
\begin{equation}
\frac{\partial L}{\partial r} = 4\pi r^2 \rho \left(\epsilon - T\frac{\partial s}{\partial t}\right),
\end{equation}
and dividing once more by $\partial r/\partial M_r$ gives
\begin{equation}
\label{heat}
\frac{\partial L}{\partial M_r} = \epsilon - T\frac{\partial s}{\partial t}.
\end{equation}
This equation is the only one that is different for a protostar than it is for a main sequence star. For a main sequence star, we simply assume that the entropy per unit mass is constant, so we drop the $\partial s/\partial t$ term. We are justified in doing this for a main sequence star, because the star is in energy equilibrium between radiative losses and internal energy generation. Thus the entropy distribution in the star changes only in response to changes in chemical composition produced by nuclear burning. That is not the case for a protostar, which is not in energy equilibrium.

This constitutes four equations in the four unknowns $r$, $P$, $T$, and $L$.  We also require functions specifying the equation of state $P(\rho)$, the opacity $\kappa_{\rm R}(\rho,T)$, the energy generation rate $\epsilon(\rho,T)$, and the entropy $s(\rho,T)$. Since radiation, degeneracy pressure, and relativistic effects are generally unimportant for protostars, the equation of state is just the usual ideal gas law
\begin{equation}
P = \frac{\rho k_B T}{\mu m_{\rm H}},
\end{equation}
where $\mu$ is the mean mass for particle in units of hydrogen masses. For constant $\mu$ the entropy is
\begin{equation}
s = \frac{k_B}{\mu m_{\rm H}} \ln \left(\frac{T^{3/2}}{\rho}\right) + \mbox{const}.
\end{equation}
For a fully ionized gas $\mu=0.61$, but in numerical calculations we generally use a numerically tabulated value of $\mu(\rho, T)$ and $s(\rho,T)$. The terms describing the opacity and nuclear energy generation rate are exactly the same as in the case of a main sequence star, with one important exception for nuclear energy generation that we will discuss below.

\subsection{Boundary Conditions}

The four structure equations require four boundary conditions to solve. Two are obvious, and are the same for protostars as main sequence stars: at $M_r=0$
\begin{eqnarray}
r(0) & = & 0\\
L(0) & = & 0.
\end{eqnarray}
The remaining two are less obvious. Thus far everything we have written down is completely identical to the case of a main sequence star, except for the time derivative in the heat equation, but the remaining two boundary conditions, describing the pressure and luminosity at the edge of the star, are different.

\paragraph{Spherical Accretion Flows.} First consider the simplest case, where we assume spherical symmetry everywhere. A main sequence star effectively has vacuum of negligible pressure outside it, but a protostar does not. It is bounded by an accretion flow, and  the pressure at the stellar surface must be sufficient to halt the flow. The accretion rate onto the star $\dot{M}$ is related to the density $\rho_i$ and velocity $v$ of the infalling material by
\begin{equation}
\dot{M} = 4\pi r^2 \rho_i v,
\end{equation}
and the ram pressure at the stellar surface is therefore
\begin{equation}
P(M) = \rho_i v^2 = \frac{\dot{M} v}{4\pi r^2},
\end{equation}
where the right hand side is to be evaluated at $r=R$. If the incoming gas is in free-fall, then we can set $v = v_{\rm ff} = \sqrt{2GM/R}$, which gives
\begin{equation}
\label{pbound1}
P(M) = \frac{\dot{M}}{4\pi} \sqrt{\frac{2 G M}{R^5}},
\end{equation}
where $M_*$ is the total stellar mass. 

The final boundary condition is on the luminosity. For a non-accreting star, in the simplest case we treat the star as radiating as a blackbody, and require that
\begin{equation}
\label{lbound2}
L(M)=4\pi R^2 \sigma_{\rm SB} T(M)^4.
\end{equation}
In more sophisticated complications we derive the luminosity from a stellar atmosphere calculation. The situation is more complex for an accreting star, because the accreting gas carries a non-negligible energy flux with it. The question therefore becomes what fraction of this energy will be radiated away at the stellar surface and what fraction will be advected or radiated into the stellar interior. We will not derive these results in detail, just sketch out the issues. The boundary condition must take the form
\begin{equation}
\label{lbound1}
L(M) = L_{\rm acc} + L_{\rm bb} - L_{\rm in},
\end{equation}
where $L_{\rm acc}=G M \dot{M}/R$ is the mechanical luminosity of the accreting gas, $L_{\rm bb} = 4\pi R^2 \sigma_{\rm SB} T(M)^4$ represents the blackbody radiation from the stellar surface, and $L_{\rm in}$ represents the inward flux of energy due to advection and radiation from the shocked gas. One way to think about $L_{\rm in}$ is that it specifies what fraction of the kinetic energy of the accreting gas escapes promptly as radiation, with the remaining portion assumed to be advected into the stellar interior with the accreting gas. The correct value of $L_{\rm in}$ is a subtle question, since it depends on the structure of the shock at the stellar surface, and on its geometry. For spherical accretion, \citet{stahler80a, stahler80b, stahler81a} show that $L_{\rm in}\approx 3 L_{\rm acc}/4$. 

\paragraph{Cold versus Hot Accretion.} Thus far we have assumed that the accretion flow is spherically symmetric, but this assumption may not be every close to correct. In particular, there is good observational evidence for T Tauri stars that accretion occurs only over a small portion of the stellar surface, likely because the star's magnetic field exerts enough pressure to prevent accretion over much of the surface. It is unknown, and a subject of great current debate, whether this is also the case during the optically-hidden main accretion phase, when the accretion rate is much higher than during the later T Tauri phase.

If the accretion is confined to a small portion of the stellar surface, this has two implications. First, the pressure boundary condition should revert to the usual vacuum one that applies to main sequence stars, since there will be no ram pressure over most of the stellar surface. We can write down this condition by integrating the equation of hydrostatic balance (equation \ref{eq:hydrobalance}) to obtain 
\begin{equation}
P(M) = \frac{G M}{R}^2 \int_{R}^{\infty} \rho \, dr.
\end{equation}
If $\kappa_{\rm R}$ changes relatively little past the stellar photosphere, then 
\begin{equation}
\int _{R}^{\infty} \rho \, dr \approx \frac{\tau_{\rm phot}}{\kappa_{\rm R, phot}},
\end{equation}
where $\tau_{\rm phot}$ is the optical depth from infinity to the photosphere and $\kappa_{\rm R, phot}$ is the opacity at the edge of the photosphere. Since the edge of the star is roughly where $\tau_{\rm phot} = 2/3$, the boundary condition becomes
\begin{equation}
\label{pbound2}
P(M) = \frac{2 G M}{3 R^2 \kappa_{\rm R, phot}}.
\end{equation}
The second implication of non-spherical accretion is that $L_{\rm in}$ might be much larger than in the spherical case. This is because, if the accretion shock covers only a small portion of the stellar surface, radiation will be able to escape out the "sides" of the shock in a way that it cannot for a fully-confined sphere. There has yet to be a fully detailed calculation of this case, and instead the usual practice in the protostellar evolution community is to parameterize the uncertainty by adopting a value of $L_{\rm in}$ that lies somewhere between the minimum possible value, corresponding to the spherical case, and the maximum possible value, in which $L_{\rm in}$ is chosen so as to set the specific entropy of the material being added to the star equal to either the specific entropy of material at the stellar surface, or the mean specific entropy of all material in the star. We refer to cases where the value of $L_{\rm in}$ is chosen equal or close to the spherical value as "hot accretion" models, because the material being added to the star is hot in this case. We refer to models where $L_{\rm in}$ is chosen so that the specific entropy of accreting material matches that of material already in the star as "cold accretion" models, since the material in this case is cold.

In the absence of a first-principles theoretical calculation, it is difficult to determine whether reality is closer to the hot or cold accretion assumption, and whether the answer to this question might be different for different stars. Approaches to settling this problem have generally relied on the empirical approach of generating synthetic tracks from hot or cold accretion assumptions and then comparing to observations to see what gives the best fit. The method by which we can generate these tracks we defer to Section \ref{ssec:protostar_numeric}.

\subsection{Deuterium Burning}

Before discussing how the structure equations can be solved numerically, it is worth delving a little further into the term $\epsilon$, representing nuclear energy generation. For a main sequence star, $\epsilon$ comes from fusion of hydrogen into helium, either via the pp-chain or the CNO cycle. However, hydrogen burning does not occur until just before the star reaches the main sequence.

There is, however, an energetically-important nuclear reaction that can occur at lower temperatures, before the star is hot enough to burn hydrogen: fusion of deuterium, via the reaction
\begin{equation}
^2\mbox{H} + \,^1\mbox{H}\, \rightarrow \, ^3\mbox{He} + \gamma.
\end{equation}
This reaction begins to occur at an appreciable rate once the temperature reaches $10^6$ K, and the reaction releases $5.5$ MeV per deuterium nucleus burned. The energy generation rate from deuterium fusion is reasonably well-approximated by \citep{kippenhahn94a}
\begin{equation}
\epsilon \approx 
\left\{
\begin{array}{ll}
0, & T < 10^6\mbox{ K} \\
4.19\times 10^7\, [\mbox{D}/\mbox{H}] \rho_0 T_6^{11.8} \mbox{ erg g}^{-1}\mbox{ s}^{-1}
\qquad & T > 10^6\mbox{ K},
\end{array}
\right.
\end{equation}
where $[\mbox{D}/\mbox{H}]$ is the ratio of D to H in the gas, $\rho_0=\rho/(1\mbox{ g cm}^{-3})$, and $T_6=T/(10^6\mbox{ K})$. For interstellar gas in the Milky Way, $[\mbox{D}/\mbox{H}]\approx 2\times 10^{-5}$, which is only slightly below the primordial abundance.

Strictly speaking the expression we have for $T>10^6$ K is only valid for temperatures near $10^6$ K, but, as we shall see, this good enough for our purposes. If we wish to run a model past the start of H burning, we need an analogous expression for it, which is the same as one used for normal main sequence stellar structure calculations.

\subsection{Numerical Solution}
\label{ssec:protostar_numeric}

We have now fully specified the equations describing our protostar. To construct a numerical model, we need to specify the accretion rate $\dot{M}$ that appears in the boundary condition equations (\ref{pbound1}) and (\ref{lbound1}) describing the pressure and luminosity at the stellar surface. In general $\dot{M}$ can be a function of time, although usually it is taken to be constant until accretion halts at some specified final stellar mass $M_*$, at which point we switch to boundary conditions (\ref{pbound2}) and (\ref{lbound2}) for the boundary pressure and luminosity.

We must also start with an initial condition, which we usually take to be a simple polytrope. This gives us initial profiles of $r$, $P$, $T$, and $L$, from which we can obtain other derived variables like $\rho$ and $s$, as a function of $M_r$. The choice of initial condition might matter a little or a lot, depending on the choice of boundary conditions, as we will see.

Given these boundary conditions, we construct the solution at each time using a shooting method in much the same way as we would for a main sequence star. We first guess a central temperature $T$ and pressure $P$, which of course also gives us the central density $\rho$ and entropy $s$. Usually a good first guess is the value of $\rho$ and $s$ at the last time step.
Then we integrate equations (\ref{mass}) - (\ref{heat}) outward in radius until we reach the outer mass shell $M_*$ (which is a function of time).

To obtain the time derivative of the entropy term that appears in the internal energy equation (\ref{heat}), we just compute the difference between the entropy $s(M_r,t)$ for mass shell $M_r$ at the current time $t$ and the value for $s(M_r,t-\Delta t)$ that we had in the previous time step. In general the solution we have constructed will not satisfy the outer boundary conditions (\ref{pbound1}/\ref{pbound2}) and (\ref{lbound1}/\ref{lbound2}), so we must modify our guesses for $T$ and $P$ in the center and try again.

We repeat this until we converge, and then we proceed to the next time step, adding new mass shells on the outside as necessary to account for new material deposited by accretion. We continue the calculation until the star's radius converges to its main sequence value. In this manner, we can generate a full evolutionary track for a given accretion rate.

\section{Evolutionary Phases for Protostars}

\begin{figure}
\includegraphics[width=\linewidth]{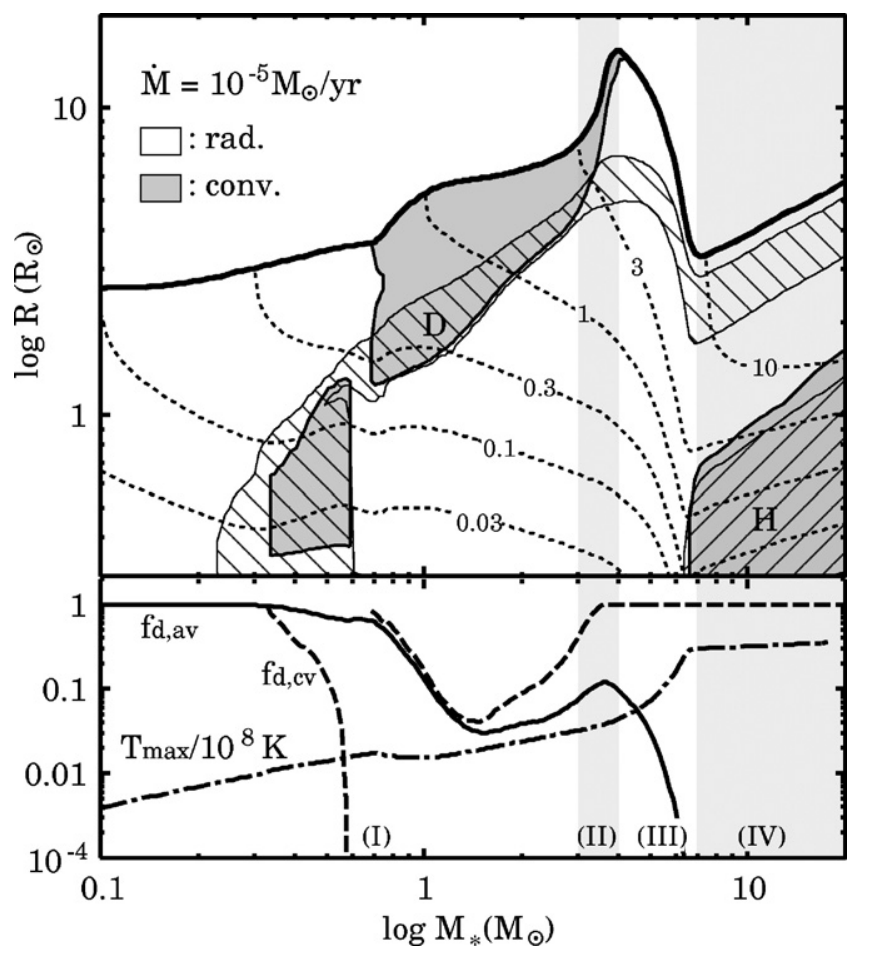}
\caption[Kippenhahn diagram for an accretion protostar]{
\label{fig:kippenhahn_hosokawa09}
Kippenhahn and composition diagrams for a protostar accreting at $10^{-5}$ $M_\odot$ yr$^{-1}$. In the top panel, the thick curve shows the protostellar radius as a function of mass, and gray and white bands show convective and radiative regions, respectively. Hatched areas show regions of D and H burning, as indicated. Thin dotted lines show the radii containing $0.1$, $0.3$, $1$, $3$, and $10$ $M_\odot$, as indicated. Shaded regions show four evolutionary phases: (I) convection, (II) swelling, (III) KH-contraction, and (IV) the main sequence. In the lower panel, the solid line shows the mean deuterium fraction in the star, normalized to the starting value, while the dashed line shows the D fraction only considering the convective parts of the star. The dot-dashed line shows the maximum temperature. Credit: \citet{hosokawa09a}, \copyright AAS. Reproduced with permission.
}
\end{figure}

We have now outlined the basic equations describing protostellar evolution, as well as the numerical method used to solve them. We will now discuss the results of these calculations. There are generally a few distinct stages though with forming stars pass, which can be read off from how the radius evolves as the star gains mass. We will use as our primary example the case of a star undergoing hot accretion at $10^{-5}$ $M_\odot$ yr$^{-1}$, as illustrated in Figure \ref{fig:kippenhahn_hosokawa09}. However, note that the ordering of these phases we describe below can vary somewhat depending on the accretion rate and the boundary conditions assumed. Moreover, for low mass stars, some of the later phases may not occur at all, or occur only after the end of accretion.

\subsection{Initial Contraction}

The initial phase of evolution is visible in Figure \ref{fig:kippenhahn_hosokawa09} as what takes place up to a mass of $\approx 0.2$ $M_\odot$ for the example shown. The first thing that happens during this phase is that the star reaches a radius that is a function solely of $M$ and $\dot{M}$. This occurs regardless of the initial radius with which we initiate the model, as long as we are using the hot accretion boundary condition. The physical reason for this behavior is easy to understand. The radius of the star is determined by the entropy profile $s(M_r)$. High entropy leads to high radius. Since the internal energy generated by the star is small compared to the accretion power when the stellar mass is low (i.e., $L_{\rm bb} \ll L_{\rm acc}$), once gas is incorporated into the star it does not lose significant energy by radiation. The only entropy it loses is due to the radiation that occurs at the shock on the star's surface. We could have guessed this result from the large value of $t_{\rm KH}$ compared to the accretion time -- in effect, this means that, once a fluid element reaches the stellar surface it will be buried and reach a nearly constant entropy quite quickly. Consequently, we can treat the material falling onto the star during this phase as having an entropy per unit mass that depends only on two factors: (1) the entropy it acquires by striking the stellar surface, and (2) how much it radiates before being buried.

The latter factor is just determined by the accretion rate. Higher accretion rates bury accreted material more quickly, leaving it with higher entropy and producing larger radii. The former depends on the velocity of the infalling material just before it strikes the stellar surface, and thus on $v_{\rm ff}\propto \sqrt{M/R}$. However, this second factor self-regulates. If at fixed $M$ the radius $R$ is very large, then $v_{\rm ff}$ is small, and the incoming material gains very little entropy in the shock. Small entropy leads to a smaller radius. Conversely, if $R$ is very small, then $v_{\rm ff}$ and the post-shock entropy will be large, and this will produce rapid swelling of the protostar. This effect means that the radius rapidly converges to a value that depends only on $M$ and $\dot{M}$.

This self-regulation does not happen if the material is assumed to accrete cold. In this case, the radial evolution of the star is determined solely by the amount of entropy that is assumed to remain in the accretion flow when it joins onto the star. As mentioned above, one common practice is to assume that the entropy of the accreting material is equal to the entropy of the gas already in the star, and, under this assumption, the choice of initial condition completely determines the subsequent evolution, since the choice of initial condition then determines the entropy content of the star thereafter.

Regardless of the boundary condition assumed, during this phase there is no nuclear burning in the star, as the interior is too cold for any such activity. Since there is no nuclear burning, and this phase generally lasts much less than the Kelvin-Helmholtz timescale on which radiation changes the star's structure, during this phase the entropy content of the star is nearly constant. This phase can therefore be referred to as the adiabatic stage in the star's evolution.

\subsection{Deuterium Ignition and Convection}

In Figure \ref{fig:kippenhahn_hosokawa09}, the next evolutionary phase begins at $\approx 0.25$ $M_\odot$, and continues to $\approx 0.7$ $M_\odot$. This stage is marked by two distinct but interrelated phenomena: the onset of nuclear burning and the onset of convection. The driving force behind both phenomena is that, as the protostar gains mass, its interior temperature rises. Recall the results of our calculation from chapter \ref{ch:protostar_form}: for a polytrope, which is not an unreasonable description of the accreting protostar, the central temperature rises with mass to the $T_c\propto M^{(2\gamma-2)/(3\gamma-2)}$. Thus even at fixed entropy the central temperature must rise as the star gains mass.

Once $T_c$ reaches $\sim 10^6$ K, deuterium will ignite at the center of the protostar. This has three significant effects. The first is that deuterium acts as a thermostat for the star's center, much as hydrogen does in a main sequence star. Because the energy generation rate is so incredibly sensitive to $T$ (rising as the 11.8 power!), any slight raise in the temperature causes it to jump enough to raise the pressure and adiabatically expand the star, reducing $T$. Thus, $T_c$ becomes fixed at $10^6$ K -- which is part of the reason we did not need an expression for $\epsilon$ that would work at higher temperatures. The star adjusts its radius accordingly, which generally requires that the radius increase as the mass rises. Thus deuterium burning temporarily halts core contraction. Both effects are visible in Figure \ref{fig:kippenhahn_hosokawa09}. The halting of core contraction is apparent from the way the dotted lines showing constant mass enclosed bend upward at $\approx 0.3$ $M_\odot$, and the nearly constant core temperature is visible from the fact that, between $\approx 0.25$ $M_\odot$ and $3-4$ $M_\odot$, a factor of more than 10 in mass, the central temperature stays within a factor of 2 of $10^6$ K.

The second effect of deuterium burning that it causes a rapid rise in the entropy at the center of the star: looking at the heat equation (\ref{heat}), we can see that if $\epsilon$ is large, then $\partial s/\partial t$ will be as well. This has the effect of starting up convection in the star. Before deuterium burning the star is generally stable against convection. That is because the entropy profile is determined by infall, and since shells that fall onto the star later arrive at higher velocities (due to the rising mass), they have higher entropy. Thus $s$ is an increasing function of $M_r$, which is the condition for convective stability. Deuterium burning reverses this, and convection follows, eventually turning much of the star convective. This also ensures the star a continuing supply of deuterium fuel, since convection will drag gas from the outer parts of the star down to the core, where they can be burned.

An important caveat here is that, although D burning encourages convection, it is not necessary for it. In the absence of D, or for very high accretion rates, the onset of convection is driven by the increasing luminosity of the stellar core as it undergoes KH contraction. This energy must be transported outwards, and as the star's mass rises and the luminosity goes up, eventually the energy that must be transported exceeds the ability of radiation to carry it. Convection results. For very high accretion rates, this effect drives the onset of convection even before the onset of D burning.

A third effect of the deuterium thermostat is that it forces the star to obey a nearly-linear mass-radius relation, and thus to obey a particular relationship between accretion rate and accretion luminosity. One can show that for a polytrope the central temperature and surface escape speed are related by
\begin{equation}
\psi = \frac{GM}{R} = \frac{1}{2}v_{\rm esc}^2 = T_n \frac{k_B T_c}{\mu m_{\rm H}},
\end{equation}
where $T_n$ is a dimensionless constant of order unity that depends only on the polytropic index. For $n=3/2$, expected for a fully convective star, $T_n = 1.86$. Plugging in this value of $T_n$, $\mu=0.61$ (the mean molecular weight for a fully ionized gas of H and He in the standard abundance ratio), and $T_c = 10^6$ K, one obtains $\psi = 2.5\times 10^{14}$ erg g$^{-1}$ as the energy yield from accretion.

\subsection{Deuterium Exhaustion and Formation of a Radiative Barrier}

The next evolutionary phase, which runs from $\approx 0.6 - 3$ $M_\odot$ in Figure \ref{fig:kippenhahn_hosokawa09}, is marked by the exhaustion of deuterium in the stellar core. Deuterium can only hold up the star for a finite amount of time. The reason is simply that there is not that much of it. Each deuterium burned provides $5.5$ MeV of energy, comparable to the $7$ MeV provided by burning hydrogen, but there are only $2\times 10^{-5}$ D nuclei per H nuclei. Thus, at fixed luminosity the "main sequence" lifetime for D burning is shorter than that for H burning by a factor of $2\times 10^{-5} \times 5.5/7 = 1.6\times 10^{-5}$.

We therefore see that, while a main sequence star can burn hydrogen for $\sim 10^{10}$ yr, a comparable pre-main sequence star of the same mass and luminosity burning deuterium can only do it for only a few times $10^5$ yr. To be more precise, the time required for a star to exhaust its deuterium is
\begin{equation}
t_{\rm D} = \frac{[\mbox{D}/\mbox{H}] \Delta E_{\rm D} M}{m_{\rm H} L} = 1.5\times 10^5 M_{0} L_1^{-1}\mbox{ yr},
\end{equation}
where $\Delta E_{\rm D}= 5.5$ MeV. Thus deuterium burning will briefly hold up a star's contraction, but cannot delay it for long. However, a brief note is in order here: while this delay is not long compared to the lifetime of a star, it is comparable to the formation time of the star. Recall that typical accretion rates are of order a few times $10^{-6}$ $\msun$ yr$^{-1}$, so a 1 $\msun$ star takes a few times $10^5$ yr to form. Thus stars may burn deuterium for most of the time they are accreting.

The exhaustion of deuterium does not mean the end of deuterium burning, since fresh deuterium that is brought to the star as it continues accreting will still burn. Instead, the exhaustion of core deuterium happens for a more subtle reason. As the deuterium supply begins to run out, the rate of energy generation in the core becomes insufficient to prevent it from undergoing further contraction, leading to rising temperatures. The rise in central temperature lowers the opacity, which is governed by a Kramers' law: $\kappa_{\rm R} \propto \rho T^{-3.5}$. This in turn makes it easier for radiation to transport energy outward. Eventually this shuts off convection somewhere within the star, leading to formation of what is called a radiative barrier.

The formation of the barrier ends the transport of D to the stellar center. The tiny bit of D left in the core is quickly consumed, and, without D burning to drive an entropy gradient, convection shuts off through the entire core. This is the physics behind the nearly-simultaneous end of central D burning and central convection that occurs near $0.6$ $M_\odot$ in Figure \ref{fig:kippenhahn_hosokawa09}. After this transition, the core is able to resume contraction, and D continues to burn as fast as it accretes. However, it now does so in a shell around the core rather than in the core.

\subsection{Swelling}

The next evolutionary phase, which occurs from $\approx 3-4$ $M_\odot$ in Figure \ref{fig:kippenhahn_hosokawa09}, is swelling. This phase is marked by a marked increase in the star's radius over a relatively short period of time. The physical mechanism driving this is the radiative barrier discussed above. The radiative barrier forms because increasing temperatures drive decreasing opacities, allowing more rapid transport of energy by radiation. The decreased opacity allows the center of the star to lose entropy rapidly, and the entropy to be transported to the outer parts of the star via radiation. The result is a wave of luminosity and entropy that propagates outward through the star.

\begin{marginfigure}
\includegraphics[width=\linewidth]{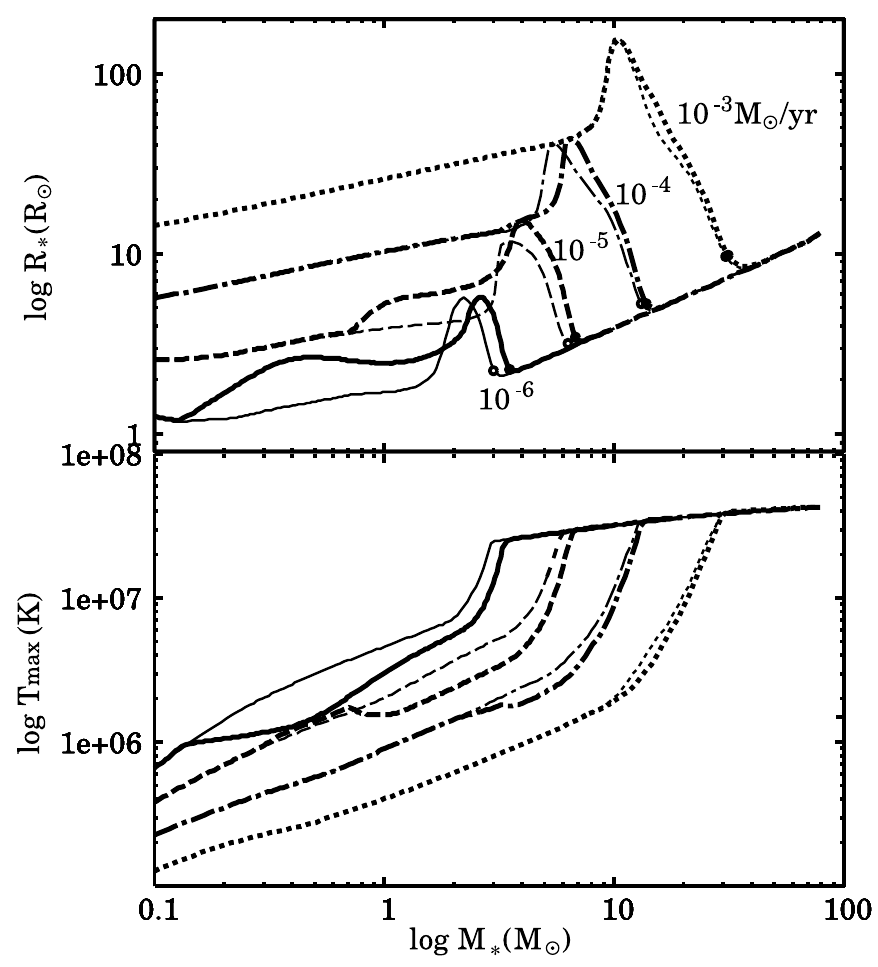}
\caption[Protostellar mass-radius relation for different accretion rates]{
\label{fig:rt_hosokawa09}
Radius versus mass (top panel) and maximum interior temperature versus mass (bottom panel) for protostars accreting at different rates. The accretion rate is indicated by the line style, as illustrated in the top panel. For each accretion rate there are two lines, one thick and one thin. The thick line is for the observed Milky Way deuterium abundance, while the thin line is the result assuming zero deuterium abundance.  Credit: \citet{hosokawa09a}, \copyright AAS. Reproduced with permission.
}
\end{marginfigure}

Once the wave of luminosity and entropy gets near the stellar surface, which is not confined by the weight of overlying material, the surface undergoes a rapid expansion, leading to rapid swelling. The maximum radius, and the mass at which the swelling phase occurs, is a strong function of the accretion rate (Figure \ref{fig:rt_hosokawa09}). However, even at very low accretion rates, swelling does not occur until the mass exceeds $1$ $M_\odot$, and thus this phase occurs only for stars more massive than the Sun.

\subsection{Contraction to the Main Sequence}

The final stage of protostellar evolution is contraction to the main sequence. Once the entropy wave hits the surface, the star is able to begin losing energy and entropy fairly quickly, and it resumes contraction. This marks the final phase of protostellar evolution, visible above $\approx 4$ $M_\odot$ in Figure \ref{fig:kippenhahn_hosokawa09}. Contraction ends once the core temperature becomes hot enough to ignite hydrogen, landing the star at least on the main sequence.

\section{Observable Evolution of Protostars}

\begin{figure}
\includegraphics[width=\linewidth]{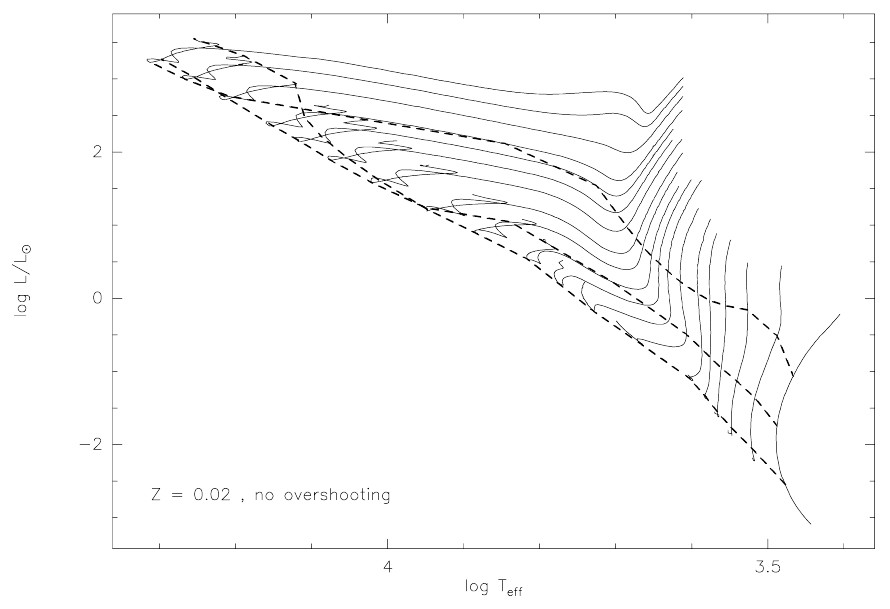}
\caption[Pre-main sequence evolutionary tracks]{
\label{fig:pms_siess00}
Solid lines show tracks taken by stars of varying masses, from $0.1$ $M_\odot$ (rightmost line) to $7.0$ $M_\odot$ (leftmost line) in the theoretical HR diagram of luminosity versus effective temperature. Stars begin at the upper right of the tracks and evolve to the lower left; tracks end at the main sequence. Dashed lines represent isochrones corresponding to $10^6$, $10^7$, and $10^8$ yr, from top right to bottom left. Credit: \citeauthor{siess00a}, A\&A, 358, 593, 2000, reproduced
with permission \copyright\, ESO.
}
\end{figure}

We have just discussed the interior behavior of an evolving protostar. While this is important, it is also critical to predict the observable properties of the star during this evolutionary sequence. In particular, we wish to understand the star's luminosity and effective temperature, which dictate its location in the Hertzsprung-Russell diagram. The required values can simply be read off from the evolutionary models (Figure \ref{fig:pms_siess00}), giving rise to a track of luminosity versus effective temperature in the HR diagram.

The two most important applications of models of this sort is in determining the mass and age distributions of young stars. The former is critical to determining the IMF, as discussed in chapter \ref{ch:imf_obs}, while the latter is critical to questions of both how clusters form, and to the problems of disk dispersal and planet formation (chapters \ref{ch:late_disk} and \ref{ch:planets}).

\subsection{The Birthline}

Before delving into the tracks themselves, we have to ask what is actually observable. As long as a star is accreting from its parent core, it will probably not be visible in the optical, due to the high opacity of the dusty gas in the core. Thus we are most concerned with stars' appearance in the HR diagram only after they have finished their main accretion phase. We refer to stars that are still accreting and thus not generally optically-observable as protostars, and those that are in this post-accretion phase as pre-main sequence stars. 

For stars below $\sim 1$ $M_\odot$, examining Figure \ref{fig:kippenhahn_hosokawa09}, we see that the transition from protostar to pre-main sequence star will occur some time after the onset of deuterium burning, either during the core or shell burning phases depending on the mass and accretion history. More massive stars will become visible only during KH contraction, or even after the onset of hydrogen burning. The lowest mass stars might be observable even before the start of deuterium burning. However, for the majority of the pre-main sequence stars that we can observe, they first become visible during the D burning phase.

\begin{marginfigure}
\includegraphics[width=\linewidth]{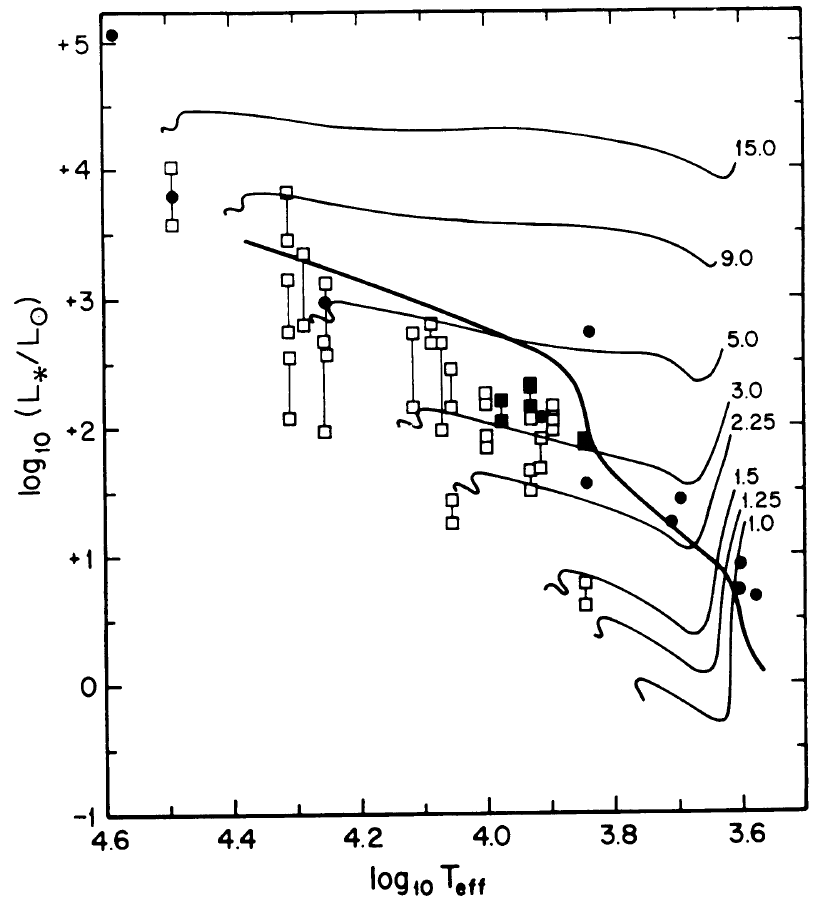}
\caption[The protostellar birthline]{
\label{fig:birthline_palla90}
Thin lines show tracks taken by stars of varying masses (indicated by the annotation, in $M_\odot$) in the theoretical HR diagram of luminosity versus effective temperature. Stars begin at the upper right of the tracks and evolve to the lower left; tracks end at the main sequence. The thick line crossing the tracks is the birthline, the point at which the stars stop accreting and become optically visible. Squares and circles represent the properties of observed young stars. Credit: \citet{palla90a}, \copyright AAS. Reproduced with permission.
}
\end{marginfigure}

Since there is a strict mass-radius relation during core deuterium burning (with some variation due to varying accretion rates), there must be a corresponding relationship between $L$ and $T$, just like the main sequence. We call this line in the HR diagram, on which protostars first appear, the birthline; it was first described by \citet{stahler83a} (Figure \ref{fig:birthline_palla90}). Since young stars are larger and more luminous that main sequence stars of the same mass, this line lies at higher $L$ and lower $T$ than the main sequence.

\subsection{The Hayashi Track}

Now that we understand what is observable, let us turn to the tracks themselves. The tracks shown in Figures \ref{fig:pms_siess00} and \ref{fig:birthline_palla90} have several distinct features. One is that, for low mass stars, the initial phases of evolution in the HR diagram are nearly vertically, i.e., at constant $T_{\rm eff}$. The vertical tracks for different masses are very close together. This vertical part of the evolution is called the Hayashi track, after its discoverer, who predicted it theoretically \citep{hayashi61a}. For low mass stars, the majority of the Hayashi track lies after the birthline, so it is directly observable.

The origin of the Hayashi track is in the physics of opacity in stellar atmospheres at low temperature. At temperatures below about $10^4$ K, hydrogen becomes neutral, and the only free electrons available come from metal atoms with lower ionization energies. Some of these electrons become bound with hydrogen atoms, forming H$^-$, and this ion is the dominant source of opacity.  Thus the opacity depends on the number of free electrons provided by metal atoms, which in turn depends extremely sensitively on the temperature.

If the temperature falls too low, the opacity will be so low that, even integrating through the rest of the star's mass, the optical depth to infinity will be $<2/3$. Since the photosphere must always be defined by a surface of optical depth unity, this effectively establishes a minimum surface temperature for the star required to maintain $\tau \approx 1$. This minimum temperature depends weakly on the star's mass and radius, but to good approximation it is simply $T_{\rm min} = T_{\rm H} = 3500$ K, where $T_{\rm H}$ is the Hayashi temperature. Low mass protostars, due to their large radii, wind up right against this limit, which is why they all contract along vertical tracks that are packed close together in $T_{\rm eff}$.

We can make this argument a bit more quantitative as follows.\footnote{This argument is taken from \citet{prialnik09a}.} Let us approximate the stellar photosphere at radius $R$ as producing blackbody emission and obeying a simple ideal gas law equation of state. In this case we have
\begin{eqnarray}
\label{eq:bb}
\log L & = & 4 \log T_R - 2 \log R + \mbox{constant} \\
\label{eq:idealgas}
\log P_R & = & \log \rho_R + \log T_R + \mbox{constant},
\end{eqnarray}
where the subscript $R$ indicates that a quantity is to be evaluated at the stellar outer radius, and we are writing things in terms of logarithms rather than powerlaw scalings for future convenience. Now let us consider a star that is a polytrope, following $P\propto K_P \rho^{(n+1)/n}$, where $n$ is the polytropic index. The polytropic constant $K_P$ is related to the stellar mass and radius by
\begin{equation}
K_P \propto M^{(n-1)/n}R^{(3-n)/n}.
\end{equation}
Thus we have
\begin{equation}
\log K_P = \left(\frac{n-1}{n}\right) \log M + \left(\frac{3-n}{n}\right)\log R + \mbox{constant},
\end{equation}
and the pressure scales with $M$ and $R$ as
\begin{equation}
\label{eq:polytrope}
\log P = \left(\frac{n-1}{n}\right) \log M + \left(\frac{3-n}{n}\right) \log R + \left(\frac{n+1}{n}\right) \log \rho + \mbox{constant}.
\end{equation}
Hydrostatic balance at the photosphere requires
\begin{equation}
\frac{dP}{dr} = \rho_R \frac{GM}{R^2} \qquad\Longrightarrow\qquad
P_R = \frac{GM}{R^2} \int_R^\infty \rho \, dr,
\end{equation}
where $P_R$ is the pressure at the photosphere and we are approximating the $GM/R^2$ is essentially constant through the photosphere. The photosphere is defined by the condition
\begin{equation}
\kappa_{\rm R} \int_R^\infty \rho\,dr \approx 1,
\end{equation}
where we are also approximating $\kappa_{\rm R}$ as constant, so putting this together we have
\begin{equation}
P_R \approx \frac{GM}{R^2\kappa_{\rm R}} \qquad\Longrightarrow\qquad
\log P_R \approx \log M - 2 \log R - \log \kappa_{\rm R}.
\end{equation}

To make further progress, we will assume that we can approximate the opacity as some powerlaw in the temperature, $\kappa_{\mathrm{R}} \propto \rho T^b$. For Kramers opacity, for example, $b=-3.5$. Substituting this into the equation for $P_R$, we have
\begin{equation}
\label{eq:opacity}
\log P_R = \log M - 2\log R - \log \rho_R - b\log T_R + \mbox{constant}.
\end{equation}
Equations (\ref{eq:bb}), (\ref{eq:idealgas}), (\ref{eq:polytrope}) and (\ref{eq:opacity}) constitute a system of four linear equations in the four unknowns $\log P_R$, $\log \rho_R$, $\log T_R$, and $\log L$. Solving this linear system yields the result
\begin{equation}
\log L = \left(\frac{9 - 2 n + b}{2-n}\right) \log T_R - \left(\frac{2n-1}{2-n}\right) \log M + \mbox{constant}.
\end{equation}
This equation describes the shape of a track in the HR diagram, because it relates $\log L$ to $\log T_R$, the photospheric temperature. To see what it implies, we can assume that young low mass stars will be fully convective thanks to D burning, so $n\approx 1.5$.

This leaves only $b$. As mentioned previously, the H$^{-}$ opacity has the property that it rises sharply with temperatures of a few thousand K, because at these temperatures collisional velocities are not high enough to dissociate H$^-$, but they are able to dissociate other atoms, which in turn produces free electrons that can yield H$^{-}$. The higher the temperature, the more free electrons available, and thus the higher the H$^{-}$ opacity. The net result is that, in this temperature range, $b$ takes on a fairly large value: $\sim 4-9$ depending on exactly where in the temperature range we are. Note that this is the opposite of the normal behavior for stellar opacities (e.g., Kramer's opacity), where the opacity falls with increasing temperature.

If we plug $b = 9$ and $n=1.5$ into the equation we have just derived, we find obtain
\begin{equation}
\log L = 30\log T_R - 4 \log M + \mbox{constant}.
\end{equation}
Using $b=4$ changes the 30 to a 20. Either way, we conclude that $\log L$ changes extremely steeply with $\log T_R$, which implies that the HR diagram track for stars with this low $T_{\rm eff}$ must be nearly vertical -- hence the Hayashi track. We also see that the location of the Hayashi tracks for stars of different masses will be slightly offset, because of the $4\log M$ term. This qualitatively explains what the numerical models produce.

\subsection{The Heyney Track}

Contraction at nearly constant $T_{\rm eff}$ continues until the star contracts enough to raise its surface temperature above $T_{\rm H}$. This increase in temperature also causes the star to transition from convective to radiative, since the opacity drops with temperature at high temperatures, and a lower opacity lets radiation rather than convection carry the energy outward.

In the HR diagram, the contraction and increase in $T_{\rm eff}$ produces a vaguely horizontal evolutionary track. This is called the Heyney track. The star continues to contract until its center becomes warm enough to allow H burning to begin.
At that point it may contract a small additional amount, but the star is essentially on the main sequence. The total time required depends on the stellar mass, but it ranges from several hundred Myr for $0.1$ $\msun$ stars to essentially zero time for very massive stars, which reach the main sequence while still accreting.

\problemset

\begin{enumerate}

\item {\bf A Simple Protostellar Evolution Model.}\\
Consider a protostar forming with a constant accretion rate $\dot{M}$. The accreting gas is fully molecular, arrives at free-fall, and radiates away a luminosity $L_{\rm acc} = f_{\rm acc} G M \dot{M}/R$ at the accretion shock, where $M$ and $R$ are the instantaneous protostellar mass and radius, and $f_{\rm acc}$ is a numerical constant of order unity. At the end of contraction the resulting star is fully ionized, all its deuterium has been burned to hydrogen, and it is in hydrostatic equilibrium. The ionization potential of hydrogen is $\psi_I = 13.6$ eV per amu, the dissociation potential of molecular hydrogen is $\psi_M=2.2$ eV per amu, and the energy released by deuterium burning is $\psi_D\approx 100$ eV per amu of total gas (not per amu of deuterium).
\begin{enumerate}
\item First consider a low-mass protostar whose internal structure is well-described by an $n=3/2$ polytrope. Compute the total energy of the star, including thermal energy, gravitational energy, and the chemical energies associated with ionization, dissociation, and deuterium burning.
\item Use your expression for the total energy to derive an evolution equation for the radius for a star. Assume the star is always on the Hayashi track, which for the purposes of this problem we will approximate as having a fixed effective temperature $T_{\rm H} = 3500$ K.
\item Numerically integrate your equation and plot the radius as a function of mass for $\dot{M} = 10^{-5}$ $\msun$ yr$^{-1}$ and $f_{\rm acc}=3/4$. As an initial condition, use $R=2.5$ $\rsun$ and $M=0.01$ $\msun$, and stop the integration at a mass of $M=1.0$ $\msun$. Plot the radius and luminosity as a function of mass; in the luminosity, include both the the accretion luminosity and the internal luminosity produced by the star.
\item Now consider two modifications we can make to allow the model to work for massive protostars. First, since massive stars are radiative, the polytropic index will be roughly $n=3$ rather than $n=3/2$. Second, the surface temperature will in general be larger than the Hayashi limit, so take the luminosity to be $L=\max[L_{\rm H}, \lsun(M/\msun)^3]$, where $L_{\rm H}=4\pi R^2 \sigma T_{\rm H}^4$ and $R$ is the stellar radius. Modify your evolution equation for the radius to include these effects, and numerically integrate the modified equations up to $M=50$ $\msun$ for $\dot{M} = 10^{-4}$ $\msun$ yr$^{-1}$ and $f_{\rm acc}=3/4$, using the same initial conditions as for the low mass case. Plot $R$ and $L$ versus $M$.
\item Compare your result to the fitting formula for the ZAMS radius of solar-metallicity stars as a function of $M$ in \citet{tout96a}\footnote{\href{http://adsabs.harvard.edu/abs/1996MNRAS.281..257T}{Tout et al., 1996, MNRAS 281, 257}}. Find the mass at which the massive star would join the main sequence. Your plots for $R$ and $L$ are only valid up to this mass, because this simple model does not include hydrogen burning.
\end{enumerate}

\item \textbf{Self-Similar Viscous Disks.}\\
Consider a protostellar disk orbiting a star, governed by the usual viscous evolution equation
\begin{displaymath}
\frac{\partial\Sigma}{\partial t} = \frac{3}{\varpi} \frac{\partial}{\partial \varpi} \left[\varpi^{1/2} \frac{\partial}{\partial \varpi} \left(\nu \Sigma \varpi^{1/2}\right)\right],
\end{displaymath}
where $\Sigma$ is the surface density, $\varpi$ is the radius in cylindrical coordinates, and $\nu$ is the viscosity. Suppose that the viscosity is linearly proportional to the radius, $\nu = \nu_1 (\varpi/\varpi_1)$.
\begin{enumerate}
\item Non-dimensionalize the evolution equation by making a change of variables to the dimensionless position, time, and surface density $x=\varpi/\varpi_1$, $T = t/t_s$, $S = \Sigma/\Sigma_1$, where $t_s = \varpi_1^2/(3\nu_1)$.
\item Use your non-dimensionalized equation to show that
\begin{displaymath}
\Sigma = \left(\frac{C}{3\pi \nu_1}\right) \frac{e^{-x/T}}{x T^{3/2}}
\end{displaymath}
is a solution of the equation for an arbitrary constant $C$.
\item Calculate the total mass in the disk in terms of $C$, $t_s$, and $t$, and calculate the time rate of change of this mass. Based on your result, give a physical interpretation of what the constant $C$ means. (Hint: what units does $C$ have?)
\item Plot $S$ versus $x$ at $T = 1, 1.5, 2$, and $4$. Give a physical interpretation of the results.\\
\end{enumerate}

\item {\bf A Simple T Tauri Disk Model.}\\
In this problem we will construct a simple model of a T Tauri star disk in terms of a few parameters: the midplane density and temperature $\rho_m$ and $T_m$, the surface temperature $T_s$, the angular velocity $\Omega$, and the specific opacity of the disk material $\kappa$. We assume that the disk is very geometrically thin and optically thick, and that it is in thermal and mechanical equilibrium.
\begin{enumerate}
\item Assume that the disk radiates as a blackbody at temperature $T_s$. Show that the surface and midplane temperatures are related approximately by
\begin{displaymath}
T_m \approx \left(\frac{3}{8}\kappa\Sigma\right)^{1/4} T_s,
\end{displaymath}
where $\Sigma$ is the disk surface density.
\item Suppose the disk is characterized by a standard $\alpha$ model, meaning that the viscosity $\nu=\alpha c_s H$, where $H$ is the scale height and $c_s$ is the sound speed. For such a disk the rate per unit area of the disk surface (counting each side separately) at which energy is released by viscous dissipation is $F_d=(9/8) \nu \Sigma \Omega^2$. Derive an estimate for the midplane temperature $T_m$ in terms of $\Sigma$, $\Omega$, and $\alpha$.
\item Calculate the cooling time of the disk in terms of the orbital period. Should the behavior of the disk be closer to isothermal or adiabatic?
\item Consider a disk with a mass of $0.03$ $\msun$ orbiting a $1$ $\msun$ star, which has $\kappa=3$ cm$^2$ g$^{-1}$ and $\alpha=0.01$. The disk runs from 1 to 20 AU, and the surface density varies with radius $\varpi$ as $\varpi^{-1}$. Use your model to express $\rho_m$, $T_m$, and $T_s$ as functions of the radius, normalized to 1 AU; i.e., derive results of the form $\rho_m = \rho_0 (\varpi/\mathrm{AU})^p$ for each of the quantities listed. Is your numerical model disk gravitationally unstable (i.e., $Q<1$) anywhere?
\end{enumerate}

\end{enumerate}

\chapter{Massive Star Formation}
\label{ch:massivestar}

\marginnote{
\textbf{Suggested background reading:}
\begin{itemize}
\item \href{http://adsabs.harvard.edu/abs/2014prpl.conf..149T}{Tan, J.~C., et al. 2014, in "Protostars and Planets VI", ed.~H.~Beuther et al., pp.~149-172} \nocite{tan14a}
\end{itemize}
\textbf{Suggested literature:}
\begin{itemize}
\item \href{http://adsabs.harvard.edu/abs/2013ApJ...766...97M}{Myers, A.~T., et al. 2013, ApJ, 766, 97} \nocite{myers13a}
\end{itemize}
}

This chapter will focus on the particular problem of massive stars. While this might seem something of a digression from our march to ever-smaller scales, we are only prepared to address massive stars now because before tackling massive stars, we first needed to develop a theory for low-mass star formation. Only with that understanding in place are we prepared to tackle the significantly more difficult problem of how massive stars form. These stars are extremely rare -- those above 10 $\msun$ constitute only about $10\%$ of all stars formed by mass, and only about $0.2\%$ by number -- but their huge energetic output gives them an importance disproportionate to their numbers. As we shall see, this energetic output also creates unique questions regarding the process by which massive stars form.

\section{Observational Phenomenology}

\subsection{Challenges}

Unfortunately, our observational knowledge of massive star formation is much more limited than our knowledge of the analogous processes governing the formation of Solar mass stars stars. The difficulty is four-fold. First, because massive stars are rare, purely on statistical grounds locations of massive star formation are likely to be much further from Earth than sites of low mass star formation. Indeed, the nearest region of massive star formation, in the Orion cloud, is 400 pc away. Many regions of study are even further, typically $1-2$ kpc. The largest clusters, where massive star formation is most active, are located in the great molecular ring at 3 kpc from the Galactic center, about 5 kpc from us. In contrast, many of the best studied regions of low mass star formation, such as the Taurus cloud, are only $\sim 100-150$ pc from Earth. The larger distance means that we can resolve only large physical scales, and that we need proportionally more telescope time to do so.

This unfortunately compounds the second challenge: crowding and confusion. Massive stars are generally found in massive star clusters. Whether this is a physical necessity or simply a result of statistics -- i.e., can massive stars only form in clusters, or is it simply improbable that a small cluster will harbor a very rare, massive star -- is a matter of hot debate. Regardless of the outcome of that debate, the clustered environment means that extreme spatial resolution is needed to avoid confusion. For example, at the center of the Orion Nebula Cluster, where the Trapezium stars are located, the stellar density is $\sim 10^5$ pc$^{-3}$ \citep{hillenbrand98a}, so the typical interstellar distance is only $0.02$ pc, or about 5000 AU. In terms of angular resolution, at the 400 pc distance to Orion this is about $10"$. The same cluster at the distance of 2 kpc has a mean angular separation between stars at its center of $2"$. Such resolutions are in reach for the highest resolution radio and sub-mm interferometers, and in the optical from HST or ground-based systems with adaptive optics, but are not far from the limits. This means that confusion is a constant worry.

The third challenge is obscuration. As we shall see, the typical region of massive star formation has a surface density of $\sim 1$ g cm$^{-2}$. For a standard Milky Way extinction curve, including the effects of ice mantles on the dust grains, this corresponds to visual extinction $A_V\approx 500$ mag. Even in the near-infrared at K band, the extinction is only a factor of $\sim 10$ smaller, so $A_K\approx 50$ mag. This means that optical and even near-IR observations are fairly useless until the tail end of the star formation process, when the vast majority of the gas has been cleared away. Only mid-IR or radio and sub-mm observations are possible during most of the star formation process. This limitation to long wavelengths of course compounds the problem of confusion, since it means that we can get high resolution only via radio interferometers.

The final problem is timescales. As discussed in Chapter \ref{ch:feedback}, feedback from massive stars rapidly destroys the environment in which they form. For example, once they become optically revealed, the disks around massive stars probably survive $\lesssim 10^5$ yr, as opposed to $>10^6$ yr for low mass stars (as we will discuss in chapter \ref{ch:late_disk}). Thus we have a very limited window in which we can observe massive star formation underway. We essentially can only see massive star formation happening when it is still in the embedded phase. In terms of our classification scheme, massive stars only have a class 0 and a class I phase, not the longer class II or class III phases.

\subsection{Massive Clumps}

\begin{figure}
\includegraphics[width=\linewidth]{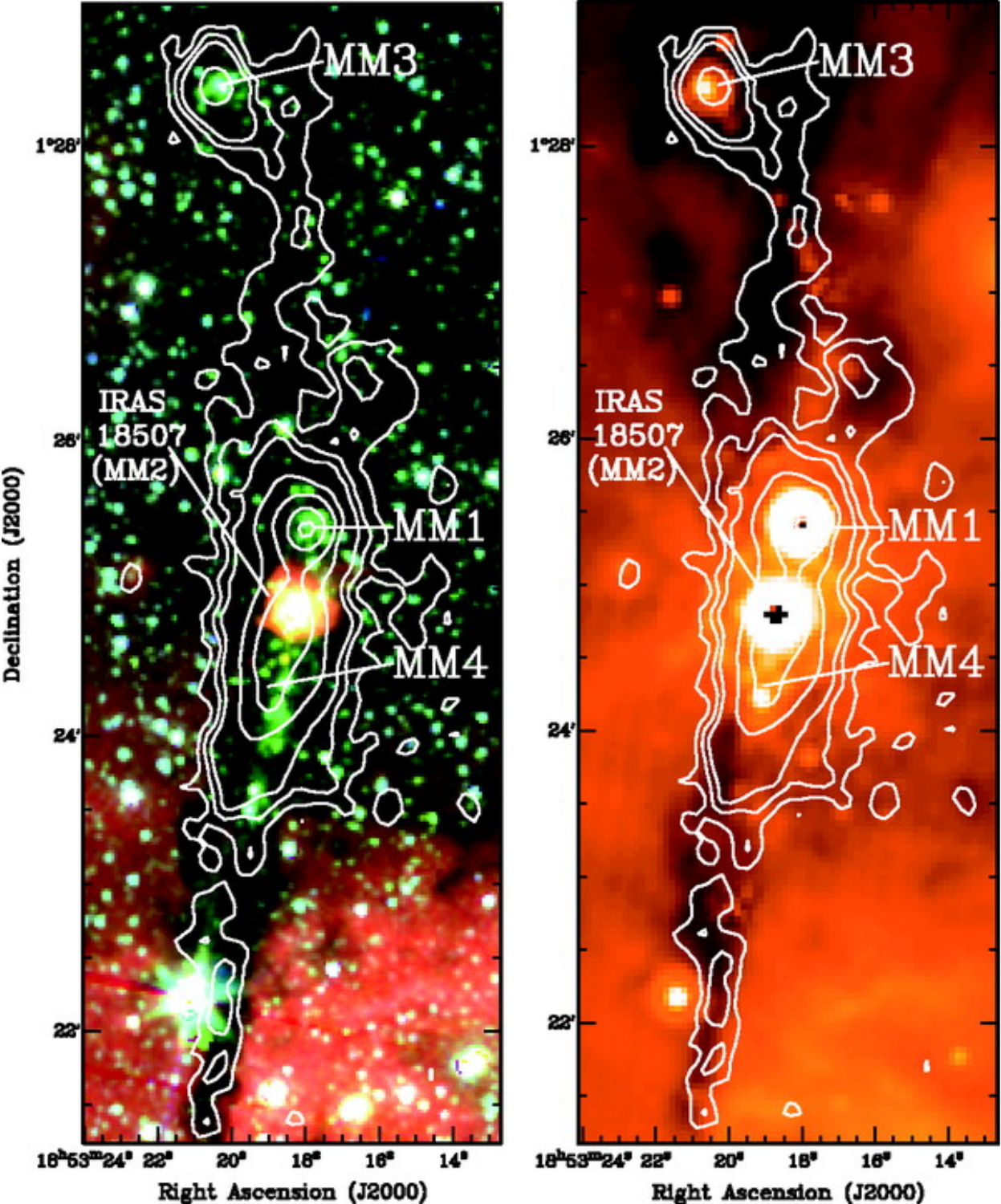}
\caption[IR and mm images of an IRDC]{
\label{fig:irdc_rathborne05}
A typical infrared dark cloud (IRDC). The left image shows \textit{Spitzer}/IRAC (near-IR), where the cloud is seen in absorption against the galactic background, while the right image shows \textit{Spitzer}/MIPS (mid-IR), where parts are seen in absorption and parts in emission. The white contours, which are the same in both panels, show mm continuum emission from cold dust. Credit: \citet{rathborne06a}, \copyright AAS. Reproduced with permission.
}
\end{figure}

Given these challenges, what do we know? Observational surveys usually find sites of massive star formation by exploiting one of three techniques. First, such sites have huge far-IR fluxes, due to the copious amounts of warm dust that are produced by an obscured massive star. Second, sites of massive star formation are characterized by having very high surface densities, such that they are opaque at near-IR wavelengths. One can also detect these regions in near-IR absorption against the galactic background, for example using the 8 $\mu$m band on \textit{Spitzer}. The classes of object discovered this way are called infrared dark clouds (IRDCs). Figure \ref{fig:irdc_rathborne05} shows an example. Third, one can look for the maser emission that often accompanies massive star formation. The maser emission comes from strong shocks in high density gas, which are probably produced by the outflows of massive stars travelling at speeds up to $\sim 1000$ km s$^{-1}$ -- the escape speed from the stellar surface. Maser emission is useful for surveys, because masers have an immense brightness temperature, making it possible to survey the sky rapidly.

The typical clump forming a massive star cluster, detected with any technique, seems to have a mass of a few thousand $\msun$, and a radius of $\sim 1-2$ pc. Combining these numbers, the surface density is $\sim 0.1-1$ g cm$^{-2}$. This much higher that the typical surface density in regions of low mass star formation, which is generally closer to $\sim 0.01 - 0.1$ g cm$^{-2}$.  Recall from our discussion of Larson's Laws in chapter \ref{ch:gmcs} that the statement that clouds have uniform surface density is equivalent to the combination of virial balance and the linewidth-size relation. The higher surface density of massive star-forming regions compared to the bulk of the material in GMCs implies either that these regions are not in virial balance, that they are not on the linewidth-size relation, or both. When we observe these regions using a molecular tracer, for the most part we find that these clumps {\it do} appear to be roughly virial, but that they are off the linewidth-size relation seen for other material in molecular clouds.

The origin of these large velocity dispersions is an interesting problem. They could be driven by gravitational collapse, of course, but that would only supply energy for one crossing time or so, and then would lead to global collapse. We know from galactic-scale surveys, however, that this gas cannot form stars rapidly any more than can the lower density material in GMCs. Otherwise the star formation rate would be too high to compared to what we observe. This suggests that these regions must be stabilized by internal feedback or disrupted by feedback in only $\sim 1$ crossing time, before they have the opportunity to convert most of the gas mass to stars.

\subsection{Massive Cores}
\label{ssec:massivecores}

\begin{marginfigure}
\includegraphics[width=\linewidth]{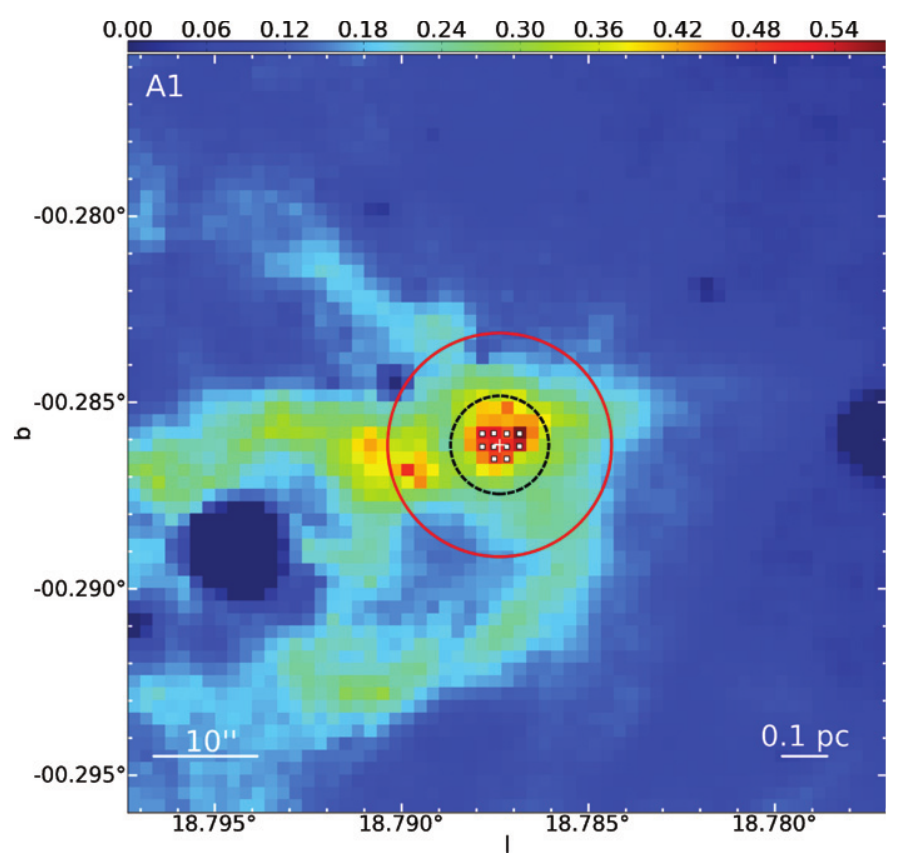}
\caption[A massive core in IR absorption]{
\label{fig:massivecore_butler12}
A massive protostellar core seen in IR absorption. Colors indicate the inferred column density in g cm$^{-2}$. Pixels marked with white dots are lower limits. The black circle shows a radius enclosing $60$ $M_\odot$, and the red circle shows the core radius inferred by fitting a core plus envelope model to the azimuthally-averaged surface density distribution. Credit: \citet{butler12a}, \copyright AAS. Reproduced with permission.
}
\end{marginfigure}

If one zooms in a bit more using an interferometer, to $\sim 0.1$ pc scales, one can find objects that are $\sim 100$ $\msun$ in mass and $\sim 0.1$ pc in radius. These are centrally concentrated, and appear to be forming stars. Their velocity dispersions are also about one is needed for them to be in virial balance, around 1 km s$^{-1}$. As with their parent clumps, such large velocity dispersions on such small scales puts these objects well off the linewidth-size relation seen in most material in GMCs. We refer to objects with these characteristics as massive cores. Figure \ref{fig:massivecore_butler12} shows an example.

In some cases we detect no mid-IR emission from massive cores, which indicates that any stars within them cannot yet be massive stars. However, even in cases with no mid-IR, there are signs of active protostellar outflows, in the form of SiO emission \citep[e.g.,][]{motte07a}. The statistics indicate that the starless phase for a massive core is at most $\sim 1000$ yr, implying that once a massive core is assembled it starts forming stars immediately, or even that star formation begins as it is being assembled.

It is instructive to perform some simple dimensional analysis for these objects. A region with a mass of $100$ $\msun$ and a radius of $0.1$ pc has a mean density of about $10^{-18}$ g cm$^{-3}$, or $n\sim 10^6$ cm$^{-3}$, and a free-fall time of $5\times 10^4$ yr. Thus we should expect one of these cores to form stars in $\sim 10^5$ yr, and to do so at an accretion rate $\dot{M}\approx M/t_{\rm ff} \approx 10^{-3}$ $\msun$ yr$^{-1}$. This is vastly higher than the expected accretion rates in the regions of low mass star formation close to Earth, and much larger than $c_s^3/G$ where $c_s$ is the thermal sound speed. 

It is also useful to phrase the accretion rate in terms of a velocity dispersion. Suppose we have a core in rough virial balance, so that
\begin{equation}
\alpha_{\rm vir} = \frac{5 \sigma^2 R}{G M} \approx 1,
\end{equation}
where the 1D velocity dispersion $\sigma$ here now includes contributions from both thermal and non-thermal motions. The density is $\rho=3 M/(4\pi R^3)$, so the free-fall time is
\begin{equation}
t_{\rm ff} = \sqrt{\frac{3\pi}{32 G \rho}} = \sqrt{\frac{\pi R^3}{8 G M}}.
\end{equation}
If the core collapses in free-fall, the accretion rate is
\begin{equation}
\dot{M} \approx \frac{M}{t_{\rm ff}} = \sqrt{\frac{8 G M^3}{\pi R^3}} = \sqrt{\frac{1000}{\pi\alpha_{\rm vir}^3}} \frac{\sigma^3}{G}.
\end{equation}
Thus, the accretion rate will be roughly $\sim 10 \sigma^3/G$.

\section{Fragmentation}

\subsection{Massive Core Fragmentation}

Given that we see these massive cores, can we understand how they turn into massive stars? The first thing that happens when one of these cores begins to collapse is that it will be subject to fragmentation. In effect, because it is so much larger than a thermal Jeans mass, a 100 $\msun$ massive core has the potential to become a small cluster rather than a single star or star system. On the other hand, both radiative heating and magnetic fields are capable of suppressing fragmentation. So what happens?

\begin{figure}
\includegraphics[width=\linewidth]{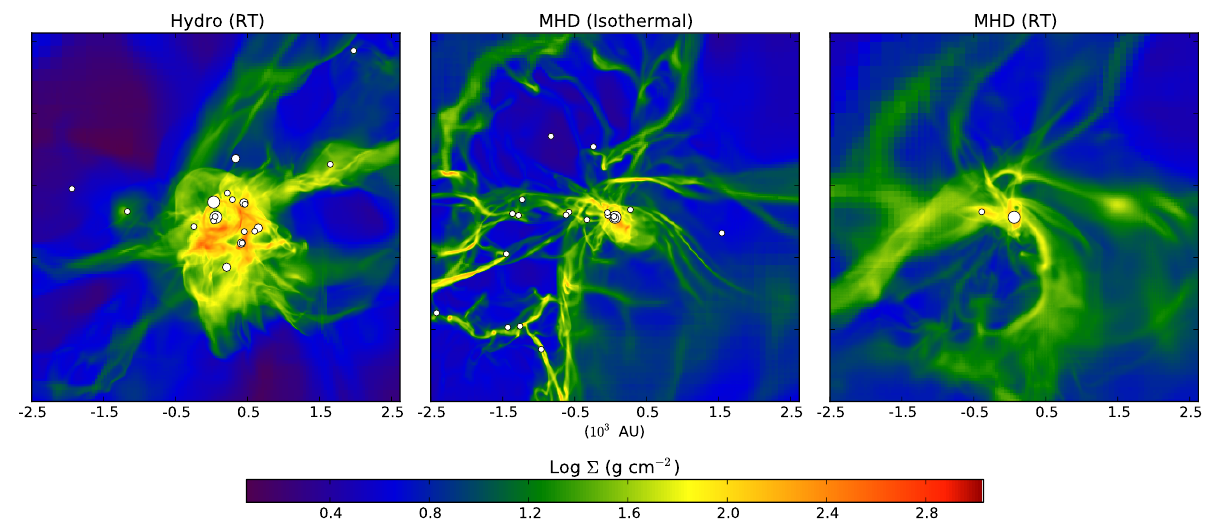}
\caption[Simulations of massive star formation with magnetic fields and radiation]{
\label{fig:msf_myers13}
Three simulations of the collapse of a 300 $M_\odot$ massive core. The color scale shows the projected gas density, and white points are stars, with the size indicating the mass. The three simulations use identical initial conditions, but different physics. The left panel uses radiative transfer but no magnetic fields, the middle uses magnetic fields but no radiation, and the right panel includes both magnetic fields and radiation. Credit: \citet{myers13a}, \copyright AAS. Reproduced with permission.
}
\end{figure}

This still a very active area of research, but recent simulations by \citet{commercon11c} and \citet{myers13a} that explore how radiative transfer and magnetic fields affect star formation suggests that a combination of the two is very effective at suppressing fragmentation of massive protostellar cores. The basic mechanism is quite analogous to the way that radiation feedback can shape the IMF overall: rapid accretion gives rise to a high accretion luminosity, which in turn heats the gas and raises the Jeans mass. Magnetic fields enhance this effect in two ways. First, by providing a convenient way of getting rid of angular momentum (as discussed in chapter \ref{ch:disks_theory}), they enhance the accretion rate. Second, they tend to stabilize the more distant, cooler parts of the core that are less heated by the radiation. These low-density regions may be Jeans unstable, but they are also magnetically subcritical and thus cannot fragment and collapse. Figure \ref{fig:msf_myers13} shows an example simulation.

\subsection{Massive Binaries}

As discussed in chapter \ref{ch:imf_obs}, massive stars are overwhelmingly members of binary or higher multiple star systems. Why this should be is obviously an interesting question, and related to the topic of fragmentation. Binaries can form in two ways. One way of making binaries is what we can call direct fragmentation: a collapsing gas core breaks up into two or more pieces during collapse. This possibility is closely related to the discussion of the IMF, in that it depends on the thermodynamics of the gas and its turbulent motions. The other possibility is disk fragmentation, in which material collapses into a disk and that disk then fragments. Direct fragmentation almost has to be the origin for wide period binaries, those with separations $\gtrsim 1000$ AU, the typical size of a protostellar disk. It could also be the origin for close ones. However, it is suggestive that the mass ratio distribution is somewhat different for close binaries than for distant ones.

There have been several numerical studies of the circumstances under which a core is expected to undergo fragmentation to produce a binary. Generally speaking, the amount of fragmentation appears to depend on the amount of initial turbulence in the core. Two important parameters controlling when and whether this happens are rate of rotation and the strength of the magnetic field in the initial cloud. A third parameter that becomes relevant in disks is the relationship between gas density and temperature.

\citet{machida08b} varied the rotation rate and magnetic field strength in clouds and found that they could draw boundaries in parameter space determining where various types of fragmentation occur. Higher rotation rates and weaker magnetic fields favor direct fragmentation, while slower rotation rates and stronger magnetic fields favor no fragmentation. Disk fragmentation appears to occur at intermediate values. Of course real cores have some level of turbulence, even if they are subsonic, and it is not entirely clear how to translate these conditions into probabilities of binary formation for turbulent cores.

The nature of disk fragmentation and its relationship with the thermal properties of the gas has been clarified in a series of papers by \citet{kratter06a} and \citet{kratter08a, kratter10a}. These authors point out that the behavior of a collapsing, rotating, non-magnetic core can be described in terms of two dimensionless numbers:
\begin{equation}
\xi \equiv \frac{\dot{M} G}{c_s^3} \qquad\qquad \Gamma = \frac{\dot{M}}{M_{*d}\Omega_{k,\rm in}} = \frac{\dot{M} \langle j\rangle_{\rm in}}{G^2 M_{*d}^3}.
\end{equation}
Here $\dot{M}$ is the rate at which matter falls onto the edge of the disk, $c_s$ is the sound speed in the disk, $M_{*d}$ is the total mass of the disk and the star it orbits, $\Omega_{k,\rm in}$ is the Keplerian angular frequency of matter entering the disk and $\langle j\rangle_{\rm in}$ is the mean specific angular momentum of matter entering the disk.

The meanings of these two dimensionless numbers are straightforward. The first, $\xi$, takes the ratio of the accretion rate to the characteristic thermal accretion rate $c_s^3/G$. This is (up to factors of order unity) the accretion rate for a singular isothermal sphere or a Bonnor-Ebert sphere, and it is also the characteristic accretion rate through an isothermal disk, as we will see in Chapter \ref{ch:disks_theory}. The second parameter, $\Gamma$, is a measure of the angular momentum content of the accretion. The quantity $\dot{M}/\Omega_{k,\rm in}$ is (neglecting a factor of $2\pi$) the amount of mass added per orbital period at the disk outer edge. Thus $\Gamma$ measures the fraction by which accretion changes the total disk plus star mass per disk orbital period. High angular momentum flows have large rotation periods, so they produce larger values of $\Gamma$ at the same total accretion rate.
\begin{marginfigure}
\includegraphics[width=\linewidth]{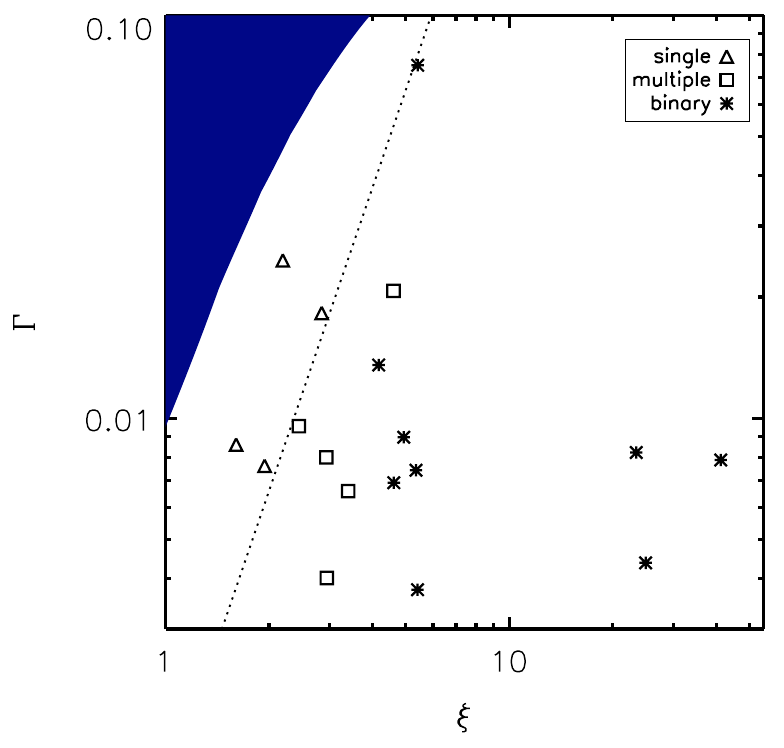}
\caption[Parameter study of disk fragmentation]{
\label{fig:diskfrag_kratter10}
Results of a series of simulations of disk fragmentation. Points show the accretion rate parameter $\xi$ and the rotation parameter $\Gamma$ for the simulations, with the type of point indicating the outcome: a single star, a multiple system, or a binary system. The shaded region is forbidden, because cores in that region are unable to collapse. Credit: \citet{kratter10a}, \copyright AAS. Reproduced with permission.
}
\end{marginfigure}

Intuitively, we expect that disk fragmentation is likely for high values of $\xi$ and low values of $\Gamma$, because both favor higher surface densities in the disk. High $\xi$ favors high disk surface density because it corresponds to matter entering the disk faster, and low $\Gamma$ favors higher surface density because it tends to make the disk more compact (since the circularization radius of the accreting material increases and $\Gamma$ does). This is exactly what a series of numerical simulations shows, as illustrated in Figure \ref{fig:diskfrag_kratter10}.

These results are very nice because they quite naturally explain why binaries are much more common among high mass stars. We showed in Section \ref{ssec:massivecores} that typical accretion rates onto massive stars are $\sim 10 \sigma^3/G$, where $\sigma$ is the velocity dispersion in the protostellar core. The parameter $\xi$ is determined by the accretion rate normalized to $c_s^3/G$ (where $c_s$ is the disk sound speed, recall), and thus we have
\begin{equation}
\xi \sim 10\left(\frac{\sigma}{c_s}\right)^3.
\end{equation}
For a massive core, the disk sound speed $c_s$ is enhanced compared to that in the core due to the radiation from the star, but much less than $\sigma$ is enhanced. Typical outer disk temperatures for massive star disks are $\sim 100$ K, corresponding to $c_s \sim 0.6$ km s$^{-1}$, whereas $\sigma \sim 1$ km s$^{-1}$, giving $\xi \gg 1$. Thus disk fragmentation is essentially inevitable.

A second effect that enhances massive star binary is N-body processing. Young clusters are born far from dynamically-relaxed, and thus there is an initial period where stars may have close encounters with one another. During this phase, encounters between binary systems, between binaries and single stars, and between three single stars can all serve to create or destroy binaries, or to modify their properties. The study of exactly how this happens is a huge topic into which we will not delve, beyond making a few general observations.

The main effects of this N-body processing are as follows: (1) wide binaries will tend to be widened and disrupted; (2) tight binaries will tend to get tighter; (3) three-body interactions may occur that will tend to preferentially keep more massive stars in binaries, thus favoring equal mass ratios. The line between close and wide binaries depends on the velocity -- binaries with orbital velocities greater than the cluster velocity dispersion are close, others are wide. All of these effects will tend to increase the binary fraction for more massive stars relative to less massive ones; the third effect does so by creating new massive binaries, and the first two favor massive binaries because a higher mass produces a higher orbital velocity, and thus a wider range of separations that can be considered close.

\section{Barriers to Accretion}

\subsection{Evolution of Massive Protostars}

Massive stars are not only somewhat different from low mass stars in terms of their parent cores, but also in their internal evolution. At accretion rates of $\sim 10^{-4}-10^{-3}$ $\msun$ yr$^{-1}$, forming a $10-100$ $\msun$ star takes of order $10^5$ yr, not that different than the time require to make a low mass star. However, massive stars are very different than low mass ones in terms of the timescales that govern their thermal evolution. As discussed in chapter \ref{ch:protostar_evol}, the characteristic timescale on which a star will contract toward the main sequence is the Kelvin-Helmholtz timescale
\begin{equation}
t_{\rm KH} = \frac{GM^2}{RL}.
\end{equation}
Evaluating this for main sequence values of $M$, $R$, and $L$, for a zero age main sequence (ZAMS) star of mass $1$ $\msun$, $t_{\rm KH} = 50$ Myr.  For a protostar where $R$ and $L$ are both larger (see chapter \ref{ch:protostar_evol}), this drops to $\sim 1$ Myr. On the other hand, to put some numbers on this for a massive star, a 50 $\msun$ ZAMS star has a radius of $10.7$ $\rsun$ and a luminosity of $3.5\times 10^5$ $\lsun$. For this star $t_{\rm KH}=20$ kyr, even without putting in a larger radius because it is pre-main sequence. Since this is less than the $\sim 100$ kyr required to form the star, we expect that massive stars will be able to reach thermal equilibrium, and thus contract to the main sequence, while forming, whereas low mass stars will not.

The rapid contraction to the main sequence has a few consequences. It means that the stars will have stronger winds, since the wind speed is linked to the Keplerian speed at the stellar surface and massive stars are able to shrink more. Similarly, the stars' comparatively small radii imply is that the effective temperature will be fairly high, so much of the light will emerge as ionizing radiation, even while the star is still forming. Finally, rapid settling means that massive protostars will put out roughly the same amount of light as main sequence stars of the same mass, since they will rapidly contract down to similar sizes. This in turn means that, unlike low mass protostars, massive protostars' luminosities come primarily from internal processes and not from accretion. The accretion luminosity of a massive star is larger than that of a low mass star because both its mass and its accretion rate are larger, but this effect is swamped by the extremely strong mass-dependence of the internal luminosity, which in the vicinity of $1$ $M_\odot$ rises as roughly $L \sim M^4$. Because of this strong mass-dependence, massive stars' accretion luminosities generally become subdominant once they reach $\sim 5-10$ $M_\odot$, depending on the accretion rate. The fact that massive protostars settle onto the main sequence while forming raises interesting problems for how they are able to keep accreting.

\subsection{Winds}

One thing that one might worry about is that the main sequence winds of a massive star, which are reasonably isotropic, might inhibit accretion. These winds may well start up while the star is still forming. However, this worry is fairly easy to dismiss. Main sequence O stars show wind speeds up to $\sim 1000$ km s$^{-1}$, with mass fluxes that are typically $\sim 10^{-7}$ $\msun$ yr$^{-1}$ or less. The mass flux is
\begin{equation}
\dot{M} = 4\pi r^2 \rho v,
\end{equation}
so the associated ram pressure is
\begin{equation}
P_{\rm wind} = \rho v^2 = \frac{\dot{M}_{\rm wind} v_{\rm wind}}{4\pi r^2}.
\end{equation}
In contrast, the accretion flow has a mass flux of $10^{-4}-10^{-3}$ $\msun$ yr$^{-1}$, and if it arrives at free-fall its ram pressure is
\begin{equation}
P_{\rm infall} = \frac{\dot{M}_{\rm acc} v_{\rm ff}}{4\pi r^2}.
\end{equation}
Thus the ratio of the ram pressures is
\begin{equation}
\frac{P_{\rm infall}}{P_{\rm wind}} = \frac{\dot{M}_{\rm acc} v_{\rm ff}}{\dot{M}_{\rm wind} v_{\rm wind}}.
\end{equation}

Since $v_{\rm wind} \approx v_{\rm ff}$ at the stellar surface, and $\dot{M}_{\rm acc}$ is larger than $\dot{M}_{\rm wind}$ by a factor of $10^3-10^4$, the ram pressure of the infall is more than enough to stop the wind. Even if the wind and the infall encounter each other further from the star, the free-fall velocity only falls off as $r^{-1/2}$, so the wind would need to be able to push the infall out to $\sim 10^6-10^8$ stellar radii before it would be able to reverse the infall. For a 50 $\msun$ ZAMS star, $10^6$ stellar radii is roughly $0.25$ pc, i.e., bigger than the initial massive core.

Thus, we generally do not expect main sequence stellar winds to inhibit accretion as long as material is left in the protostellar core. Of course the protostellar outflow carries much more momentum than the main sequence wind because it is hydromagnetically rather than radiatively driven (see chapter \ref{ch:disks_theory}). However, it is also highly collimated, and so it does not prevent accretion over $4\pi$ sr any more than protostellar outflows from lower mass stars do. It will reduce the efficiency, but not by more than low mass star outflows do.

\subsection{Ionization}

A second feedback one can worry about is ionizing radiation. Massive stars put out a significant fraction of their power beyond the Lyman limit, and this can ionize hydrogen in the envelope around them. Since when hydrogen is ionized it heats up to $\sim 10$ km s$^{-1}$, gas that is ionized may be able to escape from the massive core, which only has an escape velocity of $\sim 1$ km s$^{-1}$.

This does eventually happen, and it probably plays an important role in regulating the star formation efficiency in star clusters and on larger scales. However, one can show that, as long as the massive star is accreting quickly, this effect will not limit its ability to continue gaining mass. Problem set 5 contains a quantitative calculation of this result. Foreshadowing it here, at the accretion rates that we expect in massive cores, the ionizing radiation should all be trapped within a few stellar radii of the stellar surface. Since the escape velocity from the surface of a 50 $\msun$ ZAMS star is about $1000$ km s$^{-1}$, this gas will be trapped by the star's gravity, and will not escape. Thus ionization is an important feedback, but it is one that is likely most important after the massive star has gathered most of the mass around it and has stopped growing.

That said, this omits the fact that there is likely to be lower density within the region cleared by the protostellar outflow, so ionizing radiation may be able to escape in some directions even while the star is growing. This may eventually reduce its mass supply, and it may cause asymmetric H~\textsc{ii} regions to form, where the ionized gas is confined in certain directions (for example close to the disk) while the ionizing photons escape and drive an outflow in other directions (for example along the polar axis).

\subsection{Radiation Pressure}

By far the biggest potential worry for massive star formation is not that ionizing radiation will heat the gas enough to allow it to escape, but the pressure exerted by radiation will halt accretion. Let us go back to our picture of the structure of the envelope of dusty gas around a protostar, developed in chapter \ref{ch:protostar_form}. There is a dust destruction radius where all direct starlight is absorbed, and outside that a diffusion region. The calculation of this radius is the same as for a low mass star, except that the luminosity is not mostly due to accretion. Equating heating and cooling (and again ignoring the complication introduced by dust grain sizes smaller than $\sim 1$ $\mu$m) gives
\begin{equation}
\frac{L}{4\pi r_d^2} \pi a^2 = 4\pi a^2 \sigma_{\rm SB} T_d^4,
\end{equation}
where $r_d$ and $T_d$ are the radius and temperature at the dust destruction front. Thus
\begin{equation}
r_d = \sqrt{\frac{L_*}{16 \pi \sigma_{\rm SB} T_d^4}} = 25\mbox{ AU}\; L_{*,5}^{1/2} T_{d,3}^{-2},
\end{equation}
where $L_{5} = L/(10^5\,\lsun)$ and $T_{d,3}=T_d/(1000\mbox{ K})$. The dust destruction radius is therefore a factor of $\sim 10$ larger than it is for a low mass star.

\paragraph{Direct radiation pressure at the dust destruction front.} It is interesting to consider the force exerted by the radiation on the gas in two different regimes. One is at the dust destruction front, where the radiation still has a stellar spectrum and has not yet been down-shifted in frequency by the dust. At this front we can assume that essentially all the stellar radiation is absorbed in a thin region, so all of the momentum carried by the stellar radiation field will be transferred to the gas. Infall will reverse if this change in momentum is enough to reduce the infall velocity to zero.

Let $\dot{M}$ be the mass accretion rate onto the star. An infalling shell of material striking the dust destruction front therefore carries an inward momentum flux
\begin{equation}
\dot{p} = -\dot{M} v,
\end{equation}
where $v$ is the material's velocity, and we use the convention that $\dot{M}>0$ and $v>0$ correspond to inward motion. In comparison, the stellar radiation field carries a momentum flux
\begin{equation}
\dot{p} = \frac{L}{c}
\end{equation}
Strictly speaking this momentum is transferred to the dust grains, since they and not the gas absorb the radiation. However, the grains are coupled to the gas by collisions and magnetic fields, so they will in turn transfer any momentum they absorb to the gas.

If we let $v_0$ be the velocity of the material just before it encounters the stellar radiation field and $v_1$ be its velocity after passing through the dust destruction front, then conservation of momentum implies that
\begin{equation}
\dot{M} v_1 = \dot{M} v_0 - \frac{L}{c}.
\end{equation}
The condition that $v_1 < 0$ (i.e., that the new velocity still be inward) then requires that
\begin{equation}
\dot{M} v_0 > \frac{L}{c}
\end{equation}
If we assume that the gas is arriving at free-fall before reaching the dust destruction front, then $v_0 = \sqrt{2 G M/r_d}$, and thus the mass flux must exceed
\begin{equation}
\label{eq:mdot_min}
\dot{M} > \frac{L}{v_0 c} = \frac{L}{c} \sqrt{\frac{r_d}{2 G M}} = 8\times 10^{-5} \,\msun\mbox{ yr}^{-1} \, L_{5}^{3/2} T_{d,3}^{-1} M_{1}^{-1/2},
\end{equation}
where $M_{1}=M_*/(10\,\msun)$.

This is less than the accretion rates we inferred for massive stars based on dimensional arguments, although maybe not by quite as much as one would like. Nonetheless, this seems to imply that matter will not be stopped at the dust destruction front if it arrives as quickly as expected. More detailed evaluations of this condition by \citet{mckee03a}, who in turn build off of \citet{wolfire87a}, generally find that this is not a problem. However, there is an important caveat to mention. In deriving equation (\ref{eq:mdot_min}), we plugged in a radius $r_d$, which assumes that the direct radiation pressure encounters the gas at $r_d$. If something is able to evacuate the gas out to a radius $r > r_d$ (for example the diffuse radiation pressure we will consider momentarily), then the infall momentum is reduced as $r^{-1/2}$, while the momentum budget of the radiation remains the same. Thus direct radiation pressure cannot halt accretion by itself, but if something else begins to evacuate the region around the star, then direct radiation pressure may be able to keep it evacuated or even expand the evacuated region.

\paragraph{Diffuse radiation pressure in the envelope.} The second regime to think about this the dusty envelope, through which radiation must diffuse to escape. The radiation flux $F=L/(4\pi r^2)$, and this applies a force per unit mass to the gas
\begin{equation}
f_{\rm rad} = \frac{1}{c} \int \kappa_{\nu} F_{\nu} \, d\nu = \frac{1}{4\pi r^2 c} \int \kappa_{\nu} L_{\nu} \, d\nu,
\end{equation}
where the subscript $\nu$ indicates the frequency-dependent luminosity, flux, and opacity. Since the radiation field will be close to a black body in the envelope, we can replace the frequency integral with a Rosseland mean opacity, giving
\begin{equation}
f_{\rm rad} = \frac{\kappa_{\rm R} F}{c} = \frac{\kappa_{\rm R} L}{4\pi r^2 c}
\end{equation}

Since the opacity of the gas to the reprocessed radiation field is much less than its opacity to direct stellar radiation (i.e., $\kappa_{\rm R}$ evaluated at temperatures $T<T_d$ is much smaller than $\kappa_{\nu}$ evaluated at the peak frequency of stellar output), this force is much less than that applied at the dust destruction front. Unlike at the dust destruction front, however, this force is not applied in a quick impulse. It is applied at every radius, and thus its cumulative effect can be much stronger than that at the dust destruction front. If we think of things in terms of accelerations, the force applied in the dust envelope is much smaller, but it is applied to the gas for a much longer time, so that the total acceleration can be larger. The relevant comparison here is not to the momentum of the radiation field, but to the force of gravity exerted by the star, since we want to know whether the net acceleration is inward or outward.

The condition that gravitational force be stronger than radiative force is
\begin{equation}
\frac{G M}{r^2} > \frac{\kappa_{\rm R} L}{4\pi r^2 c},
\end{equation}
or
\begin{equation}
\frac{L}{M} < \frac{4\pi G c}{\kappa_{\rm R}}.
\end{equation}
This is just the Eddington limit calculation, or it would be if we plugged in the electron scattering opacity for $\kappa_{\rm R}$.
If we instead plug in a typical infrared dust opacity of a few cm$^2$ g$^{-1}$, we get
\begin{equation}
\left(\frac{L}{M}\right) = 1300 (\lsun/\msun) \,\kappa_{\rm R,1}^{-1},
\end{equation}
where $\kappa_{\rm R,1} = \kappa_{\rm R}/(10\mbox{ cm}^2\mbox{ g}^{-1})$.
For comparison, our standard 50 $\msun$ ZAMS star has $L/M= 7100 (\lsun/\msun)$, and thus it exceeds this limit by a factor of $\sim 5$. In fact, all ZAMS stars larger than $\sim 20$ $\msun$ exceed the limit, which would seem at face value to suggest that it should not be possible to form stars above this limit by accretion. This argument led to all sorts of contortions trying to explain how massive stars could form -- models included trying to make them only in regions of dramatically reduced dust opacity, trying to make them by collisions of lower mass stars, and various other ideas.

\begin{figure}
\includegraphics[width=\linewidth]{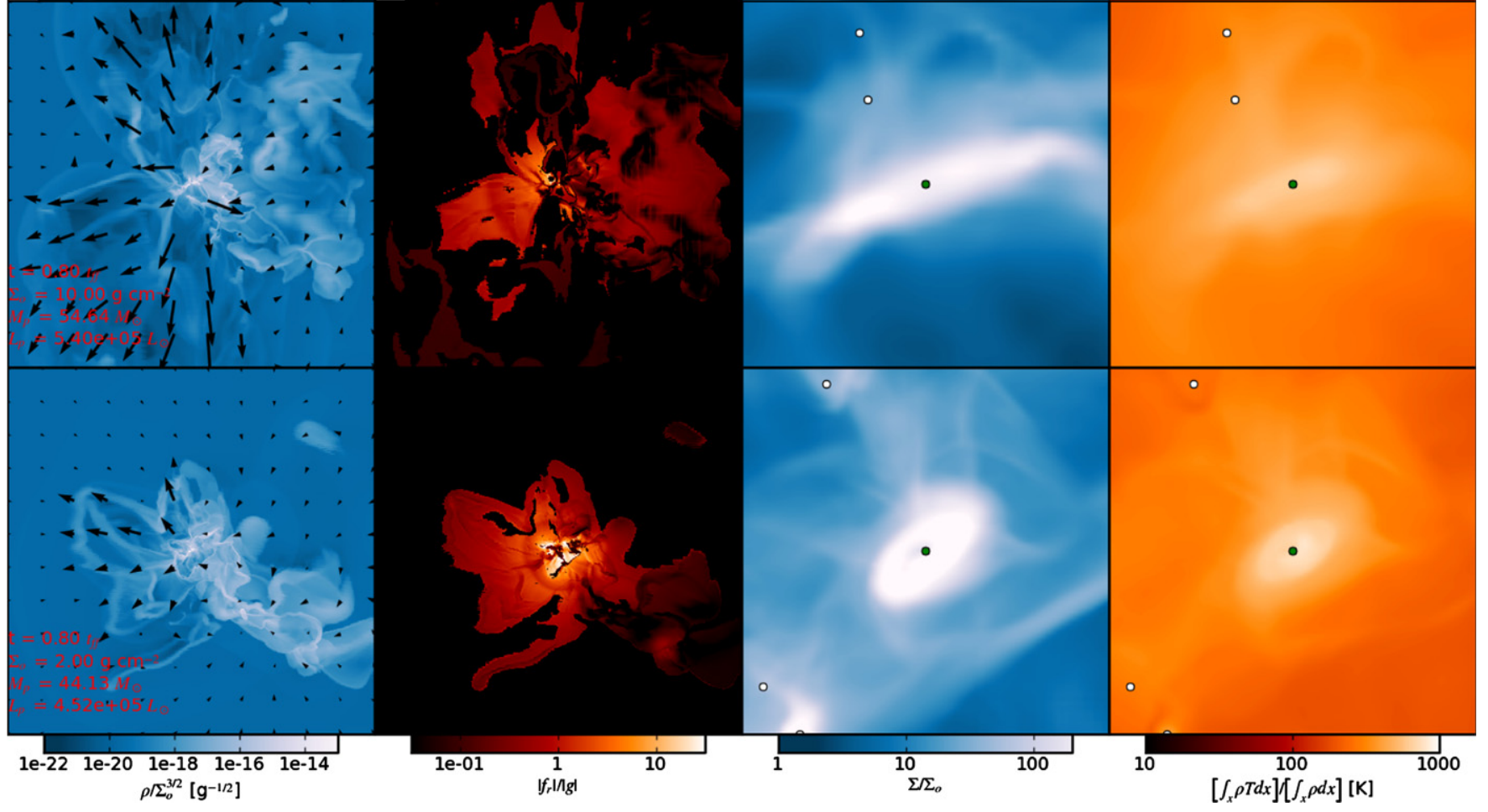}
\caption[Simulation of massive star formation with outflows]{
\label{fig:msf_cunningham11}
Two simulations of the formation of a massive star including protostellar outflows. The top row shows a simulation with an outflow, while the lower shows one without. The panels show, from left to right, normalized volume density in a slice, ratio of radiation force to gravitational force, normalized projected density, and mass-weighted mean projected temperature. Note the general absence of regions with radiation force greater than gravitational force in the simulation with winds. Credit: \citet{cunningham11a}, \copyright AAS. Reproduced with permission.
}
\end{figure}

The solution in reality is much more prosaic: the real world is not spherically symmetric, and the argument we just went through is. Multidimensional simulations show that, contrary to this naive calculation, radiation does not stop accretion in a real system. The main effect is that the ram pressure and the gas pressure can both be asymmetric, and they will conspire to be anti-correlated with one another because the radiation will escape by the path of least resistance. This asymmetry can be produced by many mechanisms; an obvious one is angular momentum, which shapes the accretion flow into a disk can concentrates ram pressure in a plane, while radiation pressure is not so confined. A shell of material held up by the radiation field turns out to be unstable, allowing radiation to escape asymmetrically and concentrating the infall. Magnetic fields or turbulence will both produce filamentary infall, again concentrating the ram pressure of the gas over a small solid angle, while allowing radiation to escape over the remaining, unoccupied solid angle. Outflows present yet another mechanism to punch a whole through which radiation can escape, as illustrated in Figure \ref{fig:msf_cunningham11}. For all these reasons, it is misleading to compare the radiation and gravitational forces averaged over $4\pi$ sr. Accretion will continue as long as there are significant patches of solid angle where gravity wins.

\chapter{The First Stars}
\label{ch:first_stars}

\marginnote{
\textbf{Suggested background reading:}
\begin{itemize}
\item \href{http://adsabs.harvard.edu/abs/2013RPPh...76k2901B}{Bromm, V. 2013, Rep.~Prog.~Phys., 76, 112901}, sections 1-5 \nocite{bromm13a}
\end{itemize}
\textbf{Suggested literature:}
\begin{itemize}
\item \href{http://adsabs.harvard.edu/abs/2011ApJ...737...75G}{Greif et al., 2011, ApJ, 737, 75} \nocite{greif11a}
\end{itemize}
}

The previous chapter focused on massive stars in the present-day Universe, and in this chapter we consider how picture changes if we go back to the early Universe. The study of primordial star formation is a major topic in contemporary astrophysics, and this chapter will not provide a comprehensive review. The goal of the chapter is instead to understand how and why the picture we have outlined thus far changes in the very early Universe, and to sketch in broad outlines how the process of star formation transitioned from that found in the early Universe to that found today. The products of such early star formation are often referred to as population III stars, following the Galactic astronomy nomenclature that metal-rich disk stars like the Sun are population I and metal-poor stars found primarily in the Galactic halo, which are presumed to be older, are population II. Population III stars would then be the oldest stars, which are completely metal-free. Unlike for most other topics covered in this book, there are almost no observations that provide useful constraints on the first stars themselves -- no population III stars have ever been observed, and for reasons we discuss below it is possible that none ever will be, except perhaps via their deaths in supernova explosions. Thus this chapter will be primarily theoretical.

\section{Cosmological Context}

We begin our discussion by setting the cosmological context for the formation of the first stars. In the modern $\Lambda$CDM cosmology\footnote{For those concerned with such details: the numerical evaluations in this section are all computed for a cosmology with Hubble constant $H_0=71$ km s$^{-1}$ Mpc$^{-1}$ and densities $\Omega_b = 0.04$, $\Omega_{\rm DM} = 0.23$, and $\Omega_\Lambda = 0.73$ for baryons, dark matter, and cosmological constant, respectively.}, the Universe begins in a nearly homogenous state, with baryons and dark matter distributed nearly uniformly. The baryonic matter consists of approximately 90\% hydrogen and 10\% helium-4 (by number, not by mass), with about 1 deuterium and helium-3 per $\sim 10^5$ H atoms, and even smaller amounts of heavier elements. As the Universe expands, gravity amplifies the tiny inhomogeneities that are present. The dark matter, which dominates the mass, collapses into halos -- virialized, self-gravitating structures -- that drag the ordinary baryonic matter into them. Collapse and virialization occur when the dark matter and baryons reach a characteristic density that is $\approx 200$ times the mean density of the Universe at whatever cosmic epoch is being considered. Once halos virialize, they obey a mass-radius relation \citep{barkana01a}
\begin{equation}
R_{\rm vir} \approx 290\,\mathrm{pc} \left(\frac{M_h}{10^6\,M_\odot}\right)^{1/3} \left(\frac{1+z}{10}\right)^{-1} \left(\frac{\Delta_c}{200}\right)^{-1/3},
\end{equation}
where $M_h$ is the mass of the halo, $z$ is the redshift, and $\Delta_c$ is the overdensity of the halo after it virializes.

The dark matter is collisionless, but the baryons that fall into a halo are not. As they fall to the center of the halo they will shock, converting their gravitational potential energy to thermal energy. After they settle into hydrostatic balance, they will be in a state of virial equilibrium, with a thermal energy equal to half their gravitational potential energy. The thermal energy per particle is $\approx k_B T$, and if we equate this with half the gravitational potential energy $G M_h m_{\rm H}/R_{\rm vir}$, we obtain
\begin{equation}
T_{\rm vir} \approx \frac{G M_h m_{\rm H}}{2R_{\rm vir} k_B} \approx 900\,K 
\left(\frac{M_h}{10^6\,M_\odot}\right)^{2/3} \left(\frac{1+z}{10}\right)^{} \left(\frac{\Delta_c}{200}\right)^{1/3},
\end{equation}
where we refer to $T_{\rm vir}$ as the virial temperature. Thus gas falling into a $\sim 10^6$ $M_\odot$ dark matter halo will be shock-heated to $\approx 1000$ K. This temperature is low enough that the gas will not be ionized, and will instead remain as neutral H and He. It is in halos like this where the first stars are thought to form.

\section{Chemistry and Thermodynamics of Primordial Gas}

\subsection{The Role of H$_2$}

The temperature of $\sim 1000$ K for primordial gas falling into dark matter halos is significant, because gas at these temperatures is much too cool to produce appreciable emission from neutral hydrogen or helium.\footnote{This statement ignores the 21 cm hyperfine transition of neutral hydrogen. However, this transition has such a low emission rate, and the photons it produces are of such low energy, that it is irrelevant for cooling.} The lowest-lying excited state of H is $E = (3/4)\cdot 13.6$ eV above ground, corresponding to a temperature $T = E/k_B = 1.2\times 10^5$ K. The lowest-lying excited states of He are at similar energies. Thus there will be no significant excitation of these states in gas at temperatures of $\sim 1000$ K, and correspondingly no cooling. Since there are essentially no heavier elements in primordial gas (excepting trace amounts of Li, which also do not provide significant cooling), this means that gas falling into early, small dark matter halos cannot cool easily. This makes the situation very different from that found in present-day star forming regions, where, as we saw in chapter \ref{ch:microphysics}, gas is able to cool on timescales many orders of magnitude smaller than the dynamical time.

If the gas could not cool at all, that would be the end of the story. Gas would fall into small halos, form hydrostatic structures, and then nothing would happen. However, there is another cooling pathway to consider: formation of H$_2$ and cooling by H$_2$ radiation. The lowest lying state of H$_2$ that can radiate, the $J=2$ rotational level\footnote{Recall from chapter \ref{ch:obscold} that transitions with $\Delta J=1$ are forbidden in homonuclear molecules due to symmetry.}, is $\approx 500$ K above ground. This makes it ineffective as a coolant in the modern Universe, where CO and C$^+$ lower the temperature well below 100 K. However, in the absence of these alternatives, H$_2$ radiation potentially provides a way of cooling the gas in a primordial halo to well below the virial temperature, thereby making it possible for the gas to collapse. This route to cooling requires that H$_2$ form, and this is a challenge. Recall from our discussion of H$_2$ formation in chapter \ref{ch:microphysics} that H$_2$ formation in the gas phase is very slow due to its symmetry, which forbids electric dipole radiation. In the present-day Universe this problem is circumvented by dust grains that act as catalysts, with H$_2$ forming on their surfaces. In primordial gas, however, there are no elements capable of forming solids under interstellar conditions, and thus no grains. How can H$_2$ form under such conditions?

The answer is that an alternative catalyst is needed, and one is available: free electrons. When the Universe recombined at $z\approx 1100$, leading the emission of the cosmic microwave background, not all electrons recombined with protons to form H~\textsc{i}. A small fraction were left free. These can catalyze the formation of H$_2$ via the reaction pathway
\begin{eqnarray}
\mathrm{H} + e^- & \rightarrow & \mathrm{H}^- + h\nu \\
\mathrm{H}^- + \mathrm{H} & \rightarrow & \mathrm{H}_2 + e^-.
\end{eqnarray}
In the first step, the H captures a free electron and radiates away excess energy -- the H$^-$ binding energy is 0.77 eV \citep{weisner64a}. Since the system is not symmetric, dipole radiation is possible, and the rate coefficient, while not particularly high, is not terribly low either: $k_{-} \approx 1.1\times 10^{-16}(T/100\,\mathrm{K})^{0.88}$ cm$^3$ s$^{-1}$ \citep{glover08a}. In the second step, H$^{-}$ interacts with neutral hydrogen to form H$_2$, and the excess binding energy goes into kinetic energy of the recoiling free electron. The rate coefficient for this reaction is been measured by laboratory experiment \citep{kreckel10a}; it reaches a maximum of $k_2 \approx 4.6\times 10^{-9}$ cm$^3$ s$^{-1}$ at around 100 K, and falls gradually to $k_2\approx 2.4\times 10^{-9}$ cm$^3$ s$^{-1}$ at 1000 K and $k_2\approx 3.0\times 10^{-10}$ cm$^3$ s$^{-1}$ at $10^4$ K.

The efficiency of this formation channel depends on two factors. One obvious one is the availability of free electrons to act as catalysts. The second is the branching ratio for H$^-$ to be destroyed by forming H$_2$, rather than by the other main mechanism of H$^-$ destruction, which is photodetachment:
\begin{equation}
\mathrm{H}^- + h\nu \rightarrow \mathrm{H} + e^{-}.
\end{equation}
This mechanism is simply the inverse of the first reaction in H$_2$ formation via $e^-$ catalysis, and can occur for any photon with an energy $>0.77$ eV, the binding energy of H$^-$. At redshifts above $z\sim 100$, the cosmic microwave background is sufficiently rich in such photons that it effectively suppresses H$_2$ formation via this channel, but at lower redshifts the photodetachment rate falls, and H$_2$ formation via the H$^-$ channel becomes the dominant formation mechanism.

\subsection{Thermal and Chemical Evolution}

\begin{figure}
\includegraphics[width=\linewidth]{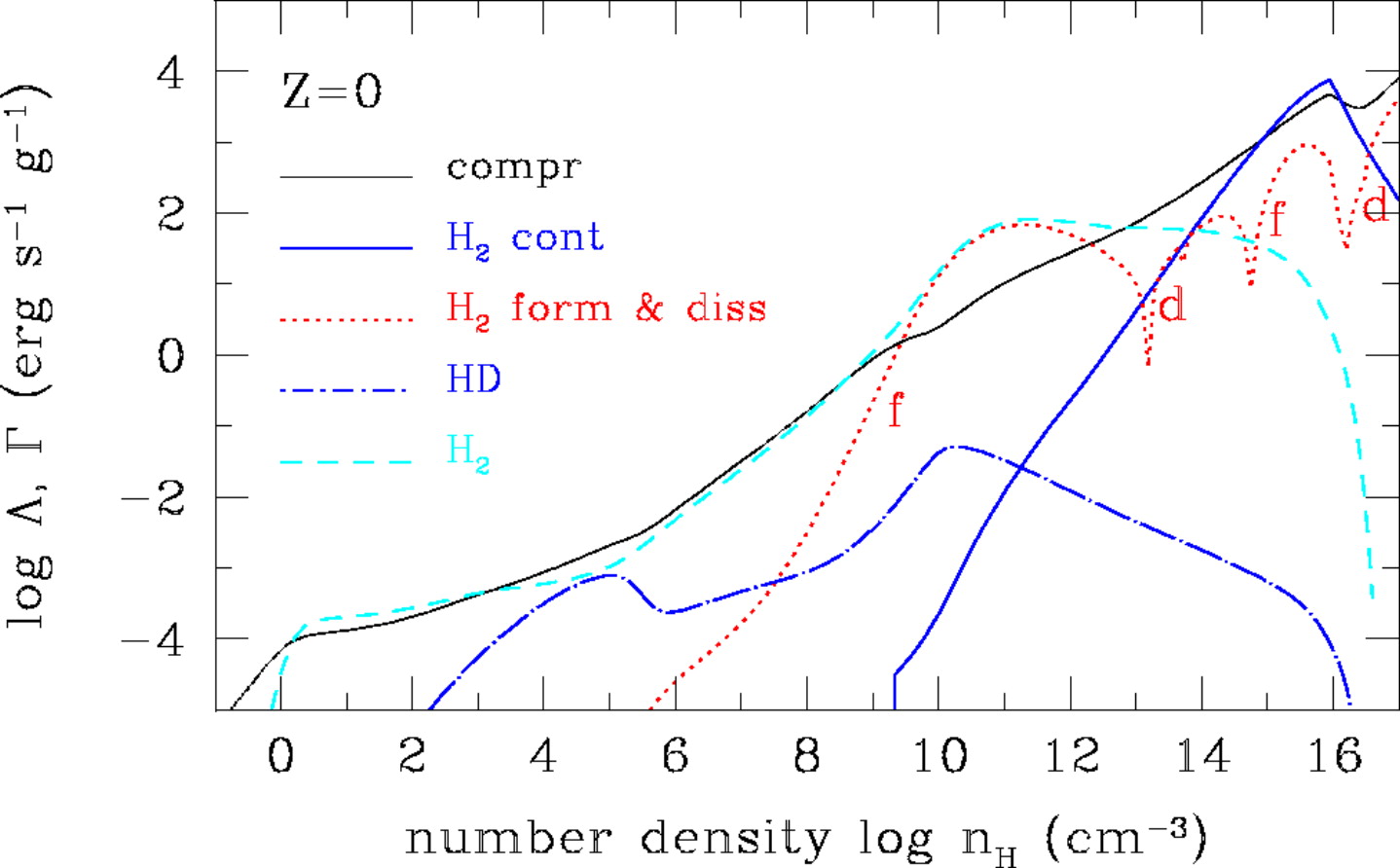}
\caption[Heating and cooling processes in primordial gas]{
\label{fig:omukai05a}
Results from a calculation of heating and cooling rates for processes operating in primordial gas of varying density, located at the center of a collapsing cloud. The processes included are heating by gravitational compression (black line), cooling by H$_2$ line (cyan) and continuum (solid blue) emission, cooling by HD emission (dot-dashed blue), and heating / cooling by collisional formation / dissociation of H$_2$ (dotted red). For each process, the value on the vertical axis indicates the rate of energy gain or less per unit mass per unit time. Credit: \citet{omukai05a}, \copyright AAS. Reproduced with permission.
}
\end{figure}

\begin{figure}
\includegraphics[width=\linewidth]{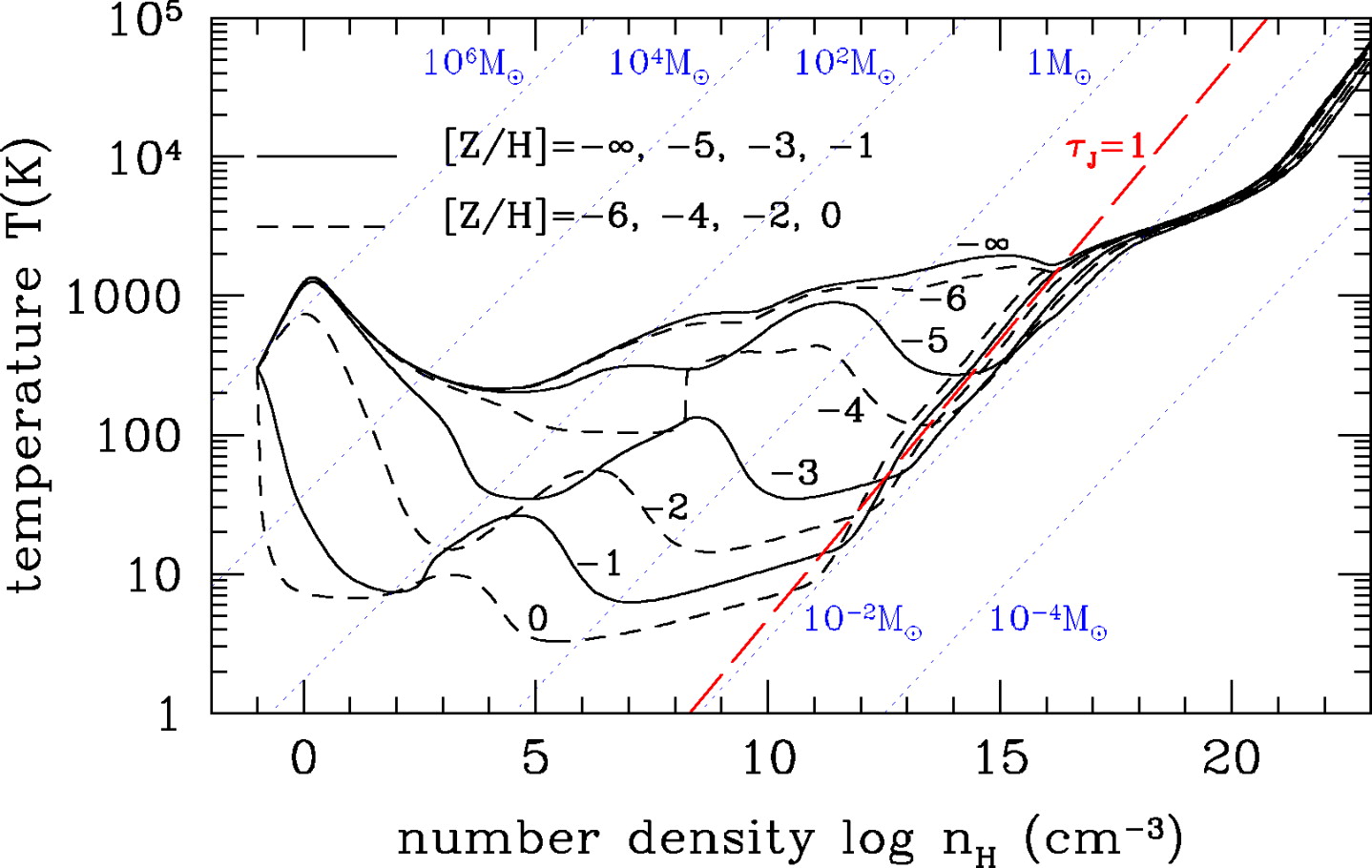}
\caption[Density-temperature evolution in primordial gas]{
\label{fig:omukai05b}
Density-temperature evolution of gas at the center of a collapsing cloud. Black dashed and solid lines show the trajectory of gas in the density-temperature plane, for different metallicities from primordial ($[{\rm Z}/{\rm H}] = -\infty$) to Solar ($[{\rm Z}/{\rm H}] = 0$). Dotted lines show loci of constant Jeans mass, as indicated. The red dashed line shows the point in density-temperature space where the center of the gas cloud becomes optically thick to its own cooling radiation. Credit: \citet{omukai05a}, \copyright AAS. Reproduced with permission.
}
\end{figure}

Armed with an understanding of how H$_2$ can form, we can now understand in rough outline the chemical and thermal processes that lead to the formation of a primordial star. These processes are summarized in Figures \ref{fig:omukai05a} and \ref{fig:omukai05b}.

Gas enters a $\sim 10^6$ $M_\odot$ halo at $z\sim 20-30$, shocks and virializes at temperatures of thousands of K. The gas begins to form a trace amount of H$_2$ via the H$^{-}$ channel, typically no more than one H$_2$ per $\sim 10^3$ H atoms. This in turn allows the gas to cool via H$_2$ line emission. As the gas cools, its density rises to maintain approximate hydrostatic balance in the halo. This continues until the gas reaches a temperature of $\approx 200$ K, about half the temperature of the lowest-lying excited state of H$_2$ that is capable of radiating; radiation by H$_2$ cannot easily cool the gas below this point, because gas at these temperatures is too cool for collisions to excite the $J=2$ state from which radiation occurs. Cooling is also slowed by critical density effects. At low densities, as gas first begins to cool in the halo, the cooling rate per unit volume increases with density as $n^2$, because the density is below the H$_2$ critical density of $\approx 10^4$ cm$^{-3}$. Once the density exceeds this value, cooling slows to increasing with density $n$. For this reason gas tends to linger at at a density of $n\sim 10^4$ cm$^{-3}$ and a temperature $T\sim 200$ K, leading to a phase known as the loitering phase of primordial star formation. During and after the loitering phase the density gradually rises as more gas cools and compresses the densest regions that have begun to collapse. The temperature gradually rises as well, because radiative emission is unable to keep up with heating from gravitational compression.

This phase ends once the density reaches $\sim 10^8$ cm$^{-3}$. At this density, another H$_2$ formation channel appears: the three-body reaction
\begin{equation}
{\rm H} + {\rm H} + {\rm H} \rightarrow {\rm H}_2 + {\rm H}.
\end{equation}
The reaction overcomes the symmetry problem by a trick of timing. When two H atoms collide, they can temporarily form an excited compound system, but because they cannot radiate they soon separate and become unbound again. However, if while they are in this short-lived compound state they are hit by a \textit{third} hydrogen atom, they can give their excess energy to the third atom, which carries it away as kinetic energy, leaving bound H$_2$ behind. Because this process effectively requires a three-way collision, its rate varies with density as $n^3$, making it extremely density-sensitive. This is why it only becomes to be significant at densities $\sim 10^8$ cm$^{-3}$. Once the density reaches this point, however, this process rapidly increases the H$_2$ fraction from $\sim 10^{-3}$ to near unity. The exact thermal and density evolution during this phase is somewhat uncertain, as there are two competing processes. The increase in H$_2$ fraction dramatically increases the cooling rate. On the other hand, each H$_2$ formation event produces an H atom recoiling with $4.5$ eV of energy. This is quickly and efficiently thermalized, providing a strong heating source. It is unclear which of these effects dominates at densities in the range $\sim 10^8-10^{12}$ cm$^{-3}$. However, once the density reaches $n\sim 10^{12}$ cm$^{-3}$, the gas must heat up fairly strongly, because the H$_2$ lines become optically thick, suppressing further radiative cooling. At this point the evolution becomes quite similar to that in present-day star formation as discussed in chapter \ref{ch:protostar_form}.

\section{The IMF of the First Stars}

We have established that the formation of the first stars happens at characteristically much higher temperatures than the formation of present-day stars, due to the absence of cooling from metal lines. We therefore turn to the question of how this will affect the properties of the stars that result, in particular their IMF. This is a question of great importance for cosmic chemical evolution, since the nucleosynthetic yield of these first stars will depend strongly on their masses.

\subsection{Fragmentation}

\marginnote{In many ways the situation for primordial star formation is simpler than for the present-day case, and this is one of them: in chapters \ref{ch:imf_th} and \ref{ch:massivestar}, we saw that radiative feedback from stars plays a critical role in regulating the temperature evolution of the gas, and thus in regulating how it fragments. For primordial star formation this effect is substantially less important because only photons that are capable of ionizing neutral hydrogen or of exciting the Lyman-Werner transitions of H$_2$ can be absorbed by the gas, and there is no dust to absorb the remaining photons and couple them to the gas. Thus radiative heating is much less important for primordial stars than for present-day ones, though ionizing photons can be important, as will be discussed in the next section.} The first question in addressing the IMF of primordial stars is how and whether the gas from which they form will fragment. Here we can bring to bear much of the same theoretical machinery we developed in chapter \ref{ch:imf_th}. Recall that, for non-isothermal gas, we expect fragmentation to be particularly pronounced at densities and temperatures where the temperature has a minimum. Primordial gas is clearly far from isothermal as it evolves, and the most obvious minimum in Figure \ref{fig:omukai05b} is that associated with the loitering phase, at a temperature of $T\approx 200$ K and a density of $n\approx 10^4$ cm$^{-3}$, when H$_2$ cooling is suppressed because the gas is too cool to excite the $J=2$ rotational level, and because the gas is beginning to exceed the critical density of that transition. The Bonnor-Ebert mass at this density and temperature is
\begin{equation}
M_{\rm BE} = 1.18 \frac{c_s^3}{\sqrt{G^3\rho}} = 380\,M_\odot \left(\frac{T}{200\,\mathrm{K}}\right)^{3/2} \left(\frac{n}{10^4\,\mathrm{cm}^{-3}}\right)^{-1/2},
\end{equation}
where the numerical evaluation of the sound speed is for primordial gas which has a mass of $1.22m_{\rm H}=2.0\times 10^{-24}$ g per H nucleus. This high mass led most early authors working on the first stars to believe that they would be quite massive, typically more than 100 $M_\odot$.

\begin{figure}
\includegraphics[width=\linewidth]{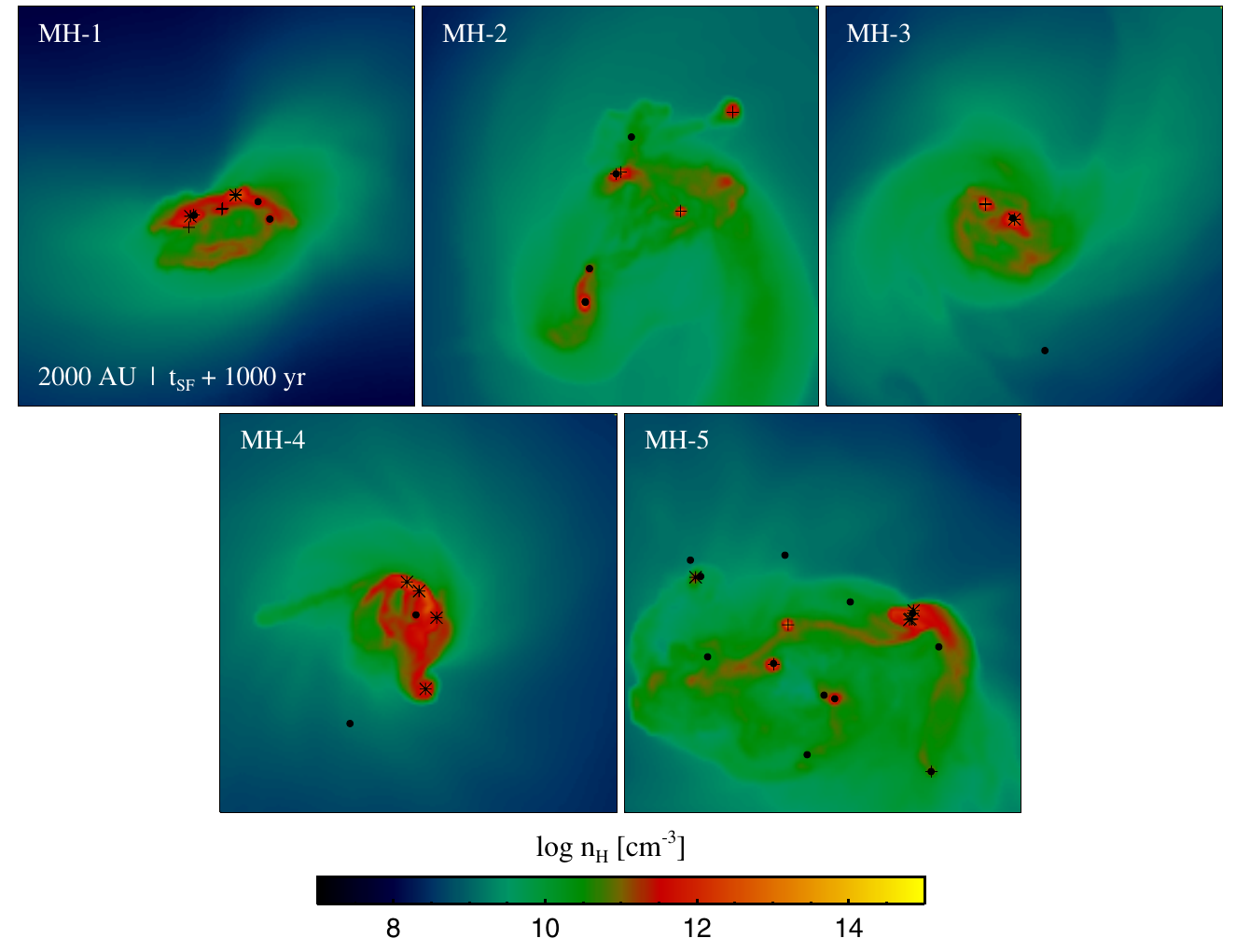}
\caption[Disk fragmentation around a primordial star]{
\label{fig:greif11a}
Results from a simulation of the formation of primordial stars. The images show the density in 2000 AU-sized regions in 5 different primordial halos, at a time 1000 years after formation of the first star in the simulation. Black dots, crosses, and stars indicate stars with masses $<1$ $M_\odot$, $1-3$ $M_\odot$, and $>3$ $M_\odot$, respectively. Credit: \citet{greif11a}, \copyright AAS. Reproduced with permission.
}
\end{figure}

Subsequent work has muddied this picture, and the true IMF of the first stars is still subject to extensive theoretical debate. The key question is whether the "protostellar core" produced during the loitering phase subsequently collapses monolithically or nearly so, or whether it fragments to much lower masses as it collapses. Fragmentation prior to the formation of a disk appears to be uncommon, but once a rotationally-supported accretion disk forms the situation changes, and becomes much more analogous the case of present-day massive stars discussed in chapter \ref{ch:massivestar}. The most commonly-seen outcome of fragmentation is formation of binaries with relatively large mass ratios \citep[e.g.,][]{stacy13a}. This would still leave the typical outcome of population III star formation extremely massive compared to the typical outcome of present-day star formation. On other hand, some simulations suggest that the disks fragment to produce small objects as well, some of which can be ejected via dynamical interactions \citep{clark11a, greif11a}; Figure \ref{fig:greif11a} shows an example. If this is the case, then, while some population III stars would be quite massive, others could be smaller than $1$ $M_\odot$.

While the true IMF of the first stars remains uncertain, the fact that we have never observed metal-free stars suggests that very few or none were formed with masses small enough such that they might still be in existence today. This in itself implies that the mass function was shifted to considerably higher values than what we observe today, since if the primordial IMF peaked at $\sim 0.2$ $M_\odot$ like the present one, there should be metal-free $0.2$ $M_\odot$ stars still in existence today. No such stars have been found.

\subsection{Feedback}

While the IMF of primordial stars at low masses depends largely on fragmentation, the IMF at high masses may well be shaped by feedback processes. Here it is instructive to compare to the case of present-day massive star formation. The absence of dust capable of absorbing non-ionizing photons means that radiation pressure is a somewhat smaller concern, though not a completely negligible one, since for very massive stars a significant fraction of the momentum budget of their radiation fields emerges in photons with energies above the hydrogen ionization threshold. Similarly, massive primordials stars will lack the fast radiatively-driven winds produced by present-day massive stars; such winds at driven by multiple resonant scattering of stellar photons off metal atoms in wind that have many closely-spaced levels, and primordial stars lack such atoms.

On the other hand, photoionization feedback seems likely to work for primordial stars in much the same way as it does for present-day ones, with the exception that the initial conditions for star formation are quite different. Moreover, feedback from non-ionizing photons with energies in the range $\approx 11-13.6$ eV, i.e., where they can be absorbed in the H$_2$ Lyman-Werner bands, can be significant as well. Analytic models and simulations suggest that these mechanisms might be able to evaporate the disks around primordial stars, limiting their maximum masses to tens of $M_\odot$ \citep{mckee08a, hosokawa11b, stacy12a}. The problem remains very much under investigation, however.

\section{The Transition to Modern Star Formation}

Not long after the first stars form, they will begin to change the environments around them, starting the process of transitioning to the modern mode of star formation that is the focus of the remainder of this book. The final topic in this chapter is how that transition happens.

\subsection{Ionization Evolution}

The first way that a transition away from population III star formation can occur is for the metal-free gas entering a halo to have a significantly higher abundance of electrons than the very low values expected to be be left over from the epoch of recombination. If this occurs, H$_2$ formation will occur significantly faster, the gas will cool earlier, and fragmentation to lower masses is much more likely. Stars resulting from this process are referred to as population III.2, while those formed in the truly primordial mode described above are referred to as population III.1.

An enhanced electron abundance could happen in several ways. One is for star formation to take place in gas that has already been photoionized by a population III.1 star. This seems unlikely while the population III.1 star is still shining, since its would likely heat the ionized gas to temperatures sufficient to prevent star formation. If the population III.1 star then explodes as a supernova this will pollute the gas with metals, leading to a population II star, as we discuss below. On the other hand, if the population III.1 star instead collapses directly to a black hole, without dispersing any metals, then any gas remaining in its halo would be ripe for population III.2 star formation.

\subsection{Chemical Evolution}
\label{ssec:chemevol}

Once a population III.1 or III.2 star explodes as a supernova, it disperses metals into the surrounding ISM, triggering a transition from population III to population II star formation. An outstanding question in this field is what level of metal enrichment is sufficient to produce this transition. Observationally, as of this writing, the most carbon-poor known star has a carbon abundance $\approx 4.5 \times 10^{-5}$ times that of the Sun \citep{caffau11a}, while the most iron-poor contains $<10^{-7}$ of the Solar iron abundance \citep{keller14a}. That both stars continue to exist today implies that it must be possible to from stars with masses well under $1$ $M_\odot$ with such low chemical abundances. In discussing this, it is helpful to consider two roles that metals have played in our discussion thus far. First, metals in the gas phase provide an important coolant for the ISM, allowing gas to cool on less than a dynamical timescale. Second, metals in the form of dust grains provide cooling at high densities where the dust and gas become collisionally-coupled, and also provide surfaces to catalyze chemical reactions, particularly H$_2$ formation.

\paragraph{Metal Line Cooling}

The distinguishing characteristic of modern star formation is the ability of star-forming gas to cool much faster than a dynamical time. Metal line cooling becomes significant when there are enough metals to make this possible, at which point they supplant H$_2$ cooling and trigger a transition to a more modern star formation mode. We can analytically estimate the metallicity required to achieve this, following the original calculation by \citet{bromm01a}, by comparing the rate of metal line cooling to the rate of heating due to adiabatic compression that we expect for baryons at the center of a virialized dark matter halo.

The calculation here is quite analogous to the one presented in Section \ref{ssec:iso_adiabat}. We
let $e$ be the thermal energy per unit mass of a particular gas parcel, and let $\Gamma$ and $\Lambda$ be the rates of change in $e$ due to heating and cooling processes, so
\begin{equation}
\frac{de}{dt} = \Gamma - \Lambda.
\end{equation}
The heating rate due to adiabatic compression is no different in the primordial case than in the present-day one: $\Gamma = -p\, (d/dt) (1/\rho) \approx -p/(\rho t_{\rm ff})$. For $t_{\rm ff} = \sqrt{3\pi/32G\rho}$, we have
\begin{equation}
\label{eq:gamma_ad_firststar}
\Gamma \approx k_B T \sqrt{\frac{32 G n}{3\pi \mu m_{\rm H}}},
\end{equation}
where $n$ is the number density of H nuclei and $\mu$ is the mean mass per H nucleus in units of $m_{\rm H}$; for near-primordial composition $\mu = 1.22$.

We must compare this to the rate of cooling due to metal lines. In the very metal-poor gas with which we are concerned, we do not expect appreciable numbers of CO or other heavy molecules to form, due to both absence of dust shielding and the long timescales required for chemical reactions when the constituent atoms are scarce. We must therefore consider cooling via atomic lines. The two most important cooling species are C$^+$ and O.\footnote{Recall from Section \ref{ssec:cochemistry} that carbon is primarily ionized under interstellar conditions because its ionization potential is smaller than that of hydrogen; in contrast, O has a higher ionization potential than H, and is largely neutral.} The former provides cooling mainly through its $158$ $\mu$m fine structure line, while the latter cools via a pair of lines at 63 and 145 $\mu$m. These transitions are generally optically thin, and their critical densities are high enough that we can treat them as being in the low-density limit. Recalling the discussion in Section \ref{ssec:molecular_lines}, and in particular Equation \ref{eq:cool_lowden}, the rate of cooling per unit mass in this limit can be written as
\begin{equation}
\Lambda = \frac{1}{\rho}n_X E A e^{-E/k_BT} \frac{n}{n_{\rm crit}} = \frac{n_X E k_{u\ell}}{\mu m_{\rm H}} e^{-E/k_B T},
\end{equation}
where here $n_X$ is the number density of the cooling species (either C$^+$ or O), $n$ is the number density of the primary collision partner (atomic hydrogen), $E$ is the energy of the cooling level, $A$ is the Einstein $A$ for the transition, $k_{u\ell}$ is the collisional de-excitation rate from the upper to the lower level, and $T$ is the temperature. Note that, in the second equality, the value of the Einstein coefficient and the number density of H have both dropped out. This is as we should expect: at low densities, cooling occurs when a hydrogen atom collides with a metal atom and excites it; the metal atom then radiates long before its next collision. Consequently, the cooling rate does not depend on exactly how long the metal atom takes to radiate (the Einstein $A$ coefficient), only on the timescale between H atoms colliding with other atoms (hence the dependence on $n_X$ but not on $n$), the collisional excitation probability (the factor $k_{u\ell} e^{-E/k_B T}$), and the energy lost per excitation ($E$).

For the lines with which we are concerned, $E/k_B = 91$ K, 228 K, and 327 K for the C$^+$ 158 $\mu$m, O 63 $\mu$m, and O 145 $\mu$m lines, respectively. The corresponding collisional de-excitation rate coefficients for collisions with H are $k_{u\ell}\approx 8\times 10^{-10}$, $5\times 10^{-10}$, and $7\times 10^{-10}$ cm$^3$ s$^{-1}$ at temperatures $\sim 1000$ K.\footnote{These rate coefficients are taken from the \href{http://home.strw.leidenuniv.nl/~moldata/}{Leiden Atomic and Molecular Database} -- \citet{schoier05a}; original data are from \citet{launay77a} and \citet{barinovs05a} for C$^+$, and from \citet{abrahamsson07a} for O.} For simplicity, let us assume that the C and O abundances are simply given by their Milky Way values scaled by a metallicity factor; specifically, we can take $n_{\rm C}/n = 2\times 10^{-4} Z'$ and $n_{\rm O}/n = 4\times 10^{-4} Z'$, where $Z'$ is the metallicity normalized to Solar. Finally, let us adopt a temperature $T \gg 300$ K, so that we can treat the $e^{-E/k_B T}$ factors for all three lines as unity; this is not strictly true, but for a rough estimate it is sufficient. Under these assumptions, and plugging in the quantities given above, we can write the cooling rate as
\begin{equation}
\Lambda = \Lambda_0 n Z',
\end{equation}
with $\Lambda_0 \approx 0.01$ erg cm$^3$ g$^{-1}$ s$^{-1}$.

Equating the heating rate $\Gamma$ and the cooling rate $\Lambda$, we find that metal-induced radiative cooling is able to match heating when
\begin{equation}
Z' = \frac{k_B T}{\Lambda_0} \sqrt{\frac{32 G}{3\pi \mu m_{\rm H} n}} = 4.5\times 10^{-4} \left(\frac{T}{1000\,\mathrm{K}}\right)\left(\frac{n}{100\,\mathrm{cm}^{-3}}\right)^{-1/2}.
\end{equation}
For the numerical evaluation we have plugged in typical densities and temperatures near the centers of virialized halos with mass $\sim 10^6$ $M_\odot$. Thus we expect that metal line cooling will become significant above metallicities of $Z'\sim 10^{-3.5}$.

\paragraph{Dust Effects}

The second channel through which metals affect the behavior of interstellar gas is by forming dust grains, which can serve as both radiators and chemical catalysts for the formation of molecules, particularly H$_2$. We defer the question of when dust becomes important as a catalyst to problem set 5, and here focus on the role of dust as a radiator. Dust is particularly important in this role because dust grains, as solid particles, can emit continuum radiation rather than being limited to emission at particular frequencies. This makes them much more efficient than gas at coupling to a radiation field, either via emission or absorption. In our discussion of the IMF and massive stars in chapters \ref{ch:imf_th} and \ref{ch:massivestar}, we focused on the latter effect, but before star formation begins in a region and produces a significant radiation field to absorb, the former effect is more important. Dust provides a potential mechanism to cool interstellar gas far beyond what would be possible with either H$_2$ or atomic line emission.

Because dust is such an efficient radiator, the bottleneck in dust cooling of the gas is usually the rate at which energy can be transmitted from the gas to the dust via collisions, not the rate at which dust can radiate it (though this can cease to be true if the dust becomes optically thick.) The grain-gas transfer rate in turn is limited by the total cross sectional area of dust grains available for collision. Since collisions are fastest when the density is high, we will focus on a high density regime when the gas has been fully converted to H$_2$, if only by three-body reactions occurring in the gas phase. To compute when dust cooling becomes important, let us consider a simplified problem\footnote{A more complete treatment of this problem may be found in \citet{schneider06a} and \citet{schneider12a}.}: suppose that we have a population of spherical dust grains with radius $a$, floating in a sea of hydrogen molecules with temperature $T$ and number density $n$. Since the velocities are grains are generally much smaller than those of individual hydrogen atoms, we can consider the grains at rest. We expect the rate at which hydrogen atoms strike a single dust grain to be of order $n\sigma v$, where $\sigma \approx \pi a^2$ is the grain cross section and $v\approx \sqrt{kT/\mu m_{\rm H}}$ is the thermal velocity of the particles; for fully-molecular gas of primordial composition, $\mu \approx 2.3$. Detailed integration over a Maxwellian distribution of gas velocities \citep{draine11a} gives a collision rate per grain
\begin{equation}
\mbox{collision rate / grain} = n \sqrt{\frac{8 k_B T}{\pi \mu m_{\rm H}}} \pi a^2,
\end{equation}
in line with this expectation.

The rate of collisions per unit gas mass is simply this multiplied by the number of dust grains per unit total (gas plus dust) mass. Let $m_{\rm gr}$ be the mass per grain, and let $\mathcal{D}$ be the ratio of dust mass to gas mass density, i.e., for every 1 g of gas, there are $\mathcal{D}$ g of dust present. Thus the rate of grain-gas collisions per unit mass is given by
\begin{equation}
\mbox{collision rate / mass} =  n \sqrt{\frac{8 k_B T}{\pi \mu m_{\rm H}}} \pi a^2 \left(\frac{\mathcal{D}}{m_{\rm gr}}\right)
\equiv n \mathcal{D} \sqrt{\frac{8 k_B T}{\pi \mu m_{\rm H}}} \mathcal{S}
\end{equation}
where we have defined the quantity $\mathcal{S} = \pi a^2/m_{\rm gr}$ as the dust cross section per unit dust mass, which is a function only of the grains themselves (their density, composition, geometry, etc.) and not of their abundance. The value of $\mathcal{S}$ is quite uncertain; we have a reasonable estimate of it for present-day interstellar dust, but it is likely to be quite different for the case relevant for the population III to population II transition, because at such early times any grains present are probably those that have condensed directly out of supernova ejecta, rather than the mix of grains produced by supernovae, AGB and red giant stars, and \textit{in situ} processing in the interstellar medium that exists in the present-day Universe. \citet{schneider12a} suggest values $\mathcal{S} \sim 10^5$ cm$^2$ g$^{-1}$, but this should be taken with considerable caution.

The rate of energy transfer should be proportional to this rate times the mean energy transfer per collision. Again integrating over a Maxwellian distribution of hydrogen atom velocities, the mean energy per hydrogen atom striking the grain is $2k_B T.$\footnote{The value $2k_B T$ is slightly higher than the canonical $(3/2) k_B T$ per particle in a Maxwellian distribution because faster-moving hydrogen atoms are more likely to collide with a grain, and thus the average kinetic energy of colliding particles is slightly higher than the average kinetic energy of all particles.}  If grain-gas collisions were perfectly elastic then there would be no energy transfer between the two, and if it were perfectly inelastic then the net energy transfer would be $2k_B T$ in the limit where the dust temperature $T_d \ll T$. We interpolate between these two extremes by writing the mean energy transfer per dust-gas collision as
\begin{equation}
\left\langle\Delta E\right\rangle = 2\alpha k_B (T - T_d),
\end{equation}
where $\alpha = 0$ corresponds to perfect elasticity, $\alpha=1$ to perfect inelasticity, and we have written the temperature dependence as proportional to $T-T_d$ to properly capture the effect that energy should flow from gas to trains for $T > T_d$ and from grains to gas for $T < T_d$, and that there should be no net energy transfer if $T=T_d$. The quantity $\alpha$ is known as the accommodation coefficient, and laboratory measurements and theoretical calculations suggest that it is of order $\sim 0.5$.

Putting this together, the rate of energy transfer from gas to grains per unit gas mass, and thus the rate of dust cooling, will be given by
\begin{equation}
\Lambda = 2 \alpha n \mathcal{D} \mathcal{S} \sqrt{\frac{8 k_B T}{\pi \mu m_{\rm H}}} k_B (T-T_d).
\end{equation}
If we now equate this cooling rate with the heating rate derived above (equation \ref{eq:gamma_ad_firststar}), we find that they become equal at a dust-to-gas ratio
\begin{equation}
\mathcal{D} = \frac{1}{\alpha \mathcal{S}} \sqrt{\frac{G}{3 n k_B T}}
\end{equation}
where for simplicity we have assumed $T \gg T_d$.

For dust cooling to be effective, it must kick in before the system becomes completely optically thick at a density $n\sim 10^{12}$ cm$^{-3}$. Thus to get the minimum value of $\mathcal{D}$ for which dust cooling matters, we will use this value of $n$. Consulting the zero-metallicity case shown in Figure \ref{fig:omukai05b}, the temperature at this density range is $T\sim 1000$ K. Plugging in these values, along with the fiducial values of $\alpha$ and $\mathcal{S}$ discussed above, we have
\begin{equation}
\mathcal{D} \approx 8\times 10^{-9}\left(\frac{0.5}{\alpha}\right)
\left(\frac{10^5\mbox{ cm}^2\mbox{ g}^{-1}}{\mathcal{S}}\right)
\left(\frac{n}{10^{12}\mbox{ cm}^{-3}}\right)^{-1/2} \left(\frac{T}{1000\mbox{ K}}\right)^{-1/2},
\end{equation}
or about $10^6$ times smaller than the Solar neighborhood value $\mathcal{D} \approx 0.01$. A further complication in making use of this value is that this estimate is phrased in terms of the dust to gas ratio, but we have little idea how to translate this into a metallicity, which is the quantity that we can measure in surviving stars. The dust-to-metals ratio in the present-day Universe is a result of a competition between grain production in supernovae and evolved stars and grain destruction (and possibly also grain growth) in the interstellar medium. It is observed to be roughly constant down to metallicities of $\sim 10\%$ of Solar, but to fall decrease below that \citep{remy-ruyer14a}. At the time of the transition from primordial to modern star formation, supernovae will have occurred, but it is not clear which other processes might have, and there is no observational guidance to be had.

\chapter{Late-Stage Stars and Disks}
\label{ch:late_disk}

\marginnote{
\textbf{Suggested background reading:}
\begin{itemize}
\item \href{http://adsabs.harvard.edu/abs/2014prpl.conf..475A}{Alexander, R., et al. 2014, in "Protostars and Planets VI", ed.~H.~Beuther et al., pp.~475-496} \nocite{alexander14a}
\end{itemize}
\textbf{Suggested literature:}
\begin{itemize}
\item \href{http://adsabs.harvard.edu/abs/2001ApJ...560..957D}{Dullemond, C.~P., Dominik, C., \& Natta, A. 2001, ApJ, 560, 957} \nocite{dullemond01a}
\item \href{http://adsabs.harvard.edu/abs/2009ApJ...700.1502A}{Andres, S.~M., et al. 2009, ApJ, 700, 1502} \nocite{andrews09a}
\end{itemize}
}

The last two chapters of this book are concerned with the fate of the left-over material from the star formation process, which is mostly collected into accretion disks around them. This chapter discusses how this material is dispersed, and the final one introduces the process by which it can begin to form planets.

\section{Stars Near the End of Star Formation}

We will begin our study of the final stages of star formation with a discussion of the stars themselves. The stars we want to study are ones that fall into the class II and class III category, and are observationally classified as either T Tauri or Herbig Ae/Be stars.\footnote{Roughly speaking, T Tauri stars are objects below $\sim 2$ $\msun$, of spectral type G0 or later, and Herbig Ae/Be stars are more massive objects of earlier spectral types. While the spectra of these two types of objects are different due to their differing surface temperatures, they appears to share a common physical nature and evolutionary history. Consequently we will discuss them as a single class in this chapter.} These stars no longer have envelopes of material around them that are sufficient to obscure the stellar photosphere. However, they are young enough that they still exhibit various signs of youth. The presence of a disk is one of these signs.

\subsection{Optical Properties}

\begin{marginfigure}
\includegraphics[width=\linewidth]{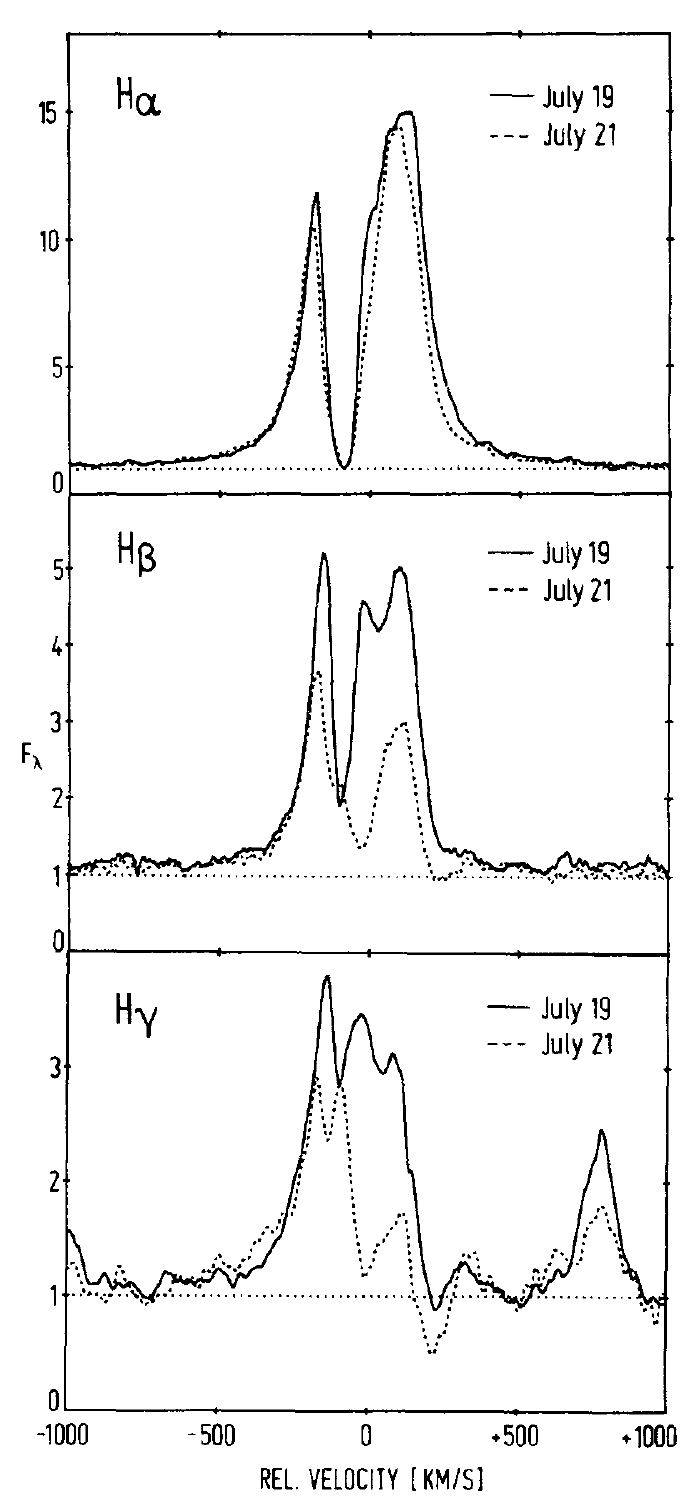}
\caption[Balmer lines of T Tauri stars]{
\label{fig:tts_appenzeller89}
Observed line profiles for three Balmer lines from the T Tauri star S Cr A taken on two nights in July, 1983. Credit: Astron.~\& Astrophys.~Rev., ``T Tauri Stars", 1, 1989, 291, \citeauthor{appenzeller89a}. With permission of Springer.
}
\end{marginfigure}

The main observational signature that has been used historically to define the T Tauri and Herbig Ae/Be classes is the presence of excess optical spectral line emission beyond that expected for a main sequence star of the same spectral class. The most prominent such line is H$\alpha$, the $n=3\rightarrow 2$ line for hydrogen, along with other hydrogen Balmer lines (Figure \ref{fig:tts_appenzeller89}). H$\alpha$ is particularly striking because in almost all main sequence objects H$\alpha$ is seen in absorption rather than emission. The strength of the H$\alpha$ emission in T Tauri stars ranges from equivalent widths (EW) of $\sim 100$ \AA\ down zero -- which still makes the stars stand out from main sequence stars, which have H$\alpha$ absorption. We divide T Tauri stars into classical ones, defined as those with H$\alpha$ EW $\gtrsim 10$ \AA, and weak-lined, those with H$\alpha$ EW $\lesssim 10$ \AA. Both classes also often show variability in their H$\alpha$ line profiles on periods of hours or days.

A large number of other optical and UV emission lines are also seen from these stars, and their strength generally correlates with that of the H$\alpha$ line. In addition to optical and ultraviolet line emission, these stars also exhibit the property of continuum veiling. What this means is that, in addition to excess line emission, these stars show excess continuum emission beyond what would be expected for a bare stellar photosphere. This excess emission arises from above the photospheric region where absorption occurs, so these photons are not absorbed. As a result they can partially or completely fill in normal photospheric absorption lines, reducing their equivalent width -- hence the name veiling.

\subsection{Infall Signatures}

The H$\alpha$ is particularly interesting, because it tells us something about the star's immediate environment. For main sequence stars, the H$\alpha$ line profile is a result of absorption at the stellar photosphere and emission from the chromosphere. At the photosphere there is a population of neutral hydrogen atoms in the $n=2$ level that absorbs photons at H$\alpha$ frequencies, producing absorption. Above that in the chromosphere is an optically thin, hot gas, which contains atoms in the $n=3$ level. Some of these emit H$\alpha$ photons, partially filling in the absorption trough, but leaving the line overall in absorption.

Producing H$\alpha$ in emission is tricky, however. The emitting material must be above the stellar photosphere, so it can fill in the absorption trough created there. This gas must be at temperatures of $5,000-10,000$ K to significantly populate the $n=3$ level. However, in order to produce enough H$\alpha$ photons to fill in the trough and produce net emission, this gas must also be dense enough for the collision rate to be high enough to force the $n=3$ level close to LTE.

Ordinary stellar chromospheres have densities that are much too low to meet this requirement. Thus H$\alpha$ emission implies the presence of material around the star at temperatures of  $5,000-10,000$ K, but at densities much higher than found in an ordinary stellar chromosphere. Moreover, the width of the H$\alpha$ emission requires that this material be moving at velocities of hundreds of km s$^{-1}$ relative to the stellar surface, i.e., comparable to the free-fall velocity. This cannot be thermal broadening, because this would require temperatures of $\sim 10^6$ K, high enough to completely ionize hydrogen. It must therefore be bulk motion.

The standard inference is that this indicates the presence of gas infalling onto the stellar surface. Such gas would provide the high densities required to produce H$\alpha$ in emission. The infall of this material would provide the requisite bulk motion. Internal shocks and the shocking of this gas against the stellar surface could easily heat gas to the required temperatures. Finally, this hot material would also produce continuum emission, explaining the continuum veiling.

\begin{marginfigure}
\includegraphics[width=\linewidth]{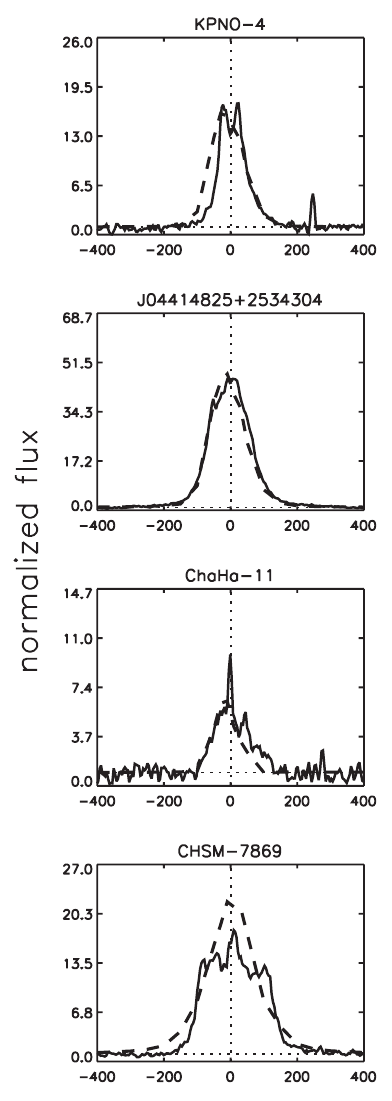}
\caption[H$\alpha$ lines from T Tauri stars compared to models]{
\label{fig:tts_ha_muzerolle05}
Comparisons between observed (solid) and model (dashed) H$\alpha$ line profiles for a sample of T Tauri stars. The $x$ axis shows velocity in km s$^{-1}$. Each model curve is a fit in which the accretion rate is one of the free parameters. Credit: \citet{muzerolle05a}, \copyright AAS. Reproduced with permission.
}
\end{marginfigure}

Quantitative radiative transfer calculations that attempt to fit the observed veiling and line emission can be used to infer the densities and velocities of the circumstellar gas, thereby constraining the accretion rate (Figure \ref{fig:tts_ha_muzerolle05}). The inferred accretion rates depend on the strength of the H$\alpha$ emission, and are typically $10^{-8}$ $\msun$ yr$^{-1}$. There is a broad range, however, running from $10^{-11}-10^{-6}$ $\msun$ yr$^{-1}$, with a very rough correlation $\dot{M}_*\propto M_*^2$.

These accretion rates are generally low enough so that accretion luminosity does not dominate over stellar surface emission. However, the estimated accretion rates are extremely uncertain, and the models used to make these estimates are very primitive. In general they simply assume that a uniform density slab of material arrives at the free-fall velocity, and covers some fraction of the stellar surface, and the accretion rate is inferred by determining the density of this material required to produce the observed spectral characteristics.

Despite this caveat, though, the H$\alpha$ line and other optical properties do seem to indicate that there must be some dense infalling material even around these stars that lack obvious envelopes. This in turn requires a reservoir of circumstellar material not in the form of an envelope, which is most naturally provided by a disk. Indeed, before the advent of space-based infrared observatories, optical indicators like this were the only real evidence we had for disks around T Tauri and Herbig Ae/Be stars.

\subsection{FU Orionis Outbursts}

\begin{figure}
\includegraphics[width=\linewidth]{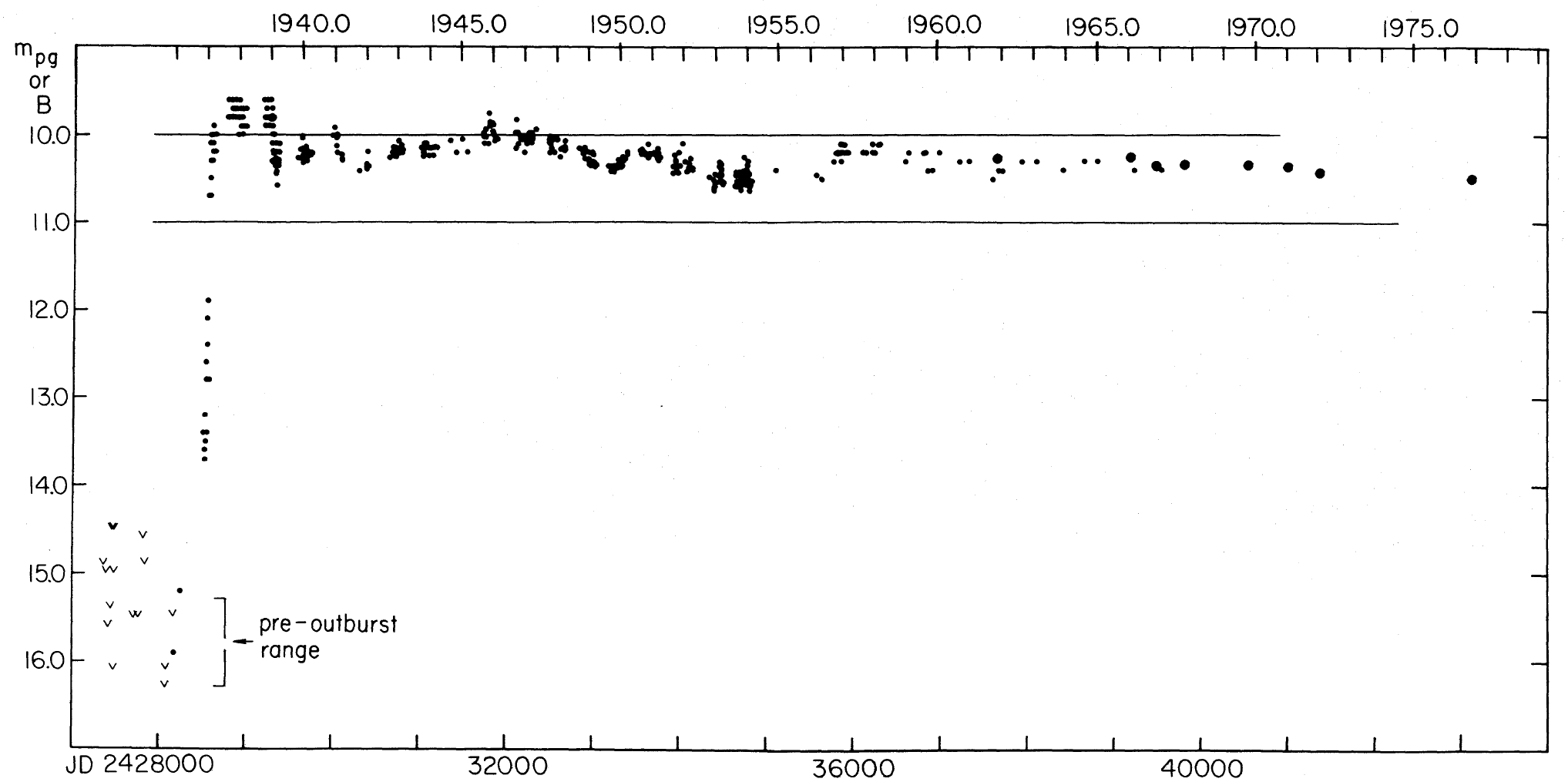}
\caption[Long-term light curve of FU Orionis]{
\label{fig:fu_ori_herbig77}
A light curve of the star FU Orionis, from the 1930s to 1970s. The $y$ axis shows the apparent magnitude in B band, or from photographic observations prior to filter standardization. Credit: \citet{herbig77a}, \copyright AAS. Reproduced with permission.
}
\end{figure}

There are many other interesting phenomena associated with these young stars, such as radio and X-ray flaring, but one in particular deserves mention both as a puzzle and a potential clue about disks. This is the FU Orionis phenomenon, named after the star FU Orionis in which it was first observed. In 1936 this star, an object in Orion, brightened by $\sim 5$ magnitudes in B band over a few months. After peaking, the luminosity began a very slow decline -- it is still much brighter today than in its pre-outburst state (Figure \ref{fig:fu_ori_herbig77}). Since then many other young stars have displayed similar behavior. When available, the spectra of these stars in the pre-outburst state generally look like ordinary T Tauri stars.

Some simple population statistics imply that this must be a periodic phenomenon. The rate of FU Ori outbursts within $\sim 1$ kpc of the Sun is roughly one per 5 years. The star formation rate in the same region is roughly 1 star per 50 years, so this implies that the mean number of FU Ori outbursts per young star is $\sim 10$.

The stellar brightening is accompanied by a rise in effective temperature, indicating the presence of hot emitting material. It is also accompanied by spectral features indicating both an outflowing wind and the presence of rapid rotation. Although a number of models have been proposed to explain exactly what is going on, and the problem is by no means solved, the most popular general idea is that outbursts like this are caused by a sudden rise in the accretion rate. For whatever reason, the disk dumps a lot of material onto the star, briefly raising the accretion rate from the tiny $10^{-8}$ $\msun$ yr$^{-1}$ typical of classical T Tauri stars up to values closer to those expected for still-embedded sources. The accreted material produces a large accretion luminosity, and the subsequent decay in the emission is associated with the cooling time of the gas that has undergone rapid infall. If this model is correct, the mystery then becomes what can set off the disk.

\section{Disk Dispersal: Observation}

We now turn to the disks that surround T Tauri and similar stars. Prior to the 2000s, we had very little direct information about such objects, since they are not visible in the optical. That changed dramatically with the launch of space-based infrared observatories, and the developed of ground-based millimeter interferometers. These new techniques made it possible to observe disks directly for the first time.

\subsection{Disk Lifetimes}

One of the most interesting properties of disks for those who are interested in planets is their lifetimes. This sets the limit on how long planets have to form in a disk before it is dispersed. In discussing disk lifetimes, it is important to be clear on how the presence or absence of a disk is to be inferred, since different techniques probe different parts and types of disks. Our discussion of disk lifetimes will therefore mirror our discussion of disk detection methods in Chapter \ref{ch:disks_obs}. In general what all these techniques have in common is that one uses some technique to survey young star clusters for disks. The clusters can be age-dated using pre-main sequence or main sequence HR diagrams, as discussed in Chapter \ref{ch:protostar_evol}. One then plots the disk fraction against age.

\begin{marginfigure}
\includegraphics[width=\linewidth]{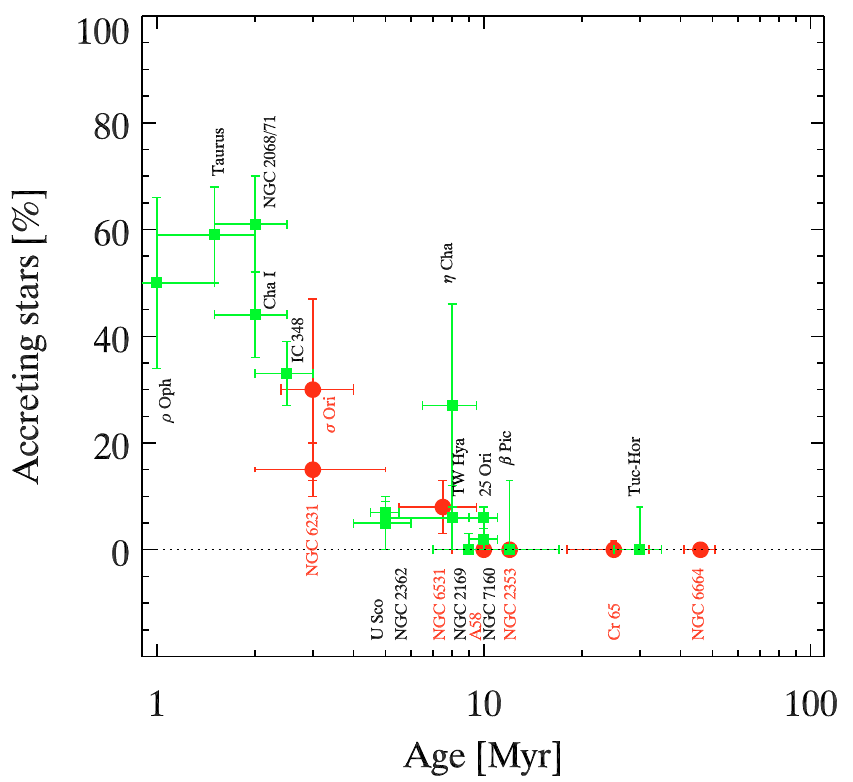}
\caption[Accreting star fraction versus cluster age]{
\label{fig:accretefrac_fedele10}
Fraction of stars that show evidence of accretion, as indicated by H$\alpha$ line emission, for clusters of different ages (indicated on the $x$ axis). The names of individual clusters are marked. Credit: \citeauthor{fedele10a}, A\&A, 510, A72, 2010, reproduced with permission \copyright\, ESO.
}
\end{marginfigure}

\begin{marginfigure}
\includegraphics[width=\linewidth]{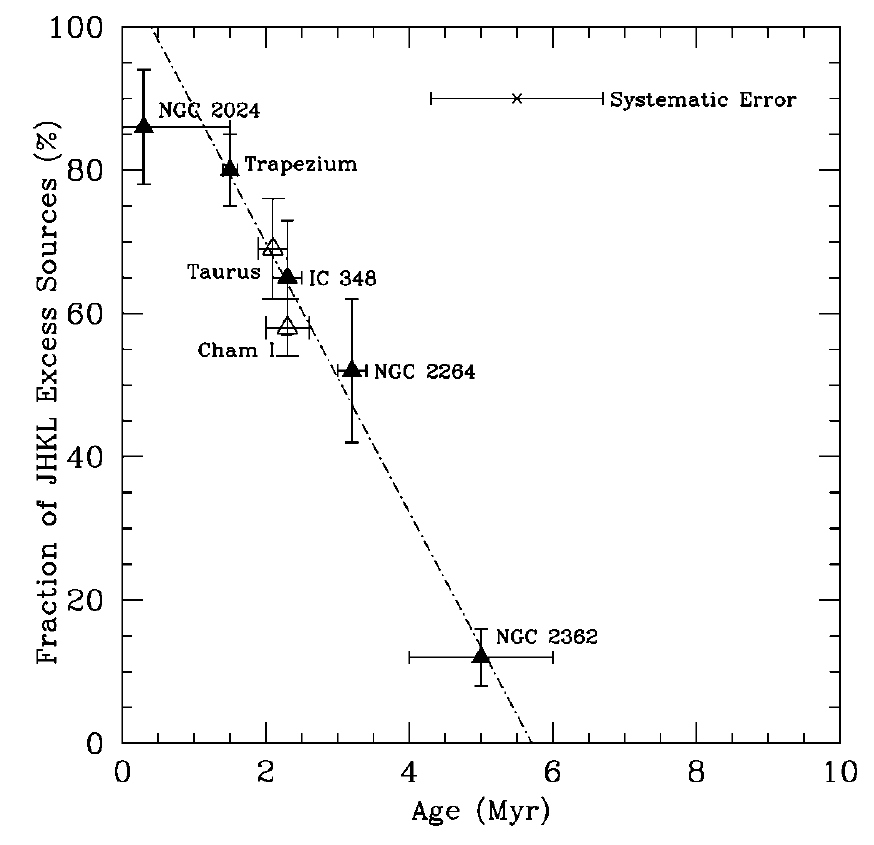}
\caption[Near infrared excess fraction versus cluster age]{
\label{fig:diskfrac_haisch01}
Fraction of stars that show near-infrared excess emission versus cluster age. The names of individual clusters are marked. Credit: \citet{haisch01a}, \copyright AAS. Reproduced with permission.
}
\end{marginfigure}

One signature of disks we have already discussed: optical line emission associated with accretion in T Tauri stars, particularly H$\alpha$. Surveys of nearby groups find that H$\alpha$ line emission usually disappears at times between 1 and 10 Myr (Figure \ref{fig:accretefrac_fedele10}). This tells us that the inner parts of disks, $\lesssim 1$ AU, which feed stars disappear over this time scale. In contrast, ground-based near infrared observations tell us about somewhat more distant parts of the disk, out to a few AU. The timescales implied by these results are very similar those obtained from the H$\alpha$: roughly half the systems loose their disks within $\sim 3$ Myr (Figure \ref{fig:diskfrac_haisch01}).

These observations are sensitive primarily to the inner disk, and the infrared techniques are generally sensitive only in cases where the dust in these regions is optically thick. Some optically thin material could still be present and would not have been detected. Observations at longer wavelengths, such as Spitzer's 24 $\mu$m band and in the mm regime from ground-based radio telescopes, probe further out in disks, at distances of $\sim 10-100$ AU. They are also sensitive to much lower amounts of material. Interestingly, unlike the shorter wavelength observations, these measurements indicate that a small but non-zero fraction of systems retain some disks out to times of $\sim 10^8$ yr. The amounts of mass needed to explain the long wavelength excess is typically only $\sim 10^{-5}$ $M_{\oplus}$ in dust. Thus in the older systems we are likely looking at an even later evolutionary phase than T Tauri disks, one in which almost all the gas and inner disk material is gone. These are debris disks, which are thought to originate from collisions between larger bodies rather than to be made up of dust from interstellar gas.

\subsection{Transition Disks}

The observation that accreting disks and inner optically thick disks disappear on a few Myr timescales, but that some fraction leave behind very small amounts of mass in the outer disk, is a very interesting one. We will discuss theoretical models for how this happens shortly. Before doing so, however, we will review what observations tell us about the transition from gaseous, accreting T Tauri disks to low-mass debris disks.

\begin{marginfigure}
\includegraphics[width=\linewidth]{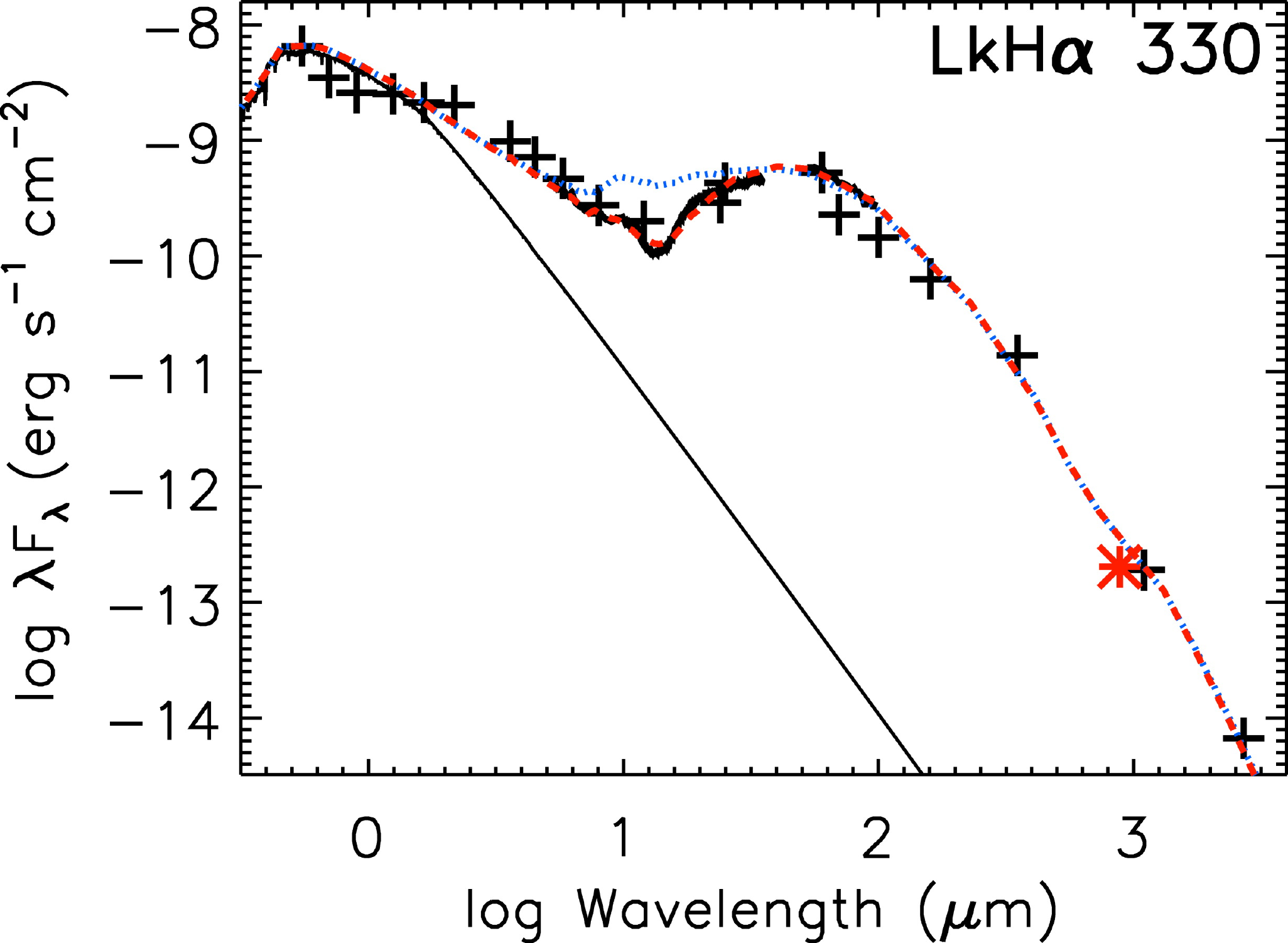}
\caption[Spectral energy distribution of LkH$\alpha$ 330]{
\label{fig:lkha330_brown08}
The spectral energy distribution of the star LkH$\alpha$ 330. Plus signs indicate measurements. The black line is a model for a stellar photosphere. The blue line is a model for a star with a disk going all the way to the central star, while the red line is a model in for a disk with a 40 AU hole in its center. Credit: \citet{brown08a}, \copyright AAS. Reproduced with permission.
}
\end{marginfigure}

This change is likely associated with an intriguing class of objects known as transition disks. Spectrally, these are defined as objects that have a significant 24 $\mu$m excess (or excess at even longer wavelengths), but little or no excess at shorter wavelengths (Figure \ref{fig:lkha330_brown08}). This spectral energy distribution (SED) suggests a natural physical picture: a disk with a hole in its center. The short wavelength emission normally comes from near the star, and the absence of material there produces the lack of short wavelength excess. Indeed, it is possible to fit the SEDs of some stars with models with holes.

\begin{marginfigure}
\includegraphics[width=\linewidth]{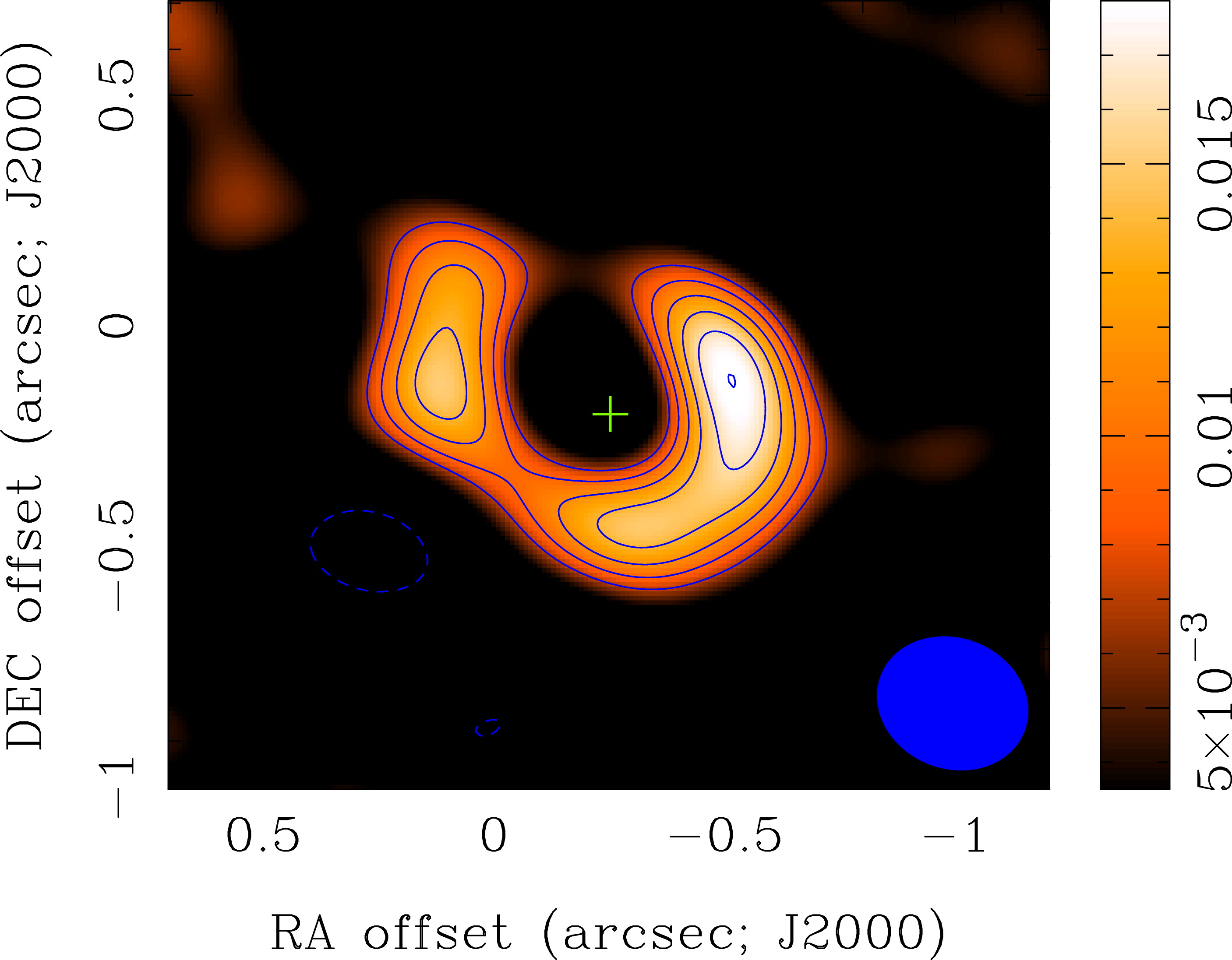}
\caption[Dust continuum image of the disk around LkH$\alpha$ 330]{
\label{fig:lkha330_image_brown08}
Dust continuum image of the disk around LkH$\alpha$ 330, taken at 340 GHz by the SMA. Colors show the detected signal, and contours show the signal to noise ratio, starting from S/N of 3 and increasing by 1 thereafter. The green plus marks the location of the star. The blue circle is the SMA beam. Credit: \citet{brown08a}, \copyright AAS. Reproduced with permission.
}
\end{marginfigure}

In the last decade it has become possible to confirm the presence of inner holes in transition disks directly, at least cases where the inferred hole is sufficiently large (Figure \ref{fig:lkha330_image_brown08}). The sizes of the holes inferred by the observations are generally very good matches to the values inferred by modelling the SEDs. The holes are remarkably devoid of dust: upper limits on the masses of small dust grains within the hole are often at the level of $\sim 10^{-6}$ $M_\odot$. The sharp edges of the holes indicate that the effect driving them is not simply the growth of dust grains to larger sizes, which should produce a more gradual transition. Instead, something else is at work. However, in some transition disks gas is still seen within the gap in molecular line emission, which also suggests that whatever mechanism is removing the dust does not necessarily get rid of all the gas as well.

\section{Disk Dispersal: Theory}

We have seen the observations suggest that disks are cleared in a few Myr. We would like to understand what mechanism is responsible for this clearing.

\subsection{Setting the Stage: the Minimum Mass Solar Nebula}

Before diving into the theoretical models, let us pause for a moment to obtain some typical numbers, which we can use below to plug in an evaluate timescales. Imagine spreading the mass in the Solar System's observed planets into an annulus that extends from each planet's present-day orbit to halfway to the next planet in each direction. Then add enough hydrogen and helium so that metal content matches that observed in the Sun. This is the mass distribution that the protoplanetary disk of the Sun must have had if all the metals in the disk wound up in planets, and if the planets were formed at their present-day locations. Neither of these assumptions is likely to be strictly true, but they are a reasonable place to begin thinking about initial conditions, and give a rough lower limit on the mass surface density of the disk from which the planets formed. We call the theoretical construct that results from this exercise the Minimum Mass Solar Nebular (MMSN). The MMSN a mass of $\sim 0.01$ $\msun$ and a surface density $\Sigma$ that varies as roughly $\varpi^{-3/2}$. The "standard" modern value for the MMSN's surface density is $\Sigma=\Sigma_0 \varpi_0^{-3/2}$, where $\Sigma_0 \approx 1700$ g cm$^{-2}$ and $\varpi_0=\varpi/\mbox{AU}$.

If solar illumination is the principal factor determining the disk temperature structure and we neglect complications like flaring of the disk, then treat the disk as a blackbody produces a temperature profile
\begin{equation}
\label{eq:TMMSN}
T=280 \varpi_0^{-1/2}\mathrm{ K}.
\end{equation}
True temperatures are probably also higher by a factor of $\sim 2$, as a result of flaring and viscous dissipation providing extra heat. The corresponding disk scale height is
\begin{equation}
H = \frac{c_g}{\Omega} = \sqrt{\frac{k_B T}{\mu m_{\rm H}}} \sqrt{\frac{\varpi^3}{GM}} = 0.03 \varpi_0^{5/4}\mbox{ AU},
\end{equation}
where $c_g$ is the gas sound speed, $\Omega$ is the angular velocity of rotation, and the numerical evaluation uses $M = M_\odot$ and $\mu=2.3$.

For solar metallicity, heavy elements will constitute roughly 2\% of the total mass, but much of this mass is in the form of volatiles that will be in the gas phase over much of the disk. For example a significant fraction of the carbon is in the form of CO and CO$_2$, and at the pressures typical of protoplanetary disks this material will not freeze out into ices until the temperature drops below $20-30$ K. Such low temperatures are found, if anywhere, in the extreme outer parts of disks. Similarly, water, which is a repository for much of the oxygen, will be vapor rather than ice at temperatures above 170 K. This temperature will be found only outside several AU.

A rough approximation to the mass in "rocks", things that are solid at any radius, and "ices", things that are solid only at comparatively low temperatures, is
\begin{eqnarray}
\Sigma_{\rm rock} & \approx & 7 \varpi_0^{-3/2}\mbox{ g cm}^{-2} \\
\Sigma_{\rm ice} & \approx &
\left\{
\begin{array}{ll}
0 & T > 170\mbox{ K}\\
23\varpi_0^{-3/2}\mbox{ g cm}^{-2} \qquad & T < 170\mbox{ K}
\end{array}
\right.
\end{eqnarray}
In other words, rocks are about $0.4\%$ of the mass, and ices, where present, are about $1.3\%$. Typical volume densities for icy material are $\sim 1$ g cm$^{-3}$, and for rocky material they are $\sim 3$ g cm$^{-3}$.

\subsection{Viscous Evolution}

Now that we have a setting, let us consider the first and most obvious mechanism for getting rid of disks: having them accrete onto their parent star. The basic process governing movement of mass in a late-stage disk is the same as during the protostellar period: viscous evolution. The difference at late stages is that there is no more mass being supplied to the disk edge, so accretion onto the star, rather than occurring in steady state, tends to drain the disk and reduce its surface density. Recall that the typical accretion rates we infer during the T Tauri phase are $\sim 10^{-9}-10^{-8}$ $\msun$ yr$^{-1}$. Since typical disk masses are $\sim 0.01$ $\msun$, this would imply that the time required to drain the disk completely into the star is $\sim 1 - 10$ Myr, not far off the observed disk dispersal lifetime.

We can make this argument more quantitative. Recall that the evolution equation for the surface density of a viscous disk is (Chapter \ref{ch:disks_theory})
\begin{equation}
\label{eq:masscons_disk_late}
\frac{\partial \Sigma}{\partial t} = \frac{3}{\varpi} \frac{\partial}{\partial \varpi} \left[\varpi^{1/2} \frac{\partial}{\partial \varpi} (\nu \Sigma \varpi^{1/2})\right],
\end{equation}
where $\nu$ is the viscosity, $\Sigma$ is the surface density, and $\varpi$ is the radius. To see how this will affect protoplanetary disks, it is useful to consider some simple cases that we can solve analytically. Let us suppose that the viscosity follows a powerlaw form $\nu=\nu_1 (\varpi/\varpi_1)^\gamma$. The equations in this case admit a similarity solution; the case $\gamma=1$ is included in problem set 4. For arbitrary $\gamma$, it is easy to verity that equation (\ref{eq:masscons_disk_late}) has the solution
\begin{equation}
\Sigma = \frac{C}{3\pi\nu_1 x^\gamma} T^{-(5-2\gamma)/(4-2\gamma)} \exp\left(-\frac{x^{2-\gamma}}{T}\right),
\end{equation}
where $C$ is a constant with units of mass over time that determines the total mass in the disk and the accretion rate, $x=\varpi/\varpi_1$, and $T$ is a dimensionless time defined by
\begin{equation}
T = \frac{t}{t_s}+1
\qquad
t_s = \frac{1}{3(2-\gamma)} \frac{\varpi_1^2}{\nu_1}.
\end{equation}
In this similarity solution, at any given time the disk surface density has two regions. For $x^{2-\gamma} \ll T$, the exponential term is negligible, and the surface density simply follows a powerlaw profile $x^{-\gamma}$. For $x^{2-\gamma} \gg T$, the exponential term imposes an exponential cutoff. As time goes on and $T$ increases, the powerlaw region expands, but its surface density also declines (at least for $\gamma < 2$, which is what most physically-motivated models produce). The quantity $t_s$ is the characteristic viscous evolution time. For times $t\ll t_s$, $T$ is about constant, so there is no evolution. Evolution becomes significant after $t>t_s$.

If we adopt a simple $\alpha$ model with constant $\alpha$, then recall that $\nu = \alpha c_g H$. For our MMSN, $T \approx 280 \varpi_0^{-1/2}$ K and $H \approx 0.03 \varpi_0^{5/4}$ K. Since $c_g\propto T^{1/2}$, this implies $\nu\propto \varpi$:
\begin{equation}
\nu \approx 5\times 10^{16} \alpha \varpi_0\mbox{ cm}^2\mbox{ s}^{-1}.
\end{equation}
Thus constant $\alpha$ for our MMSN corresponds to $\gamma = 1$. Plugging this value into the similarity solution gives $t_s = 0.024\alpha_{-2}^{-1}\mbox{ Myr}$, where $\alpha_{-2} = \alpha/0.01$. Thus we would expect the disk to drain into the star in $\sim 1$ Myr if it had values of $\alpha$ expected for the MRI.

This might seem like an appealing explanation for why disks disappear, but it faces two serious objections. The first is that, as discussed in chapter \ref{ch:disks_theory}, magnetorotational instability (MRI) seems unlikely to be able to operate everywhere in the late-stage disks. The surface will be kept ionized by stellar radiation and possibly cosmic rays, but the midplane will be too neutral for strong magnetic coupling. This should reduce the accretion rate.

A second, more serious objection is that it does not reproduce the observation that disks drain inside-out (or at least some of them do). In this model, the surface density everywhere inside the powerlaw inner region decreases with time as $t^{-3/2}$, meaning that the disk would fade uniformly rather than from the inside out. While this result is for the particular similarity solution we used, it is a generic statement that, in any model where $\alpha$ is constant with radius, the disk will tend to drain uniformly rather than inside-out. Thus something more sophisticated is needed.

\subsection{Photoevaporation Models}

One mechanism that has been proposed for disk clearing is photoevaporative winds. We will not discuss this quantitatively here, because a basic model of this process is left as an exercise in Problem Set 5. The qualitative picture is simply that the surface of the disk is heated to temperatures of $\sim 100-200$ K by stellar FUV radiation out to a fairly large region, and is heated to $\sim 10^4$ K by ionizing radiation closer in to the star. If the heated gas is far enough from the star for this temperature increase to raise its sound speed above the escape speed, it will flow away from the disk in a thermally-driven wind.

This tends to produce maximum mass loss from a region near where the sound speed equals the escape speed, since that is where there is the most gas and the radiation is most intense, but the gas can still escape. If the radiation is intense enough, a gap in the disk will open at this radius, and mass will not be able to pass through it -- any gas that gets to the gap is lost in the wind. As a result the inner disk drains viscously, and is not replenished, leaving a hole like we observe.

\subsection{Rim Accretion Models}

A second mechanism that could produce an inner hole is rim accretion. In this picture, MRI operates only on the inner rim of the disk where the gas is exposed to direct stellar radiation. Material from this rim accretes inward while the rest of the disk remains static. As the rim accretes, more disk material is exposed to stellar radiation and the MRI-active region grows. Thus the disk drains inside-out. Our treatment of this phenomenon will generally follow that set out by \citet{chiang07a}.

In this picture, we let $N_*$ be the column (in H atoms per cm$^2$) of material in the rim that is sufficiently ionized for MRI to operate. In this case the mass in the MRI-active rim at any time is
\begin{equation}
M_{\rm rim} = 4\pi N_* \mu_{\rm H} r_{\rm rim} H,
\end{equation}
where $r_{\rm rim}$ is the rim radius, H is the scale height at the rim, and $\mu_{\rm H}$ is the mass per H nucleus. The time required for this material to accrete is the usual value for viscous accretion:
\begin{equation}
t_{\rm acc} = \frac{r_{\rm rim}^2}{\nu} = \frac{r_{\rm rim}^2}{\alpha c_g H},
\end{equation}
where $c_g$ is the gas sound speed in the irradiated rim, which is presumably higher than in the shielded disk interior. Putting these together, we expect an accretion rate $\sim M_{\rm rim}/t_{\rm acc}$. \citet{chiang07a}, solving the problem a bit more exactly, get
\begin{equation}
\dot{M} \approx \frac{4\pi N_* \mu_{\rm H} \alpha c_g^3 r_{\rm rim}^2}{G M}.
\end{equation}
One can estimate $N_*$ and $c_g$ from the thermal and ionization balance of the irradiated rim, and \citet{chiang07a}'s result is $N_* \approx 5\times 10^{23}$ cm$^{-2}$ and $c_g \approx 0.9$ km s$^{-1}$, giving
\begin{equation}
\dot{M} = 1.4\times 10^{-11} \alpha_{-2} M_0^{-1} r_{\rm rim,0}^{-2} \,\msun\mbox{ yr}^{-1},
\end{equation} 
where $M_0$ is the stellar mass in units of $\msun$ and $r_{\rm rim}$ is the rim radius in units of AU.

This model nicely explains why disks will drain inside out. Moreover, it produces the additional result that any grains left in the disk that reach the rim will not accrete, and are instead blown out by stellar radiation pressure. This produces an inner hole with a small amount of gas on its way in to the star, as is required to explain the molecular observations, but with no dust.

\subsection{Grain Growth and Planet Clearing Models}

The final possible mechanism for getting rid of the disk is the formation of planets. If the dust in a disk agglomerates to form larger bodies, then the opacity per unit mass will drop dramatically, and as a result the disk will cease to produce observable infrared or millimeter emission. If the planetesimals further agglomerate into planets with significant gravitational effects, these can begin to clear the gas as well. We will therefore end this chapter with a discussion of how grains might begin to agglomerate together, starting the process of getting rid of a disk by planet formation that will be the subject of Chapter \ref{ch:planets}.

Consider a population of solid particles radius $s$, each of which individually has density $\rho_s$. The number density of particles (i.e., the number of particles per cm$^{3}$) is $n$, so the total mass density of the population of solids is
\begin{equation}
\rho_d = \frac{4}{3}\pi \rho_s s^3 n
\end{equation}
If the collection of solids has a velocity dispersion $c_s$, the mean time between collisions between them is
\begin{equation}
t_{\rm coll} = (n \pi s^2 c_s)^{-1} = \frac{4}{3} \frac{\rho_s s}{\rho_d c_s}
\end{equation}
We will see in a little while that grains of interstellar sizes will have about the same scale height as the gas. Thus, within one scale height of the disk midplane, we may take 
\begin{equation}
\rho_d \approx \frac{\Sigma_d}{H} = \frac{\Sigma_d \Omega}{c_g},
\end{equation}
where $\Sigma_d = \Sigma_{\rm rock} + \Sigma_{\rm ice}$ is the total surface density of "dust", including both rocky and icy components. 

The velocity dispersion of the solids $c_s$ depends on their sizes. In the case of small grains it will simply be the typical velocity imparted by Brownian motion in the fluid, which is
\begin{equation}
c_s = \sqrt{\frac{3}{2} \frac{k_B T}{m_s}} = \sqrt{\frac{3 \mu m_{\rm H}}{2 m_s}} c_g \approx 0.1 \varpi_0^{-1/4} s_{-4}^{-3/2} \rho_{s,0}^{-1/2}\mbox{ cm s}^{-1}
\end{equation}
where $m_s = (4/3)\pi s^3 \rho_s$ is the mass of the solid particle, $\rho_{s,0}=\rho_s/(1\mbox{ g cm}^{-3})$, $s_{-4} = s/(1\,\mu\mbox{m})$, and we have used our fiducial MMSN to estimate $c_g$. Plugging $n$ and $c_s$ into the collision time, we have
\begin{eqnarray}
t_{\rm coll} & \approx & \frac{4\sqrt{2}}{3\sqrt{3}}\frac{s\rho_s}{\Sigma_d \Omega}\sqrt{\frac{m_s}{\mu m_{\rm H}}} = \frac{8\sqrt{2\pi}}{9} \sqrt{\frac{\rho_s^3 s^5}{\Sigma_d^2\Omega^2 \mu m_{\rm H}}}
\nonumber \\
& = & (2.6, 0.6) \varpi_0^3 \rho_{s,0}^{3/2} s_{-4}^{5/2}\mbox{ yr},
\end{eqnarray}
where the two coefficients refer to the cases of rock only, or rock plus ice.

The bottom line of this calculation is that the small particles that are inherited from the parent molecular cloud will very rapidly collide with one another in the disk: a 1 $\mu$m-sized particle can expect to run into another one roughly 1 million times over the $\sim 1$ Myr lifetime of the disk. As the particles grow in size, collisions will rapidly become much less rapid, and will reach one collision per Myr at around 0.25 mm. Of course this assumes that the particles remain distributed with the same scale height as the gas, which is not a good assumption for larger particles, as we will see.

Before moving on, though, we must consider what happens when the particles collide. This is a complicated question, which is experimentally difficult enough that some groups have constructed dust-launching crossbows to shoot dust particles at one another in an attempt to answer experimentally. For very small particles, those of micron sizes, the answer is fairly easy. Such particles will be attracted to one another by van der Waals forces, and when they collide they will dissipate energy via elastic deformation. Theoretical models and experiments indicate that two particles will stick when they collide if the collisions velocity is below a critical value, and will bounce or shatter if the velocity is above that value.

For 1 $\mu$m particles, estimated sticking velocities are $\sim 1-100$ cm s$^{-1}$, depending on the composition of the body, and that this declines as $\sim s^{-1/2}$. Since this is much less than the Brownian speed, $1-10$ $\mu$m particles will very quickly grow to large sizes. Thus we have a strong theoretical prediction that grains should grow in disks. We will return to the question of how far this growth goes in the next chapter.

\chapter{The Transition to Planet Formation}
\label{ch:planets}

\marginnote{
\textbf{Suggested background reading:}
\begin{itemize}
\item \href{http://adsabs.harvard.edu/abs/2014prpl.conf..547J}{Johansen, A., et al. 2014, in "Protostars and Planets VI", ed.~H.~Beuther et al., pp.~547-570} \nocite{johansen14a}
\end{itemize}
\textbf{Suggested literature:}
\begin{itemize}
\item \href{http://adsabs.harvard.edu/abs/2010ApJ...722.1437B}{Bai, X.-N., \& Stone, J.~M. 2010, ApJ, 722, 1437} \nocite{bai10a}
\end{itemize}
}

In this final chapter, we will finish our discussion of the transition from star formation to planet formation. We have already seen in chapter \ref{ch:late_disk} that the interstellar dust grains that are captured in a star's disk will begin to collide with one another and grow, and that they will reach macroscopic size on time scales shorter than the observed disk lifetime. We now see to sharpen our understanding of how these solids will evolve. We will continue to make use of fiducial numbers from the minimum mass Solar nebula (MMSN) that we introduced in chapter \ref{ch:late_disk}.

\section{Dynamics of Solid Particles in a Disk}

\subsection{Forces on Solids}

We begin our discussion by attempting to determine the dynamics of solid particles orbiting in a protoplanetary disk. Consider such a particle. Because the mass of the disk is very small compared that of the star, we can neglect the gravitational force it exerts in the radial direction, and thus the radial gravitational acceleration felt by the particle is simply $g_\varpi = G M/\varpi^2$, where $M$ is the star's mass and $\varpi$ is the distance from the star. 

In the vertical direction we have the gravitational pull of both the star and the disk itself, and we have to think a bit more. However, one can show fairly easily that, for material distributed with the thermal scale height of the disk, the star's vertical gravity must dominate as well. The star's vertical gravitational force is 
\begin{equation}
g_{z,*} = \frac{z}{\varpi} g_\varpi = \Omega^2 z,
\end{equation}
where $z$ is the distance above the disk midplane and $\Omega$ is the angular velocity of a Keplerian orbit. We can approximate the disk as an infinite slab of surface density $\Sigma$; the gravitational force per unit mass exerted by such a slab is
\begin{equation}
g_{z,d} = 2\pi G \Sigma.
\end{equation}
The ratio of the stellar force to the disk force at a distance $H$ off the midplane, the typical disk height, is
\begin{equation}
\frac{g_{z,*}}{g_{z,d}} = \frac{\Omega^2 H}{2\pi G \Sigma} = \frac{c_g \Omega}{2\pi G \Sigma} = \frac{Q}{2},
\end{equation}
where in the last step we substituted in the Toomre $Q=\Omega c_g/(\pi G \Sigma)$ for a Keplerian disk.

Thus the vertical gravity of the disk is negligible as long as it is Toomre stable, $Q\gg 1$. For our MMSN,
\begin{equation}
\label{eq:QMMSN}
Q=55 \varpi_0^{-1/4},
\end{equation}
so unless we are {\it very} far out, stellar gravity completely dominates. As a caveat, it is worth noting that we implicitly assumed that the scale height $H$ applies to both the gas and the dust, even though we calculated it only for the gas. In fact, the motion of the dust is more complex, and, as we will see shortly, the assumption that the dust scale height is the same as that of the gas is not a good one. Nonetheless, neglecting the self-gravity of the disk is a reasonable approximation until significant gas-dust separation has occurred.

The other force on the solids that we have to consider is drag. Aerodynamic drag is a complicated topic, but we can get an estimate of the drag force for a small, slowly moving particle that is good to order unity fairly easily. Consider a spherical particle of size $s$ moving through a gas of density $\rho$ and sound speed $c_g$ at a velocity $v$ relative to the mean velocity of the gas. First note that for small particles the mean free path of a gas molecule is larger than the particle size -- Problem Set 5 contains a computation of the size scale up to which this remains the case.

For such small grains it is a reasonable approximation to neglect collective behavior of the gas and view it as simply a sea of particles whose velocity distribution does not change in response to the dust grain moving through it. If the particle is moving slowly compared to the molecules, which will be the case for most grains, then the rate at which molecules strike the grain surface will be
\begin{equation}
\mbox{collision rate} \approx 4\pi s^2 \frac{\rho}{\mu m_{\rm H}} c_g,
\end{equation}
where $\mu$ is the mean mass per molecule, so $\rho/\mu m_{\rm H}$ is the number density. This formula simply asserts that the collision rate is roughly equal to the grain area times the number density of molecules times their mean speed.

If the grain were at rest the mean momentum transferred by these collisions would be zero. However, because it is moving, collisions on the forward face happen at a mean velocity of $\sim c_g+v$, and those on the backward face have a mean velocity $\sim c_g-v$. Thus, averaging over many collisions, there will be a net momentum transfer per collision of $\mu m_{\rm H} v$. The net rate of momentum transfer, the drag force, is therefore the product of this with the collision rate:
\begin{equation}
F_D = C_D s^2 \rho v c_g,
\end{equation}
where $C_D$ is a constant of order unity.

Integrating over the Boltzmann distribution and assuming that all collisions are elastic and that the reflectance is in random directions (so-called diffuse reflection), appropriate for a rough surface, gives $C_D = 4\pi/3$. With this value of $C_D$, this formula is known as the Epstein drag law. It becomes exact in the limit $s\ll \mbox{mean free path}$, $v \ll c_g$, and for pure elastic, diffuse reflection. Larger bodies experience Stokes drag, in which the dependence changes from $s^2 \rho v c_g$ to $s^2 \rho v^2$, but we will not worry about that for now. Finally, note that solid particles will not experience significant pressure forces, since they are so much more massive than the molecules that provide pressure.

\subsection{Settling}

Now let us consider what the combination of vertical gravity and drag implies. The vertical equation of motion for a particle is
\begin{equation}
\frac{d^2 z}{dt^2} = -g_z - \frac{F_D}{\frac{4}{3}\pi s^3 \rho_s} = -\Omega^2 z - \frac{\rho c_g}{\rho_s s} \frac{dz}{dt}
\end{equation}
where $\rho_s$ is the density of the solid particle. This ODE represents a damped harmonic oscillator: the gravitational term is the linear restoring force, and the drag term is the damping term. Within one gas scale height of the midplane $\rho$ is roughly constant, $\rho \approx \Sigma/H = 3\times 10^{-9} \varpi_0^{-11/4}$ g cm$^{-3}$. For constant $\rho$ the ODE can be solved analytically:
\begin{equation}
z = z_0 e^{-t/\tau},
\end{equation}
where
\begin{equation}
\tau = 2\frac{\rho_s s}{\rho c_g} \left[1 - \left(1 - \frac{4 s^2 \rho_s^2 \Omega^2}{\rho^2 c_g^2}\right)^{1/2}\right]^{-1}.
\end{equation}

If the term in the square root is negative, which is the case when $s$ is large, the damping is not strong enough to stop particles before they reach the midplane, and they instead perform a vertical oscillation of decreasing magnitude. If it is positive, they simply drift downward, approaching the midplane exponentially. The minimum time to reach the midplane occurs when the particles are critically damped, corresponding to the case where the square root term vanishes exactly. Critical damping occurs for particles of size
\begin{equation}
s_c = \frac{\rho c_g}{2 \rho_s \Omega} = 850 \varpi_0^{-3/2} \rho_{s,0}^{-1}\mbox{ cm},
\end{equation}
where $\rho_{s,0} = \rho_s/(1\mbox{ g cm}^{-3})$.

Thus all objects smaller than $\sim 10$ m boulders will slowly drift down to the midplane without oscillating. For $s\ll s_c$, we can expand the square root term in a series to obtain
\begin{equation}
\tau \approx 4 \frac{\rho_s s}{\rho c_g} \left(\frac{s}{s_c}\right)^{-2} = \frac{\rho c_g}{\rho_s \Omega^2 s} = 270 \varpi_0^{11/4} \rho_{s,0}^{-1} s_0^{-1}\mbox{ yr},
\end{equation}
where $s_0 = s/(1\mbox{ cm})$. 

Thus 1 cm grains will settle to the midplane almost immediately, while interstellar grains, those $\sim 1$ $\mu$m in size, will take several Myr to reach the midplane. Of course these very small grains will also collide with one another and grow to larger sizes, which will let them sediment more rapidly. In practice coagulation and sedimentation occur simultaneously, and each enhances the other: growth helps particle sediment faster, and sedimentation raises the density, letting them collide more often.

\subsection{Radial Drift}

We have just considered the consequences of the forces acting on solid particles in the vertical direction. Next let us consider the radial direction. The homework includes a detailed solution to this problem for small particles, so we will not go through the calculation, just the qualitative result. The basic idea is that gas in the disk is mostly supported by rotation, but it also has some pressure support. As a result, it orbits at a slightly sub-Keplerian velocity. Solid bodies, on the other hand, do not feel gas pressure, so they can only remain in orbit at constant radius if they orbit at the Keplerian velocity. The problem is that this means that they are moving faster than the gas, and thus experience a drag force.

Problem Set 5 contains a calculation showing that the difference in velocity between the Keplerian speed and the speed with which a particle orbits is
\begin{equation}
\Delta v = \frac{n c_g^2}{2 v_K} = \eta v_K
\end{equation}
where the pressure in the disk is assumed to vary with distance from the star as $P\propto \varpi^{-n}$, $c_g$ is the gas sound speed, and the dimensionless quantity $\eta = n c_g^2/2 v_K^2$, which depends only on the local properties of the disk, has been defined for future convenience. At 1 AU for our minimum mass solar nebula model, this velocity is about $70 n$ m s$^{-1}$.

Drag takes away angular momentum, in turn causing the bodies to spiral inward. We can parameterize this effect in terms of the stopping time 
\begin{equation}
t_s = \frac{mv}{F_D},
\end{equation}
where $m$ and $v$ are the body's mass and velocity, and $F_D$ is the drag force it experiences. The stopping time is simply the characteristic time scale required for drag to stop the body.

Consider a spherical solid body of size $s$. For the Epstein law, which we discussed last time, $F_D \propto s^2$, while for Stokes drag, which describes larger bodies, $F_D \propto s^2$ at low Reynolds and $s$ at high Reynolds number. On the other hand, for a body of fixed density the mass varies as $s^3$, so the acceleration produced by drag must be a decreasing function of $s$. The stopping time is therefore an increasing function of $s$. Intuitively, big things have a lot of inertia per unit area, so they are hard to stop. Little things have little inertia per unit area, so they are easy to stop.

Now consider two limiting cases. Very small bodies will have stopping times $t_s$ much smaller then their orbital periods $t_p$, so they will always be forced into co-rotation with the gas. Since this makes their rotation sub-Keplerian, they will want to drift inward. The rate at which they can drift, however, will also be limited by gas drag, since to move inward they must also move through the gas. Thus we expect that the inward drift velocity will also decrease as the stopping time decreases, and thus as the particle size decreases. To summarize, then, for $t_s/t_p \ll 1$, we expect $v_{\rm drift} \propto s^p$, where $p$ is a positive number. Small particles drift inward very slowly, and the drift speed increases with particle size for small $s$.

Now consider the opposite limit, $t_s \gg t_p$. In this case, the drag is unable to force the solid body into co-rotation on anything like the orbital period, so the body is always in a near-Keplerian orbit, and just slowly loses angular momentum to drag. Clearly in this case the rate at which this causes the particle to drift inward will decrease as the stopping time increases, and thus as the particle size increases. Summarizing this case, then, for $t_s/t_p \gg 1$, we expect $v_{\rm drift} \propto s^{-q}$, where $q$ is a positive number.

Since the inward drift speed rises with particle size at small sizes and decreases with particle size at large sizes, there must be some intermediate size with it reaches a maximum. Conversely, the time required for drag to take away all of a body's angular momentum, so that it spirals into the star, must reach a minimum at some intermediate size. Problem set 5 contains a calculation showing that even for 1 cm pebbles the loss time is a bit shorter than the disk lifetime, and 1 cm pebbles are in the regime where $t_s \ll t_p$. The drift rate reaches a maximum for $\sim 1$ m radius objects, and for them the loss time can be as short as $\sim 100$ yr. For km-sized objects the drift rate is back down to the point where the loss time is $\sim 10^5$ to $10^6$ yr. 

\section{From Pebbles to Planetesimals}

The calculation we have just completed reveals a serious problem in how we can continue the process of growing the solids to larger sizes, forming planets and clearing away disks: it seems that once growth reaches $\sim 1$ m sizes, all those bodies should be dragged into the star in a very short amount of time. We therefore next consider how to overcome this barrier.

\subsection{Gravitational Growth}

One solution is to skip over this size range using a mechanism that allows particles to go directly from cm to km sizes, while spending essentially no time at intermediate sizes. A natural candidate mechanism for this is gravitational instability, so we begin with a discussion of whether this might work. As noted above, the gas in the MMSN is very gravitationally stable, $Q \sim 50$. However, we also saw that solids will tend to settle toward the midplane, and the solids have a much smaller velocity dispersion than the gas. The Toomre $Q$ for the solid material alone is
\begin{equation}
Q_s = \frac{\Omega c_s}{\pi G \Sigma_s}
\end{equation}
where $c_s$ and $\Sigma_s$ are the velocity dispersion and surface density of the solid material. To see what velocity dispersion is required, note that this definition of $Q$ lets use write $Q_s$ in terms of $Q_g$ as
\begin{equation}
Q_s = Q_g \left(\frac{\Sigma_g}{\Sigma_s}\right) \left(\frac{c_s}{c_g}\right) \approx(240, 60) Q_g \left(\frac{c_s}{c_g}\right),
\end{equation}
where the factors of 240 or 60 are for regions with and without solid ices, respectively.

Using equation (\ref{eq:QMMSN}) for $Q_g$ and equation (\ref{eq:TMMSN}) for the gas temperature, we have $Q_g\approx 55$ and $c_g\approx 1$ km s$^{-1}$ at 1 AU. Thus, gravitational instability for the solids, $Q_s<1$, requires that $c_s \lesssim (30, 7)$ cm s$^{-1}$, depending on whether ice is present or not. If such an instability were to occur, the characteristic mass of the resulting object would be set by the Toomre mass
\begin{equation}
M_T = \frac{4 c_s^4}{G^2 \Sigma_s} = (2\times 10^{19}, 3\times 10^{17})\mbox{ g},
\end{equation}
where the two numbers are again for the cases with and without ices in solid form. If we adopt $\rho_{\rm i,r} = (1,3)$ g cm$^{-3}$ as the characteristic densities of (icy, rocky) material, the corresponding sizes of spheres with this mass are $(20, 3)$ km. This is large enough to avoid the size range where rapid loss occurs.

To see whether this condition can be met, it is more convenient to phrase the instability criterion in terms of a density. If we use $H_s=c_s/\Omega$ in the Toomre condition, where $H_s$ is the scale height of the solids, and we take the midplane density of the solids to be $\rho_s \approx \Sigma_s/H_s$, then we have
\begin{equation}
Q_s  = \frac{\Omega^2 H_s}{\pi G \Sigma_s} \approx \frac{M_*}{\varpi^3 \rho_s},
\end{equation}
where $M_*$ is the mass of the star. A detailed stability analysis by \citet{sekiya1983} of the behavior of a stratified self-gravitating disk show that the instability condition turns out to be
\begin{equation}
\rho > 0.62\frac{M_*}{\varpi^3} = 4\times 10^{-7} M_{*,0} \varpi_0^{-3} \mbox{ g cm}^{-3},
\end{equation}
where $\rho = \rho_s + \rho_g$ is the total (gas plus solid) surface density and $M_{*,0}=M_*/\msun$. For our minimum mass solar nebula, recall that the midplane density of the gas is roughly $3\times 10^{-9}$ g cm$^{-3}$, a factor of 100 too small for instability to set in. The question then is whether the density of solids at the midplane can rise to 100 times that of the gas.

The discussion followed here closely follows that of \citet{youdin02a}. We have seen that settling causes solid particles to drift down to the midplane, and if this were the only force acting on them, then the density could rise to arbitrarily high values. However, there is a countervailing effect that will limit how high the midplane density can rise. If the midplane density of solids is large enough so that the solid density greatly exceeds the gas density, then the solid-dominated layer will rotate at the Keplerian speed rather than the sub-Keplerian speed that results from gas pressure. It is fairly straightforward to show (and a slight extension of one of the problems in Problem Set 5) that the rotation velocity required for radial hydrostatic balance is
\begin{equation}
v_{\phi} = \left(1-\eta \frac{\rho_g}{\rho}\right) v_K,
\end{equation}
where $\rho = \rho_g + \rho_s$ which approaches $v_K$ for $\rho_g \ll \rho_s$, and $(1-\eta) v_K$ for $\rho_g \gg \rho_s$. Since $\rho_s / \rho_g$ rises toward the midplane, this velocity profile has shear in it, with $v_{\phi}$ reaching a maximum at the midplane and dropping above it.

The shear can generate Kelvin-Helmholtz instability, which will in turn create turbulence that will dredge up the dust out of the midplane, halting settling and preventing the density from continuing to rise. A useful analogy to think about, which I borrow from \citet{youdin02a}, is a sandstorm in the desert. Since the midplane full of dust is trying to rotating faster than the gas-dominated layer above it, there is effectively a wind blowing above the dusty midplane layer, like a wind blowing over the desert. If the wind blows too fast, it will start picking up dust, preventing it from falling back to the desert floor.

In the case of a disk, this process will self-regulate, since reducing the amount of dust in the midplane brings its rotation velocity closer to that of the gas, thereby reducing the strength of the wind. This process of self-regulation can be calculated in terms of the condition required for KH instability. To understand how the criterion for KH instability is set, it is easiest to think about the case of a physical interface -- the results are not significantly different for a continuous medium. The most common example is a pond of water with wind blowing across its surface. Imagine that there is a small ripple in the water that causes the surface to rise a little. The wind will strike the bit of the water above the surface and try to push it horizontally. At the same time gravity will try to drag the water downward.

If the wind is strong, it will push the water horizontally faster than gravity can drag it downward. The moving water will displace the surface even more, creating a growing wave, the signature of KH instability. If it is weak, gravity will drag the ripple downward before the wind is able to displace it significantly. Thus we expect the critical condition for KH instability to involve a balance between the restoring force of gravity and the destabilizing force of shear. For a continuous medium, it turns out that the condition for instability can be stated in terms of the Richardson number
\begin{equation}
\mbox{Ri} = \frac{(g_z/\rho) (\partial \rho/\partial z)}{\left(\partial v_\phi/\partial z\right)^2} < \mbox{Ri}_c,
\end{equation}
where $z$ is the vertical distance, $g_z$ is the gravitational acceleration in the vertical direction, and the critical Richardson number for instability $\mbox{Ri}_c\approx 1/4$.\footnote{Note that the quantity in the numerator has units of one over time squared, so it is the square of a frequency. In fact, it is a frequency that is familiar from stellar structure: $(g_z/\rho) (\partial \rho/\partial z)$ is the square of the Brunt-V\"{a}is\"{a}l\"{a} frequency, the characteristic oscillation frequency for vertical displacements in a stratified medium, such as stellar atmosphere.}

The numerator here represents the stabilizing effects of gravity, which depends on both the gravitational acceleration and how quickly the density drops with height. The gravitational acceleration is
\begin{equation}
g_z = \Omega^2 z + 4\pi G \int_0^z \rho(z') \, dz',
\end{equation}
where the first term represents the gravitational pull of the star and the second represents the self-gravity of the disk. The denominator represents the amount of destabilizing velocity shear.

A reasonable approximation is that the KH instability will stop any further settling once it turns on, so the density of the solids will become as centrally peaked as possible while keeping the disk marginally stable against KH. Thus, we expect the equilibrium density profile for the solids to be the one that gives $\mbox{Ri}=1/4$. If $\rho_g(z)$ is known at a given radius, then the condition $\mbox{Ri} =1/4$ fully specifies the total density profile $\rho(z)$, since both $g_z$ and $v_\phi$ are known function of $\rho$ and $\rho_g$. Given $\rho(z)$, it is obviously trivial to deduce the density of solids $\rho_s(z)$. The equation can be solved numerically fairly easily, but we can gain additional insight by proceeding via analytic approximations.

First, note that we are interested in whether a self-gravitating layer of particles can develop at all, and that until one does then we can ignore the self-gravity of the disk in $g_z$. Thus, we can set $g_z\approx \Omega^2 z$ for our analytic approximation. If we now differentiate the velocity profile $v_{\phi}$ with respect to $z$, we get
\begin{equation}
\frac{\partial v_\phi}{\partial z} = -\eta \left(\frac{1}{\rho} \frac{\partial \rho_g}{\partial z} - \frac{\rho_g}{\rho^2}\frac{\partial \rho}{\partial z}\right) v_K.
\end{equation}
Substituting this into the condition that the Richardson is roughly $1/4$, and noting the $v_K = \varpi\Omega$, we have
\begin{eqnarray}
\mbox{Ri}_c \approx \frac{1}{4} 
& \approx &
\frac{z}{\eta^2 \varpi^2} \frac{\rho^3 (\partial \rho/\partial z)}{\left[\rho (\partial \rho_g/\partial z) - \rho_g (\partial \rho/\partial z)\right]^2} \\
& = &
\frac{z}{\eta^2 \varpi^2} \frac{\rho^3 (\partial \rho/\partial z)}{\left[\rho_s (\partial \rho_g/ \partial z) - \rho_g (\partial \rho_s/\partial z)\right]^2}.
\end{eqnarray}

Now we make our second approximation: if we focus our attention near the midplane where solids are trying to sediment out, and are being stirred up by KH instability, the density of solids should be changing much more quickly than the density of gas. In other words, we will focus our attention at heights $z$ much smaller than the gas scale height, so we can set $\partial \rho/\partial z \approx \partial \rho_s/\partial z$, and drop $\partial \rho_g/\partial z$ in comparison to $\partial \rho_s/\partial z$. Doing so gives
\begin{equation}
\mbox{Ri}_c \approx \frac{z}{\eta^2 \varpi^2} \frac{\rho^3}{\rho_g^2 (\partial\rho_s/\partial z)}
\end{equation}

To see what this implies, consider a layer of solids with scale height $H_s$ and surface density $\Sigma_s$ that marginally satisfies this equation. Plugging in $z\sim H_s$ and $\rho_s \sim \Sigma_s/H_s$ gives
\begin{equation}
\mbox{Ri}_c \sim \frac{H_s}{\eta^2 \varpi^2} \frac{(\rho_g + \Sigma_s/H_s)^3}{\rho_g^2 (\Sigma_s/H_s^2)} = \frac{(\rho_g H_s + \Sigma_s)^3}{(\eta r \rho_g)^2 \Sigma_s}
\end{equation}
Clearly this equation cannot be satisfied for arbitrarily large $\Sigma_s$, since the RHS scales as $\Sigma_s^2$ in this case. Physically, this indicates that our assumption that the KH instability can keep the Richardson number at the critical value must break down if the surface density of solids is too high. If we think about it, it makes sense that there is a maximum amount of solid material that the gas can keep aloft. To continue the sandstorm analogy, the wind can only keep a certain amount of sand aloft in the desert. It cannot pick up the entire desert.

Thus, we expect there to be a critical column density $\Sigma_p$ at which it becomes impossible to satisfy the condition that the Richardson number is $1/4$. If $\Sigma_p$ exceeds this critical value, the surface density at the midplane will rise arbitrarily, and gravitational instability becomes inevitable. For the case $\Sigma_s \gg \rho_g H_s$, this critical value is clearly given by
\begin{equation}
\Sigma_s \sim \sqrt{\mbox{Ri}_c} \eta \varpi \rho_g = 2 n \sqrt{\mbox{Ri}_c} \left(\frac{c_g}{v_K}\right)^2 \varpi \rho_g.
\end{equation}
For the conditions of our MMSN at 1 AU, using $n=1$ and $\mbox{Ri}_c=1/4$, this evaluates to $70$ g cm$^{-2}$. The numerical solution for the critical surface density is $\Sigma_s=94$ g cm$^{-2}$; the increase relative to our simple analytic estimate mostly comes from the self-gravity of the dust, which increases the shear and thus strengthens the KH instability.

This is clearly larger than the surface density of solids we have available in the MMSN, even using ices. Moreover, just increasing the total mass of the disk does not help, because $\rho_g$ will rise along with $\Sigma_s$, and thus the condition will not be any easier to meet. We therefore conclude that gravitational instability cannot be a viable mechanism to jump from cm to km sizes unless a way can be found to enhance the solid to gas ratio in the disk by a factor of $\sim 3$ in the icy part of the disk, or $\sim 10$ in the rocky part. 

\subsection{Hydrodynamic Concentration Mechanisms}

Gravitational instability by itself will not solve the problem of the meter-size barrier, but if some other mechanism can be found to increase the solid-to-gas ratio by a factor of $\sim 3-10$, then gravitational instability will take over and manufacture planetestimals. We therefore turn for the final topi in this chapter to what mechanisms might be able increase the solid to gas ratio by the required amount.

\begin{figure}
\includegraphics[width=\linewidth]{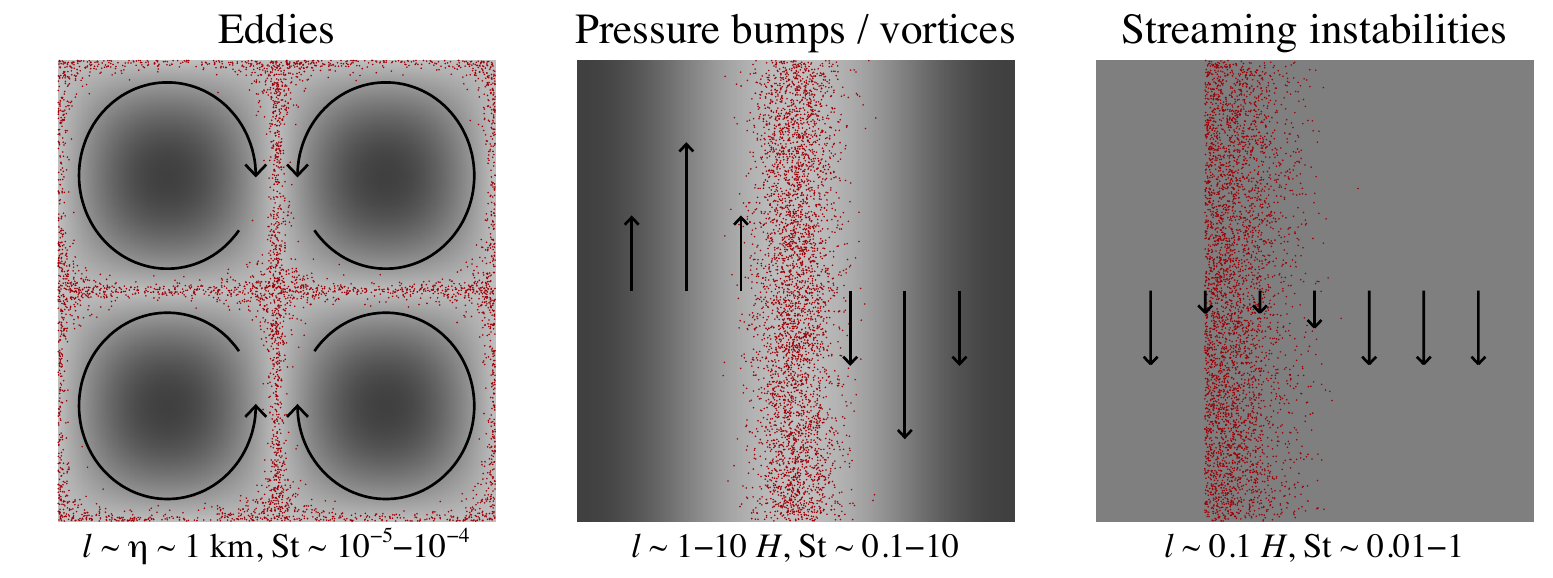}
\caption[Schematic of particle concentration by eddies in a protoplanetary disk]{
\label{fig:eddies_johansen14a}
Schematic diagram of three mechanisms to concentrate particles in a protoplanetary disk, taken from \citet{johansen14a}. The left panel shows how small-scale turbulent eddies expel particles to their outskirts. The middle panel shows how zonal flows associated with large-scale pressure bumps concentrate particles. The right panel shows concentration by streaming instabilities. In each panel, black arrows show the velocity field, and the caption indicates the characteristic length scale of the structures shown, where $H$ is the disk scale height.
}
\end{figure}

The first mechanism we will examine is concentration of small particles by eddies in a disk (Figure \ref{fig:eddies_johansen14a}). Consider a rotating eddy in a disk. By an eddy here we mean a structure where the gas moves on circular trajectories in the frame co-rotating with the disk at angular velocity $\Omega$. Suppose that the gas at some distance $r$ from the center of an eddy is rotating at some speed $v_e$. In the rotating reference frame, there are two forces acting on the gas: pressure gradients and Coriolis forces. For the eddy to remain static, the sum of these two forces must produce an acceleration per unit mass equal to the centripetal acceleration associated with the circular motion of the eddy. Specifically, we must have
\begin{equation}
2\Omega v_e - \frac{1}{\rho} \frac{dP}{dr} = -\frac{v_e^2}{r},
\end{equation}
where the first term is the Coriolis force per unit mass, the second is the pressure force per unit mass, and the right hand side is the centripetal acceleration. For a slowly-rotating eddy, $v_e /r \ll \Omega$, we can ignore the right hand side, and simply approximate that the sum of the two terms on the left is zero. Thus for slow eddies, the eddy rotation speed is given by
\begin{equation}
v_e = \frac{1}{2\rho\Omega} \frac{dP}{dr}.
\end{equation}
We see that if the eddy is associated with a pressure maximum, $dP/dr < 0$, then $v_e < 0$ as well, indicating that rotation is clockwise; eddies associated with pressure minima, $dP/dr > 0$, produce counter-clockwise rotation.

Now let us consider the dynamics of a solid particle moving through the eddy. Returning to the inertial frame, if the eddy is rotating clockwise, $v_e < 0$, then the material that is further from the star is orbiting somewhat more slowly, while the material that is closer to the star is orbiting somewhat more rapidly. This means that the material farther from the star will have a smaller velocity difference with the sub-Keplerian solids, while the material that is closer to the star will have a somewhat larger velocity difference. The drag force is therefore smaller on the far side of the eddy, and larger on the near side. The net effect is that, as solids drift from large radii inward and encounter the eddy, their rate of drift slows down, and they tend to pile up at the location of the eddy. This is a potential mechanism to raise the local ratio of solids to gas, and thus to set of gravitational instability.

The final step in this argument is to have something that provides a pressure jump and thus can produce clockwise eddies. There are a number of possible mechanisms, including a build-up of gas at the edge of a dead zone where MRI shuts off, or simply the turbulence driven by the MRI itself. Whether this actually happens in practice is still an unsolved problem, but the mechanism is at least potentially viable.

Another possible mechanism to concentrate particles is known as the streaming instability. We will not derive this rigorously, but we can describe it qualitatively. Streaming instability operates as follows: suppose that, in some region of the disk, for whatever reason, the local density of solids relative to gas is slightly enhanced. Because we are in a mid-plane layer that is at least partly sedimented, the inertia of the solids is non-negligible. Thus while we have focused on the drag force exerted by the gas on the solids, the corresponding force on gas is not entirely negligible. This force tries to make the gas rotate faster, and thus closer to Keplerian. This in turn reduces the difference in gas and solid velocities.

Now consider the implications of this: where the solid to gas ratio is enhanced, the solids force the gas to rotate closer to their velocity, which in turn reduces the drag force and thus the inward drift speed. Thus if solid particles are drifting inward, when the encounter a region of enhanced solid density, they will slow down and linger in that region. This constitutes an instability, because the slowing down of the drift enhances the solid density even further, potentially leading to a runaway instead in the gas to solid ratio. If this mechanism is able to increase the ratio enough, gravitational instability will take over and produce planetesimals.

\problemset

\begin{enumerate}

\item {\bf HII Region Trapping.}\\
Consider a star of radius $R_*$ and mass $M_*$ with ionizing luminosity $S$ photons s$^{-1}$ at the center of a molecular cloud. For the purposes of this problem, assume that the ionized gas has constant sound speed $c_i=10$ km s$^{-1}$ and case B recombination coefficient $\alphab=2.6\times 10^{-13}$ cm$^{-3}$ s$^{-1}$.
\begin{enumerate}
\item Suppose the cloud is accreting onto the star at a constant rate $\dot{M}_*$. The incoming gas arrives at the free-fall velocity, and the accretion flow is spherical. Compute the equilibrium radius $r_i$ of the ionized region, and show that there is a critical value of $\dot{M}_*$ below which $r_i \gg R_*$. Estimate this value numerically for $M_*=30$ $\msun$ and $S=10^{49}$ s$^{-1}$. How does this compare to typical accretion rates for massive stars?
\item The H~\textsc{ii} region will remain trapped by the accretion flow as long as the ionized gas sound speed is less than the escape velocity at the edge of the ionized region. What accretion rate is required to guarantee this? Again, estimate this numerically for the values given above.\\
\end{enumerate}

\item \textbf{The Transition to Grain-Mediated H$_2$ Formation.}\\
In this problem we will make some rough estimates for how the Universe transitions from H$_2$ formation being mostly by gas-phase processes, as it must in the early Universe where there are no metals, to H$_2$ formation being mostly on grain surfaces. It may be helpful for this problem to review the discussion of H$_2$ formation in Section \ref{ssec:Hchemistry}.
\begin{enumerate}
\item As a first simple example, consider atomic gas with a temperature of $100$ K immersed in a background radiation field equal to that of the Milky Way; this radiation field causes photodetachment of H$^-$ at a rate $\zeta_{\rm pd} = 2.4\times 10^{-7}$ s$^{-1}$ per H$^-$. If all H is neutral and free electrons come only from metals, then the free electron density is $n_e \approx x_{\rm C} n_{\rm H} Z$, where $x_{\rm C} \approx 10^{-4}$ is the gas-phase carbon abundance (the dominant source of free electrons) and $Z$ is the metallicity relative to Solar. Similarly, if the dust grain abundance scales linearly with metallicity, the rate coefficient for H$_2$ formation on grains is $\mathcal{R} = 3\times 10^{-17} Z$ cm$^3$ s$^{-1}$. Show that, under these assumptions, the rate of H$_2$ formation is always dominated by grain surface processes independent of the metallicity or density.
\item Now suppose that the ionization fraction of H is non-negligible, and the photodetachment rate is the same as in part (a). Determine the ionization fraction $x$ at which the rates of H$_2$ formation in the gas phase and on grain surfaces become equal. Your answer should depend on the gas density $n_{\rm H}$, temperature $T$, and metallicity $Z$. Plot the solution for $x$ as a function of metallicity for gas at temperature $T=100$ K and density $n_{\rm H} = 1$, $10$, and $100$ cm$^{-3}$.
\item In part (b), you should have found that, for a given density, there is a critical metallicity above which grain-mediated H$_2$ formation dominates regardless of the ionization fraction (except for the pathological case $x=1$). Solve for this critical metallicity as a function of density, and plot the result for $T = 100$ and $1000$ K.
\end{enumerate}

\item {\bf Disk Dispersal by Photoionization.}\\
Consider a disk around a T Tauri star of mass $M_*$ that produces an ionizing flux $\Phi$ photons s$^{-1}$. The flux ionizes the disk surface and raises the gas temperature to $10^4$ K, leading to a wind leaving the disk surface.
\begin{enumerate}
\item Close to the star the ionized gas remains bound due to the star's gravity. Estimate the gravitational radius $\varpi_g$ at which the ionized gas becomes unbound.
\item Inside $\varpi_g$, we can think of the trapped ionized gas as forming a cloud of characteristic density $n_0$. Assuming this region is roughly in ionization balance, estimate $n_0$.
\item At $\varpi_g$, a wind begins to flow off the disk surface. Because the ionizing photons are attenuated quickly as one moves away from the star, most of the mass loss comes from radii $\sim \varpi_g$. Make a rough estimate for the mass flux in the wind.
\item Evaluate the mass flux numerically for a 1 $\msun$ star with an ionizing flux of $10^{41}$ s$^{-1}$. How long would this take to evaporate a $0.01$ $\msun$ disk around this star? Given the observed lifetimes of T Tauri star disks, are photoionization-induced winds a plausible candidate for the primary disk removal mechanism?
\end{enumerate}

\item {\bf Aerodynamics of Small Solids in a Disk.}\\
Consider a solid sphere of radius $s$ and density $\rho_s$, orbiting a star of mass $M$ at a distance $\varpi$. The sphere is embedded in a protoplanetary disk, whose density and temperature where the particle is orbiting are $\rho_d$ and $T$. The gas pressure in the disk varies with distance from the star as $P\propto \varpi^{-n}$.
\begin{enumerate}
\item Because it is partially supported by gas pressure, gas in the disk orbits at a velocity slightly below the Keplerian velocity. Show that the difference between the gas velocity $v_g$ and the Keplerian velocity $v_K$ is
\begin{displaymath}
\Delta v = v_K - v_g \approx \frac{n c_g^2}{2v_K},
\end{displaymath}
where $c_g$ is the isothermal sound speed of the gas. You may assume that the deviation from Keplerian rotation is small.
\item For a particle so small that the mean free path of gas atoms is $> s$ (which is the case for grains smaller than $\sim 10$ cm), the drag force it experiences as it moves through the gas at a relative velocity $v$ is
\begin{displaymath}
F_D = \frac{4\pi}{3} s^2 \rho_d v c_g.
\end{displaymath}
This is called the Epstein drag law. We define the stopping time $t_s$ as the ratio of the particle's momentum to $F_D$; this is the time required to reduce the particle velocity by one $e$-folding. Compute $t_s$ for a particle governed by Epstein drag.
\item For small particles $t_s$ is much less than orbital period of a particle rotating at the Keplerian speed. In this case drag will force the particle's orbital velocity to match the sub-Keplerian orbital velocity of the gas, and since the particle is not supported by pressure as the disk is, it will drift inward. Estimate the equilibrium drift velocity, and the time required for the particle to drift into the star.
\item Consider a particle of size $s=1$ cm and density $\rho_s = 3$ g cm$^{-3}$ orbiting at $r=1$ AU in a protoplanetary disk of density $\rho_d=10^{-9}$ g cm$^{-3}$, temperature $T=600$ K, and pressure index $n=3$. Verify that this particle is in the regime where $t_s$ is much less than the orbital period, and then numerically evaluate the time required for the particle to drift into the star. How does this compare to the observed time scale of planet formation and disk dissipation?
\end{enumerate}

\end{enumerate}

\appendix

\chapter{Statistical Mechanics of Multi-Level Atoms and Molecules}
\label{app:multilevel_atoms}

This appendix provides a full mathematical treatment of the statistics of multi-level atoms and molecules out of thermodynamic equilibrium, including the effects of a background radiation field. This appendix is intended as a reference rather than a full derivation, and so at several points we assert results without proof. Full demonstrations of these results can be found in standard references such as \citet{rybicki86a}, \citet{shu91a}, or \citet{draine11a}.

\section{Matter-Radiation Interaction}

A general radiation field can be specified in terms of the radiation intensity $I(\nu, \mathbf{n})$ at any point in space $\mathbf{x}$ and time $t$; here $\nu$ is the frequency of radiation and $\mathbf{n}$ is a unit vector specifying the direction of radiation propagation. The intensity specifies the amount of radiant energy per unit area per unit frequency per unit solid angle. An alternative representation of the radiation field, which is more useful when dealing with problems in statistical mechanics, is the photon occupation number, defined by
\begin{equation}
n_\gamma(\nu, \mathbf{n}) = \frac{c^2}{2h\nu^3} I(\nu, \mathbf{n}).
\end{equation}
Physically, the photon occupation number is the number of quanta (photons) in a particular mode, and is dimensionless. In local thermodynamic equilibrium (LTE) at temperature $T$, the radiation intensity in all directions $\mathbf{n}$ is given by the Planck function
\begin{equation}
I(\nu,\mathbf{n}) = B_\nu(T) = \frac{2h\nu^3}{c^2}\frac{1}{e^{h\nu/k_B T} - 1}.
\end{equation}
The equivalent photon occupation number is
\begin{equation}
\label{eq:ngamma_LTE}
n_{\gamma,\rm LTE}(\nu,\mathbf{n}) = \frac{1}{e^{h\nu/k_B T} - 1}.
\end{equation}

For non-relativistic problems, the rates at which photons are emitted or absorbed by atoms undergoing a particular quantum mechanical transition does not depend upon the direction of photon propagation, and thus it is convenient to average over the direction $\mathbf{n}$. We define the directionally-averaged photon occupation number by
\begin{equation}
\langle n_\gamma\rangle(\nu) = \frac{1}{4\pi} \int n_\gamma(\nu, \mathbf{n}) \, d\Omega,
\end{equation}
where the integral is over all directions $\mathbf{n}$.

Now consider a particle of species $X$ with two quantum states that we will denote $u$ and $\ell$, with energies $E_u$ and $E_\ell$, ordered so that $E_u > E_\ell$. The states have degeneracies $g_u$ and $g_\ell$, respectively. Particles in state $u$ can spontaneously emit photons and transition to state $\ell$ with an $e$-folding timescale $A_{u\ell}$. Formally, if $n_u$ is the number density of particles in state $u$, then
\begin{equation}
\left(\frac{dn_u}{dt}\right)_{\rm spon.~emiss.} = -n_u A_{u\ell}.
\end{equation}
Particles in state $\ell$ can also absorb photons at frequencies $\nu$ near $\nu_{u\ell} = (E_u-E_\ell)/h$ and transition to state $u$, and the absorption rate is proportional to $\langle n_\gamma\rangle(\nu_{u\ell})$ and $n_\ell$, where $n_\ell$ is the number density of particles in state $\ell$. Finally, the presence of photons with frequencies near $\nu_{u\ell}$ can cause stimulated emission, whereby particles in state $u$ emit a photon and transition to state $\ell$; again, the rate at which this process occurs must be proportional to both $\langle n_\gamma\rangle(\nu_{u\ell})$ and $n_u$. We write the rates of these two processes as $(dn_u/dt)_{\rm abs.} \propto n_\ell \langle n_\gamma\rangle(\nu_{u\ell})$ and $(dn_u/dt)_{\rm stim.~emiss.} \propto -n_u \langle n_\gamma\rangle(\nu_{u\ell})$. Putting these processes together, the total rate of change of $n_u$ is given by
\begin{eqnarray}
\frac{dn_u}{dt} & = & \left(\frac{dn_u}{dt}\right)_{\rm spon.~emiss.} + \left(\frac{dn_u}{dt}\right)_{\rm stim.~emiss.} + \left(\frac{dn_u}{dt}\right)_{\rm abs.} \\
& = & -n_u A_{u\ell} - C_{u\ell} n_u \langle n_\gamma\rangle(\nu_{u\ell}) + C_{\ell u} n_\ell \langle n_\gamma\rangle(\nu_{u\ell}),
\label{eq:dnudt}
\end{eqnarray}
where the two constants of proportionality $C_{u\ell}$ and $C_{\ell u}$ are to be determined.

Consider a region where the number density of particles is so low that collisions occur negligibly often. However, the particles can still be in LTE with the radiation field. Let $n_\ell$ be the number density of particles in state $\ell$. In LTE the values of $n_u$ and $n_\ell$ must be related by the usual Boltzmann factor, so
\begin{equation}
n_u = \frac{g_u}{g_\ell} e^{-h\nu_{u\ell}/k_B T} n_\ell.
\end{equation}
The directionally-averaged photon occupation number must take on its LTE value
\begin{equation}
\label{eq:ngamma_avg_LTE}
\langle n_\gamma\rangle(\nu) = \frac{1}{e^{h\nu/k_B T} - 1}.
\end{equation}
Inserting these values of $n_u$ and $\langle n_\gamma\rangle$ into equation (\ref{eq:dnudt}), and noting that we must have $dn_u/dt = 0$ for a system in LTE, we have
\begin{equation}
-\frac{g_u}{g_\ell} e^{-h\nu_{u\ell}/k_B T} \left(A_{u\ell} + \frac{C_{u\ell}}{e^{h\nu/k_B T} - 1}\right) + \frac{C_{\ell u} }{e^{h\nu_{u\ell}/k_B T} - 1} = 0.
\end{equation}
For temperatures $T$ such that $h\nu_{u\ell} \ll k_B T$, all the exponential terms approach unity, and thus the two terms proportional to $C_{u\ell}$ and $C_{\ell u}$ are far larger than the term proportional to $A_{u\ell}$. Dropping this term, we immediately see that the equation can be satisfied only if
\begin{equation}
C_{\ell u} = \frac{g_u}{g_\ell} C_{u\ell}.
\end{equation} Conversely, for temperatures $T$ such that $h\nu_{u\ell} \gg k_B T$, the terms in the exponentials are large. We can therefore drop the $-1$ terms in the denominators, and neglect $C_{u\ell}/e^{h\nu_{u\ell}/k_B T}$ in comparison to $A_{u\ell}$. Doing so, we immediately obtain
\begin{equation}
C_{\ell u} = \frac{g_u}{g_\ell} A_{u\ell}.
\end{equation}

Inserting these results into our expressions for the rates of stimulated emission and absorption, we finally have
\begin{eqnarray}
\left(\frac{dn_u}{dt}\right)_{\rm stim.~emiss.} & = & n_u \langle n_\gamma\rangle(\nu_{u\ell}) A_{u\ell} \\
\left(\frac{dn_u}{dt}\right)_{\rm abs.} & = & \frac{g_u}{g_\ell} n_\ell \langle n_\gamma\rangle(\nu_{u\ell}) A_{u\ell}.
\end{eqnarray}

\section{Statistical Equilibrium for Multi-Level Systems}

Now let us consider some species with a series of possible quantum states. We number them $0, 1, 2, \ldots$ in order of increasing energy, so state $0$ is the ground state. We denote the energy and degeneracy of state $i$ as $E_i$ and $g_i$ respectively. We write the energy difference between any two states as $E_{ij} = E_i - E_j$,  the corresponding frequency as $\nu_{ij} = E_{ij}/h$, and we write the Einstein spontaneous emission coefficient for transitions from state $i$ to state $j$ as $A_{ij}$. The species of interest as number density $n$, and we let $n_i$ be the number density of that species in state $i$. Finally, the species of interest can undergo collisions with another species or with itself, and these can cause state transitions as well. We let $n_c$ be the number density of colliders, and we let $k_{ij}$ be the collision rate coefficient connecting any two states, so that the rate of collisionally-induced transitions from state $i$ to state $j$ is given by $n_i n_c k_{ij}$.

Given this setup, we can write out the rates of all processes that induce changes in the number density of any quantum state. Specifically, the rates of collisional transitions out of and into state $i$ are
\begin{eqnarray}
\left(\frac{dn_i}{dt}\right)_{\rm coll.~out} & = & -n_i n_c \sum_j k_{ij} \\
\left(\frac{dn_i}{dt}\right)_{\rm coll.~in} & = & n_c \sum_j n_j k_{ji}.
\end{eqnarray}
Here the first expression is a sum over the rate of collisional transitions from state $i$ to all other states, while the second is a sum over the rate of collisional transitions from all other states to state $i$. By convention we take $k_{ii} = 0$, i.e., we set the rate of collisional transitions from a state to itself to zero. The corresponding rates of transition out of and into state $i$ via spontaneous emission are
\begin{eqnarray}
\left(\frac{dn_i}{dt}\right)_{\rm spon.~emiss.~out} & = & -n_i \sum_j A_{ij} \\
\left(\frac{dn_i}{dt}\right)_{\rm spon.~emiss.~in} & = & \sum_j n_j A_{ji},
\end{eqnarray}
where we adopt the convention that $A_{ij} = 0$ for $i \leq j$, i.e., the spontaneous transition rate from a lower energy state to a higher energy one is zero. Finally, the expressions for stimulated emission- and absorption-induced transitions are
\begin{eqnarray}
\left(\frac{dn_i}{dt}\right)_{\rm stim.~emiss.~out} & = & -n_i \sum_j A_{ij} \langle n_{\gamma,ji}\rangle \\
\left(\frac{dn_i}{dt}\right)_{\rm stim.~emiss.~in} & = & \sum_j n_j A_{ji} \langle n_{\gamma,ji}\rangle \\
\left(\frac{dn_i}{dt}\right)_{\rm abs.~out} & = & -n_i \sum_j \frac{g_j}{g_i} A_{ij} \langle n_{\gamma,ij}\rangle \\
\left(\frac{dn_i}{dt}\right)_{\rm abs.~in} & = & \sum_j \frac{g_i}{g_j} n_j A_{ij} \langle n_{\gamma,ij}\rangle,
\end{eqnarray}
where for convenience we have introduced the shorthand $\langle n_{\gamma,ij}\rangle \equiv \langle n_{\gamma}\rangle(\nu_{ij})$. Note that, per our convention that $A_{ij}$ is non-zero only for $i > j$, the terms in the sums for stimulated emission are non-zero only for states $j > i$, while the terms in the sums for absorption are non-zero only for states $j < i$.

Combining all of the above expressions, we can write out the full rate of change for the number density of particles in each state $i$ as
\begin{eqnarray}
\frac{dn_i}{dt} & = & \sum_j n_j\left[n_c k_{ji} + \left(1+\langle n_{\gamma,ji}\rangle\right) A_{ji}\right] +
 \sum_j n_j \frac{g_i}{g_j} \langle n_{\gamma,ij}\rangle A_{ij}
\nonumber \\
& & \qquad {} - n_i \sum_j \left[n_c k_{ij} + \left(1+\langle n_{\gamma,ij}\rangle\right) A_{ij}\right] 
\nonumber \\
& & \qquad {} -
 n_i \sum_j \frac{g_j}{g_i} \langle n_{\gamma,ji}\rangle A_{ji}.
\label{eq:stat_eq}
\end{eqnarray}
If the system is in statistical equilibrium (but not necessarily LTE), then $dn_i/dt = 0$ for all states $i$. In this case the set of equations (\ref{eq:stat_eq}) represents a set of linear equations to be solved for the unknown number densities $n_i$. With some algebraic manipulation, one can express this system as a matrix equation
\begin{equation}
\mathbf{M} \cdot \mathbf{n} = \mathbf{n},
\end{equation}
where $\mathbf{n} = (n_0, n_1, n_2, \ldots)$ is the vector of number densities, and the matrix $\mathbf{M}$ has elements
\begin{equation}
M_{ij} = \frac{n_c k_{ji} + \left(1 + \langle n_{\gamma,ji}\rangle\right) A_{ji} + \frac{g_i}{g_j} \langle n_{\gamma,ij}\rangle A_{ij}}
{ \sum_\ell \left[n_c k_{i\ell} + \left(1 + \langle n_{\gamma,i\ell}\rangle\right) A_{i\ell} + \frac{g_\ell}{g_i} \langle n_{\gamma,\ell i}\rangle A_{\ell i}\right] }.
\end{equation}
The matrix $\mathbf{M}$ is therefore specified entirely in terms of the known rate coefficients, degeneracies, and radiation fields, and the problem of finding the level populations $\mathbf{n}$ therefore reduces to that of finding the eigenvector of $\mathbf{M}$ that has an eigenvalue of unity.

\section{Critical Densities for Multi-Level Systems}

Chapter \ref{ch:obscold} gives a derivation of the critical density for two-level systems. Armed with the formalism of the previous section, we can generalize this to many-level systems. Consider some level $i$ which has the property that it is populated primarily from below, meaning that transitions into the state via collisional excitation or radiative absorption from lower levels occur much more often than transitions into the state via radiative decays or collisional de-excitations of higher levels, or transitions out of the state to higher levels via collisions or absorptions. In this case, the time rate of change of the level population reduces to
\begin{eqnarray}
\frac{dn_i}{dt} & = & \sum_{j<i} n_j n_c k_{ji} + \sum_{j<i} n_j \frac{g_i}{g_j} \langle n_{\gamma,ij}\rangle A_{ij}
\nonumber \\
& & {} - n_i \sum_{j<i} \left[n_c k_{ij} + \left(1+\langle n_{\gamma,ij}\rangle\right)A_{ij}\right].
\end{eqnarray}
Here the first term describes collisional excitation into state $i$ from lower levels, the second describes the rate of radiative excitation into state $i$ from lower levels, and the final term describes depopulation of state $i$ via collisions, spontaneous emission, and stimulated emission.

If the system is in steady state, then $dn_i/dt = 0$, and we have
\begin{equation}
\label{eq:nsteady}
n_i = \frac{\sum_{j<i} n_j n_c k_{ji} + \sum_{j<i} n_j \frac{g_i}{g_j} \langle n_{\gamma,ij}\rangle A_{ij}}{\sum_{j<i} \left[n_c k_{ij} + \left(1 + \langle n_{\gamma,ij}\rangle\right)A_{ij}\right]}.
\end{equation}
In analogy with the case of a two-level system, we now define the critical density for state $i$ via
\begin{equation}
n_{{\rm crit},i} = \frac{\sum_{j<i} \left(1 + \langle n_{\gamma,ij}\rangle\right) A_{ij}}{\sum_{j<i} k_{ij}},
\end{equation}
i.e., the critical density is the rate of radiative de-excitation divided by the rate of collisional de-excitation. The sole differences between this and the two-level critical density defined by equation (\ref{eq:ncrit}) are that this expression sums over all states into which radiative and collisional de-excitation can occur, and that it contains an extra factor of $ \left(1 + \langle n_{\gamma,ij}\rangle\right)$ in order to properly account for enhancements in the radiative de-excitation rate due to stimulated emission.

Substituting in this definition of $n_{{\rm crit},i}$ into equation (\ref{eq:nsteady}) for the steady state population gives
\begin{equation}
n_i = \left(\frac{n_c}{n_c+n_{{\rm crit},i}}\right) \frac{\sum_{j<i} n_j k_{ji}}{\sum_{j<i} k_{ij}}
+ \left(\frac{n_{{\rm crit},i}}{n_c+n_{{\rm crit},i}}\right) \frac{\sum_{j<i} n_j \frac{g_i}{g_j} \langle n_{\gamma,ij}\rangle A_{ij}}{\sum_{j<i} \left(1+\langle n_{\gamma,ij}\rangle\right) A_{ij}}.
\end{equation}
Examining this expression, one can see that the generalized $n_{{\rm crit},i}$ plays much the same role as $n_{\rm crit}$ for a two-level system. In the limit $n_c \gg n_{{\rm crit},i}$, the first term dominates and the second is negligible. In this case the level population is simply set by collisional effects, and radiative effects become irrelevant. Given the relationships between the various collision rate coefficients $k_{ij}$ (c.f. equation \ref{eq:detailed_balance}), this implies that the level population goes to the usual Boltzmann distribution at the gas temperature $T$. Conversely, if $n_c \ll n_{{\rm crit},i}$, the first term is negligible and the second one dominates, so the level population is determined solely by the radiation field. In the absence of an external radiation field (i.e., $\langle n_{\gamma,ij}\rangle \rightarrow 0$), level $i$ becomes depopulated and thus the excitation is sub-thermal. If the radiation field follows a blackbody distribution (i.e., $\langle n_{\gamma,ij}\rangle$ has the value given by equation \ref{eq:ngamma_avg_LTE}), then one can show that the result is that the levels are populated following a Boltzmann distribution at the radiation field temperature.


\chapter{Solutions to Problem Sets}

\solutionset

\begin{enumerate}

\item \textbf{Molecular Tracers.}

\begin{enumerate}

\item The radiative de-excitation rate is
\begin{displaymath}
\left(\frac{dn_i}{dt}\right)_{\rm spon.\,emiss.} = -n_i \sum_{j<i} A_{ij} .
\end{displaymath}
The collisional de-excitation rate is
\begin{displaymath}
\left(\frac{dn_i}{dt}\right)_{\rm coll.} = -n n_i \sum_{j<i} k_{ij}.
\end{displaymath}

\item Setting the results from the previous part equal and solving, we obtain
\begin{displaymath}
 n_i \sum_{j<i} A_{ij} = n_{\rm crit} n_i \sum_{j<i} k_{ij} 
 \qquad \Longrightarrow \qquad
 n_{\rm crit} = \frac{\sum_{j<i} A_{ij}}{\sum_{j<i} k_{ij}}.
 \end{displaymath}
 
\item
Using numbers taken from the \href{http://www.strw.leidenuniv.nl/~moldata}{LAMBDA} website\footnote{Collision rates are
sufficiently uncertain that the rate coefficients listed in the
database are periodically updated as new calculations or experiments
are performed. The numerical values given here were computed using
collision rate coefficients retrieved on 10 August 2016.} for the $A_{ij}$ and $k_{ij}$ values, we have
\begin{center}
\begin{tabular}{c|c}
Line & $n_{\rm crit}$ [cm$^{-3}$] \\ \hline
CO($J=1\rightarrow 0$) & $2.2\times 10^3$ \\
CO($J=3\rightarrow 2$) & $1.9 \times 10^4$ \\
CO($J=5\rightarrow 4$) & $7.8\times 10^4$ \\
HCN($J=1\rightarrow 0$) & $1.0\times 10^6$
\end{tabular}
\end{center}

\item 
The fraction of the mass above some specified density $\rho_c$ can be obtained by integrating the PDF for mass:
\begin{equation}
f_M(\rho > \rho_0) = \frac{\int_{s_c}^\infty p_M(s) \, ds}{\int_{-\infty}^{\infty} p_M(s) \, ds}
\end{equation}
where $s_c = \ln(\rho_c/\overline{\rho})$ and the mass PDF is
\begin{equation}
p_M = \frac{1}{\sqrt{2\pi \sigma_s^2}} \exp\left[-\frac{(s+s_0)^2}{2\sigma_s^2}\right]
\end{equation}
with $s_0 = -\sigma_s^2/2$. Using the critical densities obtained in the previous part to compute $s_c$, and then to evaluate the integral, we obtain
\begin{center}
\begin{tabular}{c|c|c}
Line & $s_c$ & $f_M(n>n_{\rm crit})$ \\ \hline
CO($J=1\rightarrow 0$) & $3.1$ & 0.39 \\
CO($J=3\rightarrow 2$) & $5.2$ & 0.11 \\
CO($J=5\rightarrow 4$) & $6.6$ & 0.032 \\
HCN($J=1\rightarrow 0$) & $9.2$ & $0.0013$
\end{tabular}
\end{center}
It appears that CO($J=1\rightarrow 0$) and (to some extent) CO($J=3\rightarrow 2$) are good tracers of the bulk of the mass, while CO($J=5\rightarrow 4$) and HCN($J=1\rightarrow 0$) are better tracers of the denser parts of the cloud.

\end{enumerate}

\item \textbf{Inferring Star Formation Rates in the Infrared.}

\begin{enumerate}

\item 
This problem can be done by using the default parameters with \href{http://www.stsci.edu/~science/starburst99}{starburst99} and writing out the bolometric luminosity on a logarithmic grid from $0.1$ Myr to $1$ Gyr, for continuous star formation at a rate of 1 $\msun$ yr$^{-1}$. Taking the output luminosities, the results are
\begin{eqnarray*}
\mbox{SFR}[\msun\mbox{ yr}^{-1}] & = & 4.3\times 10^{-44} L_{\rm tot}[\mbox{erg s}^{-1}]\qquad\mbox{(10 Myr)} \\
\mbox{SFR}[\msun\mbox{ yr}^{-1}] & = & 2.9\times 10^{-44} L_{\rm tot}[\mbox{erg s}^{-1}]\qquad\mbox{(100 Myr)} \\
\mbox{SFR}[\msun\mbox{ yr}^{-1}] & = & 2.2\times 10^{-44} L_{\rm tot}[\mbox{erg s}^{-1}]\qquad\mbox{(1 Gyr)}.
\end{eqnarray*}
In comparison, the corresponding coefficient given by Kennicutt (1998) is $3.9\times 10^{-44}$, the same to within a factor of 2.

\item
The plot of the starburst99 output is shown in Figure \ref{fig:hw1sol1}. The solid line is the output with a normal IMF, and the dashed line is the output with a top-heavy IMF, for part (c).

\begin{figure}
\includegraphics[width=\linewidth]{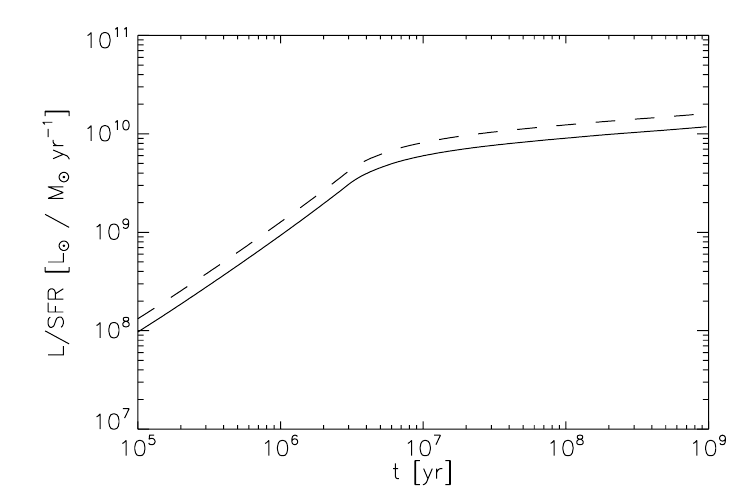}
\caption[Solution to problem set~\thesolutionset, problem~\theenumi\theenumii]{
\label{fig:hw1sol1}
Luminosity normalized by star formation rate for a normal IMF (solid line) and a top-heavy IMF (dashed line).
}
\end{figure}

\item To generate this IMF, I told starbust99 to use a 1 section IMF with a slope of $-2.3$ running from 0.5 to 100 $\msun$. At equal ages, the numbers change to
\begin{eqnarray*}
\mbox{SFR}[\msun\mbox{ yr}^{-1}] & = & 3.2\times 10^{-44} L_{\rm tot}[\mbox{erg s}^{-1}]\qquad\mbox{(10 Myr)} \\
\mbox{SFR}[\msun\mbox{ yr}^{-1}] & = & 2.1\times 10^{-44} L_{\rm tot}[\mbox{erg s}^{-1}]\qquad\mbox{(100 Myr)} \\
\mbox{SFR}[\msun\mbox{ yr}^{-1}] & = & 1.6\times 10^{-44} L_{\rm tot}[\mbox{erg s}^{-1}]\qquad\mbox{(1 Gyr)}.
\end{eqnarray*}
These are a few tens of percent lower, because the IMF contains fewer low mass stars that contribute little light. The effect is mild, but that is partly because the change in IMF is mild. These results do suggest that the IR to SFR conversion does depend on the IMF.

\end{enumerate}

\end{enumerate}

\solutionset

\begin{enumerate}

\item \textbf{The Bonnor-Ebert Sphere.}

\begin{enumerate}

\item For a uniform-density sphere with constant surface pressure, the terms that appear in the virial theorem are
\begin{eqnarray*}
\mathcal{W} & = & -\frac{3}{5} \frac{GM^2}{R} \\
\mathcal{T} & = & \frac{3}{2} M c_s^2 \\
\mathcal{T}_S & = & 4 \pi R^3 P_s.
\end{eqnarray*}
All other terms are zero. Virial equilibrium requires
\begin{eqnarray*}
0 & = & 2 (\mathcal{T} -\mathcal{T}_S) + \mathcal{W} \\
& = & 3 M c_s^2 - 8 \pi R^3 P_s - \frac{3}{5} \frac{G M^2}{R} \\
P_s & = & \frac{3 M c_s^2}{8\pi} \left[\frac{1}{R^3} - \left(\frac{G M}{5 c_s^2}\right)\frac{1}{R^4}\right].
\end{eqnarray*}
Notice that the first, positive, term in brackets dominates at large $R$, while the second, negative, one dominates at small $R$. Thus there must be a maximum at some intermediate value of $R$. To derive this maximum, we can take the derivative with respect to $R$. This gives
\begin{displaymath}
\frac{dP_s}{dR} = \frac{3 M c_s^2}{8\pi} \left[-\frac{3}{R^4} + \left(\frac{4 G M}{5 c_s^2}\right)\frac{1}{R^5}\right].
\end{displaymath}
Setting this equal to zero and solving, we find that the maximum occurs at
\begin{displaymath}
R = \frac{4 G M}{15 c_s^2}.
\end{displaymath}
Plugging this in for $P_s$, we obtain
\begin{displaymath}
P_s = \frac{10125}{2048\pi} \frac{c_s^8}{G^3 M^2} \approx 1.57 \frac{c_s^8}{G^3 M^2}.
\end{displaymath}

\item Since the gas is isothermal, we can substitute for $P$ to obtain
\begin{displaymath}
-c_s^2 \frac{1}{\rho} \frac{d}{dr}\rho = \frac{d}{dr}\phi
\end{displaymath}
The left-hand side can be re-written as
\begin{displaymath}
-c_s^2 \frac{d}{dr} \ln \rho = \frac{d}{dr}\phi,
\end{displaymath}
which makes the equation trivial to integrate:
\begin{displaymath}
-c_s^2 \ln \rho = \phi + \mathrm{const}.
\end{displaymath}
Fixing the constant of integration by the requirement that $\rho = \rho_c$ and $\phi=0$ at the origin, we have
\begin{displaymath}
\rho = \rho_c e^{-\phi/c_s^2}
\end{displaymath}

\item Substituting into the Poisson equation, we have
\begin{displaymath}
\frac{1}{r^2}\frac{d}{dr}\left(r^2 \frac{d\phi}{dr}\right) = 4 \pi G \rho_c e^{-\phi/c_s^2}
\end{displaymath}
Now define $\psi \equiv \phi/c_s^2$, giving
\begin{displaymath}
\frac{1}{r^2}\frac{d}{dr}\left(r^2 \frac{d\phi}{dr}\right) = \frac{4\pi G \rho_c}{c_s^2} e^{-\psi}.
\end{displaymath}
Finally, let
\begin{displaymath}
\xi \equiv \frac{r}{r_0},
\end{displaymath}
where 
\begin{displaymath}
r_0 = \frac{c_s}{\sqrt{4\pi G\rho_c}}.
\end{displaymath}
Substituting this in, we arrive at the desired equation:
\begin{displaymath}
\frac{1}{\xi^2}\frac{d}{d\xi}\left(\xi^2 \frac{d\psi}{d\xi}\right) = e^{-\psi}.
\end{displaymath}

\item For the purposes of numerical integration, it is most convenient to recast the problem as two first-order ODEs rather than a single second-order one. Let $\psi' = d\psi/d\xi$, and the system becomes
\begin{eqnarray*}
\frac{d\psi}{d\xi} & = & \psi' \\
\frac{d\psi'}{d\xi} & = & -2\frac{\psi'}{\xi} + e^{-\psi}.
\end{eqnarray*}
The only tricky part of the numerical solution to this system is the presence of a singularity in the equations at $\xi = 0$, which will cause numerical methods to choke. In this particular case it's not terrible to avoid this problem simply by starting the integration from a small but non-zero value of $\xi$ and setting $\psi = \psi' = 0$ at this point. However, this approach can run into problems for some equations, where the solution depends critically on the ratio of $\psi$ to $\psi'$ near the singular point. A better, more general method is to use a series expansion to solve the equation near the singularity, and then using that series expansion to numerically integrate starting from a small but non-zero value of $\xi$. Let $\psi = a_2 \xi^2 + a_3 \xi^3 + a_4 \xi^4 + \ldots$ in the vicinity of $\xi = 0$. Note that we know there is no constant or linear term due to the boundary conditions $\psi(0) = \psi'(0) = 0$. Substituting into the ODE and expanding, we obtain
\begin{displaymath}
6 a_2 + 12 a_3 \xi + O(\xi^2) = 1 + O(\xi^2).
\end{displaymath}
Since the equation must balance, we learn that $a_2 = 1/6$ and $a_3 = 0$, so the behavior of $\psi$ near $\xi = 0$ is $\psi = \xi^2/6 + O(\xi^4)$. Armed with this information, it is straightforward to integrate the equation numerically. Below is a simple Python code that can solve the problem:
\begin{verbatim}
import numpy as np
import matplotlib.pyplot as plt
from scipy.integrate import odeint

# definition of the derivatives
def derivs(y, x):
    return( [y[1], -2*y[1]/x+exp(-y[0])] )

# starting points
x0 = 1e-4
y0 = [x0**2/6, x0/3]

# solve the ode
x = np.linspace(x0, 8, 200)
ysol = odeint(derivs, y0, x)

# plot psi and exp(-psi) vs. x
plt.plot(x, ysol[:,0], lw=2, label=r'$\psi$')
plt.plot(x, np.exp(-ysol[:,0]), lw=2,
         label=r'$\rho/\rho_c$')
plt.legend(loc='upper left')
plt.xlabel(r'$\xi$')
\end{verbatim}
\begin{marginfigure}
\includegraphics[width=\linewidth]{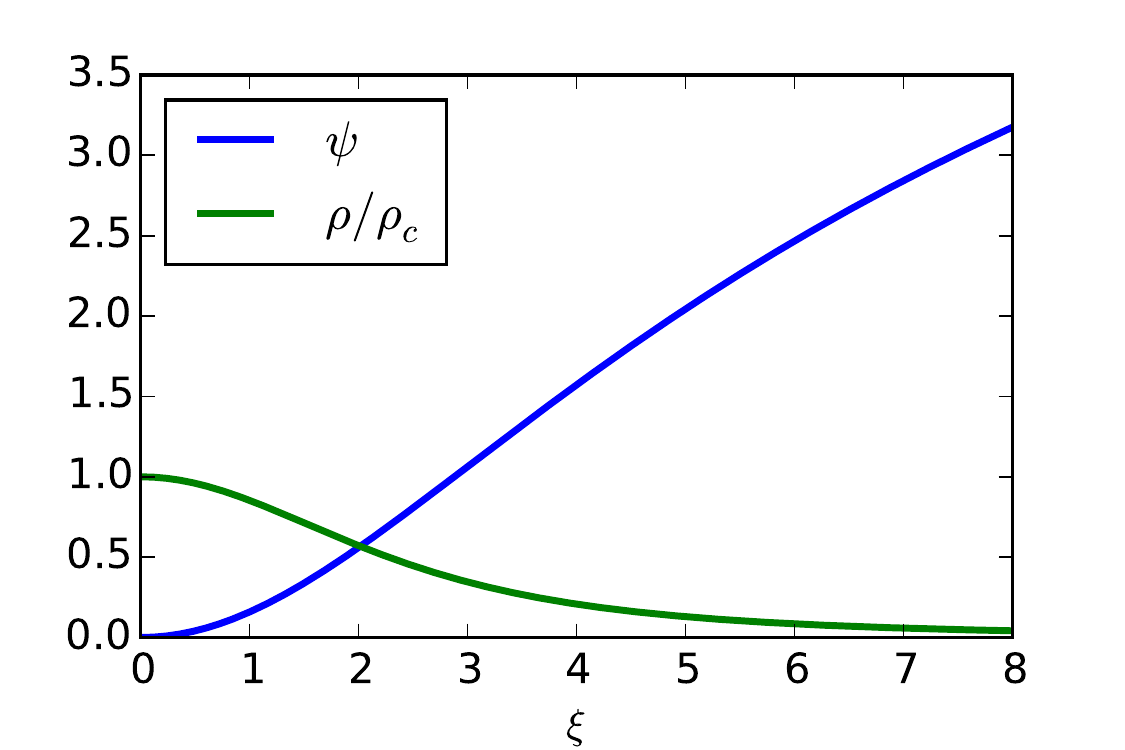}
\caption[Solution to problem set~\thesolutionset, problem~\theenumi\theenumii]{
\label{fig:hw2sol1}
Dimensionless potential $\psi$ and density $\rho/\rho_c = e^{-\psi}$ found by solving the isothermal Lane-Emden equation.
}
\end{marginfigure}
The output produced by this code is shown in Figure \ref{fig:hw2sol1}.

\item As a first step, we can substitute in the dimensionless variables from the numerical solution:
\begin{displaymath}
M = 4\pi \int_0^R \rho r^2 \, dr = 4\pi r_0^3 \rho_c \int_0^{\xi_s} e^{-\psi} \xi^2 \, d\xi
\end{displaymath}
The integral can be evaluated by plugging using the isothermal Lane-Embden equation and then using the fundamental theorem of calculus:
\begin{displaymath}
\int_0^{\xi_s} e^{-\psi} \xi^2 \, d\xi = \int_0^{\xi_s} \frac{d}{d\xi}\left(\xi^2 \frac{d\psi}{d\xi}\right) \, d\xi
= \left(\xi^2 \frac{d\psi}{d\xi}\right)_{\xi_s}.
\end{displaymath}
Note that the term coming from the endpoint at $\xi=0$ vanishes because $\xi$ and $d\psi/d\xi$ are both $0$ there. The remainder of the problem is just a matter of substitution and manipulation:
\begin{eqnarray*}
M & = & 4\pi r_0^3 \rho_c \left(\xi^2 \frac{d\psi}{d\xi}\right)_{\xi_s} \\
& = & 4\pi \frac{c_s^3}{(4\pi G\rho_c)^{3/2}} \rho_c \left(\xi^2 \frac{d\psi}{d\xi}\right)_{\xi_s} \\
& = & \frac{c_s^4}{\sqrt{4\pi G^3 \rho_c P_s/\rho_s}} \left(\xi^2 \frac{d\psi}{d\xi}\right)_{\xi_s} \\
& = & \frac{c_s^4}{\sqrt{4\pi G^3 P_s}}  \left(e^{-\psi/2}\xi^2 \frac{d\psi}{d\xi}\right)_{\xi_s}.
\end{eqnarray*}

\item Using the numerical results from above, and recalling that $\rho_c/\rho = e^{\psi}$, this is a fairly simple addition to the program. To get a bit more range on the density contrast, it is helpful to extend the range of $\xi$ a bit further than for the previous problem. A simple solution, to be executed after the previous code, is
\begin{verbatim}
# solve the ode on a slightly larger grid
x = np.linspace(x0, 1e3, 500000)
ysol = odeint(derivs, y0, x)

# Get density constrast and m
contrast = np.exp(ysol[:,0])
m = (x**2 * np.exp(-ysol[:,0]/2) * ysol[:,1]) \
    / np.sqrt(4.0*np.pi)

# Plot
plt.clf()
plt.plot(contrast, m, lw=2)
plt.xscale('log')
plt.xlabel(r'$\rho_c/\rho_s$')
plt.ylabel('m')
plt.xlim([1,1e4])
\end{verbatim}
\begin{marginfigure}
\includegraphics[width=\linewidth]{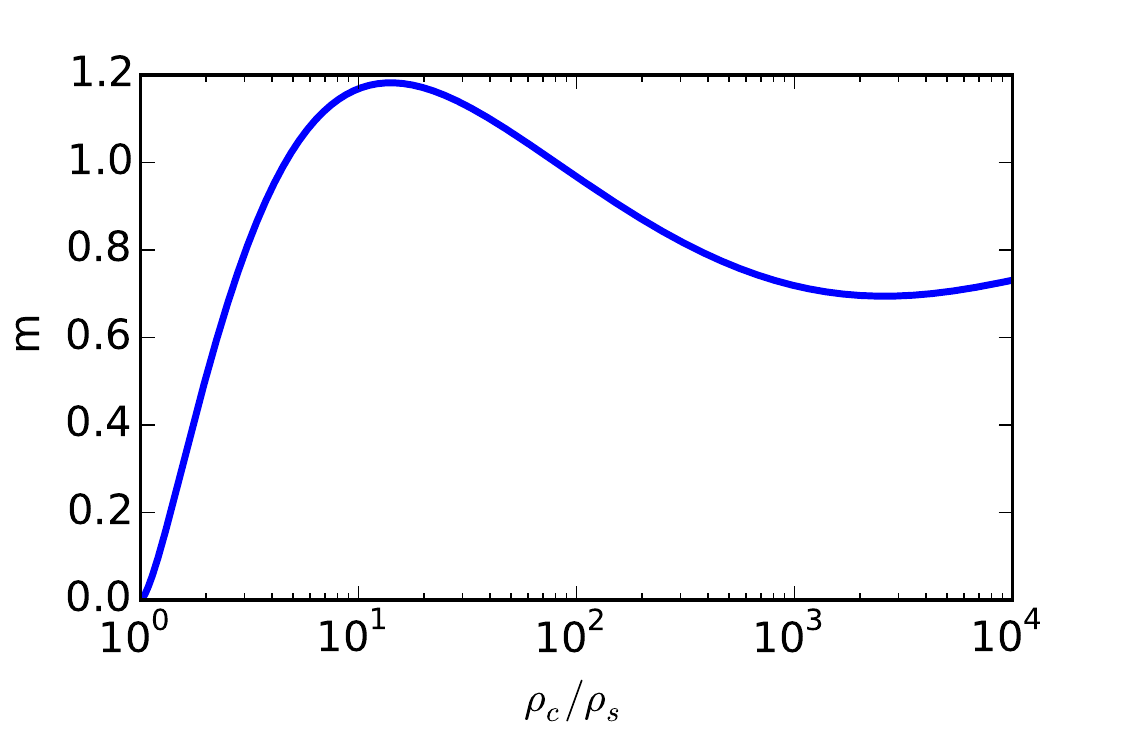}
\caption[Solution to problem set~\thesolutionset, problem~\theenumi\theenumii]{
\label{fig:hw2sol2}
Dimensionless mass $m$ versus dimensionless density contrast $\rho_c/\rho_s$ found by solving the isothermal Lane-Emden equation.
}
\end{marginfigure}
The output produced by this code is shown in Figure \ref{fig:hw2sol2}. The maximum value of $m$ (obtained via \verb=np.amax(m)=) is 1.18. The maximum is at (found via \verb=contrast[np.argmax(m)-1]=) $\rho_c/\rho_s = 14.0$.

\item The dimensionless and dimensional mass are related by
\begin{displaymath}
m = \frac{P_s^{1/2} G^{3/2} M}{c_s^4},
\end{displaymath}
so the maximum surface pressure is
\begin{displaymath}
P_{s,\rm max} = m_{\rm max}^2 \frac{c_s^8}{G^3 M^2},
\end{displaymath}
where $m_{\rm max}$ is the maximum value of $m$ produced by the numerical solution in the previous part. Plugging this in, we have
\begin{displaymath}
P_{s,\rm max} \approx 1.40 \frac{c_s^8}{G^3 M^2},
\end{displaymath}
which is only slightly different than the result we got for the uniform sphere value in part (a) -- a coefficient of 1.40 instead of 1.57.

\item The maximum mass is
\begin{displaymath}
M_{\rm BE} = m_{\rm max} \frac{c_s^4}{P_s^{1/2} G^{3/2}} \approx 1.18  \frac{c_s^4}{P_s^{1/2} G^{3/2}}
\end{displaymath}
At $T = 10$ K, a gas with $\mu=3.9\times 10^{-24}$ g has a sound speed $c_s = 0.19$ km s$^{-1}$. Plugging this in, together with the given value of $P_s$, we find $M_{\rm BE} = 0.67$ $\msun$.

\end{enumerate}

\item \textbf{Driving Turbulence with Protostellar Outflows.}

\begin{enumerate}

\item The escape speed at the stellar surface, and thus the launch velocity of the wind, is $v_w = \sqrt{2G M_*(t)/R_*}$, where $M_*(t)$ is the star's instantaneous mass. The momentum flux associated with the wind is therefore $\dot{p}_w = f \dot{M}_d v_w$. The accretion rate onto the star is $\dot{M}_* = (1-f) \dot{M}_d$. Thus at a time $t$ after the star has started accreting, we have $M_*(t) = (1-f) \dot{M}_d t$, and
\begin{displaymath}
\dot{p}_w = f (1-f)^{1/2} \dot{M}_d^{3/2} \left(\frac{2G}{R_*}\right)^{1/2} t^{1/2}.
\end{displaymath}
The time required to accrete up to the star's final mass is $t_f = M_*/\dot{M}_* = (1-f)^{-1} M_*/\dot{M}_d$, where $M_*$ is the final mass. To obtain the wind momentum per unit stellar mass, we must integrate $\dot{p}_w$ over the full time it takes to build up the star, then divide by the star's mass. Thus we have
\begin{eqnarray*}
\langle p_w\rangle & = & \frac{1}{M_*} \int_0^{(1-f)^{-1} M_*/\dot{M}_d} f (1-f)^{1/2} \dot{M}_d^{3/2} \left(\frac{2G}{R_*}\right)^{1/2} t^{1/2}\, dt \\
& = & \frac{2}{3} \frac{f}{1-f} \sqrt{\frac{2 G M_*}{R_*}}.
\end{eqnarray*}
Evaluating numerically for the given values of $f$, $M_*$, and $R_*$ gives $\langle p_w\rangle = 19$ km s$^{-1}$ $\msun^{-1}$.

\item Each outflow carries momentum $\langle p_w\rangle M_*$, and thus when it decelerates to terminal velocity $\sigma$ the mass it has swept-up must be $M_w = (\langle p_w\rangle/\sigma) M_*$. The associated kinetic energy of a single outflow is 
\begin{displaymath}
\mathcal{T}_w = \frac{1}{2} M_w \sigma^2 = \frac{1}{2} M_* \langle p_w\rangle \sigma.
\end{displaymath}
If the total star formation rate is $\dot{M}_{\rm cluster}$, then the rate at which new stars form is $\dot{M}_{\rm cluster}/M_*$. The rate of kinetic energy injection is therefore
\begin{eqnarray*}
\dot{\mathcal{T}} & = & \frac{\dot{M}_{\rm cluster}}{M_*} \mathcal{T}_w \\
& = & \frac{1}{2} \dot{M}_{\rm cluster} \langle p_w\rangle \sigma \\
& = & \frac{1}{3}\left(\frac{f}{1-f}\right) \dot{M}_{\rm cluster} \sigma \sqrt{\frac{2 G M_*}{R_*}}.
\end{eqnarray*}

\item The decay time is $L/\sigma$, to the decay rate must be the cloud kinetic energy $(3/2) M \sigma^2$ divided by this time. Thus
\begin{displaymath}
\dot{\mathcal{T}}_{\rm dec} = -\frac{3}{2} \frac{M \sigma^3}{L}.
\end{displaymath}
If we now set $\dot{\mathcal{T}}_w = -\dot{\mathcal{T}}_{\rm dec}$, we can solve for $\dot{M}_{\rm cluster}$. Doing so gives
\begin{displaymath}
\dot{M}_{\rm cluster} = \frac{9}{2}\left(\frac{1-f}{f}\right) \sqrt{\frac{R_*}{2 G M_*}} \frac{\sigma^2}{L} M.
\end{displaymath}
Using the Larson relations to evaluate this, note that $\sigma^2/L = \sigma_1^2/\mbox{pc} \equiv a_c = 3.2\times 10^{-9}$ cm s$^{-1}$ is constant, and we are left with
\begin{displaymath}
\dot{M}_{\rm cluster} = \frac{9}{2}\left(\frac{1-f}{f}\right) \sqrt{\frac{R_*}{2 G M_*}} a_c M_1 \left(\frac{L}{\mbox{pc}}\right)^2.
\end{displaymath}
Evaluating numerically for the given values of $L$ produces the results below:
\begin{center}
\begin{tabular}{l|ccc}
& $L=1$ pc & $L = 10$ pc & $L = 100$ pc \\ \hline
$\dot{M}_{\rm cluster}$ [$\msun$ yr$^{-1}$] & $1.6\times 10^{-5}$ & $1.6\times 10^{-3}$ & $1.6\times 10^{-1}$
\end{tabular}
\end{center}

\item The mass converted into stars in 1 free-fall time is $\dot{M}_{\rm cluster} t_{\rm ff}$, so the quantity we want to compute is
\begin{displaymath}
f = \frac{\dot{M}_{\rm cluster}}{M} t_{\rm ff} \equiv \frac{t_{\rm ff}}{t_*},
\end{displaymath}
where $t_*$ is the star formation timescale. From the previous part, we have
\begin{displaymath}
t_*^{-1} = \frac{\dot{M}_{\rm cluster}}{M} =  \frac{9}{2}\left(\frac{1-f}{f}\right) \sqrt{\frac{R_*}{2 G M_*}} a_c = 0.16\mbox{ Myr}^{-1}.
\end{displaymath}
The free-fall time is
\begin{eqnarray*}
t_{\rm ff} & = & \sqrt{\frac{3\pi}{32 G \rho}} = \sqrt{\frac{3\pi L^3}{32 G M}} = \sqrt{\frac{3\pi L_1^3}{32 G M_1}} \left(\frac{L}{L_1}\right)^{1/2} \\
& = & 0.81 \left(\frac{L}{L_1}\right)^{1/2}\mbox{ Myr},
\end{eqnarray*}
where $L_1 = 1$ pc. Thus we have
\begin{displaymath}
f = \frac{t_{\rm ff}}{t_*} = 0.13 \left(\frac{L}{L_1}\right)^{1/2}.
\end{displaymath}
Evaluating for $L = 1$, $10$, and $100$ pc, we get $f = 0.13$, $0.42$, and $1.3$, respectively. We therefore conclude that protostellar outflows may be a significant factor in the driving the turbulence on $\sim 1$ pc scales, and cannot be ignored there. However, they become increasingly less effective at larger size scales, and can probably be neglected at the scales of entire GMCs, $\sim 10-100$ pc.

\end{enumerate}

\item \textbf{Magnetic Support of Clouds.}

\begin{enumerate}

\item The virial ratio is (omitting constant factors of order unity)
\begin{displaymath}
\alpha_{\rm vir}\sim \frac{\sigma^2 R}{G M}.
\end{displaymath}
The Alfv\'en Mach number is the ratio of the velocity dispersion to the Alfv\'en speed
\begin{displaymath}
v_A \sim \frac{B}{\sqrt{\rho}} \sim \frac{B R^{3/2}}{M^{1/2}}.
\end{displaymath}
Thus
\begin{displaymath}
\mathcal{M}_A \sim \frac{\sigma M^{1/2}}{B R^{3/2}}.
\end{displaymath}
To rewrite this in terms of $M_\Phi$, we can eliminate $B$ from this expression by writing
\begin{displaymath}
B \sim \frac{M_\Phi G^{1/2}}{R^2},
\end{displaymath}
giving
\begin{displaymath}
\mathcal{M}_A \sim \frac{\sigma}{M_\Phi} \sqrt{\frac{MR}{G}}
\end{displaymath}
Similarly, we can eliminate $\sigma$ using the definition of the virial ratio:
\begin{displaymath}
\sigma \sim \sqrt{\alpha_{\rm vir}\frac{G M}{R}},
\end{displaymath}
and substituting this in gives
\begin{displaymath}
\mathcal{M}_A \sim \alpha_{\rm vir}^{1/2} \mu_\Phi.
\end{displaymath}

\item The expression derived in part (a) does indeed show that, if any of two of the three quantities $\mathcal{M}_A$, $\alpha_{\rm vir}$, and $\mu_\Phi$ are of order unity, the third one must be as well. Intuitively, this is because the various quantities are measures of energy ratios. Roughly speaking, $\mathcal{M}_A^2$ measures the ratio of kinetic (including thermal) energy to magnetic energy; $\alpha_{\rm vir}$ measures the ratio of kinetic to gravitational energy; and $\mu_\Phi^2$ represents the ratio of gravitational to magnetic energy. If any two of these are of order unity, then this implies that gravitational, kinetic, and magnetic energies are all of the same order. However, this in turn implies that the third dimensionless ratio should also be of order unity as well. For example, if $\mathcal{M}_A \sim \alpha_{\rm vir} \sim 1$, then this implies that kinetic energy is comparable to magnetic energy, and kinetic energy is also comparable to gravitational energy. In turn, this means that gravitational and mantic energy are comparable, in which case $\mu_\Phi \sim 1$.

\item If we have a cloud that is supported, it must have $\alpha_{\rm vir}\sim 1$. However, if the cloud is turbulent then it will naturally also go to $\mathcal{M}_A \sim 1$. This means that we are likely to measure $\mu_{\Phi}\sim 1$ even if the cloud is magnetically supercritical and not supported by its magnetic field. We would only ever expect to get $\mu_{\Phi} \gg 1$, indicating a lack of magnetic support, if the cloud were either non-virialized ($\alpha_{\rm vir} \gg 1$ or $\ll 1$) or non-turbulent.

\end{enumerate}

\end{enumerate}

\solutionset

\begin{enumerate}

\item \textbf{Toomre Instability.}

\begin{enumerate}

\item Substituting in the perturbed terms for $\Sigma$, $\vecv$, and $\phi$, the linearized equation of mass conservation is
\begin{eqnarray*}
\frac{\partial}{\partial t}\left(\Sigma_0 + \epsilon \Sigma_1\right) + 
\nabla \cdot \left[\left(\Sigma_0 + \epsilon \Sigma_1\right) \left(\vecv_0 + \epsilon \vecv_1\right)\right] & = & 0 \\
\frac{\partial}{\partial t} \Sigma_1 + \Sigma_0 \nabla \cdot  \vecv_1 + \nabla \cdot \left(\Sigma_1 \vecv_0\right) & = & 0.
\end{eqnarray*}
In going from the first line to the second, we dropped terms of order $\epsilon^2$, we used the fact that $\Sigma_0$ is constant in time to drop the term $\partial\Sigma_0/\partial t$, and we used the fact that it is constant in space (since the unperturbed state is uniform) to take the $\Sigma_0$ factor out of the divergence. Note that $\vecv_0$ and $\Sigma_1$ are not constant in space, so they cannot be taken out of the divergence.

The linearized momentum equation is
\begin{eqnarray*}
\lefteqn{\frac{\partial}{\partial t}\left(\vecv_0 + \epsilon \vecv_1\right) + \left(\vecv_0 + \epsilon \vecv_1\right)\cdot \nabla\left(\vecv_0 + \epsilon \vecv_1\right)} \qquad \\
 & = & -\frac{\nabla (\Sigma_0+\epsilon \Sigma_1)}{\Sigma_0+\epsilon\Sigma_1} c_s^2 - \nabla (\phi_0 + \epsilon\phi_1)
\nonumber \\
& & {}
- 2\veco \times \left(\vecv_0 + \epsilon \vecv_1\right) + \Omega^2 (x \ehat_x + y \ehat_y). \\
\end{eqnarray*}
To simplify this, we recall that, since the equilibrium is an exact solution, it must be the case that
\begin{displaymath}
\frac{\partial}{\partial t} \vecv_0 + \vecv_0 \cdot \nabla \vecv_0 = -c_s^2 \frac{\nabla\Sigma_0}{\Sigma_0} - \nabla \phi_0 - 2\veco\times\vecv_0 + \Omega^2 (x \ehat_x + y \ehat_y),
\end{displaymath}
and we can therefore cancel these terms. Doing so, and dropping terms of order $\epsilon^2$, we are left with
\begin{displaymath}
\frac{\partial}{\partial t} \vecv_1 + \vecv_0 \cdot\nabla \vecv_1 + \vecv_1 \cdot \nabla \vecv_0
= - \frac{\nabla \Sigma_1}{\Sigma_0} c_s^2 - \nabla \phi_1- 2  \veco\times\vecv_1.
\end{displaymath}

Finally, the linearized Poisson equation is
\begin{eqnarray*}
\nabla^2 (\phi_0 + \epsilon \phi_1) & = & 4\pi G (\Sigma_0 + \epsilon \Sigma_1) \delta(z) \\
\nabla^2 \phi_1 & = & 4\pi G \Sigma_1 \delta(z).
\end{eqnarray*}
In deriving the second line we used the fact that the unperturbed state is an exact solution to cancel $\nabla^2 \phi_0$ with $4\pi G \Sigma_0\delta(z)$.

\item First, we plug the Fourier mode trial solutions into the Poisson equation:
\begin{displaymath}
 \phi_a \nabla^2 e^{i(kx-\omega t)-|kz|} = 4\pi G \Sigma_a e^{i(kx-\omega t)} \delta(z).
\end{displaymath}
To eliminate the $\delta(z)$, we now integrate both sides in $z$ over a range $[-\zeta,\zeta]$ and evaluate in the limit $\zeta \rightarrow 0$. This gives
\begin{eqnarray*}
\lefteqn{\phi_a \int_{-\zeta}^\zeta \left(\frac{\partial^2}{\partial x^2} + \frac{\partial^2}{\partial y^2} + \frac{\partial^2}{\partial z^2}\right) e^{i(kx-\omega t)-|kz|} \, dz} \qquad\qquad\qquad
\nonumber \\
& = &  4 \pi G \Sigma_a  e^{i(kx-\omega t)} \int_{-\zeta}^\zeta \delta(z) \, dz \\
& = & 4\pi G \Sigma_a e^{i(kx-\omega t)}.
\end{eqnarray*}
To evaluate the left-hand side, note that the $\partial^2/\partial y^2$ term vanishes because there is no $y$-dependence, and the $\partial^2/\partial x^2$ term will also vanish when we take the limit $\zeta\rightarrow 0$, because the integrand is finite. Only the $\partial^2/\partial z^2$ term will survive. Thus we have
\begin{eqnarray*}
4\pi G \Sigma_a & = & \phi_a \lim_{\zeta\rightarrow 0} \int_{-\zeta}^\zeta \frac{\partial^2}{\partial z^2} e^{-|kz|} \, dz \\
& = & \phi_a \lim_{\zeta\rightarrow 0} \left[\left(\frac{d}{dz} e^{-|kz|}\right)_{z=\zeta} - \left(\frac{d}{dz} e^{-|kz|}\right)_{z=-\zeta}\right] \\
& = & -2\phi_a |k|
\end{eqnarray*}
Thus we have
\begin{displaymath}
\phi_a = -\frac{2\pi G\Sigma_a}{|k|}.
\end{displaymath}

\item As a first step, let us rewrite the terms involving $\vecv_0$ in a more convenient form; this is the Taylor expansion part. Recall that we are in a frame that is co-rotating with the disk, and where $x$ is the distance from the center of our co-rotating reference frame in the radial direction. In the lab frame, the velocity is $\vecv_0' = v_R \ehat_\phi$, and the velocity of the co-rotating reference frame at a distance $r$ from the origin is $\vecv_{\rm rot} = \Omega_0 r \ehat_\phi$. The unperturbed velocity in the rotating frame is the difference between these two, i.e.,
\begin{eqnarray*}
\vecv_0 & = & \vecv_0' - \vecv_{\rm rot} \\
& = & \left(v_R - \Omega_0 r\right) \ehat_y \\
& = & \left[\Omega_0 R - \Omega_0 \left(R + x\right)\right] \ehat_y \\
& = & -\Omega_0 x \ehat_y,
\end{eqnarray*}
where we have used the fact that $\ehat_\phi$ in the lab frame is the same as $\ehat_y$ in our co-rotating frame.

With this result in hand, we can now begin to make substitutions into the perturbed equations. The perturbed equation of mass conservation becomes
\begin{displaymath}
-i\omega \Sigma_a + i k \Sigma_0 v_{ax} = 0.
\end{displaymath}
The momentum equation becomes
\begin{eqnarray*}
\lefteqn{-i\omega \left(v_{ax} \ehat_x + v_{ay} \ehat_y\right) - \Omega_0 v_{ax} \ehat_y}
\qquad\qquad\qquad
\\
& = & -ik\frac{\Sigma_a}{\Sigma_0} c_s^2 \ehat_x - ik\phi_0 \ehat_x
- 2 \veco \times \left(v_{ax} \ehat_x + v_{ay} \ehat_y\right).
\end{eqnarray*}
Since $\veco = \Omega \ehat z$, we can write out the two components of this equation as
\begin{eqnarray*}
-i\omega v_{ax} & = & -ik c_s^2 \frac{\Sigma_a}{\Sigma_0} + i k \frac{2\pi G \Sigma_a}{|k|} + 2\Omega_0 v_{ay} \\
-i\omega v_{ay} & = & -\Omega_0 v_{ax},
\end{eqnarray*}
where we have evaluated the equation at $x=0$ and thus we have $\Omega = \Omega_0$, and in the first equation we have substituted in for $\phi_a$. We now have three equations in the three unknowns $\Sigma_0$, $v_{ax}$, and $v_{ay}$.

\item The easiest way to demonstrate the desired result is to write the system of three equations in standard form:
\begin{eqnarray*}
i k \left(\frac{2\pi G}{|k|}-\frac{c_s^2}{\Sigma_0}\right) \Sigma_a + i\omega v_{ax} + 2\Omega_0 v_{ay} & = & 0 \\
-\Omega_0 v_{ax} + i \omega v_{ay} & = & 0 \\
-i\omega \Sigma_a  + i k \Sigma_0 v_{ax} & = & 0.
\end{eqnarray*}
We can write this system as a matrix equation:
\begin{eqnarray*}
\mathbf{A} & \equiv & \left[
\begin{array}{ccc}
i k \left(\frac{2\pi G}{|k|}-\frac{c_s^2}{\Sigma_0}\right) &
i\omega &
2\Omega_0 \\
0 & -\Omega_0 & i\omega \\
-i\omega & i k \Sigma_0 & 0
\end{array}
\right] \\
\mathbf{A}
\left[
\begin{array}{c}
\Sigma_a \\
v_{ax} \\
v_{ay} 
\end{array}
\right]
& = & \left[
\begin{array}{c}
0 \\
0 \\
0 
\end{array}
\right].
\end{eqnarray*}
This matrix equation has a non-trivial solution if and only if $\mathbf{A}$ is non-invertible, i.e., it has zero determinant. Thus the condition for there to be non-trivial solutions we require
\begin{eqnarray*}
0 & = & \det(\mathbf{A}) \\
& = &
i \omega k^2 \Sigma_0 \left(\frac{2\pi G}{|k|} - \frac{c_s^2}{\Sigma_0}\right) + i \omega^3 - 2 i \omega \Omega_0^2 \\
& = & k^2 \Sigma_0 \left(\frac{2\pi G}{|k|} - \frac{c_s^2}{\Sigma_0}\right) + \omega^2 - 2\Omega_0^2 \\
\omega^2 & = & 2\Omega_0^2 - 2 \pi G\Sigma_0 |k| + k^2 c_s^2.
\end{eqnarray*}
This is the desired dispersion relation.

\item Instability requires that $\omega^2 < 0$, which requires
\begin{displaymath}
0 > 2\Omega_0^2 - 2\pi G\Sigma_0 |k| + k^2 c_s^2.
\end{displaymath}
We therefore want to find the value of $k$ that produces the minimum value of the right-hand side. The RHS is quadratic in $|k|$, and its minimum occurs at
\begin{displaymath}
|k| = \frac{\pi G \Sigma_0}{c_s^2}.
\end{displaymath}
Plugging this in, we see that the minimum value of the RHS is given by
\begin{displaymath}
2\Omega_0^2 - 2 \pi G \Sigma_0 \frac{\pi G \Sigma_0}{c_s^2} + \left(\frac{\pi G \Sigma_0}{c_s^2}\right)^2 c_s^2.
\end{displaymath}
Instability exists only if there is a value of $|k|$ that makes the RHS negative, so the condition is
\begin{eqnarray*}
2\Omega_0^2 - 2 \pi G \Sigma_0 \left(\frac{\pi G \Sigma_0}{c_s^2}\right) + \left(\frac{\pi G \Sigma_0}{c_s^2}\right)^2 c_s^2 & < & 0 \\
2\Omega_0^2 & < & \left(\frac{\pi G \Sigma_0}{c_s^2}\right) \\
\left(\frac{\sqrt{2}\Omega_0 c_s}{\pi G \Sigma_0}\right)^2 & < & 1 \\
Q & < & 1.
\end{eqnarray*}

\item The Toomre mass is
\begin{eqnarray*}
M_T = \lambda_T^2 \Sigma_0 = \frac{4 c_s^4}{G^2 \Sigma_0}.
\end{eqnarray*}
Plugging in the given values of $c_s$ and $\Sigma_0$, we obtain $M_T = 2.3\times 10^7$ $\msun$. This is a bit larger than the truncation masses reported by Rosolowsky, but only by a factor of a few.

\end{enumerate}

\item \textbf{The Origin of Brown Dwarfs.}\footnote{This problem has primarily inspired by \citet{padoan04a}.}

\begin{enumerate}

\item The Chabrier IMF is
\begin{displaymath}
\frac{dn}{d\log m} \equiv \xi(m) =
\left\{
\begin{array}{ll}
A \exp \left[-\frac{(\log m -\log m_c)^2}{2\sigma^2}\right], \quad & m<1.0\,\msun \\
B (m/\msun)^{-x}, & m>1.0\, \msun
\end{array}
\right.,
\end{displaymath}
where $m_c = 0.22$ $\msun$, $\sigma = 0.57$, $x=1.3$, $A$ is a normalization constant, and the fact that $\xi(m)$ is continuous at $m=1$ $\msun$ implies that
\begin{displaymath}
B = A \exp \left[-\frac{\log (m_c/\msun)^2}{2\sigma^2}\right].
\end{displaymath}
To compute the fraction of mass in brown dwarfs, $m<m_{\rm BD} =0.075$ $\msun$, we simply evaluate the integral of $\xi(m)$ over all masses below $m_{\rm BD}$ and divide by the integral over all masses, i.e.
\begin{displaymath}
f_{\rm BD} = \frac{\int_{m_{\rm min}}^{m_{\rm BD}} \xi(m) \, dm}{\int_{m_{\rm min}}^{m_{\rm max}} \xi(m) \, dm}.
\end{displaymath}
Note that we want to integrate with respect to $m$ and not $\log m$, because 
\begin{displaymath}
\int \frac{dn}{d\log m} \, dm \propto \int \frac{dn}{dm} m \, dm
\end{displaymath}
is the mass, which is what we want. The integrals can be evaluated analytically in terms of error functions, but it is more convenient just to evaluate them numerically from this point. Some simple python code to do so is:
\begin{verbatim}
import numpy as np
from scipy.integrate import quad

def xi(m, mc=0.22, sigma=0.57, x=1.3):
    dndlogm = np.exp( -(np.log10(m)-np.log10(mc))**2 /
                      (2.0*sigma**2) )
    idx = np.where(m > 1)[0]
    if len(idx) > 0:
        b = np.exp( -np.log10(mc)**2 / (2.0*sigma**2) )
        if type(dndlogm) is np.ndarray:
            dndlogm[idx] = b*m[idx]**-x
        else:
            dndlogm = b*m**-x
    return dndlogm

fBD = quad(xi, 0.0, 0.075)[0] / quad(xi, 0.0, 120)[0]
print("f_BD = {:f}".format(fBD))
\end{verbatim}
Using $m_{\rm min} = 0$ and $m_{\rm max} = 120~\msun$ gives $f_{\rm BD} = 0.014$.

\item The Bonnor-Ebert mass is
\begin{displaymath}
M_{\rm BE} = 1.18 \frac{c_s^4}{\sqrt{G^3 P}} = 1.18\frac{c_s^3}{\sqrt{G^3 \rho}},
\end{displaymath}
where we have taken $P = \rho c_s^2$. Solving for $\rho$, we have
\begin{displaymath}
\rho = \frac{(1.18 c_s^3)^2}{G^3 M_{\rm BE}^2}.
\end{displaymath}
Evaluating this for a gas with $\mu=2.3$, we have $c_s = \sqrt{k_B T/\mu m_{\rm H}} = 0.19$ km s$^{-1}$ and $\rho = 9.3\times 10^{-18}$ g cm$^{-3}$. This corresponds to $n_{\rm min} = \rho/\mu  m_{\rm H}=2.4\times 10^6$ molecules cm$^{-3}$.

\item First we want to derive an expression for the fraction of the mass above a given density. For a lognormal mass distribution,
\begin{displaymath}
\frac{dP}{dx} = \frac{1}{\sqrt{2\pi \sigma^2}} \exp\left[-\frac{\left(x-\overline{x}\right)^2}{2\sigma_x^2}\right],
\end{displaymath}
where $x = \ln(\rho/\overline{\rho})$, we can obtain this by integrating:
\begin{displaymath}
f(>x_0) = \int_{x_0}^\infty \frac{dP}{dx} dx = \frac{1}{2} \mbox{erfc}\left(\frac{x_0-\overline{x}}{\sqrt{2}\sigma_x}\right),
\end{displaymath}
where erfc is the complementary error function. For a lognormal turbulent density distribution, we have $\sigma_x \approx \sqrt{\ln(1+\mathcal{M}^2/4)}$ and $\overline{x} = \sigma_x^2/2$. The curve we want is the one defined implicitly by the equation $f(>x_0) = f_{\rm BD}$ with $x_0 = n_{\rm min}/\overline{n}$. Thus we wish to solve
\begin{displaymath}
\frac{1}{2} \mbox{erfc}\left[\frac{\ln (n_{\rm min}/\overline{n}) - \ln(1+\mathcal{M}^2/4)/2}{\sqrt{2\ln(1+\mathcal{M}^2/4)}}\right] = f_{\rm BD}.
\end{displaymath}
For a given $\overline{n}$ it is straightforward to solve this algebraic equation numerically to obtain $\mathcal{M}$. Some simple python code to do so is
\begin{verbatim}
from scipy.special import erfc
from scipy.optimize import brentq
import matplotlib.pyplot as plt

def resid(mach, nbar, fBD):
    nmin = 2.4e6
    x0 = np.log(nmin / nbar)
    sigmax = np.sqrt(np.log(1.0 + mach**2/4.0))
    xbar = sigmax**2 / 2.0
    return fBD - 0.5*erfc( (x0 - xbar) / (np.sqrt(2)*sigmax) )

def machsolve(nbar, fBD):
    if hasattr(nbar, '__iter__'):
        mach = np.zeros(len(nbar))
        for i, n in enumerate(nbar):
            mach[i] = brentq(resid, 1e-3, 100, args=(n, fBD))
        return mach
    else:
        return brentq(resid, 1e-3, 100, args=(nbar, fBD))

nbar = np.logspace(4,6,50)
mach = machsolve(nbar, fBD)

plt.fill_between(nbar, mach, alpha=0.5)
plt.plot([5e4], [7], 'ro')
plt.text(5.5e4, 7, 'IC 348')
plt.xscale('log')
plt.xlabel(r'$\overline{n}$')
plt.ylabel(r'$\mathcal{M}$')
\end{verbatim}
\begin{marginfigure}
\includegraphics[width=\linewidth]{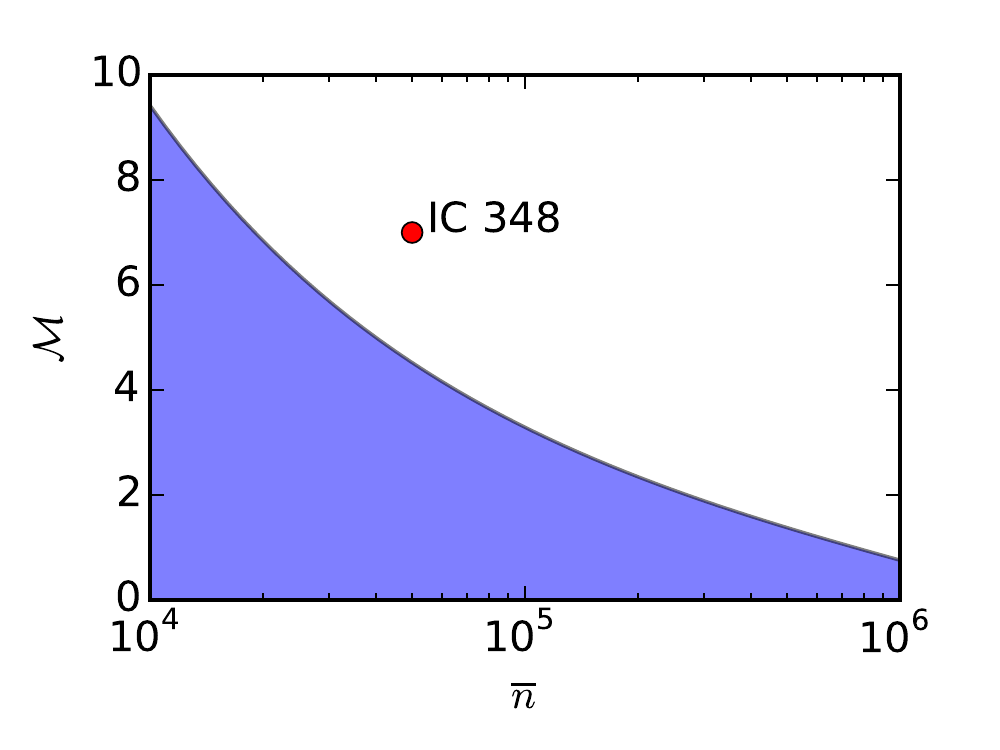}
\caption[Solution to problem set~\thesolutionset, problem~\theenumi\theenumii]{
\label{fig:hw3sol1}
Mach number $\mathcal{M}$ versus mean density $\overline{n}$, separating the region where $f(>x_0) < f_{\mathrm{BD}}$ (shaded) from the region where $f(>x_0) > f_{\mathrm{BD}}$. The red point shows the properties of IC 348.
}
\end{marginfigure}
The result is shown as Figure \ref{fig:hw3sol1}. The shaded region is the region where $f(>x_0)<f_{\rm BD}$. Clearly IC 348 (shown as the red dot in the figure) falls into the region where the mass fraction large enough to create brown dwarfs is larger than the brown dwarf mass fraction.

\end{enumerate}

\end{enumerate}
\solutionset

\begin{enumerate}

\item {\bf A Simple Protostellar Evolution Model.}

\begin{enumerate}

\item The star is a polytrope, and for a polytrope of index $n$ the gravitational energy is (e.g., see \citealt{kippenhahn94a})
\begin{displaymath}
\mathcal{W} = -\frac{3}{5-n} \frac{G M^2}{R}.
\end{displaymath}
The virial theorem tells us that the thermal energy is half the absolute value of the potential energy, so
\begin{displaymath}
\mathcal{T} = \frac{3}{2(5-n)} \frac{G M^2}{R}.
\end{displaymath}
Finally, the change in internal energy associated with dissociation, ionization, and deuterium burning is $(\psi_I  + \psi_M - \psi_D) M$. Note the opposite signs: $\psi_I$ and $\psi_M$ are positive, meaning that the final state (ionized, atomic) is higher energy than the initial one, while $\Psi_D$ is negative, indicating that the final state (all the deuterium converted to He) is a lower energy state than the initial one. Putting this all together, the total energy of the star is
\begin{displaymath}
\mathcal{E} = -\frac{3}{2(5-n)} \frac{G M^2}{R} + (\psi_I  + \psi_M - \psi_D) M.
\end{displaymath}

\item First we can compute the time rate of change of the star's energy,
\begin{displaymath}
\dot{\mathcal{E}} = \frac{3}{2(5-n)} \frac{GM}{R} \left(M\frac{\dot{R}}{R} - 2\dot{M}\right) + (\psi_I + \psi_M-\psi_D) \dot{M}.
\end{displaymath}
Now consider conservation of energy. The star's luminosity $L$ represents the rate of change of the energy "at infinity", i.e., the energy removed from the system. Since the total energy of the star plus infinity must remain constant, we require that $\dot{\mathcal{E}} + L = 0$. Writing down this condition and solving for $\dot{R}$, we obtain
\begin{displaymath}
\dot{R} = 2 R \frac{\dot{M}}{M} - \frac{2(5-n)}{3} \frac{R^2}{G M^2}\left[(\psi_I+\psi_M-\psi_D) \dot{M}+L\right]
\end{displaymath}
It is convenient to divide through by $\dot{M}$ in order to recast this as an equation for the evolution of $R$ with $M$:
\begin{displaymath}
\frac{dR}{dM} = 2 \frac{R}{M} - \frac{2(5-n)}{3} \frac{R^2}{G M^2}\left(\psi_I+\psi_M-\psi_D+\frac{L}{\dot{M}}\right).
\end{displaymath}
If we further divide by $R/M$ on both sides, we obtain
\begin{displaymath}
\frac{d\ln R}{d\ln M} = 2 - \frac{2(5-n)}{3} \frac{R}{G M}\left(\psi_I+\psi_M-\psi_D+\frac{L}{\dot{M}}\right).
\end{displaymath}
Next, we must compute the total luminosity, which contains contributions from the star's intrinsic, internal luminosity, and from the accretion luminosity. Since the star is on the Hayashi track, we can compute the intrinsic luminosity by taking its effective temperature to be fixed at $T_H$. Thus the total luminosity is
\begin{displaymath}
L = L_{\rm acc} + L_H = f_{\rm acc} \frac{G M \dot{M}}{R} + 4\pi R^2 \sigma T_H^4.
\end{displaymath}
Substituting this in, we have
\begin{displaymath}
\frac{d\ln R}{d\ln M} = 2 - \frac{2(5-n)}{3} \left[f_{\rm acc} + \left(\frac{R}{G M}\right)\left(\psi_I+\psi_M -\psi_D + \frac{4\pi R^2\sigma T_H^4}{\dot{M}}\right)\right].
\end{displaymath}
This is our final evolution equation.

\item The ODE can be integrated by standard techniques. Below is an example python program to do so, and plot the result:
\begin{verbatim}
import numpy as np
import matplotlib.pyplot as plt
from scipy.integrate import odeint

# Define some constants in cgs
G = 6.67e-8
eV = 1.6e-12
amu = 1.66e-24
sigma = 5.67e-5
Msun = 1.99e33
Rsun = 6.96e10
Lsun = 3.83e33
yr = 365.25*24.*3600.

# Problem parameters
psiI = 13.6*eV/amu
psiM = 2.2*eV/amu
psiD = 100*eV/amu
tH = 3500.0

# Default parameters
n = 1.5
facc = 0.75
Mdot = 1e-5*Msun/yr

# Define the derivative function
def dlnRdlnM(lnR, lnM, n=n, facc=facc, Mdot=Mdot):
    R = np.exp(lnR)
    M = np.exp(lnM)
    return(2.0-2.0*(5.0-n)/3.0 *
           (facc+(R/(G*M))*
            (psiI+psiM-psiD+
             4.0*np.pi*R**2*sigma*tH**4/Mdot)))

# Integrate
lnM = np.log(np.logspace(-2, 0, 500)*Msun)
lnR = odeint(dlnRdlnM, np.log(2.5*Rsun), lnM,
             args=(n, facc, Mdot))
R = np.exp(lnR[:,0])
M = np.exp(lnM)
             
# Get luminosity
L = facc*G*M*Mdot/R + 4.0*np.pi*R**2*sigma*tH**4

# Plot radius
p1,=plt.plot(M/Msun, R/Rsun, 'b', lw=2)
plt.xscale('log')
plt.xlabel(r'$M/M_\odot$')
plt.ylabel(r'$R/R_\odot$')

# Plot luminosity
plt.twinx()
p2,=plt.plot(M/Msun, L/Lsun, 'r', lw=2)
plt.ylabel(r'$L/L_\odot$')
plt.legend([p1,p2], ['Radius', 'Luminosity'],
           loc='lower right')
\end{verbatim}
\begin{marginfigure}
\includegraphics[width=\linewidth]{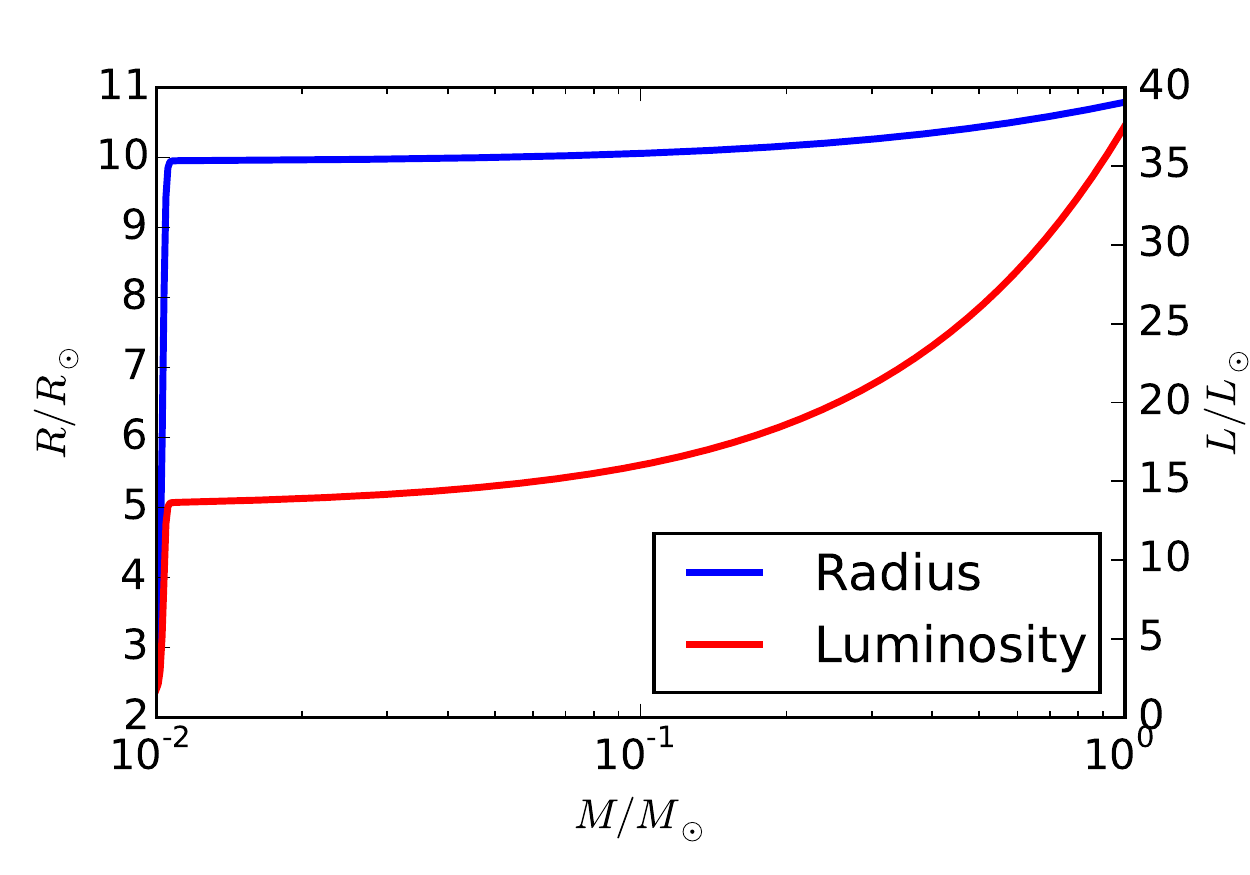}
\caption[Solution to problem set~\thesolutionset, problem~\theenumi\theenumii]{
\label{fig:hw4sol1}
Radius (blue) and luminosity (red) for the simple protostellar evolution model.
}
\end{marginfigure}
The resulting output is shown as Figure \ref{fig:hw4sol1}. Note that the radius is too large by a factor of $\sim 3$ compared to more sophisticated models, mainly due to the incorrect assumption that all the accreted deuterium burns as quickly as it accretes. In reality the D luminosity should be significantly lower, because D burning lasts longer than accretion.

\item This problem can be solved using the same basic structure as the previous part. The derivative of radius with respect to mass now becomes
\begin{eqnarray*}
\frac{d\ln R}{d\ln M} & = & 2 - \frac{2(5-n)}{3} \left[f_{\rm acc} + \left(\frac{R}{G M}\right) \cdot \right.
\\
& & \qquad \left.\left(\psi_I+\psi_M -\psi_D + \frac{\max[4\pi R^2\sigma T_H^4, \lsun (M/\msun)^3]}{\dot{M}}\right)\right].
\end{eqnarray*}
This can be integrated via a simple python program as in the previous part:
\begin{verbatim}
# Define the derivative function for the second part
def dlnRdlnM2(lnR, lnM, n=n, facc=facc, Mdot=Mdot):
    R = np.exp(lnR)
    M = np.exp(lnM)
    LH = 4.0*np.pi*R**2*sigma*tH**4
    Lstar = Lsun*(M/Msun)**3
    return(2.0-2.0*(5.0-n)/3.0 *
           (facc+(R/(G*M))*
            (psiI+psiM-psiD+np.maximum(Lstar,LH)/Mdot)))

# Integrate
Mdot2 = 1.0e-4*Msun/yr
lnM2 = np.log(np.logspace(-2, np.log10(50), 500)*Msun)
lnR2 = odeint(dlnRdlnM2, np.log(2.5*Rsun), lnM2,
              args=(3.0, facc, Mdot2))
R2 = np.exp(lnR2[:,0])
M2 = np.exp(lnM2)

# Get luminosity
L2 = facc*G*M*Mdot/R + np.maximum(
    4.0*np.pi*R2**2*sigma*tH**4,
    Lsun*(M2/Msun)**3)

# Plot radius
plt.clf()
p1,=plt.plot(M2/Msun, R2/Rsun, 'b', lw=2)
plt.xlabel(r'$M/M_\odot$')
plt.ylabel(r'$R/R_\odot$')

# Plot luminosity
plt.twinx()
p2,=plt.plot(M2/Msun, L2/Lsun, 'r', lw=2)
plt.ylabel(r'$L/L_\odot$')
plt.yscale('log')
plt.xlim([0,50])
plt.legend([p1,p2], ['Radius', 'Luminosity'], 
           loc='center right')
\end{verbatim}
\begin{marginfigure}
\includegraphics[width=\linewidth]{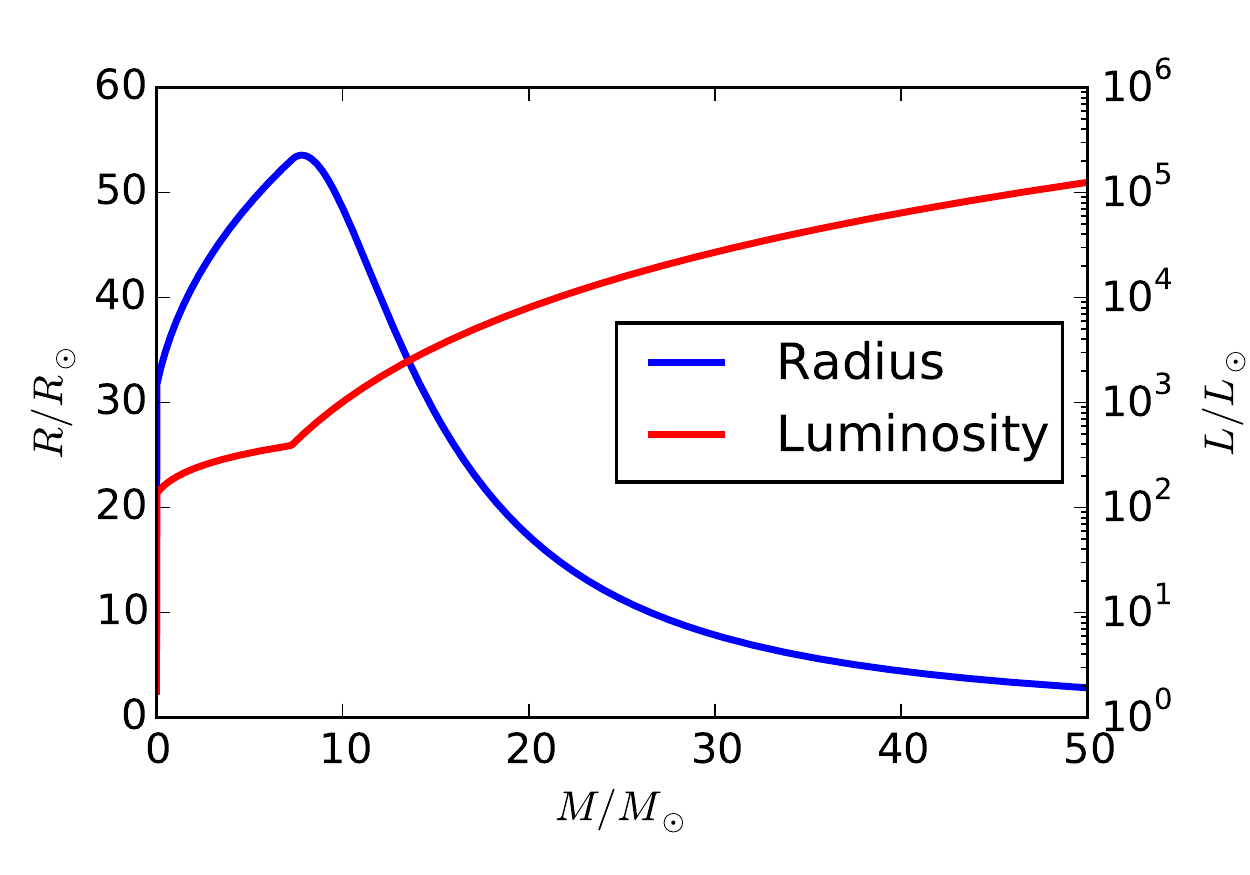}
\caption[Solution to problem set~\thesolutionset, problem~\theenumi\theenumii]{
\label{fig:hw4sol2}
Radius (blue) and luminosity (red) for the simple protostellar evolution model for a massive star.
}
\end{marginfigure}
The resulting output is shown as Figure \ref{fig:hw4sol2}.

\end{enumerate}

\item \textbf{Self-Similar Viscous Disks.}

\begin{enumerate}

\item First let us plug in the given form for $\nu$:
\begin{displaymath}
\frac{\partial\Sigma}{\partial t} = \frac{\nu_1}{\varpi_1} \frac{3}{\varpi} \frac{\partial}{\partial \varpi} \left[\varpi^{1/2} \frac{\partial}{\partial \varpi} \left(\Sigma \varpi^{3/2}\right)\right].
\end{displaymath}
Next, let's make the change of variables $\varpi = \varpi_1 x$. Note that this also implies that $\partial /\partial \varpi = (1/\varpi_1) \partial/\partial x$. With this change, we have
\begin{displaymath}
\frac{\partial\Sigma}{\partial t} = \frac{\nu_1}{\varpi_1^2} \frac{3}{x} \frac{\partial}{\partial x} \left[x^{1/2} \frac{\partial}{\partial x} \left(\Sigma x^{3/2}\right)\right].
\end{displaymath}
The third step is to make the change of variables $t = T t_s$, $\partial/\partial t = (1/t_s) \partial/\partial T$, then simplify:
\begin{eqnarray*}
\frac{3\nu_1}{\varpi_1^2} \frac{\partial\Sigma}{\partial T} & = & \frac{\nu_1}{\varpi_1^2} \frac{3}{x} \frac{\partial}{\partial x} \left[x^{1/2} \frac{\partial}{\partial x} \left(\Sigma x^{3/2}\right)\right] \\
\frac{\partial\Sigma}{\partial T} & = & \frac{1}{x} \frac{\partial}{\partial x} \left[x^{1/2} \frac{\partial}{\partial x} \left(\Sigma x^{3/2}\right)\right].
\end{eqnarray*}
The last step is to substitute in for $\Sigma$, which is trivial:
\begin{displaymath}
\frac{\partial S}{\partial T} = \frac{1}{x} \frac{\partial}{\partial x} \left[x^{1/2} \frac{\partial}{\partial x} \left(S x^{3/2}\right)\right].
\end{displaymath}
This is the non-dimensional equation we wanted.

\item We can show that the given form is a solution simply by plugging in the equivalent non-dimensional solution for $S$, which is
\begin{displaymath}
S = \frac{e^{-x/T}}{x T^{3/2}}.
\end{displaymath}
Plugging this into the two sides of the non-dimensional equation, we get
\begin{eqnarray*}
\frac{\partial S}{\partial T} & = & \left(\frac{2x - 3 T}{2 x T^{7/2}}\right) e^{-x/T} \\
\frac{1}{x} \frac{\partial}{\partial x} \left[x^{1/2} \frac{\partial}{\partial x} \left(S x^{3/2}\right)\right]
& = & 
\frac{1}{x} \frac{\partial}{\partial x} \left[\left(\frac{T-2x}{2 T^{5/2}}\right) e^{-x/T}\right] \\
& = & \left(\frac{2x - 3 T}{2 x T^{7/2}}\right) e^{-x/T}.
\end{eqnarray*}
Since the two sides match, this suffices to show that $S=e^{-x/T}/(x T^{3/2})$ is a solution. Since we have made no assumptions about the value of $\Sigma_1$ in this argument, we are free to choose its value to be whatever we want. In particular, if we choose $\Sigma_1 = C/3\pi \nu_1$, then we immediately obtain the solution
\begin{displaymath}
\Sigma = \left(\frac{C}{3\pi \nu_1}\right)\frac{1}{x T^{3/2}} e^{-x/T}.
\end{displaymath}

\item The disk mass is simply given by
\begin{eqnarray*}
M_d & = & \int_0^\infty 2\pi \varpi \Sigma \, d\varpi \\
& = & \varpi_1^2 \int_0^\infty 2\pi x \Sigma \, dx \\
& = & \left(\frac{2 C \varpi_1^2}{3 \nu_1}\right) \frac{1}{T^{3/2}} \int_0^\infty e^{-x/T} \, dx \\
& = & \frac{2 C \varpi_1^2}{3 \nu_1 T^{1/2}} \\
& = & 2 C t_s \left(\frac{t}{t_s}\right)^{-1/2}.
\end{eqnarray*}
The time rate of change of the disk mass is just
\begin{displaymath}
\dot{M}_d = -C \left(\frac{t}{t_s}\right)^{-3/2}.
\end{displaymath}
Looking at the equation, and noting that $C$ has units of mass per time, it is clear that $C$ controls the accretion rate of the disk onto the point mass in the center.

\begin{marginfigure}
\includegraphics[width=\linewidth]{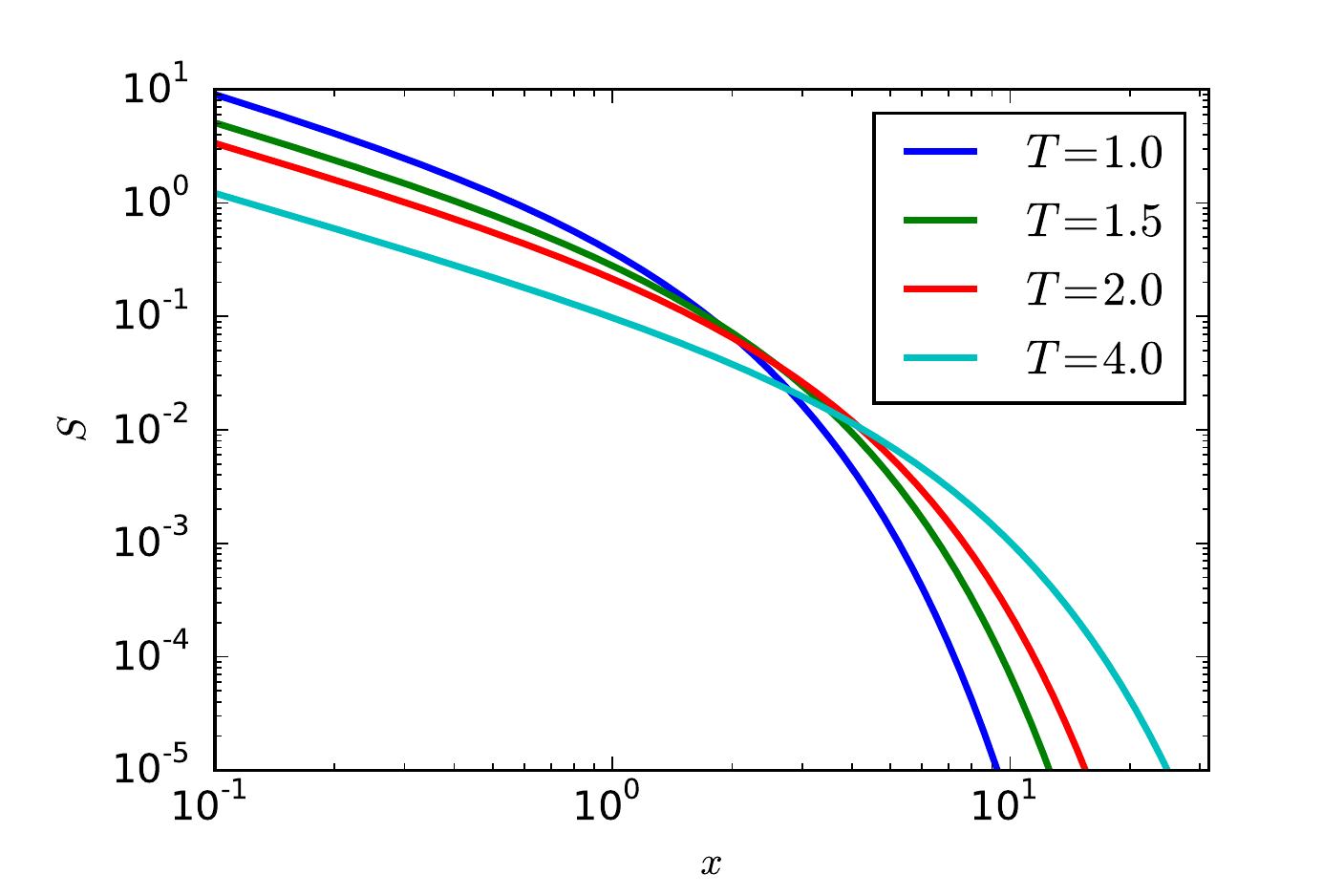}
\caption[Solution to problem set~\thesolutionset, problem~\theenumi\theenumii]{
\label{fig:hw4sol3}
$S$ versus $x$ at dimensionless times $T=1, 1.5, 2$, and 4.
}
\end{marginfigure}
\item Figure \ref{fig:hw4sol3} shows the required plot. We see that the disk surface density profile follows $S\propto x^{-1}$ for $x<T$, and is exponentially truncated at $x>T$. We can think of the inner part as the ``main disk", and the outer part as the material pushed outward by viscosity in order to compensate for the angular momentum lost as other gas moves inwards. As time passes, the inner, main disk drains onto the star, and its surface density decreases. At the same time, the outer, exponentially truncated segment of the disk grows to larger and larger radii as more and more angular momentum is extracted from the inner disk.

\end{enumerate}

\item \textbf{A Simple T Tauri Disk Model.}

\begin{enumerate}

\item The disk interior is optically thick, so the vertical radiation flux $F$ is given by the diffusion approximation:
\begin{displaymath}
F = \frac{c}{3\kappa\rho} \frac{d}{dz} E = \frac{ca}{3\kappa\rho} \frac{d}{dz} (T^4) = \frac{4 \sigma}{3\kappa\rho}  \frac{d}{dz} (T^4)
\end{displaymath}
where $E$ is the radiation energy density and $T$ is the gas temperature. In thermal equilibrium the flux does not vary with $z$, so we can re-arrange this equation and integrate from the midplane at $z=0$ to the surface at $z=z_s$:
\begin{eqnarray*}
F \int_0^{z_s} \rho \, dz & = & \frac{4\sigma}{3\kappa} \int_{T_m}^{T_s} \frac{d}{dz} T^4 \, dz \\
F \frac{\Sigma}{2} & = & \frac{4\sigma}{3\kappa} \left(T_m^4 - T_s^4\right) \\
F & \approx & \frac{8\sigma}{3\kappa\Sigma} T_m^4,
\end{eqnarray*}
where the factor of 2 in the denominator on the LHS in the second step comes from the fact that $\Sigma$ is the column density of the entire disk, and we integrated over only half of it. In the third step we assumed that $T_m^4 \gg T_s^4$, which will be true for any optically thick disk. Note that this is the flux carried away from the disk midplane in both the $+z$ and $-z$ directions -- formally the flux changes direction discontinuously at $z=0$ in this simple model, so the total flux leaving the midplane is twice this value. If the disk radiates as a blackbody, the radiation flux per unit area leaving each side of the disk surface is $\sigma T_s^4$, and this must balance the flux that is transported upward through the disk by diffusion. Thus we have
\begin{displaymath}
\frac{8\sigma}{3\kappa\Sigma} T_m^4 \approx \sigma T_s^4,
\end{displaymath}
where the expressions on either side of the equality represent the fluxes in either the $+z$ or $-z$ directions either; the total fluxes are a factor of 2 greater, but the factors of 2 obviously cancel. Solving for $T_m$ gives the desired result:
\begin{displaymath}
T_m \approx \left(\frac{3}{8}\kappa\Sigma\right)^{1/4} T_s.
\end{displaymath}

\item Equating the dissipation rate $F_d$ per unit area with the radiation rate per unit area $\sigma T_s^4$ 
\begin{eqnarray*}
\sigma T_s ^4 & = &\frac{9}{8} \nu \Sigma \Omega^2\\
T_s & = & \left(\frac{9}{8}\frac{\nu\Sigma\Omega^2}{\sigma}\right)^{1/4} \\
& = & \left(\frac{9}{8} \alpha \frac{c_s^2 \Sigma \Omega}{\sigma}\right)^{1/4}
\end{eqnarray*}
In turn, plugging this into the relation we just derived between the surface and midplane temperatures gives
\begin{eqnarray*}
T_m & \approx & \left(\frac{27}{64} \frac{\nu \kappa \Sigma^2 \Omega^2}{\sigma}\right)^{1/4} \\
& \approx & \left(\frac{27}{64} \frac{\alpha \kappa c_s^2 \Sigma^2 \Omega}{\sigma}\right)^{1/4}
\end{eqnarray*}
Substituting $c_s^2 = k_B T_m / \mu m_{\rm H}$, where $\mu$ is the mean particle mass in units of $m_{\rm H}$, and solving for $T_m$ gives
\begin{displaymath}
T_m \approx
\left(\frac{27}{64} \frac{\alpha k_B \kappa \Sigma^2 \Omega}{\sigma \mu m_{\rm H}}\right)^{1/3}.
\end{displaymath}
Note that it makes much more sense to compute $c_s$ from the midplane temperature than from the surface temperature, since the vast majority of the viscous dissipation is occurring near the midplane, not at the disk surface.\\

\item The cooling time is the thermal energy divided by the energy radiation rate. The thermal energy per unit area is
\begin{displaymath}
E_{\rm th} \approx \frac{\Sigma c_s^2}{\gamma-1} = \frac{k_B \Sigma T_m}{(\gamma-1)\mu m_{\rm H}},
\end{displaymath}
where $\gamma$ is the ratio of specific heats for the gas, which for molecular hydrogen will be somewhere between $5/3$ and $7/5$ depending on the gas temperature. The radiation rate is $2\sigma T_s^4$, so the cooling time is
\begin{eqnarray*}
t_{\rm cool} & = & \frac{E_{\rm th}}{2\sigma T_s^4} \\
& \approx & \frac{\Sigma k_B T_m}{2(\gamma-1)\mu m_{\rm H} \sigma T_s^4} \\
& \approx & \frac{3 \kappa \Sigma^2 k_B}{16(\gamma-1)\mu m_{\rm H} \sigma T_m^3} \\
& \approx & \frac{4}{9 (\gamma-1) \alpha \Omega}
\end{eqnarray*}
The orbital period is $t_{\rm orb} = 2\pi/\Omega$, so the ratio of cooling time to orbital period is
\begin{displaymath}
\frac{t_{\rm cool}}{t_{\rm orb}} \approx \frac{2}{9\pi(\gamma-1)\alpha}.
\end{displaymath}
For the typical values of $\alpha$ expected due to MRI or similar mechanisms, $\sim 0.01$ or less, this number is significantly bigger than unity, so the cooling time is longer than the orbital period. Under these conditions the disk is likely to act adiabatically rather than isothermally. Only if $\alpha$ gets quite large, $\sim 0.1$ or more, do we approach the isothermal regime.

\item Let the disk surface density be $\Sigma=\Sigma_0 (\varpi/\varpi_0)^{-1}$, and let $\varpi_0=1$ AU and $\varpi_1=20$ AU be the inner and outer radii. The mass in the disk is
\begin{displaymath}
M_{\rm disk} = \int_{\varpi_0}^{\varpi_1} \Sigma_0 \left(\frac{\varpi}{\varpi_0}\right)^{-1} 2 \pi \varpi \, d\varpi 
= 2 \pi \Sigma_0 \varpi_0 (\varpi_1 - \varpi_0),
\end{displaymath}
so 
\begin{displaymath}
\Sigma = \frac{M_{\rm disk}}{2 \pi \varpi_0 (\varpi_1 - \varpi_0)} \left(\frac{\varpi}{\varpi_0}\right)^{-1} = 2.2\times 10^3\left(\frac{\varpi}{1\mbox{ AU}}\right)^{-1}\mbox{ g cm}^{-2}.
\end{displaymath}
For a 1 $\msun$ star, the angular velocity of the orbit is
\begin{displaymath}
\Omega = \sqrt{\frac{GM}{\varpi^3}} = 2.0\times 10^{-7} \left(\frac{\varpi}{1\mbox{ AU}}\right)^{-3/2}\mbox{ s}^{-1}
\end{displaymath}
Plugging in $\kappa=3$ cm$^{-2}$ g$^{-1}$ and $\alpha=0.01$, taking $\mu=2.3$ as the mean particle mass, and plugging into the expression for $T_m$ derived in part (b) gives
\begin{displaymath}
T_m \approx 1980 \left(\frac{\varpi}{1\mbox{ AU}}\right)^{-7/6}\mbox{ K},
\end{displaymath}
and plugging this into the relation between $T_m$ and $T_s$ derived in part (a) gives
\begin{displaymath}
T_s \approx 370 \left(\frac{\varpi}{1\mbox{ AU}}\right)^{-11/12}\mbox{ K}.
\end{displaymath}
The midplane density is $\rho_m\approx \Sigma/H$, where $H$ is the scale height is $H = c_s/\Omega = \Omega^{-1}\sqrt{k_B T/\mu m_{\rm H}}$. If we use $T\approx T_m$ to compute the scale height, then we have
\begin{displaymath}
\rho_m \approx \frac{\Sigma\Omega}{\sqrt{k_B T_m/\mu m_{\rm H}}} = 1.7\times 10^{-9} \left(\frac{\varpi}{1\mbox{ AU}}\right)^{-23/12}\mbox{ g cm}^{-3}.
\end{displaymath}
Finally, the Toomre $Q$ of the disk computed using the midplane temperature (which is the most reasonable one to use, since it is the temperature of most of the mass) is
\begin{displaymath}
Q = \frac{\Omega c_s}{\pi G \Sigma} = \frac{\Omega \sqrt{k_B T_m/\mu m_{\rm H}}}{\pi G \Sigma} = 110 \left(\frac{\varpi}{1\mbox{ AU}}\right)^{-13/12}.
\end{displaymath}
This reaches a minimum value of $4.4$ at $r=20$ AU. Thus the disk is gravitationally stable.

\end{enumerate}

\end{enumerate}

\solutionset

\begin{enumerate}

\item \textbf{HII Region Trapping.}

\begin{enumerate}

\item The density profile of the accretion flow is given implicitly by
\begin{displaymath}
\dot{M}_* = 4\pi r^2 \rho v_{\rm ff},
\end{displaymath}
where $v_{\rm ff} = \sqrt{2 G M_*/r}$. Thus we have
\begin{displaymath}
\rho = \frac{\dot{M}_*}{4\pi \sqrt{2G M_*}} r^{-3/2}.
\end{displaymath}
Recombinations happen in the region between $R_*$ and $r_i$. The recombination rate per unit volume is $\alpha_B n_e n_p = 1.1 \alpha_B (\rho/\mu_{\rm H} m_{\rm H})^2$, where $\mu_{\rm H} =1.4$ is the mass per H nucleus assuming standard composition. Thus the total recombination rate within the ionized volume is
\begin{eqnarray*}
\Gamma & = & \int_{R_*}^{r_i} 4\pi r^2 (1.1\alpha_B) \left(\frac{\dot{M}_*}{4\pi \mu_{\rm H} m_{\rm H} \sqrt{2G M_*}} r^{-3/2}\right)^2 \, dr \\
& = & \frac{1.1\alpha_B \dot{M}_*^2}{8\pi \mu_{\rm H}^2 m_{\rm H}^2 G M_*} \ln \frac{r_i}{R_*}.
\end{eqnarray*}
Since this must equal the ionizing photon production rate ($\Gamma = S$), we can solve for $r_i$:
\begin{displaymath}
r_i = R_* \exp\left(\frac{8\pi \mu_{\rm H}^2 m_{\rm H}^2 G M_* S}{1.1 \alpha_B \dot{M}_*^2}\right).
\end{displaymath}
The condition that $r_i \gg R_*$ is satisfied if the term inside parentheses is $\gtrsim 1$, which in turn requires
\begin{displaymath}
\dot{M}_* \lesssim \left(\frac{8\pi \mu_{\rm H}^2 m_{\rm H}^2 G M_* S}{1.1 \alpha_B}\right)^{1/2}.
\end{displaymath}
Plugging in the given values $M_* = 30$ $\msun$ and $S=10^{49}$ s$^{-1}$, we obtain $\dot{M}_* \lesssim 7\times 10^{-5}$ $\msun$ yr$^{-1}$. This is lower (though not by a huge amount) than the typical accretion rates inferred for massive stars.

\item The escape velocity at a distance $r$ from the star is $v_{\rm esc} = \sqrt{2 G M_*/r}$. Thus the condition that $v_{\rm esc} < c_i$ at $r_i$ implies that
\begin{displaymath}
\frac{2 G M_*}{c_i^2} <  r_i = R_* \exp\left(\frac{8\pi \mu_{\rm H}^2 m_{\rm H}^2 G M_* S}{1.1 \alpha_B \dot{M}_*^2}\right).
\end{displaymath}
Solving for $\dot{M}_*$, we find
\begin{displaymath}
\dot{M}_* > \left[\frac{8\pi \mu_{\rm H}^2 m_{\rm H}^2 G M_* S}{2.2\alpha_B \ln(v_{\rm esc,*}/c_i)}\right]^{1/2},
\end{displaymath}
where $v_{\rm esc,*} = \sqrt{2 G M_*/R_*}$ is the escape speed from the stellar surface. Using $R_*=7.7$ $\rsun$ (the radius of a 30 $\msun$ ZAMS star) and plugging in the other input values gives $\dot{M}_*>2.2\times 10^{-5}$ $\msun$.

\end{enumerate}

\item \textbf{The Transition to Grain-Mediated H$_2$ Formation}\footnote{This problem was primarily inspired by \citet{glover03a}.}

\begin{enumerate}

\item To calculate the rate of H$_2$ formation catalyzed by free electrons, we first calculate the rate of formation of H$^-$:
\begin{displaymath}
\gamma({\rm H}^{-}\,{\rm form.}) = k_{-} n_e n_{\rm H} = k_{-} x_{\rm C} Z n_{\rm H}^2,
\end{displaymath}
where $k_{-} = 1.1\times 10^{-16} T_2^{0.88}$ cm$^3$ s$^{-1}$ is the rate coefficient given in Chapter \ref{ch:first_stars}. Next we must calculate the fraction of H$^-$ ions formed that yield H$_2$. The H$^-$ can be destroyed either by photodetachment or by reacting with a neutral hydrogen to form H$_2$:
\begin{displaymath}
{\rm H}^{-} + {\rm H} \rightarrow {\rm H}_2 + e^{-}.
\end{displaymath}
The rate of the latter reaction is
\begin{displaymath}
\gamma({\rm H}^{-}\rightarrow{\rm H}_2) = k_2 n_{{\rm H}^-} n_{\rm H}
\end{displaymath}
with $k_2 = 1.3\times 10^{-9}$ cm$^3$ s$^{-1}$ (the value given in class). The rate of photodetachment is
\begin{displaymath}
\gamma({\rm H}^{-}\,{\rm photo.}) = \zeta_{\rm pd} n_{{\rm H}^-}
\end{displaymath}
with $\zeta_{\rm pd} = 2.4\times 10^{-7}$ s$^{-1}$ for the Milky Way radiation field, again as given in class. Thus the branching ratio for H$^{-}$ going to H$_2$ rather than being destroyed by photodetachment is
\begin{displaymath}
\Gamma({\rm H}^{-}\rightarrow{\rm H}_2) = \frac{\gamma({\rm H}^{-}\rightarrow{\rm H}_2)}{\gamma({\rm H}^{-}\rightarrow{\rm H}_2) + \gamma({\rm H}^{-}\,{\rm photo.})} = \frac{k_2 n_{\rm H}}{k_2 n_{\rm H}+\zeta_{\rm pd}}
\end{displaymath}
Putting this together, the rate of formation of H$_2$ via free electrons is
\begin{eqnarray*}
\gamma({\rm H}_2{\rm - gas}) & = & \gamma({\rm H}^{-}\,{\rm form.}) \Gamma({\rm H}^{-}\rightarrow{\rm H}_2) \\
& = & k_{-} x_{\rm C} Z  \left[1+ \zeta_{\rm pd}/(k_2 n_{\rm H})\right]^{-1} n_{\rm H}^2. 
\end{eqnarray*}
In comparison, the rate of H$_2$ formation on grain surfaces is simply
\begin{displaymath}
\gamma({\rm H}_2{\rm - grain}) = \mathcal{R} n_{\rm H}^2 = \mathcal{R}_{\odot} Z n_{\rm H}^2,
\end{displaymath}
where $\mathcal{R}_{\odot} = 3\times 10^{-17}$ cm$^3$ s$^{-1}$ is the rate coefficient at Solar metallicity. Taking the ratio of these two, we have
\begin{eqnarray*}
\frac{\gamma({\rm H}_2{\rm - grain})}{\gamma({\rm H}_2{\rm - gas})} & = & \frac{\mathcal{R}_{\odot} \left[1+ \zeta_{\rm pd}/(k_2 n_{\rm H})\right]}{k_{-} x_{\rm C}} \\
& = & 2700 T_2^{-0.88} \left[1+ \zeta_{\rm pd}/(k_2 n_{\rm H})\right].
\end{eqnarray*}
Thus even in the limit $n_{\rm H} \rightarrow \infty$, the rate of formation on grain surfaces exceeds that in the gas phase by $\sim 3$ orders of magnitude. At densities below $\zeta_{\rm pd}/k_2 = 185$ cm$^{-3}$, where photodetachment significantly inhibits H$_2$ formation via the H$^-$ channel, grain formation wins by an even larger factor.\\

\item For gas phase formation, this calculation is exactly the same as the previous part, with two minor differences. First, the factor $x_{\rm C} Z$ in the gas-phase formation rate is replaced by $x$, since how hydrogen rather than carbon is the dominant source of free electrons. Second, all the $n_{\rm H}$ factors become $(1-x) n_{\rm H}$, since only neutral hydrogen participates in the relevant reactions. Making these changes, the H$^-$ formation rate is
\begin{displaymath}
\gamma({\rm H}^{-}\,{\rm form.}) = k_{-} x (1-x) n_{\rm H}^2,
\end{displaymath}
and the H$^-$ to H$_2$ reaction rate is
\begin{displaymath}
\gamma({\rm H}^{-}\rightarrow{\rm H}_2) = k_2 (1-x) n_{{\rm H}^-} n_{\rm H}.
\end{displaymath}
The photodetachment rate is unchanged, so the branching ratio for H$^-$ going to H$_2$ becomes
\begin{displaymath}
\Gamma({\rm H}^{-}\rightarrow{\rm H}_2) = \frac{k_2 (1-x) n_{\rm H}}{k_2 (1-x) n_{\rm H}+\zeta_{\rm pd}},
\end{displaymath}
and the gas-phase H$_2$ formation rate is therefore
\begin{displaymath}
\gamma({\rm H}_2{\rm - gas}) = k_{-} x (1-x)  \left[1+ \zeta_{\rm pd}/(k_2 (1-x) n_{\rm H})\right]^{-1} n_{\rm H}^2. 
\end{displaymath}
The grain-mediated reaction rate becomes
\begin{displaymath}
\gamma({\rm H}_2{\rm - grain}) = \mathcal{R}_{\odot} Z (1-x) n_{\rm H}^2.
\end{displaymath}
Notice that there is only one factor of $x$, because the reaction rate involves the collision of neutral hydrogen atoms with dust grains, and therefore depends on the density of gas times the density of grains. If the ionization fraction is $x$, the neutral hydrogen fraction is reduced by a factor $(1-x)$, but the grain abundance is unchanged. Setting the two rates equal, we have
\begin{displaymath}
k_{-} x \left[1+ \zeta_{\rm pd}/(k_2 (1-x) n_{\rm H})\right]^{-1} = \mathcal{R}_{\odot} Z.
\end{displaymath}
This is a quadratic in $x$, and the solution is
\begin{displaymath}
x = \frac{1}{2}\left(1+k \pm \sqrt{1-2k + k^2 - 4 k/r_{\rm pd}}\right)
\end{displaymath}
\begin{marginfigure}
\includegraphics[width=\linewidth]{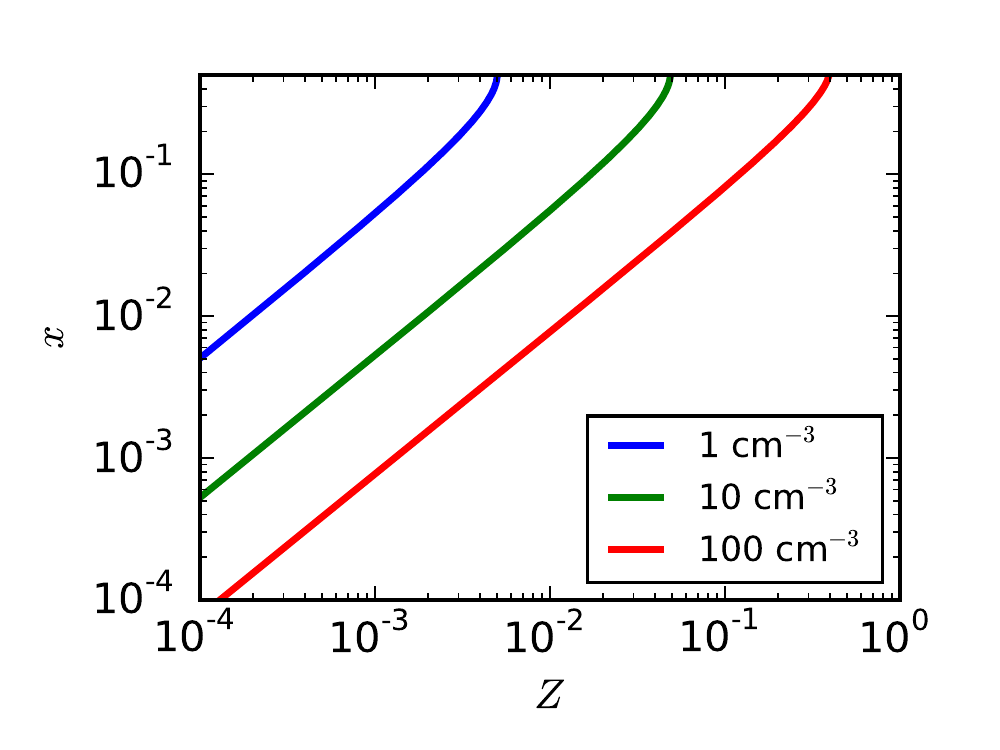}
\caption[Solution to problem set~\thesolutionset, problem~\theenumi\theenumii]{
\label{fig:hw5sol1}
$x$ versus $Z$ for $n_{\rm H} = 1$, 10, and 100 cm$^{-3}$.
}
\end{marginfigure}
where $k = \mathcal{R}_{\odot} Z/k_{-}\approx 0.27 T_2^{-0.88} Z$ is the ratio of the rate coefficients for H$_2$ formation on grains and H$^-$ formation in the gas phase, and $r_{\rm pd} = n_{\rm H} k_2 / \zeta_{\rm pd}$ is the ratio of the gas density to the critical density for H$_2$ photodetachment, $n_{\rm crit,pd} = \zeta_{\rm pd}/k_2\approx 185$ cm$^{-3}$. Note that both roots represent mathematically valid solutions, one at ionization fraction below 50\% and one at ionization fraction above 50\%. In practice, however, the low ionization fraction solution is the more physically-realistic one, since if the gas is highly ionized the temperature is unlikely to be low enough to allow formation of H$_2$. Figure \ref{fig:hw5sol1} shows the physically-realistic solution plotted for $n_{\rm H} = 1$, 10, and 100 cm$^{-3}$.

\item A solution ceases to exist when the metallicity is such that the quantity under the square root in the equation from the previous part goes to zero. Thus, grain-mediated H$_2$ formation must dominate whenever
\begin{displaymath}
1 - 2k + k^2 - 4 k r_{\rm pd}^{-1} < 0.
\end{displaymath}
Solving, this condition reduces to
\begin{displaymath}
k > 1 + \frac{2}{r_{\rm pd}} \left(1 - \sqrt{1+r_{\rm pd}}\right).
\end{displaymath}
(Note: there is another root to this equation, with a $+$ instead of a $-$ in front of the square root. However, one can easily verify that in the vicinity of this root $x>1$, which is obviously unphysical. Thus the $-$ root is the physically realistic one.) Rewriting this in terms of dimensional quantities, this is
\begin{marginfigure}
\includegraphics[width=\linewidth]{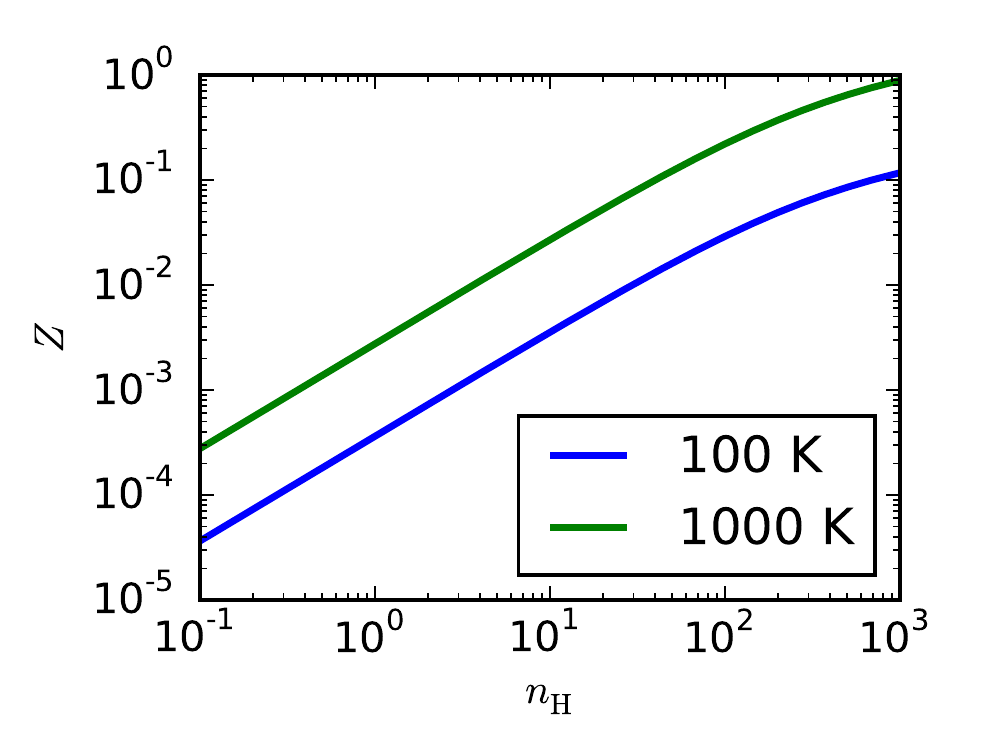}
\caption[Solution to problem set~\thesolutionset, problem~\theenumi\theenumii]{
\label{fig:hw5sol2}
$Z$ versus $n_{\rm H}$ at $T=100$ K and $1000$ K.
}
\end{marginfigure}
\begin{displaymath}
Z > 0.27 T_2^{0.88} \left[1 + \frac{2 n_{\rm crit,pd}}{n_{\rm H}} \left(1 - \sqrt{1+\frac{n_{\rm H}}{n_{\rm crit,pd}}}\right)\right].
\end{displaymath}
Figure \ref{fig:hw5sol2} shows a plot of the minimum value of $Z$ versus density $n_{\rm H}$ at $T=100$ K and 1000 K.

\end{enumerate}

\item {\bf Disk Dispersal by Photoionization.}

\begin{enumerate}

\item The gas will escape when the sound speed becomes comparable to the escape speed from the star. Thus
\begin{eqnarray*}
c_s & \approx & \sqrt{\frac{2 G M_*}{\varpi_g}}\\
\varpi_g & \approx & \frac{2 G M_*}{c_s^2} = \frac{2 G M_* \mu m_{\rm H}}{k_B T}
\end{eqnarray*}
The mean particle mass depends on whether the helium is ionized or not, but for a relatively cool star like a T Tauri star it is probably reasonable to assume that it is not, so the number of electrons equals the number of hydrogen atoms. The mean mass per particle for neutral hydrogen $\mu_{\rm H} = 1.4$, and by number hydrogen represents 93\% of all nuclei in the Milky Way, so the mean mass per particle in gas where the hydrogen is ionized is $\mu=0.61$. Plugging this in gives a sound speed $c_s = 10.7$ km s$^{-1}$.

\item Ionization balance requires that recombinations equal ionizations. If the density is $n_0$ inside $\varpi_g$, the recombination rate per unit volume is $\alpha_B n_0^2$, where $\alpha_B=2.59\times 10^{-13}$ cm$^{-3}$ s$^{-1}$ is the case B recombination coefficient, and this expression implicitly assumes that the gas is fully ionized. The recombination rate is simply this times the volume, so equating this with the ionization rate produced by the star give
\begin{eqnarray*}
\Phi & = & \frac{4}{3}\pi \varpi_g^3 \alpha_B n_0^2 \\
n_0 & = & \sqrt{\frac{3\Phi}{4\pi \alpha_B \varpi_g^3}} \\
& = & \sqrt{\frac{3\Phi k_B^3 T^3}{32\pi \alpha_B G^3 M_*^3 \mu^3 m_{\rm H}^3}}
\end{eqnarray*}

\item The wind will have a density of $\sim n_0$ and will leave a velocity $\sim c_s$, and it will be lost from an area of order $\varpi_g^2$. Thus an order of magnitude estimate for the wind mass flux is
\begin{eqnarray*}
\dot{M} & \sim & n_0 m_{\rm H} c_s \varpi_g^2 \\
& = & \sqrt{\frac{3\Phi r_g}{4\pi \alpha_B}} m_{\rm H} c_s \\
& = & \sqrt{\frac{3\Phi G M_* m_{\rm H}^2}{2\pi \alpha_B}}
\end{eqnarray*}

\item Plugging in the given numerical values gives $\dot{M} \sim 10^{-10}$ $\msun$ yr$^{-1}$. Thus it would take $\sim 100$ Myr to evaporate a $0.01$ $\msun$ star. This is much longer than the observed $\sim 2$ Myr lifetime of T Tauri disks. This indicates that photoionization by itself cannot the the primary disk removal mechanism. Instead, it is a plausible disk destruction mechanism only if it operates in tandem with some other mechanism, like accretion of the disk onto the star.

\end{enumerate}

\item {\bf Aerodynamics of Small Solids in a Disk.}

\begin{enumerate}

\item The force per unit mass in the radial direction that a parcel of gas of density $\rho$ experiences due to the combined effects of gas pressure and stellar gravity is
\begin{displaymath}
f_\varpi = -\frac{G M}{\varpi^2} - \frac{1}{\rho}\frac{\partial P}{\partial \varpi} = -\frac{GM}{\varpi^2} + \frac{n}{\rho}\frac{P}{\varpi} = -\frac{GM}{\varpi^2} + \frac{n c_g^2}{\varpi}.
\end{displaymath}
This also gives the acceleration of the gas parcel toward the star. If we equate this with the centripetal acceleration required to maintain circular motion at velocity $v_g$, then we have
\begin{displaymath}
\frac{v_g^2}{\varpi} = \frac{GM}{\varpi^2} - \frac{n c_s^2}{\varpi} = \frac{v_K^2}{\varpi} - \frac{n c_s^2}{\varpi},
\end{displaymath}
where $v_K = \sqrt{GM/\varpi}$ is the Keplerian velocity. If we solve this for $v_g$ and subtract the result from $v_K$, then we have
\begin{eqnarray*}
\Delta v = v_K - v_g & = & v_K - \sqrt{v_K^2 - n c_g^2} \\
& = & v_K \left(1 - \sqrt{1 - \frac{n c_g^2}{v_K^2}}\right) \\
& \approx & \frac{n c_g^2}{2 v_K},
\end{eqnarray*}
where the last step results from taking the Taylor expansion of the square root term in the limit $nc_g^2 \ll v_K^2$, which is equivalent to the assumption that the deviation from Keplerian rotation is small.

\item The mass of the solid particle is $m_s = (4/3)\pi s^3 \rho_s$, so its momentum is $p = (4/3)\pi s^3 \rho_s v$. Dividing this by the drag force we have
\begin{displaymath}
t_s = \frac{p}{F_D} = \frac{s}{c_s}\frac{\rho_s}{\rho_d}.
\end{displaymath}
i.e., the stopping time is just the sound crossing time of the particle's radius multiplied by the ratio of solid density to gas density.

\item In the frame co-rotating with the gas, the dust particle experiences a net radial force which contains contributions from inward stellar gravity, outward centrifugal force, and outward drag force resisting inward motion. The total force is
\begin{eqnarray*}
F & = & -\frac{GM m_s}{\varpi^2} + \frac{m_s v_g^2}{\varpi} - \frac{4\pi}{3} s^2 \rho_d v c_g \\
& = & -\frac{v_K^2}{\varpi} m_s + \left(\frac{v_K^2}{\varpi} - \frac{n c_g^2}{\varpi}\right) m_d - \frac{4\pi}{3} s^2 \rho_d v c_g \\
& = & \frac{c_s}{m_s} \left(-\frac{nc_g}{\varpi} - \frac{v}{s}\frac{\rho_d}{\rho_s}\right),
\end{eqnarray*}
where $m_s = (4/3)\pi s^3 \rho_s$ is the mass of the solid. The terminal velocity of the grain is determined by the condition that the net force be zero, so if we set the right-hand side of this equation equal to zero and solve, we find that the terminal velocity is
\begin{displaymath}
v = -n c_g\frac{s}{\varpi} \frac{\rho_s}{\rho_d}.
\end{displaymath}
The time required for the solid particle to drift into the star is roughly
\begin{displaymath}
t_{\rm drift} \approx \frac{\varpi_0}{-v} = \frac{\varpi_0^2}{n c_g s}\frac{\rho_d}{\rho_s},
\end{displaymath}

\item First let us evaluate the stopping time:
\begin{displaymath}
t_s = \frac{s}{c_g} \frac{\rho_s}{\rho_d} = \frac{s}{\sqrt{k_B T/\mu m_{\rm H}}} \frac{\rho_s}{\rho_d} = 2.1\times 10^4\mbox{ s}.
\end{displaymath}
The orbital period at 1 AU is $t_{\rm orb} = 1\mbox{ yr} = 3.1\times 10^7\mbox{ s}$, so we do have $t_s \ll t_{\rm orb}$. The timescale required for the particle to drift into the star is
\begin{displaymath}
t_{\rm drift} =  \frac{\varpi_0^2}{n c_g s}\frac{\rho_d}{\rho_s} = 1.7\times 10^{11}\mbox{ s} = 5400\mbox{ yr}.
\end{displaymath}
This is much, much smaller than the inferred timescale of $\sim 1$ Myr for planet formation and disk dissipation.

\end{enumerate}

\end{enumerate}

\backmatter


\bibliography{bib/refs} 

\begin{thebibliography}{250}
\expandafter\ifx\csname natexlab\endcsname\relax\def\natexlab#1{#1}\fi

\bibitem[{{Abdo} {et~al.}(2010){Abdo}, {Ackermann}, {Ajello}, {Baldini},
  {Ballet}, {Barbiellini}, {Bastieri}, {Baughman}, {Bechtol}, {Bellazzini},
  {Berenji}, {Bloom}, {Bonamente}, {Borgland}, {Bregeon}, {Brez}, {Brigida},
  {Bruel}, {Burnett}, {Buson}, {Caliandro}, {Cameron}, {Caraveo}, {Casandjian},
  {Cecchi}, {{\c C}elik}, {Chekhtman}, {Cheung}, {Chiang}, {Ciprini}, {Claus},
  {Cohen-Tanugi}, {Cominsky}, {Conrad}, {Dermer}, {de Palma}, {Digel}, {Silva},
  {Drell}, {Dubois}, {Dumora}, {Farnier}, {Favuzzi}, {Fegan}, {Focke},
  {Fortin}, {Frailis}, {Fukazawa}, {Funk}, {Fusco}, {Gargano}, {Gehrels},
  {Germani}, {Giavitto}, {Giebels}, {Giglietto}, {Giordano}, {Glanzman},
  {Godfrey}, {Grenier}, {Grondin}, {Grove}, {Guillemot}, {Guiriec}, {Harding},
  {Hayashida}, {Horan}, {Hughes}, {Jackson}, {J{\'o}hannesson}, {Johnson},
  {Johnson}, {Kamae}, {Katagiri}, {Kataoka}, {Kawai}, {Kerr}, {Kn{\"o}dlseder},
  {Kuss}, {Lande}, {Latronico}, {Lemoine-Goumard}, {Longo}, {Loparco}, {Lott},
  {Lovellette}, {Lubrano}, {Makeev}, {Mazziotta}, {McEnery}, {Meurer},
  {Michelson}, {Mitthumsiri}, {Mizuno}, {Monte}, {Monzani}, {Morselli},
  {Moskalenko}, {Murgia}, {Nolan}, {Norris}, {Nuss}, {Ohsugi}, {Okumura},
  {Omodei}, {Orlando}, {Ormes}, {Paneque}, {Pelassa}, {Pepe}, {Pesce-Rollins},
  {Piron}, {Porter}, {Rain{\`o}}, {Rando}, {Razzano}, {Reimer}, {Reimer},
  {Reposeur}, {Rodriguez}, {Ryde}, {Sadrozinski}, {Sanchez}, {Sander}, {Saz
  Parkinson}, {Sgr{\`o}}, {Siskind}, {Smith}, {Spandre}, {Spinelli}, {Starck},
  {Strickman}, {Strong}, {Suson}, {Takahashi}, {Tanaka}, {Thayer}, {Thayer},
  {Thompson}, {Tibaldo}, {Torres}, {Tosti}, {Tramacere}, {Uchiyama}, {Usher},
  {Vasileiou}, {Vilchez}, {Vitale}, {Waite}, {Wang}, {Winer}, {Wood}, {Ylinen},
  \& {Ziegler}}]{abdo10b}
{Abdo}, A.~A., {Ackermann}, M., {Ajello}, M., {et~al.} 2010, \apj, 710, 133

\bibitem[{{Abrahamsson} {et~al.}(2007){Abrahamsson}, {Krems}, \&
  {Dalgarno}}]{abrahamsson07a}
{Abrahamsson}, E., {Krems}, R.~V., \& {Dalgarno}, A. 2007, \apj, 654, 1171

\bibitem[{{Alexander} {et~al.}(2014){Alexander}, {Pascucci}, {Andrews},
  {Armitage}, \& {Cieza}}]{alexander14a}
{Alexander}, R., {Pascucci}, I., {Andrews}, S., {Armitage}, P., \& {Cieza}, L.
  2014, Protostars and Planets VI, 475

\bibitem[{{Andrews} {et~al.}(2009){Andrews}, {Wilner}, {Hughes}, {Qi}, \&
  {Dullemond}}]{andrews09a}
{Andrews}, S.~M., {Wilner}, D.~J., {Hughes}, A.~M., {Qi}, C., \& {Dullemond},
  C.~P. 2009, \apj, 700, 1502

\bibitem[{{Appenzeller} \& {Mundt}(1989)}]{appenzeller89a}
{Appenzeller}, I., \& {Mundt}, R. 1989, \aapr, 1, 291

\bibitem[{{Arzoumanian} {et~al.}(2011){Arzoumanian}, {Andr{\'e}}, {Didelon},
  {K{\"o}nyves}, {Schneider}, {Men'shchikov}, {Sousbie}, {Zavagno}, {Bontemps},
  {di Francesco}, {Griffin}, {Hennemann}, {Hill}, {Kirk}, {Martin}, {Minier},
  {Molinari}, {Motte}, {Peretto}, {Pezzuto}, {Spinoglio}, {Ward-Thompson},
  {White}, \& {Wilson}}]{arzoumanian11a}
{Arzoumanian}, D., {Andr{\'e}}, P., {Didelon}, P., {et~al.} 2011, \aap, 529, L6

\bibitem[{{Bai} \& {Stone}(2010)}]{bai10a}
{Bai}, X.-N., \& {Stone}, J.~M. 2010, \apj, 722, 1437

\bibitem[{{Balbus} \& {Hawley}(1991)}]{balbus91a}
{Balbus}, S.~A., \& {Hawley}, J.~F. 1991, \apj, 376, 214

\bibitem[{{Barinovs} {et~al.}(2005){Barinovs}, {van Hemert}, {Krems}, \&
  {Dalgarno}}]{barinovs05a}
{Barinovs}, {\u G}., {van Hemert}, M.~C., {Krems}, R., \& {Dalgarno}, A. 2005,
  \apj, 620, 537

\bibitem[{{Barkana} \& {Loeb}(2001)}]{barkana01a}
{Barkana}, R., \& {Loeb}, A. 2001, \physrep, 349, 125

\bibitem[{{Bastian} {et~al.}(2010){Bastian}, {Covey}, \& {Meyer}}]{bastian10a}
{Bastian}, N., {Covey}, K.~R., \& {Meyer}, M.~R. 2010, \araa, 48, 339

\bibitem[{{Bate}(2009{\natexlab{a}})}]{bate09b}
{Bate}, M.~R. 2009{\natexlab{a}}, \mnras, 392, 590

\bibitem[{{Bate}(2009{\natexlab{b}})}]{bate09a}
---. 2009{\natexlab{b}}, \mnras, 392, 1363

\bibitem[{{Bate}(2012)}]{bate12a}
---. 2012, \mnras, 419, 3115

\bibitem[{{Bigiel} {et~al.}(2010){Bigiel}, {Leroy}, {Walter}, {Blitz},
  {Brinks}, {de Blok}, \& {Madore}}]{bigiel10a}
{Bigiel}, F., {Leroy}, A., {Walter}, F., {et~al.} 2010, \aj, 140, 1194

\bibitem[{{Bigiel} {et~al.}(2008){Bigiel}, {Leroy}, {Walter}, {Brinks}, {de
  Blok}, {Madore}, \& {Thornley}}]{bigiel08a}
---. 2008, \aj, 136, 2846

\bibitem[{{Blandford} \& {Payne}(1982)}]{blandford82a}
{Blandford}, R.~D., \& {Payne}, D.~G. 1982, \mnras, 199, 883

\bibitem[{{Bochanski} {et~al.}(2010){Bochanski}, {Hawley}, {Covey}, {West},
  {Reid}, {Golimowski}, \& {Ivezi{\'c}}}]{bochanski10a}
{Bochanski}, J.~J., {Hawley}, S.~L., {Covey}, K.~R., {et~al.} 2010, \aj, 139,
  2679

\bibitem[{{Bolatto} {et~al.}(2008){Bolatto}, {Leroy}, {Rosolowsky}, {Walter},
  \& {Blitz}}]{bolatto08a}
{Bolatto}, A.~D., {Leroy}, A.~K., {Rosolowsky}, E., {Walter}, F., \& {Blitz},
  L. 2008, \apj, 686, 948

\bibitem[{{Bolatto} {et~al.}(2011){Bolatto}, {Leroy}, {Jameson}, {Ostriker},
  {Gordon}, {Lawton}, {Stanimirovi{\'c}}, {Israel}, {Madden}, {Hony},
  {Sandstrom}, {Bot}, {Rubio}, {Winkler}, {Roman-Duval}, {van Loon},
  {Oliveira}, \& {Indebetouw}}]{bolatto11a}
{Bolatto}, A.~D., {Leroy}, A.~K., {Jameson}, K., {et~al.} 2011, \apj, 741, 12

\bibitem[{{Bondi}(1952)}]{bondi52a}
{Bondi}, H. 1952, \mnras, 112, 195

\bibitem[{{Bonnor}(1956)}]{bonnor56a}
{Bonnor}, W.~B. 1956, \mnras, 116, 351

\bibitem[{{Bromm}(2013)}]{bromm13a}
{Bromm}, V. 2013, Reports on Progress in Physics, 76, 112901

\bibitem[{{Bromm} {et~al.}(2001){Bromm}, {Ferrara}, {Coppi}, \&
  {Larson}}]{bromm01a}
{Bromm}, V., {Ferrara}, A., {Coppi}, P.~S., \& {Larson}, R.~B. 2001, \mnras,
  328, 969

\bibitem[{{Brown} {et~al.}(2008){Brown}, {Blake}, {Qi}, {Dullemond}, \&
  {Wilner}}]{brown08a}
{Brown}, J.~M., {Blake}, G.~A., {Qi}, C., {Dullemond}, C.~P., \& {Wilner},
  D.~J. 2008, \apjl, 675, L109

\bibitem[{{Burkert} \& {Bodenheimer}(2000)}]{burkert00a}
{Burkert}, A., \& {Bodenheimer}, P. 2000, \apj, 543, 822

\bibitem[{{Butler} \& {Tan}(2012)}]{butler12a}
{Butler}, M.~J., \& {Tan}, J.~C. 2012, \apj, 754, 5

\bibitem[{{Caffau} {et~al.}(2011){Caffau}, {Bonifacio}, {Fran{\c c}ois},
  {Sbordone}, {Monaco}, {Spite}, {Spite}, {Ludwig}, {Cayrel}, {Zaggia},
  {Hammer}, {Randich}, {Molaro}, \& {Hill}}]{caffau11a}
{Caffau}, E., {Bonifacio}, P., {Fran{\c c}ois}, P., {et~al.} 2011, \nat, 477,
  67

\bibitem[{{Cappellari} {et~al.}(2012){Cappellari}, {McDermid}, {Alatalo},
  {Blitz}, {Bois}, {Bournaud}, {Bureau}, {Crocker}, {Davies}, {Davis}, {de
  Zeeuw}, {Duc}, {Emsellem}, {Khochfar}, {Krajnovi{\'c}}, {Kuntschner},
  {Lablanche}, {Morganti}, {Naab}, {Oosterloo}, {Sarzi}, {Scott}, {Serra},
  {Weijmans}, \& {Young}}]{cappellari12a}
{Cappellari}, M., {McDermid}, R.~M., {Alatalo}, K., {et~al.} 2012, \nat, 484,
  485

\bibitem[{{Castor} {et~al.}(1975){Castor}, {McCray}, \& {Weaver}}]{castor75a}
{Castor}, J., {McCray}, R., \& {Weaver}, R. 1975, \apjl, 200, L107

\bibitem[{{Chabrier}(2003)}]{chabrier03a}
{Chabrier}, G. 2003, \pasp, 115, 763

\bibitem[{{Chabrier}(2005)}]{chabrier05a}
{Chabrier}, G. 2005, in Astrophysics and Space Science Library, Vol. 327, The
  Initial Mass Function 50 Years Later, ed. E.~{Corbelli}, F.~{Palla}, \&
  H.~{Zinnecker} (Dordrecht: Springer), 41--+

\bibitem[{{Chakrabarti} \& {McKee}(2005)}]{chakrabarti05a}
{Chakrabarti}, S., \& {McKee}, C.~F. 2005, \apj, 631, 792

\bibitem[{Champagne(1978)}]{champagne78a}
Champagne, F.~H. 1978, J.~Fluid Mech., 86, 67

\bibitem[{{Chandrasekhar}(1939)}]{chandrasekhar39a}
{Chandrasekhar}, S. 1939, {An introduction to the study of stellar structure}
  (Chicago: The University of Chicago Press)

\bibitem[{{Chandrasekhar}(1961)}]{chandrasekhar61a}
---. 1961, {Hydrodynamic and hydromagnetic stability} (Oxford: Clarendon Press)

\bibitem[{{Chandrasekhar} \& {Fermi}(1953)}]{chandrasekhar53a}
{Chandrasekhar}, S., \& {Fermi}, E. 1953, \apj, 118, 116

\bibitem[{{Chiang} \& {Murray-Clay}(2007)}]{chiang07a}
{Chiang}, E., \& {Murray-Clay}, R. 2007, Nature Physics, 3, 604

\bibitem[{{Clark} \& {Bonnell}(2004)}]{clark04a}
{Clark}, P.~C., \& {Bonnell}, I.~A. 2004, \mnras, 347, L36

\bibitem[{{Clark} {et~al.}(2008){Clark}, {Bonnell}, \& {Klessen}}]{clark08a}
{Clark}, P.~C., {Bonnell}, I.~A., \& {Klessen}, R.~S. 2008, \mnras, 386, 3

\bibitem[{{Clark} {et~al.}(2011){Clark}, {Glover}, {Smith}, {Greif}, {Klessen},
  \& {Bromm}}]{clark11a}
{Clark}, P.~C., {Glover}, S.~C.~O., {Smith}, R.~J., {et~al.} 2011, Science,
  331, 1040

\bibitem[{{Colombo} {et~al.}(2014){Colombo}, {Hughes}, {Schinnerer}, {Meidt},
  {Leroy}, {Pety}, {Dobbs}, {Garc{\'{\i}}a-Burillo}, {Dumas}, {Thompson},
  {Schuster}, \& {Kramer}}]{colombo14a}
{Colombo}, D., {Hughes}, A., {Schinnerer}, E., {et~al.} 2014, \apj, 784, 3

\bibitem[{{Commer{\c c}on} {et~al.}(2011){Commer{\c c}on}, {Hennebelle}, \&
  {Henning}}]{commercon11c}
{Commer{\c c}on}, B., {Hennebelle}, P., \& {Henning}, T. 2011, \apjl, 742, L9

\bibitem[{{Crutcher}(2012)}]{crutcher12a}
{Crutcher}, R.~M. 2012, \araa, 50, 29

\bibitem[{{Crutcher} {et~al.}(1999){Crutcher}, {Troland}, {Lazareff},
  {Paubert}, \& {Kaz{\`e}s}}]{crutcher99b}
{Crutcher}, R.~M., {Troland}, T.~H., {Lazareff}, B., {Paubert}, G., \&
  {Kaz{\`e}s}, I. 1999, \apjl, 514, L121

\bibitem[{{Cunningham} {et~al.}(2011){Cunningham}, {Klein}, {Krumholz}, \&
  {McKee}}]{cunningham11a}
{Cunningham}, A.~J., {Klein}, R.~I., {Krumholz}, M.~R., \& {McKee}, C.~F. 2011,
  \apj, 740, 107

\bibitem[{{Da Rio} {et~al.}(2012){Da Rio}, {Robberto}, {Hillenbrand},
  {Henning}, \& {Stassun}}]{da-rio12a}
{Da Rio}, N., {Robberto}, M., {Hillenbrand}, L.~A., {Henning}, T., \&
  {Stassun}, K.~G. 2012, \apj, 748, 14

\bibitem[{{Daddi} {et~al.}(2010){Daddi}, {Elbaz}, {Walter}, {Bournaud},
  {Salmi}, {Carilli}, {Dannerbauer}, {Dickinson}, {Monaco}, \&
  {Riechers}}]{daddi10a}
{Daddi}, E., {Elbaz}, D., {Walter}, F., {et~al.} 2010, \apjl, 714, L118

\bibitem[{{Dale} {et~al.}(2014){Dale}, {Ngoumou}, {Ercolano}, \&
  {Bonnell}}]{dale14a}
{Dale}, J.~E., {Ngoumou}, J., {Ercolano}, B., \& {Bonnell}, I.~A. 2014, \mnras,
  442, 694

\bibitem[{{Dame} {et~al.}(2001){Dame}, {Hartmann}, \& {Thaddeus}}]{dame01a}
{Dame}, T.~M., {Hartmann}, D., \& {Thaddeus}, P. 2001, \apj, 547, 792

\bibitem[{{Delfosse} {et~al.}(2000){Delfosse}, {Forveille}, {S{\'e}gransan},
  {Beuzit}, {Udry}, {Perrier}, \& {Mayor}}]{delfosse00a}
{Delfosse}, X., {Forveille}, T., {S{\'e}gransan}, D., {et~al.} 2000, \aap, 364,
  217

\bibitem[{{Dobbs} {et~al.}(2014){Dobbs}, {Krumholz}, {Ballesteros-Paredes},
  {Bolatto}, {Fukui}, {Heyer}, {Low}, {Ostriker}, \&
  {V{\'a}zquez-Semadeni}}]{dobbs14a}
{Dobbs}, C.~L., {Krumholz}, M.~R., {Ballesteros-Paredes}, J., {et~al.} 2014,
  Protostars and Planets VI, 3

\bibitem[{{Draine}(2003)}]{draine03a}
{Draine}, B.~T. 2003, \araa, 41, 241

\bibitem[{{Draine}(2011)}]{draine11a}
---. 2011, {Physics of the Interstellar and Intergalactic Medium} (Princeton
  University Press: Princeton, NJ)

\bibitem[{{Draine} {et~al.}(2007){Draine}, {Dale}, {Bendo}, {Gordon}, {Smith},
  {Armus}, {Engelbracht}, {Helou}, {Kennicutt}, {Li}, {Roussel}, {Walter},
  {Calzetti}, {Moustakas}, {Murphy}, {Rieke}, {Bot}, {Hollenbach}, {Sheth}, \&
  {Teplitz}}]{draine07a}
{Draine}, B.~T., {Dale}, D.~A., {Bendo}, G., {et~al.} 2007, \apj, 663, 866

\bibitem[{{Duch{\^e}ne} \& {Kraus}(2013)}]{duchene13a}
{Duch{\^e}ne}, G., \& {Kraus}, A. 2013, \araa, 51, 269

\bibitem[{{Dullemond} {et~al.}(2001){Dullemond}, {Dominik}, \&
  {Natta}}]{dullemond01a}
{Dullemond}, C.~P., {Dominik}, C., \& {Natta}, A. 2001, \apj, 560, 957

\bibitem[{{Dunham} {et~al.}(2014){Dunham}, {Stutz}, {Allen}, {Evans},
  {Fischer}, {Megeath}, {Myers}, {Offner}, {Poteet}, {Tobin}, \&
  {Vorobyov}}]{dunham14a}
{Dunham}, M.~M., {Stutz}, A.~M., {Allen}, L.~E., {et~al.} 2014, Protostars and
  Planets VI, 195, arXiv:1401.1809

\bibitem[{{Duquennoy} \& {Mayor}(1991)}]{duquennoy91a}
{Duquennoy}, A., \& {Mayor}, M. 1991, \aap, 248, 485

\bibitem[{{Ebert}(1955)}]{ebert55a}
{Ebert}, R. 1955, Zeitschrift fur Astrophysics, 37, 217

\bibitem[{{Egusa} {et~al.}(2004){Egusa}, {Sofue}, \& {Nakanishi}}]{egusa04a}
{Egusa}, F., {Sofue}, Y., \& {Nakanishi}, H. 2004, \pasj, 56, L45

\bibitem[{{Fall} \& {Chandar}(2012)}]{fall12a}
{Fall}, S.~M., \& {Chandar}, R. 2012, \apj, 752, 96

\bibitem[{{Fall} {et~al.}(2010){Fall}, {Krumholz}, \& {Matzner}}]{fall10a}
{Fall}, S.~M., {Krumholz}, M.~R., \& {Matzner}, C.~D. 2010, \apjl, 710, L142

\bibitem[{{Fedele} {et~al.}(2010){Fedele}, {van den Ancker}, {Henning},
  {Jayawardhana}, \& {Oliveira}}]{fedele10a}
{Fedele}, D., {van den Ancker}, M.~E., {Henning}, T., {Jayawardhana}, R., \&
  {Oliveira}, J.~M. 2010, \aap, 510, A72

\bibitem[{{Federrath}(2013)}]{federrath13b}
{Federrath}, C. 2013, \mnras, 436, 1245

\bibitem[{{Federrath} \& {Klessen}(2012)}]{federrath12a}
{Federrath}, C., \& {Klessen}, R.~S. 2012, \apj, 761, 156

\bibitem[{{Fukui} {et~al.}(2008){Fukui}, {Kawamura}, {Minamidani}, {Mizuno},
  {Kanai}, {Mizuno}, {Onishi}, {Yonekura}, {Mizuno}, {Ogawa}, \&
  {Rubio}}]{fukui08a}
{Fukui}, Y., {Kawamura}, A., {Minamidani}, T., {et~al.} 2008, \apjs, 178, 56

\bibitem[{{Gao} \& {Solomon}(2004{\natexlab{a}})}]{gao04b}
{Gao}, Y., \& {Solomon}, P.~M. 2004{\natexlab{a}}, \apjs, 152, 63

\bibitem[{{Gao} \& {Solomon}(2004{\natexlab{b}})}]{gao04a}
---. 2004{\natexlab{b}}, \apj, 606, 271

\bibitem[{{Glassgold} {et~al.}(2012){Glassgold}, {Galli}, \&
  {Padovani}}]{glassgold12a}
{Glassgold}, A.~E., {Galli}, D., \& {Padovani}, M. 2012, \apj, 756, 157

\bibitem[{{Glover}(2003)}]{glover03a}
{Glover}, S.~C.~O. 2003, \apj, 584, 331

\bibitem[{{Glover} \& {Abel}(2008)}]{glover08a}
{Glover}, S.~C.~O., \& {Abel}, T. 2008, \mnras, 388, 1627

\bibitem[{{Glover} \& {Clark}(2012)}]{glover12a}
{Glover}, S.~C.~O., \& {Clark}, P.~C. 2012, \mnras, 421, 9

\bibitem[{{Glover} {et~al.}(2010){Glover}, {Federrath}, {Mac Low}, \&
  {Klessen}}]{glover10a}
{Glover}, S.~C.~O., {Federrath}, C., {Mac Low}, M., \& {Klessen}, R.~S. 2010,
  \mnras, 404, 2

\bibitem[{{Goldbaum} {et~al.}(2011){Goldbaum}, {Krumholz}, {Matzner}, \&
  {McKee}}]{goldbaum11a}
{Goldbaum}, N.~J., {Krumholz}, M.~R., {Matzner}, C.~D., \& {McKee}, C.~F. 2011,
  \apj, 738, 101

\bibitem[{{Goldreich} \& {Kwan}(1974)}]{goldreich74a}
{Goldreich}, P., \& {Kwan}, J. 1974, \apj, 189, 441

\bibitem[{{Goodman} {et~al.}(1993){Goodman}, {Benson}, {Fuller}, \&
  {Myers}}]{goodman93a}
{Goodman}, A.~A., {Benson}, P.~J., {Fuller}, G.~A., \& {Myers}, P.~C. 1993,
  \apj, 406, 528

\bibitem[{{Gould} \& {Salpeter}(1963)}]{gould63a}
{Gould}, R.~J., \& {Salpeter}, E.~E. 1963, \apj, 138, 393

\bibitem[{{Gratier} {et~al.}(2012){Gratier}, {Braine}, {Rodriguez-Fernandez},
  {Schuster}, {Kramer}, {Corbelli}, {Combes}, {Brouillet}, {van der Werf}, \&
  {R{\"o}llig}}]{gratier12a}
{Gratier}, P., {Braine}, J., {Rodriguez-Fernandez}, N.~J., {et~al.} 2012, \aap,
  542, A108

\bibitem[{{Greif} {et~al.}(2011){Greif}, {Springel}, {White}, {Glover},
  {Clark}, {Smith}, {Klessen}, \& {Bromm}}]{greif11a}
{Greif}, T.~H., {Springel}, V., {White}, S.~D.~M., {et~al.} 2011, \apj, 737, 75

\bibitem[{{Guszejnov} \& {Hopkins}(2015)}]{guszejnov15a}
{Guszejnov}, D., \& {Hopkins}, P.~F. 2015, \mnras, 450, 4137

\bibitem[{{Guszejnov} {et~al.}(2016){Guszejnov}, {Krumholz}, \&
  {Hopkins}}]{guszejnov16a}
{Guszejnov}, D., {Krumholz}, M.~R., \& {Hopkins}, P.~F. 2016, \mnras, 458

\bibitem[{{Gutermuth} {et~al.}(2011){Gutermuth}, {Pipher}, {Megeath}, {Myers},
  {Allen}, \& {Allen}}]{gutermuth11a}
{Gutermuth}, R.~A., {Pipher}, J.~L., {Megeath}, S.~T., {et~al.} 2011, \apj,
  739, 84

\bibitem[{{Haisch} {et~al.}(2001){Haisch}, {Lada}, \& {Lada}}]{haisch01a}
{Haisch}, Jr., K.~E., {Lada}, E.~A., \& {Lada}, C.~J. 2001, \apjl, 553, L153

\bibitem[{{Hansen} {et~al.}(2012){Hansen}, {Klein}, {McKee}, \&
  {Fisher}}]{hansen12a}
{Hansen}, C.~E., {Klein}, R.~I., {McKee}, C.~F., \& {Fisher}, R.~T. 2012, \apj,
  747, 22

\bibitem[{{Harper-Clark} \& {Murray}(2009)}]{harper-clark09a}
{Harper-Clark}, E., \& {Murray}, N. 2009, \apj, 693, 1696

\bibitem[{{Hartmann} {et~al.}(1998){Hartmann}, {Calvet}, {Gullbring}, \&
  {D'Alessio}}]{hartmann98a}
{Hartmann}, L., {Calvet}, N., {Gullbring}, E., \& {D'Alessio}, P. 1998, \apj,
  495, 385

\bibitem[{{Hayashi}(1961)}]{hayashi61a}
{Hayashi}, C. 1961, \pasj, 13, 450

\bibitem[{{Hennebelle} \& {Chabrier}(2008)}]{hennebelle08b}
{Hennebelle}, P., \& {Chabrier}, G. 2008, \apj, 684, 395

\bibitem[{{Hennebelle} \& {Chabrier}(2009)}]{hennebelle09a}
---. 2009, \apj, 702, 1428

\bibitem[{{Hennebelle} \& {Chabrier}(2011)}]{hennebelle11b}
---. 2011, \apjl, 743, L29

\bibitem[{{Hennebelle} \& {Fromang}(2008)}]{hennebelle08c}
{Hennebelle}, P., \& {Fromang}, S. 2008, \aap, 477, 9

\bibitem[{{Herbig}(1977)}]{herbig77a}
{Herbig}, G.~H. 1977, \apj, 217, 693

\bibitem[{{Heyer} {et~al.}(2009){Heyer}, {Krawczyk}, {Duval}, \&
  {Jackson}}]{heyer09a}
{Heyer}, M., {Krawczyk}, C., {Duval}, J., \& {Jackson}, J.~M. 2009, \apj, 699,
  1092

\bibitem[{{Heyer} \& {Brunt}(2004)}]{heyer04a}
{Heyer}, M.~H., \& {Brunt}, C.~M. 2004, \apjl, 615, L45

\bibitem[{{Hillenbrand} \& {Hartmann}(1998)}]{hillenbrand98a}
{Hillenbrand}, L.~A., \& {Hartmann}, L.~W. 1998, \apj, 492, 540

\bibitem[{{Holmberg} \& {Flynn}(2000)}]{holmberg00a}
{Holmberg}, J., \& {Flynn}, C. 2000, \mnras, 313, 209

\bibitem[{{Hopkins}(2012{\natexlab{a}})}]{hopkins12e}
{Hopkins}, P.~F. 2012{\natexlab{a}}, \mnras, 423, 2016

\bibitem[{{Hopkins}(2012{\natexlab{b}})}]{hopkins12d}
---. 2012{\natexlab{b}}, \mnras, 423, 2037

\bibitem[{{Hopkins} {et~al.}(2013){Hopkins}, {Narayanan}, {Murray}, \&
  {Quataert}}]{hopkins13c}
{Hopkins}, P.~F., {Narayanan}, D., {Murray}, N., \& {Quataert}, E. 2013,
  \mnras, 433, 69

\bibitem[{{Hopkins} {et~al.}(2011){Hopkins}, {Quataert}, \&
  {Murray}}]{hopkins11a}
{Hopkins}, P.~F., {Quataert}, E., \& {Murray}, N. 2011, \mnras, 417, 950

\bibitem[{{Hosokawa} {et~al.}(2011{\natexlab{a}}){Hosokawa}, {Offner}, \&
  {Krumholz}}]{hosokawa11a}
{Hosokawa}, T., {Offner}, S.~S.~R., \& {Krumholz}, M.~R. 2011{\natexlab{a}},
  \apj, 738, 140

\bibitem[{{Hosokawa} \& {Omukai}(2009)}]{hosokawa09a}
{Hosokawa}, T., \& {Omukai}, K. 2009, \apj, 691, 823

\bibitem[{{Hosokawa} {et~al.}(2011{\natexlab{b}}){Hosokawa}, {Omukai},
  {Yoshida}, \& {Yorke}}]{hosokawa11b}
{Hosokawa}, T., {Omukai}, K., {Yoshida}, N., \& {Yorke}, H.~W.
  2011{\natexlab{b}}, Science, 334, 1250

\bibitem[{{Hoyle}(1946)}]{hoyle46a}
{Hoyle}, F. 1946, \mnras, 106, 406

\bibitem[{{Hunter} {et~al.}(1996){Hunter}, {O'Neil}, {Lynds}, {Shaya}, {Groth},
  \& {Holtzman}}]{hunter96a}
{Hunter}, D.~A., {O'Neil}, Jr., E.~J., {Lynds}, R., {et~al.} 1996, \apjl, 459,
  L27

\bibitem[{{Imara} {et~al.}(2011){Imara}, {Bigiel}, \& {Blitz}}]{imara11b}
{Imara}, N., {Bigiel}, F., \& {Blitz}, L. 2011, \apj, 732, 79

\bibitem[{{Jappsen} {et~al.}(2005){Jappsen}, {Klessen}, {Larson}, {Li}, \& {Mac
  Low}}]{jappsen05a}
{Jappsen}, A.-K., {Klessen}, R.~S., {Larson}, R.~B., {Li}, Y., \& {Mac Low},
  M.-M. 2005, \aap, 435, 611

\bibitem[{{Jeans}(1902)}]{jeans02a}
{Jeans}, J.~H. 1902, Royal Society of London Philosophical Transactions Series
  A, 199, 1

\bibitem[{{Ji} {et~al.}(2006){Ji}, {Burin}, {Schartman}, \& {Goodman}}]{ji06a}
{Ji}, H., {Burin}, M., {Schartman}, E., \& {Goodman}, J. 2006, \nat, 444, 343

\bibitem[{{Johansen} {et~al.}(2014){Johansen}, {Blum}, {Tanaka}, {Ormel},
  {Bizzarro}, \& {Rickman}}]{johansen14a}
{Johansen}, A., {Blum}, J., {Tanaka}, H., {et~al.} 2014, Protostars and Planets
  VI, 547

\bibitem[{{Kawamura} {et~al.}(2009){Kawamura}, {Mizuno}, {Minamidani},
  {Filipovi{\'c}}, {Staveley-Smith}, {Kim}, {Mizuno}, {Onishi}, {Mizuno}, \&
  {Fukui}}]{kawamura09a}
{Kawamura}, A., {Mizuno}, Y., {Minamidani}, T., {et~al.} 2009, \apjs, 184, 1

\bibitem[{{Keller} {et~al.}(2014){Keller}, {Bessell}, {Frebel}, {Casey},
  {Asplund}, {Jacobson}, {Lind}, {Norris}, {Yong}, {Heger}, {Magic}, {da
  Costa}, {Schmidt}, \& {Tisserand}}]{keller14a}
{Keller}, S.~C., {Bessell}, M.~S., {Frebel}, A., {et~al.} 2014, \nat, 506, 463

\bibitem[{{Kennicutt} \& {Evans}(2012)}]{kennicutt12a}
{Kennicutt}, R.~C., \& {Evans}, N.~J. 2012, \araa, 50, 531

\bibitem[{{Kennicutt}(1992)}]{kennicutt92a}
{Kennicutt}, Jr., R.~C. 1992, \apj, 388, 310

\bibitem[{{Kennicutt}(1998)}]{kennicutt98a}
---. 1998, \apj, 498, 541

\bibitem[{{Kim} {et~al.}(2008){Kim}, {Hirota}, {Honma}, {Kobayashi},
  {Bushimata}, {Choi}, {Imai}, {Iwadate}, {Jike}, {Kameno}, {Kameya},
  {Kamohara}, {Kan-Ya}, {Kawaguchi}, {Kuji}, {Kurayama}, {Manabe}, {Matsui},
  {Matsumoto}, {Miyaji}, {Nagayama}, {Nakagawa}, {Oh}, {Omodaka}, {Oyama},
  {Sakai}, {Sasao}, {Sato}, {Sato}, {Shibata}, {Tamura}, \&
  {Yamashita}}]{kim08a}
{Kim}, M.~K., {Hirota}, T., {Honma}, M., {et~al.} 2008, \pasj, 60, 991

\bibitem[{{Kippenhahn} \& {Weigert}(1994)}]{kippenhahn94a}
{Kippenhahn}, R., \& {Weigert}, A. 1994, {Stellar Structure and Evolution}
  (Berlin: Springer-Verlag)

\bibitem[{{Kolmogorov}(1941)}]{kolmogorov41a}
{Kolmogorov}, A. 1941, Akademiia Nauk SSSR Doklady, 30, 301

\bibitem[{{Kolmogorov}(1991)}]{kolmogorov91a}
{Kolmogorov}, A.~N. 1991, Royal Society of London Proceedings Series A, 434, 9

\bibitem[{{Krasnopolsky} {et~al.}(2012){Krasnopolsky}, {Li}, {Shang}, \&
  {Zhao}}]{krasnopolsky12a}
{Krasnopolsky}, R., {Li}, Z.-Y., {Shang}, H., \& {Zhao}, B. 2012, \apj, 757, 77

\bibitem[{{Kratter} \& {Matzner}(2006)}]{kratter06a}
{Kratter}, K.~M., \& {Matzner}, C.~D. 2006, \mnras, 373, 1563

\bibitem[{{Kratter} {et~al.}(2008){Kratter}, {Matzner}, \&
  {Krumholz}}]{kratter08a}
{Kratter}, K.~M., {Matzner}, C.~D., \& {Krumholz}, M.~R. 2008, \apj, 681, 375

\bibitem[{{Kratter} {et~al.}(2010){Kratter}, {Matzner}, {Krumholz}, \&
  {Klein}}]{kratter10a}
{Kratter}, K.~M., {Matzner}, C.~D., {Krumholz}, M.~R., \& {Klein}, R.~I. 2010,
  \apj, 708, 1585

\bibitem[{{Kraus} \& {Hillenbrand}(2008)}]{kraus08a}
{Kraus}, A.~L., \& {Hillenbrand}, L.~A. 2008, \apjl, 686, L111

\bibitem[{{Kreckel} {et~al.}(2010){Kreckel}, {Bruhns}, {{\v C}{\'{\i}}{\v
  z}ek}, {Glover}, {Miller}, {Urbain}, \& {Savin}}]{kreckel10a}
{Kreckel}, H., {Bruhns}, H., {{\v C}{\'{\i}}{\v z}ek}, M., {et~al.} 2010,
  Science, 329, 69

\bibitem[{{Kroupa}(2001)}]{kroupa01a}
{Kroupa}, P. 2001, \mnras, 322, 231

\bibitem[{{Kroupa}(2002)}]{kroupa02c}
---. 2002, Science, 295, 82

\bibitem[{{Kruijssen}(2012)}]{kruijssen12a}
{Kruijssen}, J.~M.~D. 2012, \mnras, 426, 3008

\bibitem[{{Kruijssen} \& {Longmore}(2014)}]{kruijssen14c}
{Kruijssen}, J.~M.~D., \& {Longmore}, S.~N. 2014, \mnras, 439, 3239

\bibitem[{{Krumholz}(2011)}]{krumholz11e}
{Krumholz}, M.~R. 2011, \apj, 743, 110

\bibitem[{{Krumholz}(2013)}]{krumholz13c}
---. 2013, \mnras, 436, 2747

\bibitem[{{Krumholz}(2014)}]{krumholz14c}
---. 2014, \physrep, 539, 49

\bibitem[{{Krumholz} {et~al.}(2012{\natexlab{a}}){Krumholz}, {Dekel}, \&
  {McKee}}]{krumholz12a}
{Krumholz}, M.~R., {Dekel}, A., \& {McKee}, C.~F. 2012{\natexlab{a}}, \apj,
  745, 69

\bibitem[{{Krumholz} {et~al.}(2011{\natexlab{a}}){Krumholz}, {Klein}, \&
  {McKee}}]{krumholz11c}
{Krumholz}, M.~R., {Klein}, R.~I., \& {McKee}, C.~F. 2011{\natexlab{a}}, \apj,
  740, 74

\bibitem[{{Krumholz} {et~al.}(2012{\natexlab{b}}){Krumholz}, {Klein}, \&
  {McKee}}]{krumholz12b}
---. 2012{\natexlab{b}}, \apj, 754, 71

\bibitem[{{Krumholz} {et~al.}(2011{\natexlab{b}}){Krumholz}, {Leroy}, \&
  {McKee}}]{krumholz11b}
{Krumholz}, M.~R., {Leroy}, A.~K., \& {McKee}, C.~F. 2011{\natexlab{b}}, \apj,
  731, 25

\bibitem[{{Krumholz} \& {McKee}(2005)}]{krumholz05c}
{Krumholz}, M.~R., \& {McKee}, C.~F. 2005, \apj, 630, 250

\bibitem[{{Krumholz} \& {Tan}(2007)}]{krumholz07e}
{Krumholz}, M.~R., \& {Tan}, J.~C. 2007, \apj, 654, 304

\bibitem[{{Krumholz} {et~al.}(2014){Krumholz}, {Bate}, {Arce}, {Dale},
  {Gutermuth}, {Klein}, {Li}, {Nakamura}, \& {Zhang}}]{krumholz14e}
{Krumholz}, M.~R., {Bate}, M.~R., {Arce}, H.~G., {et~al.} 2014, Protostars and
  Planets VI, 243

\bibitem[{{Lada}(2006)}]{lada06a}
{Lada}, C.~J. 2006, \apjl, 640, L63

\bibitem[{{Lada} \& {Lada}(2003)}]{lada03a}
{Lada}, C.~J., \& {Lada}, E.~A. 2003, \araa, 41, 57

\bibitem[{{Larson}(1969)}]{larson69a}
{Larson}, R.~B. 1969, \mnras, 145, 271

\bibitem[{{Larson}(1981)}]{larson81a}
---. 1981, \mnras, 194, 809

\bibitem[{{Larson}(2005)}]{larson05a}
---. 2005, \mnras, 359, 211

\bibitem[{{Launay} \& {Roueff}(1977)}]{launay77a}
{Launay}, J.-M., \& {Roueff}, E. 1977, Journal of Physics B Atomic Molecular
  Physics, 10, 879

\bibitem[{{Leitherer} {et~al.}(1999){Leitherer}, {Schaerer}, {Goldader},
  {Delgado}, {Robert}, {Kune}, {de Mello}, {Devost}, \&
  {Heckman}}]{leitherer99a}
{Leitherer}, C., {Schaerer}, D., {Goldader}, J.~D., {et~al.} 1999, \apjs, 123,
  3

\bibitem[{{Leroy} {et~al.}(2013){Leroy}, {Walter}, {Sandstrom}, {Schruba},
  {Munoz-Mateos}, {Bigiel}, {Bolatto}, {Brinks}, {de Blok}, {Meidt}, {Rix},
  {Rosolowsky}, {Schinnerer}, {Schuster}, \& {Usero}}]{leroy13a}
{Leroy}, A.~K., {Walter}, F., {Sandstrom}, K., {et~al.} 2013, \aj, 146, 19

\bibitem[{{Li} {et~al.}(2008){Li}, {McKee}, {Klein}, \& {Fisher}}]{li08a}
{Li}, P.~S., {McKee}, C.~F., {Klein}, R.~I., \& {Fisher}, R.~T. 2008, \apj,
  684, 380

\bibitem[{{Li} {et~al.}(2005){Li}, {Mac Low}, \& {Klessen}}]{li05a}
{Li}, Y., {Mac Low}, M., \& {Klessen}, R.~S. 2005, \apjl, 620, L19

\bibitem[{{Li} {et~al.}(2014){Li}, {Banerjee}, {Pudritz}, {J{\o}rgensen},
  {Shang}, {Krasnopolsky}, \& {Maury}}]{li14a}
{Li}, Z.-Y., {Banerjee}, R., {Pudritz}, R.~E., {et~al.} 2014, Protostars and
  Planets VI, 173

\bibitem[{{Lombardi} {et~al.}(2006){Lombardi}, {Alves}, \&
  {Lada}}]{lombardi06a}
{Lombardi}, M., {Alves}, J., \& {Lada}, C.~J. 2006, \aap, 454, 781

\bibitem[{{Lopez} {et~al.}(2011){Lopez}, {Krumholz}, {Bolatto}, {Prochaska}, \&
  {Ramirez-Ruiz}}]{lopez11a}
{Lopez}, L.~A., {Krumholz}, M.~R., {Bolatto}, A.~D., {Prochaska}, J.~X., \&
  {Ramirez-Ruiz}, E. 2011, \apj, 731, 91

\bibitem[{{Lu} {et~al.}(2013){Lu}, {Do}, {Ghez}, {Morris}, {Yelda}, \&
  {Matthews}}]{lu13a}
{Lu}, J.~R., {Do}, T., {Ghez}, A.~M., {et~al.} 2013, \apj, 764, 155

\bibitem[{{Machida} {et~al.}(2008){Machida}, {Tomisaka}, {Matsumoto}, \&
  {Inutsuka}}]{machida08b}
{Machida}, M.~N., {Tomisaka}, K., {Matsumoto}, T., \& {Inutsuka}, S.-i. 2008,
  \apj, 677, 327

\bibitem[{{Masunaga} \& {Inutsuka}(2000)}]{masunaga00a}
{Masunaga}, H., \& {Inutsuka}, S.-i. 2000, \apj, 531, 350

\bibitem[{{Masunaga} {et~al.}(1998){Masunaga}, {Miyama}, \&
  {Inutsuka}}]{masunaga98a}
{Masunaga}, H., {Miyama}, S.~M., \& {Inutsuka}, S.-i. 1998, \apj, 495, 346

\bibitem[{{Matzner}(2002)}]{matzner02a}
{Matzner}, C.~D. 2002, \apj, 566, 302

\bibitem[{{Mazeh} {et~al.}(1992){Mazeh}, {Goldberg}, {Duquennoy}, \&
  {Mayor}}]{mazeh92a}
{Mazeh}, T., {Goldberg}, D., {Duquennoy}, A., \& {Mayor}, M. 1992, \apj, 401,
  265

\bibitem[{{McKee} \& {Tan}(2003)}]{mckee03a}
{McKee}, C.~F., \& {Tan}, J.~C. 2003, \apj, 585, 850

\bibitem[{{McKee} \& {Tan}(2008)}]{mckee08a}
---. 2008, \apj, 681, 771

\bibitem[{{McKee} \& {Zweibel}(1992)}]{mckee92a}
{McKee}, C.~F., \& {Zweibel}, E.~G. 1992, \apj, 399, 551

\bibitem[{{Menten} {et~al.}(2007){Menten}, {Reid}, {Forbrich}, \&
  {Brunthaler}}]{menten07a}
{Menten}, K.~M., {Reid}, M.~J., {Forbrich}, J., \& {Brunthaler}, A. 2007, \aap,
  474, 515

\bibitem[{{Motte} {et~al.}(2007){Motte}, {Bontemps}, {Schilke}, {Schneider},
  {Menten}, \& {Brogui{\`e}re}}]{motte07a}
{Motte}, F., {Bontemps}, S., {Schilke}, P., {et~al.} 2007, \aap, 476, 1243

\bibitem[{{Murray} {et~al.}(2010){Murray}, {Quataert}, \&
  {Thompson}}]{murray10a}
{Murray}, N., {Quataert}, E., \& {Thompson}, T.~A. 2010, \apj, 709, 191

\bibitem[{{Murray} \& {Rahman}(2010)}]{murray10b}
{Murray}, N., \& {Rahman}, M. 2010, \apj, 709, 424

\bibitem[{{Muzerolle} {et~al.}(2005){Muzerolle}, {Luhman}, {Brice{\~n}o},
  {Hartmann}, \& {Calvet}}]{muzerolle05a}
{Muzerolle}, J., {Luhman}, K.~L., {Brice{\~n}o}, C., {Hartmann}, L., \&
  {Calvet}, N. 2005, \apj, 625, 906

\bibitem[{{Myers} {et~al.}(2013){Myers}, {McKee}, {Cunningham}, {Klein}, \&
  {Krumholz}}]{myers13a}
{Myers}, A.~T., {McKee}, C.~F., {Cunningham}, A.~J., {Klein}, R.~I., \&
  {Krumholz}, M.~R. 2013, \apj, 766, 97

\bibitem[{{Najita} {et~al.}(2007){Najita}, {Carr}, {Glassgold}, \&
  {Valenti}}]{najita07b}
{Najita}, J.~R., {Carr}, J.~S., {Glassgold}, A.~E., \& {Valenti}, J.~A. 2007,
  Protostars and Planets V, 507

\bibitem[{{Narayanan} {et~al.}(2012){Narayanan}, {Krumholz}, {Ostriker}, \&
  {Hernquist}}]{narayanan12a}
{Narayanan}, D., {Krumholz}, M.~R., {Ostriker}, E.~C., \& {Hernquist}, L. 2012,
  \mnras, 421, 3127

\bibitem[{{Offner} {et~al.}(2014){Offner}, {Clark}, {Hennebelle}, {Bastian},
  {Bate}, {Hopkins}, {Moraux}, \& {Whitworth}}]{offner14a}
{Offner}, S.~S.~R., {Clark}, P.~C., {Hennebelle}, P., {et~al.} 2014, Protostars
  and Planets VI, 53

\bibitem[{{Offner} {et~al.}(2009){Offner}, {Hansen}, \& {Krumholz}}]{offner09b}
{Offner}, S.~S.~R., {Hansen}, C.~E., \& {Krumholz}, M.~R. 2009, \apjl, 704,
  L124

\bibitem[{{Omukai} {et~al.}(2005){Omukai}, {Tsuribe}, {Schneider}, \&
  {Ferrara}}]{omukai05a}
{Omukai}, K., {Tsuribe}, T., {Schneider}, R., \& {Ferrara}, A. 2005, \apj, 626,
  627

\bibitem[{{Ossenkopf} \& {Mac Low}(2002)}]{ossenkopf02a}
{Ossenkopf}, V., \& {Mac Low}, M.-M. 2002, \aap, 390, 307

\bibitem[{{Osterbrock} \& {Ferland}(2006)}]{osterbrock06a}
{Osterbrock}, D.~E., \& {Ferland}, G.~J. 2006, {Astrophysics of gaseous nebulae
  and active galactic nuclei} (University Science Books: Sausalito, CA)

\bibitem[{{Ostriker} \& {Shetty}(2011)}]{ostriker11a}
{Ostriker}, E.~C., \& {Shetty}, R. 2011, \apj, 731, 41

\bibitem[{{Padoan} {et~al.}(2014){Padoan}, {Federrath}, {Chabrier}, {Evans},
  {Johnstone}, {J{\o}rgensen}, {McKee}, \& {Nordlund}}]{padoan14a}
{Padoan}, P., {Federrath}, C., {Chabrier}, G., {et~al.} 2014, Protostars and
  Planets VI, 77

\bibitem[{{Padoan} \& {Nordlund}(1999)}]{padoan99a}
{Padoan}, P., \& {Nordlund}, {\AA}. 1999, \apj, 526, 279

\bibitem[{{Padoan} \& {Nordlund}(2002)}]{padoan02a}
---. 2002, \apj, 576, 870

\bibitem[{{Padoan} \& {Nordlund}(2004)}]{padoan04a}
---. 2004, \apj, 617, 559

\bibitem[{{Padoan} \& {Nordlund}(2011)}]{padoan11a}
---. 2011, \apj, 730, 40

\bibitem[{{Padoan} {et~al.}(1997){Padoan}, {Nordlund}, \& {Jones}}]{padoan97a}
{Padoan}, P., {Nordlund}, A., \& {Jones}, B.~J.~T. 1997, \mnras, 288, 145

\bibitem[{{Padoan} {et~al.}(2007){Padoan}, {Nordlund}, {Kritsuk}, {Norman}, \&
  {Li}}]{padoan07a}
{Padoan}, P., {Nordlund}, {\AA}., {Kritsuk}, A.~G., {Norman}, M.~L., \& {Li},
  P.~S. 2007, \apj, 661, 972

\bibitem[{{Palla} \& {Stahler}(1990)}]{palla90a}
{Palla}, F., \& {Stahler}, S.~W. 1990, \apjl, 360, L47

\bibitem[{{Palla} \& {Stahler}(2000)}]{palla00a}
---. 2000, \apj, 540, 255

\bibitem[{{Prialnik}(2009)}]{prialnik09a}
{Prialnik}, D. 2009, {An Introduction to the Theory of Stellar Structure and
  Evolution} (Cambridge University Press)

\bibitem[{{Pringle}(1981)}]{pringle81a}
{Pringle}, J.~E. 1981, \araa, 19, 137

\bibitem[{{Rathborne} {et~al.}(2006){Rathborne}, {Jackson}, \&
  {Simon}}]{rathborne06a}
{Rathborne}, J.~M., {Jackson}, J.~M., \& {Simon}, R. 2006, \apj, 641, 389

\bibitem[{{Rebolledo} {et~al.}(2012){Rebolledo}, {Wong}, {Leroy}, {Koda}, \&
  {Donovan Meyer}}]{rebolledo12a}
{Rebolledo}, D., {Wong}, T., {Leroy}, A., {Koda}, J., \& {Donovan Meyer}, J.
  2012, \apj, 757, 155

\bibitem[{{R{\'e}my-Ruyer} {et~al.}(2014){R{\'e}my-Ruyer}, {Madden},
  {Galliano}, {Galametz}, {Takeuchi}, {Asano}, {Zhukovska}, {Lebouteiller},
  {Cormier}, {Jones}, {Bocchio}, {Baes}, {Bendo}, {Boquien}, {Boselli},
  {DeLooze}, {Doublier-Pritchard}, {Hughes}, {Karczewski}, \&
  {Spinoglio}}]{remy-ruyer14a}
{R{\'e}my-Ruyer}, A., {Madden}, S.~C., {Galliano}, F., {et~al.} 2014, \aap,
  563, A31

\bibitem[{{Repolust} {et~al.}(2004){Repolust}, {Puls}, \&
  {Herrero}}]{repolust04a}
{Repolust}, T., {Puls}, J., \& {Herrero}, A. 2004, \aap, 415, 349

\bibitem[{{Ridge} {et~al.}(2006){Ridge}, {Di Francesco}, {Kirk}, {Li},
  {Goodman}, {Alves}, {Arce}, {Borkin}, {Caselli}, {Foster}, {Heyer},
  {Johnstone}, {Kosslyn}, {Lombardi}, {Pineda}, {Schnee}, \&
  {Tafalla}}]{ridge06a}
{Ridge}, N.~A., {Di Francesco}, J., {Kirk}, H., {et~al.} 2006, \aj, 131, 2921

\bibitem[{{Roman-Duval} {et~al.}(2010){Roman-Duval}, {Jackson}, {Heyer},
  {Rathborne}, \& {Simon}}]{roman-duval10a}
{Roman-Duval}, J., {Jackson}, J.~M., {Heyer}, M., {Rathborne}, J., \& {Simon},
  R. 2010, \apj, 723, 492

\bibitem[{{Rosolowsky}(2005)}]{rosolowsky05b}
{Rosolowsky}, E. 2005, \pasp, 117, 1403

\bibitem[{{Rybicki} \& {Lightman}(1986)}]{rybicki86a}
{Rybicki}, G.~B., \& {Lightman}, A.~P. 1986, {Radiative Processes in
  Astrophysics} (Wiley-VCH), 400

\bibitem[{{Saintonge} {et~al.}(2011{\natexlab{a}}){Saintonge}, {Kauffmann},
  {Kramer}, {Tacconi}, {Buchbender}, {Catinella}, {Fabello},
  {Graci{\'a}-Carpio}, {Wang}, {Cortese}, {Fu}, {Genzel}, {Giovanelli}, {Guo},
  {Haynes}, {Heckman}, {Krumholz}, {Lemonias}, {Li}, {Moran},
  {Rodriguez-Fernandez}, {Schiminovich}, {Schuster}, \&
  {Sievers}}]{saintonge11a}
{Saintonge}, A., {Kauffmann}, G., {Kramer}, C., {et~al.} 2011{\natexlab{a}},
  \mnras, 415, 32

\bibitem[{{Saintonge} {et~al.}(2011{\natexlab{b}}){Saintonge}, {Kauffmann},
  {Wang}, {Kramer}, {Tacconi}, {Buchbender}, {Catinella}, {Graci{\'a}-Carpio},
  {Cortese}, {Fabello}, {Fu}, {Genzel}, {Giovanelli}, {Guo}, {Haynes},
  {Heckman}, {Krumholz}, {Lemonias}, {Li}, {Moran}, {Rodriguez-Fernandez},
  {Schiminovich}, {Schuster}, \& {Sievers}}]{saintonge11b}
{Saintonge}, A., {Kauffmann}, G., {Wang}, J., {et~al.} 2011{\natexlab{b}},
  \mnras, 61

\bibitem[{{Salpeter}(1955)}]{salpeter55a}
{Salpeter}, E.~E. 1955, \apj, 121, 161

\bibitem[{{Sana} \& {Evans}(2011)}]{sana11a}
{Sana}, H., \& {Evans}, C.~J. 2011, in IAU Symposium, Vol. 272, IAU Symposium,
  ed. C.~{Neiner}, G.~{Wade}, G.~{Meynet}, \& G.~{Peters}, 474--485

\bibitem[{{Sandstrom} {et~al.}(2007){Sandstrom}, {Peek}, {Bower}, {Bolatto}, \&
  {Plambeck}}]{sandstrom07a}
{Sandstrom}, K.~M., {Peek}, J.~E.~G., {Bower}, G.~C., {Bolatto}, A.~D., \&
  {Plambeck}, R.~L. 2007, \apj, 667, 1161

\bibitem[{{Sano} \& {Stone}(2002)}]{sano02a}
{Sano}, T., \& {Stone}, J.~M. 2002, \apj, 577, 534

\bibitem[{{Schinnerer} {et~al.}(2013){Schinnerer}, {Meidt}, {Pety}, {Hughes},
  {Colombo}, {Garc{\'{\i}}a-Burillo}, {Schuster}, {Dumas}, {Dobbs}, {Leroy},
  {Kramer}, {Thompson}, \& {Regan}}]{schinnerer13a}
{Schinnerer}, E., {Meidt}, S.~E., {Pety}, J., {et~al.} 2013, \apj, 779, 42

\bibitem[{{Schmidt}(1959)}]{schmidt59a}
{Schmidt}, M. 1959, \apj, 129, 243

\bibitem[{{Schneider} {et~al.}(2011){Schneider}, {Bontemps}, {Simon},
  {Ossenkopf}, {Federrath}, {Klessen}, {Motte}, {Andr{\'e}}, {Stutzki}, \&
  {Brunt}}]{schneider11a}
{Schneider}, N., {Bontemps}, S., {Simon}, R., {et~al.} 2011, \aap, 529, A1

\bibitem[{{Schneider} {et~al.}(2012){Schneider}, {Omukai}, {Bianchi}, \&
  {Valiante}}]{schneider12a}
{Schneider}, R., {Omukai}, K., {Bianchi}, S., \& {Valiante}, R. 2012, \mnras,
  419, 1566

\bibitem[{{Schneider} {et~al.}(2006){Schneider}, {Omukai}, {Inoue}, \&
  {Ferrara}}]{schneider06a}
{Schneider}, R., {Omukai}, K., {Inoue}, A.~K., \& {Ferrara}, A. 2006, \mnras,
  369, 1437

\bibitem[{{Sch{\"o}ier} {et~al.}(2005){Sch{\"o}ier}, {van der Tak}, {van
  Dishoeck}, \& {Black}}]{schoier05a}
{Sch{\"o}ier}, F.~L., {van der Tak}, F.~F.~S., {van Dishoeck}, E.~F., \&
  {Black}, J.~H. 2005, \aap, 432, 369

\bibitem[{{Schruba} {et~al.}(2010){Schruba}, {Leroy}, {Walter}, {Sandstrom}, \&
  {Rosolowsky}}]{schruba10a}
{Schruba}, A., {Leroy}, A.~K., {Walter}, F., {Sandstrom}, K., \& {Rosolowsky},
  E. 2010, \apj, 722, 1699

\bibitem[{{Schruba} {et~al.}(2011){Schruba}, {Leroy}, {Walter}, {Bigiel},
  {Brinks}, {de Blok}, {Dumas}, {Kramer}, {Rosolowsky}, {Sandstrom},
  {Schuster}, {Usero}, {Weiss}, \& {Wiesemeyer}}]{schruba11a}
{Schruba}, A., {Leroy}, A.~K., {Walter}, F., {et~al.} 2011, \aj, 142, 37

\bibitem[{{Seifried} {et~al.}(2013){Seifried}, {Banerjee}, {Pudritz}, \&
  {Klessen}}]{seifried13a}
{Seifried}, D., {Banerjee}, R., {Pudritz}, R.~E., \& {Klessen}, R.~S. 2013,
  \mnras, 432, 3320

\bibitem[{{Shakura} \& {Sunyaev}(1973)}]{shakura73a}
{Shakura}, N.~I., \& {Sunyaev}, R.~A. 1973, \aap, 24, 337

\bibitem[{{Shu}(1991)}]{shu91a}
{Shu}, F. 1991, {Physics of Astrophysics: Volume I: Radiation} (University
  Science Books)

\bibitem[{{Shu}(1992)}]{shu92a}
{Shu}, F.~H. 1992, {Physics of Astrophysics, Vol. II} (University Science
  Books)

\bibitem[{{Shu} {et~al.}(1990){Shu}, {Tremaine}, {Adams}, \& {Ruden}}]{shu90a}
{Shu}, F.~H., {Tremaine}, S., {Adams}, F.~C., \& {Ruden}, S.~P. 1990, \apj,
  358, 495

\bibitem[{{Siess} {et~al.}(2000){Siess}, {Dufour}, \& {Forestini}}]{siess00a}
{Siess}, L., {Dufour}, E., \& {Forestini}, M. 2000, \aap, 358, 593

\bibitem[{{Smith} \& {Mac Low}(1997)}]{smith97a}
{Smith}, M.~D., \& {Mac Low}, M.-M. 1997, \aap, 326, 801

\bibitem[{{Solomon} {et~al.}(1997){Solomon}, {Downes}, {Radford}, \&
  {Barrett}}]{solomon97a}
{Solomon}, P.~M., {Downes}, D., {Radford}, S.~J.~E., \& {Barrett}, J.~W. 1997,
  \apj, 478, 144

\bibitem[{{Solomon} {et~al.}(1987){Solomon}, {Rivolo}, {Barrett}, \&
  {Yahil}}]{solomon87a}
{Solomon}, P.~M., {Rivolo}, A.~R., {Barrett}, J., \& {Yahil}, A. 1987, \apj,
  319, 730

\bibitem[{{Stacy} \& {Bromm}(2013)}]{stacy13a}
{Stacy}, A., \& {Bromm}, V. 2013, \mnras, 433, 1094

\bibitem[{{Stacy} {et~al.}(2012){Stacy}, {Greif}, \& {Bromm}}]{stacy12a}
{Stacy}, A., {Greif}, T.~H., \& {Bromm}, V. 2012, \mnras, 422, 290

\bibitem[{{Stahler}(1983)}]{stahler83a}
{Stahler}, S.~W. 1983, \apj, 274, 822

\bibitem[{{Stahler} {et~al.}(1980{\natexlab{a}}){Stahler}, {Shu}, \&
  {Taam}}]{stahler80a}
{Stahler}, S.~W., {Shu}, F.~H., \& {Taam}, R.~E. 1980{\natexlab{a}}, \apj, 241,
  637

\bibitem[{{Stahler} {et~al.}(1980{\natexlab{b}}){Stahler}, {Shu}, \&
  {Taam}}]{stahler80b}
---. 1980{\natexlab{b}}, \apj, 242, 226

\bibitem[{{Stahler} {et~al.}(1981){Stahler}, {Shu}, \& {Taam}}]{stahler81a}
---. 1981, \apj, 248, 727

\bibitem[{{Stone} {et~al.}(1998){Stone}, {Ostriker}, \& {Gammie}}]{stone98a}
{Stone}, J.~M., {Ostriker}, E.~C., \& {Gammie}, C.~F. 1998, \apjl, 508, L99

\bibitem[{{Strong} \& {Mattox}(1996)}]{strong96a}
{Strong}, A.~W., \& {Mattox}, J.~R. 1996, \aap, 308, L21

\bibitem[{{Sun} {et~al.}(2006){Sun}, {Kramer}, {Ossenkopf}, {Bensch},
  {Stutzki}, \& {Miller}}]{sun06a}
{Sun}, K., {Kramer}, C., {Ossenkopf}, V., {et~al.} 2006, \aap, 451, 539

\bibitem[{{Tafalla} {et~al.}(2004){Tafalla}, {Santiago}, {Johnstone}, \&
  {Bachiller}}]{tafalla04c}
{Tafalla}, M., {Santiago}, J., {Johnstone}, D., \& {Bachiller}, R. 2004, \aap,
  423, L21

\bibitem[{{Tamburro} {et~al.}(2008){Tamburro}, {Rix}, {Walter}, {Brinks}, {de
  Blok}, {Kennicutt}, \& {Mac Low}}]{tamburro08a}
{Tamburro}, D., {Rix}, H.-W., {Walter}, F., {et~al.} 2008, \aj, 136, 2872

\bibitem[{{Tan} {et~al.}(2014){Tan}, {Beltr{\'a}n}, {Caselli}, {Fontani},
  {Fuente}, {Krumholz}, {McKee}, \& {Stolte}}]{tan14a}
{Tan}, J.~C., {Beltr{\'a}n}, M.~T., {Caselli}, P., {et~al.} 2014, Protostars
  and Planets VI, 149

\bibitem[{{Taylor}(1964)}]{taylor64a}
{Taylor}, G.~I. 1964, {Low-Reynolds-Number Flows, National Committe for Fluid
  Mechanics Films},
  https://www.youtube.com/watch?v=51-6QCJTAjU\&list=PL0EC6527BE871ABA3\&index=7

\bibitem[{{Tielens}(2005)}]{tielens05a}
{Tielens}, A.~G.~G.~M. 2005, {The Physics and Chemistry of the Interstellar
  Medium} (Cambridge, UK: Cambridge University Press)

\bibitem[{{Tinsley}(1980)}]{tinsley80a}
{Tinsley}, B.~M. 1980, Fundamentals of Cosmic Physics, 5, 287

\bibitem[{{Tobin} {et~al.}(2012){Tobin}, {Hartmann}, {Chiang}, {Wilner},
  {Looney}, {Loinard}, {Calvet}, \& {D'Alessio}}]{tobin12a}
{Tobin}, J.~J., {Hartmann}, L., {Chiang}, H.-F., {et~al.} 2012, \nat, 492, 83

\bibitem[{{Tobin} {et~al.}(2009){Tobin}, {Hartmann}, {Furesz}, {Mateo}, \&
  {Megeath}}]{tobin09a}
{Tobin}, J.~J., {Hartmann}, L., {Furesz}, G., {Mateo}, M., \& {Megeath}, S.~T.
  2009, \apj, 697, 1103

\bibitem[{{Tomida} {et~al.}(2013){Tomida}, {Tomisaka}, {Matsumoto}, {Hori},
  {Okuzumi}, {Machida}, \& {Saigo}}]{tomida13a}
{Tomida}, K., {Tomisaka}, K., {Matsumoto}, T., {et~al.} 2013, \apj, 763, 6

\bibitem[{{Tomisaka}(1998)}]{tomisaka98a}
{Tomisaka}, K. 1998, \apjl, 502, L163+

\bibitem[{{Toomre}(1964)}]{toomre64a}
{Toomre}, A. 1964, \apj, 139, 1217

\bibitem[{{Tout} {et~al.}(1996){Tout}, {Pols}, {Eggleton}, \& {Han}}]{tout96a}
{Tout}, C.~A., {Pols}, O.~R., {Eggleton}, P.~P., \& {Han}, Z. 1996, \mnras,
  281, 257

\bibitem[{{Usero} {et~al.}(2015){Usero}, {Leroy}, {Walter}, {Schruba},
  {Garc{\'{\i}}a-Burillo}, {Sandstrom}, {Bigiel}, {Brinks}, {Kramer},
  {Rosolowsky}, {Schuster}, \& {de Blok}}]{usero15a}
{Usero}, A., {Leroy}, A.~K., {Walter}, F., {et~al.} 2015, \aj, 150, 115

\bibitem[{{van Dokkum} \& {Conroy}(2010)}]{van-dokkum10a}
{van Dokkum}, P.~G., \& {Conroy}, C. 2010, \nat, 468, 940

\bibitem[{{V{\'a}zquez} \& {Leitherer}(2005)}]{vazquez05a}
{V{\'a}zquez}, G.~A., \& {Leitherer}, C. 2005, \apj, 621, 695

\bibitem[{{Walsh} {et~al.}(2004){Walsh}, {Myers}, \& {Burton}}]{walsh04a}
{Walsh}, A.~J., {Myers}, P.~C., \& {Burton}, M.~G. 2004, \apj, 614, 194

\bibitem[{{Weaver} {et~al.}(1977){Weaver}, {McCray}, {Castor}, {Shapiro}, \&
  {Moore}}]{weaver77a}
{Weaver}, R., {McCray}, R., {Castor}, J., {Shapiro}, P., \& {Moore}, R. 1977,
  \apj, 218, 377

\bibitem[{{Weisner} \& {Armstrong}(1964)}]{weisner64a}
{Weisner}, J.~D., \& {Armstrong}, B.~H. 1964, Proceedings of the Physical
  Society, 83, 31

\bibitem[{{Wolfire} \& {Cassinelli}(1987)}]{wolfire87a}
{Wolfire}, M.~G., \& {Cassinelli}, J.~P. 1987, \apj, 319, 850

\bibitem[{{Wu} {et~al.}(2005){Wu}, {Evans}, {Gao}, {Solomon}, {Shirley}, \&
  {Vanden Bout}}]{wu05a}
{Wu}, J., {Evans}, N.~J., {Gao}, Y., {et~al.} 2005, \apjl, 635, L173

\bibitem[{{Wyder} {et~al.}(2009){Wyder}, {Martin}, {Barlow}, {Foster},
  {Friedman}, {Morrissey}, {Neff}, {Neill}, {Schiminovich}, {Seibert},
  {Bianchi}, {Donas}, {Heckman}, {Lee}, {Madore}, {Milliard}, {Rich}, {Szalay},
  \& {Yi}}]{wyder09a}
{Wyder}, T.~K., {Martin}, D.~C., {Barlow}, T.~A., {et~al.} 2009, \apj, 696,
  1834

\bibitem[{{Youdin} \& {Shu}(2002)}]{youdin02a}
{Youdin}, A.~N., \& {Shu}, F.~H. 2002, \apj, 580, 494

\bibitem[{{Zuckerman} \& {Evans}(1974)}]{zuckerman74a}
{Zuckerman}, B., \& {Evans}, N.~J. 1974, \apjl, 192, L149

\end{thebibliography}
\bibliographystyle{apj}


\printindex 

\end{document}